\documentclass[12pt,a4paper]{report}
\usepackage{comment}
\usepackage{multicol}
\usepackage{booktabs}
\usepackage{float}
\usepackage{tikz}
\usetikzlibrary{knots}
\usepackage{braids}
\usetikzlibrary{decorations.markings,hobby,knots,celtic,arrows,shapes,snakes,cd,decorations.pathreplacing,shapes.geometric,calc,shadings,decorations.pathmorphing,arrows,scopes,automata,calligraphy}
\usepackage{pgfplots}
\usepackage{qcircuit}
\usepackage{tikz-quantumgates}
\usepackage{rotating}
\usepackage{circuitikz}
\usepackage{cancel}
\usepackage{bm}
\usepackage{relsize}
\usepackage{amsmath,esint}
\usepackage{amsthm}
\usepackage{braket}
\usepackage{amssymb}
\usepackage{amstext}
\usepackage{enumerate}
\usepackage{mathtools}
\usepackage[mathscr]{euscript}
\usepackage{hyperref}
\usepackage{physics}
\usepackage{simplewick}
\usepackage{graphicx}
\usepackage{xcolor,colortbl}
\usepackage{caption}
\usepackage{subcaption}
\usepackage{nameref}
\graphicspath{{images/}}
\usepackage{color}
\usepackage[top=1.2in, bottom=1.2in, left=1in, right=1in]{geometry}
\usepackage{afterpage}
\usepackage{setspace}
\usepackage{fancyhdr}
\fancypagestyle{plain}{}
\newcommand{\eq}[1]{\begin{align}#1\end{align}}

\newcommand{\fig}[1]{\begin{figure}#1\end{figure}}
\newcommand{\tik}[1]{\begin{tikzpicture}#1\end{tikzpicture}}

\usepackage{cite}
\usepackage{multibib}

\usepackage{titlesec}
\usepackage{titletoc}
\usepackage{tocloft}
\usepackage[acronym,xindy,toc,nomain]{glossaries}
\makeglossaries

\begin{document}	
	\captionsetup[figure]{labelformat={default},labelsep=period,name={Fig.}}
	
	\begin{center}
		\fancypagestyle{empty}{}
		\pagenumbering{gobble}
	{\Large \textbf{Quantum Field Theories, Topological Materials,\\	
	and Topological Quantum Computing}}\\[12mm]
	by\\
	\large \textbf{Muhammad Ilyas}\\[12mm]
	\singlespacing
	A dissertation submitted in partial fulfillment of the\\
	requirements for the degree of\\[10mm]
	Doctor of Philosophy\\
	in\\
	Applied Physics\\[20mm]
	Dissertation Committee:\\
	Marek Perkowski, Chair\\
	Jack Straton, Co-Chair\\
	Steven Bleiler\\
	Peter Leung\\
	Aslam Khalil\\[25mm]
	
	\includegraphics[scale=0.4]{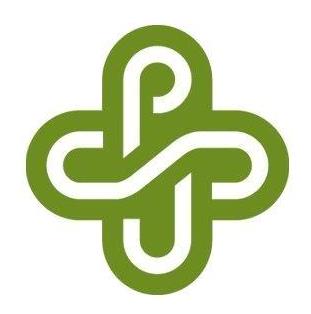}
	
	\vspace{10mm}
	
	Portland State University\\
	2021
	\end{center}
	\newpage
	\pagebreak
	\newpage\null
	\pagebreak
	
	\pagenumbering{roman}
	\setcounter{page}{1}
	\cleardoublepage
	\phantomsection
	\addcontentsline{toc}{section}{Abstract}	
	\begin{abstract}
		\phantomsection
		\thispagestyle{plain}
		A quantum computer can perform exponentially faster than its classical counterpart. It works on the principle of superposition. But due to the decoherence effect, the superposition of a quantum state gets destroyed by the interaction with the environment. It is a real challenge to completely isolate a quantum system to make it free of decoherence. This problem can be circumvented by the use of topological quantum phases of matter. These phases have quasiparticles excitations called anyons. The anyons are charge-flux composites and show exotic fractional statistics. When the order of exchange matters, then the anyons are called non-Abelian anyons. Majorana fermions in topological superconductors and quasiparticles in some quantum Hall states are non-Abelian anyons. Such topological phases of matter have a  ground state degeneracy. The fusion of two or more non-Abelian anyons can result in a superposition of several anyons. The topological quantum gates are implemented by braiding and fusion of the non-Abelian anyons. The fault-tolerance is achieved through the topological degrees of freedom of anyons. Such degrees of freedom are non-local, hence inaccessible to the local perturbations. In this dissertation, we provide a comprehensive review of the fundamentals of logic design in topological quantum computing. The braid group and knot invariants in the skein theory are discussed. The physical insight behind the braiding is explained by the geometric phases and the gauge transformation. The mathematical models for the fusion and braiding are presented in terms of the category theory and the quantum deformation of the recoupling theory. The topological phases of matter are described by the topology of band structure. The wave function of quasiparticles in the quantum Hall effect and the theory of Majorana fermions in topological superconductors are also discussed. The dynamics of the charge-flux composites and their Hilbert space are expressed through the Chern-Simons theory and the two-dimensional topological quantum field theory. The Ising and Fibonacci anyonic models for binary gates are briefly given. Ternary logic gates are more compact than their binary counterparts and naturally arise in a type of anyonic model called the metaplectic anyons. We reduced the quantum cost of the existing ternary quantum arithmetic gates and proposed that these gates can be implemented with the metaplectic anyons.
	\end{abstract}
	
	\setcounter{page}{2}
	\cleardoublepage
	\phantomsection
	\addcontentsline{toc}{section}{Table of Contents}
	\tableofcontents
	
	\cleardoublepage
	\phantomsection
	\addcontentsline{toc}{section}{List of Figures}
	\listoffigures
	
	\cleardoublepage
	\phantomsection	
	\addcontentsline{toc}{section}{List of Tables}
	\listoftables
	
	\pagebreak
	
	\cleardoublepage
	
	\pagenumbering{arabic}
	\phantomsection
	\setcounter{page}{1}

	\chapter{Introduction}\label{Intro}
		
Two of the greatest revolutions of the twentieth century were the discovery of quantum mechanics and the invention of computers. At the end of the twentieth century, these two fields merged and a new field of quantum information was born. The quantum information science ends Moore's law, according to which the computing power would double every eighteen months. This law governed silicon chip-based computers, for which the density of chips can be increased. Such computers obey the laws of classical mechanics. But we cannot reduce the physical size of chips infinitely. At the atomic level, particles behave according to the laws of quantum mechanics rather than the laws of classical mechanics.

In 1982, Richard Feynman pointed out that there is a fundamental limit with the ability of classical computers to efficiently simulate a quantum system \cite{feynman1982simulating}. He showed that some problems can be solved exponentially faster on a quantum computer using exponentially large-sized Hilbert space than they could be solved on a classical computer. David Deutsch showed that classical computers cannot efficiently simulate a quantum computer \cite{deutsch1985quantum}. Hence, a quantum computer is important for two reasons; it can perform faster, and it can answer questions about nature.

The building blocks of a classical computer are \textit{bits}. These bits are based on classical logic that has values of either $0$ or $1$. Operations on these bits are performed by a series of gates. These gates change their values and answer the operations. The classical circuits are composed in space from gates connected by wires. But quantum computers are based on quantum logic, which has values in the superposition of $0$ and $1$. The quantum gates manipulate the quantum superposition and give outputs with some probability. A ternary quantum gate is a three-valued logic design, based on the superposition of $0$,$1$, and $2$.

Many methods of encryption on a classical computer are based on difficulty in finding the prime factors of a large number. Peter Shor \cite{shor1994proceedings} invented an algorithm to find the prime factors of a number on a quantum computer with an exponential speed up. This algorithm created widespread interest in quantum computers. Many other quantum algorithms have already been proposed. Grover's search algorithm for an unstructured search \cite{grover1996fast, grover1997quantum} offers a quadratic speed up compared with a classical counterpart. These algorithms are implemented on a particular model of quantum computation.

Building a quantum computer is a great challenge due to its susceptibility to errors. The quantum superposition is destroyed due to its interaction with the environment. This process is called \textit{decoherence}. Moreover, we cannot measure the state and look for errors. In doing so we would kill the superposition. Errors can also be in the phase of a state. There are quantum error correction codes \cite{calderbank1996good, steane1996error, preskill2004lecture, calderbank1997quantum}, but a quantum system needs to be completely isolated from the environment. In 1997, Alexei Kitaev proposed a model for the fault-tolerant quantum computation \cite{kitaev2003fault}. Information is encoded in some non-local degrees of freedom of particles, hence inaccessible to local perturbations \cite{gottesman1998theory,nayak2008non,freedman2003topological}. This is done using the systems which are \textit{topological} in nature. 
The topology is a study of spaces that are continuously deformable to each other. Such spaces are called manifolds. A manifold is a space that is Euclidean flat space locally when a small patch is taken, but it has some non-Euclidean structure globally. To compare two spaces, some properties of the spaces are computed. These properties remain invariant under the continuous deformation of one space to the other. Such properties are called \textit{topological invariants}. We will discuss topology in Chapter \ref{Knot}.

The topological nature of particles can be studied through their exchange statistics. Let $\psi(\bm{r}_i,\bm{r}_j)$ be the wave function of two particles at positions $\bm{r}_i$ and $\bm{r}_j$. In three dimensions, when two particles exchange their places, the wave function gets multiplied with a phase factor. That is, 
\eq{\psi(\bm{r}_i,\bm{r}_j) = e^{i\theta}\psi(\bm{r}_j,\bm{r}_i) \label{Exchange},}
where the values $\theta = 0, 2\pi$ correspond to the exchange of bosons and $\theta = \pi$ corresponds to the exchange of fermions. The phase acquired by the wave function is $+1$ for bosons and $-1$ for fermions. Bosons are integral spin particles and obey Bose-Einstein statistics, whereas fermions are half-integer spin particles and obey Fermi-Dirac statistics. A double exchange of these particles is equivalent to no exchange. If the particles are distinguishable, then their statistics is described by the \textit{permutation group} $\mathcal{S}_N$. A group is a mathematical structure to study symmetries. The permutation group is used to study the exchange symmetry. See Appendix \ref{AbsAlg} for the definition of a group.  

Jon Magne Leinaas in 1977  \cite{leinaas1977theory} suggested that in a two-dimensional space, another statistic may occur, called \textit{fractional statistics}. For this new kind of statistics, $\theta$ has an arbitrary value between $0$ and $\pi$. Bosons and fermions do not obey fractional statistics even in two-dimensional space. Wilczek \cite{wilczek1982magnetic, wilczek1982quantum} proposed a model for the realization of the fractional statistics. He also named these particles as \textit{anyons} (neither boson nor fermion but any on) \cite{wilczek1982quantum}.

An anyon is not an elementary particle, but a collective phenomenon or a local disturbance in two-dimensional topological materials in a high magnetic field and at a very low temperature. A large number of elementary particles behave in a coordinated way to make quasiparticles. These particles can exist only inside a material, not in free space. Magnetic fluxes are attached to quasiparticles and make charge-flux composites. These quasiparticles obey fractional statistics. The quasiparticles will be discussed in Chapters \ref{QHE} and \ref{TopMaterials}. The Chern-Simons gauge theory is used as an effective field theory to describe these materials. Quasiparticles have a \textit{topological charge} which is a topological quantum number and is a generalization of the conventional charge. It is a topological invariant and changes on topological phase transition. We will discuss the topological invariants in Chapter \ref{Knot} and the topological phase transition in Chapter \ref{TopMaterials}.

The anyons are quasiparticles in quantum Hall states \cite{girvin1987quantum,moore1991nonabelions,read1999beyond} and as Majorana fermions in topological superconductors \cite{lahtinen2017short}.
Anyons are detected in laboratory \cite{camino20073, willett2009measurement,stern2006proposed,rosenow2016current,nakamura2020direct,willett2013magnetic}, and more recently \cite{rosenow2016current,nakamura2020direct}. The measurement of an anyon is done by interference as described in \cite{stern2008anyons,nayak2008non}.

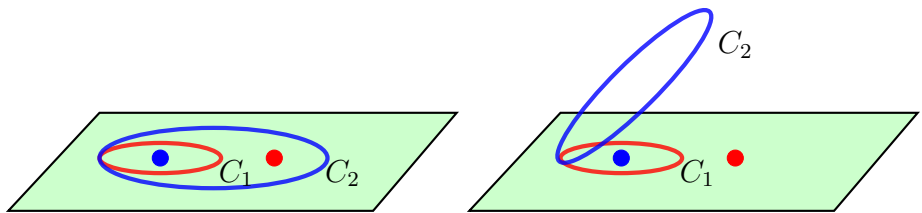
\begin{figure*}[h!]
	\centering
	\begin{tikzpicture}
		\draw[fill=green!20,thick] (-0.3,1.1) -- (4.4,1.1) -- (3.3,-0.2) -- (-1.5,-0.2)-- cycle;
		\draw [ultra thick,opacity=0.8,red] (0.5,0.5) ellipse (0.8 and 0.2);
		\draw [ultra thick,opacity=0.8,blue] (1.2,0.5) ellipse (1.5 and 0.4);
		\filldraw[blue] (0.5,0.5) circle (3pt);
		\filldraw[red] (2,0.5) circle (3pt);
		\node at (1.5,0.3) {$C_1$};
		\node at (2.9,0.3) {$C_2$};
	\end{tikzpicture}	
	\begin{tikzpicture}
		\draw[fill=green!20,thick] (-0.3,1.1) -- (4.4,1.1) -- (3.3,-0.2) -- (-1.5,-0.2)-- cycle;
		\draw [ultra thick,opacity=0.8,red] (0.5,0.5) ellipse (0.8 and 0.2);
		\draw [ultra thick,opacity=0.8,blue,rotate=45] (1.5,0.55) ellipse (1.4 and 0.3);
		\filldraw[blue] (0.5,0.5) circle (3pt);
		\filldraw[red] (2,0.5) circle (3pt);
		\node at (1.5,0.3) {$C_1$};
		\node at (2,2) {$C_2$};
	\end{tikzpicture}
	\caption[In general, two closed curves $C_1$ and $C_2$ are topologically distinct in two dimensions]{Two closed paths $C_1$ and $C_2$ are topologically distinct in two dimensions, but they can be deformed to each other in three dimensions.}
	\label{TwoPaths}
\end{figure*}

The fundamental difference between 2D and 3D is the difference in the topology of spacetime. The motion of particles makes \textit{knots and links} in spacetime. Two paths are topologically equivalent if one can be deformed to the other. In two dimensions, in general, we cannot transform one path to the other without cutting, as shown in Fig. \ref{TwoPaths}. All smoothly deformable trajectories are in the same equivalence class. Fermions and bosons do not obey the fractional statistics even in 2D, but the change in the wave function of the system in two-dimensional topological materials, when two quasiparticles are exchanged, is independent of the distance and speed of exchange. In contrast, the evolution may depend on some global characteristic of the path. Therefore, the statistics of anyons are topological. Instead of the permutation group, the double exchange of anyons is not equal to the identity. The exchange statistics of anyons is described by a \textit{braid group} in $(2+1)$-dimensional spacetime. The braid group is defined in Chapter \ref{Knot}. When the order of composition of two elements of a group does not matter then the group is \textit{Abelian}, otherwise, it is \textit{non-Abelian}.

One of the properties of the topological phases of matter is the existence of \textit{ground state degeneracy}. The degenerate ground states have a large energy gap to the excited states. The degeneracy depends on the topology of the two-dimensional system and the types of anyons present. The ground state is unique for trivial topology.
For Abelian anyons, the braid operators commute and the ground state is unique, but for non-Abelian anyons, the braiding corresponds to the evolution of the system in the degenerate ground state. The change of the system from one ground state to the other is studied using the Berry phase \cite{berry1984quantal} as described in Chapter \ref{GeoPhase}.
Let $g$ be degenerate states $\psi_a$ with $a=1,2,...,g$ of particles at positions $x_1,x_2,...,x_n$. Exchanging particles 1 and 2 may not just change the phase but may rotate it into a different state $\psi_b$. Braiding of 1 and 2 and that of 2 and 3 are given as
\eq{\psi_a \rightarrow M_{ab}\psi_b, \qquad \psi_a \rightarrow N_{ab} \psi_b \label{Exchange2},} 
where $M_{ab}$ and $N_{ab}$ are $g\cross g$ dimensional unitary matrices. For Abelian anyons, the $\theta$ in Eq. \ref{Exchange} is arbitrary and clockwise and anticlockwise exchanges commute. Which means that even the clockwise and anticlockwise exchanges may not be the same, but if we exchange particles clockwise then anticlockwise, it will be the same as if we perform anticlockwise exchange first then clockwise. In contrast, $M_{ab}$ and $N_{ab}$ in Eq. \ref{Exchange2} do not commute in general, that is $M_{ab}N_{ab}- N_{ab}M_{ab} \ne 0$ and particles obey non-Abelian statistics. 

Since the unitary evolution only depends on the topology of the path, wiggles of the path would not affect the outcome. No local perturbation can split the degeneracy, hence the system is decoherence-free. The topological nature of anyons is the source of the fault tolerance in a quantum computer.
A topological quantum computer is based on three steps; the creation of anyon-antianyon pairs from the vacuum, braiding, and fusion \cite{pachos2012introduction,nayak2008non}. Anyons can be combined by bringing them close to each other. This is called \textit{fusion}. The fusion is an inverse of the creation of the particles. The fusion of an anyon with its antiparticle gives the total topological charge zero, but the fusion of an anyon with another different type of anyon or antianyon may give the third particle or a superposition of a collection of several particles. The resultant types of particles depend on the \textit{fusion rules}. The topological charge of an anyon is assigned with respect to its fusion with other particles to get a vacuum.
There might not be a unique way to combine anyons. Different ways of the fusion of multiple anyons to get an outcome are called the \textit{fusion channels}. These fusion channels provide the basis states of the Hilbert space for quantum gates. The dimension of the Hilbert space is equal to the degeneracy of the ground state. The transformation between different fusion channels is given by \textit{$F$-matrices}.
The internal degrees of freedom of the anyons are changed by braiding and can put the system in another ground state. The phases acquired by anyons during the braiding are computed through \textit{R-matrices}.
A quantum superposition of states can be created by a suitable combination of $F$ and $R$ matrices.

From the path integral point of view, the anyon's trajectories make knots whose invariants are the probability amplitudes from an initial to a final configuration of the system of anyons. The orientations of knots correspond to the direction of particle trajectories, and the twist in a ribbon knot corresponds to the \textit{topological spin}. The topological spin is the phase due to the rotation of a topological charge around its magnetic flux attached to it. 
To specify the braiding statistics, we need the data such as; particle species, fusion rules, F-matrices, R-matrices, and topological spin.
The mathematical model for such data is the \textit{category theory} and the quantum deformation of the recoupling theory of angular momenta. 

The ternary logic gates and circuits, or the ones that consist of a combination of binary and ternary, are more compact than their binary counterparts \cite{haghparast2017towards}. There are some non-Abelian anyons for which the ternary structures naturally arise. Such anyons are called metaplectic anyons discussed in Chapter \ref{Meta}. The $F$ and $R$ matrices are computed using the recoupling theory in Chapter \ref{Rec}. In this dissertation, we proposed improved ternary arithmetic circuit designs that can be implemented with the metaplectic anyons.

This dissertation is organized as follows. In Chapter \ref{QC}, we will discuss the basics of binary and ternary logic gates in quantum computing. Chapter \ref{Knot} is on the preliminaries of knot theory and the braid group. Anyons' motion in spacetime makes braids, and the outcomes of the braids are obtained through geometric and topological phases. The geometric phases are discussed in Chapter \ref{GeoPhase}. One of the most popular physical systems for topological quantum computation is the quantum Hall effect. The existence of anyons as quasiparticles is described by Chern-Simons theory. The quantum Hall effect and topological materials are explained in Chapter \ref{QHE} and \ref{TopMaterials}, whereas the Chern-Simons theory is given in Chapter \ref{TQFT}. This chapter also includes the two-dimensional topological quantum field theory, which describes the Hilbert space of the anyon on a two-dimensional topological manifold. The knot invariant is also obtained using the partition function of the Chern-Simons theory on a topological manifold.
The concept of quantum dimensions, topological spin, F-matrices, R-matrices, and the Hilbert space can alternatively be described by the category theory as explained in Chapter \ref{Cat}. The unitary modular category and the quantum group, are also used as the mathematical models for the topological orders and topological quantum computing.

The elementary binary logic gates design in topological quantum computing is given in Chapter \ref{TQC}. It will provide a prototype theoretical model for topological quantum computing. 
The ternary gates are formed as the quantum deformation, also called q-analog, of the $F$ and $R$ matrices in the recoupling theory. The quantum deformation and the recoupling theory are discussed in Chapter \ref{Rec}. The ternary logic gates are designed using the fusion and braiding of the metaplectic anyons. These anyons are the ones with quantum dimensions equal to the square root of integers. The metaplectic anyon and ternary logic design, using the recoupling theory, are discussed in Chapter \ref{Meta}. In this chapter, we also presented our designs of the ternary adder, ternary subtractor, and ternary multiplier, and their implementation with the braiding and fusion of metaplectic anyon.

The preliminaries from the abstract algebra and topology are given in Appendices \ref{AbsAlg} and \ref{Top}. The quantum group is a deformation of the Hopf algebra. The knot invariant is calculated using the quantum group method and topological quantum field theory in Appendix \ref{HopfAlg}.
Some topics from quantum field theory are used in the main dissertation, like the concept of fields, second quantization, path integral, and gauge theory. Appendix \ref{QFT} is the basics of quantum field theory.

For this dissertation, basic textbook knowledge of the fundamentals of quantum mechanics and condensed matter physics is assumed.
	\chapter{Quantum Computing}\label{QC}

In quantum mechanics, a system is described by a state vector or a wave function that is written in the \textit{Dirac} notation as
\eq{\ket{\psi} = a_1\ket{1}+a_2\ket{2}+...+a_n\ket{n} = \sum_{i=1}^n a_i \ket{i},}
where $a_i$'s are complex numbers. The state vector $\ket{\psi}$ represents a system in a superposition of the eigenstates $\ket{i}$ of the system with probability amplitudes $a_i$. On measurement, the system will be found in an eigenstate $\ket{i}$ with a probability $\abs{a_i}^2$. The state vector is assumed to be \textit{normalized}, such that $\sum_i\abs{a_i}^2=1$.

The $\ket{\psi}$ is a vector in a Hilbert space and is written as a linear combination of the basis vectors $\ket{i}$. The Hilbert space is defined in Appendix \ref{AbsAlg}.
For each \textit{ket} $\ket{i}$, there is a dual vector \textit{bra} $\bra{i}$ in the Hilbert space. The purpose of this dual vector is to find the linear dependence of two states by computing the projections. The projections are found by \textit{inner product} or \textit{orthonormality condition} as $\bra{i}\ket{j}=\delta_{ij}$, where $\delta_{ij}=1$ when $i=j$ and $\delta_{ij}=0$ when $i\ne j$. Also, $\bra{i}\ket{\psi}=a_i$ and $\bra{\psi}\ket{\psi}=1$. The inner product works the same way as a dot product in vector analysis.
Another product, known as the \textit{outer product}, is written as $\ket{i}\bra{i}$, such that it obeys the \textit{completeness relation} $\sum_{i=1}^{n} \ket{i}\bra{i}=I$. The outer product $\ket{i}\bra{i}$ is an operator called a projection operator or a projector. Since $\ket{i}\bra{i}\ket{j}=\bra{i}\ket{j}\ket{i}$, a projector projects $\ket{j}$ along $\ket{i}$. The length of the projection is $\bra{i}\ket{j}$. In the next section, we will see that the state vector and eigenstates can be written as column vectors, so that the linear transformations on the Hilbert space are square matrices that act on the vectors as operators, see Ref. \cite{laforest2015mathematics}. 

The physical quantities or dynamical variables like position, momentum, and energy are called \textit{observables}. There is an operator corresponding to each physical quantity. The values of observables we obtain on measurement are called \textit{eigenvalues}. Mathematically, the eigenvalues $\lambda_i$ are obtained by the action of the operator on the eigenstate as 
\eq{\hat{A}\ket{i} = \lambda_i \ket{i}.}
A system in an eigenstate has a definite value of a physical quantity. But in general, a system is in a state $\ket{\psi}$ which is a superposition of the eigenstates. In this case, the eigenvalues are probabilistic or expected values in a measurement on the system. The expectation value of an operator $A$ is written as
\eq{\langle \hat{A}\rangle= \bra{\psi}\hat{A}\ket{\psi}=\sum_i \lambda_i \abs{a_i}^2.}
This equation tells that the probability of getting a particular eigenvalue $\lambda_i$ on a measurement is $\abs{\bra{i}\ket{\psi}}^2=\abs{a_i}^2$. The measurement process collapses the state vector to one of the eigenstates. Since, the eigenvalues $\lambda_i$ must be real, $\bra{\psi}\hat{A}\ket{\psi} = (\bra{\psi}\hat{A}\ket{\psi})^*$. It implies that $\hat{A}=\hat{A}^\dagger$, and $\hat{A}$ is a self-adjoint or a Hermitian operator. The adjoint operator $\hat{A}^\dagger$ is obtained by taking the transpose and complex conjugate of the matrix $\hat{A}$. Therefore, the operation $\hat{A}\ket{i}$ is equivalent to the operation $\bra{i}\hat{A}^\dagger$, where $\bra{i}$ is the adjoint of $\ket{i}$. (Compare the operators to the random variables in statistics, and computing the expectation value to finding the mean or expected value of a probability distribution.)

According to the Heisenberg uncertainty principle, \textit{complementary or conjugate variables} cannot be measured simultaneously with an absolute certainty. For conjugate variables, the eigenstates of one observable are not the eigenstates of the other observable simultaneously. So there can be a superposition of eigenvalues of one observables corresponding to each value of the other observable and vice versa. Measuring one of the conjugate variables, the other one is perturbed in an unpredictable way. Therefore, the outcomes are affected by the order in which we are observing the two quantities $A$ and $B$. The commutator quantifies the uncertainty and is written for two operators $\hat{A}$ and $\hat{B}$ as
\eq{[\hat{A},\hat{B}]=\hat{A}\hat{B}-\hat{B}\hat{A}.}
Observables corresponding to the operators involved in the commutator are also called \textit{incompatible observables}. The uncertainty principles is closely linked with the \textit{wave-particle duality} and \textit{complementarity}. A quantum object can behave as a particle in some situation and as a wave in some other situations, but it does not show both characters in the same situation. The wave function is the solution of wave equation describing the particle. This wave equation is the Scr\"odinger equation in Schr\"odinger's mechanics. The wavelength of a matter wave associated with a quantum object is the de Broglie wavelength given as $\lambda=h/p$, where $p$ is the momentum of the particle and $h$ is the Planck constant. In general, the wave funciton is not a single wave but many waves added to make a \textit{wave packet}. 

For example, a freely moving electron makes a wave packet of width $\Delta x$ spread in space, and have possible momentum values in the range $p\pm \Delta p$. As an electron can also behave as a wave, there is a probability of the electron on multiple points in the wave packet and have multiple momentum values at a single point. Measuring the position of the electron at some point gives the momentum values chosen from many values in the range $p\pm \Delta p$. On the other hand, measuring the momentum of the electron, the value of the position will be measured with an uncertainty equal to the width $\Delta x$ of the wave packet. The uncertainty for conjugate variables position and momentum is computed as $[\hat{x},\hat{p}]=i\hbar$, with $\Delta x \Delta p \ge \frac{1}{2}\abs{\langle [\hat{A},\hat{B}]\rangle}$. The values of momentum are related to the values of the position by the \textit{Fourier transform}.

Due to the wave nature of quantum objects, the wave function and eigenstates of a matter particles can also have phases associated with them, and the wave properties like diffraction and interference are observed for a particle like electron. For example, in the famous double slit experiment, the wave function of an electron can be splitted and later interfered constructively or destructively. The global phase of a wave function is canceled out while computing the probabilities and expectation values, so only the relative phases of the eigenstates have physical significance. More than one quantum objects can also be entangled. We will discuss the quantum entanglement in the next section.

The time evolution of a state is represented by the unitary time evolution operator $U(t)$ such that $U^{-1}=U^\dagger$ and $U(t)U^\dagger(t)= U^\dagger(t)U(t)=I$. It is written as $U(t)=\exp(-iHt)$, where $H$ is the Hamiltonian operator corresponds to the energy eigenvalues. The unitary evolution relates the changes in the state to the energy of the system. These changes are reversible so that the inner product is preserved. That means, the phase can be changed but the amplitude remains the same with time.
When the initial state $\ket{\psi_i}$ evolves unitarily to the final state $\ket{\psi_f}$, it is written as 
\eq{\ket{\psi_f}=U(t)\ket{\psi_i}.}
\textit{A quantum computation model involves three steps; initialization, unitary evolution, and measurement \cite{divincenzo2000physical}. The initial state is an input state and the final state is an output state \cite{Nielsen2002quantum} of a quantum gate. The evolution operator $U(t)$ corresponds to a quantum gate. The readout is a measurement in certain bases that gives a classical result.} See Ref. \cite{susskind2014quantum,zubairy2020quantum} for the theoretical minimum of quantum mechanics, \cite{orzel2010teach} for conceptual understanding, and \cite{griffiths2018introduction,reed1990quantum} for more in depth theoretical framework. For a basic study on quantum computing, see the books \cite{yanofsky2008quantum,mcmahon2007quantum,mermin2007quantum}, and for technical details, see \cite{Nielsen2002quantum,marinescu2011classical}.  
	\section{Binary Quantum Gates}
A classical bit has a value 0 or 1. A qubit is a superposition of 0 and 1, written as
\eq{\ket{\psi}= \alpha\ket{0} + \beta\ket{1}.} 
A qubit can be written in a matrix form as
\eq{\ket{\psi} = \alpha \begin{pmatrix}
		1 \\
		0
	\end{pmatrix}+ \beta \begin{pmatrix}
		0 \\
		1
	\end{pmatrix} = \begin{pmatrix}
		\alpha \\
		\beta
	\end{pmatrix},}
where $\alpha$ and $\beta$ are complex numbers. The sum of their squares is one, $\abs{\alpha}^2+\abs{\beta}^2 =1$, which means that the sum of probabilities is equal to one. A qubit can be made by any two-level quantum mechanical system. For example, a spin-half particle can be in a superposition state of spin-up state $\ket{0}$ and spin down spin state $\ket{1}$, a photon can be in a superposition of two polarization states, or an atom can be in a superposition of the ground state and the excited state. The states $\ket{0}$ and $\ket{1}$ are the eigenvectors of a system. These eigenvectors $\ket{0}$ and $\ket{1}$ provide the bases for a qubit state. These bases are orthonormal, that is $\bra{i}\ket{j}=\delta_{ij}$, where $i,j=\{0,1\}$. When $\ket{i}$ is a column vector, $\bra{i}$ is a row vector. The superposition allows us to do many calculations in parallel. For $n$ qubits, a state is written as a $2^n$-dimensional vector in a Hilbert space $\cal H$ and the qubits can be entangled.

\begin{figure}[h!]
	\centering
	\begin{subfigure}{0.4\textwidth}
		\centering
		\begin{tikzpicture}
			\qnode{0.5}{2}{$\alpha\ket{0}+\beta \ket{1}$}
			\qgateX[ibmqxA]{2}{2}
			\qnode{3.5}{2}{$\beta \ket{0} + \alpha \ket{1}$}
			\qnode{0.5}{1}{$\alpha\ket{0}+\beta \ket{1}$}
			\qgateY[ibmqxE]{2}{1}
			\qnode{3.5}{1.1}{$-i\beta \ket{0} + i\alpha \ket{1}$}
			\qnode{0.5}{0}{$\alpha\ket{0}+\beta \ket{1}$}
			\qgateZ[ibmqxC]{2}{0}
			\qnode{3.5}{0}{$\alpha\ket{0} - \beta\ket{1}$}
		\end{tikzpicture}
		\caption{}
	\end{subfigure}
	\begin{subfigure}{0.4\textwidth}
		\centering
		\begin{tikzpicture}
			\qnode{0.2}{1}{$\ket{0}$}
			\qgateH[ibmqxD]{1}{1}
			\qnode{2}{1}{$\frac{\ket{0} + \ket{1}}{\sqrt{2}}$}
			\qnode{0.2}{0}{$\ket{1}$}
			\qgateH[ibmqxA]{1}{0}
			\qnode{2}{0}{$\frac{\ket{0} - \ket{1}}{\sqrt{2}}$}
		\end{tikzpicture}
		\caption{}
	\end{subfigure}	
	\caption[Implementation of Pauli gates, Hadamard gate, and CNOT gate.]{The implementation of (a) Pauli gates, (b) Hadamard gate.}
	\label{PauliH}
\end{figure}
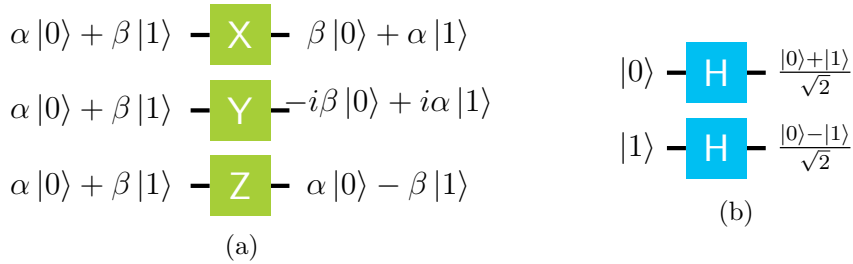

The purpose of quantum gates and circuits is to get the required output with maximum probability. Mathematically, a gate is represented by a matrix that must be unitary. The matrix elements correspond to the probabilities of getting the respective basis state. Two matrices do not commute in general.
Some elementary gates are represented by symbols $I, X, Y, Z$. These are called \textit{Pauli matrices} in physics and denoted as $\sigma_I,\sigma_x,\sigma_y, \sigma_z$. These matrices have the effect of rotating the qubit about the z-axis by an angle $\theta$.
\eq{I = \begin{pmatrix}
		1 & 0\\
		0 & 1
	\end{pmatrix} \ , \ X =
	\begin{pmatrix}
		0 & 1\\
		1 & 0
	\end{pmatrix} \ , \ Y =
	\begin{pmatrix}
		0 & -i\\
		i & 0
	\end{pmatrix} \ , \ Z =
	\begin{pmatrix}
		1 & 0\\
		0 & -1
	\end{pmatrix},}
where $X$ is the NOT gate, $Z$ is the phase gate and $Y$ is the phase and NOT gate together, that is $Y = iXZ$. These matrices have properties that $X^2=Y^2=Z^2=I$ and $XY=iZ, \ YZ=iX, \ ZX=iY$. The superposition is created by the \textit{Hadamard gate},
\eq{H\ket{\psi} = \frac{1}{\sqrt{2}}\begin{pmatrix}
		1 & 1 \\
		1 & -1
	\end{pmatrix} \begin{pmatrix}
		\alpha \\
		\beta
	\end{pmatrix} = \frac{1}{\sqrt{2}} \begin{pmatrix}
		\alpha + \beta \\
		\alpha - \beta
	\end{pmatrix}.}
The implementation of these gates is shown in Fig. \ref{PauliH} (a) and (b).
Another example of one-qubit gates is \textit{phase gate} that can be written as
\eq{P(\phi)= \begin{pmatrix}
		1 & 0\\
		0 & e^{i\phi}
	\end{pmatrix}.}
Applying on the state ket $\ket{\psi}$, we get
\eq{P\ket{\psi} = \alpha \ket{0}+e^{i\phi} \beta \ket{1}=\begin{pmatrix}
	\alpha\\
	e^{i\phi}\beta
\end{pmatrix}.}
When $\phi=\pi$ we have the Pauli matrix $Z$, that is $P(\pi)=Z$. Other examples of phase gates are $S$ and $T$ gates written as
\eq{S=P(\pi/2)=\begin{pmatrix}
	1 & 0\\
	0 & i
\end{pmatrix}, \qquad T = e^{i\pi/8}\begin{pmatrix}
	e^{-i\pi/8} & 0\\
	0 & e^{i\pi/8}
\end{pmatrix}=\begin{pmatrix}
	1 & 0\\
	0 & e^{i\pi/4}
\end{pmatrix}.}
We can see $T^2=S$. $T$ is also known as $\pi/8$ gate.

\begin{figure}
	\centering
	\begin{tikzpicture}
		\draw [very thick,cyan,dashed] (2,0) arc(0:180:2cm and 0.7cm);
		\draw [very thick,cyan,dashed] (0,2) arc(90:-270:0.7cm and 2cm);
		\draw[ultra thick,-{latex[scale=3.0]},olive] (0,0)--(0.9,1.6);
		
		\draw[very thick,->,blue] (0,0)--(2.5,0);
		\draw[very thick,->,blue] (0,0)--(-2.5,0);
		\draw[very thick,->,red] (0,0)--(0,2.5);
		\draw[very thick,->,red] (0,0)--(0,-2.5);
		
		\draw[very thick,->,teal] (0,0)--(-1.6,-1.6);
		\draw [very thick,cyan] (0,2) arc(90:270:0.7cm and 2cm);
		\draw [very thick,cyan] (2,0) arc(0:-180:2cm and 0.7cm);
		\draw[very thick,->,teal] (0,0)--(1.6,1.6);
		\draw [very thick,cyan] (0,0) ellipse (2 and 2);
		
		\node [above left,red] at (0,2) {$\ket{1}$};
		\node [below left,red] at (0,-2) {$\ket{0}$};
		\node [below right,blue] at (1.9,0) {$\frac{1}{\sqrt{2}}[\ket{0}+i\ket{1}]$};
		\node [above left,blue] at (-2,0) {$\frac{1}{\sqrt{2}}[\ket{0}-i\ket{1}]$};
		\node [above right,teal] at (1.5,1.5) {$\frac{1}{\sqrt{2}}[\ket{0}-\ket{1}]$};
		\node [below left,teal] at (-1.5,-1.5) {$\frac{1}{\sqrt{2}}[\ket{0}+\ket{1}]$};
		
		\draw[thick,dashed,olive] (0.05,-0.05)--(1.4,-0.5);
		\draw [thick,dashed,olive] (0,2) arc(90:-14:1.45cm and 2cm);
		\draw [thick] (0,1) arc(90:65:1cm and 2cm);
		\node [above right] at (0,1) {$\theta$};
		\draw [thick] (-0.15,-0.2) arc(190:350:0.3cm and 0.2cm);
		\node [above right] at (-0.5,-0.8) {$\phi$};
		\node [olive] at (1.2,2) {$\ket{\psi}$};	
	\end{tikzpicture}
	\caption{One-qubit state can be represented by a Bloch sphere.}
	\label{Bloch}	
\end{figure}
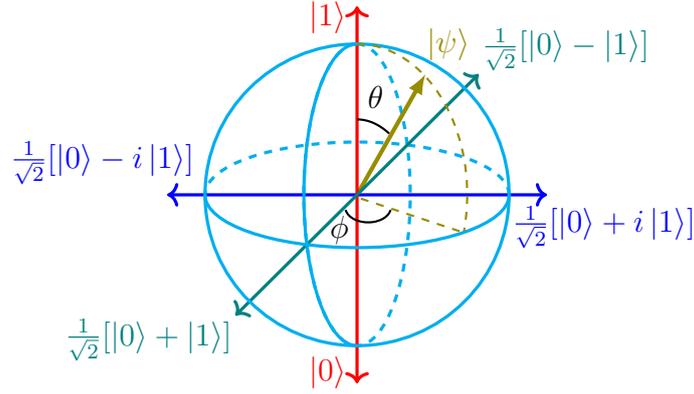

The one-qubit state can be represented by the Bloch sphere as shown in Fig. \ref{Bloch}. The general state on the Bloch sphere is given by \cite{Nielsen2002quantum}
\eq{\ket{\psi} = \cos \frac{\theta}{2} \ket{0} + e^{i\phi}\sin\frac{\theta}{2} \ket{1} = \begin{pmatrix}
		\cos\frac{\theta}{2} \\
		e^{i\phi}\sin\frac{\theta}{2}
	\end{pmatrix}, \ \text{where} \ 0 \le \theta \le \pi, \ \ 0 \le \phi \le 2\pi .}
The operators that rotate the state on the Bloch sphere can be written as
\eq{&R_x(\theta) \equiv e^{-i\frac{\theta}{2} X} = \cos \frac{\theta}{2} I + i\sin\frac{\theta}{2} X = \begin{pmatrix}
		\cos \frac{\theta}{2} & -i\sin\frac{\theta}{2}\\
		-i\sin\frac{\theta}{2} & \cos \frac{\theta}{2}
	\end{pmatrix},\nonumber \\
	&R_y(\theta) \equiv e^{-i\frac{\theta}{2} Y} = \cos \frac{\theta}{2} I + i\sin\frac{\theta}{2} Y = \begin{pmatrix}
		\cos \frac{\theta}{2} & -\sin\frac{\theta}{2}\\
		\sin\frac{\theta}{2} & \cos \frac{\theta}{2}
	\end{pmatrix},\nonumber \\
	&R_z(\theta) \equiv e^{-i\frac{\theta}{2} Z} = \cos \frac{\theta}{2} I + i\sin\frac{\theta}{2} Z = \begin{pmatrix}
		e^{-i \frac{\theta}{2}} & 0\\
		0 & e^{i \frac{\theta}{2}}
	\end{pmatrix}.}
These three rotations on a Bloch sphere are combined into a general rotation as
\eq{R_{\hat{n}}(\theta) = \exp(-i \frac{\theta}{2} \hat{n}\cdot \vec{\sigma}) = \cos \frac{\theta}{2} I - i\sin\frac{\theta}{2}\Big(n_xX + n_yY + n_zZ\Big),}
where $\hat{n}$ is a unit vector in three dimensions and $\vec{\sigma}$ is the three-component vector of Pauli matrices. $R_{\hat{n}}(\theta)$ is the effect of rotation on the state around the unit vector $\hat{n}$. Let there exist real numbers $\alpha$, $\beta$, $\gamma$ and $\delta$ such that a general unitary operation on a qubit \cite{Nielsen2002quantum} can be written as
\eq{U = e^{i\alpha}R_z(\beta)R_y(\gamma)R_z(\delta) = \begin{pmatrix}
e^{i(\alpha - \beta/2 - \delta/2)}\cos\frac{\gamma}{2} & -e^{i(\alpha - \beta/2 + \delta/2)}\sin\frac{\gamma}{2} \\
e^{i(\alpha + \beta/2 - \delta/2)}\sin\frac{\gamma}{2} & e^{i(\alpha + \beta/2 + \delta/2)}\cos\frac{\gamma}{2}
\end{pmatrix}.}
The Bloch sphere representation is limited to a single qubit state only.

Two-qubit states can be \textit{separable or inseparable}. The separable states are written for independent composite systems, whereas the inseparable states are entangled state. The separable state, also called product state, can be factorized into two separate states and written as the \textit{tensor product} of the two states as
\eq{\ket{\psi} =& \ket{\phi_1} \otimes \ket{{\phi_2}}  
	= \Big(\alpha_1 \ket{0} + \beta_1 \ket{1} \Big)\otimes \Big(\alpha_2 \ket{0} + \beta_2 \ket{1}\Big) \nonumber \\
	=& \alpha_1 \alpha_2 \ket{0}\ket{0} + \alpha_1 \beta_2\ket{0}\ket{1} 
	+\beta_1 \alpha_2 \ket{1}\ket{0} + \beta_1 \beta_2 \ket{1}\ket{2} \nonumber \\
	=& \alpha \ket{00} + \beta\ket{01} + \gamma\ket{10} + \delta \ket{11},}
where $\abs{\alpha}^2 + \abs{\beta}^2 + \abs{\gamma}^2 + \abs{\delta}^2 = 1$. The tensor product notation $\otimes$ means that we multiply each term of the first vector to each term of the second vector. The dimension of the Hilbert space of the product state is a product of the dimensions of the two systems, that is $\mathcal{H}=\mathcal{H}_1\otimes \mathcal{H}_2$. Since the gates are operators and we write them as matrices, the tensor product of two operators is expressed in such a way that multiply each entry of the first matrix to all entries of the second matrix. As an example, let us consider two matrices $A$ and $B$ as
 $$A=\begin{pmatrix}
 	a_1 & a_2\\
 	a_3 & a_4
 \end{pmatrix}, \qquad B=\begin{pmatrix}
 b_1 & b_2\\
 b_3 & b_4
\end{pmatrix}.$$
The tensor product of $A$ and $B$ is given as
\eq{A\otimes B = \begin{pmatrix}
		a_1\begin{pmatrix}
			b_1 & b_2\\
			b_3 & b_4
		\end{pmatrix} & a_2 \begin{pmatrix}
			b_1 & b_2\\
			b_3 & b_4
		\end{pmatrix}\\
		a_3\begin{pmatrix}
			b_1 & b_2\\
			b_3 & b_4
		\end{pmatrix} & a_4 \begin{pmatrix}
			b_1 & b_2\\
			b_3 & b_4
		\end{pmatrix}
	\end{pmatrix}=\begin{pmatrix}
		a_1b_1 & a_1b_2 & a_2b_1 & a_2b_2\\
		a_1b_3 & a_1b_4 & a_2b_3 & a_2b_4\\
		a_3b_1 & a_3b_2 & a_4b_1 & a_4b_2\\
		a_3b_3 & a_3b_4 & a_4b_3 & a_4b_4
	\end{pmatrix}.}
When $A\ket{a}=\alpha \ket{a}$ and $B\ket{b}=\beta \ket{b}$ then the following rules are defined for the tensor product
\eq{&(A\otimes B)(\ket{a}\otimes \ket{b}) = A\ket{a} \otimes B\ket{b},\nonumber\\	
	&(\ket{a}+\ket{b})\otimes \ket{c}= \ket{a}\otimes \ket{c}+\ket{b}\otimes \ket{c},\nonumber\\
	&\ket{a}\otimes(\ket{b}+ \ket{c})= \ket{a}\otimes \ket{b}+\ket{a}\otimes \ket{c}.}
We also have the notations $\ket{a}\otimes \ket{b}\equiv \ket{a}\ket{b} \equiv \ket{ab}$. If $\ket{a}$ and $\ket{b}$ are column vectors of two elements each, then $\ket{ab}$ is a column vector of four elements. Corresponding operators $A$ and $B$ would become a four by four matrix $A\otimes B$. It can be generalized to $n$-dimensional Hilbert space. \textit{The number of states is increased exponentially with the increase of the number of qubits.}

\begin{figure}[h!]
\begin{subfigure}{0.25 \textwidth}
	\centering
	\begin{tikzpicture}
		\qnode{0}{1}{$\ket{a}$}
		\qnode{0}{0}{$\ket{b}$}
		\qgateCNC[ibmqx]{b}{1}{1}
		\qgateCNX[ibmqx]{t}{1}{0}
		\qnode{2}{1}{$\ket{a}$}
		\qnode{2.25}{0}{$\ket{a\oplus b}$}
	\end{tikzpicture}
\caption{}
\end{subfigure}
\begin{subfigure}{0.2\textwidth}
	\centering
	\begin{tikzpicture}
		\draw[ultra thick,cyan] (0,0)--(0,1);
		\qgateCNC[ibmqx]{b}{0}{1}
		\qgateU[ibmqxA]{0}{0}{U}
	\end{tikzpicture}
	\caption{}
\end{subfigure}
\begin{subfigure}{0.25\textwidth}
	\begin{tikzpicture}
		\draw[ultra thick,cyan] (0,0)--(0,1);	
		\qwire[ibmqx]{0}{1}
		\qwire[ibmqx]{0}{0}
		\qnode{0}{1}{\Large\textcolor{cyan}{\textbf{$\cross$}}}
		\qnode{0}{0}{\Large\textcolor{cyan}{\textbf{$\cross$}}}
		\qnode{-1}{1}{$\ket{a}$}
		\qnode{-1}{0}{$\ket{b}$}
		\qnode{1}{1}{$\ket{b}$}
		\qnode{1}{0}{$\ket{a}$}
	\end{tikzpicture}
	\caption{}
\end{subfigure}
\begin{subfigure}{0.2\textwidth}
	\begin{tikzpicture}
		\qgateCNC[ibmqx]{b}{1}{1}
		\qgateCNX[ibmqx]{t}{1}{0}
		\qgateCNC[ibmqx]{t}{2}{0}
		\qgateCNX[ibmqx]{b}{2}{1}
		\qgateCNC[ibmqx]{b}{3}{1}
		\qgateCNX[ibmqx]{t}{3}{0}
	\end{tikzpicture}
	\caption{}
\end{subfigure}
\caption[Controlled-NOT, Controlled-U, SWAP gate, and its physical realization.]{(a) CNOT gate (b) Controlled-U gate and (c) SWAP gate (d) Physical realization of SWAP gate.}
\label{TwoQubit}
\end{figure}
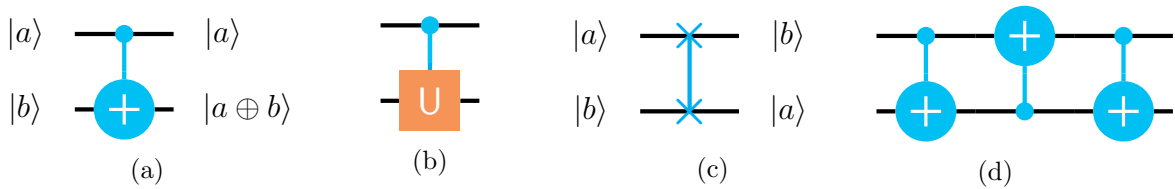

A typical example of a two-qubit gate is the \textit{controlled-NOT or CNOT gate} shown in Fig. \ref{TwoQubit} (a). This gate flips the second qubit when the first qubit state is $\ket{1}$. The first qubit is called \textit{control qubit} and the second qubit is called \textit{target qubit}. This gate is a classical analog of exclusive-OR gate based on exclusive-OR logic represented by $\oplus$. The symbol $\oplus$ is defined such that the output state $\ket{x\oplus y}$ will give a value 0 when both inputs are either zero or 1, but the output value will be 1 when one of the inputs is 1.
A two-qubit state
\eq{\ket{\psi} &= \alpha \ket{00} + \beta\ket{01} + \gamma\ket{10} + \delta \ket{11}}
is changed by the operation of CNOT gate as
\eq{CNOT\ket{\psi} &= \begin{pmatrix}
		1 & 0 & 0 & 0 \\
		0 & 1 & 0 & 0 \\
		0 & 0 & 0 & 1 \\
		0 & 0 & 1 & 0 
	\end{pmatrix}
	\begin{pmatrix}
		\alpha \\
		\beta \\
		\gamma \\
		\delta
	\end{pmatrix} = \begin{pmatrix}
		\alpha \\
		\beta \\
		\delta \\
		\gamma
	\end{pmatrix} \nonumber\\
	&= \alpha \ket{00} + \beta\ket{01} + \delta\ket{10} + \gamma \ket{11}.}
The second qubit remains the same when the first qubit is $\ket{0}$, whereas $X$ gate is applied to the target qubit when the control state is $\ket{1}$. But in general, there can be any one-qubit gate at the place of $X$ as shown in Fig. \ref{TwoQubit} (b). In that case, we can write the controlled-U (CU) gate as
\eq{CU =\begin{pmatrix}
		1 & 0 & 0 & 0\\
		0 & 1 & 0 & 0\\
		0 & 0 & u_1 & u_2\\
		0 & 0 & u_3 & u_4
\end{pmatrix}.}

Another two-qubit gate is a SWAP gate that swaps the states of input qubits. The SWAP gate and its physical realization are shown in Fig. \ref{TwoQubit} (c) and (d). It can also be written in matrix notation as
\eq{SWAP=\begin{pmatrix}
		1 & 0 & 0 & 0\\
		0 & 0 & 1 & 0\\
		0 & 1 & 0 & 0\\
		0 & 0 & 0 & 1
	\end{pmatrix}.}

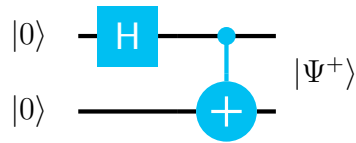
\begin{figure}[h!]
	\centering
	\begin{tikzpicture}
		\qnode{-1}{1}{$\ket{0}$}
		\qnode{-1}{0}{$\ket{0}$}
		\qgateH[ibmqxE]{0}{1}
		\qwire[ibmqx]{0}{0}
		\qgateCNC[ibmqx]{b}{1}{1}
		\qgateCNX[ibmqx]{t}{1}{0}
		\qnode{2}{0.5}{$\ket{\Psi^+}$}
	\end{tikzpicture}
	\caption{The entanglement generating gate.}
	\label{BellState}
\end{figure}

The computation cannot be performed by a single qubit only. A system should consist of several qubits and should have the capability to entangle these qubits. One qubit is a superposition of two basis states, but when there is a quantum correlation among two systems, then we say that these systems are entangled. The entangled state is non-separable and cannot mathematically be factorized into two separate superposition states. The Bell's state
\eq{\ket{\Psi^+} = \alpha_{00}\ket{00} + \alpha_{11}\ket{11},}
with $\abs{\alpha_{00}}^2 + \abs{\alpha_{11}}^2 =1$, is an example of the entangled state. It is non-local, that is, the information of only one qubit is not accessible locally when two states are far apart. Classically, the first qubit can be in state $\ket{0}$ or $\ket{1}$, so can be the second qubit. Therefore, there are four possibilities of values on measurement. But an entangled state like this Bell's state would give $\ket{00}$ or $\ket{11}$ with the probabilities $\abs{\alpha_{00}}^2$ and $\abs{\alpha_{11}}^2$. On measurement, if the first qubit collapse to the state $\ket{0}$ $(\ket{1})$ then the second qubit is forced to collapse to $\ket{0}$ $(\ket{1})$. The state is maximally entangled when $\abs{\alpha_{00}}^2 = \abs{\alpha_{11}}^2 = \frac{1}{2}$. The entangled state is created in a process that can be shown as in Fig. \ref{BellState}.

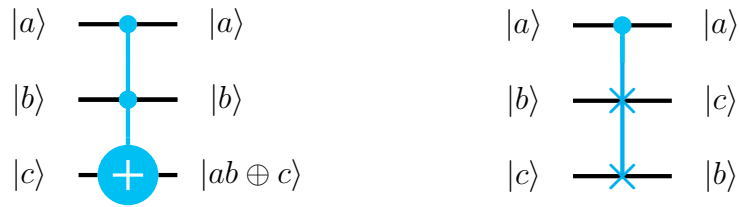
\begin{figure}[h!]
	\centering
	\begin{subfigure}{0.4\textwidth}
		\begin{tikzpicture}
			\qnode{0}{2}{$\ket{a}$}
			\qnode{0}{1}{$\ket{b}$}
			\qnode{0}{0}{$\ket{c}$}
			\qgateCNC[ibmqx]{b}{1}{2}
			\qgateCNC[ibmqx]{b}{1}{1}
			\qgateCNX[ibmqx]{t}{1}{0}
			\qgateCNC[ibmqx]{t}{1}{1}
			\qnode{2}{2}{$\ket{a}$}
			\qnode{2}{1}{$\ket{b}$}
			\qnode{2.25}{0}{$\ket{ab\oplus c}$}
		\end{tikzpicture}
	\end{subfigure}
	\begin{subfigure}{0.2\textwidth}
		\begin{tikzpicture}
			\draw[ultra thick,cyan] (0,0)--(0,2);
			\qgateCNC[ibmqx]{b}{0}{2}	
			\qwire[ibmqx]{0}{1}
			\qwire[ibmqx]{0}{0}
			\qnode{0}{1}{\Large\textcolor{cyan}{\textbf{$\cross$}}}
			\qnode{0}{0}{\Large\textcolor{cyan}{\textbf{$\cross$}}}
			\qnode{-1}{2}{$\ket{a}$}
			\qnode{-1}{1}{$\ket{b}$}
			\qnode{-1}{0}{$\ket{c}$}
			\qnode{1}{2}{$\ket{a}$}
			\qnode{1}{1}{$\ket{c}$}
			\qnode{1}{0}{$\ket{b}$}
		\end{tikzpicture}
	\end{subfigure}
	\caption[Toffoli and Fredkin gates.]{(a) Toffoli and (b) Fredkin gates.}
	\label{Toffoli}
\end{figure}

Toffoli and Fredkin gates shown in Fig. \ref{Toffoli} are examples of three-qubit gates. Toffoli gate, also known as controlled-controlled-NOT or CCNOT, consists of two control qubits and one target qubit. When the first and second qubits will be in state $\ket{1}$, then the NOT gate $X$ will be applied to the third gate, but nothing will happen in all other cases. For the Fredkin gate, a SWAP gate is applied on the second and third qubit when the first one is in state $\ket{1}$, nothing will happen otherwise. Therefore, this gate is also known as the controlled-SWAP gate.

Gates are drawn from left to right in a diagram but mathematically they appear in order from right to left. For example, the gate in Fig. \ref{BellState} is written as
\eq{\ket{\Psi^+} = CNOT(H\otimes I)\ket{00}.}

The set of elementary gates used to perform all kinds of computations is called the \textit{universal set of gates} \cite{barenco1995elementary}. There are several universal sets of gates. Hadamard, $\pi/8$, and a CNOT gate can make one of the sets of universal quantum gates \cite{Nielsen2002quantum}. The number of gates used to implement a circuit is called \textit{quantum cost} of the circuit.
Quantum algorithms are used to solve a particular problem. The most popular quantum algorithms are Shor's factoring algorithm \cite{lavor2003shor} and Grover search algorithm \cite{lavor2003grover}. The former is an exponential speedup over classical factoring algorithm, and the latter is a quadratic speedup.

The gates and circuits in quantum computing are made reversible.
The reversible circuits are the ones with the same number of outputs as the number of inputs, and there is also a one-to-one correspondence between input and output states. Landauer \cite{landauer1961irreversibility} proved that the minimum energy dissipation for the processing of information is  $KT \ln 2$.  
Bennett et al. \cite{bennett1973logical} proposed that the energy dissipation can be avoided if the information processing is made \textit{reversible}.

The conditions necessary for constructing a quantum computer are known as \textit{DiVincenzo' criteria} \cite{divincenzo2000physical}. Five requirements for quantum computations are scalability of the physical system, ability to initialize the qubits in a particular basis state, long decoherence time, universal set of gates, and measurement capability. Two more conditions are added for quantum communication, which are the ability to interconvert stationary and flying qubits, and the ability to transmit the qubits between two locations.
Some physical systems used to construct a quantum computer are: ion trap, neutral atoms trapped in an optical lattice, superconductors, quantum dots, nitrogen vacancy centers in diamond, optical, and topological.

\section{Ternary Quantum Gates}
In quantum technologies, hybrid circuits are sometimes employed, which are a combination of binary and multivalued circuits. Such gates and corresponding circuits may be advantageous in some ways, such as the reduction in inputs and outputs, reduction in the quantum cost, and the complexity of interconnects. Ternary logic is the most popular multi-value logic. The basic unit of information for multivalued logic is called a \textit{qudit} and that of ternary logic is called a \textit{qutrit}.
Khan and Perkowski \cite{khan2007quantum} showed that the ternary logic needs fewer gates comparing with its respective binary system. A binary quantum system requires $n_2 = \log_2N$ qubits for a Hilbert space of dimensions $N$. On the other hand, an $m$-valued quantum system requires $n_m = \log_mN$ qudits, we have 
\eq{n_m = \log_mN = \frac{\log_2N}{\log_2m} = \frac{n_2}{\log_2m}.}
Therefore, an $m$-valued quantum system requires $1/\log_2m$ times the memory of its binary counterpart. Hence, a logarithmic reduction in the number of qudits for an $m$-valued logic. It is also shown in \cite{khan2007quantum} that $m=3$ is the most favorable choice. Haghparast et al. \cite{haghparast2017towards} proved that the ternary is $37\%$ more compact than binary. Therefore, by using ternary logic gates, we can reduce the cost of circuits and make them more efficient.

In ternary quantum logic, the state $\ket{0}$, $\ket{1}$ and $\ket{2}$ are computational bases. A state can be in a superposition of these three basis states, and is written as 
\eq{\phi = \alpha \ket{0} + \beta \ket{1} + \gamma\ket{2},}
with $\alpha$, $\beta$ and $\gamma$ being complex numbers such that $\abs{\alpha}^2 + \abs{\beta}^2 + \abs{\gamma}^2 = 1$. The state vector $\ket{\psi}$ is a three-dimensional column vector
\eq{\ket{\psi}=\begin{pmatrix}
		\alpha\\\beta\\\gamma
	\end{pmatrix} =\alpha\begin{pmatrix}
		1\\0\\0
	\end{pmatrix}+\beta\begin{pmatrix}
		0\\1\\0
	\end{pmatrix}+\gamma\begin{pmatrix}
		0\\0\\1
	\end{pmatrix}.} 

The elementary gates for ternary logic are $3\cross 3$ unitary matrices \cite{khan2006design,khan2007quantum,di2011elementary,shah2010design,giesecke2006ternary}. The qutrit gates and their symbols are shown in Fig. \ref{1QutritGates}, where $Z_3(+1)$ shifts the qutrit state by 1 and $Z_3(+2)$ gate shifts the qutrit state by 2. $Z_3(01)$, $Z_3(12)$ and $Z_3(02)$ permute the states $\ket{0}$ and $\ket{1}$, $\ket{1}$ and $\ket{2}$, and $\ket{0}$ and $\ket{2}$ respectively \cite{giesecke2006ternary,khan2007quantum}. 

\begin{figure}[h!]
	\centering
	\begin{tikzpicture}
		\qgateU[ibmqxA]{0}{0}{+1}
		\qnode{0}{-1.5}{$\begin{pmatrix}
				0 & 0 & 1\\
				1 & 0 & 0\\
				0 & 1 & 0
			\end{pmatrix}$}
	\end{tikzpicture}
	\begin{tikzpicture}
		\qgateU[ibmqxA]{0}{0}{+2}
		\qnode{0}{-1.5}{$\begin{pmatrix}
				0 & 1 & 0\\
				0 & 0 & 1\\
				1 & 0 & 0
			\end{pmatrix}$}
	\end{tikzpicture}
	\begin{tikzpicture}
		\qgateU[ibmqxD]{0}{0}{01}
		\qnode{0}{-1.5}{$\begin{pmatrix}
				0 & 1 & 0\\
				1 & 0 & 0\\
				0 & 0 & 1
			\end{pmatrix}$}
	\end{tikzpicture}
	\begin{tikzpicture}
		\qgateU[ibmqxD]{0}{0}{12}
		\qnode{0}{-1.5}{$\begin{pmatrix}
				1 & 0 & 0\\
				0 & 0 & 1\\
				0 & 1 & 0
			\end{pmatrix}$}
	\end{tikzpicture}
	\begin{tikzpicture}
		\qgateU[ibmqxD]{0}{0}{02}
		\qnode{0}{-1.5}{$\begin{pmatrix}
				0 & 0 & 1\\
				0 & 1 & 0\\
				1 & 0 & 0
			\end{pmatrix}$}
	\end{tikzpicture}	
	\caption{Quantum one-qutrit gates.}
	\label{1QutritGates}
\end{figure}

Analogous to the Hadamard in binary, there is a Chrestenson transform that creates a superposition state from the bases states and is written as \cite{al2002multiple,moraga2014some}
\eq{CH=\frac{1}{\sqrt{3}}\begin{pmatrix}
		1 &1 & 1\\
		1 & \omega & \omega^*\\
		1 & \omega^* & \omega 
\end{pmatrix},}
where $\omega= \exp(2\pi i/3)$. The $\omega$ is a cube root of unity, that means that it is equal to unity if we raise it to the cubic power.

A two-qutrit state is written as 
\eq{\ket{\psi} =& \ket{\phi_1} \otimes \ket{\phi_2}
	= \big(\alpha_1 \ket{0} + \beta_1 \ket{1} + \gamma_1\ket{2} \big) \otimes \big(\alpha_2 \ket{0} + \beta_2 \ket{1} + \gamma_2\ket{2}\big) \nonumber \\
	=& \alpha_1 \alpha_2 \ket{00} + \alpha_1 \beta_2 \ket{01} + \alpha_1 \gamma_2 \ket{02} + \beta_1 \alpha_2 \ket{10} + \beta_1 \beta_2\ket{11} + \beta_1 \gamma_2\ket{12} +\nonumber\\
	& \gamma_1 \alpha_2\ket{20} + \gamma_1\beta_2\ket{21} + \gamma_1\gamma_2\ket{22}.}
The two-qutrit gates analogous to CNOT gate in Fig. \ref{TwoQubit} is such that the $U$ is applied when the controlling qutrit is at $\ket{2}$ otherwise the second qutrit does not change. Here, $U$ is one of the five one-qutrit gates in Fig. \ref{1QutritGates}. The Toffoli gate is implemented in such a way that $U$ is applied on the third qutrit when both the controlling qutrit at $\ket{2}$. The one-qutrit gates are called \textit{shift gates} and the two-qutrit gates are referred to as \textit{Muthukrishnan-Stroud (MS) gates} \cite{muthukrishnan2000multivalued}. The shift gates and MS gates have quantum cost unity \cite{asadi2020efficient,monfared2017design}. We will further discuss ternary gates and circuits in Chapter \ref{Meta}.

The physical realization of ternary logic was suggested by Ref. \cite{muthukrishnan2000multivalued} for an ion-trap quantum computer, by Ref. \cite{morisue1989novel,morisue1998josephson} for a Josephson junction, by Ref. \cite{smith2013quantum} for cold atoms, and by Ref. \cite{malik2016multi} for entangled photons. Some circuit architectures are better described by the multi-valued logic. In certain systems containing the non-Abelian anyons, called metaplectic anyons, qutrits naturally appear. We will discuss ternary gates with metaplectic anyons in Chapter \ref{Meta}.

\section{Computational Complexity}
Is a quantum computer more powerful than a classical computer? The answer is in the computational complexity theory. To discuss the computational complexity, we will first talk about the questions such as: what is computable? And which problems are solvable and how efficiently?
In 1936, Alan Turing proposed a thought machine to answer these questions \cite{turing1937computable}. A classical computer is based on a Turing machine model, so we will use these two interchangeably. Turing showed that there exists a universal Turing machine that can simulate any problem that is solvable on any other hardware. This machine can truly capture the algorithmic process. Alonzo Church simultaneously arrived at a similar result \cite{church1936unsolvable}.  

The classical \textit{Church-Turing thesis} can be stated as, \textit{Any computation process or algorithm that can be devised by a mathematician can be effectively implemented on a Turing machine}.

After the invention of the transistor, several other models were suggested. So for the comparison of the models, efficiency became an important factor. The Church-Turing thesis was modified and strengthened with the word efficient. According to the \textit{strong Church-Turing thesis}: \textit{Any computation that can be performed by any physical machine can be simulated efficiently by a Turing machine.}

The Church-Turing thesis and the strong Church-Turing thesis were stated for the deterministic Turing machine. Later, some algorithms were devised for which a solution could be found with some bounded probability \cite{pathak2013elements}. One of the examples is the Solovay-Strassen algorithm, which appeared in 1977 \cite{solovay1977fast} for finding whether an integer is a prime or not. It did not answer with certainty but with some probability. These kinds of algorithms could not be efficiently implemented on a deterministic Turing machine. A probabilistic Turing machine can make a random choice at each step. The modified strong Church-Turing thesis states that: \textit{Any computation that can be performed by any physical machine can be simulated efficiently by a probabilistic Turing machine.}

Since nature is quantum mechanical, another challenge to the strong Church-Turing thesis came from quantum theory. Quantum theory is not based on probabilities, but on probability amplitudes, so complex numbers are used. And in addition, there is a measurement problem in quantum theory. The quantum Turing machine should have circuits that preserve the inner product. The quantum circuit model is equivalent to the quantum Turing machine \cite{pathak2013elements}.

In 1985, David Deutsch \cite{deutsch1985quantum} proposed a model of the Turing machine that is based on the laws of quantum theory. Now the strong Church-Turing thesis can be stated as: 
\textit{A quantum Turing machine can efficiently simulate any realistic model of computation.}

Now the question of what is computable is answered regarding what is computable on a Turing machine. There is a problem of uncomputability or undecidability called the \textit{halting problem}.
On a computer, an algorithm runs for several steps and gives out an answer, yes or no. For simple programs, we know that a computer will halt after some number of steps or run forever. But for some complex algorithms, it is not easy to know if it will halt. Alan Turing proved that it is impossible to have an algorithm that can tell if an arbitrary algorithm will eventually halt. It is related to G\"odel's two incompleteness theorems, according to which a formal axiomatic system cannot be complete and consistent at the same time. In every formal system, some statements cannot be proved or disproved by using the same system. The undecidability brought great philosophical consequences to the field of computing \cite{lloyd2006programming,penrose1990emperor}.

Apart from the halting problem, to classify the difficulty of solving different problems on a Turing machine, the idea of complexity classes is introduced. These classes share some common features when one talks about the resources of time and space needed to solve a problem. The resource of time is related to the number of steps needed for an algorithm to solve a problem.

The complexity class \textbf{P} is the set of problems that can be solved on a classical computer in polynomial time, whereas when the solution of a problem can be checked on the classical computer, then the problem is in the complexity class \textbf{NP}. For example, there is no fast way to solve the problem of finding factors of a large number, so it is not in the class \textbf{P}. But if someone gave the factors, then it can easily be checked if it is the right solution. Therefore, the factor finding problem is in the class \textbf{NP}. The \textbf{P} is a subset of the \textbf{NP}. Whether $\textbf{P} = \textbf{NP}$ is an open problem. It is one of the million-dollar problems of Clay Mathematics Institute.

When a problem needs small space resources but no limit on time, then it falls into the complexity class of \textbf{PSPACE}. Another class is \textbf{BPP} when the problem can be solved in polynomial time if a bounded probability of error is allowed. The \textbf{BQP} is defined as the set of problems that can be solved on quantum computers with some bounded probability of error is allowed \cite{aaronson2013quantum}. This complexity class lies somewhere between the classes \textbf{P} and \textbf{PSPACE} \cite{Nielsen2002quantum}.
It implies \cite{kaye2007introduction} that
\eq{\textbf{P}\subseteq \textbf{BPP}\subseteq \textbf{BQP}\subseteq \textbf{PSPACE}, \qquad \text{and} \qquad \textbf{P}\subseteq \textbf{NP} \subseteq \textbf{PSPACE}. \label{ComplexitySubset}}
This is shown in the Fig. \ref{CClasses}.
There are many more complexity classes, see \cite{aaronson2013quantum,yanofsky2008quantum}. So far, there is no proof that the containments in Eq. \ref{ComplexitySubset} are strict, but it is widely believed that \textbf{P$\ne$NP}, \textbf{NP$\ne$PSPACE}, and \textbf{BPP$\ne$BQP} \cite{kaye2007introduction}. It is a non-trivial problem to prove that the quantum computer is more powerful than the classical computer, despite evidences in the favor of this proposition \cite{Nielsen2002quantum}. 
\begin{figure}[h!]
	\centering
	\begin{tikzpicture}[scale=0.9]
		\draw [ultra thick,blue] (-0.2,0.2) ellipse (3 and 2.5);
		\draw [ultra thick,blue] (1,-1.1) ellipse (0.5 and 0.7);
		\draw [ultra thick,blue] (0,-1) ellipse (1.6 and 1.2);
		\draw [ultra thick,red] (-0.2,-0.5) ellipse (2.2 and 1.7);
		\draw [ultra thick,blue] (0.7,0.2) ellipse (1.4 and 2.1);
		\node at (1,-1.1) {\textbf{P}};
		\node at (-1,-1.2) {\textbf{BPP}};
		\node at (-1.5,0) {\textbf{BQP}};
		\node at (1,1.4) {\textbf{NP}};
		\node at (-1.5,1.5) {\textbf{PSPACE}};
	\end{tikzpicture}
	\caption[Complexity classes.]{Complexity classes \cite{kaye2007introduction}.}
	\label{CClasses}
\end{figure}

To analyze the cost of an algorithm with the increase of the input size, three notations are used in computer science. When two functions $f(n)$ and $g(n)$ are defined on non-negative integers, we say $f(n)$ is in the class of functions $O(g(n))$ or $f(n)$ in $O(g(n))$ if there are positive integers $c$ and $N$ such that for all values of $n\ge N$, $f(n)\le cg(n)$. That is $g(n)$ is an upper bound on $f(n)$ up to a constant factor for large $n$. $O$ is useful for worse case scenario or upper bound of the resources consumed by an algorithm. 
$\Omega$ notation is used for a lower bound on resources. A function $f(n)$ is $\Omega(g(n))$ if there are positive integers $c$ and $N$ such that for all values of $n\ge N$, $f(n) \ge cg(n)$.
We say that $f(n)$ is $\Theta(g(n))$ if it is both $O(g(n))$ and $\Omega(g(n))$.
As an example, let $24n + 4[\log n] +8$ be gates required to perform a task. For large $n$, only the first term is important. The function $2n$ is in the class $O(2n^2)$ since $2n\le 2n^2$ for large $n$.

The calculations show that there is a significant speedup in solving problems on a quantum computer. Peter Shor \cite{shor1994algorithms} invented an algorithm to find prime factors of a number on the quantum computer in $O((logN)^3)$ steps. The same problem on a classical computer takes is $O(N^{1/3})$ steps. Therefore, it is an exponential speed-up solving the prime factors finding a problem on a quantum computer.
Simon's period finding problem falls into \textbf{BQP} class \cite{pathak2013elements} and the Grover search algorithm has a quadratic speedup comparing with the classical search. See \cite{Nielsen2002quantum,kaye2007introduction} for a detailed analysis of the complexity of different quantum computing algorithms.
	\chapter{Topology and Knot Theory}\label{Knot}
The knot theory is of fundamental importance in topological quantum computing. Topological quantum gates are made up of knots, links, and braids. The integral along the worldlines of quasiparticles in topological materials gives knot invariants. The geometric phases are associated with such invariants. The knot theory is studied as a branch of topology.

\section{Topology}
The equivalence of two spaces in Euclidean geometry is shown by comparing their lengths and angles, but angles and lengths are irrelevant in topology. 
Instead, imagine that the spaces are made up of a stretchable and moldable material so that we can continuously deform one space to the other without tearing. For example, a sphere cannot be turned into a torus without tearing a hole, so they are topologically different. A hole or a handle in a topological space is called a \textit{genus}. From this point of view, a disk is equivalent to a rectangle or a square but different from an annulus. A curve and a straight line are equivalent shapes, but both are different from a closed curve. A closed curve is equivalent to a circle. A torus is equivalent to a coffee cup as each has one hole in it, and we can smoothly deform one into the other. Some examples of topological spaces are shown in Fig. \ref{TopDiagrams}. For further knowledge on topology, see Appendix \ref{Top}. 
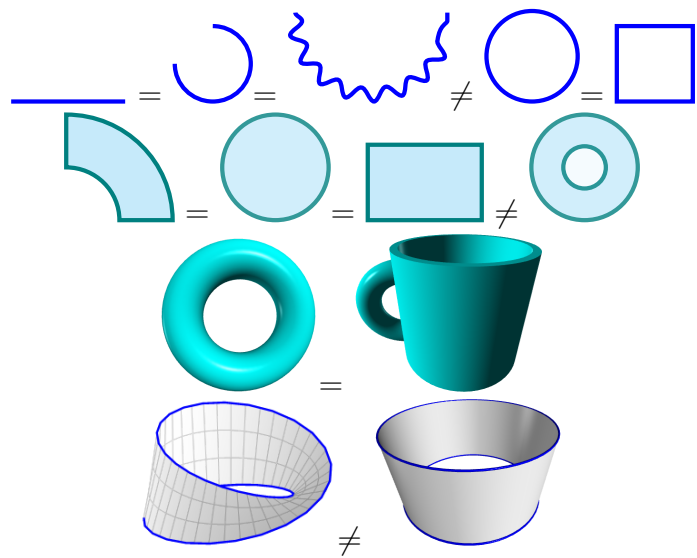
\begin{figure}[h!]
	\centering
	\tik{\draw [blue,ultra thick] (0,0)--(1.5,0);}$\ =$
	\tik{\draw[ultra thick,blue] (0,-0.5) arc (90:-180:0.5);}$=$
	\tik{\draw [decorate,ultra thick,blue, decoration={snake}] (0,0) arc (-180:0:1);}$\ne$
	\tik{\draw[blue,ultra thick] (0,0) circle[radius=0.6];}$=$
	\tik{\draw [ultra thick,blue] (0,0) rectangle (1,1);}
	
	\tik{[scale=0.7]\draw[teal,ultra thick,fill = cyan!20] (1,0) arc (0:90:1)--(0,2) arc (90:0:2)--cycle;} $=$
	\tik{\draw [teal,ultra thick,fill=cyan!20,opacity=0.8] (0,0) ellipse (0.7 and 0.7);}$=$
	\tik{\draw [teal,ultra thick,fill = cyan!20] (0,0) rectangle (1.5,1);}$\ \ne$
	\tik{[scale=0.7]\draw [teal,ultra thick,fill=cyan!20,opacity=0.8] (0,0) ellipse (1 and 1);		
		\draw [teal,ultra thick,fill=white,opacity=0.8] (0,0) ellipse (0.4 and 0.4);}
	
	\includegraphics[scale=0.21]{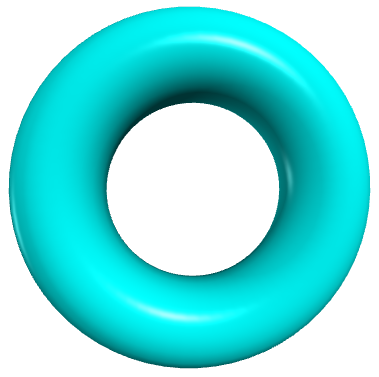}$=$
	\includegraphics[scale=0.25]{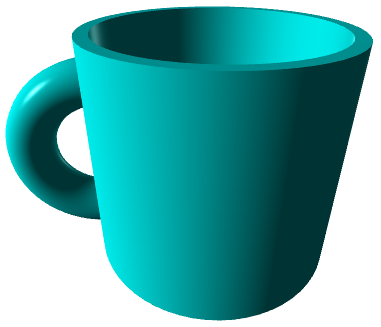}
	
	\includegraphics[scale=0.45]{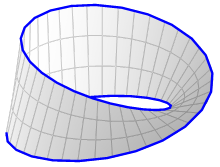}$\ne$
	\includegraphics[scale=0.25]{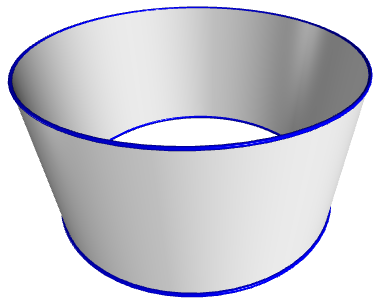}
	\caption{Equivalence of topological spaces.}
	\label{TopDiagrams}
\end{figure}
Topology is the study of properties that are preserved under continuous deformation in such a way that the dimension of the diagram should not change. The continuous deformation is called \textit{homeomorphism}. The topological properties that characterize the equivalence of two shapes under homeomorphism are called \textit{topological invariants}. These invariants can be numbers, or certain properties of the topological spaces like connectedness, compactness, homotopy group, homology group, or cohomology group \cite{armstrong2013basic,nakahara2003geometry,nash1988topology,isham1999modern,frankel2011geometry}. A genus is a topological invariant, but in some cases, it is not a very useful one. 

\begin{figure}[h!]
	\centering
	\includegraphics[scale=0.65]{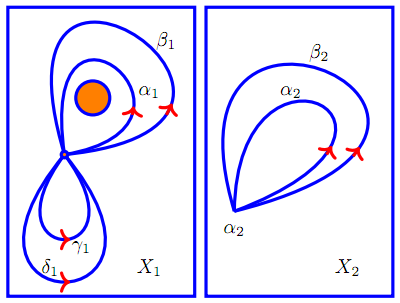}
	\caption[Homotopy group.]{In the region $X_1$, the loops $\beta_1$ and $\alpha_1$ can be deformed to each other but they cannot be deformed to $\delta_1$ or $\gamma_1$. In the region $X_2$, both the loops can be shrunk to identity.}
	\label{Homotopy}
\end{figure}

The first homotopy group provides an intuition for anyonic statistics and braids.
Consider two regions $X_1$ and $X_2$ in Euclidean space as shown in Fig. \ref{Homotopy}. Imagine that any loop in $X_2$ can be shrunk to a point, but when there is a hole as in $X_1$, a loop cannot be shrunk to a point. From the Fig. \ref{Homotopy}, $\alpha _1$ can be deformed to $\beta _1$, and $\gamma_1$ can be deformed to $\delta_1$, but the loops $\alpha_1$ and $\beta_1$ cannot be deformed to $\gamma_1$ and $\beta_1$. 
When two loops can be continuously deformed to each other then they are in the same equivalence class or a homotopic class \cite{nash1988topology}. 

Spaces are distinguished by working with equivalent classes rather than loops. This suggests that the holes are determined by using these equivalence classes. The group structure on these equivalence classes is called the \textit{first homotopy group or a fundamental group} represented by $\pi _1(X)$. See Appendix \ref{AbsAlg} for the formal definition of a group. The group axioms of a homotopy group are described below.
The composition of two group elements corresponds to two loops that start at the same point and are combined to make the third one. In Fig. \ref{Homotopy}, $\gamma_1$ and $\beta_1$ loops are combined to make $\tilde {\gamma}$ that can be written as $\tilde {\gamma} = \gamma_1 \ \beta_1$. This loop first goes along the $\beta_1$ then along the $\gamma_1$. The inverse $\beta_1^{-1}$ is given by a loop that goes in the opposite direction. The identity loop is the one that stays at some point all the time. The loop $\varepsilon = \beta_1 \ \beta_1^{-1}$ is not an identity but homotopic to the identity \cite{nash1988topology}. When these loops are physically made by the motion of particles on a two-dimensional space then they make braids in the third dimension which is time.

\section{Knot Invariants}
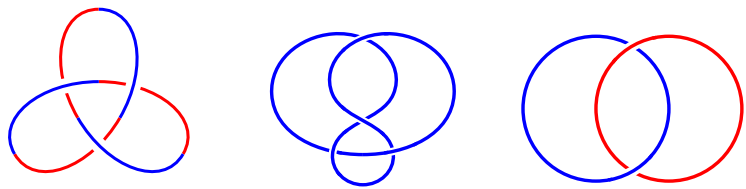
\begin{figure}[h!]
	\centering
	\scalebox{0.8}{
	\begin{tikzpicture}[use Hobby shortcut,every path/.style={line width=1mm,white,double=red, double distance=.5mm},scale=0.8]
		\def\nfoil{3}
		\draw ([closed]0,2)
		foreach \k in {1,...,\nfoil} {
			.. ([blank=soft]90+360*\k/\nfoil-180/\nfoil:-.5) .. (90+360*\k/\nfoil:2)
		};
		\draw[use previous Hobby path={invert soft blanks,disjoint},double=blue];
	\end{tikzpicture} \qquad
	\begin{tikzpicture}[use Hobby shortcut]
		\begin{knot}[consider self intersections=true,ignore endpoint intersections=false,flip crossing=3,only when rendering/.style={ultra thick,blue}]
			\strand ([closed,ultra thick,blue]0,0) .. (1.5,1) .. (.5,2) .. (-.5,1) .. (.5,0) .. (0,-.5) .. (-.5,0) .. (.5,1) .. (-.5,2) .. (-1.5,1) .. (0,0);
		\end{knot}
		\path (0,-.7);
	\end{tikzpicture} \ \qquad
	\begin{tikzpicture}[scale=1.2]
		\begin{knot}[flip crossing=2]
			\strand [red,ultra thick](2,0) circle[radius=1cm];
			\strand[blue,ultra thick,->] (1,0) circle[radius=1cm];
		\end{knot}
	\end{tikzpicture}}
	\caption{Trefoil knot, figure 8 knot, and the Hopf link.}
	\label{Knots}
\end{figure}

A knot is a closed loop embedded in the three-dimensional space. A link is a disjoint union of more than one loop. 
A knot diagram is a projection of a knot into the plane $\mathbb{R}^2$ such that the points are segments and double points are under-crossings and over-crossings.  A circle is an \textit{unknot or a trivial link}. The simplest non-trivial link is a \textit{Hopf link}. For example, the trefoil, figure 8 knot, and Hopf link are shown in Fig. \ref{Knots}. For further study on knots, see \cite{kauffman2001knots,adams2004knot,baez1994gauge}.

\begin{figure}[h!]
	\centering
	\begin{subfigure}{0.3\textwidth}
		\centering
		\begin{tikzpicture}[scale = 0.5]
			\draw[blue,knot,ultra thick] (1,4) .. controls +(0,-2) .. (0,1.7);
			\draw[knot=blue,ultra thick] (1,0) .. controls +(0,2) .. (0,2.3);
			\draw [blue,ultra thick] (0,1.7) to [curve through={(-0.4,2)}](0,2.3);
			\node [left] at (2.5,2) {=};
			\draw[red,knot,ultra thick] (3,0) -- (3,4);
			\node [] at (4,2) {=};
			\draw[blue,knot,ultra thick] (5,4) .. controls +(0,-2) .. (6,1.7);
			\draw[knot=blue,ultra thick] (5,0) .. controls +(0,2) .. (6,2.3);
			\draw [blue,ultra thick] (6,1.7) to [curve through={(6.4,2)}](6,2.3);
		\end{tikzpicture}
		\caption{}
	\end{subfigure}
	\begin{subfigure}{0.3\textwidth}
		\centering
		\begin{tikzpicture}[scale = 0.5]
			\draw[knot=blue,ultra thick] (1,0) .. controls +(0,1) and +(0,-1) .. (0,2) .. controls +(0,1) and +(0,-1) .. (1,4);
			\draw[red,knot,ultra thick] (0,0) .. controls +(0,1) and +(0,-1) .. (1,2) .. controls +(0,1) and +(0,-1) .. (0,4);
			\node [left] at (2.5,2) {=};
		\end{tikzpicture}
		\begin{tikzpicture}[scale = 0.5]
			\draw[blue,ultra thick] (1,0) -- (1,4);
			\draw[red,knot,ultra thick] (0,0) -- (0,4);
			\node [left] at (2,2) {=};
		\end{tikzpicture}
		\begin{tikzpicture}[scale = 0.5]
			\draw[red,knot,ultra thick] (0,0) .. controls +(0,1) and +(0,-1) .. (1,2) .. controls +(0,1) and +(0,-1) .. (0,4);
			\draw[knot=blue,ultra thick] (1,0) .. controls +(0,1) and +(0,-1) .. (0,2) .. controls +(0,1) and +(0,-1) .. (1,4);
		\end{tikzpicture}
		\caption{}
	\end{subfigure} 
	\begin{subfigure}{0.3\textwidth}
		\centering
		\begin{tikzpicture}[scale = 0.5]
			\draw[blue,knot,ultra thick] (1,0) -- (1,4);
			\draw[red,knot,ultra thick] (0,0) -- (3,4);
			\draw[teal,knot,ultra thick] (3,0) -- (0,4);
			\node [left] at (4,2) {=};
		\end{tikzpicture}
		\begin{tikzpicture}[scale = 0.5]
			\draw[blue,knot,ultra thick] (2,0) -- (2,4);
			\draw[red,knot,ultra thick] (0,0) -- (3,4);
			\draw[teal,knot,ultra thick] (3,0) -- (0,4);
		\end{tikzpicture}
		\caption{}
	\end{subfigure}
	\caption[Reidemeister moves.]{Reidemeister moves: (a) The move I is undoing a twist in the strand, (b) the move II separates two unbraided strands, and (c) the move III slides a strand under a crossing.}
	\label{Reid}
\end{figure}
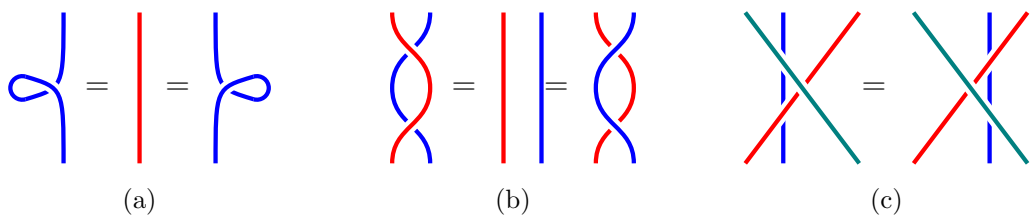

Two knot diagrams are equivalent if we can bend, stretch and smoothly deform one to the other without cutting. In knot theory, the equivalence of two knots is called \textit{ambient isotopy}. Other than stretching and bending, a simple way of showing that two knots are isotopic to each other is by a finite number of the \textit{Reidemeister moves} shown in Fig. \ref{Reid}. These moves are always permitted but not always sufficient to show the isotopy of two knots. When a knot is modified by applying these moves on a small portion of the diagram while keeping the rest of the diagram fixed, we get the formulas called \textit{Skein Relations}.

The \textit{knot invariants} are a set of rules that give the same output for two equivalent knots. That is, these invariants should not change under an ambient isotopy. Therefore, different knots are distinguished by their knot invariants. 
The knot invariants have their merits and limitations. The \textit{knot polynomials} are among several knot invariants assigned to knots and relatively easy to calculate. 
The \textit{Jones polynomial} \cite{jones1997polynomial} is of particular interest to us because of its connection to physics. This connection was first explored by Edward Witten in 1989 in his seminal paper \cite{witten1989quantum}. Physically, the trajectories of anyons in spacetime make knots. The knot invariants of trajectories are calculated by the path integral approach to the Chern-Simons theory. The present form of Jones polynomial is due to Kauffman who formulated it in a simpler way \cite{kauffman2001knots}. 

\subsection{Kauffman Bracket}
The Kauffman bracket is a polynomial invariant of unoriented link. The normalized version of Kauffman bracket yields the Jones polynomial when framing of a knot or a link is also considered. A Kaufman bracket $\langle L \rangle$ of a knot or a link $L$ assigns to each crossing a number that is either $A$ or $B$ as in diagrammatic Eq. \ref{SkeinKauff}. The values of variables $A$ and $B$ are to be computed in present section. This \textit{Kauffman skein relation} is used recursively until we get a resulting link diagram to have no crossings. Thus it consists of a finite set of unlinks or circles \cite{kauffman1990invariant,kauffman2001knots}. The Kauffman skein relation is written as
\eq{
	\begin{tikzpicture}[scale = 0.5]
		\draw [teal, ultra thick] (1,-1)--(0,0)--(1,1);
		\draw [teal, ultra thick] (2,-1)--(3,0)--(2,1);
		\draw[blue,knot,ultra thick] (1.9,-0.8) -- (1.1,0.8);
		\draw[red,knot,ultra thick] (1.1,-0.8) -- (1.9,0.8);
		\node [left] at (5,0) {$ = A $};
	\end{tikzpicture}
	\begin{tikzpicture}[scale = 0.5]
		\draw [teal, ultra thick] (1,-1)--(0,0)--(1,1);
		\draw [teal, ultra thick] (2,-1)--(3,0)--(2,1);
		\draw [red,ultra thick] (1.1,-0.8) to [curve through={(1.3,0)}](1.1,0.8);
		\draw [blue,ultra thick] (1.9,-0.8) to [curve through={(1.7,0)}](1.9,0.8);
		\node [left] at (5,0) {$ + B $};
	\end{tikzpicture}
	\begin{tikzpicture}[scale = 0.5]
		\draw [teal, ultra thick] (1,-1)--(0,0)--(1,1);
		\draw [teal, ultra thick] (2,-1)--(3,0)--(2,1);
		\draw [red,ultra thick] (0.8,0.5) to [curve through={(1.4,0.2)}](2.2,0.5);
		\draw [blue,ultra thick] (0.8,-0.5) to [curve through={(1.4,-0.2)}](2.2,-0.5);
	\end{tikzpicture}.
	\label{SkeinKauff}}
Here the variables $A$ and $B$ are assigned according to the convention such that when the first strand goes over the second while going in upward direction, we call it a positive crossing or an overcrossing,. In contrast, when the first strand goes below the second while going in upward direction, we call it a negative crossing or an undercrossing. The variables $A$ and $B$ on the right hand side will exchange their places comparing with the case of overcrossing. Now we will look for the invariance of Kauffman bracket under the Reidemeister moves. The Kauffman bracket is not an invariant under the Reidemeister move I. We will see later that it is an invariant when one considers a strand not as a string but as a \textit{ribbon or frame}. First, we will discuss the invariance of Kauffman bracket under the moves II and III. For the move II, we can write an equation as
\eq{
	\begin{tikzpicture}[scale = 0.5]
		\draw [teal, ultra thick] (1,-1)--(0,0)--(1,1);
		\draw [teal, ultra thick] (2,-1)--(3,0)--(2,1);
		\draw [red,knot,ultra thick] (1.1,-0.8) to [curve through={(1.7,0)}](1.1,0.8);
		\draw [blue,knot,ultra thick] (1.9,-0.8) to [curve through={(1.3,0)}](1.9,0.8);
	\end{tikzpicture} \nonumber
	&\begin{tikzpicture}[scale = 0.5]
		\node [left] at (0,0) {$ = A $};
		\draw [teal, ultra thick] (1,-1)--(0,0)--(1,1);
		\draw [teal, ultra thick] (2,-1)--(3,0)--(2,1);
		\draw [red,knot,ultra thick] (1.1,-0.8) to [curve through={(1.7,0)}](1.9,0.8);
		\draw [blue,knot,ultra thick] (1.9,-0.8) to [curve through={(1.3,0)}](1.1,0.8);
	\end{tikzpicture}
	\begin{tikzpicture}[scale = 0.5]
		\node [left] at (0,0) {$ + B $};
		\draw [teal, ultra thick] (1,-1)--(0,0)--(1,1);
		\draw [teal, ultra thick] (2,-1)--(3,0)--(2,1);
		\draw [red,knot,ultra thick] (0.9,0.7) to [curve through={(1.4,0.5)}](2.1,0.7);
		\draw [blue,knot,ultra thick] (1.1,-0.8) to [curve through={(1.3,-0.8)}](1.7,0);
		\draw [blue,knot,ultra thick] (1.9,-0.8) to [curve through={(1.7,-0.8)}](1.3,0);
		\draw [blue,ultra thick] (1.3,0) to [curve through={(1.5,0.1)}](1.7,0);
	\end{tikzpicture}\\
	&\begin{tikzpicture}[scale = 0.5]
		\node [left] at (0,0) {$ = A \Bigg[B$};
		\draw [teal, ultra thick] (1,-1)--(0,0)--(1,1);
		\draw [teal, ultra thick] (2,-1)--(3,0)--(2,1);
		\draw [red,ultra thick] (1.1,-0.8) to [curve through={(1.3,0)}](1.1,0.8);
		\draw [blue,ultra thick] (1.9,-0.8) to [curve through={(1.7,0)}](1.9,0.8);
	\end{tikzpicture}
	\begin{tikzpicture}[scale = 0.5]
		\node [left] at (0,0) {$ + A $};
		\draw [teal, ultra thick] (1,-1)--(0,0)--(1,1);
		\draw [teal, ultra thick] (2,-1)--(3,0)--(2,1);
		\draw [red,ultra thick] (0.8,0.5) to [curve through={(1.4,0.2)}](2.2,0.5);
		\draw [blue,ultra thick] (0.8,-0.5) to [curve through={(1.4,-0.2)}](2.2,-0.5);
		\node [left] at (4,0) {$\Bigg]$};
	\end{tikzpicture}
	\begin{tikzpicture}[scale = 0.5]
		\node [left] at (0,0) {$ + B \Bigg[A$};
		\draw [teal, ultra thick] (1,-1)--(0,0)--(1,1);
		\draw [teal, ultra thick] (2,-1)--(3,0)--(2,1);
		\draw [blue, ultra thick] (1.5,0) circle (0.25);
		\draw [red,ultra thick] (0.8,0.6) to [curve through={(1.4,0.4)}](2.2,0.6);
		\draw [blue,ultra thick] (0.8,-0.6) to [curve through={(1.4,-0.4)}](2.2,-0.6);
	\end{tikzpicture}
	\begin{tikzpicture}[scale = 0.5]
		\node [left] at (0,0) {$ + B $};
		\draw [teal, ultra thick] (1,-1)--(0,0)--(1,1);
		\draw [teal, ultra thick] (2,-1)--(3,0)--(2,1);
		\draw [red,ultra thick] (0.8,0.5) to [curve through={(1.4,0.2)}](2.2,0.5);
		\draw [blue,ultra thick] (0.8,-0.5) to [curve through={(1.4,-0.2)}](2.2,-0.5);
		\node [left] at (4,0) {$\Bigg]$};
	\end{tikzpicture}\nonumber\\
	&\begin{tikzpicture}[scale = 0.5]
		\node [left] at (0,0) {$ = AB$};
		\draw [teal, ultra thick] (1,-1)--(0,0)--(1,1);
		\draw [teal, ultra thick] (2,-1)--(3,0)--(2,1);
		\draw [red,ultra thick] (1.1,-0.8) to [curve through={(1.3,0)}](1.1,0.8);
		\draw [blue,ultra thick] (1.9,-0.8) to [curve through={(1.7,0)}](1.9,0.8);
	\end{tikzpicture}
	\begin{tikzpicture}[scale = 0.5]
		\node [left] at (0,0) {$+A^2 $};
		\draw [teal, ultra thick] (1,-1)--(0,0)--(1,1);
		\draw [teal, ultra thick] (2,-1)--(3,0)--(2,1);
		\draw [red,ultra thick] (0.8,0.5) to [curve through={(1.4,0.2)}](2.2,0.5);
		\draw [blue,ultra thick] (0.8,-0.5) to [curve through={(1.4,-0.2)}](2.2,-0.5);
	\end{tikzpicture}
	\begin{tikzpicture}[scale = 0.5]
		\node [left] at (0,0) {$+ BAd$};
		\draw [teal, ultra thick] (1,-1)--(0,0)--(1,1);
		\draw [teal, ultra thick] (2,-1)--(3,0)--(2,1);
		\draw [red,ultra thick] (0.8,0.5) to [curve through={(1.4,0.2)}](2.2,0.5);
		\draw [blue,ultra thick] (0.8,-0.5) to [curve through={(1.4,-0.2)}](2.2,-0.5);
	\end{tikzpicture}
	\begin{tikzpicture}[scale = 0.5]
		\node [left] at (0,0) {$ + B^2$};
		\draw [teal, ultra thick] (1,-1)--(0,0)--(1,1);
		\draw [teal, ultra thick] (2,-1)--(3,0)--(2,1);
		\draw [red,ultra thick] (0.8,0.5) to [curve through={(1.4,0.2)}](2.2,0.5);
		\draw [blue,ultra thick] (0.8,-0.5) to [curve through={(1.4,-0.2)}](2.2,-0.5);
	\end{tikzpicture}\nonumber\\
	&\begin{tikzpicture}[scale = 0.5]
		\node [left] at (0,0) {$ = AB$};
		\draw [teal, ultra thick] (1,-1)--(0,0)--(1,1);
		\draw [teal, ultra thick] (2,-1)--(3,0)--(2,1);
		\draw [red,ultra thick] (1.1,-0.8) to [curve through={(1.3,0)}](1.1,0.8);
		\draw [blue,ultra thick] (1.9,-0.8) to [curve through={(1.7,0)}](1.9,0.8);
	\end{tikzpicture}
	\begin{tikzpicture}[scale = 0.5]
		\node [left] at (0,0) {$ + (A^2 + B^2 + BAd)$};
		\draw [teal, ultra thick] (1,-1)--(0,0)--(1,1);
		\draw [teal, ultra thick] (2,-1)--(3,0)--(2,1);
		\draw [red,ultra thick] (0.8,0.5) to [curve through={(1.4,0.2)}](2.2,0.5);
		\draw [blue,ultra thick] (0.8,-0.5) to [curve through={(1.4,-0.2)}](2.2,-0.5);
	\end{tikzpicture},
	\label{MoveIIKauff}}
where the $d$ is a real number assigned to an unknot or a circle. For the Reidemeister move II, only the first term in Eq. \ref{MoveIIKauff} must survive. Therefore, we must have the following conditions
\eq{AB =1, \qquad A^2 + B(Ad+B) = 0. \label{KnotConditions}} 
From the first condition in Eq. \ref{KnotConditions}, we have $B= A^{-1}$. The second condition would become 
\eq{
	A^2 + B(Ad+B) &= A^2 + A^{-1}(Ad + A^{-1}) = A^2 +d + A^{-2}, \nonumber \\
	\Rightarrow d &= -(A^2 + A^{-2}).
	\label{LoopNumber}}
From this last relation we can get another relation which will be useful later
\eq{Ad+B = -A(A^2 + A^{-2}) + A^{-1} = -A^3. \label{TwistPhase1}}
The Kauffman bracket is invariant under the Reidemeister move III as shown in Eq. \ref{MoveIIIKauff},
\eq{
	\begin{tikzpicture}[scale = 0.5]
		\draw [teal, ultra thick] (1,-1)--(0,0)--(1,1);
		\draw [teal, ultra thick] (2,-1)--(3,0)--(2,1);
		\draw[blue,knot,ultra thick] (1.09,-0.8) -- (1.09,0.8);
		\draw[red,knot,ultra thick] (2.2,-0.6) -- (0.8,0.6);
		\draw[green,knot,ultra thick] (0.8,-0.6) -- (2.2,0.6);
	\end{tikzpicture}
	\begin{tikzpicture}[scale = 0.5]
		\node [left] at (0,0) {$ = A $};
		\draw [teal, ultra thick] (1,-1)--(0,0)--(1,1);
		\draw [teal, ultra thick] (2,-1)--(3,0)--(2,1);
		\draw[blue,knot,ultra thick] (1.15,-0.9) -- (1.15,0.9);
		\draw [red,knot,ultra thick] (0.9,-0.5) to [curve through={(1.5,0)}](0.9,0.5);
		\draw [green,knot,ultra thick] (2.2,-0.5) to [curve through={(1.7,0)}](2.2,0.5);
	\end{tikzpicture}
	\begin{tikzpicture}[scale = 0.5]
		\node [left] at (0,0) {$ + B $};
		\draw [teal, ultra thick] (1,-1)--(0,0)--(1,1);
		\draw [teal, ultra thick] (2,-1)--(3,0)--(2,1);
		\draw[blue,knot,ultra thick] (1.2,-0.8) -- (1.2,0.8);
		\draw [red,knot,ultra thick] (0.8,0.5) to [curve through={(1.4,0.3)}](2.2,0.5);
		\draw [green,knot,ultra thick] (0.8,-0.5) to [curve through={(1.4,-0.3)}](2.2,-0.5);
	\end{tikzpicture}
	\begin{tikzpicture}[scale = 0.5]
		\node [left] at (0,0) {$ = A $};
		\draw [teal, ultra thick] (1,-1)--(0,0)--(1,1);
		\draw [teal, ultra thick] (2,-1)--(3,0)--(2,1);
		\draw[blue,knot,ultra thick] (1.8,-0.9) -- (1.8,0.9);
		\draw [red,knot,ultra thick] (0.8,-0.5) to [curve through={(1.3,0)}](0.8,0.5);
		\draw [green,knot,ultra thick] (2.2,-0.5) to [curve through={(1.5,0)}](2.2,0.5);
	\end{tikzpicture}
	\begin{tikzpicture}[scale = 0.5]
		\node [left] at (0,0) {$ + B $};
		\draw [blue, ultra thick] (0,0) -- (1,1);
		\draw [teal, ultra thick] (1,-1)--(0,0)--(1,1);
		\draw [teal, ultra thick] (2,-1)--(3,0)--(2,1);
		\draw[blue,knot,ultra thick] (1.9,-0.8) -- (1.9,0.8);
		\draw [red,knot,ultra thick] (0.8,0.5) to [curve through={(1.4,0.3)}](2.2,0.5);
		\draw [green,knot,ultra thick] (0.8,-0.5) to [curve through={(1.4,-0.3)}](2.2,-0.5);
	\end{tikzpicture}
	\begin{tikzpicture}[scale = 0.5]
		\node [left] at (0,0) {$ = $};
		\draw [teal, ultra thick] (1,-1)--(0,0)--(1,1);
		\draw [teal, ultra thick] (2,-1)--(3,0)--(2,1);
		\draw[blue,knot,ultra thick] (1.9,-0.8) -- (1.9,0.8);
		\draw[red,knot,ultra thick] (2.2,-0.6) -- (0.8,0.6);
		\draw[green,knot,ultra thick] (0.8,-0.6) -- (2.2,0.6);
	\end{tikzpicture}.\label{MoveIIIKauff}}
The invariance of Kauffman bracket under the Reidemesister move I is shown in Eq. \ref{MoveIKauff},
\eq{
	\begin{tikzpicture}[scale = 0.5]
		\draw [teal, ultra thick] (1,-1)--(0,0)--(1,1);
		\draw [teal, ultra thick] (2,-1)--(3,0)--(2,1);
		\draw [blue,knot,ultra thick] (1.1,0.8) .. controls (1.2,-0.5) and (2.3,-0.5) .. (2.3,0);
		\draw [blue,knot,ultra thick] (1.1,-0.8) .. controls (1.2,0.5) and (2.3,0.5) .. (2.3,0);
	\end{tikzpicture}
	&\begin{tikzpicture}[scale = 0.5]
		\node [left] at (0,0) {$ = A$};
		\draw [teal, ultra thick] (1,-1)--(0,0)--(1,1);
		\draw [teal, ultra thick] (2,-1)--(3,0)--(2,1);
		\draw [red, ultra thick] (2.1,0) circle (0.4);
		\draw [blue,knot,ultra thick] (1,0.8) to [curve through={(1.5,0)}](1,-0.8);
	\end{tikzpicture}
	\begin{tikzpicture}[scale = 0.5]
		\node [left] at (0,0) {$ + B$};
		\draw [teal, ultra thick] (1,-1)--(0,0)--(1,1);
		\draw [teal, ultra thick] (2,-1)--(3,0)--(2,1);
		\draw [blue,knot,ultra thick] (1.1,0.8) .. controls (1.2,-0.8) and (2.3,1) .. (2.3,0);
		\draw [blue,knot,ultra thick] (1.1,-0.8) .. controls (1.2,0.8) and (2.3,-1) .. (2.3,0);
	\end{tikzpicture}
	\begin{tikzpicture}[scale = 0.5]
		\node [left] at (0,0) {$ = Ad$};
		\draw [teal, ultra thick] (1,-1)--(0,0)--(1,1);
		\draw [teal, ultra thick] (2,-1)--(3,0)--(2,1);
		\draw[blue,knot,ultra thick] (1.5,-0.8) -- (1.5,0.8);
	\end{tikzpicture}
	\begin{tikzpicture}[scale = 0.5]
		\node [left] at (0,0) {$ + B$};
		\draw [teal, ultra thick] (1,-1)--(0,0)--(1,1);
		\draw [teal, ultra thick] (2,-1)--(3,0)--(2,1);
		\draw[blue,knot,ultra thick] (1.5,-0.8) -- (1.5,0.8);
	\end{tikzpicture}\nonumber \\
	&\begin{tikzpicture}[scale = 0.5]
		\node [left] at (0,0) {$ = (Ad+B)$};
		\draw [teal, ultra thick] (1,-1)--(0,0)--(1,1);
		\draw [teal, ultra thick] (2,-1)--(3,0)--(2,1);
		\draw[blue,knot,ultra thick] (1.5,-0.8) -- (1.5,0.8);
	\end{tikzpicture}
	\begin{tikzpicture}[scale = 0.5]
		\node [left] at (0,0) {$ = -A^{3}$};
		\draw [teal, ultra thick] (1,-1)--(0,0)--(1,1);
		\draw [teal, ultra thick] (2,-1)--(3,0)--(2,1);
		\draw[blue,knot,ultra thick] (1.5,-0.8) -- (1.5,0.8);
	\end{tikzpicture}.\label{MoveIKauff}}
In the last line of Eq. \ref{MoveIKauff}, Eq. \ref{TwistPhase1} is used. 
The Kauffman bracket polynomial is based on the skein relations which are written as 
\eq{&\langle L \rangle = A \langle L_A \rangle + A^{-1}\langle L_B \rangle\nonumber\\
	&\langle L \cup O \rangle = d \langle L \rangle = -(A^2+A^{-2}) \langle L \rangle \nonumber\\
	&\langle O \rangle = 1 \label{SkeinKauff2}}
where $L$ is a link and $O$ refers to an unknot or a trivial link. The first relation is the same as \ref{SkeinKauff}, the second relation tells that if a knot or a link is a union of a knot and an unknot then the resultant knot would be $d$ times that knot. The third relation implies that an isolated unknot is assigned a value of 1.

As we can see, the Kauffman bracket is not an invariant under the Reidemeister move I. When two knot diagrams are the same under only Reidemeister moves II, and III then the diagrams are called \textit{regular isotopic} but not ambient isotopic. The smoothing out of the twist gives a factor of $-A^3$. This factor will be compensated for by taking a twist in a framed or ribbon strand as in Fig. \ref{RibbonTwist}. A ribbon strand has some width rather than a line.
Hence a ribbon is related to a string with a factor $-A^{3}$ multiplied. In quantum theory, it is related to the phase accumulated by a particle with a spin when it does a $2\pi$ rotation.

\fig{[h!]
	\centering
	\begin{tikzpicture}[use Hobby shortcut,scale=2]
		\begin{knot}[
			consider self intersections=true,
			ignore endpoint intersections=false,
			flip crossing=2,
			only when rendering/.style={
			}
			]
			\strand [ultra thick,blue](0.34,0.9)..(0.5,0.8).. (0.7,0.9) .. (0.7,1.05) .. (0.33,1)..(0.2,0.6);
			\strand [ultra thick,blue](0.27,0.7).. (0.45,0.57).. (0.95,0.95) .. (0.5,1.37) .. (0.25,1.26)..(0,0.75);
			\strand [ultra thick,blue] (0,1) .. (0,1.8);
			\strand [ultra thick,blue] (0.2,1.3) .. (0.2,1.8);
			\strand [ultra thick,blue] (0,0) .. (0,0.8);
			\strand [ultra thick,blue] (0.2,0) .. (0.2,0.6);
			\draw [ultra thick,blue] (0,1.8) -- (0.2,1.8);
			\draw [ultra thick,blue] (0,0) -- (0.2,0);
		\end{knot}
		\path (0,0);
	\end{tikzpicture}
	\begin{tikzpicture}[scale=2]
		\node [] at (-0.2,1) {$ \rightarrow $};
		\draw[ultra thick,blue] (0,0) -- (0,0.2);
		\draw[ultra thick,blue] (0.2,0) -- (0.2,0.2);
		\draw[ultra thick,knot=blue] (0.2,0.2) -- (0,0.6);
		\draw[ultra thick,knot=blue] (0,0.2) -- (0.2,0.6);		
		\draw[ultra thick,blue] (0,0.6) -- (0,1.2);
		\draw[ultra thick,blue] (0.2,0.6) -- (0.2,1.2);
		\draw[ultra thick,knot=blue] (0.2,1.2) -- (0,1.6);
		\draw[ultra thick,knot=blue] (0,1.2) -- (0.2,1.6);		
		\draw[ultra thick,blue] (0,1.6) -- (0,1.8);
		\draw[ultra thick,blue] (0.2,1.6) -- (0.2,1.8);
		\draw[ultra thick,blue] (0,0) -- (0.2,0);
		\draw[ultra thick,blue] (0,1.8) -- (0.2,1.8);	
	\end{tikzpicture}
	\begin{tikzpicture}[scale=2]
		\node [] at (-0.4,1) {$ =-A^3 $};
		\draw[ultra thick,blue] (0,0) -- (0,1.8)--(0.2,1.8)--(0.2,0)--cycle;	
	\end{tikzpicture}
	\caption{Straightening a loop in a ribbon gives a twist factor.}
	\label{RibbonTwist}}
\subsection{Jones Polynomial}
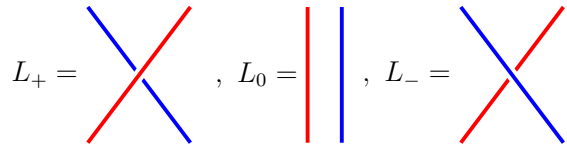
\begin{figure}[h!]
	\centering
	\scalebox{0.9}{
	\begin{tikzpicture}[scale = 0.5]
		\node [left] at (0,2) {$L_+ = $};
		\draw[blue,knot,ultra thick] (3,0) -- (0,4);
		\draw[red,knot,ultra thick] (0,0) -- (3,4);
	\end{tikzpicture}
	\begin{tikzpicture}[scale = 0.5]
		\node [left] at (0,2) {$, \ L_0 = $};
		\draw[blue,ultra thick] (1,0) -- (1,4);
		\draw[red,knot,ultra thick] (0,0) -- (0,4);
	\end{tikzpicture}
	\begin{tikzpicture}[scale = 0.5]
		\node [left] at (0,2) {$, \ L_- = $};
		\draw[red,knot,ultra thick] (0,0) -- (3,4);
		\draw[blue,knot,ultra thick] (3,0) -- (0,4);
	\end{tikzpicture}}
	\caption{Skein relations for Jones polynomial}
	\label{SkeinJones}
\end{figure}

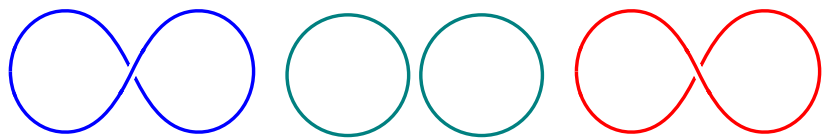
\begin{figure}[h!]	
	\centering
	\scalebox{0.8}{
	\begin{tikzpicture}[use Hobby shortcut]
		\begin{knot}[
			consider self intersections=true,
			ignore endpoint intersections=false,
			flip crossing=1,
			only when rendering/.style={}		]
			\strand [blue,ultra thick] ([closed]0,0) .. (1,-1) .. (2,0) .. (3,1) .. (4,0) .. (3,-1) .. (2,0)..(1,1)..(0,0);
		\end{knot}
		\path (0,0);
	\end{tikzpicture} \ \ \
	\begin{tikzpicture}
		\begin{knot}[flip crossing=2]
			\strand [teal, ultra thick](1,0) circle[radius=1];
			\strand[teal,ultra thick] (-1.2,0) circle[radius=1];
		\end{knot}
	\end{tikzpicture}
	\begin{tikzpicture}[use Hobby shortcut]
		\begin{knot}[consider self intersections=true,
			ignore endpoint intersections=false,
			flip crossing=1,
			only when rendering/.style={}]
			\strand [red,ultra thick] ([closed]0,0) .. (1,1) .. (2,0) .. (3,-1) .. (4,0) .. (3,1) .. (2,0)..(1,-1)..(0,0);
		\end{knot}
		\path (0,0);
	\end{tikzpicture}}
	\caption{Skein relations for unlinked circles.}
	\label{ThreeLinks}
\end{figure}

The Kauffman bracket is not an invariant under all the Reidemeister moves. We need to account for the twist factor or the self linking. This twist is also called a \textit{writhe}. Now we will also assign an orientation to the knot diagrams. If $w_+$ is an overcrossing and $w_-$ is an undercrossing for an oriented knot or link, then the writhe is given by $w(L) = w_+ - w_-$. We can construct a quantity as 
\eq{V_L(A) = (-A^3)^{-w (L)} \langle L \rangle.}
This is called \textit{Jones polynomial}. It is an invariant under all three Reidemeister moves. Two knots are equivalent if they have the same value of the Jones polynomial. A twist that contributes a factor $(-A^{-3})$ would get canceled with $(-A^3)^{w(L)}$. With the change of variable $t^{1/2} = A^{-2}$, the Jones polynomial agrees with the original form in Jones' paper \cite{jones1997polynomial}. The Jones polynomial is an invariant under orientation change.
The skein relations for Jones polynomial can be written as
\eq{-t^{-1} V(L_+) + (t^{1/2} - t^{1/2})V(L_0) + t V(L_-) = 0,}
where $L_0,L_-,L_+$ are shown in Fig. \ref{SkeinJones}.
The writhe is $+1$ for $L_+$, $-1$ for $L_-$, and $0$ for $L_0$. The Jones polynomial is an invariant for a knot's mirror image when $t$ is replaced by $t^{-1}$. The Jones polynomial for a Hopf link is $V(\text{Hopf})= -t^{5/2}-t^{1/2}$. The polynomial invariant for the knot in Fig. \ref{ThreeLinks} is given as follows. Since $L_+$ and $L_-$ are isotopic to an unknot, we have $V(L_+)= V(L_-)=1$, that is the unknot is assigned the value of the Jones polynomial as 1. Therefore,
\eq{-t^{-1} V(L_+) + &(t^{1/2} - t^{1/2})V(L_0) + t V(L_-) = 0\\
	&V(L_0)= \frac{t^{-1} -t}{t^{1/2} - t^{-1/2}}.}
The knot invariants are calculated in terms of topological quantum field theory (TQFT) and quantum group in Appendix \ref{HopfAlg}. The TQFT is discussed in Chapter \ref{TQFT} and the quantum group is defined in Appendix \ref{HopfAlg}. 
\section{Braid Group}
The trajectories of $N$ particles from their initial position at a time $t_i$ to the final position at a time $t_f$ are in one-to-one correspondence with the elements of the braid group $\cal{B}_N$. 
The time direction is taken vertically upward. The trajectories of particles are equivalence classes of all those trajectories which can be continuously deformed into each other. 
Assume that the particle number is fixed, which is the same as saying that there are no loops inside the braids. We will discuss in Chapter \ref{TQFT} that the loops correspond to the creation and annihilation of particles.

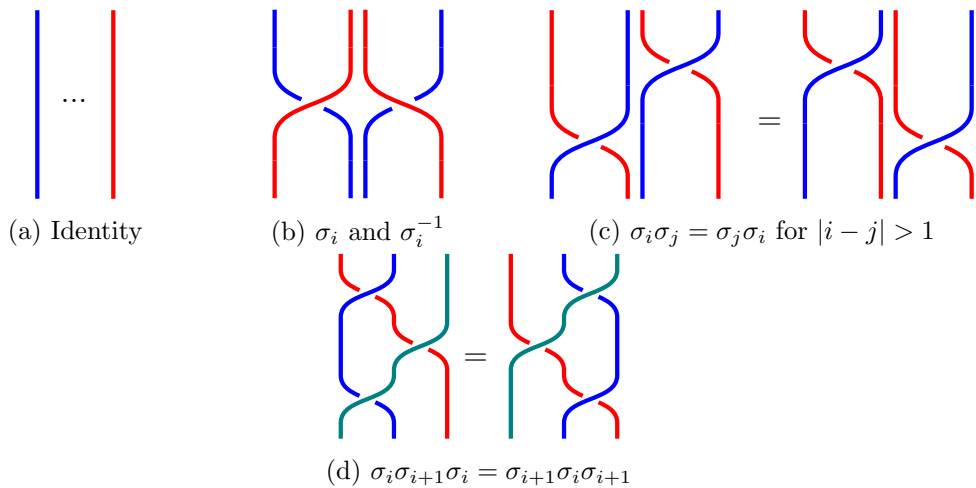
\begin{figure}[h!]
	\centering
	\begin{subfigure}{0.2\textwidth}
		\centering
		\begin{tikzpicture}[rotate=180]
			\draw[ultra thick,red] (1,0) -- (1,-2.5);
			\draw[ultra thick,blue] (2,0) -- (2,-2.5);
			\node [left] at (1.2,-1.3) {$...$};
		\end{tikzpicture}
		\caption{Identity}
	\end{subfigure}
	\begin{subfigure}{0.25\textwidth}
		\centering
		\begin{tikzpicture}
			\braid[ultra thick, style strands={1}{blue},style strands={2}{red},style strands={3}{green}]
			s_1^{-1};
			\draw[ultra thick,blue] (1,0) -- (1,0.5);
			\draw[ultra thick,red] (2,0) -- (2,0.5);
			\draw[ultra thick,red] (1,-1.5) -- (1,-2);
			\draw[ultra thick,blue] (2,-1.5) -- (2,-2);
		\end{tikzpicture}
		\begin{tikzpicture}
			\braid[ultra thick, style strands={1}{red},style strands={2}{blue},style strands={3}{green}]
			s_1;
			\draw[ultra thick,red] (1,0) -- (1,0.5);
			\draw[ultra thick,blue] (2,0) -- (2,0.5);
			\draw[ultra thick,blue] (1,-1.5) -- (1,-2);
			\draw[ultra thick,red] (2,-1.5) -- (2,-2);
		\end{tikzpicture}
		\caption{$\sigma_i$ and $\sigma_i^{-1}$}
	\end{subfigure}
	\begin{subfigure}{0.4\textwidth}
		\centering
		\begin{tikzpicture}
			\braid[ultra thick, style strands={1}{red},style strands={2}{blue},style strands={3}{green}]
			s_1^{-1};
			\draw[ultra thick,red] (1,0) -- (1,1);
			\draw[ultra thick,blue] (2,0) -- (2,1);
		\end{tikzpicture}
		\begin{tikzpicture}
			\braid[ultra thick, style strands={1}{red},style strands={2}{blue},style strands={3}{green}]
			s_1^{-1};
			\draw[ultra thick,blue] (1,-2.5) -- (1,-1.5);
			\draw[ultra thick,red] (2,-2.5) -- (2,-1.5);
			\node [left] at (3,-1.5) {$=$};
		\end{tikzpicture}
		\begin{tikzpicture}
			\braid[ultra thick, style strands={1}{red},style strands={2}{blue},style strands={3}{green}]
			s_1^{-1};
			\draw[ultra thick,blue] (1,-2.5) -- (1,-1.5);
			\draw[ultra thick,red] (2,-2.5) -- (2,-1.5);
		\end{tikzpicture}
		\begin{tikzpicture}
			\braid[ultra thick, style strands={1}{red},style strands={2}{blue},style strands={3}{green}]
			s_1^{-1};
			\draw[ultra thick,red] (1,0) -- (1,1);
			\draw[ultra thick,blue] (2,0) -- (2,1);
		\end{tikzpicture}
		\caption{$\sigma_i\sigma_j = 
			\sigma_j \sigma_i$ for $\abs{i-j}>1$}
	\end{subfigure}
	\begin{subfigure}{0.3\textwidth}
		\centering
		\begin{tikzpicture}[scale=0.7]
			\braid[ultra thick, style strands={1}{red},style strands={2}{blue},style strands={3}{teal}]
			s_1^{-1} s_2^{-} s_1^{-1};
			\node [left] at (4,-2) {$=$};
		\end{tikzpicture}
		\begin{tikzpicture}[scale=0.7]
			\braid[ultra thick, style strands={1}{red},style strands={2}{blue},style strands={3}{teal}]
			S_2^{-1} s_1^{-1} s_2^{-1}; 
		\end{tikzpicture}
		\caption{$\sigma_i\sigma_{i+1}\sigma_i=\sigma_{i+1}\sigma_i\sigma_{i+1}$}
	\end{subfigure}
	\caption{The braid group; generators and their properties.}
	\label{Braiding1}
\end{figure}

Let the braiding of the first and the second strand be represented by  $\sigma_1$ and braiding of the second and the third strands be represented by $\sigma_2$ and so on.
The braid group generators are shown in Fig. \ref{Braiding1}. (a) is the identity element of the braid group. It consists of all straight strands. As shown in (b), the clockwise exchange of strands $i$ and $i+1$ is represented by the generator $\sigma_i$, whereas counterclockwise exchange is represented by the inverse $\sigma_i^{-1}$. 
The group composition of two braids is given by stacking the strands on top of each other. For a non-Abelian group, the multiplication is noncommutative, and the order of stacking matters in this case. This is because of the degeneracy of the ground state, as discussed in Chapter \ref{GeoPhase}. The braid group generators also satisfy two conditions shown in Fig. \ref{Braiding1} (c) and (d).
\eq{
	\sigma_i\sigma_j &= \sigma_j\sigma_i \ \text{for} \ \abs{i-j} > 1,\\
	\sigma_i\sigma_{i+1}\sigma_i &= \sigma_{i+1}\sigma_i\sigma_{i+1}.} 
The second relation is the famous Yang-Baxter equation.

\subsection{Matrix Representation of the Braid Group}
The generators of a braid group above can be written in the matrix representation as follows. Suppose, $\sigma_i$ represents a positive crossing when two stands are going upward in such a way that the first strand is at the top and the second strand is at the bottom. The $\sigma_i^{-1}$ represents a negative crossing when the second strand is at the top and the first is at the bottom while going upward. See \cite{pourkia2018new} for further details.

\begin{figure}[h!]
	\centering
	\begin{subfigure}{0.35\textwidth}
		\centering
		\begin{tikzpicture}
			\braid[ultra thick, style strands={1}{blue},style strands={2}{red},style strands={3}{green}]
			s_1^{-1};
		\end{tikzpicture}\qquad
		\begin{tikzpicture}
			\braid[ultra thick, style strands={1}{red},style strands={2}{blue},style strands={3}{green}]
			s_1;
		\end{tikzpicture}
		\caption{}
	\end{subfigure}
	\begin{subfigure}{0.35\textwidth}
		\centering
		\begin{tikzpicture}
			\braid[ultra thick, style strands={1}{red},style strands={2}{blue},style strands={3}{green}]
			s_1s_1^{-1};
			\draw[ultra thick,red] (3,-2.5) -- (3,0);
			\draw[ultra thick,blue] (4,-2.5) -- (4,0);
			\node[] at (2.5,-1.5) {$=$};
		\end{tikzpicture}
		\caption{}
	\end{subfigure}
	\caption[Generators of the braid group and their composition.]{(a) The generator $\sigma_1$ and $\sigma^{-1}$ of the braid $\mathcal{B}_2$, (b) the composition of $\sigma$ and $\sigma^{-1}$.}
	\label{B2}
\end{figure}
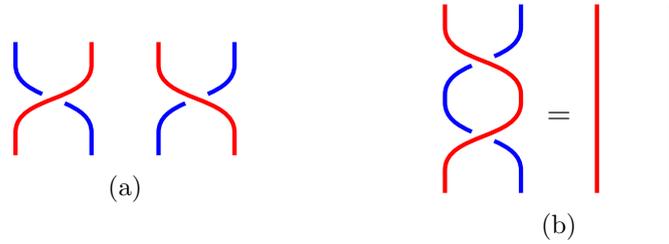

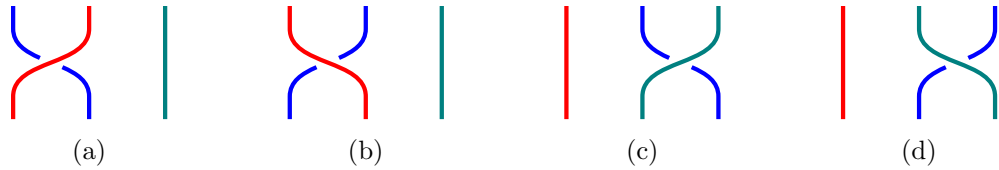
\begin{figure}[h!]
	\begin{subfigure}{0.22\textwidth}
		\centering	
		\begin{tikzpicture}
			\braid[ultra thick, style strands={1}{blue},style strands={2}{red},style strands={3}{teal}]
			s_1^{-1};
			\draw[ultra thick,teal] (3,-1.5) -- (3,0);
		\end{tikzpicture}
		\caption{}
	\end{subfigure}
	\begin{subfigure}{0.22\textwidth}
		\centering
		\begin{tikzpicture}
			\braid[ultra thick, style strands={1}{red},style strands={2}{blue},style strands={3}{teal}]
			s_1;
			\draw[ultra thick,teal] (3,-1.5) -- (3,0);
		\end{tikzpicture}
		\caption{}
	\end{subfigure}
	\begin{subfigure}{0.22\textwidth}
		\centering
		\begin{tikzpicture}
			\braid[ultra thick, style strands={1}{red},style strands={2}{blue},style strands={3}{teal}]
			s_2^{-1};	
		\end{tikzpicture}
		\caption{}
	\end{subfigure}
	\begin{subfigure}{0.22\textwidth}
		\centering
		\begin{tikzpicture}
			\braid[ultra thick, style strands={1}{red},style strands={2}{teal},style strands={3}{blue}]
			s_2;
		\end{tikzpicture}
		\caption{}
	\end{subfigure}
	\caption{The generators $\sigma_1,\sigma_1^{-1},\sigma_2,\sigma_2^{-1}$ in  $\mathcal{B}_3$}
	\label{B3}
\end{figure}

To find the braid matrices, first consider the case of a two-strand braid group $\mathcal{B}_2$. For positive crossing, that is $\sigma$, we assigned a value $\psi_{12}= t$ and $\psi_{21}= b$. Similarly, for negative crossing, that is $\sigma^{-1}$, we assign $\psi_{21}^{-1} = t^{-1}$ and $\psi_{12}^{-1} = b^{-1}$. When there is no strand connecting the upper and lower points, the corresponding values are zero. Therefore, $\psi_{11}= \psi_{22}^{-1} = 0$ and $\psi_{11}^{-1}= \psi_{22}^{-1} = 0$. We can write the $\sigma$ matrices as
\eq{\sigma_1 =\begin{pmatrix}
		0&t\\
		b&0
	\end{pmatrix}, \qquad
	\sigma_1^{-1} =\begin{pmatrix}
		0&b^{-1}\\
		t^{-1}&0
	\end{pmatrix}.}
For a braid in $\mathcal{B}_n$, we would have $n\cross n$ matrix representation. $\mathcal{B}_2$ and its inverse are shown in Fig. \ref{B2} (a).
To check whether these are the inverse of each other, compose $\sigma$ with $\sigma^{-1}$. Since we get $\sigma_1\sigma_1^{-1}= I_2$, the braid gives an identity which is also the Reidemeister move I as shown in Fig. \ref{B2} (b).
We can extend the procedure for writing the braid matrices to the case of $\mathcal{B}_3$. On similar lines, we can write the $\sigma$ and $\sigma^{-1}$ matrices for three strands shown in Fig. \ref{B3}.
\eq{
	&\sigma_1 =\begin{pmatrix}
		0&t&0\\
		b&0&0\\
		0&0&1
	\end{pmatrix}, \
	\sigma_2 =\begin{pmatrix}
		1&0&0\\
		0&0&t\\
		0&b&0
	\end{pmatrix}, \nonumber\\ 
	&\sigma_1^{-1} =\begin{pmatrix}
		0&b^{-1}&0\\
		t^{-1}&0&0\\
		0&0&1
	\end{pmatrix},\ \sigma_2^{-1} =\begin{pmatrix}
		1&0&0\\
		0&0&b^{-1}\\
		0&t^{-1}&0
	\end{pmatrix}.}
It is easy to check that 
$\sigma_1 \sigma_1^{-1} = \sigma_2\sigma_2^{-1} = I_3$. 
The Yang-Baxter equation from the Fig. \ref{Braiding1} is given by
\eq{\sigma_1 \sigma_2 \sigma_1 = \sigma_2\sigma_1 \sigma_2 = \begin{pmatrix}
		0&0&t^2\\
		0&bt&0\\
		b^2&0&0
	\end{pmatrix}.}

\section{Knot and Physics}
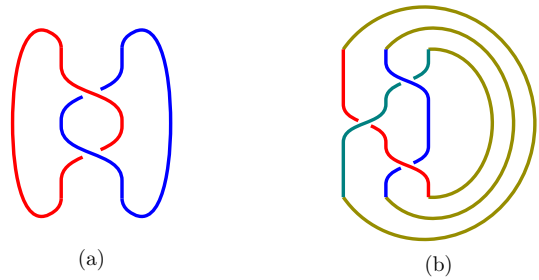
\begin{figure}[h!]
	\centering
	\scalebox{0.8}{
	\begin{subfigure}{0.35\textwidth}
		\centering
		\begin{tikzpicture}[use Hobby shortcut,scale=1]
			\begin{knot}[
				consider self intersections=true,
				ignore endpoint intersections=false,
				flip crossing=2,
				only when rendering/.style={}]
				\strand [ultra thick,red](1,-2.5)..(0.9,-2.7)..(0.2,-1.25)..(0.9,0.2)..(1,0);
				\strand [ultra thick,blue](2,-2.5)..(2.1,-2.7)..(2.8,-1.25)..(2.1,0.2)..(2,0);
				
			\end{knot}
			\braid[ultra thick, style strands={1}{red},style strands={2}{blue}]
			s_1 s_1 s;
		\end{tikzpicture}	
		\caption{}
	\end{subfigure}
	\begin{subfigure}{0.35\textwidth}
		\centering
		\begin{tikzpicture}[use Hobby shortcut,scale=0.7]
			\begin{knot}[
				consider self intersections=true,
				ignore endpoint intersections=false,
				flip crossing=2,
				only when rendering/.style={}]
				\strand [ultra thick,olive](3,0)..(3.2,0)..(4.5,-1.8).. (3.2,-3.5)..(3,-3.5);
				
				\strand [ultra thick,olive](2,0)..(3,0.5)..(5,-1.8).. (3,-4)..(2,-3.5);
				\strand [ultra thick,olive](1,0)..(3,1)..(5.5,-1.8).. (3,-4.5)..(1,-3.5);
			\end{knot}
			\braid[ultra thick, style strands={1}{red},style strands={2}{blue},style strands={3}{teal}]
			S_2^{1} s_1^{-1} s_2^{1};
		\end{tikzpicture}
		\caption{}
	\end{subfigure}}
	\caption{The trace closure and plat closure.}
	\label{Closure}
\end{figure}

A knot or a link can be obtained from a braid by the closure. There are two types of closures. The \textit{trace closure} is identifying the top ends of the strands to their corresponding bottom ends by keeping track of each strand. So that each colored strand at the top is connected to the same color at the bottom. Whereas the \textit{plat closure} is to identify each strand at the top to the adjacent end at the bottom without knowing the paths in the middle. That is, to identify the rightmost strand at the top to the rightmost at the bottom without bothering about its color. These two types of closure are shown in Fig. \ref{Closure} (a) and (b). Physically, the strands are paths of moving particles in a time direction that is taken upward. A closure corresponds to taking the trace of multi-particle input and output states. The braiding causes the change in the phase of the wave function. These phases are geometric phases discussed in Chapter \ref{GeoPhase}. If we get the same phase for the motion of an anyon clockwise and anticlockwise around the other anyon then the braid group is an Abelian or commutative, otherwise it is a non-Abelian group.

The amplitude of the process in quantum physics is related to the Kauffman bracket. For example, a Hopf link in Fig. \ref{Closure} (a) is corresponding to the creation of two pairs of particles and antiparticles in a $(2+1)$-dimensional spacetime, their braiding, and then fusion. The particles may or may not be fused to the vacuum. It will have some probability amplitude associated with the fusion. This probability amplitude corresponds to taking the trace of a braid.
The quantum amplitude for these processes does not depend on the geometry, speed of evolution, or the distance between particles, but depends only on the topology of the worldlines. The same quantum amplitude occurs when two paths have the same topology. That is, the quantum amplitude is a knot invariant, we will see that in Chapter \ref{TQFT}. Edward Witten made the connection between the knot theory and quantum physics \cite{witten1989quantum}. He won the field medal for this work with Vaughn Jones in 1990.
	\chapter{Geometric Phases in Quantum Physics}\label{GeoPhase}
	
Anyons are charge-flux composites that arise in a highly correlated system, such as the quantum Hall state. The interaction between these charges and fluxes is through braiding. Topological quantum gates are implemented by braiding anyons on a two-dimensional manifold. The quantum gates are the unitary operators that change a system from one state to the other, as explained in Chapter \ref{QC}. The state-space of a topological quantum gate is the ground state degeneracy of the system. This braiding and the ground state degeneracy can be understood through geometric phases \cite{bohm2013geometric}, such as Aharonov-Bohm phase \cite{aharonov1959significance} and Berry phase \cite{berry1984quantal}.

\section{A Charged Particle in a Magnetic Field}
Classically, the motion of a nonrelativistic charge particle in a magnetic field is described by the Lorentz equation given as
\eq{m\bm{\ddot{x}}_i = q \big(\bm{E} + (\bm{v}\cross \bm{B}\big)).}
The electric field $\bm{E}$ and the magnetic field $\bm{B}$ can be written in terms of vector potential $\textbf{A} = (A_x,A_y,A_z)$ and scalar potential $\phi(x)$ as
\eq{\bm{E}= - \frac{\partial}{\partial t}\bm{A} - \nabla \phi , \ \ \bm{B}= \nabla \cross \bm{A}.\label{EM}}
In quantum mechanics, the momentum $p$ is used instead of $v$ and its status is raised to an operator $\bm{p}= -i\hbar \nabla$. 
The Schr\"odinger's equation for a charged particle moving in an electromagnetic field can be written as
\eq{i\hbar \frac{\partial \psi(x)}{\partial t} = \Big[\frac{1}{2m}\big(\bm{p} - q \bm{A}\big)^2 + q \phi\Big]\psi(x).}
Let $\psi_0(x)$ be an eigenstate of the Hamiltonian when there is no vector potential. The wave function of the particle in the presence of vector potential is related to $\psi_0$ as 
\eq{\psi(x) = \exp({i\frac{q}{\hbar} \int \bm{A}\cdot d\bm{x}}) \psi_0(x). \label{WaveFuncA}}
The wave function in a magnetic field will get a phase $\phi=\frac{q}{\hbar} \int \bm{A}\cdot d\bm{x}$ other than the dynamical phase. A dynamical phase is the one that a wave function gets during the time evolution. The momentum operator $(\bm{p}-q\bm{A})$ appears as a combination of $\bm{p} = -i\hbar \nabla$ and the vector potential $\bm{A}$.

\subsection{Gauge Transformation}
Let us transform the vector potential as
\eq{\bm{A} \rightarrow \bm{A}^{'} = \bm{A} + \nabla \Lambda, \ \phi \to \phi' = \phi- \frac{\partial \Lambda}{\partial t}, \label{GaugeTransf}} 
where $\Lambda$ is a scalar function. The transformation in Eq. \ref{GaugeTransf} is called \textit{gauge transformation}. Now the Schr\"odinger equation will be written in terms of $\bm{A}',\psi'$, and $\phi'$. The wave function in Eq. \ref{WaveFuncA} can be written as
\eq{\psi(x)\to \psi'(x) &= \exp(i\frac{q}{\hbar}\int \bm{A}'\cdot d\bm{x})\psi_0(x)\nonumber\\
	&= \exp(i\frac{q}{\hbar}\Lambda(x))\psi(x).}
As a result of the gauge transformation, the wave function gets an additional phase of $\exp(i\frac{q}{\hbar} \Lambda(x))$. The fields $\bm{E}$ and $\bm{B}$ will remain invariant under this transformation. The physical quantities are modulus squared, so the complex phases do not appear.
This gauge transformation is \textit{local} as $\Lambda (x)$ is a function of $x$. The global gauge transformation is not as significant. It is independent of the position and is corresponding to the transformation of the whole system. 
The gauge transformation should be thought of as a phase transformation rather than some scale transformation as the name implies.

Quantum field theory of the electromagnetic field is the Abelian gauge field. Yang and Mills in 1954 pointed out that this phase function $\Lambda (x)$ can be a matrix instead of just a number. Since matrices are non-commutative in general, this phase transformation is non-Abelian. The Yang-Mills field is a non-Abelian gauge field. The Chern-Simons theory, which is the effective field theory for the topological phases, also used to compute the knot polynomial, is a type of non-Abelian gauge field in two dimensions. The gauge theory is discussed in Appendix \ref{QFT}.

\subsection{Aharonov-Bohm Effect}
In 1959, Yakir Aharonov and David Bohm \cite{aharonov1959significance} suggested that in quantum mechanics, the vector potential is not just a mathematical artifact, but it leads to detectable results. The effect of vector potential can be observed in a region where $\bm{B}= 0$ but $\bm{A} \ne 0$. They proposed an experiment shown in Fig. \ref{AB1}.
Suppose an infinitely long solenoid having a current through it produces a magnetic field along the z-axis. Since according to the right-hand rule the magnetic field is along the axis of the solenoid and zero outside, it can be taken as a tube of magnetic flux. The electron beam from the source $S$ is separated into two parts as shown in Fig. \ref{AB1}. The two parts of the beam combined at the screen make an interference pattern. 
The phase we get with the wave function of a charged particle in a magnetic field is given in the Eq. \ref{WaveFuncA}.
The phase acquired by the evolution of the wave function around a loop $C$ can be derived as
\eq{\phi = \frac{q}{\hbar} \oint\bm{A}\cdot d\bm{r} = \frac{q}{\hbar}\int_S \nabla \cross \bm{A} \cdot d\bm{s} = \frac{q}{\hbar} \int_S \bm{B}\cdot d\bm{s} = \frac{q}{\hbar}\Phi,}
where $d\bm{r}$ is a segment of the loop $C$ and $S$ is the surface enclosed by $C$, $d\bm{s}$ is the surface area element, and $\Phi$ is the total flux through $S$. This phase is gauge invariant, i.e. it is independent of the choice of $\bm{A}$ provided that it gives the same $\bm{B}$.
This phase is topological, as it does not depend on the shape of the path around the flux. Also, it remains invariant under the deformation of the surface that makes $\Phi$ fixed.
The two paths in Fig. \ref{AB1} are facing different relative vector potentials, hence interference fringes are modulated by the magnetic flux in the coil which is affected by the change of the electric current through the coil. Therefore, the choice of potentials instead of fields is not merely a convenience but a necessity. The electromagnetic field needs to be described in terms of an abstract four-dimensional vector $A_\mu = (\bm{A}, \phi)$.
\fig{[h!]
	\centering
	\tik{
		\draw[ultra thick,blue,->] (-3,-1) .. controls(0,2) and (1.5,1.5)..(3,1.1);
		\node [cylinder,draw=black,thick,aspect=1.5,minimum height=3cm,minimum width=1cm,shape border rotate=90,cylinder uses custom fill, cylinder body fill=green!30,cylinder end fill=red!5,opacity=0.8] at (0.5,0){};
		\draw[ultra thick,->,red] (0.5,-1.5)--(0.5,2);
		\draw[ultra thick,blue,->] (-3,-1) .. controls(-1,-1) and (1.5,-1.5)..(3,0.9);
		\draw [snake=coil, segment amplitude=12pt, segment length=4pt](0.5,-1.2)--(0.5,1.3);
		\draw[snake=coil,segment aspect=0,segment amplitude=15pt,ultra thick,red] (3.2,1) -- (4.8,1);
		\node [above] at (0.5,2){$\Phi$};
	}
	\caption[Aharonov-Bohm Effect]{Aharonov-Bohm Effect: switching on and off the flux in the flux tube, causes a shift in the interference fringes.}
	\label{AB1}}

\subsection{Anyon and Aharonov-Bohm Effect}
An anyon is a quasiparticle having fractional charge and fractional statistics. We can think of these particles as a composite of charge $q$ and flux $\Phi$. These composites arise in two-dimensional physical systems \cite{girvin1987quantum,wilczek1982quantum}. We will explain the formation of these composites in Chapters \ref{QHE} and \ref{TQFT}.
Let us exchange two anyons in a 2-dimensional space. The movement of anyons around each other in the $(2+1)$-dimensional space, is described by the braid group, see Chapter \ref{Knot}. The charge 1 going around the flux of 2 gets the Aharanov-Bohm phase $e^{iq \Phi}$. At the same time, the flux of 1 going around the charge of 2 and gets the phase $e^{iq \Phi}$, as in Fig. \ref{AnyonAB}. Therefore, the system gets a total phase $e^{2iq \Phi}$. This phase depends on the number of times one charge circulates the other, but it does not depend on the shape of the path, provided that the adiabaticity condition is satisfied. The \textit{adiabaticity condition} dictates that the charges must be moved slowly enough so that the system is not perturbed drastically from the ground state. The number of times a charge circulates another charge is called the \textit{winding number}. The phase will be written as $e^{imq\Phi}$ when the winding number is $m$. The statistical angle $\phi = q \Phi$ on an exchange corresponds to the phase shift of their wave function.
The $2\pi$ rotation of an anyon around itself gives a phase of $e^{iq \Phi}$ due to the charge ring around it. The spin-statistics theorem \cite{pauli1940connection} says that if $s$ is the effective spin of an anyon taken counterclockwise, we get a phase $e^{i2\pi s}$. Thus, we have a non-trivial spin $s= \frac{q \Phi}{2\pi}$ \cite{pachos2012introduction}. 
The \textit{topological spin} or a twist of an anyon is the rotation of the charge around its own flux. The phase due to the topological spin and the phase due to the braiding are related to each other. We will discuss that again later in Chapters \ref{TQFT}, \ref{Cat} and Appendix \ref{HopfAlg}.

\fig{[h!]
	\centering\tik{
		\draw[fill=red!10] (-0.5,1.3) -- (6.5,1.3) -- (5.2,-1.3) -- (-2.1,-1.3)-- cycle;
		\draw [ultra thick,fill=cyan] (0,0) ellipse (0.8 and 0.4);
		\draw [ultra thick,blue] (2.5,0) ellipse (2.6 and 1);
		\draw [ultra thick,fill=cyan] (4,0) ellipse (0.8 and 0.4);
		\node [cylinder,draw=black,thick,aspect=1.5,minimum height=2cm,minimum width=1cm,shape border rotate=90,cylinder uses custom fill, cylinder body fill=green!30,cylinder end fill=red!5] at (0,0.6){};
		\draw[ultra thick,->,red] (0,1)--(0,2);
		\draw[ultra thick,->,red] (0.2,1)--(0.2,2);
		\draw[ultra thick,->,red] (-0.2,1)--(-0.2,2);
		\node [cylinder,draw=black,thick,aspect=1.5,minimum height=2cm,minimum width=1cm,shape border rotate=90,cylinder uses custom fill, cylinder body fill=green!30,cylinder end fill=red!5,opacity=0.8] at (4,0.6){};
		\draw[ultra thick,->,red] (4,1)--(4,2);
		\draw[ultra thick,->,red] (3.8,1)--(3.8,2);
		\draw[ultra thick,->,red] (4.2,1)--(4.2,2);
		\draw[->,bend left,ultra thick] (4.5,2) to (4.8,1);
		\node [above] at (0,2){$\Phi$};
		\node [above] at (4,2){$\Phi$};
		\node [] at (-0.8,-0.5){$q_1$};
		\node [] at (3.2,-0.5){$q_2$};
		\node [] at (4.2,-1){$C$};
	}\caption{Anyons moving around each other acquire Aharanov-Bohm phase.}
	\label{AnyonAB}}
\section{Berry Phase}
The Aharanov-Bohm phase is a special case of the geometric phases when the system has time reversal symmetry \cite{berry2010geometric,cohen2019geometric}. It appears when the underlying geometry is changed by the magnetic field in an abstract way. A more general geometric phase, acquired by a wave function during the evolution in parametric space, is called the Berry phase. As an example, consider a spin-1/2 particle in a magnetic field that is oriented in a particular direction. By slowly varying the magnetic field orientation and bringing it back to the initial value, the system will come back to the initial state up to an overall phase with the wave function of the particle \cite{pachos2012introduction}.

Let us compute the geometric phase in quantum mechanics for a general situation described by two variables $\bm{r}$ and $\bm{R}(t)$. Let $\bm{r}$ describe a fast motion and $\bm{R}(t)$ be a variable that describes a slow motion. The slow variable describes a parameter that varies slowly with time and modulates the fast variable. For example, the motion of electrons of the atoms in a diatomic molecule is described by the fast variable $r$ and the vibratory motion of atoms is described by a slower variable $\bm{R}(t)$.
We suppose that the system returns to the original state after completing a loop in parametric space.
The adiabaticity condition should be satisfied, which means that the motion of the system in the parametric space should be slow enough so that the system does not go to the excited state.
The Schr\"odinger equation with state vector $\psi(t)$ can be written as
\eq{ i\hbar \frac{\partial }{\partial t} \ket{\psi (t)} = H(\bm{R(t)}) \ket{\psi (t)}.} 
Let $\ket{n, \bm{R}}$ be an eigenstate of the Hamiltonian that has energy eigenvalue as $E_n(\bm{R})$.
As the $\bm{R}(t)$ is slowly varying, at an instant of time $t$ we can take $\ket{n, \bm{R}(t)}$ as a basis vector, therefore we can write
\eq{H(\bm{R}(t)) \ket{n, \bm{R}(t)} = E_n(\bm{R}(t))\ket{n, \bm{R}(t)}.}
The solution of Schr\"odinger equation is given by
\eq{\ket{\psi(t)} = e^{i\gamma_n (t)} \exp \Big[-\frac{i}{\hbar} \int_{0}^{t}dt^{'} E_n(\bm{R}(t^{'}))\Big] \ket{n,\bm{R}(t)}. \label{BerrySol}}
The phase in brackets is the dynamical phase that depends on time, whereas $\gamma_n(t)$ is the Berry phase that depends on the geometry of parametric space. 
In 1984, Berry pointed out that $\gamma_n(t)$ has deep physical meaning and cannot be ignored \cite{berry1984quantal}. Consider a situation when the slow variable $\bm{R}(t)$ returns to the starting point $(\bm{R}(t)=\bm{R}(T))$ at a time $t=T$ after completing a turn in the parametric space in a closed path $C$. If we put the solution \ref{BerrySol} in the Schr\"odinger equation, take derivative, and cancel then exponentials on both sides, and then apply $\bra{n,\bm{R(t')}}$, we get
\eq{\gamma_n(C) = i \int_{0}^{t} dt'\bra{n, \bm{R}(t')} \frac{d}{dt'} \ket{n, \bm{R}(t')} = i\oint_{C} d\bm{R}\cdot\bra{n, \bm{R}} \nabla_R \ket{n, \bm{R}} \label{BerryPhase}.}
One of the examples of the Berry phase is the evolution of a system from one ground state to the other in topological materials. The degeneracy corresponds to the parametric space $\bm{R}(t)$. This idea is used in topological quantum computation in Chapter \ref{TQC}. In addition to the quantum computation, the Berry phase is also used to classify the topological materials, as discussed in Chapter \ref{TopMaterials}. 
The quantum analog of the geometric phase in polarization optics was discussed by Pancharatnam in 1956 \cite{pancharatnam1956generalized}. The non-adiabatic generalization of the Berry phase was proposed by Aharonov-Anandan \cite{aharonov1987phase}. 

\begin{figure}[h!]
	\centering
	\begin{subfigure}{0.4\textwidth}
		\centering
		\includegraphics[scale=0.7]{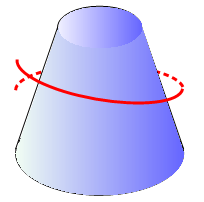}
		\caption{}
	\end{subfigure}
	\begin{subfigure}{0.4\textwidth}
		\centering
		\includegraphics[scale=0.2]{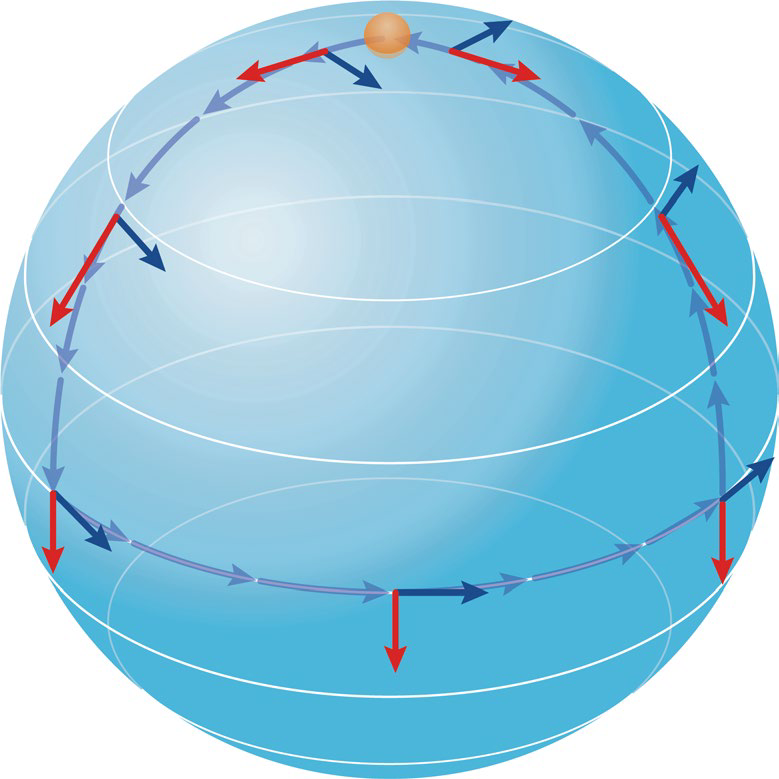}
		\caption{}
	\end{subfigure}
	\caption[Parallel transport and the holonomy on a fiber bundle.]{(a) Parallel transport on curved space. (b) Geometric phase is the holonomy on a fiber bundle. This diagram is taken from \cite{cohen2019geometric}.}
	\label{ParallelTrans}
\end{figure}

The topological aspect of the Berry phase is \textit{holonomy} on a fiber bundle. The holonomy is a failure of \textit{parallel transport} around some manifold \cite{baez1994gauge,bohm2013geometric} as shown in Fig. \ref{ParallelTrans} (a). A conic manifold is drawn because a cylinder is a trivial bundle, so the geometric phase may not appear.
A fiber bundle is a topological space that is a product space locally, but globally it may have a different structure than the product space. See Appendix \ref{Top} for the concept of the fiber bundle and parallel transport.

The \textit{parallel transport} is moving a vector on a manifold such that the direction of the vector keeps pointing in the same direction. In a plane or a flat space, the initial direction and the final direction would coincide. But in a curved space, the initial and the final directions are different in general. This can be visualized if we bring the vector back after making a complete loop and match it with the initial direction, as shown in Fig. \ref{ParallelTrans} (b). If we start from the North Pole, parallel transporting the blue and red arrows while keeping the direction pointing in one direction, when we come back to the North Pole, the final direction of the vector is different when comparing with the starting one.

Two vectors at two different positions at the manifold belong to different tangent spaces, so they are not related. To compare the two vectors, a \textit{covariant derivative} is taken. It involves an extra term with the flat space derivative. The failure of the covariant derivatives along two different paths gives the \textit{curvature}. In our case, we are not moving a vector in a physical space but in a space where vector potential and the magnetic field strength play the role of connection and curvature respectively. This is explained in terms of the gauge theory on fiber bundle in Appendix \ref{QFT}. 

\subsection{Anyons on a Torus}\label{AnyonTorus}
\begin{figure}[h!]
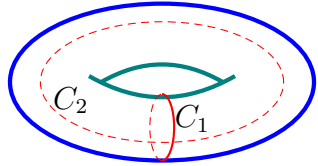

	\centering
	\tik{[scale=0.8]\draw[bend left,ultra thick,teal] (-1,-0) to (1,-0);
		\draw[bend right,ultra thick,teal] (-1.2,0.1) to (1.2,0.1);	
		\draw[rotate=0,ultra thick,blue] (0,0) ellipse (2.5 and 1.3);
		\draw[densely dashed,red] (0,0) ellipse (2 and 1);	
		\draw[densely dashed,red] (0,-1.3) arc (270:90:.2 and 0.550);
		\draw[thick,red] (0,-1.3) arc (-90:90:.2 and .550);
		\node [] at (0.5,-0.6) {$C_1$};
		\node [] at (-1.5,-0.3) {$C_2$};}
	\caption{Two non-trivial paths on a torus.}
	\label{PathsTorus}
\end{figure}

The system may have multiple types of anyons categorized according to their topological charge. As we discussed in \ref{Intro}, anyons are charge-flux composite found in topological materials. When two anyons are brought close to each other, their fusion may result in another anyon or a superposition of several anyons. These two anyons may annihilate to vacuum if they are antiparticles to each other. The ground state somehow knows what types of anyons can be created. Let there be two paths $C_1$ and $C_2$ on a torus along meridian and longitude, as shown in Fig. \ref{PathsTorus}.
Let $T_1$ and $T_2$ be operators correspond to the creation of anyon-antianyon pair from the vacuum and carrying around meridian and longitude respectively. 
$T_2^{-1}T_1^{-1}T_2T_1$ is two particles created, braided around each other and then re-annihilated. Since the operators $T_1$ and $T_2$ are implemented with some time-dependent Hamiltonian \cite{nayak2008non}, they are unitary. These two operators do not commute with each other, we have 
\eq{T_2T_1 = e^{-2i\theta} T_1T_2.}
Therefore, the system has ground state degeneracy. As $T_1$ is unitary, its eigenvalues must have a unit modulus, that is, they are just complex numbers. The operation of $T_1$ on a state $\alpha$ can be written as 
\eq{T_1\ket{\alpha} = e^{i\alpha}\ket{\alpha}.}
where $\alpha$ is the space of possible ground states. $T_2\ket{\alpha}$ must also be a ground state since $T_2$ commutes with $H$. Therefore, we can write
\eq{T_1(T_2 \ket{\alpha}) = e^{2i\theta}e^{i\alpha} (T_2 \ket{\alpha}).}
Let us call this new ground state $\ket{\alpha+2\theta} = T_2\ket{\alpha}$. 
On similar lines, we can generate more ground states. Consider a system where anyons have a statistical phase $\theta = \pi p/m$, where $p$ and $m$ are relatively prime so that $p/m$ is an irreducible fraction. The ground states can be written as
\eq{\ket{\alpha},\ket{\alpha+2\pi p/m}, \ket{\alpha+4\pi p/m},...,\ket{\alpha+2\pi (m-1)/m}.}
The phase $\alpha+2\pi = \alpha$ so that we are back to the original state. Now we have $m$ independent ground states. Since anyons have the statistical angle $\theta = \pi p/m$, the charge-flux composite will get $(q,\Phi) = (\pi p/m,1)$. When there is a fusion of $n$ elementary anyons then we have $\ket{n} = (q=n\pi p/m,\Phi =n) = (n\pi p/m,n)$. When there are $m$ anyons, we have $\ket{m} = (\pi p,m)$.
Now if we braid $\ket{n} = (n \pi p/m ,n)$ around one of these $\ket{m} = (\pi p/m)$, we obtain a net phase of $2\pi p$ which is equivalent to no phase at all. 
Hence, the cluster of $m$ elementary anyons is equivalent to a vacuum. In this way, we have $m$ species of anyon and $m$ different ground states on torus \cite{nayak2008non}. The subspace used to implement the topological gates depends on a particular model of anyons and also on the number of anyons present.

In the case of an annulus instead of a torus, the $T_1$ operator corresponds to a particle moving along a circular loop and $T_2$ to the particle moving from the inside edge to the outside edge. The degeneracy is $2$. On similar bases, the degeneracy for the higher genus space is $m^g$, where $g$ is \textit{genus}. The genus is a handle in a topological space. The integral along the path followed by a particle gives a phase accumulated by the particle. It is an example of the \textit{Wilson line} operator. We can also explain the degeneracy using punctures on a topological surface, as discussed in Chapter \ref{TQFT}. The above discussion is related to the gauge invariance and the coupling constant in Chern-Simons field theory \cite{nayak2008non} and the number of anyons in the fractional quantum Hall effect as discussed in Chapter \ref{QHE},.
The total charge and flux of a fusion of two particles must be zero: that is, we should get the vacuum. Therefore, the antianyon must have charge $-q$ and phase $\Phi$. The phase of an anyon moving clockwise around another anyon is the same as an antianyon moving clockwise around another antianyon. However, the phase of an anyon around an antianyon is $-2\phi$. The fusion of a particle with its antiparticle gives the vacuum, but when two particles are pushed together, we get a charge $2q$ and flux $2\Phi$. Now, the phase of exchanging these two particles is $\phi= 4q\Phi/\hbar$.
	\chapter{Quantum Hall Effect}\label{QHE}	
The physical systems for topological quantum computation are particles called non-Abelian anyons \cite{nayak2008non,stern2008anyons}. The non-Abelian anyons are found as quasiparticles in fractional quantum Hall effect \cite{girvin1987quantum}, and as Majorana fermions in topological superconductors \cite{kitaev2001unpaired}. The fractional quantum Hall effect and topological superconductors are examples of the non-Abelian states of matter that are explained by topology. In this chapter, we will discuss the quantum Hall effect. The topological aspect of the quantum Hall effect and Majorana fermions in topological superconductors will be discussed in the next Chapter \ref{TopMaterials}.

\section{Classical Hall Effect}
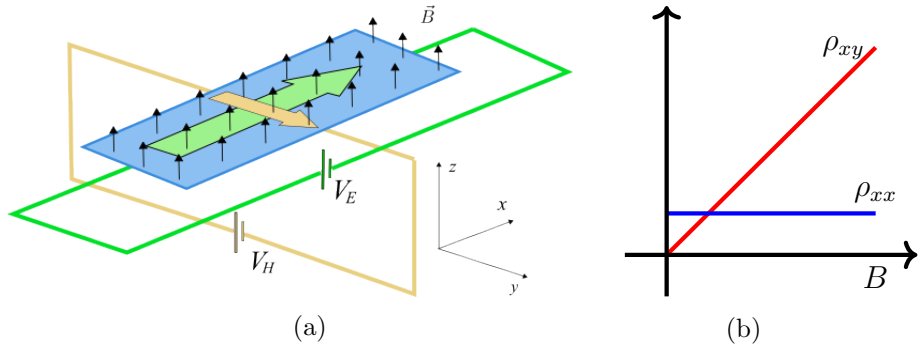
\begin{figure}[h!]
	\centering
	\begin{subfigure}{0.5\textwidth}
		\includegraphics[scale=0.5]{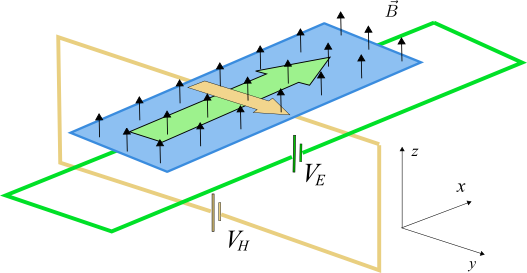}
		\caption{}
	\end{subfigure}
	\begin{subfigure}{0.2\textwidth}
		\begin{tikzpicture}[scale=1.1]
			\draw[ultra thick,red] (0,0)--(2.5,2.5);
			\draw[ultra thick,blue] (0,0.5)--(2.5,0.5);
			\draw[ultra thick,->] (-0.5,0)--(3,0);
			\draw[ultra thick,->] (0,-0.5)--(0,3);		
			\node [left] at (2.5,2.5) {$\rho_{xy}$};
			\node [above] at (2.5,0.5) {$\rho_{xx}$};
			\node [below] at (2.5,0) {$B$};
		\end{tikzpicture}
		\caption{}
	\end{subfigure}
	\caption[Classical Hall effect.]{(a) The experimental configuration for the Hall effect \cite{vaid2012quantum} (b) Resistivity as a function of magnetic field.}
	\label{Hall}
\end{figure}

The Hall effect was discovered in 1879 by Edwin Hall, about eighteen years before the discovery of the electron. Let a magnetic field $\bm{B}$ be applied along the $z$-direction to a metal conductor as shown in Fig. \ref{Hall} (a). The conventional current under the influence of the electric field is along the $x$-direction whereas the electrons move in the -$x$-direction. Under the influence of the magnetic field $\bm{B}$, electrons move in the $y$-direction due to the Lorentz force $\bm{F} = q(\bm{E} + \bm{v} \cross \bm{B})$. As a result, the positive and negative charges are accumulated on the opposite sides of the metal sheet. There is a net electric field between two edges and a potential difference called \textit{Hall voltage}. In a steady-state, the Hall voltage and the Hall conductance are given by
\eq{V_H= \frac{IB}{ned}, \qquad \rho = \frac{B}{en},}
where $I,e,n,d$ respectively are the current, electron charge, electron density, and the thickness of the material.
In classical Hall effect, the Hall voltage increases linearly with the magnetic field whereas the longitudinal voltage remains constant as shown in diagram \ref{Hall} (b). 

\begin{figure}
	\centering
	\includegraphics[scale=0.6]{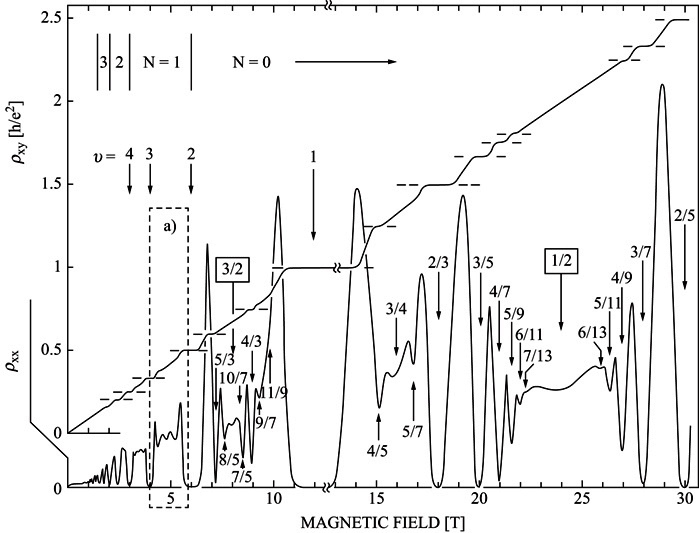}
	\caption[Resistivity as a function of magnetic field in the quantum Hall effect.]{Resistivity as a function of magnetic field in the quantum Hall effect \cite{eisenstein1990fractional}.}
	\label{QHEGraph}
\end{figure}

\section{Integer Quantum Hall Effect}
Almost a century later, the quantum Hall effect was discovered in 1980 by Klaus von Klitzing et al. \cite{klitzing1980new}. Electrons are bound to move on a two-dimensional surface of the 50 nm thin layer metal-oxide-semiconductor-field-effect transistor (MOSFET). After lowering the temperature to $\sim 4 K$, the transversal resistivity is measured in a very strong magnetic field $(\sim 10 T)$ along the $z$-direction \cite{stormer1999nobel}. 
The plateaus appear in the transverse Hall conductivity with simultaneous disappearance of the longitudinal conductivity with the magnetic field as shown in Fig. \ref{QHEGraph}. The Nobel Prize was awarded to Klaus von Klitzing in 1985 for the discovery of the quantum Hall effect.

\subsection{Two-Dimensional Electrons in a Magnetic Field}
\begin{figure}[h!]
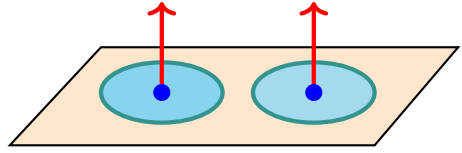

	\centering
	\tik{
		\draw[fill=orange!20,thick] (-0.3,1.1) -- (4.4,1.1) -- (3.3,-0.2) -- (-1.5,-0.2)-- cycle;
		\draw [ultra thick,opacity=0.8,teal,fill=cyan!50] (0.5,0.5) ellipse (0.8 and 0.4);
		\draw [ultra thick,opacity=0.8,teal,fill=cyan!40] (2.5,0.5) ellipse (0.8 and 0.4);	
		\draw[red,ultra thick,->] (0.5,0.5) to (0.5,1.7);
		\draw[red,ultra thick,->] (2.5,0.5) to (2.5,1.7);
		\filldraw[blue] (0.5,0.5) circle (3pt);
		\filldraw[blue] (2.5,0.5) circle (3pt);
	}
	\caption{Two-dimensional electrons move in circular paths around the magnetic flux tubes make the Landau levels.}
	\label{LandauLevels}
\end{figure}

The motion of an electron in a magnetic field was first studied by Landau \cite{landau1958quantum}. The quantum mechanical wave function for a two-dimensional electron in the presence of a magnetic field can be derived as follows \cite{cristofano1993topological,girvin1987quantum}. The spin gets polarized by the magnetic field, so we can consider a spinless electron in the $x,y$ plane. Let the magnetic field be applied perpendicular to the plane in the $z$-direction.
The wave function is periodic in the $y$-direction with a length $L_y$ and extended in the $x$-direction with a length $L_x$. We will explain the effect of impurities later, but for the moment, let us ignore the disorders. It will simplify the discussion without the loss of generality. The Hamiltonian for a charge particle in a magnetic field is
\eq{H= \frac{\big(\bm{p}-e\bm{A}\big)^2}{2m},}
where $\bm{A}$ is the vector potential. For the Schr\"odinger equation for an electron in a two-dimensional surface, we can choose the Landau gauge that is written as
$A_x = -B\hat{x}\hat{y}, \ A_y = 0, \ \bm{B}= \nabla \cross \bm{A} = B\hat{z}$,  so that we have 
\eq{\frac{1}{2m}\big(p_x^2 + (p_y+eBx)^2 \big)\psi = E\psi,}
where $p_x = -i\hbar\frac{\partial}{\partial x}$ and $p_y= -i\hbar\frac{\partial}{\partial y}$. 
Let us introduce the \textit{magnetic length} $l \equiv (\hbar /eB)^{1/2}$,
that is the minimum radius allowed by the uncertainty principle for an electron moving in a magnetic field. We also choose $\psi(x,y) = e^{ik_yx}\phi_{k_y}(x)$ because the Hamiltonian is translationally invariant in the $y$-direction. Therefore, we can write
\eq{\left\{ \frac{p_x^2}{2m}+ \frac{1}{2}m\omega_c^2(k_y l^2 + x)^2\right\}\phi_{k_y}(x) = E\phi_{k_y}(x),}
where $\omega_c = eB/m$ is the \textit{cyclotron frequency}. $\phi$ satisfies the harmonic oscillator equation with the potential shifted by $x= -k_yl^2$, that is a Gaussian wave packet located at $x= -k_yl^2$. The states are extended in the $x$ direction but confined in the $y$-direction. The eigenvalue of energy is given by 
\eq{E_n = (n +\frac{1}{2})\hbar \omega_c.}
The energy $E_n$ corresponds to the quantized eigenstates known as the \textit{Landau levels}. This situation is shown in Fig. \ref{LandauLevels}.

For the periodic wave function in the $y$-direction, we have $k_y = \frac{2\pi}{L_y}$. Since the $x$ ranges over $L_x$, so that $k_x$ ranges over $L_x/l^2$.
It implies that the number of states in the lowest Landau level is 
\eq{N_\phi =\frac{L_xL_y}{2\pi l^2}= \frac{ABe}{h}= \frac{AB}{\Phi_0},}
where $A$ is the area of the sample $A=L_xL_y$ and $\Phi_0 = h/e$ is the \textit{magnetic flux quantum}. The number of states in a Landau level is equal to the number of flux quanta incidents on the plane. So we have a massive degeneracy, as an electron can move in a circular path around magnetic flux at many places on the two-dimensional sheet. The number of filled Landau levels is given by  
$n= \nu eB/\hbar $ \cite{cristofano1993topological,nayak2008non}.
This $\nu$ is called \textit{filling fraction}. It is a ratio of the number of electrons to the number of flux quanta present in the system, that is, $N_e = \nu N_\phi$, where $N_\phi = \Phi/\Phi_0$.

In the case that the filled Landau levels are an integer number, then the chemical potential lies somewhere between filled levels and empty levels. The \textit{chemical potential} is the energy required to insert one more electron into the system. Now we need energy for the excitation of an electron from an occupied level to an empty level, we say that the system is \textit{incompressible} and the excitations are gapped. At this value of the integer filling fraction, the Hall resistivity and conductivity are quantized as
\eq{\sigma_{xy} = \nu\frac{e^2}{h}, \qquad \rho_{xy}= \frac{1}{\nu} \frac{h}{e^2} \label{resistivity},}
whereas the longitudinal resistance is zero. When there are disorders in the sample, the degeneracy is lifted, that is the degenerate levels split in energy, and we have bands. The value of $\nu$ in these experiments is highly accurate (up to one part in six million \cite{halperin1986quantized}) that makes the QHE one of the most accurate methods for the resistance measurement \cite{avron2003topological}.  

\subsection{Explanation of Integer Quantum Hall Effect}
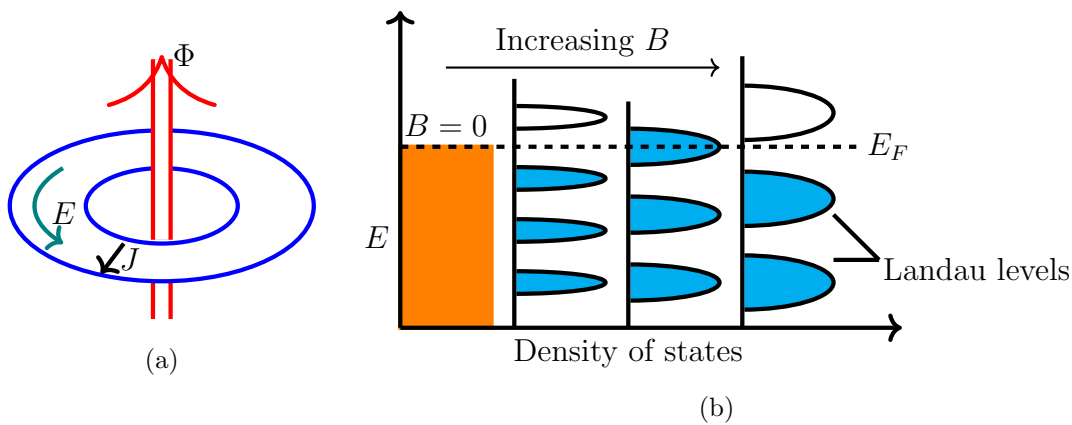
\begin{figure}[h!]
	\centering
	\begin{subfigure}{0.3\textwidth}
		\centering
		\begin{tikzpicture}
			\draw[ultra thick,red,double distance = 5] (0,-1.5) -- (0,-1);
			\draw [ultra thick,blue] (0,0) ellipse (1 and 0.5);
			\draw [ultra thick,blue] (0,0) ellipse (2 and 1);
			\draw[ultra thick,red,->,double distance = 5] (0,-0.45) -- (0,2);
			\draw[ultra thick,teal,->] (-1.3,0.5) .. controls(-1.8,0.2) and (-1.8,-0.2)..(-1.3,-0.5);
			\draw[ultra thick,->](-0.5,-0.5)--(-0.8,-0.9);
			\node [right] at (-1.6,-0.1) {$E$};
			\node [right] at (0,2) {$\Phi$};
			\node [right] at (-0.7,-0.7) {$J$};		
		\end{tikzpicture}
		\caption{}
	\end{subfigure}
	\begin{subfigure}{0.6\textwidth}
		\centering
		\begin{tikzpicture}[scale=0.3]
			\draw [ultra thick,fill=cyan] (15,2) ellipse (4 and 1.2);
			\draw [ultra thick,fill=cyan] (15,5.7) ellipse (4 and 1.2);
			\draw [ultra thick] (15,9.5) ellipse (4 and 1.2);
			\filldraw[ultra thick,->,white] (10,0) rectangle (15,15);	
			\draw [ultra thick,fill=cyan] (10,2) ellipse (4 and 0.8);
			\draw [ultra thick,fill=cyan] (10,5) ellipse (4 and 0.8);
			\draw [ultra thick,fill=cyan] (10,8) ellipse (4 and 0.8);
			\filldraw[ultra thick,->,white] (5,0) rectangle (10,15);	
			\draw [ultra thick,fill=cyan] (5,2) ellipse (4 and 0.5);
			\draw [ultra thick,fill=cyan] (5,4.3) ellipse (4 and 0.5);
			\draw [ultra thick,fill=cyan] (5,6.6) ellipse (4 and 0.5);
			\draw [ultra thick] (5,9.3) ellipse (4 and 0.5);	
			\filldraw[ultra thick,->,white] (0,0) rectangle (5,15);
			\filldraw[ultra thick,->,orange] (0,0) rectangle (4,8);
			\draw[ultra thick,->] (0,0) --(22,0);
			\draw[ultra thick,->] (0,0) --(0,14);
			\draw[ultra thick] (15,0) --(15,12);
			\draw[ultra thick] (10,0) --(10,10);
			\draw[ultra thick] (5,0) --(5,11);
			\draw[ultra thick,dashed] (0,8) --(20,8);	
			\node [left] at (0,4) {$E$};
			\node [above] at (2,8) {$B=0$};
			\node [above] at (8,11.5) {Increasing $B$};
			\node [below] at (10,0) {Density of states};
			\node [right] at (20,8) {$E_F$};
			\node [right] at (20.7,2.6) {Landau levels};
			\draw[thick,->] (2,11.5) --(14,11.5);
			\draw[ultra thick] (19,5) --(21,3);
			\draw[ultra thick] (19,3) --(21,3);
		\end{tikzpicture}
		\caption{}
	\end{subfigure}	
	\caption[Laughlin's argument]{(a) Laughlin's argument: the adiabatic increase of the flux creates electric field $\bm{E}$ and causes current flow $J$ in the radial direction. (b) Oscillations of density of states at the Fermi level.}
	\label{LaughlinArgu}
\end{figure}

To explain the quantum Hall effect, Laughlin proposed a thought experiment \cite{laughlin1981quantized}. The Laughlin's argument was later refined by Halperin \cite{halperin1982quantized}.
Consider the system where a disorder is only in the bulk (material other than the boundary) of the annulus, but free of disorder in the outside regions. The eigenstate in the lowest Landau level is characterized by angular momentum $m$
\eq{\phi_m \sim z^m e^{-\abs{z}/4l^2}}
where $z=x+iy$ is the representation of the position in complex coordinates. The wave function is localized in the region of disorders (impurities). When the flux is changed, the wave function moves outwards in a circle. When the flux is changed by an integral number of quantum flux $\phi_0$, angular momentum is changed by an integer from $m$ to the next eigenstate with $m+1$. While the wave function goes outward radially, it can get pushed into the disordered region. But one electron will be taken out from the disordered region, so the flux only affects the extended states wrapped around the annulus, as in Fig. \ref{LaughlinArgu} (a). The presence of disorder only affects the number of extended states in the way that there are fewer states than could be without the disorder. By the gauge invariance, the system is back to itself up to a geometric phase \cite{avron2003topological}. The gauge transformation is hinted at the topological nature of the quantum Hall effect \cite{avron2003topological}. The deformation of the sample does not change the Hall conductivity, so it is a topological invariant \cite{girvin1987quantum}. This is discussed in the next Chapter \ref{TopMaterials}. 

Halperin further explained the integer quantum Hall effect using the semiclassical approach and the band diagram \cite{halperin1986quantized}.
In the Landau levels, there are many independent states with the same energy, and each of these states has a corresponding wave function. The electrons move in the cyclotron motion at various places in the sample. The number of these states depends on the strength of the magnetic field. The stronger the magnetic field, the tighter the orbits are and more orbits can fit without overlapping. The quantum Hall effect is observed only in two dimensions, because the motion in the third dimension, in the direction of the magnetic field, can add any amount of energy to the energy of Landau levels, and hence gaps are filled up.

The impurities are responsible for the formation of plateaus and vanishing of the longitudinal resistance at the plateaus. In the presence of an impurity, independent states in a given Landau level are no longer equal in energy, that is, the degeneracy is lifted. They form bands. The electrons moving in circular orbits around disorders are in the localized states. Whereas the extended states are responsible for the current flowing through the sample.
When the magnetic field is increased, more states get pushed into the localized regions. Therefore, the longitudinal resistance keeps on increasing. When vacancies in the localized states become exhausted, then the longitudinal resistance becomes zero, until the magnetic field is at a value that the next Landau level starts filling. At this stage, the Hall resistance stays constant. With increasing the strength of the magnetic field, the \textit{Fermi energy level} keeps on moving from band to band, as shown in Fig. \ref{LaughlinArgu} (b). Electrons scatter to a new Landau level when the magnetic field reaches a value such that the Fermi level overlaps with the Landau level. The ratio of Hall resistances at two plateaus is a ratio of integers because for any given Hall voltage the current is proportional to the number of occupied sub-bands of the extended states. An integral number of sub-bands is filled at the plateau. 

\begin{figure}[h!]
	\centering
	\begin{subfigure}{0.35\textwidth}
		\centering
		\begin{tikzpicture}[scale=0.8]
			\draw[ultra thick,blue,fill=cyan!20] (-0.5,0) rectangle (3,4.8);
			\draw[ultra thick,red,->] (1,2.5) -- (1,2.6);
			\draw [ultra thick,red,->] (1.5,2.5) ellipse (0.5 and 0.5);
			\draw [ultra thick,red,->] (3,0) arc (270:90:0.5);
			\draw [ultra thick,red,->] (3,1.2) arc (270:90:0.5);
			\draw [ultra thick,red,->] (3,2.4) arc (270:90:0.5);
			\draw [ultra thick,red,->] (3,3.6) arc (270:90:0.5);			
			\draw [ultra thick,red,->] (-0.5,1.2) arc (90:-90:0.5);
			\draw [ultra thick,red,->] (-0.5,2.4) arc (90:-90:0.5);
			\draw [ultra thick,red,->] (-0.5,3.6) arc (90:-90:0.5);
			\draw [ultra thick,red,->] (-0.5,4.8) arc (90:-90:0.5);	
		\end{tikzpicture}
		\caption{}
	\end{subfigure}
	\begin{subfigure}{0.5\textwidth}
		\centering
		\begin{tikzpicture}[scale=0.8]
			\draw [ultra thick,blue] plot [smooth] coordinates {(1,5)(2,3)(4.5,2.7)(7,3)(8,5)};
			\draw [ultra thick,blue] plot [smooth] coordinates {(1,5)(2,2)(4.5,1.5)(7,2)(8,5)};
			\draw [ultra thick,blue] plot [smooth] coordinates {(1,5)(2,1)(4.5,0.3)(7,1)(8,5)};
			\draw[ultra thick] (0,0) --(0,5);
			\draw[ultra thick] (0,0) --(10,0);
			\draw[ultra thick,dashed,red] (0,2) --(10,2);
			\filldraw[red] (1.6,2) circle (3pt);
			\filldraw[red] (2,2) circle (3pt);	
			\filldraw[red] (7,2) circle (3pt);
			\filldraw[red] (7.4,2) circle (3pt);
			\node [right] at (10,2) {$E_F$};
			\node [right] at (10,0) {$x$};
			\node [left] at (0,3) {$E$};
			\draw[thick] (7.6,1.9) --(7.9,1.5);
			\node [] at (9.2,1.1) {Edge states};
			\node [above] at (4.5,0.3) {$\nu=0$};
			\node [] at (4.5,1.75) {$\nu=1$};
			\node [above] at (4.5,2.7) {$\nu=2$};
		\end{tikzpicture}
		\caption{}
	\end{subfigure}	
	\caption[The edge modes on quantum Hall bar.]{Edge modes: (a) the material remains an insulator in the bulk but there is a flow of current on the edges, (b) the Fermi level in the bulk is lower than that on the edges.}
	\label{EdgeModes}
\end{figure}
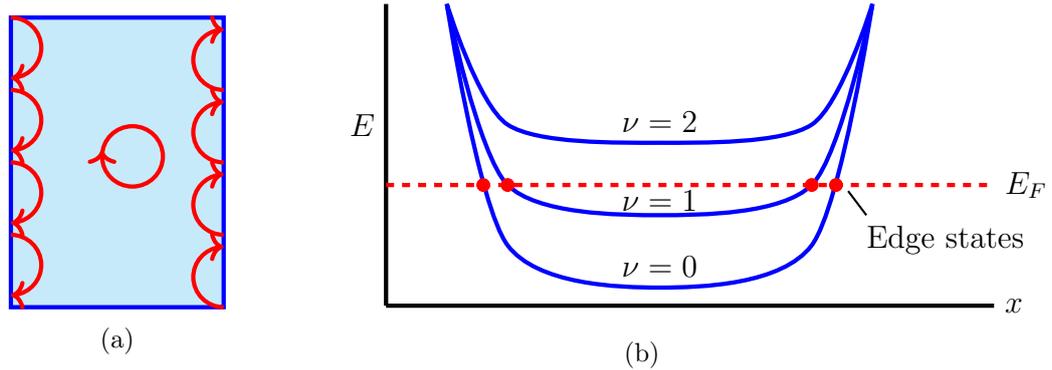

As electrons try to move in a circular path in the magnetic field and the applied electric field pushes them towards the right, they follow the cycloid motion. The drift velocity towards the right is directly proportional to the strength of the electric field and inversely proportional to the strength of the magnetic field. As the magnetic field is increased, more electrons deflect towards edges, hence the Hall voltage must be increased. There is a movement of charges along the edges, but the bulk remains an insulator. The bulk of the quantum Hall system is gapped, but the energy gap is reduced at the edges, hence there are excitations at the edges as shown in Fig. \ref{EdgeModes} (a) and (b). The edge states are topologically protected, which means when we deform the sample, the resistance remains the same. This happens unless the topology of the sample remains the same. The current at the edges of the QHE has chiral symmetry, that is, the edge modes are moving in one direction on one side and the other on the other side. The chiral symmetry is broken in topological insulators. The quantum Hall states are called the \textit{topologically ordered} phases, whereas the topological insulators are the \textit{symmetry protected topological phases} \cite{wen1995topological,wen2017colloquium}. For a discussion on how topology can explain the topological materials, see Chapter \ref{TopMaterials}. 

\section{Fractional Quantum Hall Effect}
The integer quantum Hall effect is observed in an impure sample, and no interaction among electrons is considered.
In 1982 Dan Tsui and Horst Stormer \cite{tsui1982two} discovered plateaus in quantum Hall conductivity at the filling fraction $\nu= p/q$. Hence, the Hall resistance is 
\eq{R_H = \frac{h}{e^2}\frac{p}{q},} 
where $p$ and $q$ are integers.
Tsui et al. used a more pure interface of GaAs-GaAlAs heterojunction with a stronger magnetic field and a lower temperature. They observed that the plateaus in $\rho_{xy}$ appear at fractional filling factor at $\nu = \frac{1}{3}, \frac{2}{3},$ etc. Later on plateaus at filling factors $\frac{4}{3}, \frac{5}{3}, \frac{2}{5}, \frac{4}{5}, \frac{2}{7}$, etc were also observed. The fractional quantum Hall effect is a counterintuitive phenomenon, creating particles with a charge smaller than the charge of an individual electron without splitting the electron. For the FQHE, the mutual Coulomb interactions are taken into account, which would make the gapless excitations of the ground state impossible. Electrons are moving around the magnetic flux and also around each other and at the same time avoiding each other due to the Coulomb repulsion \cite{wen2016introduction}. Therefore, we have to consider many-body wave functions for $N$ electrons with a strong correlation among them and obeying the Fermi statistics. The theory of fractional quantum Hall effect was proposed by Bob Laughlin. Dan Tsui, Horst Stormer, and Bob Laughlin were awarded the Nobel Prize for FQHE in 1998.

\subsection{Quasiparticles}
The fractional filling factor corresponds to the partially filled Landau levels. The quasiparticle excitations above the ground state are possible. A bump with a positive charge due to the perturbation in the uniform density is a quasihole. These are the flux-charge composites. Attachment of flux to the charges is also explained by Chern-Simons theory in Chapter \ref{TQFT}. 

The formation of the quasiparticle is explained as follows \cite{stormer1999nobel}.
Classically, 2D electrons are like charged billiard balls on a table, but quantum mechanically, they are smeared out on the two-dimensional plane. There is a uniform probability of finding electrons here and there in 2D and the material behaves like a featureless liquid. Their motion among electrons is still correlated. The perpendicular magnetic field creates tiny whirlpools in the liquid called \textit{vortices}. These are not elementary particles but quasiparticles. The vortices are formed by the flux quantum attached to the charge in such a way that the charge is displaced away from the center. Each electron is at the center of the vortex and part of the pool at the same time. The correlation of their mutual positions is energetically beneficial. It is also to satisfy the Pauli exclusion principle. There is one flux quantum per electron for IQHE. The vortices can be characterized by flux quanta attached. At a stronger magnetic field, there are more fluxes per electron and push more electrons away to reduce the Coulomb energy. The quasiparticles cause current while moving through the material.

Another way to imagine the vortices is as new entities called \textit{composite particles}. In this picture, the flux is considered as a shield to the other particles leads to the removal of mutual interaction from the problem \cite{stormer1999nobel}. This changes the character from fermion to boson and back. In this way, they can condense to the liquid \cite{marinescu2011classical}.
When a composite particle encircling the other composite particle, the wave function is multiplied by $-1$ and an odd number of times. An extra twist is created due to flux. Therefore, an odd number of flux and electron make a \textit{composite boson} and an even number of fluxes make a composite fermion.
Three fluxed are attached per electron for $1/3$ fractional Hall state. It is a composite boson. They move in apparently zero magnetic fields and condensed into a new ground state \cite{stormer1999nobel}.

\subsection{Laughlin's Wave Function}
\begin{figure}[h!]
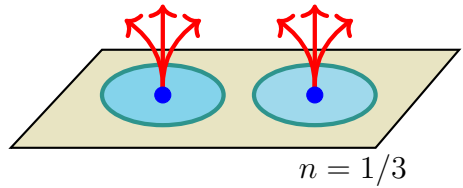

	\centering
	\tik{
		\draw[fill=olive!20,thick] (-0.3,1.1) -- (4.4,1.1) -- (3.3,-0.2) -- (-1.5,-0.2)-- cycle;
		\draw [ultra thick,opacity=0.8,teal,fill=cyan!50] (0.5,0.5) ellipse (0.8 and 0.4);
		\draw [ultra thick,opacity=0.8,teal,fill=cyan!40] (2.5,0.5) ellipse (0.8 and 0.4);
		\draw[red,bend right,ultra thick,->] (0.5,0.5) to (0,1.5);
		\draw[red,bend right,ultra thick,->] (2.5,0.5) to (2,1.5);
		\draw[red,bend left,ultra thick,->] (0.5,0.5) to (1,1.5);
		\draw[red,bend left,ultra thick,->] (2.5,0.5) to (3,1.5);
		\draw[red,ultra thick,->] (0.5,0.5) to (0.5,1.7);
		\draw[red,ultra thick,->] (2.5,0.5) to (2.5,1.7);
		\filldraw[blue] (0.5,0.5) circle (3pt);
		\filldraw[blue] (2.5,0.5) circle (3pt);
		\node [] at (3,-0.5) {$n=1/3$};
	}
	\caption[Quasiparticles as flux-charge composites.]{Quasiparticles as flux-charge composites. Three fluxes are attached to each charge.}
	\label{Quasiparticle}
\end{figure}

Laughlin \cite{laughlin1983anomalous} proposed a trial wave function for the filling fraction $1/3$, which was later generalized to the filling fraction with odd denominators.
Let two particles' unnormalized wave function in the lowest Landau level be written as
\eq{\Psi \sim (z_1+z_2)^M(z_1-z_2)^me^{-(\abs{z_1}^2 +\abs{z_2}^2)/4l_B^2},}
where $m$ is the relative momentum of particles and $M$ is the angular momentum of the center of mass. The many-particle wave funciton would look like
\eq{\Psi(z_1,...,z_n)= f(z_1,...,z_N)e^{-\sum_{i=1}^{N}\abs{z_i}^2/4l^2_B}.}
The wave function must be antisymmetric, as the particles are fermions.
The Laughlin's proposal for the form of the trial function is   
\eq{\Psi_m= \prod_{i<j}^{N}(z_i-z_j)^m \exp(-\sum_{j=1}^{N}z_j^* z_j).}
The most probable position of a particle is at $(z_,...z_N)$ with the probability given as $\abs{\Psi(z_1,...,z_N)}^2$. 
The number of flux quanta $N_\phi = mN_e$ corresponding to the filling fraction $\nu = N_e/N_\phi = 1/m$. The case of $m=3$ is shown in Fig. \ref{Quasiparticle}.
The wave function is symmetric corresponds to even $m$ whereas it is antisymmetric in case that $m$ is odd.
The wave function is the eigenstate of angular momentum. There are $N$ pairings, so the maximum exponent that a single coordinate can have is $Nm$. Each pairing has angular momentum $m$. The maximum exponent is equal to the flux quanta. The fractional charge is also explained by the gauge invariance. By the Laughlin argument in Fig. \ref{LaughlinArgu}, each insertion of flux $\phi_0$ pumps a charge $e^* = e/3$ from the inner boundary of the annulus to the outer. The Fermi level is located at a $1/3$ filled band.

The wave function describing a quasihole at a position $w$ is written as
\eq{\Psi_m=\prod_{k=1}^{N}(z_k-w)^m  \prod_{i<j}^{N}(z_i-z_j)^m \exp(-\sum_{j=1}^{N}z_j^* z_j).}
The quasihole is a lack of electron density at the position $w$. For $M$ quasiholes, we can write
\eq{\Psi_m=\prod_{l=1}^{M}\prod_{k=1}^{N}(z_k-w_l)^m  \prod_{i<j}^{N}(z_i-z_j)^m \exp(-\sum_{j=1}^{N}z_j^* z_j).}
Now $w$ is just a parameter, otherwise, if it is a dynamical variable then the wave function is just the original wave function with an extra electron at position $w$. A single quasihole has a charge of $+e/m$, hence $m$ holes act like a deficit of one electron. The quasiparticle excitations are emergent anyons. These anyons have a charge $e/m$ and a statistical angle $\theta = \pi/m$ for a filling fraction $\nu = 1/m$ \cite{arovas1984fractional,nayak2008non}. The calculations show that these anyons are Abelian \cite{wilczek1990fractional}. Even number denominator quantum Hall states cannot be described by the Laughlin wave function.

\subsection{Hierarchy Theory of Composite Fermion}
The Laughlin's wave function approach can explain quantum Hall states with the filling fraction $\nu = 1/m$ with $m$ odd integer. After the discovery of the Laughlin state with $\nu= 1/3$, there were many other states discovered with filling fractions $\nu= 2/3, 2/7, 3/5,...,$ etc. The hierarchical model was presented by Haldane \cite{haldane1983fractional} and Halperin \cite{halperin1984statistics} to account for the filling fraction $\nu= p/q$. They wrote the Laughlin's wave function in a slightly different form as
\eq{\phi(w_1,...,w_M) = \prod_{\alpha < \beta} (w_\alpha -w_\beta)^{1/m +p}, }
where $p$ is an even integer. The corresponding quasiparticles condense into the Laughlin's wave function with the filling fraction
\eq{\nu = \frac{1}{m \pm 1/p},} 
where $\pm$ is for whether quasiparticles condense or quasiholes. Any odd denominator can be obtained by repeating the procedure so that we can write
\eq{\nu = \frac{1}{m+ \frac{\alpha_1}{p_1+ \frac{\alpha_2}{... \frac{...}{+\frac{\alpha_n}{p_n}}}}},}
with $m= 1,3,5,...$, $\alpha_i = \pm 1$, and $p_i = 2,4,6,...$.
The state with filling fraction $\nu = [m, p_1,...,p_n]$ occurs only if its parent state $\nu = [m,p_1,...,p_{n-1}]$ also occurs. The Laughlin state $\nu = 1/3$ is a ground state and a parent state of quaisparticles that are excited and condensed into higher order fractions of a daughter state with filling factor $\nu = 2/5$. There can be further daughter states and this process is repeated.  

This approach \cite{jain1989composite} can also be used to explain the quantum Hall states with even denominator filling factors. 
The FQHE with the filling factor $\nu = 1/m$ under the magnetic field $B$ is equivalent to the IQHE of the composite fermions in a reduced magnetic field $B^* = 1/m B$. Assume a general filling fraction written as
\eq{\nu= \frac{n}{2pn \pm 1},}
where $p$ and $n$ are integers. Let $\rho$ be the electron number density, and we take the effective magnetic field as $B^* = B - 2p\rho \phi_0$, then the filling fraction of composite fermions is $\nu^* = \frac{\rho\phi_0}{B^*}$.
So $\nu$ and $\nu^*$ are related as
\eq{\nu= \frac{\nu^*}{2p \nu^* \pm 1}.}
Hence the condition for IQHE for composite fermions is $\nu^* = n$.

\section{Non-Abelian Quantum Hall States}
The discussion above can only explain the odd denominator filling fraction states, but the even denominator filling fraction states were also observed. The filling fraction with $\nu=5/2$ state was first discovered in 1987 \cite{willett1987observation}.
The state at a filling factor $1/2$ is not like others. There are two times as many vortices as charges. That makes them composite fermions (CFs), so do not condense to the lowest energy state. The behavior of the system is changed. They fill up the lowest-lying states, similar to electrons fill at $B=0$. The magnetic field is used up making the composite fermions, so these quasiparticles move effectively in a zero field. Their mass arises from the interaction instead of the property of the individual electron. There is no corresponding plateau, instead the Hall line is featureless. The odd denominator states are Bose-condensed, but $1/2$ state is a Fermi sea. The composite fermions generate their own Landau levels, that is IQHE of CFs \cite{stormer1999nobel}.

\subsection{Moore-Read State $\nu=5/2$}
This state can be thought of as two filled Landau levels for both spins up and spin down electron, that is $5/2=2+1/2$. Two filled levels are inert, therefore it should behave like $1/2$, but it does not. This state belongs to the first excited Landau level.

Hansson et al. \cite{hansson2009quantum} proposed a wave function for the Moore-Read Pfaffian state \cite{moore1991nonabelions}, also expressed the composite fermion Jain state \cite{jain1989composite}. See also 
series of papers on this \cite{hansson2007conformal, hansson2007composite, bergholtz2007microscopic, bergholtz2008quantum}. The wave function of the Moore-Read state with an even number of particles with the filling fraction $\nu = 1/m$
\eq{\Psi = Pf(\frac{1}{z_i-z_j})\prod_{i<j}(z_i -z_j)^m \prod_{i=1}^N e^{-\abs{z_i}^2/(4l^2)},}
where $Pf$ is a \textit{Pfaffian} which has properties that for any antisymmetric matrix $M_{ij}$, the determinant is zero when $N$ is odd but non-zero when $N$ is even. When $N$ is even, its determinant is written in terms of the object called Pfaffian,
\eq{\det(M) = Pf(M)^2.}
The partition of $N$ into $N/2$ pairs gives
\eq{Pf(M) = \mathcal{A}[M_{12}M_{34}...M_{N-1, N}],}
where $\mathcal{A}$ means antisymmetrise the argument. Therefore, we have
\eq{Pf(M) = \frac{1}{2^{N/2}(N/2)!} \Sigma_{\sigma} \text{sign}(\sigma) \prod_{k=1}^{N/2}M_{\sigma(2k-1),\sigma(2k)},}
where sign($\sigma$) is the signature of $\sigma$ and sum is over $\sigma \in S_N$ symmetric group.
For example, for four particles
\eq{Pf(\frac{1}{z_i -z_j}) = \frac{1}{z_1 -z_2}\frac{1}{z_3 -z_4} + \frac{1}{z_1 -z_3}\frac{1}{z_4 -z_2} + \frac{1}{z_1 -z_4}\frac{1}{z_2 -z_3}.}
The number of terms grows rapidly with $N$. This wave function is for a boson when $m$ is odd and fermion when $m$ is even. For $m=1$, the wave function does not vanish when a pair of particles coincide, since the zero of $(z_1-z_2)$ is compensated by the Pfaffian.
For $p=1$ we would have the smallest quasihole. Now the wave function becomes
\eq{\Psi_{qh}(w) = \big[\prod_{i=1}^{N}(z_i -w)\big]\Psi^m_{MR.}}
The resulting charge for this object is $e^* = e\nu = e/m$.
These quasiholes are non-Abelian anyons. The explicit calculation for four anyons was done by Nayak and Wilczek in \cite{nayak19962n}. They found that the dimension of the Hilbert space of $2n$ anyons is $2^{n-1}$, which implies there are $2^{n-1}$ possible ground states for $2n$ quasiholes. The Moore-Read state corresponds to the spin-polarized p-wave Cooper pairing in the BCS theory of superconductors for a fixed number of composite fermions \cite{read2000paired}. The quasiparticles in topological superconductors are Majorana fermions. We will discuss p-wave topological superconductors and Majorana fermion in the next Chapter \ref{TopMaterials}. If the quasiparticles are moved around each other, the state of the whole system changes in a way that depends only on the topology of the exchange. If we move one of the quasiholes around the others in a closed path, this quasihole would get a geometric phase. This is an example of the non-Abelian Berry phase \cite{bonderson2011plasma}. It occurs in the second Landau level in the presence of Coulomb repulsion. 

For the Laughlin's state, there is only a unique ground state and the quasiparticles are Abelian anyons. The non-Abelian anyons in the Moore-Read state can be fused to more than one outcome.
There are many ways to rearrange the electron in partially filled Landau levels, hence the system is degenerate.
Degeneracy of the ground state characterizes different phases of the quantum Hall system. It is a topological invariant and does not change under a small perturbation \cite{wen1990ground}. 

An extension of Moore-Read quantum Hall state is proposed by Read and Rezayi \cite{read1999beyond} at a filling fraction of $\nu=12/5$. Moore and Read observed \cite{moore1991nonabelions} that the conformal blocks of certain correlation functions in conformal field theories (CFT) can be directly interpreted as the wave functions of electrons in the lowest Landau level. For brief introduction to conformal field theory, See \cite{gaberdiel2006two,simmons2017conformal,gaberdiel2000introduction} and for somewhat detailed version of the theory, see \cite{schellekens1995conformal,moore1989classical,ginsparg1988applied,blumenhagen2009introduction,francesco2012conformal}. For an experimental setup to observe anyons and the evidence for their existence see \cite{stern2006proposed,rosenow2016current,nakamura2020direct,willett2013magnetic}. There are non-Abelian states other than the quantum Hall effect \cite{stern2010non}.

\section{Detection of Non-Abelian Anyons}\label{AnyonDetect}

\begin{figure}[h!]
	\centering
	\begin{tikzpicture}[scale=0.8]
		\draw[ultra thick, blue,->](-1.5,0)--(3,0)--(3,4.5);
		\draw[ultra thick, blue,->](0,-1.5)--(0,3)--(4.5,3);
		\filldraw[ultra thick,cyan,rotate=45](-0.5,-0.1) rectangle (0.5,0.1);
		\begin{scope}[shift={(3,3)}]
			\filldraw[ultra thick,cyan,rotate=45](-0.5,-0.1) rectangle (0.5,0.1);
		\end{scope}
		\begin{scope}[shift={(3.1,-0.1)}]
			\draw[ultra thick,cyan,rotate=45](-0.5,-0.1) rectangle (0.5,0.1);
		\end{scope}
		\begin{scope}[shift={(-0.1,3.1)}]
			\draw[ultra thick,cyan,rotate=45](-0.5,-0.1) rectangle (0.5,0.1);
		\end{scope}
		\begin{scope}[shift={(1.5,3)}]
			\filldraw[ultra thick,gray](-0.5,-0.2) rectangle (0.5,0.2);
		\end{scope}
		\node[] at (-0.7,-0.7) {$BS_1$};
		\node[] at (3.6,3.7) {$BS_2$};
		\node[] at (-0.5,3.5) {$M_2$};
		\node[] at (3.7,-0.3) {$M_1$};
		\node[above] at (1.5,3.1) {$\phi$};
	\end{tikzpicture}
	\caption{Mach-Zehnder interferometer.}
	\label{Mach}
\end{figure}
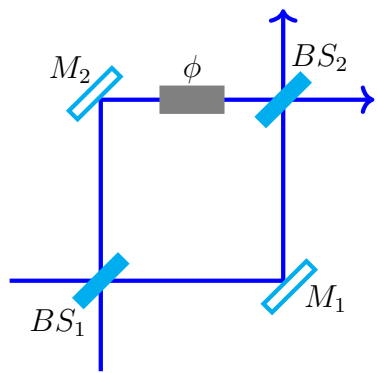
The non-Abelian anyons are detected by using the Mach-Zehnder interferometer shown in Fig. \ref{Mach}. Let a photon be incident at the beam splitter $BS_1$. There is an equal probability of reflection and transmission. $M_1$ and $M_2$ are reflectors. The two paths are combined at the second beam splitter $BS_2$. In case the two paths are of equal length, the photon transmitted at the first beam splitter will also be transmitted at the second beam splitter. If we want to control the probability of reflection and transmission at the second beam splitter $BS_2$, we need to change the length of one of the paths. That can be done by inserting a phase delay $\phi$ through inserting a glass plate. By tuning the phase delay $\phi$, the interference pattern at the outputs can be varied.

\begin{figure}[h!]
	\centering
	\includegraphics[scale=0.6]{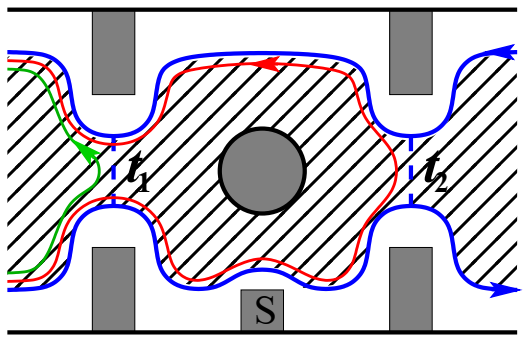}
	\caption[Detection of non-Abelian anyons on a quantum Hall bar.]{Detection of non-Abelian anyons on a quantum Hall bar. The diagram is taken from Ref. \cite{nayak2008non}.}
	\label{AnyonDetection1}
\end{figure}

Since the braiding changes the ground state and fusion channels, the Mach-Zehnder interferometer is used to detect the non-Abelian anyons on a quantum Hall bar \cite{feldman2021fractional,bonderson2007non,sarma2006topological,nayak2008non} shown in Fig. \ref{AnyonDetection1}. In our case, instead of the incident photon, a test anyon is introduced. The constrictions in gray color in the Fig. are electrodes that play the role of the beam splitters. The phase is inserted by the existence of the anyons. This phase is the Aharonov-Bohm phase. An inserted anyon will change the interference at two output states. The two paths of a test anyon are made to interfere. The longitudinal resistivity $\sigma_{xx}$ is measured. The relative phase $\phi$ can be changed either by changing the magnetic field or area through $S$. By braiding the test anyons, the final ground state is different when the anyon is non-Abelian. The tunneling probability amplitudes $t_1$ and $t_2$ are controlled by the voltages between the top and the bottom edges at two electrodes. The result will be different when there is a non-Abelian anyon present. The interference can be written as
\eq{G_L\propto \abs{t_1}^2+\abs{t_2}^2+2Re{(t_1^*t_2e^{i\phi})}.}
This is an example of the measurement based on interference. For projective measurement, the topological charge of an anyon is measured by fusion with another anyon.

\begin{figure}[h!]
	\centering
	\includegraphics[scale=0.6]{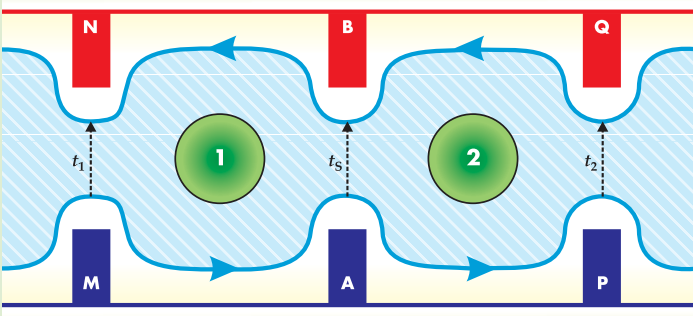}
	\caption[Physical implementation of the NOT gate on a quantum Hall bar.]{Physical implementation of the NOT gate on a quantum Hall bar. The diagram is taken from Ref. \cite{sarma2006topological}.}
	\label{AnyonDetection2}
\end{figure}

Now let there be two quasiholes at locations 1 and 2 as in Fig. \ref{AnyonDetection2}. The tunneling amplitudes $t_1$ and $t_2$ are changed by changing the voltages at $M,N$ and $P,Q$. The relative phase depends on the state of the qubit. The longitudinal resistance can be written as $\sigma_{xx}=\abs{t_1\pm it_2}^2$, where the plus sign is for state the $\ket{0}$ and minus for the state $\ket{1}$. The measurement will project the state on one of the eigenstates.
The tunneling between the edges is used to flip the state. As a braiding of non-Abelian anyons changes the ground state, the tunneling at constrictions $A$ and $B$ will transform the state from $\ket{0}$ to $\ket{1}$. The change of the conductivity is a sign of a non-Abelian quantum Hall state.	
	\chapter{Topological Materials}\label{TopMaterials}
	
One of the goals of condensed matter physics is to predict the properties of many-body systems. The physical prediction cannot be tested by solving the Schr\"odinger's equation for a large number of particles. The correlation among observable properties is obtained by using the effective field theories containing degrees of freedom at low-energy excitations.  
Another theme of condensed matter physics is to characterize the systems at the phase transition and classify the different states of matter. A phase of matter is the one in which many-body systems are assembled and arranged in some particular way, which gives the matter some distinct properties.
In this chapter, we will discuss how topology explains the topological states of matter, such as topological insulators and topological superconductors. The Majorana fermions in topological superconductors are non-Abelian anyons used to perform topological quantum computation.

\section{Landau Theory of Symmetry Breaking}

Two states of the many-body systems are in different phases when they manifest different symmetries. 
The symmetry is described by group theory. The new phase would have a new symmetric group which is a subgroup of the original one. 
As an example, on lowering the temperature, liquid crystallizes and continuous translation symmetry is broken to discrete translation symmetry. The spontaneous symmetry breaking theory also provides information about the low-energy excitations in the ordered phase. This theory can distinguish the continuous and discrete symmetry and the global and local symmetries. The symmetry-breaking theory demands that the two phases are not continuously connected and have different symmetries.

The classification of phases is based on macroscopic properties governed by the conservation laws and the broken symmetries. In a given phase, the physical properties are the function of some parameters that change smoothly, like temperature, coupling constant, or magnetic field. The phase transition is signaled by discontinuities in the functional dependence of these parameters. These parameters are called the \textit{order parameters}.
For example, the phase transition for magnetic material, from paramagnetic to ferromagnetic, is characterized by the magnetization in a specific direction. The system was rotationally invariant before the phase transition. But after the phase transition, spins point in a specific direction and so the rotational symmetry is broken. Here, the magnetization is the \textit{order parameter}. It is non-zero in a low symmetric ferromagnetic phase and vanishes for the paramagnetic phase.
The theory that describes the phase transition based on the order parameter is called the \textit{Landau theory of symmetry breaking} \cite{ginzburg1950theory}.

The \textit{critical exponents} describe the behavior of a system near a critical point. These exponents give information about the behavior of the thermodynamic quantities at the phase transition. The critical exponents depend on the order parameters, symmetry of interactions, and spatial dimensions. When the critical exponents, describing the phase transition, are the same for different physical systems, these systems are in the same \textit{universality class}. 
The length scale on which the fluctuations of order parameters are correlated is known as \textit{correlation length}.

The \textit{Mermin-Wagner theorem} states that the symmetry of a system cannot be broken in dimensions $d\le 2$. But there are other mechanisms for the phase transition like a low-energy distortion of the order parameter called a topological defect, dislocations in periodic crystals, or vortices in superfluid liquid. These distortions cannot be eliminated by the change of the order parameter. 
In this case, the two phases may have the same symmetries. The transition between the two phases is a change of topological properties. The topological material exists at a very low temperature. The phase transition by symmetry breaking is topologically trivial. For more on topology in condensed matter, see \cite{monastyrsky2006topology}.

\section{Topology of Band Structure}
In the early 1980s, the fractional quantum Hall effect \cite{tsui1982two} opened a new paradigm in condensed matter physics. The study of the quantum Hall phase revealed that the classification based on the theory of symmetry breaking is not complete. Despite being separate phases, the different FQH phases have the same symmetry. These phases are characterized by topological properties and the phase transition is called \textit{topological phase transition}. This kind of phase transition does not break any symmetry but changes the topology.  
The topological spaces are characterized by topological invariants. These invariants are some properties of manifolds that remain the same for the equivalent manifolds. See Appendix \ref{Top} for an introduction to topological spaces and topological invariants.

An atom has various discrete energy levels called atomic orbitals. When more than one atom join and make molecules, their atomic orbitals split into different molecular energy levels. In solids, there are a large number of atoms arranged in such a way that each atom can influence other neighboring atoms. Their energy levels split into a large number of different discrete energy levels. These energy levels are close to each other and make the \textit{energy bands}. The topmost empty or partially filled band is called \textit{conduction band} and the completely occupied band below the conduction band is called \textit{valence band}. The conduction and valence bands are close to each other in metals, but there is a large energy gap between the two in insulators. These are also called \textit{trivial insulators} to differentiate from the topological insulators. The energy gap is small for semiconductors. The highest occupied state or the energy of the electron with the highest energy is called a \textit{Fermi level}. The Fermi level has a fifty percent probability of being occupied at equilibrium. A \textit{semimetal} is the one for which the valence and conduction bands touch each other at some points. There is no well-defined \textit{Fermi surface} in semimetals \cite{pariari2020atoms}. The Fermi surface is the highest occupied state in the \textit{reciprocal space}. When the state in the bands is arranged in momentum space, then it is called the reciprocal space.

In the integral quantum Hall effect (IQHE), electrons have a cyclotron motion in the magnetic field. In the presence of an impurity, the energy levels split into sub-bands called Landau levels, as we discussed in the last chapter. The energy gap between the $N$th filled energy band and the $(N+1)$th empty band makes the bulk an insulator. When we increase the magnetic field, the degeneracy of Landau levels increases, and the Landau levels pass through the Fermi level as shown in Fig. \ref{LaughlinArgu}. There is an oscillation of the density of states at the Fermi level, that causes the oscillation in the electronic properties of the material, such as the electrical resistance. Electrons at edges drift towards the right, causing a current flow. There is a finite resistance, unlike the trivial insulators.

Now we will discuss the topology of the \textit{Brillouin zone} that is a unit cell in the momentum space or reciprocal space \cite{ando2013topological}. See Ref. \cite{cayssol2021topological} for detailed discussion on topological and geometrical aspects of band theory. When the atoms are arranged in a periodic lattice, the pattern of the first Brillouin zone is repeated. Therefore, only the first Brillouin zone is usually studied. 

According to \textit{Bloch's theorem}, the solutions of Schr\"odinger equation in a system of periodic potential $V(\textbf{k})= V(\textbf{k+a})$, are in the plane wave form, and there exist energy eigenstates $\ket{u_n(\textbf{k})}$ such that
\eq{\mathcal{H}\ket{\psi_n(\textbf{k})} = E_{\textbf{k}}\ket{\psi_n(\textbf{k})}, \qquad \ket{\psi_n(\textbf{k})} = e^{i\textbf{k}\cdot \textbf{r}}\ket{u_n(\textbf{k})},}
where $\ket{u_n(\textbf{k})} = \ket{u_n(\textbf{k+a})}$ and
$\textbf{k}$ values are inside the Brillouin zone (BZ) and $n$ are the labels for different bands. The eigenfunctions $\ket{u_n(\textbf{k})}$ are the \textit{Bloch states}. The Fermi energy $E_F$ lies between the empty conduction and the filled valence band. The ground state of many body systems is determined by the filled Bloch states.
The reduced Schr\"odinger equation in the band insulator is
\eq{H(\textbf{k})\ket{u_n(\textbf{k})} = E_n(\textbf{k})\ket{u_n(\textbf{k})},}
where $H(\textbf{k})= e^{-i\textbf{k}\cdot \textbf{r}}\mathcal{H} e^{-i\textbf{k}\cdot \textbf{r}}$ and $E_n(\textbf{k})$ are eigenvalues. The magnetic unit cell satisfy periodic boundary conditions, which corresponds to the topology of a torus. Two Bloch states in the same band, but different $\textbf{k}$, do not have to be orthogonal. The eigenvectors of $H(\textbf{k})$ and $H(\textbf{k}')$ now may have a non-zero overlap.

The quantized value of the Hall conductivity is explained by the concept of the \textit{fiber bundle} in topology. This is just like the \textit{gauge transformation}. 
The change of wave function of an electron by some phase, corresponding to the transformation of the vector potential as $\textbf{A}\rightarrow \textbf{A}+\nabla\phi$ and the magnetic flux $F=\nabla\cross \textbf{A}$. In terms of the fiber bundle, the vector potential corresponds to the \textit{connection}, and the strength of the magnetic field is the \textit{curvature}. The fiber bundle, connection, and curvature will be discussed in Appendix \ref{Top}.
Using the same idea, we can have the Berry phase and Berry curvature as we did in Chapter \ref{GeoPhase}. 

The bands $n$ collectively form the band structure. There is a phase ambiguity in the Bloch wave function. The Bloch wave function can be transformed as 
\eq{\ket{u_n(\textbf{k})} \rightarrow e^{i\phi(\textbf{k})}\ket{u_n(\textbf{k})}.}
Corresponding to this phase change, there will be a connection and curvature, analogous to the vector potential and the field strength in electromagnetism. The Berry connection and Berry curvature on the Brillouin zone are written as
\eq{\textbf{A}^n(\textbf{k}) = i\bra{u_n(\textbf{k})} \nabla_k \ket{u_n(\textbf{k})}, \qquad F^n(\textbf{k}) = \nabla_k \cross \textbf{A}^n(\textbf{k}).}
The \textit{Gauss-Bonnet theorem} relates the geometry and topology by the relation
\eq{ \int_M K dA = 2\pi \chi,}
where $\chi = 2(1-g)$ is the Euler characteristic, and $g$ is the genus, $A$ is an area of a region in $M$, and $K$ is the \textit{Gaussian curvature} given as $K=k_1k_2$, with $k_1$ is a curvature while going in one direction and $k_2$ is for the other direction. A flat surface has the Gaussian curvature zero. A saddle has the Gaussian curvature negative, as one of the $k_1$ or $k_2$ is negative. A sphere has a positive Gaussian curvature. A torus has the negative curvature on some points and positive on some other points.

In 1982, Thouless, Kohmoto, Nightingale, den Nijs \cite{thouless1982quantized} made a theoretical discovery that the filling fraction $\nu$ is a topological invariant now known as the \textit{TKNN invariant}. The integration of field strength over the whole BZ defines the topological invariant called the \textit{Chern number} and is written as
\eq{\nu_n = \frac{1}{2\pi} \int_{BZ} dk_xdk_y F_{xy}^{(n)}(\textbf{k}).}
This is the Chern number of the $n$th band. The total Chern number of the occupied bands $\nu = \sum_n \nu_n$. This is the same number as in the quantum Hall resistivity 
\eq{\sigma_{xy} = \frac{e^2}{h}\nu.}
This integer $\nu$ is called the \textit{first Chern number} and is a topological invariant. This topological invariant characterizes IQH. Hence, the Hall conductance is an integral multiple of $e^2/h$.
The time-reversal symmetry, $F_{xy}(-\textbf{k}) = -F_{xy}(-\textbf{k})$, leads to the negative sign with the Chern number. Therefore, the time reversal symmetry is broken in the quantum Hall state.
The line integral of $A^n(\textbf{k})$ along a closed path $C$ in the momentum space is the Berry phase $\exp(\oint_C d\textbf{k}\cdot \textbf{A}^n (\textbf{k}))$. This is a gauge invariant but not a topological invariant \cite{sato2017topological}.

The edge states, at the interface of the quantum Hall state and the vacuum or trivial insulator, are the consequence of the topological classification of the gapped states. The gap must be closed to change the $\nu$ at the boundary. The number of edge channels is determined by the bulk-boundary correspondence, as described in the previous Chapter. The number of edge modes intersecting the Fermi energy is related to the change in bulk topological invariant $\nu$ across the interface. The $N$ filled Landau levels have $N$ number of edge channels at the interface \cite{pariari2020atoms}. 

\subsection{Topological Insulators and Semimetals}
The symmetry of the Hamiltonian is important in determining the role of the topology of the occupied bands. Without imposing symmetry, any deformation of the Hamiltonian is possible as long as the gap is not closed \cite{sato2017topological}. 
\textit{Kramer's theorem} states that all eigenstates of a $\mathcal{T}$ invariant Hamiltonian of the spin-1/2 system have a twofold degeneracy. 
The time reversal operator obeys $\mathcal{T}^2=-1$ and $\bra{\mathcal{T}u}\ket{\mathcal{T}v}=\bra{u}\ket{v}$. Therefore, $\bra{u}\ket{\mathcal{T}u}=0$, that means they are orthogonal. $\ket{u}$ and $\mathcal{T}\ket{u}$ have the same energy, so there is a degeneracy. The Bloch Hamiltonian is given by
\eq{\mathcal{T}H(\textbf{k})\mathcal{T}= H(-\textbf{k}).}
Now $\ket{u_n(\textbf{k})}$ and $\mathcal{T}\ket{u_n(\textbf{k})}$ are the eigenstates of $H(\textbf{k})$ and $H(-\textbf{k})$, and the phase ambiguity can be written as
\eq{\ket{u_n(\textbf{k})} = e^{i\phi_n(\textbf{k})}\mathcal{T}\ket{u_n(\textbf{k})}.}
The quantum spin Hall (QSH) states, also called topological insulators (TIs), have the time-reversal symmetry due to a large spin-orbit coupling. Spin-orbit coupling is the relativistic effect when the spin of moving charges interacts with its motion. In atoms, it leads to a shift in the energy of two orbits having opposite spins. It is a kind of Zeeman splitting due to the motion in a magnetic field in the electron's frame of reference and the magnetic moment of the electron due to its spin in the electric field of the nucleus. Its correction is a fine structure and is detectable in spectral lines in fine structure. 

An analogous effect in a bulk crystal is called \textit{Rashba effect} also known as Bychkov-Rashba effect \cite{manchon2015new}. Moving charges create a magnetic field. In the absence of the external magnetic field, there is some kind of fictitious effective magnetic field acting in the upward direction on the up spins and in the downward direction on the down spins. This magnetic field is due to the motion of electrons moving in an electric field. When electron motion is along the $x$-axis, they feel a magnetic field along the $y$-axis. This field is called the \textit{Rashba field} written as 
\eq{H_R = \alpha (\bm{\sigma} \cross \textbf{p})\cdot \hat{z},}
where $\alpha$ is the Rashba coupling, $\bm{\sigma}$ is the Pauli matrix-vector and $\textbf{p}$ is the momentum. This equation is a two-dimensional version of the Dirac Hamiltonian. The Rashba field bends the trajectory of an electron. The direction of distortion depends on the direction of angular momentum. This effect is used to get the terms in the model Hamiltonian \cite{bernevig2013topological}. The momenta are locked to the perpendicular spin to preserve the time-reversal symmetry. The upward and downward spins move in the opposite direction. The total charge flow is zero as left moving and right moving charges are equal, but there is a spin current. As there is no dissipation of energy by these two channels, this spin current can be a useful application in integrated circuit technology. The related field of study is called \textit{spintronics}. 

\begin{figure}[h!]
	\centering
	\includegraphics[scale=0.8]{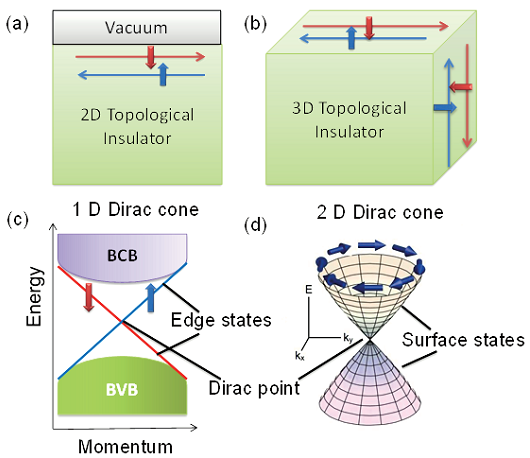}
	\caption[Edge states in 2D and 3D topological materials.]{Edge state (a) and (b), and band diagrams (c) and (d), for 2D and 3D topological insulators. The diagram is taken from the Ref. \cite{bhardwaj2020topological}.}
	\label{TopInsulator}
\end{figure}

The QSH state is realized as the superposition of the two systems. These kinds of materials cannot be characterized by the TKNN invariant, because there are equal and opposite invariants for the two channels, so $\nu=0$.
Kane Mele \cite{kane2005z} model introduced a $\mathbb{Z}_2$ invariant, that has the value 0 or 1.
The edge states are explained by the bulk-boundary correspondence. There are gapped energy states in the bulk, but gapless at states at the edges. At the interface with the trivial insulators, the topological invariant changes. The energy gap closes and the surface states appear. The topology changes from the non-trivial to the trivial at the interface, as shown in Fig. \ref{TopInsulator} (a) and (c). In 3D topological insulators, there are surface states, as in Fig. \ref{TopInsulator} (b),  described by four $\mathbb{Z}_2$ topological invariants \cite{pariari2020atoms}. 

There are relatively recent discoveries of the materials known as Dirac semimetals and Weyl semimetals, realized in graphene and 3D topological materials. Dirac semimetal exists at the phase transition from trivial insulators to topological insulators. The bulk gap can be tuned by chemical doping or external pressure. At the critical point, the valence band and the bulk conduction band touch at a special point in the momentum space called the \textit{Dirac node} \cite{neto2009electronic}. The Dirac node is shown in Fig. \ref{TopInsulator} (d). These materials are classified as topological if the node points are topologically protected due to the bulk band structure \cite{bhardwaj2020topological}. As the electrons can easily be excited from the valence band to the conduction band at the Dirac point, these materials are good absorbers of light and have high electron mobility. Just like graphene, these are 3D topological materials and the surface states in these materials behave like metals. The charge carrier concentration is three orders of magnitude less than that in the metals. The dynamics of the charge carriers are governed by the Dirac-type equation, with $c$ and $\textbf{p}$ in the Dirac equation are replaced by $V_F$ and $\textbf{k}$. But there are two differences from the three-dimensional relativistic fermions; the velocity is two orders of magnitude less than the speed of light, and these materials have charge carriers constrained to two dimensions \cite{pariari2020atoms}. See Appendix \ref{QFT} for Dirac equation and Dirac and Weyl massive fermions. These semimetals host Dirac and Weyl massless fermions.
\subsection{Classification of Topological Materials}
Topological states of matter include integer and fractional quantum Hall states, spin liquids, topological insulators, and topological superconductors. These are characterized by the topological invariant, which is calculated from the ground state wave function. 

The topological materials have common characteristics such as ground state degeneracy, energy gap separating the ground state from the excited state in the bulk, and edge modes at the surface or boundary. The edge modes are robust against perturbations. If one phase can be deformed to the other without closing the energy gap, then those materials are in the same equivalent class. The idea is to change the Hamiltonian adiabatically so that the system remains in the ground state. The collapse of the gap signals the \textit{topological quantum phase transition}. Different equivalence classes have different topologies.

There are several models to characterize topological materials. The two well-known categories are based on long-range entanglement and short-range entanglement. The first one is known as the topological order. The quantum Hall state falls in this category. These kinds of phases have fractional statistics, and edge modes have chiral symmetry. The chiral symmetry means that the edge current flows only in one direction on one side and in the opposite direction on the other side. These phases are described by the effective field theory, which is topological quantum field theory. There are strongly interacting fermionic or bosonic systems. The IQHE is a state of free fermions of topological order characterized by the Chern number. The FQHE consists of fractional excitation and has braiding statistics. String-net condensation for $2+1$D topological order is also in this category \cite{levin2005string}. These have gapped edges and are classified by unitary modular fusion categories. The group theory is the mathematical foundation of symmetry and symmetry breaking, the modular category is the foundation of the topological order. See Chapter \ref{Cat} on category theory. Topological order is used in topological quantum computing. See \cite{wen2016introduction,wen2017colloquium} for a brief introduction to the topological order.

The second category is the one with short-range entanglement. The quantum spin Hall effect is the model for such phases \cite{kane2005z}. These are also called \textit{symmetry-protected topological} (SPT) states \cite{wen1995topological}. Topological insulators and superconductors \cite{qi2011topological} fall in this category \cite{lahtinen2017short}.
These are robust against perturbations that respect the time-reversal symmetry and $U(1)$ symmetry, whereas the topological orders are robust against any kind of perturbation. The edge modes are flowing in the absence of a magnetic field. These edge modes break the chiral symmetry, that is, the same edge can have modes in both directions.
The topological insulators are both insulators and conductors, insulators in the bulk and conductors at the edges. These edges are one-dimensional for a two-dimensional material and are two-dimensional surface states for a three-dimensional topological insulator. 

For further details on the classification of the topological phases of matter, see 
\cite{wen2017colloquium,bernevig2013topological} and review articles \cite{chiu2016classification,bansil2016colloquium,qi2011topological,pariari2020atoms,hasan2010colloquium}
	\section{Superconductivity}
Non-Abelian anyons are the physical systems needed for topological quantum computation. These anyons exist as quasiparticles in fractional quantum Hall effect and Majorana fermions in topological superconductors. The conventional superconductors are known as \textit{s-wave} superconductors. Here, s stands for spin-singlet state, analogous to atomic configuration. The Cooper pairs are formed by two electrons with opposite spins and opposite momenta. The p-wave superconductors are spin-triplet superconductors and are known as unconventional superconductors or chiral superconductors. These materials show topological properties. Read and Green \cite{read2000paired} theoretically proposed that the anyons in p-wave superconductors correspond to the quasiparticle in non-Abelian quantum Hall states \cite{moore1991nonabelions}. See \cite{liu2014creating,sola2020majorana,sarma2015majorana} for a brief discourse on the Majorana fermions and their use in topological quantum computation. 

Superconductivity was discovered by Heike Kamerlingh Onnes in 1911. Such materials are explained by Bardeen-Cooper-Schrieffer's (BCS) theory of superconductivity \cite{bardeen1957theory}. Superconducting materials show zero resistance and expel the magnetic field. Below a critical temperature $T_c$, the material shows superconducting behavior. The \textit{Cooper pairs} are formed below the critical temperature. In such pairs, electrons have equal and opposite momenta and opposite spin, so they are not restricted by the Pauli exclusion principle. The pairs make a bosonic state and the material condenses to the superconducting phase. The thermal energy of the pair is less than their binding energy, so the paired state is stable. High-temperature superconductivity at temperature 90 K was discovered in 1986 by George Bednorz and Alex Muller \cite{bednorz1986possible}. Recently, in 2020, the first room-temperature superconductor is found at a very high-pressure \cite{snider2020room}. See \cite{campbell1987superconductivity} for a brief introduction to superconductivity and \cite{sahu1990bcs} for the basic derivation of BCS theory. For further study of the field, see the books by Tinkham, M \cite{tinkham2004introduction}, Steven M. Girvin \cite{girvin2019modern} and G. E. Volovik \cite{volovik2003universe}.

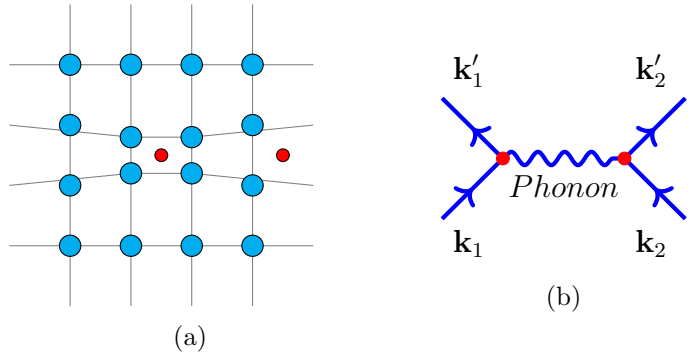
\begin{figure}
	\centering
	\begin{subfigure}{0.3\textwidth}		
		\begin{tikzpicture}[scale=0.8]
			\draw[gray] (0,4)--(5,4);
			\draw[gray] (0,3)--(2,2.8)--(3,2.8)--(5,3);
			\draw[gray] (0,2)--(2,2.2)--(3,2.2)--(5,2);
			\draw[gray] (0,1)--(5,1);
			\draw[gray] (4,0)--(4,5);
			\draw[gray] (3,0)--(3,5);
			\draw[gray] (2,0)--(2,5);
			\draw[gray] (1,0)--(1,5);			
			\draw[fill=cyan] (4,4) circle (5pt);
			\draw[fill=cyan] (4,3) circle (5pt);
			\draw[fill=cyan] (4,2) circle (5pt);
			\draw[fill=cyan] (4,1) circle (5pt);
			\draw[fill=cyan] (3,4) circle (5pt);
			\draw[fill=cyan] (3,2.8) circle (5pt);
			\draw[fill=cyan] (3,2.2) circle (5pt);
			\draw[fill=cyan] (3,1) circle (5pt);
			\draw[fill=cyan] (2,4) circle (5pt);
			\draw[fill=cyan] (2,2.8) circle (5pt);
			\draw[fill=cyan] (2,2.2) circle (5pt);
			\draw[fill=cyan] (2,1) circle (5pt);
			\draw[fill=cyan] (1,4) circle (5pt);
			\draw[fill=cyan] (1,3) circle (5pt);
			\draw[fill=cyan] (1,2) circle (5pt);
			\draw[fill=cyan] (1,1) circle (5pt);
			\draw[fill=red] (2.5,2.5) circle (3pt);
			\draw[fill=red] (4.5,2.5) circle (3pt);
		\end{tikzpicture}
		\caption{}
	\end{subfigure}
	\centering
	\begin{subfigure}{0.3\textwidth}
		\centering
		\tikzset{middlearrow/.style={decoration={markings,mark= at position 0.5 with {\arrow{#1}},},postaction={decorate}}}
		\begin{tikzpicture}[scale=0.8]
			\draw[ultra thick,blue,middlearrow={>}] (1,1)--(0,2);
			\draw[ultra thick,blue,middlearrow={>}] (0,0)--(1,1);	
			\draw [decorate,ultra thick,blue, decoration={snake}] (1,1) --(3,1);
			\draw[ultra thick,blue,middlearrow={>}] (4,0)--(3,1);
			\draw[ultra thick,blue,middlearrow={>}] (3,1)--(4,2);
			\filldraw[red] (1,1) circle (3pt);
			\filldraw[red] (3,1) circle (3pt);
			\node[above right] at (0,2) {$\textbf{k}'_1$};
			\node[below right] at (0,0) {$\textbf{k}_1$};	
			\node[below=2] at (2,1) {$Phonon$};	
			\node[above right] at (3,2) {$\textbf{k}'_2$};
			\node[below right] at (3,0) {$\textbf{k}_2$};
		\end{tikzpicture}
		\caption{}
	\end{subfigure}
	\caption[Interaction between two electrons in a Cooper pair.]{(a) The atomic lattice is deformed by the existence of one electron so that the other electron is attracted to the deformed lattice.}
	\label{BCSLattice}
\end{figure}

A rough pictorial view of the Cooper pairs is shown in Fig. \ref{BCSLattice} (a). A moving electron deforms the lattice formed by the positive ions. The lattice remains deformed for some time after the passing electron due to the large mass of the nucleus. The other passing electron effectively gets attracted to this deformation. These two electrons remain connected during their motion in the lattice. The attraction must be greater than the Coulomb repulsion. The bandgap in superconductors is due to the electron-electron interactions. It is the energy required to break the pairs and make separate electrons. When temperature increases, the gap decreases and becomes zero at the transition temperature. In the field-theoretic approach, two electrons approach each other, exchange phonon, and then scattered as shown in Fig. \ref{BCSLattice} (b). The interaction between two electrons is attractive when the frequency of phonon is below \textit{Debye frequency} $\omega_D$.
The Cooper pairs cannot be formed at high temperatures because the lattice is not formed due to the thermal vibrations. High–temperature superconductors cannot be explained by the BCS theory, since that mainly deals with the lattice deformation. The superconductivity can be similar to superfluidity in which the viscosity of the liquid becomes zero below some critical temperature \cite{volovik2003universe}.

According to the Landau-Ginsberg theory of symmetry breaking, superconductivity is the second-order phase transition in which the order parameter is different for the ordered and disordered states. The value is changed at the critical temperature \cite{tinkham2004introduction}. Landau-Ginsberg's theory is a macroscopic theory, whereas the BCS theory is a microscopic theory when the interaction between electrons is attractive in the presence of the Fermi sea.

The resistance is due to an interaction of electrons with the lattice. This interaction is caused by lattice vibrations and defects or impurities. Resistance of a superconducting material drops to zero at the critical temperature $T_c$. Below the critical temperature, the lattice is perfectly periodic, hence zero resistance.

The supercurrent in the bulk creates its own magnetic field, which opposes the applied one. Therefore, the applied field does not penetrate the material. This is called \textit{Meissner effect}. Due to this effect, superconducting material can levitate in a magnetic field. This effect was surmised long before BCS in 1935 by London. The material shows \textit{perfect diamagnetism} which is different from the conventional diamagnetism by the surface current. The applied field decay exponentially inside the material at \textit{penetration depth}. The penetration depth is temperature-dependent.
There are two types of materials distinguished by their response to the external magnetic field. \textit{Type-I} superconductors are the ones for which the superconductivity is lost when the magnetic field is increased up to a critical value. In contrast, the \textit{type-II} superconductors have two critical values of the field. By the increase of field to some critical value, the field penetrates the material in discrete threads and the material remains superconducting. The threads of the magnetic field are tubes of flux quantum and make vortices called \textit{Abrikosov vortices}. Further increase of the field to another critical value, the superconductivity is lost, and the material becomes a normal material \cite{tinkham2004introduction}.

When two or more superconductors are connected by some non-superconducting or insulating materials, the supercurrent flow across the junction even without any applied voltage. This effect is known as \textit{Josephson effect}. It is due to the tunneling of the Cooper pairs. The quantum mechanical circuits such as superconducting qubit and SQUID are based on the Josephson effect. Another effect on the junction is \textit{proximity effect}. When a superconductor is brought into contact with a normal metal, the Cooper pairs at the boundary are not destroyed abruptly but carried over to the normal metal, where they decohere by scattering events. Depending on the materials and their quality, the Cooper pairs can persist up to hundreds of microns inside the normal metal. A related concept is when electrons from the normal metal enter into a superconductor, at the superconductor-normal metal contact, two electrons can enter into the superconductor and make a Cooper pair. A single electron transfer is forbidden due to the energy gap, but when one electron is transferred from the normal metal to the superconductor, a hole is reflected in the normal materials. This effect is called the \textit{Andreev reflection} \cite{tinkham2004introduction}. It is important in our discussion for the realization of Majorana fermions.

\subsection{BCS Theory of Superconductivity}
As discussed in \cite{sahu1990bcs,leggett1975theoretical,cooper1956bound}, consider a pair of electrons in a metal with one electron has momentum $\textbf{k}$ and spin up whereas the other has momentum $\textbf{-k}$ with spin down at $T=0$. Let the two electrons be coupled by the effective attractive interaction and make the \textit{Cooper pairs}. These pairs interact in the filled Fermi sea and have less energy than without pairing.
More such pairs reduce energy further and the new state of matter is formed \cite{bardeen1957theory}.

For theoretical description of the superconducting state, the method of second quantization is used. The second quantization is discussed in Appendix \ref{QFT}. The wave function of non-interacting $N$-particles system is written as $\phi_i(\textbf{r}_i, s_i), i= 1,2...,N$ where $\textbf{r}_i$ are positions of particles and $s_i$ are spins. In second quantization, particle number $n_1,n_2...$ are the state variables.
Let a number operator be written as $\hat{n}_i=\hat{c}_i^\dagger\hat{c}_i$, where $\hat{c}_i^\dagger$ is the creation operator and $\hat{c}_i$ is the annihilation operator for the state $i$. The eigenvalues of the number operator are the particle numbers. It obeys anti-commutation relation for fermion, 
\eq{\left\{c_i, c_j^\dagger\right\} = c_ic_j^\dagger + c_j^\dagger c_i = \delta_{ij}.} 
Let $\ket{0}$ be the uncorrelated vacuum state. The wave function of BCS theory at $T=0$ ground state is written as a product of pair states as 
\eq{\Psi = \prod_k (u_k + v_kc^\dagger_{k\uparrow}c^\dagger_{-k\downarrow}) \ket{0} \label{Pair},}
where $c^\dagger_{k\uparrow}c^\dagger_{-k\downarrow}$ is a creation operator for the superconducting pair state, $v_k$ probability amplitude that pair state is occupied, and $u_k$ is probability amplitude for the state to be unoccupied. We have $\abs{u_k}^2 + \abs{v_k}^2 = 1$, that means the total probability for a state at $\textbf{k}$ must be equal to 1.
When $({\textbf{k}\uparrow},{-\textbf{k}\downarrow})$ be occupied state and $({\textbf{k}\uparrow}^{'},{-\textbf{k}\downarrow}^{'})$ is the unoccupied state initially. Here $\textbf{k}^{'}$ and $\textbf{k}$ differ by some momentum $\textbf{q}$. The probability of the system being in initial state is $u_kv_{k^{'}}$. Let the first state gets unoccupied and the second gets occupied, then probability that this happens is $u^*_kv^*_{k^{'}}v_ku_{k^{'}}$. The Eq. \ref{Pair} is a state with a superposition of different number of particles. 

Now the Cooper pair Hamiltonian $\hat{H}$ for the interacting electron system, corresponding to the wave function in Eq. \ref{Pair} can be written as
\eq{\hat{H}\ket{\Psi} = E_{CP}\ket{\Psi},}
where $\hat{H}= \hat{H}_0+\hat{V}_{int}$ with 
$\hat{H}_0= \sum_{\textbf{k}\sigma} \xi_\textbf{k} \hat{c}^\dagger_{\textbf{k}\sigma}\hat{c}_{\textbf{k}\sigma}$. So that we have
\eq{\hat{H}= \sum_{\textbf{k}\sigma} \xi_\textbf{k} \hat{c}^\dagger_{\textbf{k}\sigma}\hat{c}_{\textbf{k}\sigma} + \sum_{\textbf{k}\textbf{k}'}V_{\textbf{k}\textbf{k}'}\hat{c}^\dagger_{\textbf{k}\uparrow}\hat{c}^\dagger_{-\textbf{k} \downarrow}\hat{c}_{-\textbf{k}'\downarrow}\hat{c}_{\textbf{k}'\uparrow},}
with $\xi_\textbf{k} = \epsilon(\textbf{k})-\mu=\hbar^2k^2/2m-\mu$ is the kinetic energy of an electron with respect to the chemical potential $\mu$. The chemical potential is the change of the energy of the system due to the change in particle number. $\sigma$ is a spin index that can be $\uparrow$ or $\downarrow$. The \textit{mean-field approximation} is to take the average of the interactions on a one body make many-body problem to one-body problem. By using the mean-field approximation, we have
$\Delta_\textbf{k} = -\sum_{\textbf{k}'}V_{\textbf{k}\textbf{k}'}\langle \hat{c}_{-\textbf{k}'\downarrow}\hat{c}_{\textbf{k}'\uparrow}\rangle$. Now we can write the BCS Hamiltonian in the form
\eq{\hat{H}_{B\hat{c}S}^{MF}= \sum_{\textbf{k}\sigma} \xi_\textbf{k} \hat{c}^\dagger_{\textbf{k}\sigma}\hat{c}_{\textbf{k}\sigma} + \sum_{\textbf{k}\textbf{k}'}[\Delta_\textbf{k} \hat{c}^\dagger_{\textbf{k}\uparrow}\hat{c}^\dagger_{-\textbf{k} \downarrow} + \Delta_\textbf{k}^*\hat{c}_{-\textbf{k}'\downarrow}\hat{c}_{\textbf{k}'\uparrow}]\label{HBCS},}
where $\Delta$ is the Cooper pair energy gap. It is the energy needed to break the Cooper pairs into single electrons. Therefore, the $\Delta$ is binding energy of the Cooper pairs. It can serve as a complex order-parameter for the theory of superconductivity.

\section{Topological Superconductors and Majorana\\ Fermion}
There are examples when some particles were predicted first but discovered later. Majorana fermion is one such example. Majorana in 1937 \cite{majorana1937nuovo} proposed a solution to the Dirac equation which predicts an antiparticle that is identical to its own particle, that is, it has charge-conjugation and particle-hole symmetry. Since $\psi$ and $\psi^*$ satisfy the Dirac equation, as we discussed in Appendix \ref{QFT}, therefore, $\psi= \psi^*$ can be imposed without any contradiction. Dirac's equation involving imaginary solution. Majorana's solutions to the Dirac equation are real numbers. Therefore, the self-conjugated Dirac particle is the Majorana particle \cite{wilczek2009majorana}.
There are proposals for a particle to be a Majorana particle. See a discussion in ref \cite{wilczek2009majorana}. The emergent particles in condensed matter systems can support Majorana fermions \cite{wilczek2009majorana}. These particles are not elementary particles and cannot move at the speed of light and do not obey the Lorentz invariance of the Dirac equation. Therefore, these are distinct from the original proposal by Majorana. These particles obey two conditions to be the Majorana particles; they obey the Dirac equation, and they are their own antiparticles. These two conditions are met in topological superconductors. They are found at the boundaries of the topological superconductors and the spin-liquids.

The Cooper pairs in the conventional superconductors have an even parity spin-singlet state, with $S=0$ and $l=0$. Such materials are called s-wave superconductors. These are antisymmetric in spin wave function but symmetric in angular momentum. An extension of this theory is proposed for spin-triplet pairing $S=1$ and $l=1$. Now, the spin wave function is symmetric, but the Pauli exclusion principle forces the parity of angular momentum to be odd. Reversing the motion flips the orbital angular momentum to break time-reversal symmetry. Such materials are called the p-wave superconductors. In two dimensions, these are known as $p_x\pm ip_y$ superconductors. These materials are ferromagnetic like where momentum circulates. These are also known as chiral superconductors. Such superconductors host topological phases with Majorana excitations
at their boundaries and the defects.

The Hamiltonian \ref{HBCS} can be diagonalized by using the \textit{Bogoliubov transformation} in which the fermionic operators are written as a linear combination of electron creation and annihilation operators
\eq{\hat{\gamma}_{\textbf{k}'\uparrow} = u^*_\textbf{k}\hat{c}_{\textbf{k}'\uparrow}+ v_\textbf{k}\hat{c}^\dagger_{\textbf{-k}'\downarrow}\nonumber \\
	\hat{\gamma}_{\textbf{-k}'\downarrow} = u_\textbf{k}\hat{c}^\dagger_{\textbf{-k}'\downarrow}- v^*_\textbf{k}\hat{c}_{\textbf{k}'\uparrow},}
with
$u_\textbf{k}= \frac{1}{2}(1+\xi_k/E_\textbf{k}), \ v_\textbf{k}= \frac{1}{2}(1-\xi_\textbf{k}/E_\textbf{k})$, and $u_\textbf{k}v_\textbf{k}=\Delta_\textbf{k}/(2E_\textbf{k})$, where $E_\textbf{k}=\sqrt{\xi_\textbf{k}^2 + \abs{\Delta_\textbf{k}}^2}$ is the quasiparticles excitation energy \cite{stanescu2016introduction}. The s-wave pairing implies that $\Delta_{-\textbf{k}}= \Delta_\textbf{k}$, $u_{-\textbf{k}} = u_\textbf{k}$, and $v_{-\textbf{k}}=v_\textbf{k}$. 
Now BCS Hamiltonian becomes
\eq{\hat{H}_{BC}^{MF}= \sum_{\textbf{k}\sigma} E_{\textbf{k}}\hat{\gamma}^\dagger_{\textbf{k}'\sigma}\hat{\gamma}_{\textbf{k}'\sigma}+const.}
The $\hat{\gamma}^\dagger$ and $\hat{\gamma}$ are creation and annihilation operators for quasiparticles. The quasiparticle operators $\hat{\gamma}$ satisfy the relation $\hat{\gamma}^\dagger_E=\hat{\gamma_{-E}}$. At $E=0$, $\hat{\gamma}^\dagger_0=\hat{\gamma_0}$. This is the definition of Majorana fermion. These operators are a combination of electrons and holes. At the Fermi surface $\abs{u_\textbf{k}}\approx \abs{v_\textbf{k}}$, the quasiparticles are a superposition of holes and electrons. Deep inside the Fermi sea, the Bogoliubov quasiparticles are hole-like with $\abs{u_\textbf{k}}\approx 0$ and $\abs{v_\textbf{k}}\approx 1$, whereas above the $E_F$ with $\abs{u_\textbf{k}}\approx 1$ and $\abs{v_\textbf{k}}\approx 0$ the quasiparticle are particle-like. The system is in \textit{strong pairing} regime when the Cooper pair of bounded electron formed over a length $\zeta$, whereas the \textit{weak pairing} regime is when the Cooper pair size is infinite in real space \cite{alicea2012new}.

Using the fermion anticommutation relation, the first term in the BCS Hamiltonian can be written as
$\frac{1}{2}\sum_{\textbf{k}\sigma}[\xi_\textbf{k}\hat{c}^\dagger_{\textbf{k}\sigma}\hat{c}_{\textbf{k}\sigma}- \xi_{-\textbf{k}}\hat{c}_{\textbf{k}\sigma}\hat{c}^\dagger_{\textbf{k}\sigma}+\xi_\textbf{k}]$
and $\Delta_\textbf{k} \hat{c}^\dagger_{\textbf{k}\uparrow}\hat{c}^\dagger_{-\textbf{k} \downarrow}= \frac{1}{2}[\Delta_\textbf{k} \hat{c}^\dagger_{\textbf{k}\uparrow}\hat{c}^\dagger_{-\textbf{k} \downarrow} - \Delta_\textbf{k} \hat{c}^\dagger_{-\textbf{k}\downarrow}\hat{c}^\dagger_{\textbf{k} \uparrow}]$. Similar relations can be written for the complex conjugate terms. Introducing the four-component spinor $\Psi^\dagger = (\hat{c}^\dagger_{\textbf{k}\uparrow},\hat{c}^\dagger_{\textbf{k}\downarrow},\hat{c}_{-\textbf{k}\uparrow},\hat{c}_{-\textbf{k}\downarrow})$. In this \textit{Nambu spinor} representation we have
\eq{\hat{H}_{BCS}=\frac{1}{2}\sum_{\textbf{k}}\Psi^\dagger_\textbf{k}H_{BdG}(\textbf{k})\Psi_\textbf{k} + const.,}
where $H_{BdG}$ is the \textit{Bogoliubov-de Gennes Hamiltonian}. It is a $4 \cross 4$ matrix with eigenvalues $E_\textbf{k}= \pm \sqrt{\xi_\textbf{k}^2+\abs{\Delta_\textbf{k}}^2}$. This Hamiltonian has particle-hole symmetry, that means changing particle to antiparticle and vice versa would not affect the equation. That hints at a 2-fold degeneracy. 

To find the topological invariant for the \textit{chiral superconductors}, writing the Hamiltonian in a two-dimensional representation as
\eq{\hat{H}=\frac{1}{2}\sum_{\textbf{k}}\Psi^\dagger_\textbf{k}H_{BdG}(\textbf{k})\Psi_\textbf{k},}
\eq{H_{BdG}(\textbf{k})=\begin{pmatrix}
		H_k & \Delta_k \\
		\Delta_k^\dagger & -H_k^*
	\end{pmatrix} \qquad \Psi = \begin{pmatrix}
		u_k\\
		v_k
	\end{pmatrix}.}
The const. is ignored. $H_k = \frac{k^2}{2m} - \mu$ is a single-particle Hamiltonian, whereas the diagonal terms are pairing interactions. The component $u_k, v_k$ of the wave function are particle and hole-like wave funciton. $\Delta_k = \psi(k)= \psi(k)$ for spin-singlet and $\Delta_k = d_z(k)=-d_z(-k)$ for spin-triplet. We have $\Delta_k = \Delta'_k(k_x +ik_y)$, where $\Delta'$ is taken real.
In the basis of Pauli matrices $\bm{\sigma}=(\sigma_1,\sigma_2,\sigma_3)$ we can write
\eq{\mathcal{H}_k = \bm{\sigma} \cdot \textbf{d}_k, \qquad \textbf{d}_k= [\Delta'k_x, -\Delta'k_y, \frac{k^2}{2m} - \mu].}
It resembles the spin Hamiltonian of ferromagnetism, with magnetization represented by the vector $\textbf{d}_k$. The Chern number is written as
\eq{C= \frac{1}{4\pi}\int \textbf{n}_k \cdot (\frac{\partial\textbf{n}_k}{\partial k_x}\cross \frac{\partial\textbf{n}_k}{\partial k_y}) dk_xdk_y, \qquad \textbf{n}_k = \frac{\textbf{d}_k}{\abs{\textbf{d}_k}}.} 
The Chern number $\abs{C}$ is the number of times the vector $\textbf{n}_k$ sweeps the unit sphere as $\textbf{k}$ covers the momentum space.
\begin{figure*}[h!]
	\centering
	\includegraphics[scale=0.6]{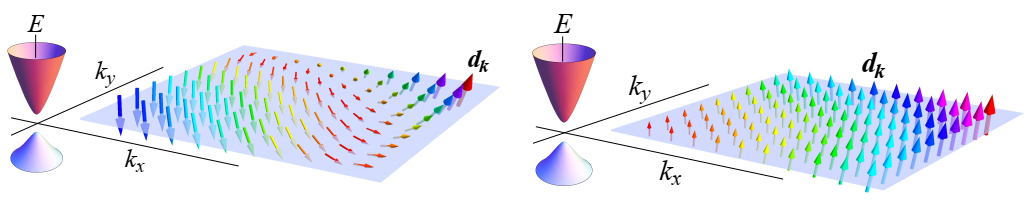}
	\caption[Chern number of topological superconductors.]{Topological superconductor with Chern number $C=\pm 1$ and the conventional superconductor with Chern number $C=0$. This diagram is taken directly from Ref. \cite{culcer2020transport}.}
	\label{SuperChern}
\end{figure*}

\subsection{Kitaev's Toy Model}
\begin{figure*}[h!]
	\centering
	\begin{tikzpicture}[scale=0.5]	
		\draw [teal,ultra thick,fill=cyan!20] (1,0) ellipse (1.5 and 0.5);
		\draw [teal,ultra thick,fill=cyan!20] (5,0) ellipse (1.5 and 0.5);
		\draw [teal,ultra thick,fill=cyan!20] (9,0) ellipse (1.5 and 0.5);
		\draw [teal,ultra thick,fill=cyan!20] (13,0) ellipse (1.5 and 0.5);
		\draw[ultra thick, blue] (0,0)--(2,0);
		\draw[ultra thick, blue] (4,0)--(6,0);
		\draw[ultra thick, blue] (8,0)--(10,0);
		\draw[ultra thick, blue] (12,0)--(14,0);
		\fill[red] (0,0) circle(6pt);
		\fill[red] (2,0) circle(6pt);
		\fill[red] (4,0) circle(6pt);
		\fill[red] (6,0) circle(6pt);
		\fill[red] (8,0) circle(6pt);
		\fill[red] (10,0) circle(6pt);
		\fill[red] (12,0) circle(6pt);
		\fill[red] (14,0) circle(6pt);
		\node[] at (-4,0) {$Trivial$};	
	\end{tikzpicture}\\
	
	\vspace{10pt}
	
	\begin{tikzpicture}[scale=0.5]
		\draw [teal,ultra thick,fill=cyan!20] (1,0) ellipse (1.5 and 0.5);
		\draw [teal,ultra thick,fill=cyan!20] (5,0) ellipse (1.5 and 0.5);
		\draw [teal,ultra thick,fill=cyan!20] (9,0) ellipse (1.5 and 0.5);
		\draw [teal,ultra thick,fill=cyan!20] (13,0) ellipse (1.5 and 0.5);
		\draw[ultra thick, blue] (2,0)--(4,0);
		\draw[ultra thick, blue] (6,0)--(8,0);
		\draw[ultra thick, blue] (10,0)--(12,0);
		\fill[red] (0,0) circle(6pt);
		\fill[red] (2,0) circle(6pt);
		\fill[red] (4,0) circle(6pt);
		\fill[red] (6,0) circle(6pt);
		\fill[red] (8,0) circle(6pt);
		\fill[red] (10,0) circle(6pt);
		\fill[red] (12,0) circle(6pt);
		\fill[red] (14,0) circle(6pt);
		\node[] at (-3,0) {$Topological$};
	\end{tikzpicture}
	\caption[Kitaev's toy model.]{When Majorana fermions are paired at the same site, we get the trivial phase. But by tuning the pairing potential and the chemical potential, Majorana fermions can be paired at adjacent sites to get the topological phase.}
	\label{KitaveToy}
\end{figure*}
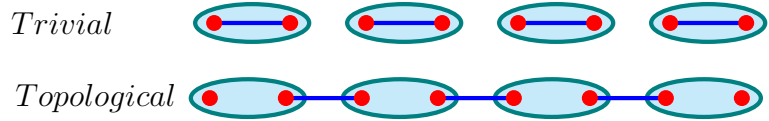

To realize the spinless p-wave superconductor, Kitaev proposed a 1D toy model \cite{kitaev2001unpaired}.
\eq{H = -\frac{t}{2}\sum_{j=0}^{N-1}(c_j^\dagger c_{j+1} + c^\dagger_{j+1}c_j) - \frac{\Delta}{2}\sum_{j=0}^{N-1}(c_j^\dagger c_{j+1}^\dagger + c_{j+1}c_j) - \mu \sum_{j=0}^{N} c_j^\dagger c_j,}
where $t$ is nearest-neighbor hopping strength, $\Delta$ is p-wave pairing amplitude, and $\mu$ is the chemical potential. The chemical potential couples Majorana at the same lattice site, whereas $\Delta$ and $t$ couple two Majorana on nearest sites. Fermion operators and Majorana operators obey the following relations  
\eq{c_j = \frac{1}{2} (\gamma^B_{j} + i\gamma^A_{,j}), \qquad c_j^\dagger = \frac{1}{2} (\gamma^B_{j} - i\gamma^A_{j}),}
\eq{&\left\{\gamma^A,\gamma^B\right\}= \delta_{AB}, \qquad \gamma^\dagger = \gamma, \qquad \gamma^2 =1,\\
	&\left\{c_i,c_j\right\}=\left\{c_i^\dagger,c_j^\dagger\right\}=0, \qquad \left\{c_i,c_j^\dagger \right\}=\delta_{ij}.}
When the strength of pairing equals the hopping strength, we get
\eq{H = -\Delta \frac{i}{2}\sum_{j=0}^{N-1} \gamma^B_{j}\gamma^A_{j+1} +\mu \frac{i}{2}\sum_{j=0}^N \gamma^B_{j}\gamma^A_{j+1}.}
For $\Delta = t= 0$ and $\mu < 0$, only the chemical potential term remains. The coupling between neighboring sites is less than the coupling between the same site. All Majorana fermions are paired as shown in Fig. \ref{KitaveToy} (top). It is a trivial phase, we have
\eq{H = \mu \frac{i}{2}\sum_{j=0}^N \gamma^B_{j}\gamma^A_{j}.}
For $\mu=0$ and $t=\Delta$, we have
\eq{H = -\frac{i}{2}\Delta \sum_{j=0}^{N-1} \gamma^B_{j}\gamma^A_{j+1}.}
This is the case in Fig. \ref{KitaveToy} (bottom). There are Majorana fermions at adjacent sites. By adjusting the $\Delta$ and $\mu$, we can have the phase transition between the trivial and topological superconductors. In terms of the fermion operators, the Hamiltonian is given as
\eq{H=\Delta \sum_{x=1}^{N-1}(c_x^\dagger c_x -\frac{1}{2}),}
zero-energy mode at each end. These can be used as a qubit. 

\subsection{Semiconductor Heterostructure}
The p-wave superconductors are a theoretical construct. They are rare in nature. Fu and Kane proposed that they can be created with the combination of s-wave superconductors and topological insulators \cite{fu2008superconducting}. Later, a combination of s-wave superconductors with semiconductors was also proposed \cite{sau2010generic,alicea2010majorana}. For further details on materials such as topological insulators, topological superconductors, and heterostructures, and their detection techniques, see the reviews \cite{culcer2020transport,zhu2021majorana}.

\begin{figure*}[h!]
	\centering
	\begin{subfigure}{0.4\textwidth}
		\includegraphics[scale=0.34]{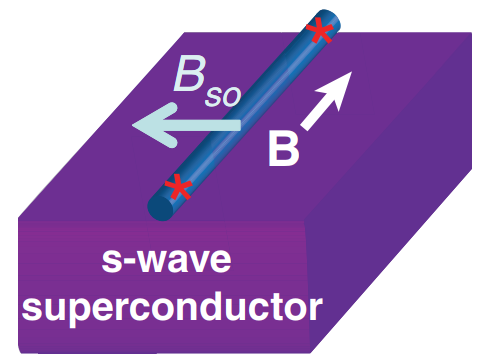}
		\caption{}
	\end{subfigure}
	\begin{subfigure}{0.4\textwidth}
		\includegraphics[scale=0.45]{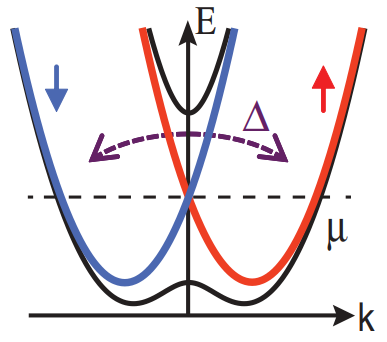}
		\caption{}
	\end{subfigure}	
	\caption[Physical realization of Majorana fermions on a nanowire.]{(a) A semiconductor nanowire is combined with the s-wave superconductor. A magnetic field is applied along the nanowire that caused the Zeeman splitting. (b) Blue and red curves correspond to the sub-bands in the presence of the Rashba spin-orbit coupling, and black curves are Zeeman gapped states obtained after the application of external magnetic field. These diagrams are taken directly from Ref. \cite{zhu2021majorana}.}
	\label{Nanowire}
\end{figure*}

The idea is to put a nanowire of material, with a large spin-orbit coupling such as InAs or InSb, in contact with the s-wave superconductor and apply the external magnetic field to the semiconductor. The strong spin-orbit coupling locks the spin-up and spin-down modes with orbital momentum. The external magnetic field will cause the Zeeman splitting and try to align both the modes to one direction and make the spinless configuration as shown in Fig. \ref{Nanowire} (a). The bandgap is introduced by the Zeeman splitting, as shown in Fig. \ref{Nanowire} (b). When this material is proximatized with the s-wave superconductors, due to the Andreev reflection, the electron tunnel to the superconductor, and a hole is reflected at the boundary. This makes an electron-hole pair at the boundary of the nanowire. For a suitable value of the chemical potential, superconducting gap, and magnetic field, the system turned to the Kitaev chain. Majorana zero modes appear when the chemical potential is within the Zeeman gap, as represented by the dashed line in Fig. \ref{Nanowire}. See \cite{alicea2012new} for details. The evidence for Majorana fermion is detected \cite{mourik2012signatures}. The Hamiltonian for the nanowire can be written as 
\eq{H = H_{wire} + H_\Delta,}
where $H_\Delta$ is associated with the spin-singlet pairing entering from the s-wave superconductor due to the proximity effect. The $H_{wire}$ and $H_\Delta$ can be written as
\eq{H_{wire}= \int dx\psi^\dagger\Big(-\frac{\hbar^2\partial_x^2}{2m}-\mu-i\hbar \alpha \hat{\textbf{e}}.\bm{\sigma} \partial_x - \frac{g\mu_B B_z}{2}\sigma^z\Big)\psi,}
\eq{H_\Delta = \int dx (\abs{\Delta} e^{i\phi}\psi_{\downarrow x} \psi_{\uparrow x}+h.c.),}
where $\psi^\dagger$ adds an electron with the chemical potential $\mu$ and effective mass $m$, $\alpha$ is the spin-orbit coupling strength and $\bm{\sigma}=(\sigma^x,\sigma^y,\sigma^z)$ is the vector of Pauli matrices, and $\hat{\textbf{e}}$ is a unit vector in $(x,y)$ plane along which the spin-orbit coupling is favored. The last term corresponds to the Zeeman coupling due to the magnetic field $B_z$. The eigenvalues of the Hamiltonian is now given by
\eq{\epsilon_{\pm}(k) = \frac{k}{2m}-\mu \pm \sqrt{(\alpha k)^2 +B_z}.}
When $B_z=\Delta =0$ there is no spinless regime. This is shown as red and blue curves in Fig. \ref{Nanowire}. For nonzero $B_z$, there is a bandgap as shown by black lines. For nonzero $\Delta$, the material becomes p-wave in the lower band, with $k$ and $-k$. The realization of the topological phase requires that $\abs{\Delta}<g\mu_B\abs{B_z}/2$ and $B_z> \sqrt{\Delta^2+\mu^2}$. The gap vanish when $B_z= \sqrt{\Delta^2+\mu^2}$. Below this $B_z$ there is no spinless state. The edges of these systems support Majorana modes, since time-reversal symmetry is broken by the Zeeman term. The Majorana modes obey particle-hole symmetry, since quasiparticle at $E$ and quasihole at energy $-E$ are equal. Therefore, at zero energy there is degeneracy and zero energy is needed to switch from one state to another. Hence, these are also known as \textit{Majorana zero mode (MZMs)}. The degeneracy leads to non-Abelian statistics. More general no-Abelian anyons can be obtained by the proximity of superconductors and quantum Hall states, see \cite{barkeshli2012topological,clarke2013exotic,lindner2012fractionalizing,cheng2012superconducting,vaezi2013fractional}.

\section{Majorana Fermion and Topological Quantum Computation}
Suppose a zero-mode $\gamma$ is localized in a vortex core. A single zero-mode $\gamma$ cannot define the creation and annihilation operator. The zero mode is self conjugate, so it satisfies one condition of Majorana fermion
\eq{\gamma^\dagger = \gamma} 
But if $\gamma^\dagger$ is a creation operator and $\gamma$ is an annihilation operator, but they are both equal, there is a contradiction. Therefore, $\gamma^\dagger$ cannot be considered as a creation operator. 
This difficulty is resolved by considering a pair of vortices. Let $\gamma_1$ and $\gamma_2$ be Majorana modes for vortices 1 and 2 that obey anticommutation relation $\left\{\gamma_i, \gamma_j \right\} = 2 \delta_{ij}$. The fermion creation and annihilation operators $c^\dagger$ and $c$ can be written in terms of the Majorana operators as
\eq{c^\dagger = \frac{\gamma_1 + i \gamma_2}{2} , \ \ c = \frac{\gamma_1 - i \gamma_2}{2}.}
These operators satisfy the relation
\eq{\left\{c^\dagger, c\right\} = 1.} 
The separated vortices are necessary for the creation and annihilation operators, which leads to a non-local correlation between the vortices and results in drastic changes to statistics \cite{ivanov2001non,read2000paired}. The degeneracy leads to non-Abelian statistics.

On exchanging two bosons or fermions, the final state remains the same as the initial state. But in general, the final state wave function in 2D p-wave superconductors is different from the initial state due to the existence of Majorana zero mode particles. Therefore, the Majorana particles can be used as a physical system for topological quantum computation.

For a particle $\gamma$ to be a Majorana fermion, it must satisfy the properties as $\gamma$ is fermionic, $\gamma^2 = 1$, $[H,\gamma]=0$. The second condition implies that the corresponding operator is self-adjoint, and the third condition implies that it is a zero mode. They come in pairs, so for $2n$ isolated vortices, there is a $2^n$-fold degeneracy. As the $\gamma$ operators are real, exchanging them will change their signs. The fermion parity must be conserved, which implies that $\gamma_1$ and $\gamma_2$ must pick opposite signs, that is, 
$\gamma_1 \rightarrow - \gamma_2, \ \gamma_2 \rightarrow \gamma_1$. This transformation is generated by the unitary operator
\eq{U=e^{i\theta}e^{\frac{\pi}{4}\gamma_1\gamma_2}.}

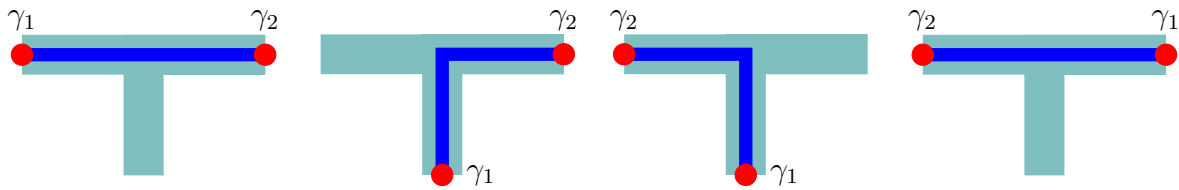
\begin{figure}[h!]
	\centering
	\tikzset{middlearrow/.style={decoration={markings,mark= at position 0.6 with {\arrow{#1}},},postaction={decorate}}}
	\begin{tikzpicture}[scale=0.8]
		\draw[teal!50,line width=15](0,-2)--(0,0)--(2,0)--(-2,0);
		\draw[line width=5,blue] (-2,0)--(2,0);
		\filldraw[red] (-2,0) circle (5pt);
		\filldraw[red] (2,0) circle (5pt);
		\node[above=5] at (-2,0) {$\gamma_1$};
		\node[above=5] at (2,0) {$\gamma_2$};
	\end{tikzpicture}
	\begin{tikzpicture}[scale=0.8]
		\draw[teal!50,line width=15](0,-2)--(0,0)--(2,0)--(-2,0);
		\draw[line width=5,blue] (0,-2)--(0,0)--(2,0);
		
		\filldraw[red] (2,0) circle (5pt);
		\filldraw[red] (0,-2) circle (5pt);
		\node[right=5] at (0,-2) {$\gamma_1$};
		\node[above=5] at (2,0) {$\gamma_2$};
	\end{tikzpicture}
	\begin{tikzpicture}[scale=0.8]
		\draw[teal!50,line width=15](0,-2)--(0,0)--(2,0)--(-2,0);
		\draw[line width=5,blue] (0,-2)--(0,0)--(-2,0);
		
		\filldraw[red] (-2,0) circle (5pt);
		\filldraw[red] (0,-2) circle (5pt);
		\node[above=5] at (-2,0) {$\gamma_2$};
		\node[right=5] at (0,-2) {$\gamma_1$};
	\end{tikzpicture}
	\begin{tikzpicture}[scale=0.8]
		\draw[teal!50,line width=15](0,-2)--(0,0)--(2,0)--(-2,0);
		\draw[line width=5,blue](-2,0)--(2,0);			
		\filldraw[red] (-2,0) circle (5pt);
		\filldraw[red] (2,0) circle (5pt);
		\node[above=5] at (-2,0) {$\gamma_2$};
		\node[above=5] at (2,0) {$\gamma_1$};
	\end{tikzpicture}
	\caption[Braiding two Majorana fermions.]{Schematics for the braiding of two Majorana fermion \cite{alicea2011non}. The Majorana particles as red spheres at the edges of 1D nanowire on conventional superconductor.}
	\label{MajoranaBraiding}
\end{figure}

\begin{figure}[h!]
	\centering
	\tikzset{middlearrow/.style={decoration={markings,mark= at position 0.6 with {\arrow{#1}},},postaction={decorate}}}
	\begin{tikzpicture}[scale=0.8]
		\draw[teal!50,line width=15](0,-2)--(0,0)--(2,0)--(-2,0);
		\draw[line width=5,blue] (-2,0)--(-0.3,0);
		\draw[line width=5,blue] (0.3,0)--(2,0);			
		\filldraw[red] (-2,0) circle (5pt);
		\filldraw[red] (-0.3,0) circle (5pt);		
		\filldraw[red] (0.3,0) circle (5pt);
		\filldraw[red] (2,0) circle (5pt);
		\node[above=5] at (-2,0) {$\gamma_1$};
		\node[above=5] at (-0.3,0) {$\gamma_2$};
		\node[above=5] at (0.3,0) {$\gamma_3$};
		\node[above=5] at (2,0) {$\gamma_4$};
	\end{tikzpicture}
	\begin{tikzpicture}[scale=0.8]
		\draw[teal!50,line width=15](0,-2)--(0,0)--(2,0)--(-2,0);
		\draw[line width=5,blue] (-2,0)--(-0.3,0);
		\draw[line width=5,blue] (0,-2)--(0,0)--(2,0);		
		\filldraw[red] (-2,0) circle (5pt);
		\filldraw[red] (-0.4,0) circle (5pt);			
		\filldraw[red] (2,0) circle (5pt);
		\filldraw[red] (0,-2) circle (5pt);
		\node[above=5] at (-2,0) {$\gamma_1$};
		\node[above=5] at (-0.4,0) {$\gamma_2$};
		\node[right=5] at (0,-2) {$\gamma_3$};
		\node[above=5] at (2,0) {$\gamma_4$};
	\end{tikzpicture}
	\begin{tikzpicture}[scale=0.8]
		\draw[teal!50,line width=15](0,-2)--(0,0)--(2,0)--(-2,0);
		\draw[line width=5,blue] (-2,0)--(0,0)--(0,-2);
		\draw[line width=5,blue] (0,0)--(2,0);
		\filldraw[red] (-2,0) circle (5pt);
		\filldraw[red] (0,0) circle (5pt);			
		\filldraw[red] (2,0) circle (5pt);
		\filldraw[red] (0,-2) circle (5pt);
		\node[above=5] at (-2,0) {$\gamma_1$};
		\node[above=5] at (0,0) {$\gamma_2$};
		\node[right=5] at (0,-2) {$\gamma_3$};
		\node[above=5] at (2,0) {$\gamma_4$};
	\end{tikzpicture}
	\begin{tikzpicture}[scale=0.8]
		\draw[teal!50,line width=15](0,-2)--(0,0)--(2,0)--(-2,0);
		\draw[line width=5,blue] (-2,0)--(0,0)--(0,-2);		
		\draw[line width=5,blue] (0.4,0)--(2,0);
		\filldraw[red] (-2,0) circle (5pt);
		\filldraw[red] (0.4,0) circle (5pt);			
		\filldraw[red] (2,0) circle (5pt);
		\filldraw[red] (0,-2) circle (5pt);
		\node[above=5] at (-2,0) {$\gamma_1$};
		\node[right=5] at (0,-2) {$\gamma_3$};
		\node[above=5] at (0.2,0) {$\gamma_2$};
		\node[above=5] at (2,0) {$\gamma_4$};
	\end{tikzpicture}
	\caption[Braiding more than two Majorana fermions.]{Braiding of more than two Majorana fermions \cite{litinski2017combining}.}
	\label{MajoranaBraiding2}
\end{figure}
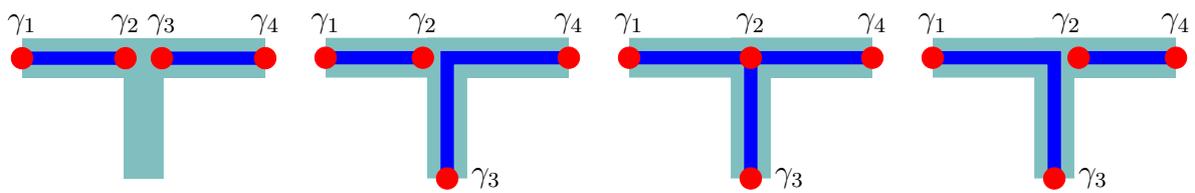

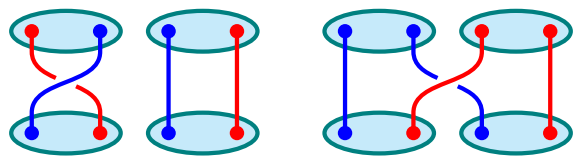
\begin{figure*}[h!]
	\centering
	\begin{tikzpicture}[scale=0.9]
		\draw [teal,ultra thick,fill=cyan!20] (1.5,0) ellipse (0.8 and 0.3);
		\draw [teal,ultra thick,fill=cyan!20] (1.5,-1.5) ellipse (0.8 and 0.3);
		\draw [teal,ultra thick,fill=cyan!20] (3.5,0) ellipse (0.8 and 0.3);
		\draw [teal,ultra thick,fill=cyan!20] (3.5,-1.5) ellipse (0.8 and 0.3);
		\braid[teal,ultra thick, style strands={1}{red},style strands={2}{blue},style strands={3}{teal}]
		s_1^{-1};
		\draw[ultra thick,blue] (3,-1.5) -- (3,0);
		\draw[ultra thick,red] (4,-1.5) -- (4,0);
		\fill[red] (1,0) circle(3pt);
		\fill[blue] (2,0) circle(3pt);
		\fill[blue] (1,-1.5) circle(3pt);
		\fill[red] (2,-1.5) circle(3pt);
		\fill[blue] (3,0) circle(3pt);
		\fill[red] (4,0) circle(3pt);
		\fill[blue] (3,-1.5) circle(3pt);
		\fill[red] (4,-1.5) circle(3pt);	
	\end{tikzpicture}\qquad
	\begin{tikzpicture}[scale=0.9]
		\draw [teal,ultra thick,fill=cyan!20] (0.5,0) ellipse (0.8 and 0.3);
		\draw [teal,ultra thick,fill=cyan!20] (0.5,-1.5) ellipse (0.8 and 0.3);
		\draw [teal,ultra thick,fill=cyan!20] (2.5,0) ellipse (0.8 and 0.3);
		\draw [teal,ultra thick,fill=cyan!20] (2.5,-1.5) ellipse (0.8 and 0.3);
		\braid[teal,ultra thick, style strands={1}{blue},style strands={2}{red},style strands={3}{teal}]
		s_1^{-1};
		\draw[ultra thick,red] (3,-1.5) -- (3,0);
		\draw[ultra thick,blue] (0,-1.5) -- (0,0);	
		\fill[blue] (0,0) circle(3pt);
		\fill[blue] (1,0) circle(3pt);	
		\fill[blue] (0,-1.5) circle(3pt);
		\fill[red] (1,-1.5) circle(3pt);	
		\fill[red] (2,0) circle(3pt);
		\fill[blue] (2,-1.5) circle(3pt);	
		\fill[red] (3,0) circle(3pt);	
		\fill[red] (3,-1.5) circle(3pt);		
	\end{tikzpicture}
	\caption[The change of state on braiding of the Majorana fermions.]{Majorana braiding of the same pair does not change the ground state, whereas the braiding of a Majorana from one pair with the one from other pair change the state.}
	\label{MajoranaChangeState}
\end{figure*}

This is braiding of Ising anyons for $\theta = \pi/8$, and we have $U=\exp(-i\frac{\pi}{4} \gamma_1\gamma_2)=(1+\gamma_1\gamma_2)/\sqrt{2}$ \cite{ivanov2001non,sarma2015majorana}. The braiding of two and more than two Majorana particles on \textit{T-junction} \cite{alicea2011non,litinski2017combining} are shown in Fig. \ref{MajoranaBraiding} and in Fig. \ref{MajoranaBraiding2}. The braiding of two Majorana fermions does not change the ground state, but only get an Abelian phase with their wave function. Whereas, a braiding of a particle from one pair to that of from the other pair may put the system in another ground state. This is shown in Fig. \ref{MajoranaChangeState}. The fusion and braiding matrices of Ising anyons are discussed in Chapter \ref{TQC}. In the Ising model, there are three particles $I,\sigma, \psi$. Here the $\sigma$ is a Majorana fermion, whereas the $\psi$ is an electron.
	\chapter{Topological Quantum Field Theories}\label{TQFT}
	
Quantum field theory can be used to describe the quasiparticles in condensed matter physics. The topological phases are defined as the materials whose low energy effective field theory is the topological quantum field theory (TQFT). The TQFT has observable properties, such as the correlation function, that are invariant under the smooth deformation of spacetime manifold in which is the system lives \cite{nayak2008non}. Low-energy means that the energy of our system should be low enough that the gap between the ground state and excited state should not be filled.
A quantum field is an entity in spacetime that can show a wavelike behavior whose amplitude can be a scalar, a vector, a complex number, or a tensor.

Energy and mass are interconvertible according to the relativity theory. At high energies, particles can be created and destroyed. Quantum field theory is formulated by combining special relativity with quantum mechanics, and therefore accounts for the creation and annihilation of particles. In relativity, time is dealt with equal footings as a space dimension, therefore, we use four spacetime coordinates $(t,\bm{x}) = (t,x,y,z)=(x^0,x^1,x^2,x^3)$, three space and one time. The observable quantities are defined as the four-vectors. For example, $x^\mu=(t,\bm{x}),\ p^\mu=(E,\bm{p}), \ A^\mu= (\phi,\bm{A}), \ j^\mu = (\rho,\bm{j})$, and $\partial_\mu \equiv \partial/\partial x^\mu = (\partial/\partial t,\nabla)$, where the boldface letters mean the three space components. The dynamics of the fields are studied by using the Lagrangian formulation. The quantum field is obtained by quantization of a classical field. The theory related to this quantization is called the quantum field theory. The classical fields are promoted to the field operators whose application on the vacuum corresponds to the creation of particles at a particular spacetime point. All the particles are excitations in their corresponding fields. The basics of quantum field theory are given in Appendix \ref{QFT}, see \cite{tom2015quantum,mcmahon2008quantum} for further knowledge. 

There are two ways to formulate a quantum field theory; canonical quantization, and Feynman path integrals. We will only talk about the path integral approach in this chapter. For second quantization, see Appendix \ref{QFT}. In the path integral approach, we compute the amplitude of having a final configuration of a field such that an initial configuration of the field is given. This probability amplitude is called the \textit{partition function} of the theory. See Appendix \ref{QFT} for the derivation of the partition function.
The fundamental formula in quantum field theory is given by
\eq{Z[A] = \int dA e^{iS(A)/\hbar}=\int dA e^{i\int d^4x \mathcal{L}(A)},\label{ZCS}}
where $S$ is the action given as $S= \int dt L = \int dt dx \mathcal{L}$ and $L$ and $\mathcal{L}$ are the Lagrangian and Lagrangian density.
The Euler-Lagrange equation gives the equation of motion, which is found by minimizing the action with respect to the functional. A functional is a function of a function. Here the action is a function of the field, which is a function of spacetime. A quantum particle follows all possible paths between two points. A phase is associated with each path. The integral is a weighted sum of the contributions from all the paths. 

The action of a charge particle in an electromagnetic field is written as \cite{pachos2012introduction}
\eq{S[A] = \frac{1}{2} \int d^4x (\bm{E}^2 +\bm{B}^2 +\bm{A \cdot J} + A_0\rho),} 
where $\bm{E},\bm{B},\bm{A}$ and $\rho$ are the electric field, magnetic field, vector potential and charge density. We used the notation  $x= (t,\bm{x})=(x^0,x^1,x^2,x^3)$ and $A=(A_0,\bm{A})$.
The equations of motion are Maxwell's equations that give the interaction between charge particles and the electromagnetic field. The Maxwell's equations are written as
\eq{\nabla \cdot \bm{E} = \rho, \ \nabla \cross \bm{B} - \partial_t \bm{E} = \bm{J},\nonumber\\
	\nabla \cdot \bm{B} = 0, \ \nabla \cross \bm{E} - \partial_t \bm{B} = 0.}
If we transform the vector potential as $\bm{A}\to \bm{A}' = \bm{A}+ \nabla \lambda,\ A_0 \to A'_0 = A_0 + \partial_t \lambda$, the action and $\bm{E}$ and $\bm{B}$ will remain invariant. This transformation is called the \textit{gauge transformation}. 
Now let our system be confined to three spacetime dimensions, that is, two spatial and one time dimension. The coordinates and the field can be written as $x= (t,\bm{x}) = (x^0,x^1,x^2)$ and $A=(A_0,\bm{A}) = (A_0,A_1,A_2)$. Now the action in $(2+1)$-dimensions is given by
\eq{S[A] = \frac{1}{2} \int d^4x (\bm{E}^2 +\bm{B}^2 +\bm{A\cdot J} + A_0\rho + \frac{k}{2\pi} \varepsilon^{\mu\nu\rho}a_\mu \partial_\nu a_\rho) \label{TopAction},}
The indices $\mu,\nu,\rho$ take the values from 0 to 3. $\varepsilon$ is the Levi-Civita symbol. It is zero when more than one index are equal, 1 when the indices are in a cyclic permutation, and -1 otherwise. The last term can be written simply as $\frac{k}{2\pi}\int a\mathcal{D}a$. This term is independent of the coordinate reparameterization and the gauge transformation, hence this action is topological in nature. Its topological nature will be apparent later. The Euler-Lagrange equation of motion for the action in Eq. \ref{TopAction} will involve the potential $a$, all other terms decay exponentially at larger distances \cite{pachos2012introduction}.

\section{Abelian Chern-Simons Theory}
The Chern-Simons theory \cite{chern1974characteristic} is an effective field theory that describes the low-energy degrees of freedom in topological materials. It is a gauge theory valid for energies much smaller than the energy gap between the ground state and the excited state. 
When a system is confined to two dimensions, the path integrals involve the action and the Lagrangian is independent of the metric. The metric is related to distances in space. In topology, the distances are irrelevant. 
The Chern-Simons action is written as 
\eq{S_{CS} = \frac{k}{4\pi} \varepsilon^{\mu \nu \rho}a_\mu \partial_\nu a_\rho,}
where $a_\mu$ is a $U(1)$ gauge field  and $k$ is a constant called the \textit{level} of the theory. It is the coupling constant of the Chern-Simons field. Let us have a coordinate transformation of $a_\mu(x)$ as $a_\mu(x) = \frac{\partial x^\mu}{\partial x'^\nu}a_\nu(x')$. 
Under this transformation, the line integral $\int dx^\mu a_\mu = \int dx^\mu \frac{\partial x^\mu}{\partial x'^\nu}a_\nu(x')$ will remain invariant. Hence, the action is independent of the spacetime metric.

In the presence of the external electromagnetic field and the quasiparticles we have  
\eq{S= S_{CS}-\frac{k}{2\pi} k \varepsilon^{\mu \nu \rho}a_\mu \partial_\nu a_\rho + a_\mu J_\mu.}
In QFT, the equation of motion is calculated by using the relation $J_\mu = \partial \mathcal{L}/\partial A_\mu = \frac{\delta S_{CS}[a]}{\delta a_i}$. Therefore, we can have 
\eq{J_\mu = \frac{1}{2\pi} \varepsilon^{\mu \nu \rho} \partial_\nu a_\rho.} 
Integrating out the $a_\mu$ field, we can get a conductance from this current as 
\eq{\sigma_{xx}=0, \qquad \sigma_{xy} = \frac{1}{k}e^2/h.}
This is the quantum Hall conductance. Comparing with the Maxwell's theory, the counterpart for four-vector charge density is $J_\mu = (\rho,\mathbf{J})$, magnetic field $b= \partial_2 a_1 - \partial_1 a_2$, and electric field $e^i= \partial_0 a_i - \partial_i a_0$, we have
\eq{\rho = \frac{k}{2\pi}b, \qquad J^i = \frac{k}{2\pi} \varepsilon ^{ij} e^j.}
The charge from these equations can be obtained as
\eq{Q= k \int d^2x b.}
It implies that the charge of the source field is proportional to the flux of gauge field $a_\mu$. Hence, the Chern-Simons theory ties $a_\mu$ flux to the charge of the source field. The contribution of $e^{iS_{CS}}$ is the Aharanov-Bohm phase. The attachment of the flux to the charge causes a change in the statistics, which becomes fractional now. The Aharanov-Bohm phase of a moving charge around another causes a phase of $e^{iq\Phi}$, where $q$ is the charge and $\Phi$ is the flux. For the Chern-Simons theory, the charge is given by $q= k\Phi$. For this charge, the Aharanov-Bohm phase would be $e^{-i\frac{q^2}{k}}$. For the rotation of $2\pi$, the phase should be equal to $e^{2\eta\pi}$, therefore we have
\eq{2\pi\eta = -\frac{q^2}{k},}
where $\eta = -\frac{q^2}{2\pi k}$ is not necessarily an integer, hence we have fractional statistics. The Chern-Simons theory gives the Hall conductance of $\nu$ filled Landau levels if we identify the Chern-Simons level as $k=e^2\nu/\hbar$. Therefore, the Chern-Simons theory explains the quantum Hall effect, see Chapter \ref{QHE}. The Abelian Chern-Simons term implements the Abelian anyonic statistics. The Chern-Simons field has no dynamics and no local degrees of freedom.

\section{Non-Abelian Chern-Simons Theory}

By promoting the $a_\mu$ to the vector of matrices, the field $a_\mu$ takes the values in the Lie algebra which are non-commutative in general. The Lie algebra will be discussed in Appendix \ref{AbsAlg}. The generators in the fundamental representation of $SU(2)$ are $i\sigma_x, i\sigma_y, i\sigma_z$. The Lie algebra gauge field is
\eq{a_\mu(x) = a_\mu^k(x) \sigma_k \frac{i}{2},}
where $i/2$ is a convention. 
For $a_\mu$ being a matrix-valued quantity, the Chern-Simons action is
\eq{S_{CS} = \frac{\hbar k}{4\pi} \int_{M}d^3x \epsilon^{\alpha \beta \gamma} Tr \Bigg[a_\alpha \partial_\beta a_\gamma - \frac{2i}{3} a_\alpha a_\beta a_\gamma\Bigg].}
The second term appears because of the non-commutativity of $a_\mu$. This term is zero in the case of Abelian $a_\mu$. When the gauge group is $SU(2)$ at the level $k$, we call such theory a $SU(2)_k$ theory. By the same reasoning as given for the Abelian case, this action is independent of the spacetime metric. Topological nature is also apparent when the action is written in differential geometric form, as in the Ref. \cite{witten1989quantum}.

In the case of a non-Abelian gauge field, the gauge transformation is 
\eq{a_\mu \rightarrow Ua_\mu U^{-1} - i \frac{\hbar}{q} U\partial_\mu U^{-1},}
where $U:M\to G$ is a function on the manifold $M$. It has values in a group $G$. The Chern-Simons theory is not invariant under the gauge transformation. The \textit{winding number} $m$ classifies the gauge transformation as 
\eq{S_{CS} \rightarrow S_{CS} + 2\pi k m.}
It implies that the non-Abelian Chern-Simons theory is parameterized by the coupling constant that can only have integer values. The integer, winding number, counts how many times the gauge transformation covers the group $G$ as $x$ covers the manifold $M$.
The Chern-Simons action is invariant under a small gauge transformation, that is when $m=0$. It is not invariant under a large gauge transformation when $m\ne 0$. On a manifold, the Chern-Simons action is invariant up to a surface term. It is not invariant under the gauge transformation that does not vanish on the boundary. So there is a bulk-boundary correspondence. For example, on the torus, $m=0$ corresponds to the contractible loops which do not wind around the longitude or meridian. However, $e^{iS_{CS}}$ is an invariant when $k$ is an integer. A small perturbation to the Hamiltonian cannot change the value of $k$.
There are gapless edge modes whose dynamics can be determined from the properties of the bulk.

The Chern-Simon theory is an example of topological quantum field theories \cite{witten1989quantum}. The bulk-boundary correspondence is analogous to the conformal field theory (CFT) and topological field theory (TQFT) correspondence \cite{freidel20042d,felder2000conformal}. 
Let the boundary of a 3-manifold has punctures and the Wilson lines start and end on these punctures. A Hilbert space is associated with each boundary. A state is given by the Chern-Simons functional integral. The properties of these associated Hilbert spaces are also given by the conformal blocks of the $SU(2)_k$ Wess-Zumino-Witten conformal field theory on the boundary with punctures \cite{kaul1994chern}. In other words, the topological field theory in the bulk has a correspondence with the conformal field theory on the boundary \cite{witten1989quantum}.

\section{Knot Invariant and Chern-Simons Theory}
The quantity of our interest in this section is the Wilson loop operator. The expectation value of this operator gives the probability amplitude of the creation of particles from the vacuum at one spacetime point, traveling, and then annihilation at another spacetime point. This is the path integral approach in quantum field theory as we discussed in Appendix \ref{QFT}, except that the path integral in Chern-Simon theory depends on the topology of the path. We can take the trace of the gauge group in any representation. The field $a_\mu$ is a matrix, hence $a_\mu(x)$ and $a_\mu(x^{'})$ do not commute. 
Thus, if we choose $k$ as an integer, then the functional integral becomes
\eq{Z(M) = \int_M {\cal D}a_\mu(x) e^{iS_{CS}/\hbar}.}
The $Z(M)$ is a manifold invariant, that is a smooth deformation would not change its value. The $Z[M]$ depends only on the topology of 4-manifold, see the Section \ref{2TQFT}.

Physically, the knots are the path integrals along trajectories of particles. To compute a knot invariant from the Chern-Simons theory, let each particle be defined in some representation of a group. The representation of a group is defined in Appendix \ref{AbsAlg}. Different representations correspond to different particle types. A quantity called \textit{Wilson loop} operator is defined as 
\eq{W = Tr \Big[P\exp(i \frac{q}{\hbar} \oint_L dl^\mu a_\mu)\Big],}
where $P$ is the path ordering operator, analogous to the time ordering in quantum mechanics. The quantity inside the integral is called the holonomy. The trace of the holonomy is the Wilson loop \cite{baez1994gauge}. The holonomy is the mismatch of a `parallel transported' vector along a loop in some curved space with the original vector. The holonomy is related to the geometric phases we discussed in Chapter \ref{GeoPhase}.

The insertion of the Wilson loop operator inside the path integral gives the knot invariant of a link $L$. See  \cite{witten1989quantum} for further explanation. The link is to be thought of as embedded inside three manifold $S^3$. Similar to the path integral formulation in Chapter \ref{QFT}, a link invariant of two links $L_1$ and $L_2$ is written as
\eq{\text{Knot Invariant} = \langle W(L)\rangle = \frac{Z(S^3, L_1,L_2)}{Z(S^3)} = \frac{\int_{S^3} {\cal D}a_\mu(x)W_{L_1}W_{L_2} e^{iS_{CS}/\hbar}}{\int_{S^3} {\cal D}a_\mu(x) e^{iS_{CS}/\hbar}}.}
The expectation value of a loop inside $S^3$ for an unknot is given by \cite{witten1989quantum}
\eq{\langle W(L)\rangle = \frac{Z_{CS}(S^3;C)}{Z_{CS}(S^3)}= dim_q\alpha_i.}
This quantity is a single loop, as we saw in the skein relations of the Kauffman bracket in Chapter \ref{Knot}. On the right-hand side, we have the quantum dimension of the particle. This number is defined in the context of the Hilbert space in Chapter \ref{TQC}.
When $C$ is a link that is a disjoint union of the knot components $C_1,...,C_m$, then the Wilson loop is written as a product of the Wilson loops for each component. 
\eq{W(L) = W(C_1,...,C_m) = \prod_{j=1}^{m} W(C_j)}
The knot invariant would be the Jones polynomial when the framing is also considered. If $w(L)$ is writhe in a link, then the link invariant in $U(1)$ representation can be written as $\langle L \rangle = e^{i\pi w(L)/k}$. The writhe is defined in Chapter \ref{Knot} in the context of knot theory. In Appendix \ref{HopfAlg}, $SU(N)$ representation and framing are considered to calculate the knot invariant.

Let us calculate the Wilson loop around two different cyclic paths $L_1$, $L_2$ on torus along the longitude and meridian as
\eq{W_j = \exp(iq/\hbar\oint_{L_i} \vec{dl}\cdot \vec{a})= e^{iq w_i/\hbar}.}
This is the non-Abelian analog of the Aharanov-Bohm effect. The gauge transformation would shift the gauge field and the $w_i$. The phase will also shift correspondingly, so that for an integral $k$ we will get the same phase $w_i$ as for zero shift of the gauge field. This reminds us of the effect of the flux change in the quantum Hall effect, see Fig. \ref{LaughlinArgu}. As we know, $e^Ae^B = e^Be^A e^{[A,B]}$. This holds when $[A, B]$ is a number. Therefore, we have  
\eq{W_1W_2 = e^{i\theta}W_2W_1 \label{WilsonQG},}
where the $\theta$ is a statistical angle of the theory. This result is similar to the one we discussed in Chapter \ref{GeoPhase} on geometric phase and anyons. The operators $W$ correspond to the $T$ operators, which create particles, move them around the two loops, and fuse them later. We saw in Section \ref{AnyonTorus} that the degeneracy of the ground state is related to the genus and the number of types of particles. Eq. \ref{WilsonQG} is also related to the quantum group symmetry in Chern-Simons theory \cite{ho1996quantum,guadagnini1990braids}. When the framing is also considered, the knot invariant for a manifold $M$ is calculated in Appendix \ref{HopfAlg} on the same lines as derived in the seminal paper by Edward Witten \cite{witten1989quantum}.
	\section{Two-Dimensional Topological Quantum Field Theory}\label{2TQFT}
Consider the path integral on a closed 3-manifold $M$. A manifold is a surface locally looking like Euclidean space, but globally it may have a complicated structure. The fundamental formula in quantum field theory, as written in Eq. \ref{ZCS} is given by 
\eq{Z(M) = \int_{A_{| \Sigma_1 = A_1}}^{A_{| \Sigma_2 = A_2}} {\cal {D}}A \exp(iS[A]/\hbar),}
where $S$ is the action, and $Z(M)$ is called the partition function. It is the probability amplitude of getting a final configuration of the field given an initial configuration. The link invariant $Z(M)$ calculated in terms of the Chern-Simons theory is often not well-defined. Therefore, the link invariant in the case of a two-dimensional topological quantum field theory (TQFT) can be advantageous where the links are labeled by particle types.
The TQFT is a set of rules which gives output as a complex number such that input is given as a labeled link embedded in a 3-manifold. In this way, the Hilbert space can be defined on a topological manifold. The TQFT is a field theory that depends on the topological properties, not on the geometric properties. It implies that the TQFT gives us $Z(M)$ which is independent of the metric \cite{bartlett2005categorical}, therefore it is a \textit{topological invariant}.

\subsection{Atiyah's Definition of TQFT}
First, we will consider a spacetime manifold without particles. It is still non-trivial. Particles, and their motion around, will be added to this manifold later on. Here we have a $(d+1)$ dimensional manifold $M$ is considered and $\Sigma$ is a $d$-dimensional oriented slice of the manifold $M$. This definition can be generalized to any dimension.  In our case, we will consider $d=2$ and $\Sigma$ slice can be considered as a space of dimensions $d$ at some fixed time.  
Atiyah's definition of $d$-dimensional TQFT \cite{atiyah1990geometry} consists of the following axioms.

\textbf{Axiom 1:}
A vector space $V(\Sigma)$ is associated with a $d$-dimensional space $\Sigma$, which depends only on the topology of $\Sigma$.
Sometimes this $V$ is also called $H$ for Hilbert space. For example, when $\Sigma$ is a torus, then due to degenerate ground states, we have a non-trivial Hilbert space $V(\Sigma)$. As the degeneracy depends on the particle type, space $V(\Sigma)$ also depends on the particle type. 

\textbf{Axiom 2:}
The Hilbert space of two disjoint Hilbert spaces of $\Sigma_1$ and $\Sigma_2$ is the tensor product of the spaces of each $\Sigma$
\eq{V(\Sigma_1 \cup \Sigma_2) = V(\Sigma_1) \otimes V(\Sigma_2)}
This implies 
$V(\emptyset) = \mathbb{C}$ so $\emptyset \cup \Sigma = \Sigma$ and $\mathbb{C} \otimes V(\Sigma) = V(\Sigma)$. 

\textbf{Axiom 3:}
If the manifold $M$ is a $(d+1)$-dimensional manifold with boundary $\partial M = \Sigma$ , then associated with this manifold is a particular element of the vector space $V(\Sigma)$ given by
\eq{Z(M) \in V(\partial M)}
This association depends only on the topology of the manifold. Here $\partial M$ can be thought of as a time slice of the system at some fixed time, and $V(\partial M)$ some possible Hilbert space of the ground state. The interior of the rest of the manifold, other than a boundary, is the spacetime history of the system. $Z(M)$ is a wave function that is picked out by this spacetime history.
For a closed manifold, we have $\partial M = \emptyset$, so $Z(M) \in \mathbb{C}$. 

\textbf{Axiom 4:}
Reversing the orientation of the surface $\Sigma$ gives a dual vector space $V^*$, that is, bras turn into kets. 
\eq{V(\Sigma^*) = V^*(\Sigma)}

\subsection{Cobordism}
As we discussed in Appendix \ref{Top}, a topological surface can be imagined as the gluing of smaller manifolds. This is called \textit{surgery} in topology. In TQFT, the knot invariant of a manifold can be computed from the knot invariant of the smaller pieces. We need to study how to glue the pieces to make the manifold, or equivalently, how to break the manifold into pieces. 

\begin{figure}[h!]
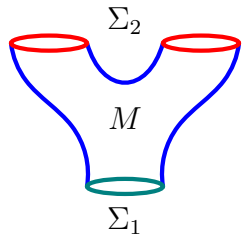

	\centering
	\tik{
		[use Hobby shortcut,scale=1]
		\begin{knot}[
			consider self intersections=true,
			ignore endpoint intersections=false,
			flip crossing=2,
			only when rendering/.style={}]
			\strand [ultra thick,knot=blue](0.5,-0.4).. (0.7,0.4).. (1.3,1).. (1.5,1.5);
			\strand [ultra thick,knot=blue](-0.5,-0.4).. (-0.7,0.4)..(-1.3,1).. (-1.5,1.5);
		\end{knot}	
		\draw[ultra thick,blue](-0.5,1.5) .. controls(-0.3,0.8) and (0.3,0.8)..(0.5,1.5);
		\draw [ultra thick,teal] (0,-0.4) ellipse (0.5 and 0.1);
		\draw [ultra thick,red] (-1,1.5) ellipse (0.5 and 0.1);
		\draw [ultra thick,red] (1,1.5) ellipse (0.5 and 0.1);
		\node at (0,0.5) {$M$};
		\node at (0,1.8) {$\Sigma_2$};
		\node at (0,-0.85) {$\Sigma_1$};}
	\caption{Cobordism, $\partial M = \Sigma_1^*\cup \Sigma_1$}
	\label{Cob}
\end{figure}

We are taking the time direction upward. Let the \textit{in-boundary} at an initial time be called $\Sigma_1$, and at a later time we have $\Sigma_2$, the \textit{out-boundary}. The evolution between these two surfaces is a manifold, say $M$. We can see that when $\Sigma$ is 2-dimensional, then $M$ is 3-dimensional. This can be extended to any number of dimensions when $\Sigma$ is a $d-1$-dimensional manifold and $M$ is a $d$-dimensional manifold, but we will only discuss the case of a two-dimensional $\Sigma$. 
If the disjoint union of $\Sigma_1$ and $\Sigma_2$ is the boundary of a manifold $M$, that is $\partial M = \Sigma_1 \cup \Sigma_2$, then we say that $M$ is a \textit{cobordism} between $\Sigma_1$ and $\Sigma_2$, and $\Sigma_1$ and $\Sigma_2$ are cobordant, as shown in Fig. \ref{Cob}. In abstract algebra, an interval between two points $a$ and $b$ is given as $I=[a,b]$. Correspondingly, the interval between the initial time and the final time we call $I$ so that the cobordism can be written as $M=\Sigma\cross I$. The algebra used for the cobordism is called \textit{Frobenius algebra} \cite{kock2004frobenius}.

The composition of a cobordism $M$ and a cobordism $M'$ correspond to the passage of time followed by another passage of time. This is equivalent to the total passage of time given by the cobordism $MM^{'}$. No effect occurs on the state if topology does not change with time. This means TQFT has no local degrees of freedom. $Z$ is unitary means that the time evolution operators for a process and its time reversal process are complex conjugates of each other. In other words, the time evolution is unitary if topology does not change. In Chapter \ref{Cat}, we will discuss that TQFT is a functor from the category of manifolds to the category of vector spaces, where the particle are objects in a category and the cobordisms are morphisms.

\subsection{Hilbert Space on a Two-Manifold}
\begin{figure}[h!]
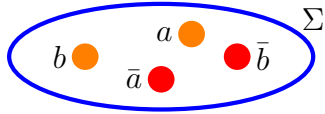

	\centering
	\tik{\draw [ultra thick,blue] (2,3) ellipse (2 and 0.7);
		\fill[orange] (1,3) circle(5pt);
		\fill[red] (2,2.7) circle(5pt);
		\fill[orange] (2.4,3.3) circle(5pt);
		\fill[red] (3,3) circle(5pt);
		\node [] at (4,3.5) {$\Sigma$};
		\node [left=3] at (1,3) {$b$};
		\node [left=3] at (2,2.7) {$\bar{a}$};
		\node [left=3] at (2.4,3.3) {$a$};
		\node [right=3] at (3,3) {$\bar{b}$};
	}
	\caption{Time slice of a manifold with particles.}
	\label{TimeSlice}
\end{figure}

Let us have a two-dimensional manifold as a disc $\Sigma$. This manifold is topologically equivalent to a sphere with one hole. Let time be in the upward direction. When there are no holes on the two-dimensional manifold $\Sigma$, the spacetime history will make a cylindrical manifold $M$ whose boundary is $\Sigma$. 
In Fig. \ref{TimeSlice}, we have holes or punctures on a two-dimensional boundary $\Sigma$ of a manifold $M$. In this section, puncture, hole or punctured hole are used interchangeably. When we move these punctures around each other, the trajectories or worldlines of the punctures will make braids in $(2+1)$-dimensional spacetime, and the spacetime history will have a nontrivial topology as shown in Fig. \ref{Worldlines}. Imagine an orientation is assigned to each punctured (not shown in Figs.) so that the punctures with one type of orientation make a vector space and the punctures with opposite orientation make a dual vector space.

Now we will identify the physical picture of TQFT with the anyonic model. Here the boundary $\partial M= \Sigma$ can be thought of as a slice of the system at an instant of time, and $V(\partial M)$ as a possible Hilbert space of the ground state.
In Fig. \ref{Worldlines}, worldlines correspond to creation, fusion, and braiding. It is the evolution of the system that changes the system from one ground state to the other in general. Sky-blue strands correspond to the creation of particle-antiparticle pairs, the green strand represents the fusion of particles to the vacuum. The orange loop corresponds to the creation and fusion of a particle-antiparticle pair.
We can also observe that the state at $\Sigma$ and the state at $\Sigma'$ look similar locally, but globally they are different in general. We also see that a manifold can be a composition of two separate pieces, or we can cut the manifold at some particular instant of time. The identification of the in-boundary to the out-boundary is given by taking the trace. This identification will make knots and links of the worldlines.
The opposite orientations of the punctured holes correspond to the directed trajectories of particles and antiparticles. In quantum field theory, a trajectory of an antiparticle is in the backward time direction comparing with that of a particle. When a particle is its own antiparticle, the corresponding vector space and dual vector space are equal, therefore we do not need directed trajectories. The \textit{twist} of a puncture corresponds to the topological spin obtained when an anyon rotates around its own magnetic flux.

\fig{[h!]\centering
	\begin{tikzpicture}
		\draw[ultra thick,knot=cyan](0,0) .. controls(-0.3,-2) and (1.5,-1.5)..(0.9,-0.3);
		\draw[ultra thick,knot=cyan](2,-0.2) .. controls(2,-2) and (2.5,-1.5)..(3,0);
		\draw [ultra thick,blue] (1.5,0) ellipse (2 and 0.7);
		\draw[ultra thick,blue] (-0.5,0)--(-0.5,3);
		\draw[ultra thick,blue] (3.5,0)--(3.5,3);
		\draw[ultra thick,knot=black](0,0) .. controls(0,1) and (2,1)..(2,2.7);
		\draw[ultra thick,knot=cyan](0,3) .. controls(0,2) and (1.5,1.5)..(1.5,3.2);
		\draw[ultra thick,knot=teal](2,-0.2) .. controls(0,1) and (0,1.5)..(1,1.3);
		\draw[ultra thick,knot=violet](1,-0.4) .. controls(1,1) and (4,2)..(3,3);
		\draw[ultra thick,knot=teal](3,0) .. controls(2,0.8) ..(1.3,1.2);
		\draw[ultra thick,knot=orange](2.5,1.5) .. controls(2,2) and (2,2.5)..(2.6,2);
		\draw[ultra thick,knot=orange](2.9,1.7) .. controls(3.5,1.3) and (3.5,0.6)..(2.5,1.5);
		\fill[orange] (0,0) circle(5pt);
		\fill[red] (1,-0.4) circle(5pt);
		\fill[red] (2,-0.2) circle(5pt);
		\fill[orange] (3,0) circle(5pt);
		\fill[orange] (0,3) circle(5pt);
		\fill[red] (1.5,3.2) circle(5pt);
		\fill[red] (2,2.7) circle(5pt);
		\fill[orange] (3,3) circle(5pt);
		\draw [ultra thick,blue] (1.5,3) ellipse (2 and 0.7);
		\draw[ultra thick,blue](-0.5,0) .. controls(-0.3,-3) and (3,-3)..(3.5,0);
		\node [] at (4.8,3.5) {$\Sigma'= \partial (M\cup M')$};
		\node [] at (4.4,0) {$\Sigma = \partial M$};
		\node [] at (0.1,1.6) {$M'$};
		\node [] at (1.5,-1.5) {$M$};
		\node [] at (0.3,3) {$b$};
		\node [] at (2.3,2.7) {$\bar{a}$};
		\node [] at (2.8,3.3) {$a$};
		\node [] at (1.1,3) {$\bar{b}$};
	\end{tikzpicture}
	\caption{Worldlines of particles in the manifold.}
	\label{Worldlines}}

Let us have two punctures on a \textit{Riemann sphere} which is a complex manifold. The gluing axiom of topological quantum field theory \cite{atiyah1990geometry} states that we can glue the two punctures when they have opposing orientations. If we imagine one puncture as a particle, then the other hole must be considered as an antiparticle. It is explained in \ref{GeoPhase} that the ground state degeneracy for an $m$ number of particles on a torus is $m^g$, where $g$ is a genus. A genus is a handle in a manifold. From Fig. \ref{PunturesTori}, we get a genus-one torus $T^2$ when we glue the two opposing particles together. Therefore, the dimension of Hilbert space on a torus is equal to the number of particles types.
\fig{[h!]
	\centering
	\tik{\draw[bend right,ultra thick,red] (-1,0) to (1,0);
		\draw [ultra thick,blue] (0,0.1) ellipse (1 and 1);
		\draw [ultra thick,red] (-0.5,0.5) ellipse (0.2 and 0.2);
		\draw [ultra thick,red] (0.5,0.5) ellipse (0.2 and 0.2);
		\node [below] at (-0.5,0.3) {$a$};
		\node [below] at (0.5,0.3) {$\bar{a}$};
		\draw[ultra thick,->] (1.2,0)--(1.8,0);
	}
	\tik{[use Hobby shortcut,scale=1]
		\draw[bend right,ultra thick,red] (-1,0) to (1,0);
		\draw [ultra thick,blue] (0,0) ellipse (1 and 1);
		\begin{knot}[
			consider self intersections=true,
			ignore endpoint intersections=false,
			flip crossing=2,
			only when rendering/.style={}]
			\strand [ultra thick,knot=blue](-0.8,0.2) ..(-0.7,0.5).. (-0.7,1.3).. (0,1.7).. (0.7,1.3).. (0.7,0.5).. (0.8,0.2);
			\strand [ultra thick,knot=blue](-0.2,0.2).. (-0.3,0.5).. (-0.3,1.2).. (0,1.4).. (0.3,1.2).. (0.3,0.5).. (0.1,0.2);
		\end{knot}		
		\draw[ultra thick,red] (0,1.37) to (0,1.73);}
	\caption{Punctures on tori equivalent to the types of particles present}
	\label{PunturesTori}
}
This idea can be generalized to $n$ punctures and higher genus torus, written as $n$-torus or $T^n$. Any Riemann surface can be formed by the composition of three punctured Riemann spheres, which is also called \textit{pants} \cite{nayak2008non,fuchs2002tft} shown in Fig. \ref{Cob}. If two of them are fused, a two-punctured sphere will result.
Since the opposite orientations of punctures in TQFT are the opposite charges in anyonic models, two punctures on a sphere with labels $a$ and $\bar{a}$ should have the same topological charge to fuse into the vacuum. The fusion of two particles requires that for $k$ charges, there are $k+1$ different possible allowed boundary conditions. These charges can be identified as $j=0,1/2,...,k/2$ and the corresponding anyonic model is $SU(2)_k$ model with $k+1$ quasiparticles \cite{elitzur1989remarks,nayak2008non}.

	\chapter{Category Theory}\label{Cat}
	
So far, we have looked at things from the geometric point of view. Category theory is the algebraic approach for the computation of $F$-symbols and $R$-symbols used in Chapter \ref{TQC} and \ref{Meta}.
Anyons are simple objects in the categories whose trajectories are modeled by the morphisms of certain unitary modular categories (UMC). UMC is considered as a computing system and morphisms are circuits for computation \cite{rowell2018mathematics}. 

\section{Category and Functor}

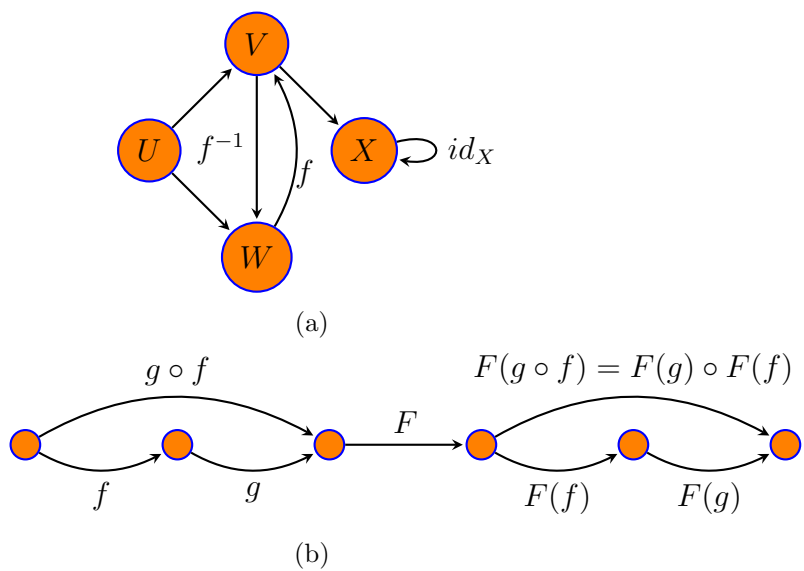
\begin{figure}[h!]
	\centering
	\begin{subfigure}{0.5\textwidth}
		\centering
		\begin{tikzpicture}[->,node distance=2cm,thick,>=stealth,shorten >=1pt]
			\node[circle,draw=blue,fill=orange] (U) {$U$};
			\node[circle,draw=blue,fill=orange,above right of =U] (V) {$V$};
			\node[circle,draw=blue,fill=orange,below right of =U] (W) {$W$};
			\node[circle,draw=blue,fill=orange,above right of=W] (X) {$X$};
			\path (U) edge (V);
			\draw (U) edge (W);
			\draw (V) edge node [left] {$f^{-1}$} (W);
			\draw [above right] (V) edge (X);
			\draw [loop right] (X) edge node {$id_X$} (X);
			\draw [bend right]  (W) edge node [right=3,below] {$f$} (V);
		\end{tikzpicture}\\
		\caption{}
	\end{subfigure}\\
	\begin{subfigure}{0.5\textwidth}
		\centering
		\begin{tikzpicture}[->,node distance=2cm,thick,>=stealth,shorten >=1pt,scale=0.7]
			\node[circle,draw=blue,fill=orange] (U) {};
			\node[circle,draw=blue,fill=orange,right of =U] (V) {};
			\node[circle,draw=blue,fill=orange, right of =V] (W) {};
			\path [bend right](U) edge node [below] {$f$} (V);
			\draw [bend right](V) edge node [below] {$g$} (W);
			
			\node[circle,draw=blue,fill=orange,right of = W] (A) {};
			\node[circle,draw=blue,fill=orange,right of =A] (B) {};
			\node[circle,draw=blue,fill=orange, right of =B] (C) {};
			
			\path [bend left](U) edge node [above] {$g\circ f$} (W) ;
			\path [bend left](A) edge node [above] {$F(g\circ f) = F(g)\circ F(f)$} (C) ;
			
			\path  (W) edge node [above] {$F$} (A) ;
			\path [bend right] (A) edge node [below] {$F(f)$} (B);
			\draw [bend right] (B) edge node [below] {$F(g)$} (C);
		\end{tikzpicture}
		\caption{}
	\end{subfigure}		
	\caption[Categories and functors.]{(a) A category theory, (b) a functor $F$. }
	\label{CatFunc}
\end{figure}

A \textit{category} ${\cal C}$ consists of two elements: \textit{objects} and \textit{morphisms} (or \textit{arrows}), such that 
\begin{itemize}
	\item Each morphism $f$ is assigned with a pair of objects $(U,V)$ in which case $U$ is called the domain (or source) of $f$ whereas $V$ is called the codomain (or target) of the morphism $f$. This is written as $f: U\rightarrow V$.
	\item There is a rule that if there are morphisms $f: U\rightarrow V$ and $g: V\rightarrow W$, then their composition is $g\circ f: U \rightarrow W$, that is, the codomain of $f$ is equal to the domain of $g$.
	\item This composition must be associative $(f\circ g)\circ h = f\circ (g\circ h)$. 
	\item Finally, there exists an identity morphism $id \circ f = f = f \circ id$ for each object.
\end{itemize}
The category theory is pictorially shown in Fig. \ref{CatFunc} (a).
One typical example of a category is the \textbf{Set} for which the objects are sets and the morphisms are functions $f:X \rightarrow Y$ which are composed ordinarily. But in general, objects are taken merely as labels that may not have substructures.
There are categories in which objects are sets and morphisms are functions preserving the structure of the objects. One such example is the \textbf{Grp} in which case the objects are groups and the morphisms are group homomorphisms. When these objects are Abelian groups, then the category is represented as \textbf{Ab}. The \textbf{Vect} categories are the ones for which the objects are vector spaces and the morphisms are linear maps. The \textbf{Hilb} is the category with objects that are the Hilbert spaces and the morphisms are linear maps. 

A mapping from one category to the other is called a \textit{functor}. A functor $F:\mathcal{C} \to \mathcal{D}$ from a category $\mathcal{C}$ to a category $\mathcal{D}$ is a rule such that
\begin{itemize}
	\item To each object $U$ of $\mathcal{C}$, there is associated an object $V$ of $\mathcal{D}$ and each morphism $f:a \to b$ in $\mathcal{C}$, there is associated a corresponding morphism $F(f):F(a) \to F(b)$ in $\mathcal{D}$.
	\item The association must preserve the composition and the units, that is, $F(fg) = F(f)F(g)$ and $F(1_a) = 1_{F(a)}$, where $1_a$ is a unit object in $\mathcal{C}$ and $1_{F(a)}$ is a unit object in $\mathcal{D}$. 
\end{itemize} 
A functor can be depicted as in Fig. \ref{CatFunc}. 
The products of categories are defined in the same way as the Cartesian product ${\cal C} \cross {\cal D}$. The objects in this case are $(a,b)$ where $a \in {\cal C}$ and $b \in {\cal D}$ and maps are $(f,f_2)$ when $f_1 \in {\cal C}$ and $f_2 \in {\cal D}$.
For further knowledge on category theory, see 
\cite{bakalov2001lectures,bartlett2005categorical,turaev2016quantum,fuchs2002tft,heunen2019categories}.

\section{Monoidal or Tensor Categories}
A \textit{monoid} is a set $X$ with some binary operation defined on it. Groups, rings, modules, and algebra are examples of monoids.  
A monoid $X$ can be expressed as follows. For $A,B \ \in \ X$ and $(A,B) \in X \cross X$, a product $A \circ B$  is defined in the way that $A \circ (B \circ C) = (A \circ B) \circ C$. For all $A \ \in \ X$, there is a unit element for which $1 \circ A = A \circ 1 = A$.
A monoid is also armed with two operations, $\mu : X\cross X \rightarrow X, \qquad \eta : 1 \rightarrow X$.

A process of defining a similar binary operation on a category is called the \textit{categorification}. In a category theory, numbers are replaced with vectors of dimension $n$, and equality is replaced with isomorphism. The bifunctors $\oplus$ and $\otimes$ are the categorifications of $+$ and $\cross$ of a ring. The ring is defined in Appendix \ref{AbsAlg}. A \textit{monoidal category} is a category with the following additional data:
\begin{itemize}
	\item A bifunctor $\otimes : {\cal C} \cross {\cal C} \to {\cal C}$.
	\item Associativity isomorphism of the functor\\ $\alpha_{UVW}:(U \otimes V)\otimes W \rightarrow U \otimes (V \otimes W)$.
	\item Unit object $\lambda_v: \bm{1}\otimes V \rightarrow V, \qquad \rho_V: V\otimes \bm{1} \rightarrow V$.
\end{itemize}

The tensor or monoidal category is called \textit{strict} if $U \otimes (V\otimes W) = (U \otimes V) \otimes W$ and $1\otimes U = U \otimes 1 = U$. For a strict category, we can write tensor products without bothering with parenthesis. The opposite category $\mathcal{C}^{op}$ is obtained by reversing the arrows.

The monoidal categories can be understood by drawing graphs as in Fig. \ref{Hom1}. The morphisms $f$ are represented by a rectangle, whereas the source object and the target objects are represented by arrows. In Fig. \ref{Hom1} $(a)$, morphism $f$ is from the object $U$ to $V$. The Fig. \ref{Hom1} $(b)$ is the identity morphism, which is equivalent to doing nothing to the object. In $(c)$, we have the composition of two morphisms, that is, if $f: U\to V$ and $g: V\to W$ then $f\circ g: U\to W$. In $(d)$, the tensor product of two morphisms is shown, that is, $f\otimes g: U_1\otimes U_2 \to V_1 \otimes V_2$, where $f:U_1\to V_1$ and $g:U_2\to V_2$. 

\subsection{Deformation of Diagrams and Duality}
The further generalization to the arbitrary category can be done by introducing the duals of the vector spaces. This duality would help to examine the diagrams which are equivalent under the deformation by keeping their endpoints fixed. 
Let there be an object $V^{*}$ in ${\cal C}$ for each object $V$ in ${\cal C}$. There are two morphisms called \textit{right duals}
\eq{e_V:V^* \otimes V\rightarrow 1 \qquad i_V:I \rightarrow V \otimes V^*.}
The right duals are shown graphically in Fig. \ref{Duals} (a). The structure is called \textit{duality} when there are the following two composites 
\eq{V &= 1 \otimes V \xrightarrow{i_V \otimes id_V} V\otimes V^* \otimes V \xrightarrow{i_d \otimes e_V} V \otimes 1 = V, \\
	V^* &= V^* \otimes 1 \xrightarrow{id_{V^*} \otimes i_V} V^* \otimes V \otimes V^* \xrightarrow{e_V \otimes id_{V^*}} 1 \otimes V^*= V^*.}
When $e_i$ is a basis for $V$ then $e^i$ is its dual basis.
Duals can be incorporated in the diagrams by representing the duals as arrows pointing down, whereas usual arrows are pointing up. For example, $f:V^* \rightarrow W$ is shown in Fig. \ref{Deform}.
By using the duality, we can see that the graphical calculus remains invariant under deformation.
Duals of morphisms can also be defined. If $f: U\rightarrow V$ then its dual is $f^* : U^* \rightarrow V^*$.

Suppose $f: V\rightarrow V$ and an object $V$ is in a monoidal category ${\cal C}$. The \textit{trace} $\Tr f \in \mathbb{C}$ is defined as in Fig. \ref{Trace} $(a)$. The \textit{dimension} of $V$ is given by the trace of the identity, that is, $dim V = \Tr id_V$ shown in Fig. \ref{Trace} $(b)$.
The trace and dimension should behave like those in \textbf{Vect}
\eq{&(a) \ \Tr(f\otimes g) = \Tr f \Tr g \ \ (b) \ \Tr(f^*) = \Tr f, \nonumber \\
	&(c) \ dim(V \otimes W) = dimV dimW \ \ (d) \ dimV^* = dimV.}
\begin{figure}[h!]
	\centering
	\begin{subfigure}{0.15\textwidth}
		\centering
		\begin{tikzpicture}		
			\node [] at (0.5,2.5) {$ f $};
			\draw[->,ultra thick,blue] (0.5,1) -- (0.5,2);
			\draw[->,ultra thick,blue] (0.5,3)--(0.5,4);
			\draw [ultra thick,red] (0,2) rectangle (1,3);
			\node [above] at (0.5,4) {$ V $};
			\node [below] at (0.5,1) {$ U $};
		\end{tikzpicture}
		\caption{}
	\end{subfigure}
	\begin{subfigure}{0.2\textwidth}
		\centering
		\begin{tikzpicture}		
			\node [] at (0,2.5) {$ id_V $};
			\draw[->,ultra thick,blue] (0,1) -- (0,2);
			\draw[->,ultra thick,blue] (0,3)--(0,4);
			\draw [ultra thick,red] (-0.5,2) rectangle (0.5,3);
			\node [above] at (0,4) {$ V $};
			\node [below] at (0,1) {$ V $};
			\node [] at (1,2.5) {$ = $};	
			\draw[->,ultra thick,blue] (1.5,1) -- (1.5,4);
			\node [above] at (1.5,4) {$ V $};
			\node [below] at (1.5,1) {$ V $};
		\end{tikzpicture}
		\caption{}
	\end{subfigure}
	\begin{subfigure}{0.4\textwidth}
		\centering
		\begin{tikzpicture}		
			\node [] at (0,2.5) {$ f \circ g $};
			\draw[->,ultra thick,blue] (0,1) -- (0,2);
			\draw[->,ultra thick,blue] (0,3)--(0,4);
			\draw [ultra thick,red] (-0.5,2) rectangle (0.5,3);
			\node [above] at (0,4) {$ W $};
			\node [below] at (0,1) {$ U $};
			\node [] at (1.5,2.5) {$ = $};
			\draw[->,ultra thick,blue] (2.5,3.5)--(2.5,4);	
			\draw[->,ultra thick,blue] (2.5,2.3)--(2.5,2.8);
			\draw[->,ultra thick,blue] (2.5,1) -- (2.5,1.6);
			\draw [ultra thick,red] (2,2.8) rectangle (3,3.5);
			\draw [ultra thick,red] (2,1.6) rectangle (3,2.3);
			\node [] at (2.5,3.1) {$ g $};
			\node [] at (2.5,1.9) {$ f $};
			\node [above] at (2.5,4) {$ W $};
			\node [right] at (2.7,2.55) {$ V $};
			\node [below] at (2.5,1) {$ U $};
		\end{tikzpicture}
		\caption{}
	\end{subfigure}\\
	\begin{subfigure}{0.45\textwidth}
		\centering
		\begin{tikzpicture}		
			\node [] at (0,2.5) {$ f \otimes g $};
			\draw[->,ultra thick,blue] (0,1) -- (0,2);
			\draw[->,ultra thick,blue] (0,3)--(0,4);
			\draw [ultra thick,red] (-0.5,2) rectangle (0.5,3);
			\node [above] at (0,4) {$ V_1\otimes V_2 $};
			\node [below] at (0,1) {$ U_1 \otimes U_2 $};
			\node [] at (1,2.5) {$ = $};
			\draw[->,ultra thick,blue] (1.7,3)--(1.7,4);	
			\draw[->,ultra thick,blue] (2.3,1) -- (2.3,2);
			\draw[->,ultra thick,blue] (2.3,3)--(2.3,4);	
			\draw[->,ultra thick,blue] (1.7,1) -- (1.7,2);
			\draw [ultra thick,red] (1.5,2) rectangle (2.5,3);
			\node [] at (2,2.5) {$ f\otimes g $};
			\node [above] at (1.7,4) {$ V_1 $};
			\node [below] at (1.7,1) {$ U_1 $};
			\node [above] at (2.3,4) {$ V_2 $};
			\node [below] at (2.3,1) {$ U_2 $};
			\node [] at (3,2.5) {$ = $};		
			\draw[->,ultra thick,blue] (4,3)--(4,4);	
			\draw[->,ultra thick,blue] (4,1) -- (4,2);
			\draw [ultra thick,red] (3.5,2) rectangle (4.5,3);
			\node [] at (4,2.5) {$ f $};
			\node [above] at (4,4) {$ V_1 $};
			\node [below] at (4,1) {$ U_1 $};		
			\draw[->,ultra thick,blue] (5.5,3)--(5.5,4);	
			\draw[->,ultra thick,blue] (5.5,1) -- (5.5,2);
			\draw [ultra thick,red] (5,2) rectangle (6,3);
			\node [] at (5.5,2.5) {$ g $};
			\node [above] at (5.5,4) {$ V_2 $};
			\node [below] at (5.5,1) {$ U_2 $};
		\end{tikzpicture}
		\caption{}
	\end{subfigure}
	\caption[Graphical calculus for the morphisms in a category theory.]{Graphical calculus for the morphisms in a category theory. (a) $f:U\rightarrow V$, (b) $id_V: V\rightarrow V$, (c) composition of $f:U\rightarrow V$ and $g: V\rightarrow W$ is $f\circ g:  U\rightarrow W$, (d) tensor product of $f:U_1\rightarrow V_1$ and $g: U_2\rightarrow V_2$ is $f\otimes g:U_1\otimes U_2 \rightarrow V_1 \otimes V_2$ }
	\label{Hom1}
\end{figure}
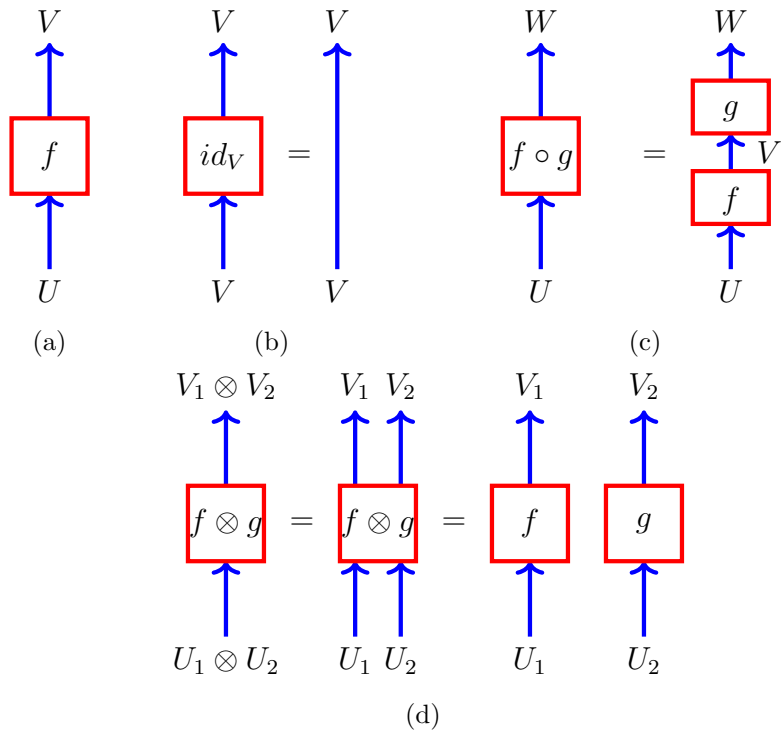

\begin{figure}[h!]
	\centering
	\begin{subfigure}{0.4\textwidth}
		\centering
		\begin{tikzpicture}[scale=0.7]
			\draw[ultra thick,->,red] (2,0) .. controls(1.8,1.5) and (0.2,1.5)..(0,0);
		\end{tikzpicture}
		\begin{tikzpicture}[scale=0.7]
			\draw[ultra thick,->,blue] (2,0) .. controls(1.8,-1.5) and (0.2,-1.5)..(0,0);
		\end{tikzpicture}
		\caption{}
	\end{subfigure}
	\begin{subfigure}{0.4\textwidth}
		\centering
		\begin{tikzpicture}[scale=0.7]
			\draw[ultra thick,->,red] (0,0) .. controls(0.2,1.5) and (1.8,1.5)..(2,0);
		\end{tikzpicture}
		\begin{tikzpicture}[scale=0.7]
			\draw[ultra thick,->,blue] (0,0) .. controls(0.2,-1.5) and (1.8,-1.5)..(2,0);
		\end{tikzpicture}
		\caption{}
	\end{subfigure}	
	\caption[Left and right duals.]{ (a) Right dual: $e_V:V^*\cross V\rightarrow I, \ i_V:I\rightarrow V\cross V^*$, (b) left dual $e'_V:V\otimes ^*V \to I, \ i'_V:I \to ^*V\otimes V$.}
	\label{Duals}
\end{figure}

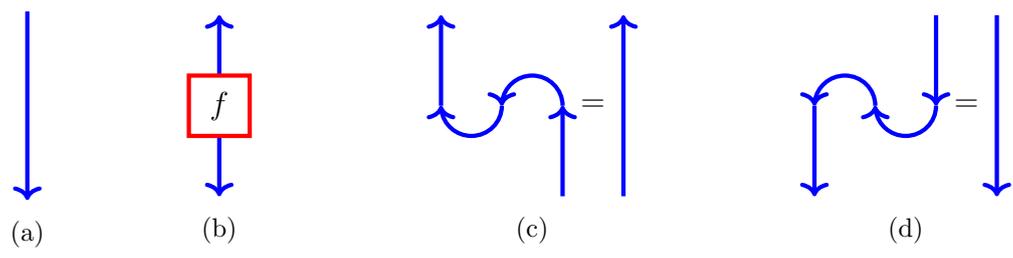
\begin{figure}[h!]
	\centering
	\begin{subfigure}{0.1\textwidth}
		\centering	
		\begin{tikzpicture}
			\draw[->,ultra thick,blue] (0,3.5) -- (0,1);
		\end{tikzpicture}
		\caption{}
	\end{subfigure}
	\begin{subfigure}{0.2\textwidth}
		\centering	
		\begin{tikzpicture}[scale=0.8]			
			\node [] at (0,2.5) {$ f $};
			\draw[->,ultra thick,blue] (0,2) -- (0,1);
			\draw[->,ultra thick,blue] (0,3)--(0,4);
			\draw [ultra thick,red] (-0.5,2) rectangle (0.5,3);
		\end{tikzpicture}
		\caption{}
	\end{subfigure}
	\begin{subfigure}{0.3\textwidth}
		\centering
		\begin{tikzpicture}[scale=0.8]
			\draw [ultra thick,->,blue] (0,0) arc (0:-180:0.5);
			\draw [ultra thick,->,blue] (1,0) arc (0:180:0.5);
			\draw[ultra thick,->,blue] (-1,0) -- (-1,1.5);
			\draw[ultra thick,->,blue] (1,-1.5) -- (1,0);
			\node [] at (1.5,0) {$ = $};
			\draw[->,ultra thick,blue] (2,-1.5) -- (2,1.5);
		\end{tikzpicture}
		\caption{}
	\end{subfigure}
	\begin{subfigure}{0.3\textwidth}
		\centering
		\begin{tikzpicture}[scale=0.8]
			\draw [ultra thick,->,blue] (0,0) arc (0:180:0.5);
			\draw [ultra thick,->,blue] (1,0) arc (0:-180:0.5);
			\draw[ultra thick,->,blue] (-1,0) -- (-1,-1.5);
			\draw[ultra thick,->,blue] (1,1.5) -- (1,0);
			\node [] at (1.5,0) {$ = $};
			\draw[->,ultra thick,blue] (2,1.5) -- (2,-1.5);
		\end{tikzpicture}
		\caption{}
	\end{subfigure}
	\caption[Pictorial representation of morphisms containing duals.]{Pictorial representation of morphisms containing duals; (a) $id_{V^*}$, (b) $f:V^*\rightarrow W$, (c) $(id_V \otimes e_V)(i_V\otimes id_V) = id_V$, (d) $(e_V \otimes id_{V^*})(id_{V^*}\otimes i_V) = id_V^*$.}
	\label{Deform}
\end{figure}

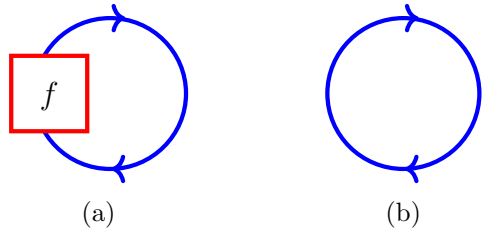
\begin{figure}[h!]
	\centering
	\begin{subfigure}{0.2\textwidth}
		\centering
		\begin{tikzpicture}	
			\draw [ultra thick,blue] (0,0) ellipse (1 and 1);
			\draw[->,ultra thick,blue] (0.1,1) -- (0.2,1);
			\draw[-<,ultra thick,blue] (0.1,-1) -- (0.2,-1);
			\draw [ultra thick,red,fill = white] (-1.3,-0.5) rectangle (-0.3,0.5);
			\node [below] at (-0.8,0.3) {$f $};
		\end{tikzpicture}
		\caption{}
	\end{subfigure}\qquad
	\begin{subfigure}{0.2\textwidth}
		\centering
		\begin{tikzpicture}
			\draw [ultra thick,blue] (0,0) ellipse (1 and 1);
			\draw[->,ultra thick,blue] (0.1,1) -- (0.2,1);
			\draw[-<,ultra thick,blue] (0.1,-1) -- (0.2,-1);
		\end{tikzpicture}
		\caption{}
	\end{subfigure}
	\caption[Trace of the morphisms.]{Trace of the morphisms: (a) $\Tr f$, (b) $\Tr id_V$.}
	\label{Trace}	
\end{figure}

\subsection{Braided Tensor Category}
\begin{figure}[h!]
	\centering
	\begin{subfigure}{0.2\textwidth}
		\centering
		\begin{tikzpicture}
			\draw[ultra thick,->,blue] (1,2) -- (1,2.2);		
			\draw[ultra thick,->,red] (0,2) -- (0,2.2);
			\draw[ultra thick,knot=red] (1,1) -- (0,2);
			\draw[blue,ultra thick,knot=blue] (0,1) -- (1,2);
			\node [below] at (0,1) {$V$};
			\node [below] at (1,1) {$W$};
			\node [] at (0.5,0) {$ \sigma_{V,W}$};
		\end{tikzpicture}
		\begin{tikzpicture}
			\draw[ultra thick,->,red] (1,2) -- (1,2.2);		
			\draw[ultra thick,->,blue] (0,2) -- (0,2.2);
			\draw[ultra thick,knot=red] (0,1) -- (1,2);
			\draw[ultra thick,knot=blue] (1,1) -- (0,2);		
			\node [below] at (0,1) {$V$};
			\node [below] at (1,1) {$W$};
			\node [] at (0.5,0) {$ \sigma^{-1}_{V,W}$};
		\end{tikzpicture}
		\caption{}
	\end{subfigure}
	\begin{subfigure}{0.4\textwidth}
		\centering
		\begin{tikzpicture}[scale=0.7]
			\draw[ultra thick,->,red] (1,4) -- (1,4.3);		
			\draw[ultra thick,->,blue] (0,4) -- (0,4.3);
			\draw[ultra thick,knot=red] (0,3) -- (1,4);
			\draw[ultra thick,knot=blue] (1,3) -- (0,4);		
			\draw[ultra thick,blue] (1,2) -- (1,3);		
			\draw[ultra thick,red] (0,2) -- (0,3);
			\draw[ultra thick,knot=red] (1,1) -- (0,2);
			\draw[ultra thick,knot=blue] (0,1) -- (1,2);				
			\node [] at (2,2.5) {$=$};
			\draw[ultra thick,->,blue] (3,1) -- (3,4.3);
			\draw[ultra thick,->,red] (4,1) -- (4,4.3);
			\node [above] at (0,4.3) {$V$};
			\node [above] at (1,4.3) {$W$};
			\node [left] at (0,2.5) {$W$};
			\node [right] at (1,2.5) {$V$};
			\node [below] at (0,1) {$V$};
			\node [below] at (1,1) {$W$};
			\node [above] at (3,4.3) {$V$};
			\node [above] at (4,4.3) {$W$};
			\node [below] at (3,1) {$V$};
			\node [below] at (4,1) {$W$};
			\draw[ultra thick,->,red] (7,4) -- (7,4.3);		
			\draw[ultra thick,->,blue] (6,4) -- (6,4.3);
			\draw[ultra thick,knot=blue] (7,3) -- (6,4);
			\draw[ultra thick,knot=red] (6,3) -- (7,4);				
			\draw[ultra thick,blue] (7,2) -- (7,3);		
			\draw[ultra thick,red] (6,2) -- (6,3);
			\draw[ultra thick,knot=blue] (6,1) -- (7,2);
			\draw[ultra thick,knot=red] (7,1) -- (6,2);						
			\node [] at (5,2.5) {$=$};
			\node [above] at (6,4.3) {$V$};
			\node [above] at (7,4.3) {$W$};
			\node [below] at (6,1) {$W$};
			\node [below] at (7,1) {$V$};
		\end{tikzpicture}
		\caption{}
	\end{subfigure}\qquad
	\begin{subfigure}{0.3\textwidth}
		\centering
		\begin{tikzpicture}[scale=0.7]
			\draw[ultra thick,->,red] (1,4) -- (1,4.3);		
			\draw[ultra thick,->,teal] (0,4) -- (0,4.3);
			\draw[ultra thick,->,blue] (2,4) -- (2,4.3);
			\draw[ultra thick,blue] (2,3) -- (2,4);
			\draw[ultra thick,knot=red] (0,3) -- (1,4);
			\draw[ultra thick,knot=teal] (1,3) -- (0,4);
			\draw[ultra thick,blue] (1,2) -- (2,3);		
			\draw[ultra thick,red] (0,2) -- (0,3);
			\draw[ultra thick,teal] (2,1) -- (2,2);
			\draw[ultra thick,knot=blue] (0,1) -- (1,2);
			\draw[ultra thick,knot=red] (1,1) -- (0,2);
			\draw[ultra thick,knot=teal] (2,2) -- (1,3);
			\node [] at (3,2.5) {$=$};
			\node [above] at (0,4.2) {$U$};
			\node [above] at (1,4.2) {$V$};
			\node [above] at (2,4.2) {$W$};
			\node [below] at (0,1) {$W$};
			\node [below] at (1,1) {$V$};
			\node [below] at (2,1) {$U$};
			\draw[ultra thick,->,red] (5,4) -- (5,4.3);		
			\draw[ultra thick,->,teal] (4,4) -- (4,4.3);
			\draw[ultra thick,->,blue] (6,4) -- (6,4.3);
			\draw[ultra thick,blue] (5,3) -- (6,4);
			\draw[ultra thick,knot=red] (6,3) -- (5,4);
			\draw[ultra thick,knot=teal] (4,3) -- (4,4);
			\draw[ultra thick,knot=blue] (4,2) -- (5,3);
			\draw[ultra thick,knot=teal] (5,2) -- (4,3);		
			\draw[ultra thick,red] (6,2) -- (6,3);
			\draw[ultra thick,knot=red] (5,1) -- (6,2);
			\draw[ultra thick,knot=blue] (4,1) -- (4,2);		
			\draw[ultra thick,knot=teal] (6,1) -- (5,2);
			\node [above] at (4,4.2) {$U$};
			\node [above] at (5,4.2) {$V$};
			\node [above] at (6,4.2) {$W$};
			\node [below] at (4,1) {$W$};
			\node [below] at (5,1) {$V$};
			\node [below] at (6,1) {$U$};
		\end{tikzpicture}
		\caption{}
	\end{subfigure}
	\caption{The braiding in a braided tensor category.}
	\label{Braiding}
\end{figure}
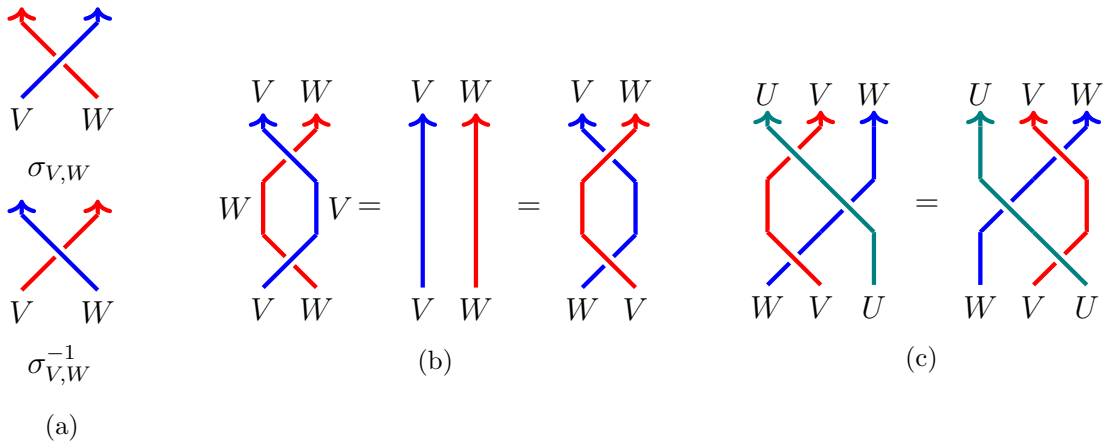

The discussion above is sufficient for a planar deformation. For the three-dimensional deformation, we have to introduce the concept of braiding. Two types of crossings in braiding, `under' and `over', cannot be deformed into each other. For every pair of objects in the monoidal category $({\cal C}, \otimes, 1)$, a natural isomorphism is assigned\\$\sigma_{V,W} :V\otimes W \to W \otimes V$, such that for any three objects $U,V,W$ in ${\cal C}$ we have
\eq{\sigma_{U,V\otimes W} = (id_V \otimes \sigma_{U,W}) (\sigma_{U,V} \otimes id_W),\nonumber\\ 
	\sigma_{U \otimes V, W} = (\sigma_{U,W} \otimes id_V)(id_U \otimes \sigma_{V,W}).}
The natural isomorphism means that it does not matter whether the braiding is done before or after the maps. The braiding is graphically shown in Fig. \ref{Braiding} (a).

To see how $\sigma^{-1}_{V,W}$ is really the inverse of $\sigma_{V,W}$, we will compose $\sigma_{V,W}$ with its inverse, and the result must be the identity. This is shown in Fig. \ref{Braiding} $(b)$. The result is the \textit{second Reidemeister move} we discussed in the Chapter on knot theory \ref{Knot}. 
$\sigma_{V,W}$ has to satisfy the following relation which is famously called the Yang-Baxter equation,
\eq{(&id_W \otimes \sigma_{U,V}) \circ (\sigma_{U,W} \otimes id_V) \circ (id_U \otimes \sigma_{V,W}) \nonumber\\
	&= (\sigma_{W,V} \otimes id_U)\circ (id_V \otimes \sigma_{U,W})\circ (\sigma_{U,V} \otimes id_W).}
This is also the \textit{third Reidemeister move}. It is shown graphically in Fig. \ref{Braiding} (c). A braided tensor category is called a \textit{symmetric tensor category} when $\sigma$ satisfies $\sigma_{W,V}\sigma_{V,W} = id_{V\otimes W}$. i.e. $\sigma_{V\otimes W} = \sigma^{-1}_{W\otimes V}$.
For a symmetric category, it does not matter whether strands are crossing above or below one another. 

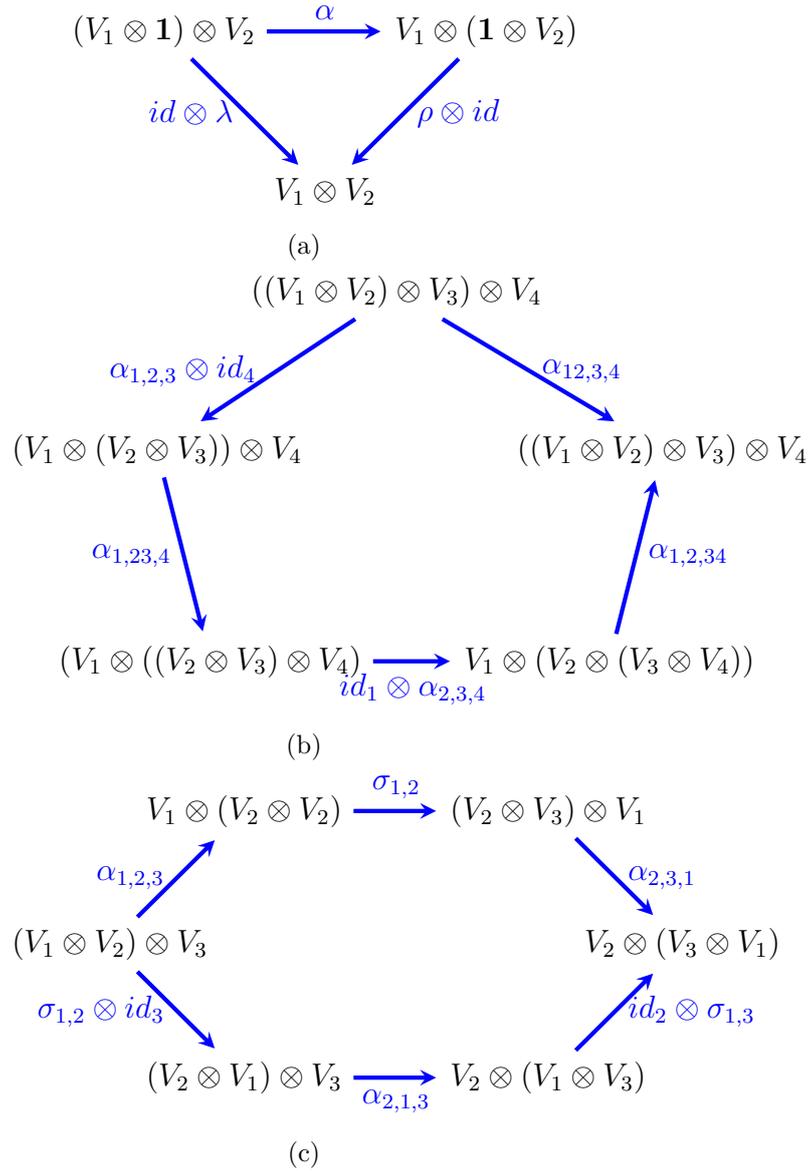
\begin{figure}[h!]
	\centering
	\begin{subfigure}{0.4\textwidth}
		\centering
		\begin{tikzpicture}[->,node distance=3cm,thick,>=stealth,shorten >=1pt,scale=0.5]
			\node[] (A) {$(V_1\otimes \bm{1}) \otimes V_2$};
			\node[below right of =A] (B) {$V_1\otimes V_2$};
			\node[above right of=B] (C) {$V_1\otimes (\bm{1} \otimes V_2)$};
			\draw [blue,ultra thick] (A) edge node [above,blue] {$\alpha$} (C);
			\draw (A) [blue,ultra thick] edge node [left,blue] {$id\otimes \lambda$} (B);
			\draw (C) [blue,ultra thick] edge node [right,blue] {$\rho \otimes id$} (B);
		\end{tikzpicture}
		\caption{}
	\end{subfigure}\\
	\begin{subfigure}{0.5\textwidth}
		\centering
		\begin{tikzpicture}[->,node distance=3.5cm,thick,>=stealth,shorten >=1pt,scale=0.7]
			\node[] (A) at (6,3) {$((V_1\otimes V_2)\otimes V_3)\otimes V_4$};
			\node[] (B) at (1.5,0) {$(V_1\otimes (V_2\otimes V_3))\otimes V_4$};
			\node[] (C) at (2.5,-4) {$(V_1\otimes ((V_2\otimes V_3)\otimes V_4)$};		
			\node[] (D) at (10,-4) {$V_1\otimes (V_2\otimes (V_3\otimes V_4))$};
			\node[] (E) at (11,0) {$((V_1\otimes V_2)\otimes V_3)\otimes V_4$};
			
			\draw [blue,ultra thick] (A) edge node [left=3,blue] {$\alpha_{1,2,3}\otimes id_4$} (B);
			\draw [blue,ultra thick] (B) edge node [left,blue] {$\alpha_{1,23,4}$} (C);
			\draw [blue,ultra thick] (C) edge node [below,blue] {$id_1\otimes \alpha_{2,3,4}$} (D);
			\draw [blue,ultra thick] (D) edge node [right,blue] {$\alpha_{1,2,34}$} (E);
			\draw [blue,ultra thick] (A) edge node [right,blue] {$\alpha_{12,3,4}$} (E);			
		\end{tikzpicture}
		\caption{}
	\end{subfigure}\\
	\begin{subfigure}{0.5\textwidth}
		\centering
		\begin{tikzpicture}[->,node distance=2.5cm,thick,>=stealth,shorten >=1pt]
			\node[] (A) {$(V_1\otimes V_2) \otimes V_3$};
			\node[above right of=A] (B) {$V_1\otimes (V_2 \otimes V_2)$};		
			\node[right of =B,node distance=4cm] (C) {$(V_2\otimes V_3)\otimes V_1$};
			\node[below right of =C] (D) {$V_2\otimes (V_3\otimes V_1)$};
			\node[below right of =A] (E) {$(V_2\otimes V_1)\otimes V_3$};
			\node[right of =E,node distance=4cm] (F) {$V_2\otimes (V_1\otimes V_3)$};
			
			\draw [blue,ultra thick] (A) edge node [left,blue] {$\alpha_{1,2,3}$} (B);
			\draw [blue,ultra thick] (B) edge node [above,blue] {$\sigma_{1,2}$} (C);
			\draw [blue,ultra thick] (C) edge node [right,blue] {$\alpha_{2,3,1}$} (D);
			
			\draw [blue,ultra thick] (A) edge node [left,blue] {$\sigma_{1,2}\otimes id_3$} (E);
			\draw [blue,ultra thick] (E) edge node [below,blue] {$\alpha_{2,1,3}$} (F);
			\draw [blue,ultra thick] (F) edge node [right,blue] {$id_2\otimes \sigma_{1,3}$} (D);
		\end{tikzpicture}
		\caption{}
	\end{subfigure}
	\caption[Triangle, pentagon and hexagon identities]{(a) Triangle Identity, (b) pentagon identity, (c) hexagon identity}
	\label{Identities}
\end{figure}

The data $(\mathcal{C},\otimes,\bm{1},\alpha,\lambda,\rho)$ constitute a tensor category iff they satisfy the two axioms; the triangle axiom and the pentagon axiom.
The data $(\mathcal{C},\otimes,\bm{1},\alpha,\lambda,\rho,\sigma)$ constitute a braided tensor category iff they satisfy the  three axioms; the triangle axiom, the pentagon axiom and the hexagon axiom shown in Fig. \ref{Identities}. These identities are also discussed in Chapter \ref{TQC} with reference to braiding and fusion of anyons. 

\subsection{Ribbon Category}
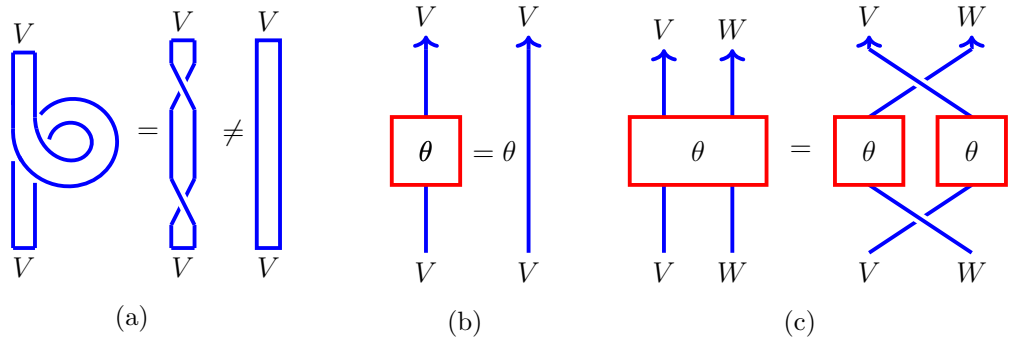
\begin{figure}[h!]
	\centering
	\begin{subfigure}{0.2\textheight}
		\centering
		\scalebox{0.9}{
		\begin{tikzpicture}[use Hobby shortcut,scale=1.6]
			\begin{knot}[
				consider self intersections=true,
				ignore endpoint intersections=false,
				flip crossing=2,
				only when rendering/.style={
				}
				]
				\strand [ultra thick,blue] (0.2,1.2) .. (0.25,1) .. (0.7,0.9) .. (0.7,1.05) .. (0.33,1);
				\strand [ultra thick,blue] (0,1.1) ..(0.55,0.57).. (0.95,0.96) .. (0.5,1.37) .. (0.25,1.23);
				\strand [ultra thick,blue] (0,1) .. (0,1.8);
				\strand [ultra thick,blue] (0.2,1.2) .. (0.2,1.8);
				\strand [ultra thick,blue] (0,0) .. (0,0.8);
				\strand [ultra thick,blue] (0.2,0) .. (0.2,0.6);
				\draw [ultra thick,blue] (0,1.8) -- (0.2,1.8);
				\draw [ultra thick,blue] (0,0) -- (0.2,0);
				\node [below] at (0.1,0) {$V$};
				\node [above] at (0.1,1.8) {$V$};
			\end{knot}
			\path (0,0);
		\end{tikzpicture}
		\begin{tikzpicture}[scale=1.7]
			\node [] at (-0.2,1) {$ = $};
			\draw[ultra thick,blue] (0,0) -- (0,0.2);
			\draw[ultra thick,blue] (0.2,0) -- (0.2,0.2);
			\draw[ultra thick,knot=blue] (0,0.2) -- (0.2,0.6);
			\draw[ultra thick,knot=blue] (0.2,0.2) -- (0,0.6);
			\draw[ultra thick,blue] (0,0.6) -- (0,1.2);
			\draw[ultra thick,blue] (0.2,0.6) -- (0.2,1.2);
			\draw[ultra thick,knot=blue] (0,1.2) -- (0.2,1.6);
			\draw[ultra thick,knot=blue] (0.2,1.2) -- (0,1.6);
			\draw[ultra thick,blue] (0,1.6) -- (0,1.8);
			\draw[ultra thick,blue] (0.2,1.6) -- (0.2,1.8);
			\draw[ultra thick,blue] (0,0) -- (0.2,0);
			\draw[ultra thick,blue] (0,1.8) -- (0.2,1.8);
			\node [below] at (0.1,0) {$V$};
			\node [above] at (0.1,1.8) {$V$};	
		\end{tikzpicture}
		\begin{tikzpicture}[scale=1.7]
			\node [] at (-0.2,1) {$ \ne $};
			\draw[ultra thick,blue] (0,0) -- (0,1.8)--(0.2,1.8)--(0.2,0)--cycle;
			\node [below] at (0.1,0) {$V$};
			\node [above] at (0.1,1.8) {$V$};
		\end{tikzpicture}}
		\caption{}
	\end{subfigure}\qquad
	\begin{subfigure}{0.15\textwidth}
		\centering
		\scalebox{0.9}{
		\begin{tikzpicture}
			\draw[ultra thick,->,blue] (5.5,4) -- (5.5,4.2);
			\draw[ultra thick,knot=blue] (5.5,3) -- (5.5,4);
			\draw[ultra thick,knot=blue] (5.5,1) -- (5.5,2);
			\node [] at (5.5,2.5) {$ \theta $};
			\node [above] at (5.5,4.2) {$ V $};
			\node [below] at (5.5,1) {$ V $};
			\draw [ultra thick,red] (5,2) rectangle (6,3);
			\node at (6.5,2.5) {$= \theta$};
			\draw[ultra thick,knot=blue] (7,1) -- (7,4);
			\draw[ultra thick,->,blue] (7,4) -- (7,4.2);
			\node [] at (5.5,2.5) {$ \theta $};
			\node [above] at (7,4.2) {$ V $};
			\node [below] at (7,1) {$ V $};
		\end{tikzpicture}}
		\caption{}
	\end{subfigure}\qquad
	\begin{subfigure}{0.3\textwidth}
		\centering
		\scalebox{0.9}{
		\begin{tikzpicture}	
			\draw[->,ultra thick,blue] (2.5,3)--(2.5,4);	
			\draw[ultra thick,blue] (2.5,1) -- (2.5,2);
			\draw[->,ultra thick,blue] (3.5,3)--(3.5,4);	
			\draw[ultra thick,blue] (3.5,1) -- (3.5,2);
			\draw [ultra thick,red] (2,2) rectangle (4,3);
			\node [] at (3,2.5) {$ \theta $};
			\node [above] at (2.5,4) {$ V $};
			\node [below] at (2.5,1) {$ V $};
			\node [above] at (3.5,4) {$ W $};
			\node [below] at (3.5,1) {$ W $};
			\node [] at (4.5,2.5) {$ = $};
			\draw[ultra thick,->,blue] (7,4) -- (7,4.2);		
			\draw[ultra thick,->,blue] (5.5,4) -- (5.5,4.2);
			\draw[ultra thick,knot=blue] (5.5,3) -- (7,4);
			\draw[ultra thick,knot=blue] (7,3) -- (5.5,4);
			\draw[ultra thick,knot=blue] (5.5,1) -- (7,2);
			\draw[ultra thick,knot=blue] (7,1) -- (5.5,2);
			\node [] at (7,2.5) {$ \theta $};
			\node [above] at (7,4.2) {$ W $};
			\node [below] at (7,1) {$ W $};
			\node [] at (5.5,2.5) {$ \theta $};
			\node [above] at (5.5,4.2) {$ V $};
			\node [below] at (5.5,1) {$ V $};
			\draw [ultra thick,red] (5,2) rectangle (6,3);
			\draw [ultra thick,red] (6.5,2) rectangle (7.5,3);
		\end{tikzpicture}}
		\caption{}
	\end{subfigure}
	\caption{Representation of a twist in a ribbon category.}
	\label{Twist}
\end{figure}

The \textit{first Reidemeister move} is to straighten out the twist in a strand, as shown in Fig. \ref{Twist} (a). The Reidemeister moves 1, 2, and 3 are for unframed tangles.
To show the relation between duality and braiding, we need two more arrows, as given by 
\eq{e^{'}_V: V\otimes ^*V \rightarrow 1, \qquad i^{'}_V : 1 \rightarrow ^*V \otimes V.}
These are called the \textit{left duals} and are shown graphically in Fig. \ref{Duals} (b). These identities are used to swap the inputs of $i_V$ and $e_V$.
To get rid of the twist, a natural family of isomorphisms is introduced $\theta_V: V\xrightarrow{}V$, such that 
\eq{&\theta_{V\otimes W} = \sigma_{W,V}\sigma_{V,W}(\theta_V \theta_W), \label{GetRidTwistA}\nonumber \\
	&(\theta_V\otimes id_{V^*})i_V = (id_V\otimes \theta_{V^*})i_V, \nonumber \\
	&\theta_1 = id, \nonumber \\
	&\theta_{V^*} = (\theta_V)^*.}
The first of these relations is shown in Fig. \ref{Twist} (c). The $\theta$ is the actual $2\pi$ twist in a ribbon. For a balanced category, we can write $^*V = V^*$, which means the left, and the right duals are equal.

Now we will introduce the \textit{ribbon category} $(\mathcal{C},\otimes,1,\sigma,\theta)$ which is a braided monoidal category with a compatible twist.  
A ribbon category is a rigid, balanced, and braided tensor category. By compensating for the twist, the ribbon diagrams remain invariant under 3d isotopy. Two ribbon diagrams are isotopic when they are topologically deformable into each other.

\section{Semisimple Ribbon Categories}
Suppose an object $U$ of a category does not contain nontrivial subobjects, then the object is called \textit{simple}. An Abelian category is \textit{semisimple} if any object in ${\cal C}$ is isomorphic to the direct sum of simple objects $V_i$ \cite{bakalov2001lectures}.
\eq{V = \bigoplus_{i\in I} N_iV_i, \qquad V_i \otimes V_j \simeq \bigoplus_{k \in \Pi_{\cal C}} N_{i,j}^k V_k.}
These formulas are called the fusion rules, and $N_{ij}^k$ is called the fusion coefficient. These are positive integers and are given by
\eq{N_{i,j}^k = dim(Hom_{\cal C}(V_k,V_i\otimes V_j))= dim(Hom(\bm{1},V_i\otimes V_j\otimes V_k)),}
\eq{N_{ij}^k = N_{ji}^k=N_{ik^*}^{j^*}=N_{i^*j^*}^{k^*}, \ N_{ij}^0 = \delta_{ij^8}.}
We also have
\eq{&\theta_V = \theta_i id_{V_i}, \ dimV_i = d_i, \ \theta_0 = 1, \ \theta_i = \theta_{i^*}, \nonumber \\
	& d_0 = 1, d_i = d_{i^*}, \ d_id_j = \sum_k N_{ij}^kd_k.}
The fusion matrix $N_i$ of $V$ is defined as $(N_i)_{k,j} = N_{i,j}^k$ with the non-negative entries. The largest eigenvalue is called \textit{Frobenus-Perron dimension} of ${\cal C}$.

A \textit{fusion category} is a semisimple rigid category with a unit object \cite{etingof2005fusion}, a finite-dimensional morphism, and a finite number of simple objects. The braided fusion category with a ribbon structure is called a \textit{ribbon fusion}, or a \textit{premodular category}. A premodular category is a \textit{modular category} if the S-matrix of ${\cal C}$ is defined as follows 
\eq{S_{ij} = \Tr(\sigma_{V_j,V_{i}}\circ \sigma_{V_{i},V_j}) \qquad \text{for} \ i,j \in \Pi_{\cal C},}
\eq{S_{ij}:V_j\otimes V_j \to V_i \otimes V_i^*.}
The $S$-matrix is the basis change between the components of the links. This $S$-matrix is a topological analogue of the $S$-matrix in the scattering theory we will discuss in the Appendix \ref{QFT}. A modular category is a semisimple ribbon category if the matrix $S$ is invertible and $\mathcal{C}$ has only a finite number of isomorphism classes of objects.
The $S$-matrix is an invariant of a Hopf link colored by $a$ and $b$, which are the entries of the $S$-matrix. 
A single loop can be thought of as a loop linking vacuum, which give $S_{a0} = S_{0a}$.
For the normalization, we take the loop of vacuum linking vacuum $S_{00}$, so
\eq{d_a = \frac{S_{a0}}{S_{00}}.}
The total quantum dimension is a number $D = \sqrt{\sum_{a \in \Pi_{\cal C}}d_a^2}$. It is an invariant of the category and is called the quantum order. 
The columns of the $S$-matrix can be seen to be simultaneous eigenvectors for the fusion matrix so that S diagonalizes all $N_a$ simultaneously. This leads to \textit{Verlinde formula} \cite{verlinde1988fusion}.
\eq{N_{ab}^c = \sum_x \frac{S_{ax} S_{bx}S_{cx}}{S_{1x}}.}
See \cite{witten1989quantum,dijkgraaf1988modular} for the proof of the Verlinde formula. By the Verlinde formula, $S$ matrix contains the information about the braiding as well as the fusion.

Let, $V_{ab}^c$ be $Hom(a\otimes b,c)$, the Hilbert space that is assigned to a surface when there are three objects. When there are four simple objects in a semisimple category, the F-matrix can be written as
\eq{F_d^{abc}:Hom((a\otimes b)\otimes c,d)\to Hom(a\otimes(b\otimes c),d)}

\subsection{Modular Transformation}
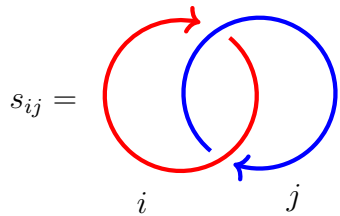
\begin{figure}[h!]
	\centering
	\begin{tikzpicture}
		\node [] at (-2.3,-1) {$ s_{ij} = $};
		\draw[ultra thick, ->,red] (0.15,-0.1) arc (50:-285:1);
		\draw[ultra thick, ->,blue] (-0.1,-1.6) arc (230:-110:1);
		\node [below] at (-1,-2) {$ i $};
		\node [below] at (1,-1.9) {$ j $};
	\end{tikzpicture}
	\caption{Modular transformation between two colored links}
	\label{SMatrix}
\end{figure}

The S-matrix is the change of basis between the two links shown in Fig. \ref{SMatrix}. 
This transformation is called the \textit{modular transformation}. The topological twist $\theta_a$ is encoded in a diagonal matrix $T= (\delta_{ab} \theta_a)$. $S_{ab}$ is an invariant of the Hopf link of two charges colored by $a$ and $b$. The $S$ and $T$ are called the \textit{modular data} of the category.

The $S$ and $T$ matrices define a projective representation of a modular group or mapping class group (MCG).
The mapping class group on a torus is $SL(2,\mathbb{Z})$, the special linear group on $\mathbb{Z}$. The Lie algebra and special linear groups are discussed in the Appendix \ref{AbsAlg}. The mapping class group is obtained from the deformation of spaces. Taking an example of a torus, which is the identification of a sheet at two opposite edges. The modular transformation of the torus is cutting the torus, putting one twist, called the \textit{Dehn twist}, and then joining again. A torus remains invariant under non-trivial transformation of the modular group.

A torus is made up of a lattice in the form of a parallelogram in complex plane with two opposite sides identified. The parallelogram is generated by two complex numbers $w_1$ and $w_2$. Modular parameter $\tau=w_1/w_2 \in \mathbb{H}$, where $\mathbb{H}=\{z\in \mathbb{C}|z>0\}$ is the complex upper half plane. We can denote the lattice as $L(w_1,w_2)$. Let another torus denoted by a lattice $L(u_1,u_2)$ with the modular parameter $\tau =u_1/u_2$. The two tori are equal if their modular parameters coincide, that is when they are related by the modular transformation as
\eq{\begin{pmatrix}
		w_1\\
		w_2
	\end{pmatrix}=\begin{pmatrix}
		a&b\\
		c&d
	\end{pmatrix}\begin{pmatrix}
		u_1\\
		u_2
	\end{pmatrix}, \qquad \text{where} \ \begin{pmatrix}
		a&b\\
		c&d
	\end{pmatrix}\in \text{PSL}(2,\mathbb{Z}).}
See \cite{schreivogl2013generalized} for detailed discussion. The most general modular transformation has the form
\eq{\tau \to \frac{a\tau + b}{c\tau +d}, \qquad a,b,c,d \in \mathbb{Z}; ad-bc =1.}
This is also known as global conformal transformation or M\"obious transformation \cite{francesco2012conformal}. The modular transformation has generators $S: \tau \to -\frac{1}{\tau}$ and $T:\tau \to \tau+1$. The matrices $S$ and $T$ obey $S^2=1,(ST)^3=1$. These matrices represent the rescaling of coordinates, where $S$ represents the inversion in the unit circle followed by reflection around the imaginary axis, and $T$ represents a unit translation. The projective special linear group $\text{PSL}(2,\mathbb{Z})$ is a quotient group $SL(2,\mathbb{Z})/\mathbb{Z}_2$, see Appendix \ref{AbsAlg} for the definition of the quotient group. $\text{PSL}(2,\mathbb{Z})$ is a group of $2\cross 2$ matrices with integer coefficients and unit determinant. Some authors write the modular matrix as a bigger group $SL(2,\mathbb{Z})$ instead of $\text{PSL}(2,\mathbb{Z})$. The quotient $\mathbb{Z}_2$ is due to the fact that the above transformation is unaffected by taking $a,b,c,d$ negative.

The modular transformation is an alternative to the fusion coefficients $N_{ab}^c$. This transformation encodes the braiding statistics and fusion rule of anyons \cite{nayak2008non}.
A braid is a special case of MCG when there are $n$ punctures. The non-Abelian Berry phase is obtained from the transformations $T$ and $S$ on a torus. A ground state subspace form a unitary representation of MCG \cite{zhu2020quantum}.
For example, let there be charges in an Ising anyon model given as $\ket{1}_m,\ket{\sigma}_m,\ket{\psi}_m$. In this case, the $S$ matrix would be 3 by 3, and the effect of its application would be to change the basis to a new basis $\ket{1}_l,\ket{\sigma}_l,\ket{\psi}_l$. The $S$ matrix is a change of basis, so this matrix should be unitary.

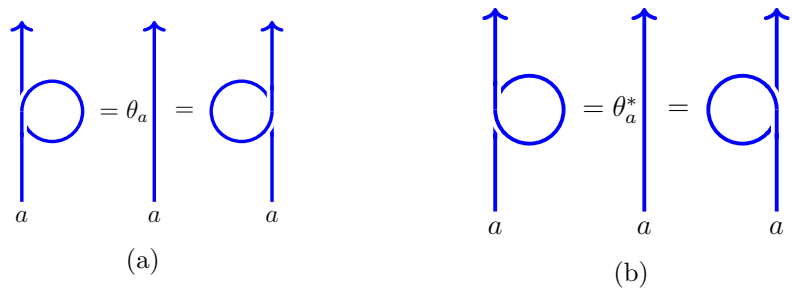
\begin{figure}[h!]
	\centering
	\begin{subfigure}{0.3\textwidth}
		\centering
		\scalebox{0.8}{
		\begin{tikzpicture}[use Hobby shortcut,scale=1]
			\begin{knot}[consider self intersections=true,
				ignore endpoint intersections=false,
				flip crossing=2,
				only when rendering/.style={}]		
				\strand [ultra thick,blue] (0,1) ..(0.5,0.5).. (1,1) .. (0.5,1.5) .. (0,1);
				\strand [ultra thick,blue] (0,-0.5) .. (0,1);
				\strand [ultra thick,blue,->] (0,1) .. (0,2.5);		
				\node [below] at (0,-0.5) {$a$};
			\end{knot}
			\path (0,0);
		\end{tikzpicture}
		\begin{tikzpicture}
			\node at (0,1) {$=\theta_a$};
			\node at (1,1) {$=$};
			\draw[ultra thick,blue,->] (0.5,-0.5) -- (0.5,2.5);
			\node [below] at (0.5,-0.5) {$a$};
		\end{tikzpicture}
		\begin{tikzpicture}[use Hobby shortcut,scale=1]
			\begin{knot}[consider self intersections=true,
				ignore endpoint intersections=false,
				flip crossing=2,
				only when rendering/.style={}]
				\strand [ultra thick,blue,->] (0,1) .. (0,2.5);
				\strand [ultra thick,blue] (0,-0.5) .. (0,1);
				\strand [ultra thick,blue] (0,1) ..(-0.5,0.5).. (-1,1) .. (-0.5,1.5) .. (0,1);
				\node [below] at (0,-0.5) {$a$};
			\end{knot}
			\path (0,0);
		\end{tikzpicture}}
		\caption{}
	\end{subfigure}\qquad \qquad
	\begin{subfigure}{0.3\textwidth}
		\centering
		\scalebox{0.9}{
		\begin{tikzpicture}[use Hobby shortcut,scale=1]
			\begin{knot}[consider self intersections=true,
				ignore endpoint intersections=false,
				flip crossing=2,
				only when rendering/.style={}]
				\strand [ultra thick,blue] (0,-0.5) .. (0,1);
				\strand [ultra thick,blue,->] (0,1) .. (0,2.5);
				
				\strand [ultra thick,blue] (0,1) ..(0.5,0.5).. (1,1) .. (0.5,1.5) .. (0,1);
				\node [below] at (0,-0.5) {$a$};
			\end{knot}
			\path (0,0);
		\end{tikzpicture}
		\begin{tikzpicture}
			\node at (0,1) {$=\theta^*_a$};
			\node at (1,1) {$=$};
			\draw[ultra thick,blue,->] (0.5,-0.5) -- (0.5,2.5);
			\node [below] at (0.5,-0.5) {$a$};
		\end{tikzpicture}
		\begin{tikzpicture}[use Hobby shortcut,scale=1]
			\begin{knot}[consider self intersections=true,
				ignore endpoint intersections=false,
				flip crossing=2,
				only when rendering/.style={}]
				
				\strand [ultra thick,blue] (0,1) ..(-0.5,0.5).. (-1,1) .. (-0.5,1.5) .. (0,1);
				\strand [ultra thick,blue] (0,-0.5) .. (0,1);
				\strand [ultra thick,blue,->] (0,1) .. (0,2.5);		
				\node [below] at (0,-0.5) {$a$};
			\end{knot}
			\path (0,0);
		\end{tikzpicture}}
		\caption{}
	\end{subfigure}
	\caption[Removing a twist is equivalent to adding a phase.]{Removing a twist is equivalent to adding a phase (a) $\theta_a = e^{2\pi i h_a}$ and (b) $\theta^*_a = e^{-2\pi ih_a}$.}
	\label{TwistPhase}
\end{figure}

\begin{figure}[h!]
	\centering
	\begin{tikzpicture}		
		\draw[ultra thick,knot=blue] (0.2,0)--(0.2,0.5);
		\draw[ultra thick,knot=blue](-0.2,0)--(-0.2,0.5);
		\draw[ultra thick,knot=blue](0.2,0.5)--(-0.2,1);
		\draw[ultra thick,knot=blue](-0.2,0.5)--(0.2,1);
		\draw[ultra thick,knot=blue](0.2,1)--(-0.2,1.5);
		\draw[ultra thick,knot=blue](-0.2,1)--(0.2,1.5);
		\draw[ultra thick,knot=blue](-0.2,1.5)--(-0.2,2);
		\draw[ultra thick,knot=blue](0.2,1.5)--(0.2,2);
		\draw[ultra thick,knot=blue](-0.2,2)--(-1,3);
		\draw[ultra thick,knot=blue](0.2,2)--(1,3);
		
		\draw[ultra thick,knot=blue](-1,3)--(-1,3.5);
		\draw[ultra thick,knot=blue](1,3)--(1,3.5);
		
		\draw[ultra thick,blue](0,2.3)--(0.6,3)--(0.6,3.5);
		\draw[ultra thick,blue](0,2.3)--(-0.6,3)--(-0.6,3.5);
		\draw[ultra thick,knot=blue](0.6,3.5)--(1,4);
		\draw[ultra thick,knot=blue](1,3.5)--(0.6,4);
		\draw[ultra thick,knot=blue](-1,3.5)--(-0.6,4);
		\draw[ultra thick,knot=blue](-0.6,3.5)--(-1,4);
		\draw[ultra thick,knot=blue](0.6,4)--(1,4.5);
		\draw[ultra thick,knot=blue](1,4)--(0.6,4.5);
		\draw[ultra thick,knot=blue](-1,4)--(-0.6,4.5);
		\draw[ultra thick,knot=blue](-0.6,4)--(-1,4.5);
		\draw[ultra thick,blue](-1,4.5)--(-1,5);
		\draw[ultra thick,blue](-0.6,4.5)--(-0.6,5);
		\draw[ultra thick,blue](1,4.5)--(1,5);
		\draw[ultra thick,blue](0.6,4.5)--(0.6,5);
		\draw[ultra thick,blue](-0.2,0)--(0.2,0);
		\draw[ultra thick,blue](-1,5)--(-0.6,5);
		\draw[ultra thick,blue](0.6,5)--(1,5);
		\node at (1.5,2.5) {$=$};
	\end{tikzpicture}
	\begin{tikzpicture}[scale=0.83]	
		\draw[ultra thick,knot=blue](-0.2,0)--(-0.2,0.5);
		\draw[ultra thick,knot=blue](0.2,0)--(0.2,0.5);
		\draw[ultra thick,knot=blue](-0.2,0.5)--(-1,1.5);
		\draw[ultra thick,knot=blue](0.2,0.5)--(1,1.5);
		\draw[ultra thick,blue](0,0.8)--(0.6,1.5);
		\draw[ultra thick,blue](0,0.8)--(-0.6,1.5);
		
		\draw[ultra thick,knot=blue](0.6,1.5)--(-1,3.5);	
		\draw[ultra thick,knot=blue](1,1.5)--(-0.6,3.5);	
		\draw[ultra thick,knot=blue](-0.6,1.5)--(1,3.5);
		\draw[ultra thick,knot=blue](-1,1.5)--(0.6,3.5);
		
		\draw[ultra thick,knot=blue](0.6,3.5)--(-1,5.5);	
		\draw[ultra thick,knot=blue](1,3.5)--(-0.6,5.5);
		
		\draw[ultra thick,knot=blue](-1,3.5)--(0.6,5.5);
		\draw[ultra thick,knot=blue](-0.6,3.5)--(1,5.5);
		
		\draw[ultra thick,knot=blue](-0.6,5.5)--(-0.6,6);
		\draw[ultra thick,knot=blue](-1,5.5)--(-1,6);
		\draw[ultra thick,knot=blue](0.6,5.5)--(0.6,6);
		\draw[ultra thick,knot=blue](1,5.5)--(1,6);
		\draw[ultra thick,blue](-0.2,0)--(0.2,0);
		\draw[ultra thick,blue](-1,6)--(-0.6,6);
		\draw[ultra thick,blue](0.6,6)--(1,6);
	\end{tikzpicture}
	\caption{Braid-twist correspondence.}
	\label{BraidTwist}
\end{figure}
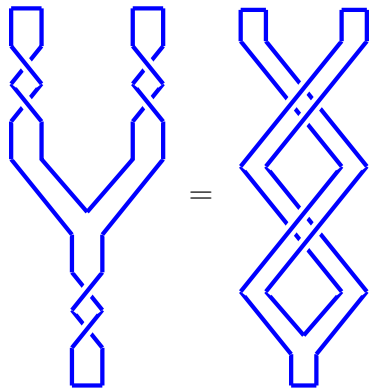

The topological spin or a twist factor $\theta_a = e^{2 \pi i h_a}$ is a twist when an anyon is rotated by $2\pi$, where $h_a$ is a \textit{scaling dimension} or conformal dimension. See Fig. \ref{TwistPhase}. It is an integer for bosons and gives a spin factor identity. It is a half-integer for fermions that would give the spin factor as $-1$. For vacuum, $h_0 = 0$. The exchange of two particles without twisting is governed by the braid group for anyons. In general, the left over right exchange is not equal to the right over the left exchange. The product of two left over right exchanges is called \textit{monodromy}. It returns the excitations to their original position but results in a nontrivial effect on the state of the system. This effect is given by the ribbon equation pictorially shown in Fig. \ref{BraidTwist},
\eq{e^{i \theta_{ab}^{c}} = \frac{e^{2 \pi i h_c}}{e^{2 \pi i h_a} e^{2 \pi i h_b}}.}
This is the same as the full twist of fusion products combined with the full twist of charges in the opposite direction. The effect of monodromy introduces the phase factor 
\eq{\frac{\theta_c}{\theta_a \theta_b} = e^{2 \pi i(h_c - h_a -h_b)}= [R^{ab}_c]^2.}
This is shown in Fig. \ref{BraidTwist}. This relation is used to derive the entries of the $R$-matrix for Fibonacci and Ising anyons. The trace of monodromy on $a$ and $b$ gives $S_{ab}$.

The classification of topological phases of matter is actually the classification of modular categories \cite{bruillard2015classification,bruillard2016classification}. For example, the Fibonacci and Ising categories given below are the phases of matter in which Fibonacci and Ising anyons are present \cite{rowell2009classification}. These two types of anyonic models are the most widely used in topological quantum computing. We will discuss these types of anyons in Chapter \ref{TQC}.

\subsubsection*{Fibonacci Modular Tensor Category}
Anyon types: $1,\tau$\\
Fusion rules: $\tau^2 = 1+ \tau$\\
Quantum dimensions: $d_1 = 1$, $d_\tau = \phi$\\
Twists: $\theta_1=1$, $\theta_\tau = e^{4i\pi/5}$\\
Braiding: $R_1^{\tau \tau}= e^{-4i\pi/5}$, $R^{\tau\tau}_\tau = e^{3i\pi/5}$\\
S-matrix: $S= \frac{1}{\sqrt{2+\phi}} \begin{pmatrix}
	1 & \phi\\
	\phi & 1
\end{pmatrix}$\\
F-matrices: $F^{\tau\tau\tau}_\tau = \begin{pmatrix}
	\phi^{-1} & \phi^{-1/2}\\
	\phi^{-1/2} & -\phi^{-1}
\end{pmatrix}$

\subsubsection*{Ising Modular Tensor Category}
Anyon types: $1,\sigma,\psi$\\
Fusion rules: $\sigma^2 = 1+ \psi, \sigma \psi = \psi \sigma = \sigma, \psi^2 = 1$\\
Quantum dimensions: $d_1 = 1$, $d_\sigma = \sqrt{2}, d_\psi = 1$\\
Twists: $\theta_1=1$, $\theta_\sigma = e^{i\pi/8}, \theta_\psi = -1$\\
Braiding: $R_1^{\sigma\sigma}= e^{-i\pi/8}, R^{\psi \psi}_1=-1, R^{\psi\sigma}_\sigma = -i, R^{\sigma\sigma}_\psi = e^{3i\pi/8}$\\
S-matrix: $S= \frac{1}{2} \begin{pmatrix}
	1 & \sqrt{2} & 1\\
	\sqrt{2}&0 & -\sqrt{2}\\
	1 & -\sqrt{2} & 1
\end{pmatrix}$\\
F-matrices: $F^{\sigma\sigma\sigma}_\sigma = \frac{1}{2} \begin{pmatrix}
	1 & 1\\
	1 & -1
\end{pmatrix}, F^{\psi\sigma\psi}_\sigma = (-1), F^{\sigma\psi\sigma}_\psi =(-1)$

\section{Category Theory and Topological Quantum\\ Computation}
There are three steps to implement quantum computation: the creation of anyons from the vacuum, braiding these anyons by moving them around, and measuring the anyon type of the pair of neighboring particles.
In category theory, the first step is the implementation of $Hom(\textbf{1}, X^{\otimes n})$ and the third step is $Hom(X^{\otimes n}, \textbf{1})$. After the braiding, neighboring anyons may have a different total charge besides vacuum. The computing result is a probability distribution on anyon types obtained by repeating the same process polynomially many times. Usually, the probability of neighboring pairs of anyons returning to vacuum is taken as the computing answer, which approximates some topological invariant of links obtained from the braiding process \cite{rowell2018mathematics}. 

A group of particles in the configuration space makes a quantum state in the Hilbert space. The objects in a category correspond to particles, and dual objects correspond to antiparticles. The twist is a topological spin of an anyon. A loop is a trace of a worldline and gives the dimension of a particle. If an object $X$ has $d_X=1$, then it is an Abelian anyon, whereas the object is a non-Abelian anyon when the dimension $d_X>1$.  
The decompositions are called the fusion rules.
The comparison of the anyon model to the category theory is given in \cite{wang2010topological,rowell2018mathematics}.

\section{Categorical Aspect of TQFT}
The functorial view of the quantum field theory is a mapping from geometry to algebra. More specifically, from a geometric and dynamical structure of spacetime to the algebraic structure of observables and states. 
The TQFT is a symmetric monoidal functor, can be written as \cite{bartlett2005categorical} 
\eq{Z = \textbf{nCob} \rightarrow \textbf{Vect} \label{SymMonFunc},}
where $\textbf{nCob}$ is a category whose objects are closed $n-1$-dimensional manifold $\Sigma$ and whose morphisms $M:\Sigma_1 \to \Sigma_2$ are cobordism, that is, $n$-dimensional manifolds having an input boundary $\Sigma_1$ and an output boundary $\Sigma_2$. The concept of cobordism is explained in Chapter \ref{TQFT}.
Physically, that $Z$ is a functor means $Z(MM^{'}) = Z(M)Z(M^{'})$, and when $Z$ is monoidal means that the two non-interacting systems have a state-space which is the tensor product of the states of individual systems. $Z$ is a symmetric functor when we are dealing with a bosonic statistic.

	\chapter{Topological Quantum Computation}\label{TQC}
Topological quantum computing is a fault-tolerant quantum computing, proposed by Alexei Kitaev \cite{kitaev2003fault}, manifested by manipulating quantum information using anyons. A quantum computation model involves three main steps; initialization, unitary evolution, and measurement \cite{divincenzo2000physical}.
In quantum theory, the time evolution of a state is represented by the unitary time evolution operator $U(t)$. When the initial state $\ket{\psi_i}$ evolves unitarily to the final state, it is written as $\ket{\psi_f}=U(t)\ket{\psi_i}$.
The initial state is an input and the final state is an output of a quantum gate, and the readout is a measurement in a certain basis to get a classical result \cite{Nielsen2002quantum}. See Chapter \ref{QC} for further details.
The gate operation $U(t)$ is equivalent to the rotation of states in the Hilbert space. Analogous to conventional quantum computing, topological quantum computing has three steps; the creation of pairs of anyons from the vacuum, their braiding, and their fusion. The result of fusion corresponds to the measurement, and the braiding corresponds to the unitary transformation $\Psi_f = (Braid) \Psi_i$. The braids cause rotation within degenerate N-particles space. The change of state by braiding can be explained by geometric phases. The braid group and the geometric phase are discussed in Chapters \ref{Knot} and \ref{GeoPhase}.
To see how topological quantum computing is a fault-tolerant quantum computing, let two spacetime histories $\ket{1}$ and $\ket{2}$ in Fig. \ref{Histories} have time reversed states $\bra{1}$ and $\bra{2}$. By the Kauffman bracket we have $\bra{1}\ket{1} = \bra{2}\ket{2} = d^2$, $\bra{2}\ket{1} = d$. The number $d$ is assigned to a loop as we discussed in context of the Kauffman bracket in Chapter \ref{Knot}. So $\ket{1}$ and $\ket{2}$ are distinct states as long as $\abs{d} \ne 1$ \cite{simon2010quantum}. The states $\ket{1}$ and $\ket{2}$ locally look the same, but the outcomes of their fusion with $\ket{1}$ or $\ket{2}$ are different. Therefore, disturbing one of the particles would not affect the outcome if the topology of a spacetime trajectory is not changed. The state $\ket{1}$ can also be interpreted as the creation of two pairs of anyons and the state $\bra{1}$ as a fusion of the two pairs of anyons.

For a basic introduction on topological quantum computation, see \cite{field2018introduction,pachos2012introduction}. The topological gates with Ising anyons, using the Kauffman version of the recoupling theory \cite{kauffman1994temperley}, are proposed by \cite{fan2010braid}. For the implementation of gates with Ising anyons in the quantum Hall phase, see \cite{georgiev2008towards,ahlbrecht2009implementation} and for gates with Fibonacci anyons, see \cite{hormozi2007topological}.

\begin{figure}[h!]
	\centering
	\begin{tikzpicture}[scale=0.5]
		\draw [ultra thick,fill=cyan!20] (-4.5,0) ellipse (4 and 0.7);
		\braid[ultra thick,rotate=180, style strands={1}{red},style strands={2}{blue},style strands={3}{red},style strands={4}{blue},style strands={5}{red},style strands={6}{blue},style strands={7}{red},style strands={8}{blue}]s_5s_7S_4^{-1}s_1s_3S_6^{-1}s_2;
		\draw [ultra thick,fill=cyan!20] (-4.5,7.9) ellipse (4 and 0.7);
		\fill[blue] (-8,0) circle(6pt);
		\fill[red] (-7,0) circle(6pt);
		\fill[blue] (-6,0) circle(6pt);
		\fill[red] (-5,0) circle(6pt);
		\fill[blue] (-4,0) circle(6pt);
		\fill[red] (-3,0) circle(6pt);
		\fill[blue] (-2,0) circle(6pt);
		\fill[red] (-1,0) circle(6pt);
		\node [] at (-8.5,4) {\textcolor{black}{\small \begin{turn}{90}Time\end{turn}}};
		\node [] at (1.2,8) {Measure};
		\node [left] at (1.5,4) {Braid};
		\node [] at (1.2,0) {Initialize};	
	\end{tikzpicture}
	\caption[Creation, fusion and braiding of anyons.]{Creation of anyon pairs, braiding them by dragging around each other, then fusing them and getting the result of fusion \cite{simon2010quantum}}
	\label{ThreeSteps}	
\end{figure}
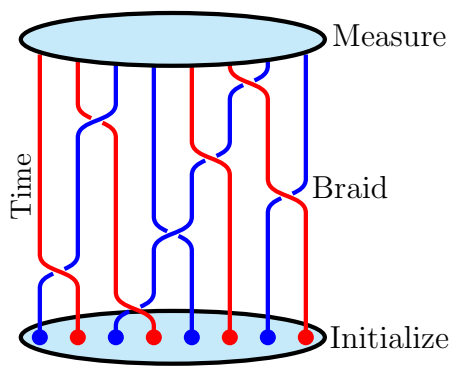

\begin{figure}[h!]
	\centering
	\begin{tikzpicture}[scale=0.9]
		\draw [ultra thick,fill=cyan!30,opacity=0.5] (0,0) ellipse (1.7 and 0.7);
		\path[draw,ultra thick,red,postaction={decorate,decoration={markings,mark=between positions 0.15 and 1 step 0.25 with {\arrow[blue]{};}}}] (-1,0) .. controls(0,3) ..(1,0);
		\path[draw,ultra thick,red,postaction={decorate,decoration={markings, mark=between positions 0.15 and 1 step 0.25 with {\arrow[blue]{};}}}](-0.2,-0.2) .. controls(0,2.2) ..(0.5,0.3);
		\fill[blue] (-1,0) circle(5pt);
		\fill[blue] (1,0) circle(5pt);
		\fill[red] (-0.2,-0.2) circle(5pt);
		\fill[red] (0.5,0.3) circle(5pt);
		\path[draw,ultra thick,red,postaction={decorate,decoration={markings, mark=between positions 0.15 and 1 step 0.25 with {\arrow[blue]{};}}}] (-1,-2.5) .. controls(0,-5) ..(1,-2.5);
		\path[draw,ultra thick,red,postaction={decorate,decoration={markings, mark=between positions 0.15 and 1 step 0.25 with {\arrow[blue]{};}}}] (-0.2,-2.8) .. controls(0,-4.5) ..(0.5,-2.2);
		\draw [ultra thick,fill=cyan!30,opacity=0.5] (0,-2.5) ellipse (1.7 and 0.7);
		\fill[blue] (-1,-2.5) circle(5pt);
		\fill[blue] (1,-2.5) circle(5pt);
		\fill[red] (-0.2,-2.8) circle(5pt);
		\fill[red] (0.5,-2.2) circle(5pt);
		\node [left] at (-2,0) {$\bra{1}$};
		\node [left] at (-2,-2.5) {$\ket{1}$};
	\end{tikzpicture}\qquad
	\begin{tikzpicture}[scale=0.9]
		\draw [ultra thick,fill=cyan!30,opacity=0.5] (0,0) ellipse (1.7 and 0.7);
		\path[draw,ultra thick,red,postaction={decorate,decoration={markings, mark=between positions 0.15 and 1 step 0.25 with {\arrow[blue]{};}}}] (-1,0) .. controls(-0.5,3) ..(-0.2,-0.3);
		
		\path[draw,ultra thick,red,postaction={decorate,decoration={markings, mark=between positions 0.15 and 1 step 0.25 with {\arrow[blue]{};}}}] (0.2,0.2) .. controls(0.5,3) ..(1,0);
		\fill[blue] (-1,0) circle(5pt);
		\fill[red] (-0.2,-0.3) circle(5pt);	
		\fill[red] (0.2,0.2) circle(5pt);
		\fill[blue] (1,0) circle(5pt);	
		\path[draw,ultra thick,red,postaction={decorate,decoration={markings, mark=between positions 0.15 and 1 step 0.25 with {\arrow[blue]{};}}}] (-1,-2.3) .. controls(-1,-5) ..(-0.3,-2.8);	
		\path[draw,ultra thick,red,postaction={decorate,decoration={markings, mark=between positions 0.15 and 1 step 0.25 with {\arrow[blue]{};}}}] (0.2,-2.1) .. controls(0,-5) ..(0.9,-2.6);
		\draw [ultra thick,fill=cyan!30,opacity=0.5] (0,-2.5) ellipse (1.7 and 0.7);
		\fill[blue] (-1,-2.3) circle(5pt);
		\fill[red] (-0.3,-2.8) circle(5pt);	
		\fill[red] (0.2,-2.2) circle(5pt);
		\fill[blue] (0.9,-2.6) circle(5pt);
		\node [left] at (-2,0) {$\bra{2}$};
		\node [left] at (-2,-2.5) {$\ket{2}$};
	\end{tikzpicture}\qquad
\scalebox{0.9}{
	\begin{tikzpicture}[scale=0.7]
		\path[draw,ultra thick,red,postaction={decorate,decoration={markings,
				mark=between positions 0.15 and 1 step 0.25 with {\arrow[blue]{};}}}] (-1,0) .. controls(0,-3) ..(1,0);
		\path[draw,ultra thick,red,postaction={decorate,decoration={markings,
				mark=between positions 0.15 and 1 step 0.25 with {\arrow[blue]{};}}}] (-0.2,-0.2) .. controls(0,-2.2) ..(0.5,0.3);
		\draw [ultra thick,fill=cyan!30,opacity=0.5] (0,0) ellipse (2 and 0.8);
		\path[draw,ultra thick,red,postaction={decorate,decoration={markings,
				mark=between positions 0.15 and 1 step 0.25 with {\arrow[blue]{};}}}] (-1,0) .. controls(0,3) ..(1,0);
		\path[draw,ultra thick,red,postaction={decorate,decoration={markings,
				mark=between positions 0.15 and 1 step 0.25 with {\arrow[blue]{};}}}] (-0.2,-0.2) .. controls(0,2.2) ..(0.5,0.3);	
		\fill[teal] (-1,0) circle(5pt);
		\fill[teal] (1,0) circle(5pt);
		\fill[teal] (0.5,0.3) circle(5pt);
		\fill[teal] (-0.2,-0.2) circle(5pt);
		\node [left] at (-2,0) {$\bra{1}\ket{1}=$};
		\node [right] at (2.5,0) {$=d^2$};
		\path[draw,ultra thick,red,postaction={decorate,decoration={markings,
				mark=between positions 0.15 and 1 step 0.25 with {\arrow[blue]{};}}}] (-1,-5) .. controls(-1,-8) ..(-0.2,-5.5);
		\path[draw,ultra thick,red,postaction={decorate,decoration={markings,
				mark=between positions 0.15 and 1 step 0.25 with {\arrow[blue]{};}}}] (0.5,-5) .. controls(0,-8) ..(1,-5);
		\draw [ultra thick,fill=cyan!30,opacity = 0.5] (0,-5.1) ellipse (2 and 0.8);
		\path[draw,ultra thick,red,postaction={decorate,decoration={markings,
				mark=between positions 0.15 and 1 step 0.25 with {\arrow[blue]{};}}}] (-1,-5) .. controls(0,-2) ..(1,-5);
		\path[draw,ultra thick,red,postaction={decorate,decoration={markings,
				mark=between positions 0.15 and 1 step 0.25 with {\arrow[blue]{};}}}] (-0.2,-5.5) .. controls(0,-2.5) ..(0.5,-5);
		\fill[teal] (-1,-5) circle(5pt);
		\fill[teal] (-0.2,-5.5) circle(5pt);
		\fill[teal] (0.5,-5) circle(5pt);		
		\fill[teal] (1,-5) circle(5pt);
		\node [left] at (-2,-5) {$\bra{1}\ket{2}=$};
		\node [right] at (2.5,-5) {$=d$};
	\end{tikzpicture}}
	\caption[States of spacetime histories.]{States of spacetime histories \cite{simon2010quantum}.}
	\label{Histories}
\end{figure}
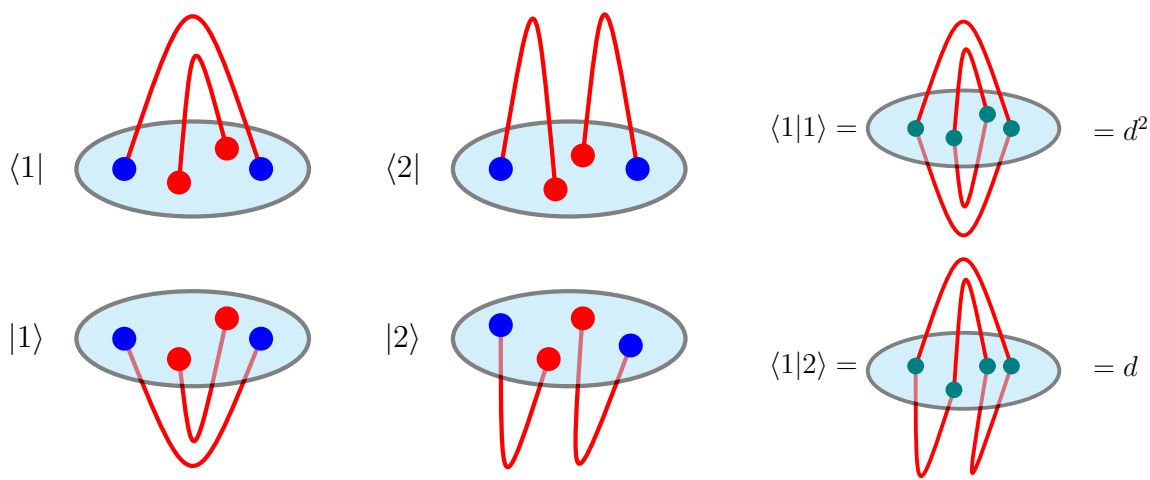

\section{Hilbert Space}
The types of anyons are categorized by quantum numbers attached to them called the \textit{topological charges}. When the charges are created from the vacuum, their total charge must be zero. Therefore, the value of the topological charge is assigned with respect to its fusion with other anyons. The topological charge zero is assigned to the vacuum. The vacuum is also called a \textit{trivial charge}.
For example, in Ising anyon model, there are three charges $\left\{\textbf{1},\sigma,\psi \right\}$. In some anyonic models, the vacuum is represented by $0$ or $I$. The $\textbf{1}$ represents a vacuum, or a trivial particle, whereas the $\sigma$ and $\psi$ are nontrivial particles. Since the anyons are created from the vacuum, to conserve the total charge, they must be fused to vacuum. The number of ways these anyons are fused to vacuum is called the \textit{fusion channels or fusion trees}. The number of the fusion channels is equal to the ground state degeneracy of the system. The fusion space is a shared property of a collection of non-Abelian anyons regardless of where they are located. Therefore, local perturbations do not affect the degeneracy of the system. In Chapter \ref{TQFT}, we identified the anyons on a two-dimensional manifold as the punctures or holes on a sphere.

Let several anyons be created from the vacuum, and let us consider a subtree consisting of two anyons $a$ and $b$. A fusion tree diagram is equivalent to the anyons' creation tree if the time direction is reversed.
When these anyons are Abelian, they fused to only one outcome and the dimension of Hilbert space for two anyons is one. But when the particles $a$ and $b$ are non-Abelian, there is more than one fusion outcome, that is $a\cross b = \sum_c N_{ab}^c$.
Their Hilbert space is denoted as $V_{ab}^{c}$. The dimension of Hilbert space is given as $N_{ab}^c = dim(V_{ab}^c)$. The numbers $N_{ab}^c$ are also called the \textit{fusion rules}. These fusion rules appear in conformal field theory and category theory in the form as $\phi_a\cross \phi_b = \sum_c N_{ab}^c\phi_c$. The fusion rules put some restrictions on what types of anyons a particular anyonic model can have. As the fusion of two non-Abelian anyons can result in several anyons, in general, $N_{ab}^c$ could have more values than one. Most of the anyonic models are built by considering only the two fusion outcomes; vacuum or another anyon, that is $N_{ab}^c=0,1$. When the fusion of $a$ and $b$ gives the topological charge $c$ then $N_{ab}^c=1$, but when $a$ and $b$ cannot be fused to $c$ then $N_{ab}^c=0$.

The dimension of Hilbert space increases with the number of anyons, analogous to the addition of spins $1/2 \otimes 1/2 = 0 \oplus 1 $. This analogy is not exact, because the anyons are not elementary particles, but they have internal degrees of freedom. The dimension of Hilbert space of $N$ particles of type $a$ is roughly $\sim d_a^N$, where $d$ is the dimension of Hilbert space of one anyon and is called the \textit{quantum dimension}. It is a measure of how much the Hilbert space is increased by adding this anyon. Therefore, it is an asymptotic degeneracy per particle. It needs not to be an integer. The quantum dimension of vacuum is one. In terms of knot theory, the $d$ is a number assigned to a loop. For the process $a\cross b = \sum_c N_{ab}^c c$, the quantum dimension can be defined to satisfy $d_ad_b = \sum_c N_{ab}^c d_c$.

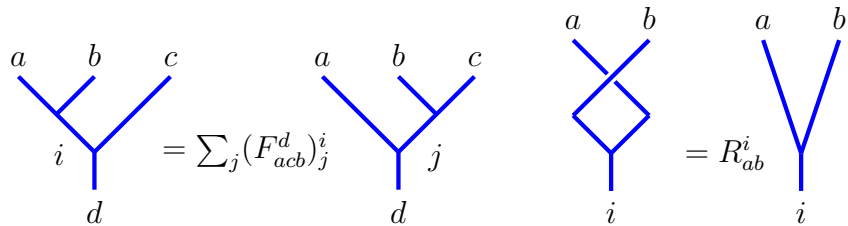
\begin{figure}[h!]
	\centering
	\begin{tikzpicture}[scale=0.5]
		\draw [blue, ultra thick] (2,0) -- (2,1); 
		\draw [blue, ultra thick] (2,1) -- (4,3);
		\draw [blue, ultra thick] (2,1) -- (0,3);
		\draw [blue, ultra thick] (1,2) --(2,3);
		\node [above] at (0,3) {$a$};
		\node [above] at (2,3) {$b$};
		\node [above] at (4,3) {$c$};
		\node [below] at (2,0) {$d$};
		\node [below left] at (1.5,1.5) {$i$};
		\node [] at (6,1) {$= \sum_{j}(F_{acb}^d)^i_j $};
		\draw [blue, ultra thick] (10,0) -- (10,1); 
		\draw [blue, ultra thick] (10,1) -- (12,3);
		\draw [blue, ultra thick] (10,1) -- (8,3);
		\draw [blue, ultra thick] (11,2) --(10,3);		
		\node [above] at (8,3) {$a$};
		\node [above] at (10,3) {$b$};
		\node [above] at (12,3) {$c$};
		\node [below] at (10,0) {$d$};
		\node [below right] at (10.5,1.5) {$j$};
	\end{tikzpicture}\qquad
	\begin{tikzpicture}[scale=0.5]
		\draw [blue, ultra thick] (1,0) -- (1,1);
		\draw [blue, ultra thick] (1,1) -- (0,2);
		\draw [blue, ultra thick] (1,1) -- (2,2);
		\draw [blue, ultra thick] (0,2) -- (2,4);
		\draw [blue, ultra thick] (2,2) -- (1.1,2.9);
		\draw [blue, ultra thick] (0.9,3.1) -- (0,4);
		\node [below] at (1,0) {$i$};
		\node [above] at (0,4) {$a$};
		\node [above] at (2,4) {$b$};
		\node [] at (4,1) {$= R_{ab}^{i}$};
		\draw [blue, ultra thick] (6,0) -- (6,1);
		\draw [blue, ultra thick] (6,1) -- (7,4);
		\draw [blue, ultra thick] (6,1) -- (5,4);
		\node [below] at (6,0) {$i$};
		\node [above] at (5,4) {$a$};
		\node [above] at (7,4) {$b$};
	\end{tikzpicture}
	\caption{F and R moves.}
	\label{FRMatrices}
\end{figure}

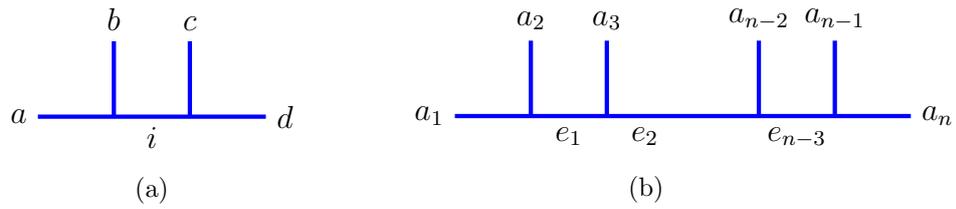
\begin{figure}[h!]
	\centering
	\begin{subfigure}{0.4\textwidth}
		\centering		
		\begin{tikzpicture}
			\draw[ultra thick,blue] (0,0)--(3,0);
			\draw[ultra thick,blue] (1,0)--(1,1);
			\draw[ultra thick,blue] (2,0)--(2,1);
			\node [left] at (0,0) {$ a $};
			\node [above] at (1,1) {$ b $};
			\node [below] at (1.5,0) {$ i $};
			\node [above] at (2,1) {$ c $};
			\node [right] at (3,0) {$ d $};
		\end{tikzpicture}
		\caption{}
	\end{subfigure}
	\begin{subfigure}{0.4\textwidth}
		\centering
		\begin{tikzpicture}
			\draw[ultra thick,blue] (0,0)--(6,0);
			\draw[ultra thick,blue] (1,0)--(1,1);
			\draw[ultra thick,blue] (2,0)--(2,1);
			\draw[ultra thick,blue] (4,0)--(4,1);
			\draw[ultra thick,blue] (5,0)--(5,1);
			\node [left] at (0,0) {$ a_1 $};
			\node [above] at (1,1) {$ a_2 $};
			\node [below] at (1.5,0) {$ e_1 $};
			\node [above] at (2,1) {$ a_3 $};
			\node [below] at (2.5,0) {$ e_2 $};
			\node [above] at (4,1) {$ a_{n-2} $};
			\node [below] at (4.5,0) {$ e_{n-3} $};
			\node [above] at (5,1) {$ a_{n-1} $};
			\node [right] at (6,0) {$ a_n $};
		\end{tikzpicture}
		\caption{}
	\end{subfigure}
	\caption[Fusion space of multiple anyons.]{Fusion space of anyons \cite{pachos2012introduction} for (a) four anyons, (b) $n$ anyons.}
	\label{FusionSpace}
\end{figure}

Let there be a situation when the fusion outcome of three non-Abelian anyons $a,b,c$ is $d$. (It can be a subtree of another fusion tree in which $d$ can be fused with another anyon and the outcome is vacuum or some other anyons).
There are more different fusion channels than one. For example, $a$ can be fused to $b$ first, then their outcome $i$ is fused with $c$ to get $d$. Or $b$ can be fused to $c$ first, then their fusion outcome $j$ can be fused with $a$ to get $d$. The fusion channels $i$ and $j$ make two sets of basis. The transformation between these bases $i$ and $j$ is given by $F$-symbols or $F$-moves. As for the non-Abelian anyons, $i$ and $j$ occur in more ways than one, the $F$-moves between different $i$'s and $j$'s will be a matrix called the \textit{$F$-matrix} shown in Fig. \ref{FRMatrices} (a). The vector space for these anyons can be written as $V_{abc}^d=\bigoplus_i V_{ab}^i \otimes V_{ic}^d=\bigoplus_j V_{bc}^j \otimes V_{ja}^d$. The fusion diagrams which can be continuously deformed into each other are equivalent and represent the same state.
For a diagram of $n$ anyons shown in Fig. \ref{FusionSpace}, the dimension of Hilbert space can be written in terms of the fusion rules as $N_{a_1a_2}^{e_1}N_{e_1a_3}^{e_2}...N_{e_{n-3}a_{n-1}}^{a_n}$.

Now let us braid two anyons $a$ and $b$ by exchanging their places. Abelian anyons get a complex phase that depends on the types of anyons and on whether the exchange is clockwise or counterclockwise. But it does not depend on the order of exchange. Therefore, for a Abelian anyon we have the $R$-move given as $R_{ab} = e^{i\theta_{ab}}$. But for non-Abelian anyons, the $R$-move is a matrix $R_{ab}^{i} = e^{i \theta_{ab}^{i}}$ and it depends on the order of the exchange \cite{lahtinen2017short}. Topologically equivalent braids have the same outcomes. The braid matrix is shown in Fig. \ref{FRMatrices} (b).

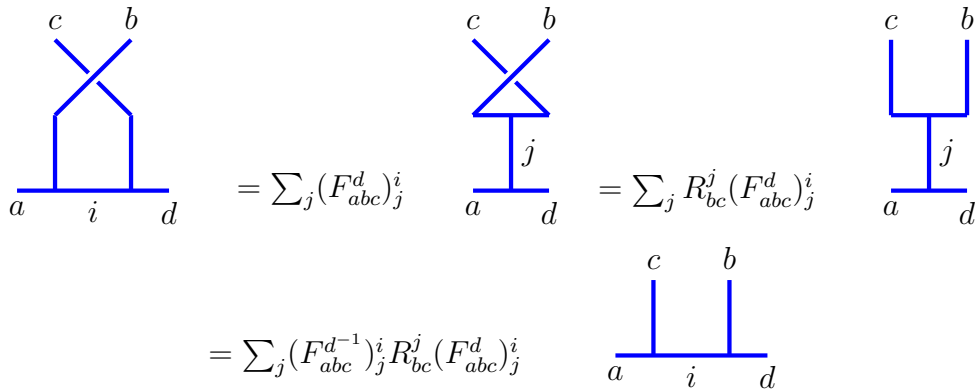
\begin{figure}[h!]
	\centering
	\begin{tikzpicture}[scale=1]
		\draw[ultra thick,blue] (0.5,0)--(2.5,0);
		\draw[ultra thick,blue] (1,0)--(1,1);
		\draw[ultra thick,blue] (2,0)--(2,1);
		\node [below] at (0.5,0) {$ a $};
		\node [above] at (1,2) {$ c $};
		\node [below] at (1.5,0) {$ i $};
		\node [above] at (2,2) {$ b $};
		\node [below] at (2.5,0) {$ d $};
		\draw[ultra thick,knot=blue] (2,1)--(1,2);
		\draw[ultra thick,knot=blue] (1,1)--(2,2);
		\node [] at (4.5,0) {$= \sum_{j}(F_{abc}^d)^i_j $};
		\draw[ultra thick,blue] (6.5,0)--(7.5,0);
		\draw[ultra thick,blue] (7,0)--(7,1);	
		\draw[ultra thick,knot=blue] (7.5,1)--(6.5,2);
		\draw[ultra thick,knot=blue] (6.5,1)--(7.5,2);
		\draw[ultra thick,blue] (6.5,1)--(7.5,1);
		\node [below] at (6.5,0) {$ a $};
		\node [above] at (6.5,2) {$ c $};
		\node [right] at (7,0.5) {$ j $};
		\node [above] at (7.5,2) {$ b $};
		\node [below] at (7.5,0) {$ d $};
		\node [right] at (8,0) {$= \sum_{j}R_{bc}^j(F_{abc}^d)^i_j $};
		\draw[ultra thick,blue] (12,0)--(13,0);
		\draw[ultra thick,blue] (12.5,0)--(12.5,1);	
		\draw[ultra thick,knot=blue] (12,1)--(12,2);
		\draw[ultra thick,knot=blue] (13,1)--(13,2);
		\draw[ultra thick,blue] (12,1)--(13,1);
		\node [below] at (12,0) {$ a $};
		\node [above] at (12,2) {$ c $};
		\node [right] at (12.5,0.5) {$ j $};
		\node [above] at (13,2) {$ b $};
		\node [below] at (13,0) {$ d $};
	\end{tikzpicture}\\
		\begin{tikzpicture}		
		\node [right] at (13.5,0) {$= \sum_{j}(F_{abc}^{d^{-1}})^i_jR_{bc}^j(F_{abc}^d)^i_j $};
		\draw[ultra thick,blue] (19,0)--(21,0);
		\draw[ultra thick,blue] (19.5,0)--(19.5,1);
		\draw[ultra thick,blue] (20.5,0)--(20.5,1);
		\node [below] at (19,0) {$ a $};
		\node [above] at (19.5,1) {$ c $};
		\node [below] at (20,0) {$ i $};
		\node [above] at (20.5,1) {$ b $};
		\node [below] at (21,0) {$ d $};
	\end{tikzpicture}
	\caption{Braiding for the superposition of the two fusion channels.}
	\label{BMatrix}
\end{figure}

The fusion channel of a pair of anyons cannot be changed by $R$-move only. In other words, the system would not evolve from one ground state to the other by exchanging the two anyons of the same pair. To change their fusion channel, we need to braid $b$ and $c$. For this braiding, we have to transform $i$ bases to the $j$ bases by an $F$-move. Therefore, having only two anyons is not enough for making a topological qubit.
To see the effect of the exchange of $b$ and $c$ in the basis $i$, first an $F$-matrix is applied to change the basis from $i$ to $j$, then an $R$ matrix is applied, and then an $F^{-1}$-matrix is applied to change the basis back to $i$. This process is shown in Fig. \ref{BMatrix} and can be written as
\eq{B_{ab} = {F_{acb}^d}^{-1} R_{ab} F_{acb}^d.}
The matrix $B$ creates the superposition of the fusion channels, with a distinct phase factor for different fusion channels.

\begin{figure}[h!]
	\centering
	\begin{tikzpicture}[]
		\node (P0) at (1.5,1) {$\tik{[scale = 0.3]
				\draw [blue, ultra thick] (0,0) -- (0,1); 
				\draw [blue, ultra thick] (0,1) -- (3,4);
				\draw [blue, ultra thick] (0,1) -- (-3,4);
				\draw [blue, ultra thick] (-2,3) --(-1,4);
				\draw [blue, ultra thick] (-1,2) --(1,4);
				\node [above] at (-3,4) {1};
				\node [above] at (-1,4) {2};
				\node [above] at (1,4) {3};
				\node [above] at (3,4) {4};
				\node [below] at (0,0) {5};
				\node [below] at (-2,3) {a};
				\node [below] at (-1,2) {b};}$};
		\node (P1) at (4.5,3) {$\tik{[scale = 0.3]
				\draw [blue, ultra thick] (0,0) -- (0,1); 
				\draw [blue, ultra thick] (0,1) -- (3,4);
				\draw [blue, ultra thick] (0,1) -- (-3,4);
				\draw [blue, ultra thick] (-2,3) --(-1,4);
				\draw [blue, ultra thick] (2,3) --(1,4);
				\node [above] at (-3,4) {1};
				\node [above] at (-1,4) {2};
				\node [above] at (1,4) {3};
				\node [above] at (3,4) {4};
				\node [below] at (0,0) {5};
				\node [below] at (-2,3) {a};
				\node [below] at (2,3) {c};}$};
		\node (P2) at (7.5,1) {$\tik{[scale = 0.3]
				\draw [blue, ultra thick] (0,0) -- (0,1); 
				\draw [blue, ultra thick] (0,1) -- (3,4);
				\draw [blue, ultra thick] (0,1) -- (-3,4);
				\draw [blue, ultra thick] (1,2) --(-1,4);
				\draw [blue, ultra thick] (2,3) --(1,4);
				\node [above] at (-3,4) {1};
				\node [above] at (-1,4) {2};
				\node [above] at (1,4) {3};
				\node [above] at (3,4) {4};
				\node [below] at (0,0) {5};
				\node [below] at (1,2) {d};
				\node [below] at (2,3) {c};}$};
		\node (P3) at (3,-2) {$\tik{[scale = 0.3]
				\draw [blue, ultra thick] (0,0) -- (0,1); 
				\draw [blue, ultra thick] (0,1) -- (3,4);
				\draw [blue, ultra thick] (0,1) -- (-3,4);
				\draw [blue, ultra thick] (0,3) --(-1,4);
				\draw [blue, ultra thick] (-1,2) --(1,4);
				\node [above] at (-3,4) {1};
				\node [above] at (-1,4) {2};
				\node [above] at (1,4) {3};
				\node [above] at (3,4) {4};
				\node [below] at (0,0) {5};
				\node [below] at (0,3) {e};
				\node [below] at (-1,2) {b};}$};
		\node (P4) at (6,-2) {$\tik{[scale = 0.3]
				\draw [blue, ultra thick] (0,0) -- (0,1); 
				\draw [blue, ultra thick] (0,1) -- (3,4);
				\draw [blue, ultra thick] (0,1) -- (-3,4);
				\draw [blue, ultra thick] (1,2) --(-1,4);
				\draw [blue, ultra thick] (0,3) --(1,4);
				\node [above] at (-3,4) {1};
				\node [above] at (-1,4) {2};
				\node [above] at (1,4) {3};
				\node [above] at (3,4) {4};
				\node [below] at (0,0) {5};
				\node [below ] at (0,3) {e};
				\node [below ] at (1,2) {d};}$};
		\draw [ultra thick,red,->] (P0) -- (P1);
		\draw [ultra thick,red,->] (P1) -- (P2);
		\draw [ultra thick,red,->] (P0) -- (P3);
		\draw [ultra thick,red,->] (P3) -- (P4);
		\draw [ultra thick,red,->] (P4) -- (P2);
		\node at (2.5,2.5) {$(F^5_{a34})_b^c$};
		\node at (6,2.5) {$(F^5_{12c})_a^d$};
		\node at (1.5,-0.6) {$(F^b_{123})_a^e$};
		\node at (4.5,-2.5) {$(F^5_{1e4})_b^d$};
		\node at (7.4,-0.7) {$(F^d_{234})_e^c$};
	\end{tikzpicture}
	\caption{The pentagon identity.}
	\label{Pentagon}
\end{figure}
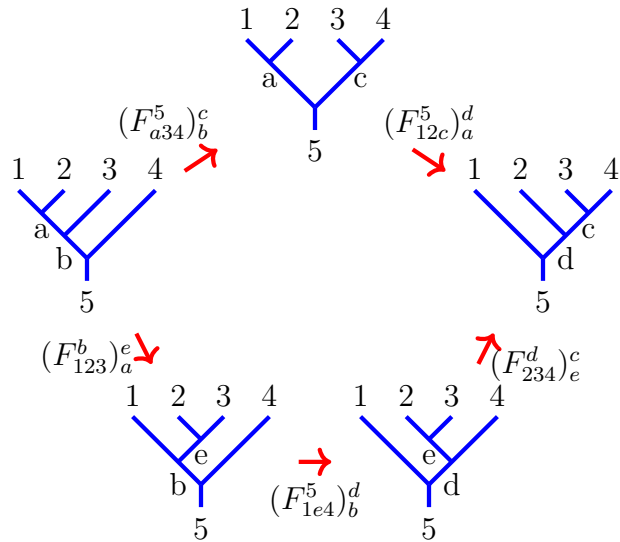

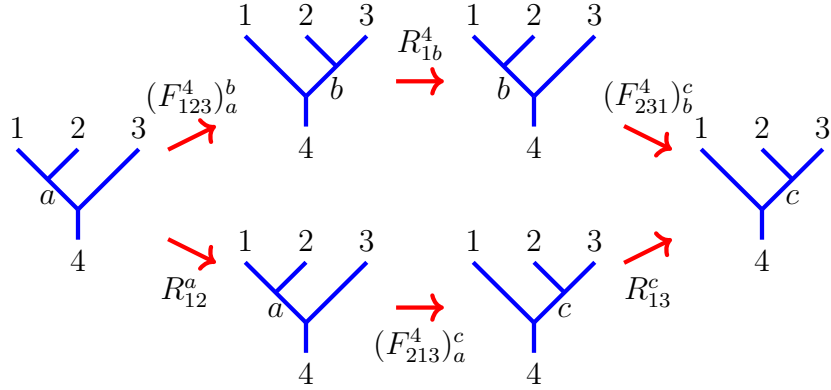
\begin{figure}[h!]
	\centering
	\begin{tikzpicture}[]
		\node (P1) at (0,0) {$\tik{[scale = 0.4]
				\draw [blue, ultra thick] (0,0) -- (0,1); 
				\draw [blue, ultra thick] (0,1) -- (-2,3);
				\draw [blue, ultra thick] (0,1) -- (2,3);
				\draw [blue, ultra thick] (-1,2) --(0,3);
				\node [above] at (-2,3) {$1$};
				\node [above] at (0,3) {$2$};
				\node [above] at (2,3) {$3$};			
				\node [below] at (-1,2) {$a$};
				\node [below] at (0,0) {$4$};}$};
		\node (P2) at (3,-1.5) {$\tik{[scale = 0.4]
				\draw [blue, ultra thick] (0,0) -- (0,1); 
				\draw [blue, ultra thick] (0,1) -- (-2,3);
				\draw [blue, ultra thick] (0,1) -- (2,3);
				\draw [blue, ultra thick] (-1,2) --(0,3);
				\node [above] at (-2,3) {$1$};
				\node [above] at (0,3) {$2$};
				\node [above] at (2,3) {$3$};			
				\node [below] at (-1,2) {$a$};
				\node [below] at (0,0) {$4$};}$};
		\node (P3) at (6,-1.5) {$\tik{[scale = 0.4]
				\draw [blue, ultra thick] (0,0) -- (0,1); 
				\draw [blue, ultra thick] (0,1) -- (-2,3);
				\draw [blue, ultra thick] (0,1) -- (2,3);
				\draw [blue, ultra thick] (1,2) --(0,3);
				\node [above] at (-2,3) {$1$};
				\node [above] at (0,3) {$2$};
				\node [above] at (2,3) {$3$};			
				\node [below] at (1,2) {$c$};
				\node [below] at (0,0) {$4$};}$};		
		\node (P4) at (9,0) {$\tik{[scale = 0.4]
				\draw [blue, ultra thick] (0,0) -- (0,1); 
				\draw [blue, ultra thick] (0,1) -- (-2,3);
				\draw [blue, ultra thick] (0,1) -- (2,3);
				\draw [blue, ultra thick] (1,2) --(0,3);
				\node [above] at (-2,3) {$1$};
				\node [above] at (0,3) {$2$};
				\node [above] at (2,3) {$3$};			
				\node [below] at (1,2) {$c$};
				\node [below] at (0,0) {$4$};}$};
		\node (P5) at (3,1.5) {$\tik{[scale = 0.4]
				\draw [blue, ultra thick] (0,0) -- (0,1); 
				\draw [blue, ultra thick] (0,1) -- (-2,3);
				\draw [blue, ultra thick] (0,1) -- (2,3);
				\draw [blue, ultra thick] (1,2) --(0,3);
				\node [above] at (-2,3) {$1$};
				\node [above] at (0,3) {$2$};
				\node [above] at (2,3) {$3$};			
				\node [below] at (1,2) {$b$};
				\node [below] at (0,0) {$4$};}$};
		\node (P6) at (6,1.5) {$\tik{[scale = 0.4]
				\draw [blue, ultra thick] (0,0) -- (0,1); 
				\draw [blue, ultra thick] (0,1) -- (-2,3);
				\draw [blue, ultra thick] (0,1) -- (2,3);
				\draw [blue, ultra thick] (-1,2) --(0,3);
				\node [above] at (-2,3) {$1$};
				\node [above] at (0,3) {$2$};
				\node [above] at (2,3) {$3$};			
				\node [below] at (-1,2) {$b$};
				\node [below] at (0,0) {$4$};}$};
		\draw [ultra thick,red,->] (P1) -- (P2);
		\draw [ultra thick,red,->] (P1) -- (P5);
		\draw [ultra thick,red,->] (P2) -- (P3);
		\draw [ultra thick,red,->] (P3) -- (P4);
		\draw [ultra thick,red,->] (P5) -- (P6);
		\draw [ultra thick,red,->] (P6) -- (P4);
		\node at (1.5,1.3) {$(F^4_{123})_a^b$};
		\node at (1.4,-1.3) {$R^a_{12}$};
		\node at (4.5,2) {$R^4_{1b}$};
		\node at (4.5,-2) {$(F^4_{213})_a^c$};
		\node at (7.5,1.3) {$(F^4_{231})_b^c$};
		\node at (7.5,-1.3) {$R^c_{13}$};
	\end{tikzpicture}
	\caption{The hexagon identity.}
	\label{Hexagon}
\end{figure}

$F$ and $R$ have to satisfy some consistency conditions, which restrict the multiplicity of possible models to finitely many. Such conditions are called pentagon and Hexagon equations \cite{turaev2016quantum} due to their geometric interpretation. These conditions are also studied algebraically in Chapter \ref{Cat} on category theory. 
In the Fig. \ref{Pentagon}, the fusion process of four anyons $1,2,3,4$ is written as
\eq{{(F_{12c}^5)}_a^d {(F_{a34}^5)}_b^c = \sum_e {(F_{234}^d)}_e^c {(F_{1e4}^5)}_b^d {(F_{123}^b)}_a^e \label{EqPenta}.}
By the addition of braiding $R$-matrices, we get the relation pictorially shown in Fig. \ref{Hexagon}
\eq{\sum_b {(F_{231}^4)}_b^c R_{1b}^4 {(F_{123}^4)}_a^b = R_{13}^c{(F_{213}^4)}_a^c R_{12}^a \label{EqHexa}.}

The \textit{twist factor} or spin factor of an anyon is a phase corresponding to the rotation of a charge around its own magnetic flux. It can be thought of as a twist in a framed ribbon, as discussed in Chapter \ref{Knot}. This factor is written as $\theta_a = e^{2 \pi i h_a}$ when an anyon is rotated by $2\pi$, where $h_a$ is the \textit{topological spin} of an anyon $a$. It is an integer for bosons and gives spin factor identity. It is a half-integer for fermions that would give the spin factor $-1$. Its value is between 0 and 1 for an anyon. For vacuum, $h_0 = 0$. The twist factor is also discussed in Chapter \ref{Cat} and is shown in Fig. \ref{TwistPhase}. Through the ribbon equation, the spin factor can also be used to derive the entries of an $R$-matrix for a particular anyonic model. For example, when two anyons $a$ and $b$ are fused to $c$, the spin factor is given by the ribbon equation pictorially shown in Fig. \ref{BraidTwist} \cite{pachos2012introduction,eliens2014diagrammatics}
\eq{[R_{ab}^c]^2=\frac{\theta_c}{\theta_a \theta_b}= \frac{e^{2 \pi i h_c}}{e^{2 \pi i h_a} e^{2 \pi i h_b}} = e^{2 \pi i(h_c - h_a -h_b)}.}
This is interpreted as the full twist of fusion product $c$ combined with the full twist of the charges $a$ and $b$ in the opposite direction, and is equal to the double exchange of the $a$ and $b$.

\section{Topological Qubit}
Now we will summarize what we discussed in the last section and build a qubit from this discussion. A pair of non-Abelian anyons cannot be used directly as a qubit, because the two states belong to different topological charge sectors $\ket{ab;i_1}$ and $\ket{ab;i_2}$, and cannot be superposed by braiding.
Let three anyons $a$, $b$ and $c$ be fused to $d$. The first two are fused to $i$, then their outcome is fused with the third gives $d$.
We can write the two different fusion channels as the two states of our qubit as
\eq{\ket{i} = \ket{a,b \rightarrow i} \ket{i,c \rightarrow d},}
where the tensor product symbol is omitted. Alternatively, when the last two anyons $b$ and $c$ are fused to $j$, the $j$ is fused with $c$ to make $d$. $i$ and $j$ are two sets of bases. This is shown in the Fig. \ref{FRMatrices}. A qubit can also be formed as
\eq{\ket{j} = \ket{b,c \rightarrow j}\ket{j,a \rightarrow d}.}
The change of basis is performed by using the $F$-matrices as
\eq{\ket{i} = \sum_j (F_{abc}^d)_j^i \ket{j}.}
The $(F^d_{abc})_j^i$ are the matrix elements of $F_{abc}^d$ summed over $j$. $F$ and $R$ matrices are obtained for a particular anyon model from the solution of the pentagon and hexagon Eqs. \eqref{EqPenta} and \eqref{EqHexa} \cite{pachos2012introduction}. Ising and Fibonacci anyons are the most popular systems to make the topological quantum computing logic gates. These anyons are found as quasiparticles in non-Abelian fractional quantum Hall effect and topological superconductors.

A possible error in topological quantum computation is due to the braiding or fusion with some unattended anyon in the system. This error can be avoided by carefully accounting for the charges participating in the encoding. Another type of error could be due to the energy of the system, such that the gap between the ground state and the excited state gets filled. This error can be minimized by keeping the system at a very low temperature. The measurement of outcome is either interference or the projective measurement as discussed in Section \ref{AnyonDetect}, see also \cite{nayak2008non,bonderson2007non}.

\subsection{Example 1: Fibonacci Anyon}

The simplest non-Abelian anyon model consists of only two particles $\textbf{1}$ and $\tau$ \cite{trebst2008short,wang2010topological,delaney2016local}. The fusion rules for this anyon model are
\eq{\tau \cross \tau = \textbf{1} + \tau}
The basis states can be written as 
\eq{\ket{0} = \ket{\tau,\tau \rightarrow \textbf{1}}, \qquad \ket{1} = \ket{\tau,\tau \rightarrow \tau}}
The dimension is the different number of ways the fusion of all anyons can result in topological charge $\textbf{1}$ or $\tau$. In the fusion outcome of two $\tau$ we get $\textbf{1}$ with probability $p_0 = 1/\phi^2$ and $\tau$ with probability $p_1 = \phi/\phi^2 = 1/\phi$ \cite{delaney2016local}.
The dimension grows as the Fibonacci series, in which the next number is a sum of the last two numbers. The quantum dimension $d_\tau = \phi = (1+ \sqrt{5})/2$ is the golden mean.
\eq{\tau \cross \tau \cross \tau &= \textbf{1} + 2\tau \nonumber \\
	\tau \cross \tau\cross \tau\cross \tau &= 2 \cdot 1 + 3 \cdot \tau \nonumber \\
	\tau \cross \tau\cross \tau\cross \tau\cross \tau &= 3 \cdot 1 + 5 \cdot \tau \nonumber}
As we discussed above, no amount of braiding can change one qubit state to the other. Therefore, we need more than two $\tau$ particles for the qubit. Also, topological quantum computation has no tensor product structure. That means, if three $\tau$ anyons are used to make a qubit then the six anyons have only the five dimensional fusion space. Therefore, only a subspace is used to encode a qubit. Three Fibonacci anyons are required for the qubit and the fusion of four particles results in the vacuum \cite{lahtinen2017short}. $F$ and $R$ matrices for this model are obtained by consistency conditions, as in Ref. \cite{pachos2012introduction} are given as  
\eq{[F_{\tau\tau 1}^\tau]^\tau_\tau = \begin{pmatrix}
		\frac{1}{\phi} & \frac{1}{\sqrt{\phi}} \\
		\frac{1}{\sqrt{\phi}} & -\frac{1}{\phi}
	\end{pmatrix} \label{FFib}}
\eq{R_{\tau\tau} = \begin{pmatrix}
		R^1_{\tau\tau} & 0 \\
		0 & R^\tau_{\tau \tau}
	\end{pmatrix} = \begin{pmatrix}
		e^{4\pi i/5} & 0 \\
		0 & -e^{3\pi i/5}
	\end{pmatrix} \label{RFib}}
Quantum computing with Fibonacci anyons is done as follows.
The fusion of two $\tau$ particles gives either $\bm{1}$ or $\tau$. These two orthogonal states are represented as $\ket{(\bullet, \bullet)_\mathbf{1}}$ and $\ket{(\bullet, \bullet)_\tau}$. The addition of the third $\tau$ particle to the state $\ket{(\bullet, \bullet)_\mathbf{1}}$ gives $\tau$ and is denoted as $\ket{((\bullet, \bullet)_\mathbf{1}, \bullet)_\tau} \equiv \ket{0}$. But when the third particle is added to the state $\ket{(\bullet, \bullet)_\mathbf{1}}$, we get either $\bm{1}$ or $\tau$. These states are represented as $\ket{((\bullet,\bullet)_\tau,\bullet)_\tau}\equiv\ket{1}$ and
$\ket{((\bullet,\bullet)_\tau,\bullet)_{\bm{1}}}\equiv \ket{N}$, here $\ket{N}$ stands for the non-computational state. The amplitude in this state is considered as the \textit{leakage error} \cite{nayak2008non,hormozi2007topological}. The states $\ket{((\bullet, \bullet)_\mathbf{1}, \bullet)_\tau} \equiv \ket{0}$ and $\ket{((\bullet,\bullet)_\tau,\bullet)_\tau}\equiv\ket{1}$ are the basis states for a qubit and which are interchanged by an $F$ matrix \eqref{FFib}. The braiding of these particles is represented by an $R$ matrix \eqref{RFib}. These basis states and the fusion of Fibonacci anyons is shown in Fig.\ref{FusionSpaceFib}. The set of gates required to build any kind of circuit is called the universal quantum gate set. The Fibonacci anyonic model is the universal for quantum computing. These kinds of anyons are proposed to be found in the Read-Rezayi state $\nu = 12/5$ which is a very fragile state, so other anyon models are also under consideration \cite{nayak2008non}.

\begin{figure}[h!]
	\centering
	\begin{subfigure}{0.3\textwidth}
		\centering
		\begin{tikzpicture}
			\draw[ultra thick,blue] (0,0)--(3,0);
			\draw[ultra thick,blue] (1,0)--(1,1);
			\draw[ultra thick,blue] (2,0)--(2,1);
			\node [left] at (0,0) {$ \tau $};
			\node [above] at (1,1) {$ \tau $};
			\node [below] at (1.5,0) {$ i $};
			\node [above] at (2,1) {$ \tau $};
			\node [right] at (3,0) {$ \tau $};
		\end{tikzpicture}
		\caption{}
	\end{subfigure}
	\begin{subfigure}{0.4\textwidth}
		\centering
		\begin{tikzpicture}
			\draw[ultra thick,blue] (0,0)--(6,0);
			\draw[ultra thick,blue] (1,0)--(1,1);
			\draw[ultra thick,blue] (2,0)--(2,1);
			\draw[ultra thick,blue] (4,0)--(4,1);
			\draw[ultra thick,blue] (5,0)--(5,1);
			\node [left] at (0,0) {$ \tau $};
			\node [above] at (1,1) {$\tau $};
			\node [below] at (1.5,0) {$ e_1 $};
			\node [above] at (2,1) {$ \tau $};
			\node [below] at (2.5,0) {$ e_2 $};
			\node [above] at (4,1) {$ \tau $};
			\node [below] at (4.5,0) {$ e_{n-3} $};
			\node [above] at (5,1) {$ \tau $};
			\node [right] at (6,0) {$ \tau $};
		\end{tikzpicture}
		\caption{}
	\end{subfigure}	
	\caption[Fusion space of Fibonacci anyons.]{(a) Two orthogonal qubit states, $i$ would be either $1$ or $\tau$. (b) fusion space for $n$ Fibonacci anyons.}
	\label{FusionSpaceFib}
\end{figure}
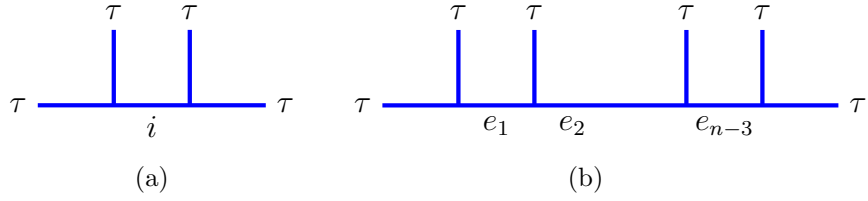

\begin{figure}[h!]
	\centering
	\begin{tikzpicture}
		\node at (-1.5,0) {$\ket{0} = \ket{((\bullet,\bullet)_1,\bullet)_\tau} =$};
		\draw [ultra thick,blue] (1.9,0) ellipse (1.5 and 0.5);
		\draw [ultra thick,blue] (1.55,0) ellipse (1 and 0.3);
		\draw [ultra thick,blue] (1.55,0) ellipse (1 and 0.3);
		\node[red] at (1,0) {$\bullet$};
		\node[red] at (2,0) {$\bullet$};
		\node[red] at (3,0) {$\bullet$};
		\node[below] at (2.65,0) {$\bm{1}$};
		\node[below] at (3.55,0) {$\tau$};
		\node at (4,0) {$=$};
		\draw [blue, ultra thick] (5,-1) -- (5,-0.5); 
		\draw [blue, ultra thick] (5,-0.5) -- (6,0.5);
		\draw [blue, ultra thick] (5,-0.5) -- (4,0.5);
		\draw [blue, ultra thick] (4.5,0) --(5,0.5);
		\node [above] at (6,0.5) {$\tau$};
		\node [above] at (4,0.5) {$\tau$};
		\node [above] at (5,0.5) {$\tau$};
		\node [below] at (5,-1) {$\tau$};
		\node [below] at (4.5,0) {$\bm{1}$};
	\end{tikzpicture}\\
	\begin{tikzpicture}
		\node at (-1.5,0) {$\ket{1} = \ket{((\bullet,\bullet)_\tau,\bullet)_\tau} =$};
		\draw [ultra thick,blue] (1.9,0) ellipse (1.5 and 0.5);
		\draw [ultra thick,blue] (1.55,0) ellipse (1 and 0.3);
		\draw [ultra thick,blue] (1.55,0) ellipse (1 and 0.3);
		\node[red] at (1,0) {$\bullet$};
		\node[red] at (2,0) {$\bullet$};
		\node[red] at (3,0) {$\bullet$};
		\node[below] at (2.65,0) {$\tau$};
		\node[below] at (3.55,0) {$\tau$};
		\node at (4,0) {$=$};
		\draw [blue, ultra thick] (5,-1) -- (5,-0.5); 
		\draw [blue, ultra thick] (5,-0.5) -- (6,0.5);
		\draw [blue, ultra thick] (5,-0.5) -- (4,0.5);
		\draw [blue, ultra thick] (4.5,0) --(5,0.5);
		\node [above] at (6,0.5) {$\tau$};
		\node [above] at (4,0.5) {$\tau$};
		\node [above] at (5,0.5) {$\tau$};
		\node [below] at (5,-1) {$\tau$};
		\node [below] at (4.4,0) {$\tau$};
	\end{tikzpicture}\\
	\begin{tikzpicture}
		\node at (-1.5,0) {$\ket{N} = \ket{((\bullet,\bullet)_\tau,\bullet)_{\bm{1}}}=$};
		\draw [ultra thick,blue] (1.9,0) ellipse (1.5 and 0.5);
		\draw [ultra thick,blue] (1.55,0) ellipse (1 and 0.3);
		\node[red] at (1,0) {$\bullet$};
		\node[red] at (2,0) {$\bullet$};
		\node[red] at (3,0) {$\bullet$};
		\node[below] at (2.65,0) {$\tau$};
		\node[below] at (3.55,0) {$\bm{1}$};
		\node at (4,0) {$=$};
		\draw [blue, ultra thick] (5,-1) -- (5,-0.5); 
		\draw [blue, ultra thick] (5,-0.5) -- (6,0.5);
		\draw [blue, ultra thick] (5,-0.5) -- (4,0.5);
		\draw [blue, ultra thick] (4.5,0) --(5,0.5);
		\node [above] at (6,0.5) {$\tau$};
		\node [above] at (4,0.5) {$\tau$};
		\node [above] at (5,0.5) {$\tau$};
		\node [below] at (5,-1) {$\bm{1}$};
		\node [below] at (4.4,0) {$\tau$};
	\end{tikzpicture}
	\caption[The orthogonal states of three Fibonacci particles.]{Orthogonal states of three Fibonacci particles \cite{nayak2008non}}
	\label{FibQubit}
\end{figure}
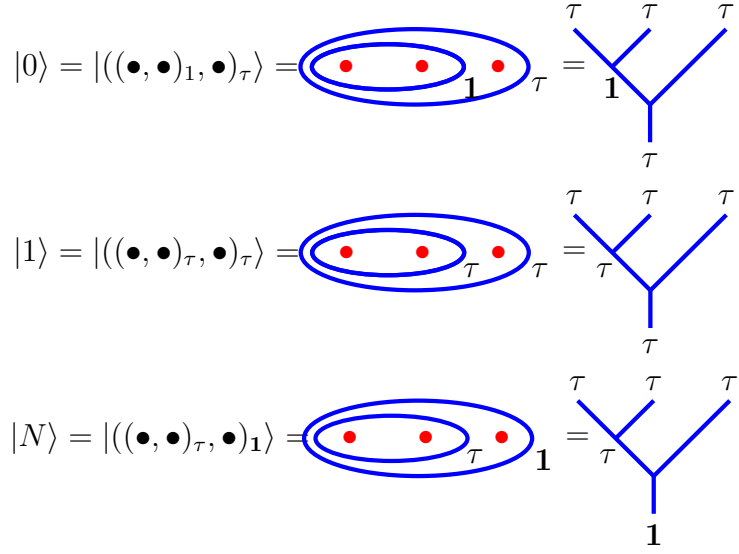

\subsection{Example 2: Ising Anyon}
This model has three anyons $\textbf{1}$, $\sigma$ and $\psi$.
The fusion rules for these anyons are;
\eq{\sigma \cross \sigma = \textbf{1} + \psi, \qquad \psi \cross \psi = \textbf{1}, \qquad \psi \cross \sigma = \sigma}
Two basis states can be written as
\eq{\ket{0}=\ket{\sigma, \sigma \rightarrow \textbf{1}}, \qquad \ket{1}=\ket{\sigma, \sigma \rightarrow \psi}}
Since two fusions belong to different topological charge sectors, at least three anyons are needed that can fuse to $\sigma$ in two different ways. For every added $\sigma$ the dimension of fusion space doubles, hence for $2N$ anyons the dimension is $2^{N-1}$. This model is non-universal, so non-topological schemes are also devised in addition to topological computation. The $F$ and $R$ matrices for this model \cite{lahtinen2017short} are,
\eq{F_{\sigma \sigma \sigma}^\sigma = \frac{1}{\sqrt{2}}\begin{pmatrix}
		1 & 1 \\
		1 & -1
	\end{pmatrix},\qquad
	R_{\sigma \sigma} = e^{-i \pi/8} \begin{pmatrix}
		1 & 0 \\
		0 & i
	\end{pmatrix},}
where $R_{\sigma \sigma}^1 = e^{-i\pi/8}$ and $R_{\sigma \sigma}^\psi = e^{i3\pi/8}$. The topological charges are labeled as $0,1/2, 1$ corresponds to $\textbf{1}, \sigma, \psi$. The total charge of two particles with charges $1/2$ is either $1$ or $0$. The total charge is $0$ when both particles have charges $1$, but the total charge is $1/2$ when one particle has charge $1$ and other has $1/2$. The quantum dimensions, $d_1=d_\psi =1$ and $d_\sigma=\sqrt{2}$, for these anyons are computed in Ref. \cite{pachos2012introduction} using the fusion rules.   

Let us consider four particles of charge $1/2$ with a total charge of $0$. The first two are fused either to $0$ or $1$. In case it is zero, then the total of the third and fourth must be zero. If the total of first and second is 1 then the total of third and fourth must be equal to 1. In this way, we have two states of four $1/2$ quasiparticles. There are $2^{n-1}$ states for $2n$ particles \cite{nayak19962n,nayak2008non}. 
When particles of the same pair, say $i$, are braided, only the phase is changed, but when a particle of pair $i$ is braided with the particle of other pair $j$, a NOT gate is applied \cite{nayak2008non}. Taking both particles of $i$ around both particles in $j$ then the basis state is multiplied by $+1$ if $j$ has a charge $0$, but it gets multiplied by $-1$ if it has a charge $1$. Six $\sigma$ anyons are required for two-qubit encoding. See Ref. \cite{nayak2008non,lahtinen2017short} for the implementation of CNOT and phase gates. The Ising anyon model is implemented by using the quantum Hall state $\nu = 5/2$ and Majorana particle in topological superconductors, see Chapters \ref{QHE} and \ref{TopMaterials}.

	\chapter{Recoupling Theory}\label{Rec}
	
The fusion of quasiparticles is similar to the recoupling theory of addition of angular momentum in quantum mechanics \cite{rose1995elementary,varshalovich1988quantum,biedenharn1981angular}. Ternary logic gates are designed using the q-deformed version of the recoupling theory. Therefore, for the intuition, we will discuss the quantum deformation and the recoupling of angular momenta and then discuss the quantum deformation of recoupling theory \cite{biedenharn1995quantum,kirillov1988representations,kirillov1991clebsch}. The quantum deformed quantities are also called q-analogs.

\section{Quantum Deformation}
In classical mechanics, states on phase space make a manifold represented by say $(q,p)$. Physical quantities are observables that are the functions of $(q,p)$. The Abelian algebra formed for these observables and associated geometry is commutative.
In quantum mechanics, due to the Heisenberg uncertainty principle, there is no arbitrary precision of the quantities. Algebra is non-commutative and classical mechanics is the limiting case when Planck constant $h \rightarrow 0$. Therefore, quantum mechanics is a kind of deformation of classical mechanics. 
In quantum mechanics, the commuting classical observables are replaced with the noncommuting Hermitian operators. Hence we can say that we deform classical algebra and the deformation parameter is $h$. 
The noncommutativity of the variables $X$ and $Y$ in the deformed space is written as 
\eq{XY = qYX,}
where $q$ is a complex number in general. It is called the \textit{deformation parameter}. Let $q$ be a number different from $1$, and $h$ be a number different from $0$. If we take $x= qx_0$ or $x=x_0+h$ we can have the classical values when $h\to 0$ or $q\to 1$. These two are related as $q= e^h$ \cite{jaganathan2000introduction}. For $q \to 1$, we would get back the classical commuting variables. 
Let us choose $q = e^{i\theta}$. Consider an example when the operators $T_\alpha$ and $G_{\theta/\alpha}$ are acting on a function $\psi(x)$ of real variable $x$, such that 
\eq{T_\alpha \psi(x) = \psi(x+\alpha), \qquad G_{\theta/\alpha} \psi(x) = e^{i \theta x / \alpha} \psi(x).}
When we apply both $T_\alpha$ and $G_{\theta/\alpha}$ operators, we get
\eq{T_\alpha G_{\theta/\alpha} \psi(x) = e^{i\theta(x+\alpha)} \psi(x+\alpha) = e^{i\theta} G_{\theta/\alpha} T_\alpha\psi(x).}
With the fixed value of variables $\theta$ and $\alpha$, $T_\alpha$ and $G_{\theta/\alpha}$ become noncommuting variables that can be written as
\eq{T_\alpha G_{\theta/\alpha} = e^{i\theta}G_{\theta/\alpha} T_\alpha.}
The q-analogs of an algebra is discussed in Appendix \ref{HopfAlg}. See also  \cite{jaganathan2000introduction}. 

\subsection{q-Analogs}

An anyon or a pair of anyons interacts through braiding in the plane deformed by the existence of the fields of the other anyons in an abstract way. This braiding interaction may cause the twist factor that is related to the topological spin of an anyon and involves the parameter $q$. Algebraically, we can think of $q$ as a small perturbation of the usual mathematical objects. The $q$ can be generic or a \textit{root of unity}. The root of unity is defined as when a complex number $q$ is raised to some power $n$ so that it equals 1 for that power, then we say that this complex number is $n$th root of unity. 
Now the $q$-analog quantities give the corresponding classical quantities for the limiting case when $q \rightarrow 1$.
For $n\in \mathbb{Z}$ we define what is called a \textbf{$q$-integer}
\eq{[n]_q= \frac{q^n - q^{-n}}{q-q^{-1}},}
which is the Laurent polynomial equal to $n \in \mathbb{Z}$ for $q \rightarrow 1$. 
It is done by taking the limit $q \to 1$ and applying L'Hospital's rule.
The $q$-analog of natural numbers is as follows, $[0]_q =0,\ [1]_q = 1,\ [2]_q = \frac{q^2-q^{-2}}{q-q^{-1}}$, and so on.
We can get \textbf{$q$-factorial} $[n]!$ that can be written as
\eq{[n]! &= [n][n-1]...[1] \\
	&=\frac{q^n - q^{-n}}{q-q^{-1}}\cdot \frac{q^{n-1} - q^{-(n-1)}}{q-q^{-1}}...\frac{q^2 - q^{-2}}{q-q^{-1}}\cdot 1}
Sometimes, we take $[n]_q = \frac{1-q^n}{1-q}$, as in \cite{jaganathan2000introduction,le2015quantum}.
In that case the $q$-analog of natural numbers and $q$-factorials are written as $[0]_q =0,\ [1]_q = 1,\ [2]_q = 1+q,\ [3]_q = 1+q+q^2$,
\eq{[n]!=1\cdot (1+q)(1+q+q^2)...(1+q+...+q^{n-1}).}
	\section{Recoupling of Angular Momenta}
We will briefly present the quantum theory of angular momentum to get the idea of the $SU(2)_k$ anyonic model will be discussed in the next section. This model is the quantum deformation of recoupling theory. For more detailed study, see \cite{biedenharn1981angular,varshalovich1988quantum,rose1995elementary}.
Let us consider two systems with angular momentum operators $J_1$ and $J_2$ with eigenvalues $j_1$ and $j_2$. These two systems may be the orbital angular momentum of two different particles or they may be the spin and orbital angular momentum of a single particle. The z-components of the angular momenta $J_z$ have allowed eigenvalues $-j_1 \le m_1 \le +j_1$ with $2j_1 + 1$ values, and $-j_2 \le m_2 \le +j_2$ with $2j_2 + 1$ states. The combined system is written as $j_1 \otimes j_2$.

It is like the vector spaces $V_1$ and $V_2$, with dimensions $2j_2 +1$ each, are combined as $V_1 \otimes V_2$ with dimensions $(2j_1 +1)(2j_2 +1)$. The total angular momentum operator is acting on $V_1 \otimes V_2$. This operator constitutes $SU(2)$ Lie algebra.
Two quantum numbers are needed to specify an individual system and four quantum numbers to specify the combined system. 
For the whole system 
\eq{J^2 = (J_1 +J_2)^2, \qquad J_z = J_{1z} + J_{2z}.}
These operators are applied to states as
\eq{J^2\ket{j,m} &= j(j+1)\ket{j,m}\nonumber \\
	J_z\ket{j,m} &=m\ket{j,m} \nonumber\\
	J_\pm\ket{j,m} &= \sqrt{(j\mp m)(j\pm m+1)}\ket{j,m\pm 1}}
where $J_+ = J_x+iJ_y$ and $J_- = J_x -iJ_y$.
The uncoupled states $\ket{j_1j_2m_1m_2}$ are the eigenstates of operators $\left\{J_1^2,J_{1z},J^2_2, J_{2z}\right\}$ and the coupled state $\ket{j_1j_2jm}$ are the eigenstate of operators $\left\{J_1^2,J^2_2,J^2, J^2_z\right\}$.
The coupling of $j_1$ and $j_2$ is the construction of eigenfunctions of $J^2$ and $J_z$
We will write the total angular momentum basis $\ket{j_1j_2;jm}$ in terms of tensor product basis $\ket{j_1m_1}\ket{j_2m_2}$. That is done by expressing the coupled state in terms of the uncoupled state.
\eq{&\ket{(j_1 j_2)jm}= \sum_{m_1 = -j_1}^{j_1}\sum_{m_2 = -j_2}^{j_2} \ket{j_1m_1}\ket{j_2m_2}\bra{j_1m_1j_2m_2}\ket{jm}\nonumber\\
	&=\sum_{j_1j_2m_1m_2}C_{mm_1m_2}^{jj_1j_2}\ket{j_1m_1}\ket{j_2m_2}.}
where $j=\abs{j_1-j_2},...,\abs{j_1+j_2}, \ m = -j,...,j$.
The coefficients $C_{mm_1m_2}^{jj_1j_2}$
\eq{\sum_{j_1j_2m_1m_2}C_{mm_1m_2}^{jj_1j_2} = \bra{j_1m_1;j_2m_2}\ket{j_1j_2;jm} = \bra{j_1m_1;j_2m_2}\ket{jm}}
are called Clebsch-Gordon coefficients (CGC). These are non-zero only when $m = m_1 + m_2$ and $\abs{j_1-j_2} \le j \le j_1 +j_2$.

\begin{figure*}[h!]
	\centering
	\begin{subfigure}{0.5\textwidth}
		\centering
		\begin{tikzpicture}[scale=0.7]
			\draw [blue, ultra thick] (2,0) -- (2,1); 
			\draw [blue, ultra thick] (2,1) -- (4,3);
			\draw [blue, ultra thick] (2,1) -- (0,3);
			\draw [blue, ultra thick] (1,2) --(2,3);
			\node [above] at (0,3) {$j_1$};
			\node [above] at (2,3) {$j_2$};
			\node [above] at (4,3) {$j_3$};
			\node [below] at (2,0) {$j$};
			\node [below left] at (1.5,1.5) {$j_{12}$};
		\end{tikzpicture}
		\begin{tikzpicture}[scale=0.7]	
			\draw [blue, ultra thick] (10,0) -- (10,1); 
			\draw [blue, ultra thick] (10,1) -- (12,3);
			\draw [blue, ultra thick] (10,1) -- (8,3);
			\draw [blue, ultra thick] (11,2) --(10,3);		
			\node [above] at (8,3) {$j_1$};
			\node [above] at (10,3) {$j_2$};
			\node [above] at (12,3) {$j_3$};
			\node [below] at (10,0) {$j$};
			\node [below right] at (10.5,1.5) {$j_{23}$};
		\end{tikzpicture}
		\caption{}
	\end{subfigure}
	\begin{subfigure}{0.2\textwidth}
		\centering	
		\begin{tikzpicture}[scale=1.5]
			\draw[ultra thick, blue] (0,-1) -- (0,0.8);
			\draw[ultra thick, blue] (0,-1) -- (1,-0.6);
			\draw[ultra thick, blue] (0,-1) -- (-1,-0.6);
			\draw[ultra thick,dashed, blue] (-1,-0.6) -- (1,-0.6);
			\draw[ultra thick, blue] (-1,-0.6) -- (0,0.8);
			\draw[ultra thick, blue] (1,-0.6) -- (0,0.8);
			\node [left] at (0.05,0) {$j_3$};
			\node [] at (0.5,-0.45) {$j_1$};
			\node [] at (0.5,-1) {$j_{12}$};
			\node [] at (-0.5,-1) {$j_2$};
			\node [] at (0.55,0.3) {$j$};
			\node [] at (-0.55,0.4) {$j_{23}$};
		\end{tikzpicture}
		\caption{}
	\end{subfigure}
	\caption[Recoupling of three angular momenta and triangle condition.]{(a) Recoupling of three angular momenta (b) the triangle condition. }
	\label{TriCond}
\end{figure*}
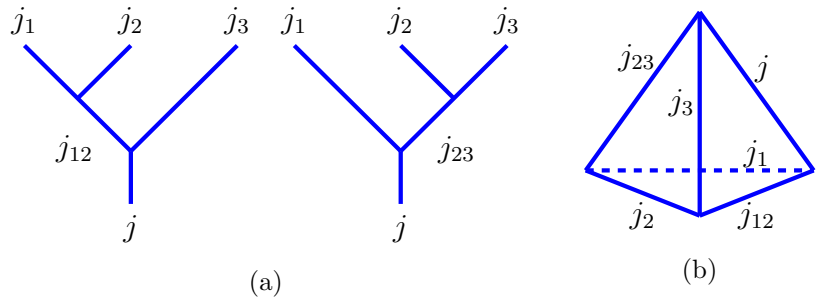

The total angular momentum can have value $j = j_1 + j_2, j_1+j_2 -1,...,\abs{j_1 -j_2}$.
Any of the numbers $j_1,j_2,j$ can have values that are greater than or equal to the difference of the other two and less than or equal to the sum of the other two. This condition is called \textit{triangle condition} and is represented by $\Delta(j_1j_2j)$.
We can also write the total angular momentum basis in terms of the product basis by using the \textit{Wigner's 3j-symbols}. In that case, the coefficients are called \textit{Wigner coefficients}.
The Wigner 3j-symbol is zero unless the triangle condition is satisfied. 
The CGC are related to the 3j-symbols as
\eq{\bra{j_1m_1;j_2m_2}\ket{j_1j_2;jm} = (-1)^{-j_1+j_2 -m}\sqrt{2j+1}
	\begin{Bmatrix}
		j_1&j_2&j \\
		m_1&m_2&-m
	\end{Bmatrix}.}
On similar lines, we can couple three angular momenta
$J_1,J_2,J_3$ whose total angular momentum $J= J_1+J_2+J_3$. There are two coupling schemes as
\eq{(J_1+J_2)+J_3 = J_{12}+J_3 = J \qquad J_1+(J_2 +J_3) = J_1+J_{23} = J.} 
The total coupling may be done by 
first coupling $j_1$ and $j_2$ to $j_{12}$ and then $j_{12}$ and $j_3$ to $J$
\eq{\ket{(j_1j_2)j_{12}j_3;jm}
	= \sum_{m_{12} = -j_{12}}^{j_{12}}\sum_{m_3 = -j_3}^{j_3} \ket{(j_1j_2);j_{12}m_{12}}\ket{j_3m_3} 
	\bra{j_{12}j_3;m_{12}m_3}\ket{j_{12}j_3;jm}.}
Alternatively, we can first combine $j_2$ and $j_3$ to get $j_{23}$ and next $j_{23}$ can be combined with $j_1$ to make $J$
\eq{\ket{j_1((j_2 j_3)j_{23});jm} 
	= \sum_{m_1 = -j_1}^{j_1}\sum_{m_{23} = -j_{23}}^{j_{23}}	
	\ket{j_1m_1}
	\ket{j_2j_3;j_{23}m_{23}} \bra{j_1 j_{23};m_1m_{23}}\ket{j_1j_{23};jm}.}
These two coupling schemes are shown in Fig. \ref{TriCond} (a). The coupling scheme results in a complete orthonormal bases for the $(2j_1 + 1)(2j_2 + 1)(2j_3+1)$-dimensional space spanned by $\ket{j_1,m_1}\ket{j_2,m_2}\ket{j_3,m_3}$, $m_1 = -j_1,...,j_1, m_2 = -j_2,...,j_2$; $m_3 = -j_3,...,j_3$.

The angular momenta in two coupling schemes are related by a unitary transformation. The matrix elements of this unitary transformation are known as \textit{recoupling coefficients}. These coefficients are independent of $m$ and so we have

\eq{\ket{((j_1j_2)j_{12}j_3)jm} &= \sum_{j_{23}} \ket{(j_1(j_2j_3)j_{23})jm}\bra{(j_1(j_2j_3)j_{23})j}\ket{((j_1j_2)j_{12}j_3)j}.}
These coefficients can be written in terms of \textit{Wigner 6j-symbols},
\eq{\bra{(j_1(j_2j_3)j_{23})j}\ket{((j_1j_2)j_{12}j_3)j} = (-1)^{j_1+j_2+j_3+j} \sqrt{(2j_{12} +1)(2j_{23} +1)}\begin{Bmatrix}
		j_1 & j_2 & j_{12} \\
		j_3 & j & j_{23}
	\end{Bmatrix}.}
The 6j-symbols have a symmetry that permutation of columns or rows leaves it invariant. Similar to the 3j-symbols, 6j-symbols are not matrices. The Racah coefficients \cite{racah1942theory} are related to the recoupling coefficients as
\eq{W(j_1 j_2 j_3J; j_{12}j_{23}) = \frac{\bra{(j_1(j_2j_3)j_{23})j}\ket{((j_1j_2)j_{12}j_3)j}}{\sqrt{(2J_{12}+1)(2J_{23}+1)}}.}
Therefore, the Racah coefficients are related to the Wigner 6j-symbols by
\eq{\begin{Bmatrix}
		j_1 & j_2 & j_{12} \\
		j_3 & j & j_{23}
	\end{Bmatrix} = (-1)^{j_1+j_2+j_3+j} W(j_1j_2j_3j;j_{12}j_{23})}
If $a \equiv j_1, \ b \equiv j_2, \ c \equiv j_3, \ d \equiv j, \ e \equiv j_{12}, \ f \equiv j_{23}$, we have the triangle condition as
\eq{\Delta(abc) = \sqrt{\frac{(a+b-c)!(a-b+c)!(-a+b+c)!}{(a+b+c+1)!}}.}
The right hand side is zero unless the triangle condition is satisfied. This condition is satisfied by each side of the quadrilateral in Fig. \ref{TriCond} (b). The Racah coefficient is a product of four of these factors
\eq{W(abcd;ef) = \Delta(abe)\Delta(cde)\Delta(acf)\Delta(bdf) \omega(abcd;ef),}
where $$\omega(abcd;ef) = \sum_z \frac{(-1)^{z+\beta_1}(z+1)!}{(z-\alpha_1)!(z-\alpha_2)!(z-\alpha_3)!(z-\alpha_4)!(\beta_1 - z)!(\beta_2 - z)!(\beta_3 - z)!},$$
$\alpha_1 = a+b+e, \ \alpha_2 = c+d+e, \ \alpha_3 = a+c+f \ \alpha_4 = b+d+f$
$\beta_1 = a+b+c+d, \ \beta_2 = a+d+e+f, \ \beta_3 = b+c+e+f$.
The sum over $z$ is finite over the range $max(\alpha_1,\alpha_2,\alpha_3,\alpha_4) \le z \le min(\beta_1,\beta_2,\beta_3)$
See \cite{rose1995elementary,aquilanti2009combinatorics,santos2017couplings} for detailed derivation of the above equation and coupling and recoupling of angular momenta. 

\section{$SU(2)_k$ Anyon Model}
The F-symbols in topological quantum computation can be computed using the $SU(2)_k$ model \cite{bonderson2007non} and Temperley-Lieb recoupling theory \cite{kauffman1994temperley} also called the $JK_k$ anyon model \cite{levaillant2015universal}. Where the $k$ is called the level of the theory. It is the coupling constant of Chern-Simons theory and is related to the number of particles present as we discussed in Chapter \ref{TQC}. These theories are the quantum analog of theory of addition of angular momentum. The $SU(2)_k$ is the q-deformed version of $SU(2)$ with $q=\exp(i2\pi/(k+2))$, for $q$ at the root of unity. The detailed derivation of the parameter $q$ in this form is given in \cite{witten1989quantum}. The anyon's fusion amplitudes would be written as the recoupling coefficients. The $F$-symbols and $R$-symbols are obtained by using these two models at level $k=4$ gave identical values. The charges are half integral in $SU(2)_k$ model, but are integral in $JK_k$ model. For $F$ and $R$ symbols in the $JK_4$ model, see \cite{levaillant2015universal}. In our work, we will use the $SU(2)_4$ model. The topological data for this model are given as
\eq{[n]_q = \frac{q^{n/2}-q^{-n/2}}{q^{1/2} - q^{-1/2}}, \qquad {\mathcal{C}} = \left\{0, 1/2,..., k/2 \right\}, \qquad j_1 \times j_2 = \sum_{j = |{j_1-j_2}|}^{\min{j_1+j_2, k-j_1-j_2}},}

\eq{[F_j^{j_1,j_2,j_3}]_{j_{12},j_{23}} = (-1)^{j_1 + j_2 + j_3 +j} \sqrt{[2j_{12}+1]_q [2j_{23}+1]_q}
	\begin{Bmatrix} 
		j_1 & j_2 & j_{12}\\ 
		j_3 & j & j_{23}\\ 
	\end{Bmatrix}_q,}
where
\eq{\begin{Bmatrix} 
		j_1 & j_2 & j_{12}\\ 
		j_3 & j & j_{23}\\ 
	\end{Bmatrix}=\Delta(j_1,j_2,j_3)\Delta(j_{12},j_3,j)\Delta(j_2,j_3,j_{23})\Delta(j_1,j_{23},j) \nonumber\\ \times \sum_z \frac{(-1)^z[z+1]_q !}{[z-j_1-j_2 -j_{12}]_q![z-j_{12}-j_3 -j]_q![z-j_2-j_3 -j_{23}]_q![z-j_1-j_{23} -j]_q!} \nonumber\\ \times \frac{1}{[j_1 +j_2 +j_3 +j -z]_q![j_1 +j_{12} +j_3 +j_{23} -z]_q![j_2 +j_{12} +j +j_{23} -z]_q!},}

$$ \Delta(j_1,j_2,j_3) = \sqrt{\frac{[-j_1 +j_2 +j_3]_q![j_1-j_2+j_3]_q! [j_1 +j_2 -j_3]_q!}{[j_1 + j_2 +j_3 +1]_q!}}, \qquad [n]_q! \equiv \prod_{m=1}^n [m]_q,$$

$$ R_j^{j_1,j_2} = (-1)^{j-j_1-j_2} q^{\frac{1}{2}[j(j+1) - j_1(j_1+1) -j_2(j_2+1)]},$$

$$ d_j = [2j+1]_q = \frac{\sin \big[\frac{(2j+1)\pi}{k+2}\big]}{\sin \big(\frac{\pi}{k+2}\big)}, \quad {\mathcal{D}} = \frac{\sqrt{\frac{k+2}{2}}}{\sin \big(\frac{\pi}{k+2}\big)},$$


$$ \theta_j = q^{j(j+1)} = e^{i 2\pi \frac{j(j+1)}{k+2}}, \qquad S_{j_1j_2} = \sqrt{\frac{2}{k+2}} \sin \big[ \frac{(2j_1 +1)(2j_2 +1) \pi}{k+2} \big].$$

Where $d$, $\mathcal{D}$, and $S_{j_1,j_2}$ are quantum dimension, total quantum dimension and topological $S$-matrix respectively. The data in $SU(2)_k$ theory are used to compute the $[F_j^{j_1,j_2,j_3}]_{j_{12},j_{23}}$ and $R^{j_1j_2}_j$ matrices \cite{cui2015universal}.
For $SU(2)_4$, the charges are $0,1,2,3,4$, and the fusion rules are 
$$0\cross 0 = 0, \ 0\cross 1 = 1, \ 0\cross 2 = 2, \ 0\cross 3 = 3, \ 0\cross 4 = 4,$$ 
$$1\cross 1 = 0+2, \ 1\cross 2 = 1+3, \ 1\cross 3 = 2+4, \ 1\cross 4 = 3,$$ 
$$2\cross 2 = 0+2+4, \ 2\cross 3 = 1+3, \ 2\cross 4 = 2,$$ 
$$3\cross 3 = 0+2, \ 3\cross 4 = 1, \ 4\cross 4 = 0$$

The $F$ and $R$ symbols are calculated using these fusion rules and from the $F$ and $R$ matrices, the $\sigma$ matrices are obtained in the next Chapter \ref{Meta} \cite{cui2015universal,levaillant2015universal}. The $F$ and $R$ symbols for $SU(2)_k$ anyonic model are given as\\
$[F^{abc}_d]_{ef}=1$ when any of the $a,b,c,d$ is zero, \\
$F_4^{114} = F_4^{123} = F_3^{124} = F_4^{132} = F_3^{133} = F_2^{134} = F_4^{141} = F_3^{142} = F_2^{143} = F_1^{144} = F_4^{213} = F_3^{214} = F_4^{222} = F_2^{224} = F_4^{231} = F_1^{234} = F_3^{241} = F_2^{242} = F_1^{243} = F_4^{312} = F_3^{313} = F_2^{314} = F_4^{321} = F_1^{324} = F_3^{331} = F_1^{331} = F_4^{334} = F_2^{341} = F_1^{342} = F_4^{343} = F_3^{344} = F_4^{411} = F_3^{412} = F_2^{413} = F_1^{414} = F_3^{421} = F_2^{422} = F_1^{423} = F_2^{431} = F_1^{432} = F_4^{433} = F_3^{434} = F_1^{441} = F_3^{443} = -1$,\\
$F_1^{111} = F_3^{131} = F_1^{313} = F_3^{333} = \begin{pmatrix}
	-\frac{1}{\sqrt{3}} & \sqrt{\frac{2}{3}} \\
	\sqrt{\frac{2}{3}} & \frac{1}{\sqrt{3}}
\end{pmatrix},$\\
$F_2^{112} = F_1^{122} = F_3^{122} = F_2^{132} = F_2^{211} = F_2^{213} = F_1^{221} = F_3^{221} = F_1^{223} = F_2^{231} = F_2^{312} = F_1^{322} = \begin{pmatrix}
	-\frac{1}{\sqrt{2}} & \sqrt{\frac{1}{2}} \\
	\sqrt{\frac{1}{2}} & \frac{1}{\sqrt{2}}
\end{pmatrix},$\\
$F_3^{113} = F_1^{133} = F_3^{311} = F_1^{331} = \begin{pmatrix}
	-\frac{2}{\sqrt{3}} & \sqrt{\frac{1}{3}} \\
	\sqrt{\frac{1}{3}} & \frac{2}{\sqrt{3}}
\end{pmatrix},$\\
$F_2^{121} = F_1^{212} = \begin{pmatrix}
	-\frac{1}{2} & \frac{\sqrt{3}}{2} \\
	\frac{\sqrt{3}}{2} & \frac{1}{2}
\end{pmatrix},$\\
$F_2^{123} = F_3^{212} = F_1^{232} = F_2^{321} = \begin{pmatrix}
	-\frac{\sqrt{2}}{3} & \frac{1}{2} \\
	\frac{1}{2} & \frac{\sqrt{2}}{3}
\end{pmatrix},$\\
$F_3^{223} = F_2^{233} = F_3^{322} = F_2^{332} = \begin{pmatrix}
	\frac{1}{\sqrt{2}} & -\sqrt{\frac{1}{2}} \\
	-\sqrt{\frac{1}{2}} & -\sqrt{\frac{1}{2}}
\end{pmatrix},$\\
$F_3^{232} = F_2^{323} = \begin{pmatrix}
	\frac{1}{2} & -\frac{\sqrt{3}}{2} \\
	-\frac{\sqrt{3}}{2} & -\frac{1}{2}
\end{pmatrix},$\\
$F_2^{222} = \begin{pmatrix}
	\frac{1}{2} & -\frac{1}{\sqrt{2}} & \frac{1}{2} \\
	-\frac{1}{\sqrt{2}} & 0 & \frac{1}{\sqrt{2}} \\
	\frac{1}{2} & \frac{1}{\sqrt{2}} & \frac{1}{2}
\end{pmatrix},$\\
$R_0^{00} = R_1^{01} = R_2^{02} = R_3^{03} = R_4^{04} = R_1^{10} = R_2^{20} = R_3^{30} = R_4^{40} = R_0^{44} = 1,$
$R_0^{11} = e^{3i\pi/4},$
$R_2^{11} = e^{i \pi/12},$
$R_1^{12} = R_1^{21} = R_2^{22} = R_3^{23} = R_3^{32} = e^{2i\pi/3},$\\
$R_3^{12} = R_3^{21} = e^{i\pi/6},$
$R_2^{13} = R_2^{31} = e^{7i\pi/12},$
$R_4^{13} = R_4^{31} = e^{i\pi/4},$
$R_3^{14} = R_3^{41} = i,$
$R_0^{22} = e^{-2i\pi/3},$\\
$R_4^{22} = e^{i\pi/3},$
$R_1^{23} = R_1^{32} = e^{-5i\pi/6},$
$R_2^{24} = R_2^{42} = -1,$
$R_0^{33} = e^{-i\pi/4},$
$R_2^{33} = e^{-11i\pi/12},$
$R_1^{34} = R_1^{43} = -i$.

	\chapter{Ternary Logic Design with Metaplectic Anyons}\label{Meta}
	
Quantum computation is performed with metaplectic anyons which are simple objects in weakly integral categories. The term metaplectic is for a braid group that is in the metaplectic representation. These representations are the symplectic analog of spinor representation. \cite{cui2015universal,bocharov2016efficient}.

The metaplectic anyons can be studied from the category theory. Anyons are simple objects in a unitary modular category. See Chapter \ref{Cat} for the introduction to category theory and the use of category theory in topological quantum computation. A category is integral when the quantum dimension or Frobenius-Perron dimension of a simple object is an integer, whereas a category is called weakly integral if the squares of the quantum dimensions of all the simple objects are integer. Weakly integral categories are a class of metaplectic categories \cite{bruillard2016classification,hastings2014metaplectic}.
There are five anyons $\left\{1,Z,X,X^{'},Y\right\}$ in the theory of metaplectic anyons with fusion rules, quantum dimensions, and topological twists given as
\eq{X \otimes X &= 1 +Y, \nonumber\\
	Y \otimes Y &= 1 + Z + Y,\nonumber\\
	X \otimes Z &= X',\nonumber\\
	X \otimes X' &= Z+Y,} 
\eq{d_1 = d_Z = 1, \ d_X = d_{X'}= \sqrt{3}, \ d_{Y} = 2,}
\eq{\theta_0 = \theta_4 = 1, \ \theta_1 = \theta_3 = e^{i\pi/4}, \ \theta_2 = e^{i2\pi/3}.}

There is a non-Abelian boson quasiparticle $Z$. These theories also have a fundamental particle $X$. This particle is also a non-Abelian. It is a vortex for the $Z$ boson. This $X$ particle is fused with another $X$ particle to give $Y_i$ or vacuum, where $i = 1,2,...,r$ and $r= (m-1)/2$. These non-Abelian particles $Y_i$ have the quantum dimension $2$. When $X$ and $Z$ are fused, the result is the particle $X'$ \cite{hastings2013metaplectic,hastings2014metaplectic}.
A collection of $N$ quasiparticles $X$ at a fixed position has an $n_N$-dimensional degenerate subspace with $n_N \sim m^{N/2}$.
The proposed metaplectic anyon systems are the quantum Hall effect and Majorana zero modes \cite{barkeshli2012topological,clarke2013exotic,lindner2012fractionalizing,cheng2012superconducting,vaezi2013fractional}.

\section{One-Qutrit Braiding Gates}
Let us consider four $X$ anyons. The first two of the four are fused to $c_{12}$ and the last two are fused to $c_{34}$ as shown in Fig. \ref{OneTwoQutrit} (a).
With the constraint $c_{14} = Y$, we get three fusion trees \cite{cui2015universal,bocharov2016efficient}
\eq{(c_{12},c_{34}) \in \left\{-(YY),(\textbf{1}Y),(Y\textbf{1})\right\}.}
These are corresponding to the three states of qutrit $\ket{0},\ket{1},\ket{2}$. The minus sign is just to make the algebra nicer later. 
Let $\sigma_1$ be a braid matrix for the first two particles, and $\sigma_2$ corresponds to a braid of the second with the third, and $\sigma_3$ is a braid matrix for the third and fourth as shown in Fig. \ref{Sigma13} and \ref{Sigma2}.
The associated Hilbert space is represented by $V_y^{\epsilon \epsilon \epsilon \epsilon}$, for $\epsilon=X$. Under the basis $\left\{-\ket{YY},\ket{\bm{1}Y}, \ket{Y\bm{1}}\right\}$, the generators of the braid group ${\mathcal{B}}_4$ for the representation $V_y^{\epsilon \epsilon \epsilon \epsilon}$ are
\eq{&\sigma_1 = \gamma \begin{pmatrix}
		1 & 0 & 0\\
		0 & \omega & 0\\
		0 & 0 & 1
	\end{pmatrix}, \qquad \sigma_3 = \gamma \begin{pmatrix}
		1 & 0 & 0\\
		0 & 1 & 0\\
		0 & 0 & \omega
\end{pmatrix},}
\eq{\sigma_2 &= \frac{\gamma^3}{\sqrt{3}} \begin{pmatrix}
		1 & \omega & \omega\\
		\omega & 1 & \omega\\
		\omega & \omega & 1
	\end{pmatrix} = \gamma \begin{pmatrix}
		\frac{1}{2} + \frac{\sqrt{3}i}{6} & -\frac{1}{2} + \frac{\sqrt{3}i}{6} & -\frac{1}{2} + \frac{\sqrt{3}i}{6} \\
		-\frac{1}{2} + \frac{\sqrt{3}i}{6} & \frac{1}{2} + \frac{\sqrt{3}i}{6} & -\frac{1}{2} + \frac{\sqrt{3}i}{6} \\
		-\frac{1}{2} + \frac{\sqrt{3}i}{6} & -\frac{1}{2} + \frac{\sqrt{3}i}{6} & \frac{1}{2} + \frac{\sqrt{3}i}{6}
\end{pmatrix},}
where $\omega = e^{2 \pi i/3}$ and $\gamma = e^{\pi i/12}$.
Ignoring the $\gamma$ in front, let us define \cite{cui2015universal}
$p = \sigma_1 \sigma_2 \sigma_1$, $q = \sigma_2 \sigma_3 \sigma_2$, 
\eq{p^2 = -\begin{pmatrix}
		0 & 1 & 0\\
		1 & 0 & 0\\
		0 & 0 & 1
	\end{pmatrix}, \qquad
	q^2 = -\begin{pmatrix}
		0 & 0 & 1\\
		0 & 1 & 0\\
		1 & 0 & 0
	\end{pmatrix}, \qquad
	-(q^2pq^2)^2 = \begin{pmatrix}
		1 & 0 & 0\\
		0 & 0 & 1\\
		0 & 1 & 0
	\end{pmatrix}, \nonumber \\
	(q^2pq^2)^2Z^*((q^2pq^2)^2)^* = \begin{pmatrix}
		0 & 0 & 1\\
		1 & 0 & 0\\
		0 & 1 & 0
	\end{pmatrix}, \qquad 
	(q^2pq^2)^2Z((q^2pq^2)^2)^* = \begin{pmatrix}
		0 & 1 & 0\\
		0 & 0 & 1\\
		1 & 0 & 0
	\end{pmatrix}.\label{ZGates}}
These gates correspond to one-qutrit gates $Z_3(+1)$, $Z_3(+2)$, $Z_3(01)$, $Z_3(12)$, $Z_3(02)$ in conventional quantum computing discussed in Chapter \ref{QC}. The phase gate $Z=\sigma_1\sigma_3^{-1}=\sigma_1\sigma_3^2$ and the ternary Hadamard gate $H=q^2pq^2$ can be written in matrix form as
\eq{Z = \begin{pmatrix}
		1 & 0 & 0\\
		0 & \omega & 0\\
		0 & 0 & \omega^2
	\end{pmatrix}, \
	H = \frac{1}{\sqrt{3}i}\begin{pmatrix}
		1 & 1 & 1\\
		1 & \omega & \omega^2\\
		1 & \omega^2 & \omega
\end{pmatrix}.}

\section{Two-Qutrit Braiding Gates}
The two-qutrit model would consist of eight $X$ anyons with the final fusion outcome $Y$ as shown in Fig. \ref{OneTwoQutrit} (b). 
The braid matrices for two qutrits are written as $\sigma_1,\sigma_2,\sigma_3,\sigma_4,\sigma_5,\sigma_6,\sigma_7$.
Let us define
\eq{s_1 = \sigma_2\sigma_1\sigma_3\sigma_2, \ s_2 = \sigma_4\sigma_3\sigma_5\sigma_4, \ s_3 = \sigma_6\sigma_5\sigma_7\sigma_6.}
From these matrices, we can calculate a matrix
\eq{\Lambda (Z) = s_1^{-1}s_2^2 s_1 s_3^{-1}s_2^2s_3.}
The two-qutrit encoding is obtained when restricting the vector space $V_y^{\epsilon \epsilon \epsilon \epsilon \epsilon \epsilon \epsilon \epsilon}$ to nine dimensional subspace $V_y^{\epsilon \epsilon \epsilon \epsilon}\otimes V_y^{\epsilon \epsilon \epsilon \epsilon} \subset V_y^{\epsilon \epsilon \epsilon \epsilon \epsilon \epsilon \epsilon \epsilon}$
with $c_{14}=c_{58}= Y$. This nine-dimensional restriction of the $\Lambda (Z)$ is the Controlled-Z gate (CZ).
The SUM gate is a generalization of the CNOT gate \cite{cui2015universal,bocharov2016efficient}. It is related to CZ as
\eq{SUM = (I\otimes H)CZ(I \otimes H^{-1}).\label{SumGate}}
This $SUM$ gate will be combined with the topological charge measurement to build arithmetic circuits in the next section.
\begin{figure}[t!]
	\centering
	\begin{subfigure}{0.3\textwidth}
		\centering
		\begin{tikzpicture}[scale=0.5]
			\draw [blue, ultra thick] (3,0) -- (3,1); 
			\draw [blue, ultra thick] (3,1) -- (0,4);
			\draw [blue, ultra thick] (3,1) -- (6,4);
			\draw [blue, ultra thick] (1,3) -- (2,4);
			\draw [blue, ultra thick] (5,3) --(4,4);
			\node [below left] at (1.5,2.5) {$c_{12}$};
			\node [below right] at (4.5,2.5) {$c_{34}$};
			\node [below] at (3,0) {$c_{14}$};
			\node [above] at (0,4) {X};
			\node [above] at (2,4) {X};
			\node [above] at (4,4) {X};
			\node [above] at (6,4) {X};
		\end{tikzpicture}
		\caption{}
	\end{subfigure}
	\begin{subfigure}{0.5\textwidth}
		\centering
		\begin{tikzpicture}[scale = 0.3]
			\draw [blue, ultra thick] (7,0) -- (7,1);
			\draw [blue, ultra thick] (7,1) -- (0,8);
			\draw [blue, ultra thick] (7,1) -- (14,8);%
			\draw [blue, ultra thick] (3,5) --(6,8);
			\draw [blue, ultra thick] (11,5) --(8,8);%
			\draw [blue, ultra thick] (1,7) --(2,8);
			\draw [blue, ultra thick] (5,7) --(4,8);
			\draw [blue, ultra thick] (9,7) --(10,8);
			\draw [blue, ultra thick] (13,7) --(12,8);%
			\node [above] at (0,8) {X};
			\node [above] at (2,8) {X};
			\node [above] at (4,8) {X};
			\node [above] at (6,8) {X};%
			\node [above] at (8,8) {X};
			\node [above] at (10,8) {X};
			\node [above] at (12,8) {X};
			\node [above] at (14,8) {X};%
			\node [below left] at (2,6) {$c_{12}$};
			\node [below right] at (4,6) {$c_{34}$};%
			\node [below left] at (10,6) {$c_{56}$};
			\node [below right] at (12,6) {$c_{78}$};%
			\node [below left] at (5,3) {$c_{14}$};
			\node [below right] at (9,3) {$c_{58}$};%
			\node [below] at (7,0) {Y};
		\end{tikzpicture}
		\caption{}
	\end{subfigure}
	\caption[One-qutrit and two-qutrit gates.]{(a) One-qutrit gate (b) Two-qutrit gate \cite{bocharov2016efficient}.}
	\label{OneTwoQutrit}
\end{figure}
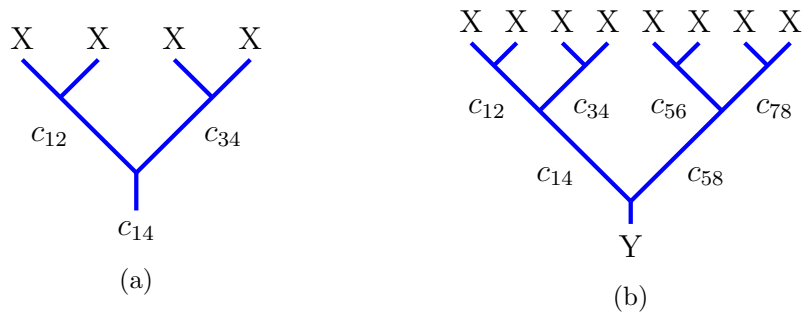
\begin{figure}[t!]
	\centering
	\begin{subfigure}{0.4\textwidth}
		\centering
		\begin{tikzpicture}[scale = 0.3]
			\draw [blue, ultra thick] (3,0) -- (3,1); 
			\draw [blue, ultra thick] (3,1) -- (6,4);
			\draw [blue, ultra thick] (3,1) -- (0,4);
			\draw [blue, ultra thick] (1,3) --(2,4);
			\draw [blue, ultra thick] (5,3) --(4,4);
			\node [below] at (3,0) {$b$};
			\node [above] at (0,4) {$a$};
			\node [above] at (2,4) {$a$};
			\node [above] at (4,4) {$a$};
			\node [above] at (6,4) {$a$};
			\node [below left] at (2,2.5) {$x_i$};
			\node [below right] at (4,2.5) {$y_i$};
		\end{tikzpicture}
		$=R^{aa}_{x_i}$
		\begin{tikzpicture}[scale = 0.3]
			\draw [knot=blue, ultra thick] (2,4) --(0,6);
			\draw [knot=blue, ultra thick] (0,4) -- (2,6);
			\draw [blue, ultra thick] (3,0) -- (3,1); 
			\draw [blue, ultra thick] (3,1) -- (6,4);
			\draw [blue, ultra thick] (3,1) -- (0,4);	
			\draw [blue, ultra thick] (1,3) --(2,4);	
			\draw [blue, ultra thick] (5,3) --(4,4);
			\node [below] at (3,0) {$b$};
			\node [above] at (2,6) {$a$};
			\node [above] at (0,6) {$a$};
			\node [above] at (4,4) {$a$};
			\node [above] at (6,4) {$a$};
			\node [below left] at (2,2.5) {$x_i$};
			\node [below right] at (4,2.5) {$y_i$};
		\end{tikzpicture}
		\caption{}
	\end{subfigure}
	\begin{subfigure}{0.4\textwidth}
		\centering
		\begin{tikzpicture}[scale = 0.3]
			\draw [blue, ultra thick] (3,0) -- (3,1); 
			\draw [blue, ultra thick] (3,1) -- (6,4);
			\draw [blue, ultra thick] (3,1) -- (0,4);
			\draw [blue, ultra thick] (1,3) --(2,4);
			\draw [blue, ultra thick] (5,3) --(4,4);
			\node [below] at (3,0) {$b$};
			\node [above] at (0,4) {$a$};
			\node [above] at (2,4) {$a$};
			\node [above] at (4,4) {$a$};
			\node [above] at (6,4) {$a$};
			\node [below left] at (2,2.5) {$x_i$};
			\node [below right] at (4,2.5) {$y_i$};
		\end{tikzpicture}
		$=R^{aa}_{y_i}$
		\begin{tikzpicture}[scale = 0.3]
			\draw [knot=blue, ultra thick] (6,4) --(4,6);
			\draw [knot=blue, ultra thick] (4,4) -- (6,6);
			\draw [blue, ultra thick] (3,0) -- (3,1); 
			\draw [blue, ultra thick] (3,1) -- (6,4);
			\draw [blue, ultra thick] (3,1) -- (0,4);	
			\draw [blue, ultra thick] (1,3) --(2,4);	
			\draw [blue, ultra thick] (5,3) --(4,4);
			\node [below] at (3,0) {$b$};
			\node [above] at (2,4) {$a$};
			\node [above] at (0,4) {$a$};
			\node [above] at (4,6) {$a$};
			\node [above] at (6,6) {$a$};
			\node [below left] at (2,2.5) {$x_i$};
			\node [below right] at (4,2.5) {$y_i$};
		\end{tikzpicture}
		\caption{}
	\end{subfigure}
	\caption[One-qutrit braid matrices $\sigma_1$ and $\sigma_3$.]{One-qutrit braid matrices (a) $\sigma_1$ and (b) $\sigma_3$.}
	\label{Sigma13}
\end{figure}
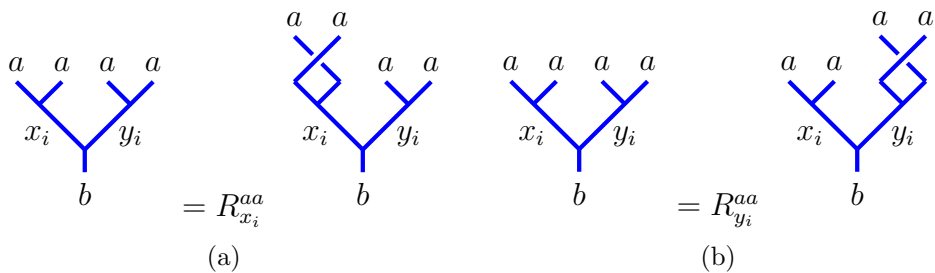

\begin{figure}[h!]
	\centering
	\begin{tikzpicture}[scale = 0.3]
		\draw [blue, ultra thick] (3,0) -- (3,1); 
		\draw [blue, ultra thick] (3,1) -- (6,4);
		\draw [blue, ultra thick] (3,1) -- (0,4);
		\draw [blue, ultra thick] (1,3) --(2,4);
		\draw [blue, ultra thick] (5,3) --(4,4);
		\node [below] at (3,0) {$b$};
		\node [above] at (0,4) {$a$};
		\node [above] at (2,4) {$a$};
		\node [above] at (4,4) {$a$};
		\node [above] at (6,4) {$a$};
		\node [below left] at (2,2.5) {$x_i$};
		\node [below right] at (4,2.5) {$y_i$};
	\end{tikzpicture}
	$=\sum_c[F^{aay_i}_b]_{x_ic}$	
	\begin{tikzpicture}[scale = 0.3]	
		\draw [blue, ultra thick] (3,0) -- (3,1); 
		\draw [blue, ultra thick] (3,1) -- (6,4);
		\draw [blue, ultra thick] (3,1) -- (0,4);
		\draw [blue, ultra thick] (4,2) --(2,4);
		\draw [blue, ultra thick] (5,3) --(4,4);
		\node [below] at (3,0) {$b$};
		\node [above] at (0,4) {$a$};
		\node [above] at (2,4) {$a$};
		\node [above] at (4,4) {$a$};
		\node [above] at (6,4) {$a$};
		\node [below right] at (3,1.5) {$c$};
		\node [below right] at (4,2.5) {$y_i$};
	\end{tikzpicture}\\
	$=\sum_{cd}[F^{aay_i}_b]_{x_ic}[F^{aaa}_c]^{-1}_{dy_i}$
	\begin{tikzpicture}[scale = 0.3]	
		\draw [blue, ultra thick] (3,0) -- (3,1); 
		\draw [blue, ultra thick] (3,1) -- (6,4);
		\draw [blue, ultra thick] (3,1) -- (0,4);
		\draw [blue, ultra thick] (4,2) --(2,4);
		\draw [blue, ultra thick] (3,3) --(4,4);
		\node [below] at (3,0) {$b$};
		\node [above] at (0,4) {$a$};
		\node [above] at (2,4) {$a$};
		\node [above] at (4,4) {$a$};
		\node [above] at (6,4) {$a$};
		\node [below right] at (3,1.5) {$c$};
		\node [below left] at (3.45,3.3) {$d$};
	\end{tikzpicture}
	$=\sum_{cd}[F^{aay_i}_b]_{x_ic}[F^{aaa}_c]^{-1}_{dy_i}R^{aa}_d$
	\begin{tikzpicture}[scale = 0.3]
		\draw [blue, ultra thick] (3,0) -- (3,1); 
		\draw [blue, ultra thick] (3,1) -- (6,4);
		\draw [blue, ultra thick] (3,1) -- (0,4);
		\draw [blue, ultra thick] (4,2) --(2,4);
		\draw [blue, ultra thick] (3,3) --(4,4);
		\draw [blue, ultra thick] (2,4) --(4,6);
		\draw [blue, ultra thick] (4,4) --(3.2,4.8);
		\draw [blue, ultra thick] (2.8,5.2) --(2,6);
		\node [below] at (3,0) {$b$};
		\node [above] at (0,4) {$a$};
		\node [above] at (2,6) {$a$};
		\node [above] at (4,6) {$a$};
		\node [above] at (6,4) {$a$};
		\node [below right] at (3,1.5) {$c$};
		\node [below left] at (3.45,3.3) {$d$};
	\end{tikzpicture}
	$=\sum_{cde}[F^{aay_i}_b]_{x_ic}[F^{aaa}_c]^{-1}_{dy_i}R^{aa}_d [F^{aaa}_c]_{de}$
	\begin{tikzpicture}[scale = 0.3]
		\draw [blue, ultra thick] (3,0) -- (3,1); 
		\draw [blue, ultra thick] (3,1) -- (6,4);
		\draw [blue, ultra thick] (3,1) -- (0,4);
		\draw [blue, ultra thick] (4,2) --(2,4);
		\draw [blue, ultra thick] (5,3) --(4,4);
		\draw [blue, ultra thick] (2,4) --(4,6);
		\draw [blue, ultra thick] (4,4) --(3.2,4.8);
		\draw [blue, ultra thick] (2.8,5.2) --(2,6);
		\node [below] at (3,0) {$b$};
		\node [above] at (0,4) {$a$};
		\node [above] at (2,6) {$a$};
		\node [above] at (4,6) {$a$};
		\node [above] at (6,4) {$a$};
		\node [below right] at (4.5,3) {$e$};
		\node [below right] at (3,1.5) {$c$};
	\end{tikzpicture}\\
	$=\sum_{cdef}[F^{aay_i}_b]_{x_ic}[F^{aaa}_c]^{-1}_{dy_i}R^{aa}_d [F^{aaa}_c]_{de}
	[F^{aae}_b]^{-1}_{fc}$
	\begin{tikzpicture}[scale = 0.3]
		\draw [blue, ultra thick] (3,0) -- (3,1); 
		\draw [blue, ultra thick] (3,1) -- (6,4);
		\draw [blue, ultra thick] (3,1) -- (0,4);
		\draw [blue, ultra thick] (1,3) --(2,4);
		\draw [blue, ultra thick] (5,3) --(4,4);
		\draw [blue, ultra thick] (2,4) --(4,6);
		\draw [blue, ultra thick] (4,4) --(3.2,4.8);
		\draw [blue, ultra thick] (2.8,5.2) --(2,6);
		\node [below] at (3,0) {$b$};
		\node [above] at (0,4) {$a$};
		\node [above] at (2,6) {$a$};
		\node [above] at (4,6) {$a$};
		\node [above] at (6,4) {$a$};
		\node [below right] at (4,2.5) {$e$};
		\node [below left] at (2,2.5) {$f$};
	\end{tikzpicture}
	\caption[One-qutrit braid matrix $\sigma_2$.]{One-qutrit braid matrix $\sigma_2$ \cite{cui2015universal}.}
	\label{Sigma2}
\end{figure}
\pagebreak
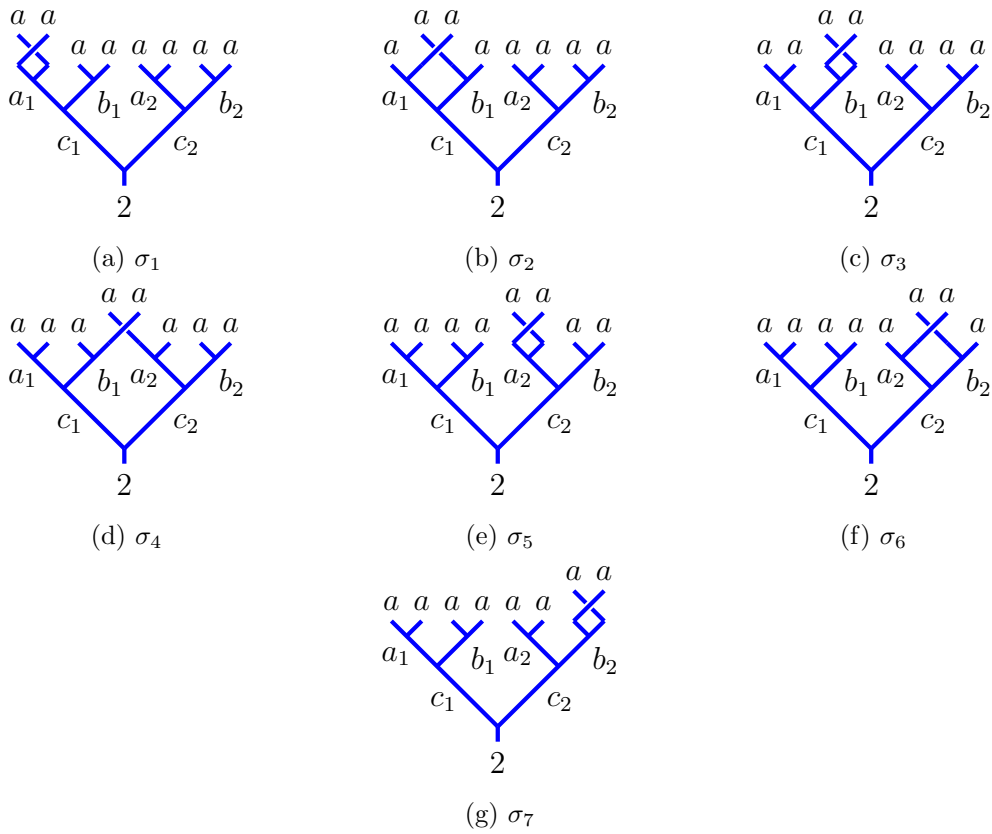
\begin{figure}[h!]
	\centering
	\begin{subfigure}{0.3\textwidth}
		\centering
		\begin{tikzpicture}[scale = 0.2]
			\draw [blue, ultra thick] (7,0) -- (7,1); 
			\draw [blue, ultra thick] (7,1) -- (0,8);
			\draw [blue, ultra thick] (7,1) -- (14,8);%
			\draw [blue, ultra thick] (3,5) --(6,8);
			\draw [blue, ultra thick] (11,5) --(8,8);%
			\draw [blue, ultra thick] (1,7) --(2,8);
			\draw [blue, ultra thick] (5,7) --(4,8);
			\draw [blue, ultra thick] (9,7) --(10,8);
			\draw [blue, ultra thick] (13,7) --(12,8);%
			\draw [blue, ultra thick] (2,8) --(1.2,8.8);
			\draw [blue, ultra thick] (0.8,9.2) --(0,10);
			\draw [blue, ultra thick] (0,8) --(2,10);%
			\node [above] at (0,10) {$a$};
			\node [above] at (2,10) {$a$};
			\node [above] at (4,8) {$a$};
			\node [above] at (6,8) {$a$};%
			\node [above] at (8,8) {$a$};
			\node [above] at (10,8) {$a$};
			\node [above] at (12,8) {$a$};
			\node [above] at (14,8) {$a$};%
			\node [below left] at (2,7) {$a_1$};
			\node [below right] at (4.5,7) {$b_1$};%
			\node [below left] at (10,7) {$a_2$};
			\node [below right] at (12.5,7) {$b_2$};%
			\node [below left] at (5,4) {$c_1$};
			\node [below right] at (9.5,4) {$c_2$};%
			\node [below] at (7,0) {2};%
		\end{tikzpicture}
		\caption{$\sigma_1$}
	\end{subfigure}
	\begin{subfigure}{0.3\textwidth}
		\centering
		\begin{tikzpicture}[scale = 0.2]
			\draw [blue, ultra thick] (7,0) -- (7,1); 
			\draw [blue, ultra thick] (7,1) -- (0,8);
			\draw [blue, ultra thick] (7,1) -- (14,8);%
			\draw [blue, ultra thick] (3,5) --(6,8);
			\draw [blue, ultra thick] (11,5) --(8,8);%
			\draw [blue, ultra thick] (1,7) --(2,8);
			\draw [blue, ultra thick] (5,7) --(4,8);
			\draw [blue, ultra thick] (9,7) --(10,8);
			\draw [blue, ultra thick] (13,7) --(12,8);%
			\draw [blue, ultra thick] (4,8) --(3.2,8.8);
			\draw [blue, ultra thick] (2.8,9.2) --(2,10);
			\draw [blue, ultra thick] (2,8) --(4,10);%
			\node [above] at (0,8) {$a$};
			\node [above] at (2,10) {$a$};
			\node [above] at (4,10) {$a$};
			\node [above] at (6,8) {$a$};%
			\node [above] at (8,8) {$a$};
			\node [above] at (10,8) {$a$};
			\node [above] at (12,8) {$a$};
			\node [above] at (14,8) {$a$};%
			\node [below left] at (2,7) {$a_1$};
			\node [below right] at (4.5,7) {$b_1$};%
			\node [below left] at (10,7) {$a_2$};
			\node [below right] at (12.5,7) {$b_2$};%
			\node [below left] at (5,4) {$c_1$};
			\node [below right] at (9.5,4) {$c_2$};%
			\node [below] at (7,0) {2};%
		\end{tikzpicture}
		\caption{$\sigma_2$}
	\end{subfigure}
	\begin{subfigure}{0.3\textwidth}
		\centering
		\begin{tikzpicture}[scale = 0.2]
			\draw [blue, ultra thick] (7,0) -- (7,1); 
			\draw [blue, ultra thick] (7,1) -- (0,8);
			\draw [blue, ultra thick] (7,1) -- (14,8);%
			\draw [blue, ultra thick] (3,5) --(6,8);
			\draw [blue, ultra thick] (11,5) --(8,8);%
			\draw [blue, ultra thick] (1,7) --(2,8);
			\draw [blue, ultra thick] (5,7) --(4,8);
			\draw [blue, ultra thick] (9,7) --(10,8);
			\draw [blue, ultra thick] (13,7) --(12,8);%
			\draw [blue, ultra thick] (6,8) --(5.2,8.8);
			\draw [blue, ultra thick] (4.8,9.2) --(4,10);
			\draw [blue, ultra thick] (4,8) --(6,10);%
			\node [above] at (0,8) {$a$};
			\node [above] at (2,8) {$a$};
			\node [above] at (4,10) {$a$};
			\node [above] at (6,10) {$a$};%
			\node [above] at (8,8) {$a$};
			\node [above] at (10,8) {$a$};
			\node [above] at (12,8) {$a$};
			\node [above] at (14,8) {$a$};%
			\node [below left] at (2,7) {$a_1$};
			\node [below right] at (4.5,7) {$b_1$};%
			\node [below left] at (10,7) {$a_2$};
			\node [below right] at (12.5,7) {$b_2$};%
			\node [below left] at (5,4) {$c_1$};
			\node [below right] at (9.5,4) {$c_2$};%
			\node [below] at (7,0) {2};%
		\end{tikzpicture}
		\caption{$\sigma_3$}
	\end{subfigure}
	\begin{subfigure}{0.3\textwidth}
		\centering
		\begin{tikzpicture}[scale = 0.2]
			\draw [blue, ultra thick] (7,0) -- (7,1); 
			\draw [blue, ultra thick] (7,1) -- (0,8);
			\draw [blue, ultra thick] (7,1) -- (14,8);%
			\draw [blue, ultra thick] (3,5) --(6,8);
			\draw [blue, ultra thick] (11,5) --(8,8);%
			\draw [blue, ultra thick] (1,7) --(2,8);
			\draw [blue, ultra thick] (5,7) --(4,8);
			\draw [blue, ultra thick] (9,7) --(10,8);
			\draw [blue, ultra thick] (13,7) --(12,8);
			\draw [blue, ultra thick] (8,8) --(7.2,8.8);
			\draw [blue, ultra thick] (6.8,9.2) --(6,10);
			\draw [blue, ultra thick] (6,8) --(8,10);%
			\node [above] at (0,8) {$a$};
			\node [above] at (2,8) {$a$};
			\node [above] at (4,8) {$a$};
			\node [above] at (6,10) {$a$};%
			\node [above] at (8,10) {$a$};
			\node [above] at (10,8) {$a$};
			\node [above] at (12,8) {$a$};
			\node [above] at (14,8) {$a$};%
			\node [below left] at (2,7) {$a_1$};
			\node [below right] at (4.5,7) {$b_1$};%
			\node [below left] at (10,7) {$a_2$};
			\node [below right] at (12.5,7) {$b_2$};%
			\node [below left] at (5,4) {$c_1$};
			\node [below right] at (9.5,4) {$c_2$};%
			\node [below] at (7,0) {2};%
		\end{tikzpicture}
		\caption{$\sigma_4$}
	\end{subfigure}
	\begin{subfigure}{0.3\textwidth}
		\centering
		\begin{tikzpicture}[scale = 0.2]
			\draw [blue, ultra thick] (7,0) -- (7,1); 
			\draw [blue, ultra thick] (7,1) -- (0,8);
			\draw [blue, ultra thick] (7,1) -- (14,8);%
			\draw [blue, ultra thick] (3,5) --(6,8);
			\draw [blue, ultra thick] (11,5) --(8,8);%
			\draw [blue, ultra thick] (1,7) --(2,8);
			\draw [blue, ultra thick] (5,7) --(4,8);
			\draw [blue, ultra thick] (9,7) --(10,8);
			\draw [blue, ultra thick] (13,7) --(12,8);%
			\draw [blue, ultra thick] (10,8) --(9.2,8.8);
			\draw [blue, ultra thick] (8.8,9.2) --(8,10);
			\draw [blue, ultra thick] (8,8) --(10,10);%
			\node [above] at (0,8) {$a$};
			\node [above] at (2,8) {$a$};
			\node [above] at (4,8) {$a$};
			\node [above] at (6,8) {$a$};%
			\node [above] at (8,10) {$a$};
			\node [above] at (10,10) {$a$};
			\node [above] at (12,8) {$a$};
			\node [above] at (14,8) {$a$};%
			\node [below left] at (2,7) {$a_1$};
			\node [below right] at (4.5,7) {$b_1$};%
			\node [below left] at (10,7) {$a_2$};
			\node [below right] at (12.5,7) {$b_2$};%
			\node [below left] at (5,4) {$c_1$};
			\node [below right] at (9.5,4) {$c_2$};%
			\node [below] at (7,0) {2};%
		\end{tikzpicture}
		\caption{$\sigma_5$}
	\end{subfigure}
	\begin{subfigure}{0.3\textwidth}
		\centering
		\begin{tikzpicture}[scale = 0.2]
			\draw [blue, ultra thick] (7,0) -- (7,1); 
			\draw [blue, ultra thick] (7,1) -- (0,8);
			\draw [blue, ultra thick] (7,1) -- (14,8);%
			\draw [blue, ultra thick] (3,5) --(6,8);
			\draw [blue, ultra thick] (11,5) --(8,8);%
			\draw [blue, ultra thick] (1,7) --(2,8);
			\draw [blue, ultra thick] (5,7) --(4,8);
			\draw [blue, ultra thick] (9,7) --(10,8);
			\draw [blue, ultra thick] (13,7) --(12,8);%
			\draw [blue, ultra thick] (12,8) --(11.2,8.8);
			\draw [blue, ultra thick] (10.8,9.2) --(10,10);
			\draw [blue, ultra thick] (10,8) --(12,10);%
			\node [above] at (0,8) {$a$};
			\node [above] at (2,8) {$a$};
			\node [above] at (4,8) {$a$};
			\node [above] at (6,8) {$a$};%
			\node [above] at (8,8) {$a$};
			\node [above] at (10,10) {$a$};
			\node [above] at (12,10) {$a$};
			\node [above] at (14,8) {$a$};%
			\node [below left] at (2,7) {$a_1$};
			\node [below right] at (4.5,7) {$b_1$};%
			\node [below left] at (10,7) {$a_2$};
			\node [below right] at (12.5,7) {$b_2$};%
			\node [below left] at (5,4) {$c_1$};
			\node [below right] at (9.5,4) {$c_2$};%
			\node [below] at (7,0) {2};%
		\end{tikzpicture}
		\caption{$\sigma_6$}
	\end{subfigure}
	\begin{subfigure}{0.3\textwidth}
		\centering
		\begin{tikzpicture}[scale = 0.2]
			\draw [blue, ultra thick] (7,0) -- (7,1); 
			\draw [blue, ultra thick] (7,1) -- (0,8);
			\draw [blue, ultra thick] (7,1) -- (14,8);%
			\draw [blue, ultra thick] (3,5) --(6,8);
			\draw [blue, ultra thick] (11,5) --(8,8);%
			\draw [blue, ultra thick] (1,7) --(2,8);
			\draw [blue, ultra thick] (5,7) --(4,8);
			\draw [blue, ultra thick] (9,7) --(10,8);
			\draw [blue, ultra thick] (13,7) --(12,8);%
			\draw [blue, ultra thick] (14,8) --(13.2,8.8);
			\draw [blue, ultra thick] (12.8,9.2) --(12,10);
			\draw [blue, ultra thick] (12,8) --(14,10);%
			\node [above] at (0,8) {$a$};
			\node [above] at (2,8) {$a$};
			\node [above] at (4,8) {$a$};
			\node [above] at (6,8) {$a$};%
			\node [above] at (8,8) {$a$};
			\node [above] at (10,8) {$a$};
			\node [above] at (12,10) {$a$};
			\node [above] at (14,10) {$a$};%
			\node [below left] at (2,7) {$a_1$};
			\node [below right] at (4.5,7) {$b_1$};%
			\node [below left] at (10,7) {$a_2$};
			\node [below right] at (12.5,7) {$b_2$};%
			\node [below left] at (5,4) {$c_1$};
			\node [below right] at (9.5,4) {$c_2$};%
			\node [below] at (7,0) {2};%
		\end{tikzpicture}
		\caption{$\sigma_7$}
	\end{subfigure}
	\caption{Two-qutrit braid matrices.}
	\label{TwoQutritSigma}
\end{figure}

\pagebreak

	\section{Ternary Arithmetic Circuits}\label{Arith}

The most important challenge in circuit design is reducing the number of gates. The more the gates, the harder it is to implement the circuit.
The MS gates in conventional quantum computing are designed by keeping the controlling value 2. Since, in topological quantum computation, any anyon can be braided to another anyon at any stage of the implementation, we can have a controlling value of $0,1,2$. Therefore, we can create a more general methodology of designing topological circuits presented here. We redesigned the qutrit arithmetic circuits that can be implemented with one-qutrit and two-qutrit gates made by metaplectic anyons \cite{bocharov2015improved} described in the last section. The universal set of gates cannot be made by braiding alone, it is to be combined with the topological charge measurement \cite{cui2015universal}.

The gates that can be obtained by braiding alone are the Clifford gates, whereas the non-Clifford gates cannot be implemented by braiding alone. In this work, we have the Clifford gate $SUM$ that can be implemented by braiding alone, whereas $C_c(X)$ is a non-Clifford gate that is implemented by the measurement of the topological charge.
As in \ref{QC}, one-qutrit ternary gates are represented as $Z_3(+1)$, $Z_3(+2)$, $Z_3(01)$, $Z_3(12)$, and $Z_3(02)$, where the first two are \textit{increment gates} and the last three are \textit{permutation gates}.
The non-Clifford gate, that is $C_c(X)$, applies $X$ when the controlling value is $c=0,1,2$, where $X$ can be a permutation or increment gate. The gates $C_c(X)$ are $9\cross 9$ matrices can be written as $diag(I_3,I_3,X)$ for control $\ket{2}$, $diag(I_3,X,I_3)$ for control $\ket{1}$, and $diag(X,I_3,I_3)$ for control $\ket{0}$, where $I_3$ is $3\cross 3$ identity matrix. In Ref. \cite{bocharov2015improved}, the $SUM$ gates are called soft-controlled whereas the $C_c(U)$ are called hard-controlled gates.

The measurement can be projective or based on interference \cite{cui2015universal,nayak2008non,stern2008anyons}.
In case of interferometric measurement, a probe charge is sent through the paths around some region. The interference between different paths is related to the total topological charge in that region. The charge of the region can be found by measuring the charge of the probe. This kind of measurement can distinguish charge of fusion channel from an overall collection. This is non-demolition measurement but the fusion channels evolve in non-universal manner.
Local measurement for topological charge of a single quasiparticle is performed by bringing two charges close to each other and find their charge from their fusion outcome.
Let the measurement $\mathcal{M}_1 = \left\{\Pi_1,\Pi'_1\right\}$ correspond to the topological charge measurement of the first pair of anyons spanned by $\ket{\textbf{1}Y}$ and its orthogonal complements $\ket{-YY},\ket{Y\textbf{1}}$. The topological charge of the first pair of anyons is found by this measurement. If it is $\textbf{1}$ or $Y$ then the second pair is still in a coherent superposition of $\textbf{1}$ and $Y$. This measurement allows us to find whether an anyon is trivial or not.
For the hard-controlled gates $C_c(U)$, braiding gate is applied only when the controlling value of topological charge is 'c', the one mentioned on the gate, otherwise go to previous step or start over. The process is repeated several times until we get the required result.
The braiding supplemented with the projective measurement provides the universal set of gates for anyonic quantum computation \cite{cui2015universal}.

\begin{figure}[h!]
	\centering
	\begin{tikzpicture}
		\draw[ultra thick,cyan] (0,0)--(0,1);
		\qnode{0}{-0.8}{$SUM_1$}
		\qgateCNC*[ibmqxD]{b}{0}{1}
		\qgateU[ibmqxD]{0}{0}{+1}
	\end{tikzpicture}\qquad
	\begin{tikzpicture}
		\draw[ultra thick,cyan] (0,0)--(0,1);
		\qnode{0}{-0.8}{$SUM_2$}
		\qgateCNC*[ibmqxD]{b}{0}{1}
		\qgateU[ibmqxD]{0}{0}{+2}
	\end{tikzpicture}\qquad
	\begin{tikzpicture}
		\draw[ultra thick,cyan] (0,0)--(0,1);
		\qnode{0}{-0.8}{$C_c(X)$}
		\qgateCNC[ibmqx]{b}{0}{1}
		\qgateU[ibmqxA]{0}{0}{X}
		\qnode{-0.2}{0.8}{$c$}
	\end{tikzpicture}\qquad
\begin{tikzpicture}
	\draw[ultra thick,cyan] (0,0)--(0,2);
	\qnode{0}{-0.8}{$C(SUM)$}
	\qgateCNC[ibmqxD]{b}{0}{2}
	\qgateCNC*[ibmqxD]{b}{0}{1}
	\qgateU[ibmqxA]{0}{0}{X}
	\qnode{-0.2}{1.8}{$c$}
\end{tikzpicture}
	\caption[Graphical representation of two-qutrit ternary gates.]{The graphical representation of two-qutrit ternary gates \cite{bocharov2015improved}.}
	\label{GTGates}
\end{figure}
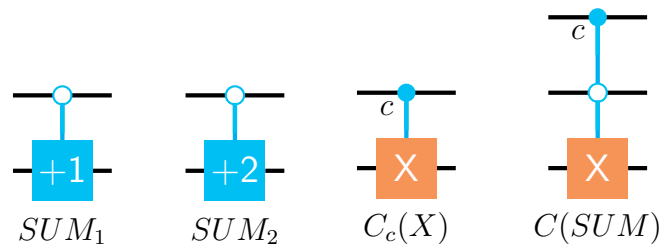

As discussed in previous section, the Clifford gate $SUM = (I\otimes H)\Lambda (Z)(I \otimes H^{-1})$ in Eq. \ref{SumGate} is a generalization of the CNOT gate. It can also be written as $SUM= \ket{0}\bra{0} \otimes I + \ket{1}\bra{1} \otimes X + \ket{2}\bra{2} \otimes X^2$. Here, $X$ is an increment gate $Z_3(+1)$ or $Z_3(+2)$. The two-qutrit $9\cross 9$ matrix for the $SUM$ is written as $diag(I_3, X, X^2)$. Let us call this gate $SUM_1$. But if we use $\Lambda (Z^{-1})$ in Eq. \ref{SumGate}, we get the matrix form as $diag(I_3, X^2, X)$. Let us represent this form as $SUM_2$. The braiding implementation of $SUM_1$ and $SUM_2$ is equivalent. The $SUM_1$ and $SUM_2$ are used for designing the two-qutrit braiding gates $Z_3(+1)$ and $Z_3(+2)$. These gates are shown in Fig. \ref{GTGates} and their matrices are shown in Eq. \ref{SUMMatrix}. We can also note that the matrix $diag(I_3, X^2, X)$ is the square of the matrix $diag(I_3, X, X^2)$. When control is the second qutrit and target is the first qutrit, then the $SUM$ gates are written in the form $SUM=I \otimes \ket{0}\bra{0} + X \otimes \ket{1}\bra{1} + X^2 \otimes \ket{2}\bra{2}= (H\otimes I)\Lambda (Z)(H^{-1} \otimes I)$.

Two qutrits can be swapped by the $SWAP$ gate, as discussed in \ref{QC}. This gate can be formed by braiding alone \cite{bocharov2016efficient} with the use of the permutation gates. One of the realizations of the $SWAP$ gate $SWAP: \ket{i,j} \to \ket{j,i}$ is obtained by using the gate $Z_3(12)$ \cite{bocharov2016efficient} and can be written as
\eq{SWAP = (Z_3(12) \otimes I)SUM_{1,2}SUM_{2,1}SUM_{2,1}SUM_{1,2}, \label{Swap}}
where $SUM_{j,k}$ is a two-qutrit $SUM$ gate applied to $k$th qutrit when $j$th qutrit is the control qutrit. Two other non-Clifford gates used in this paper are Honer gate and controlled-SUM gate given as
\eq{&Horner = \Lambda(\Lambda(X)): \ket{i,j,k} \to \ket{i,j,ij+k}\nonumber\\
	&C(SUM) = C_c(SUM): \ket{i,j,k} \to \ket{i,j,j\delta_{i,c}+k}}
These gates are generalization of Toffoli gate \cite{bocharov2015improved,bocharov2017factoring}.

In Fig. \ref{GTGates}, the non-Clifford gates are represented by the filled circles at the controlling values with $c=0,1,2$, whereas for the $SUM$ gates, the hollow circles are drawn at the controlling values. To avoid cluttering, labels of the control is omitted when its value is 2.
The increment gates will be represented by $+1$ and $+2$ and the permutation gates will be represented by $01$, $12$, and $02$. The gates in the boxes are related to the sum or product column of truth tables. Blue and orange colors of gates correspond to Clifford gates and non-Clifford gates respectively. A circuit is read from left to right but when it is written as matrices it is read from right to left, the same as the matrix multiplication.

\eq{SUM_1=\begin{pmatrix}
		1 & 0 & 0 & & & & & & \\
		0 & 1 & 0 & & & & & &\\
		0 & 0 & 1 & & & & & &\\
		& & & 0 & 0 & 1 & & &\\
		& & & 1 & 0 & 0 & & &\\
		& & & 0 & 1 & 0 & & &\\
		& & & & & & 0 & 1 & 0\\
		& & & & & & 0 & 0 & 1\\
		& & & & & & 1 & 0 & 0\end{pmatrix}, \ SUM_2=\begin{pmatrix}
		1 & 0 & 0 & & & & & & \\
		0 & 1 & 0 & & & & & &\\
		0 & 0 & 1 & & & & & &\\
		& & & 0 & 1 & 0 & & &\\
		& & & 0 & 0 & 1 & & &\\
		& & & 1 & 0 & 0 & & &\\
		& & & & & & 0 & 0 & 1\\
		& & & & & & 1 & 0 & 0\\
		& & & & & & 0 & 1 & 0\end{pmatrix}\label{SUMMatrix}}

\subsection{Ternary Adder}
The adder circuit is the most important arithmetic circuit used in almost all circuits, especially in algorithms such as Grover, Shor, and HHL algorithms.
Binary adder circuits are proposed by  Ref. \cite{draper2004logarithmic,cuccaro2004new,vedral1996quantum} and their ternary counterparts are given in Ref. \cite{khan2007quantum,khan2004quantum,monfared2017design}. 
The adder circuit of Ref. \cite{haghparast2017towards} consists of 14 MS and shift gates and the circuit from Ref. \cite{deibuk2015design} obtained using the genetic algorithm, has 13 MS and shift gates. In Ref. \cite{asadi2020efficient}, half adder is designed by 5 MS gates and one shift gate. An output that remains unused and thrown out is called the garbage output. Most of these designs used the Toffoli gate for their implementation, but the Toffoli gate cannot be built by braiding alone \cite{cui2015universal}. Our circuit design for half adder consist of four gates. Only one braiding gate is used to implement the sum and three non-Clifford gates are used for the implementation of carry. The constant inputs and the garbage outputs are the same for our designs as in the existing designs.

When we add two one-digit numbers, then we get the half adder, whereas the full adder circuit adds three one-digit numbers. The third digit can be a carry from the previous half adder.   
The truth table for a half adder is shown in Table \ref{TruthHalfAdd} and the circuit realization is shown in Fig. \ref{Half}. Let us discuss the cases when there is a nonzero carry. For example, the case when the input $A$ has value 1 and input $B$ has value 2. For the first gate, the control value is not 2 so it would not be applied. For the second gate, control value is 2 but third input is zero. The second gate will also remain ineffective. Third gate will be applied and it will give the carry 1. At the fourth gate, withing the box, $S$ will be zero as $B=2$ will add 2 to $A=1$.

When $A=2$ and $B=2$, the first gate will change the third qutrit from 0 to 2. At the second gate, third qutrit will be changed to 1. The third gate will not be applied. The fourth gate will add 2 to $A=2$ and give the sum $S=1$. The garbage bits at the end of the computation will be ignored.

\begin{figure}[h!]
	\centering
	\begin{minipage}{0.4\textwidth}
		\centering
		\begin{tikzpicture}[scale=1]
			\draw[fill=red!20,opacity=0.5] (4.6,0.6) rectangle (5.8,2.6);
			\qnode{0.2}{2}{$A$}
			\qnode{0.2}{1}{$B$}
			\qnode{0.2}{0}{$0$}
			\qwire[ibmqx]{1}{1}
			\draw[ultra thick,cyan] (1.3,0)--(1.3,2);			
			\qgateU[ibmqxA]{1}{0}{+1}
			\qgateCNC[ibmqx]{b}{1}{2}
			\qgateCNC*[ibmqx]{b}{1}{1}
			\qwire[ibmqx]{2}{2}
			\draw[ultra thick,cyan] (2.6,0)--(2.6,1);			
			\qgateU[ibmqxA]{2}{0}{12}
			\qgateCNC[ibmqx]{b}{2}{1}
			\draw[ultra thick,cyan] (3.9,0)--(3.9,2);			
			\qgateU[ibmqxA]{3}{0}{+1}
			\qgateCNC[ibmqx]{b}{3}{2}
			\qgateCNC[ibmqx]{b}{3}{1}
			\qnode{2.8}{2.3}{$1$}
			\qwire[ibmqx]{4}{0}		
			\draw[ultra thick,cyan] (5.2,1)--(5.2,2);
			\qgateU[ibmqxD]{4}{2}{+1}
			\qgateCNC*[ibmqx]{t}{4}{1}			
			\qnode{4.7}{2}{$S$}
			\qnode{4.7}{1}{$g$}
			\qnode{4.85}{0}{$c_{out}$}
		\end{tikzpicture}
		\captionof{figure}{Ternary half adder circuit realization.}
		\label{Half}
	\end{minipage}
	\begin{minipage}{0.4\textwidth}
		\centering			
		\begin{tabular}{|>{\columncolor{green!20}}c|>{\columncolor{green!20}}c||>{\columncolor{orange!40}}c|>{\columncolor{orange!20}}c|}
			\hline
			\rowcolor{cyan!50}
			$A$ & $B$ & $S$ & $c_{out}$\\
			\hline\hline
			0 & 0 & 0 & 0\\
			0 & 1 & 1 & 0\\
			0 & 2 & 2 & 0\\
			1 & 0 & 1 & 0\\
			1 & 1 & 2 & 0\\
			1 & 2 & 0 & 1\\
			2 & 0 & 2 & 0\\
			2 & 1 & 0 & 1\\
			2 & 2 & 1 & 1\\
			\hline
		\end{tabular}
		\captionof{table}{Truth table of the ternary half adder.}
		\label{TruthHalfAdd}
	\end{minipage}	
\end{figure}

The full adder adds three qutrits $A$, $B$, $C$ as shown in Fig. \ref{Full}. The truth table for the ternary full adder is given in Table \ref{TruthFullAdd}. The sum of $A$ and $B$ is obtained that is added to the third input qutrit $C$ to get the output $\bm{S}$. The input $C$ can be a carry from the previous sum of two qutrits. The garbage outputs $g_1$ and $g_2$ are ignored.

The addition of two-qutrit numbers and its circuit realization are shown in Fig. \ref{2-adder} (a), (b). A half adder and a full adder can be used. 
The first qutrit $A_0$ of $A_0B_0$ is added by the first half adder and the first digit of the sum $\bm{S}_0$ and $g_1$ are obtained. Their carry $c_0$ is to be added with the sum of the second qutrits $A_1$ and $B_1$. This $c_0$ corresponds to the input $C$ of the full adder. The addition of $A_1+B_1 +c_0$ gives the second digits of the output as $\bm{S}_1$ and a carry $\bm{c}_{out}$. The $SWAP$ gate in Eq. \ref{Swap} is used to exchange the qutrits $S$ and $c_0$, and the garbage qutrits are thrown out.

\begin{figure}[h!]
	\centering
	\begin{minipage}{0.65\textwidth}
		\centering
			\begin{tikzpicture}[scale=1]
				\draw[fill=red!20,opacity=0.5] (3.3,1.7) rectangle (4.5,3.6);
				\draw[fill=red!20,opacity=0.5] (8.5,0.7) rectangle (9.7,3.6);
				\qnode{-0.7}{3}{$A$}
				\qnode{-0.7}{2}{$B$}
				\qnode{-0.7}{1}{$C$}
				\qnode{-0.7}{0}{$0$}
				\qwire[ibmqx]{0}{3}
				\qwire[ibmqx]{0}{1}
				\draw[ultra thick,cyan] (0,0)--(0,3);
				\qgateU[ibmqxA]{0}{0}{+1}
				\qgateCNC[ibmqx]{b}{0}{3}
				\qgateCNC*[ibmqx]{b}{0}{2}
				\qwire[ibmqx]{1}{3}				
				\qwire[ibmqx]{1}{1}			
				\draw[ultra thick,cyan] (1.3,0)--(1.3,2);				
				\qgateCNC[ibmqx]{b}{1}{2}
				\qgateU[ibmqxA]{1}{0}{12}
				\qwire[ibmqx]{2}{1}
				\draw[ultra thick,cyan] (2.6,0)--(2.6,3);			
				\qgateCNC[ibmqx]{b}{2}{3}
				\qgateCNC[ibmqx]{b}{2}{2}
				\qgateU[ibmqxA]{2}{0}{+1}
				\qnode{1.8}{3.3}{$1$}
				\qwire[ibmqx]{3}{0}
				\qwire[ibmqx]{3}{1}
				\draw[ultra thick,cyan] (3.9,2)--(3.9,3);
				\qgateU[ibmqxD]{3}{3}{+1}
				\qgateCNC*[ibmqx]{t}{3}{2}
				\qwire[ibmqx]{4}{2}
				\draw[ultra thick,cyan] (5.2,0)--(5.2,3);			
				\qgateCNC[ibmqx]{b}{4}{3}
				\qgateCNC*[ibmqx]{b}{4}{1}
				\qgateU[ibmqxA]{4}{0}{+1}
				\qwire[ibmqx]{5}{2}
				\draw[ultra thick,cyan] (6.5,0)--(6.5,3);			
				\qgateCNC[ibmqx]{b}{5}{3}
				\qgateCNC[ibmqx]{b}{5}{1}
				\qgateU[ibmqxA]{5}{0}{+2}
				\qwire[ibmqx]{6}{2}
				\draw[ultra thick,cyan] (7.8,0)--(7.8,3);			
				\qgateCNC[ibmqx]{b}{6}{3}
				\qgateCNC[ibmqx]{b}{6}{1}
				\qgateU[ibmqxA]{6}{0}{+1}
				\qnode{5.8}{3.3}{$1$}
				\qwire[ibmqx]{7}{0}
				\qwire[ibmqx]{7}{2}
				\draw[ultra thick,cyan] (9.1,1)--(9.1,3);
				\qgateU[ibmqxD]{7}{3}{+1}
				\qgateCNC*[ibmqx]{t}{7}{1}
				\qnode{7.7}{3}{$S$}
				\qnode{7.7}{2}{$B$}
				\qnode{7.7}{1}{$C$}
				\qnode{7.9}{0}{$c_{out}$}
			\end{tikzpicture}
			\captionof{figure}[Ternary full adder circuit realization.]{Ternary full adder circuit realization.}
			\label{Full}
	\end{minipage}
	\begin{minipage}{0.3\textwidth}
		\centering
		\scalebox{0.8}{
		\begin{tabular}{|>{\columncolor{green!20}}c|>{\columncolor{green!20}}c|>{\columncolor{green!20}}c||>{\columncolor{orange!40}}c|>{\columncolor{orange!20}}c|}
				\hline
				\rowcolor{cyan!50}
				$A$ & $B$ & $C$ & $S$ & $c_{out}$\\
				\hline\hline
				0 & 0 & 0 & 0 & 0\\
				0 & 0 & 1 & 1 & 0\\
				0 & 0 & 2 & 2 & 0\\
				0 & 1 & 0 & 1 & 0\\
				0 & 1 & 1 & 2 & 0\\
				0 & 1 & 2 & 0 & 1\\
				0 & 2 & 0 & 2 & 0\\
				0 & 2 & 1 & 0 & 1\\
				0 & 2 & 2 & 1 & 1\\
				1 & 0 & 0 & 1 & 0\\
				1 & 0 & 1 & 2 & 0\\
				1 & 0 & 2 & 0 & 1\\
				1 & 1 & 0 & 2 & 0\\
				1 & 1 & 1 & 0 & 1\\
				1 & 1 & 2 & 1 & 1\\
				1 & 2 & 0 & 0 & 1\\
				1 & 2 & 1 & 1 & 1\\
				1 & 2 & 2 & 2 & 1\\
				2 & 0 & 0 & 2 & 0\\
				2 & 0 & 1 & 0 & 1\\
				2 & 0 & 2 & 1 & 1\\
				2 & 1 & 0 & 0 & 1\\
				2 & 1 & 1 & 1 & 1\\
				2 & 1 & 2 & 2 & 1\\
				2 & 2 & 0 & 1 & 1\\
				2 & 2 & 1 & 2 & 1\\
				2 & 2 & 2 & 0 & 2\\
				\hline
		\end{tabular}}
		\captionof{table}{Truth table for ternary full adder.}
		\label{TruthFullAdd}
	\end{minipage}	
\end{figure}

\begin{figure}[h!]
	\centering
	\begin{subfigure}{0.2\textwidth}
		\centering
		\begin{tabular}{ccc}
			\rowcolor{green!20}
			&$c_0$&\\
			\rowcolor{cyan!20}
			&$A_1$&$A_0$\\
			\rowcolor{cyan!20}
			$+$&$B_1$&$B_0$\\
			\hline
			\rowcolor{orange!20}
			$c_{out}$&$S_1$&$S_0$
		\end{tabular}
		\caption{}
	\end{subfigure}
	\begin{subfigure}{0.75\textwidth}
		\centering
		\begin{tikzpicture}[scale=1]
			\draw[fill=red!20,opacity=0.5] (3.3,4.7) rectangle (4.5,6.6);			
			\qnode{-0.7}{6}{$A_0$}
			\qnode{-0.7}{5}{$B_0$}
			\qnode{-0.7}{4}{$0$}
			\qwire[ibmqx]{0}{5}
			\draw[ultra thick,cyan] (0,4)--(0,6);			
			\qgateU[ibmqxA]{0}{4}{+1}
			\qgateCNC[ibmqx]{b}{0}{6}
			\qgateCNC*[ibmqx]{b}{0}{5}
			\qwire[ibmqx]{1}{6}
			\draw[ultra thick,cyan] (1.3,4)--(1.3,5);			
			\qgateU[ibmqxA]{1}{4}{12}
			\qgateCNC[ibmqx]{b}{1}{5}
			\draw[ultra thick,cyan] (2.6,4)--(2.6,6);			
			\qgateU[ibmqxA]{2}{4}{+1}
			\qgateCNC[ibmqx]{b}{2}{6}
			\qgateCNC[ibmqx]{b}{2}{5}
			\qnode{1.8}{6.3}{$1$}
			\qwire[ibmqx]{3}{4}
			\draw[ultra thick,cyan] (3.9,5)--(3.9,6);
			\qgateU[ibmqxD]{3}{6}{+1}
			\qgateCNC*[ibmqx]{t}{3}{5}			
			\qnode{3.7}{6}{$\textbf{S}_0$}
			\qnode{3.7}{5}{$g_1$}
			\qnode{3.7}{4.2}{$c_0$}
			
			===================================
			
			\draw[fill=red!20,opacity=0.5] (3.3,1.7) rectangle (4.5,3.6);
			\draw[fill=red!20,opacity=0.5] (9.8,0.7) rectangle (11,3.6);
			\qnode{-0.7}{3}{$A_1$}
			\qnode{-0.7}{2}{$B_1$}
			\qnode{3.8}{1}{$C$}
			\qnode{-0.7}{0}{$0$}
			
			\qwire[ibmqx]{4}{3}
			\qwire[ibmqx]{4}{2}
			\qwire[ibmqx]{4}{0}		
			
			\draw[ultra thick,cyan] (0,0)--(0,3);
			\qgateU[ibmqxA]{0}{0}{+1}
			\qgateCNC[ibmqx]{b}{0}{3}
			\qgateCNC*[ibmqx]{b}{0}{2}
			\qwire[ibmqx]{1}{3}				
			\draw[ultra thick,cyan] (1.3,0)--(1.3,2);				
			\qgateCNC[ibmqx]{b}{1}{2}
			\qgateU[ibmqxA]{1}{0}{12}
			\draw[ultra thick,cyan] (2.6,0)--(2.6,3);			
			\qgateCNC[ibmqx]{b}{2}{3}
			\qgateCNC[ibmqx]{b}{2}{2}
			\qgateU[ibmqxA]{2}{0}{+1}
			\qnode{1.8}{3.3}{$1$}
			\qwire[ibmqx]{3}{0}

			\draw[ultra thick,cyan] (3.9,2)--(3.9,3);
			\qgateU[ibmqxD]{3}{3}{+1}
			\qgateCNC*[ibmqx]{t}{3}{2}
			\qwire[ibmqx]{5}{2}
			\draw[ultra thick,cyan] (6.5,0)--(6.5,3);			
			\qgateCNC[ibmqx]{b}{5}{3}
			\qgateCNC*[ibmqx]{b}{5}{1}
			\qgateU[ibmqxA]{5}{0}{+1}
			\qwire[ibmqx]{6}{2}
			\draw[ultra thick,cyan] (7.8,0)--(7.8,3);			
			\qgateCNC[ibmqx]{b}{6}{3}
			\qgateCNC[ibmqx]{b}{6}{1}
			\qgateU[ibmqxA]{6}{0}{+2}
			\qwire[ibmqx]{7}{2}
			\draw[ultra thick,cyan] (9.1,0)--(9.1,3);			
			\qgateCNC[ibmqx]{b}{7}{3}
			\qgateCNC[ibmqx]{b}{7}{1}
			\qgateU[ibmqxA]{7}{0}{+1}
			\qnode{6.8}{3.3}{$1$}
			\qwire[ibmqx]{8}{0}
			\qwire[ibmqx]{8}{2}
			\draw[ultra thick,cyan] (10.4,1)--(10.4,3);
			\qgateU[ibmqxD]{8}{3}{+1}
			\qgateCNC*[ibmqx]{t}{8}{1}
			\qnode{8.7}{3}{$\textbf{S}_1$}
			\qnode{8.7}{2}{$g_2$}
			\qnode{8.7}{1}{$g_3$}
			\qnode{8.8}{0}{$c_{out}$}
			\draw[ultra thick, red] (4.5,4)--(5.2,4)--(5.2,1)--(5.9,1);
		\end{tikzpicture}
		\caption{}
	\end{subfigure}
	\caption[Ternary two-qutrit addition and its circuit realization.]{Ternary two-qutrit (a) addition and (b) circuit realization by using one half adder and one full adder.}
	\label{2-adder}
\end{figure}
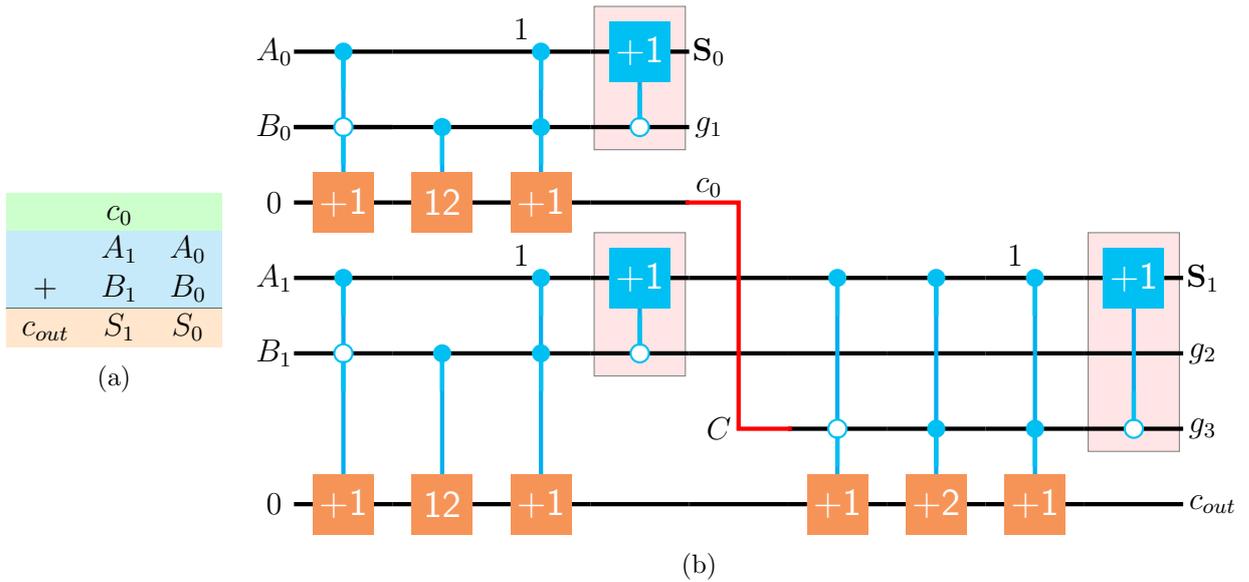

\begin{figure*}[h!]
	\centering

\end{figure*}

\subsection{Ternary Subtractor}
A ternary subtractor gives an output as a difference between two inputs. The subtractor circuit takes two inputs $A$ and $B$ and one ancilla. The difference between two inputs and the borrow is obtained at the output. For ternary subtractor in conventional quantum computing, see Ref. \cite{monfared2017design}. The half subtractor truth table is shown in Table \ref{TruthHalfSub} and the circuit realization is shown in Fig. \ref{HalfSub}. Let us discuss the case when $A=1,B=2$. Since the controlling, value needs to be at 0, but we have $B=2$, the first $C(SUM)$ gate would not be applied. For the second gate, the controlling value is 2, therefore the MS permutation gate would be applied, but the value at the third qutrit is 0 so this gate will remain ineffective. The gate for the carry will give 1, because the controlling values are 2. The fourth gate changes $A$ from 1 to 2. Therefore, at the output, we get a difference $D=2$ and the borrow $b_{out} = 1$. This design of half subtractor circuit consists of only one braiding gate and three non-Clifford gates.

\begin{figure}[h!]
	\centering
	\begin{minipage}{0.4\textwidth}
		\centering
		\begin{tikzpicture}[scale=1]
			\draw[fill=red!20,opacity=0.6] (3.3,0.5) rectangle (4.5,2.6);
			\qnode{-0.7}{2}{$A$}
			\qnode{-0.7}{1}{$B$}
			\qnode{-0.7}{0}{$0$}
			\draw[ultra thick,cyan] (0,0)--(0,2);			
			\qgateU[ibmqxA]{0}{0}{+1}
			\qgateCNC[ibmqx]{b}{0}{2}
			\qgateCNC*[ibmqx]{b}{0}{1}
			\qnode{-0.2}{2.3}{$0$}
			\draw[ultra thick,cyan] (1.3,0)--(1.3,1);			
			\qwire[ibmqx]{1}{2}
			\qwire[ibmqx]{1}{1}
			\qgateCNC[ibmqx]{b}{1}{1}
			\qgateU[ibmqxA]{1}{0}{12}
			\draw[ultra thick,cyan] (2.6,0)--(2.6,2);			
			\qwire[ibmqx]{2}{2}
			\qwire[ibmqx]{2}{1}
			\qgateCNC[ibmqx]{b}{2}{2}
			\qgateCNC[ibmqx]{b}{2}{1}
			\qgateU[ibmqxA]{2}{0}{+1}
			\qnode{1.8}{2.3}{$1$}
			\draw[ultra thick,cyan] (3.9,1)--(3.9,2);
			\qgateU[ibmqxD]{3}{2}{+2}
			\qgateCNC*[ibmqx]{t}{3}{1}
			\qwire[ibmqx]{3}{0}
			\qnode{3.7}{2}{$D$}
			\qnode{3.7}{1}{$g$}
			\qnode{3.85}{0}{$b_{out}$}
		\end{tikzpicture}
		\captionof{figure}{Ternary half subtractor circuit realization.}
		\label{HalfSub}
	\end{minipage}
	\begin{minipage}{0.4\textwidth}
		\centering		
		\begin{tabular}{|>{\columncolor{green!20}}c|>{\columncolor{green!20}}c||>{\columncolor{orange!40}}c|>{\columncolor{orange!20}}c|}
			\hline
			\rowcolor{cyan!50}
			A & B & D & $b_{out}$ \\
			\hline\hline
			0 & 0 & 0 & 0\\
			0 & 1 & 2 & 1\\
			0 & 2 & 1 & 1\\
			1 & 0 & 1 & 0\\
			1 & 1 & 0 & 0\\
			1 & 2 & 2 & 1\\
			2 & 0 & 2 & 0\\
			2 & 1 & 1 & 0\\
			2 & 2 & 0 & 0\\
			\hline
		\end{tabular}
	\captionof{table}{Truth table for ternary half subtractor.}
		\label{TruthHalfSub}
	\end{minipage}	
\end{figure}

When taking the difference of two numbers and a borrow is needed, then have a full subtractor. It has three inputs $A$, $B$, and $C$, where $C$ is the borrow-in. The truth table for full subtractor is shown in Table \ref{TruthFullSub} and the circuit realization is shown in Fig. \ref{FullSub}. 

\begin{figure}[h!]
	\centering
	\begin{minipage}{0.7\textwidth}
		\centering
		\begin{tikzpicture}[scale=1]
			\draw[fill=red!20,opacity=0.5] (3.3,0.4) rectangle (5.8,3.6);
			\qnode{-0.7}{3}{$A$}
			\qnode{-0.7}{2}{$B$}
			\qnode{-0.7}{1}{$C$}
			\qnode{-0.7}{0}{$0$}
			\qwire[ibmqx]{0}{3}
			\qwire[ibmqx]{0}{1}
			\draw[ultra thick,cyan] (0,0)--(0,2);
			\qgateU[ibmqxA]{0}{0}{+1}
			\qgateCNC[ibmqx]{b}{0}{2}
			\qwire[ibmqx]{1}{1}			
			\draw[ultra thick,cyan] (1.3,0)--(1.3,3);				
			\qgateCNC[ibmqx]{b}{1}{3}
			\qgateCNC[ibmqx]{b}{1}{2}
			\qgateU[ibmqxA]{1}{0}{+1}
			\qnode{0.8}{3.3}{$0$}
			\qnode{0.8}{2.3}{$1$}
			\qwire[ibmqx]{2}{1}
			\draw[ultra thick,cyan] (2.6,0)--(2.6,3);			
			\qgateCNC[ibmqx]{b}{2}{3}
			\qgateCNC[ibmqx]{b}{2}{2}
			\qgateU[ibmqxA]{2}{0}{+2}
			
			\qwire[ibmqx]{3}{0}
			\qwire[ibmqx]{3}{1}
			\draw[ultra thick,cyan] (3.9,2)--(3.9,3);
			\qgateU[ibmqxD]{3}{3}{+2}
			\qgateCNC*[ibmqx]{t}{3}{2}			
			\qwire[ibmqx]{4}{0}
			\qwire[ibmqx]{4}{2}
			\draw[ultra thick,cyan] (5.2,1)--(5.2,3);
			\qgateU[ibmqxD]{4}{3}{+2}
			\qgateCNC*[ibmqx]{t}{4}{1}			
			
			\qwire[ibmqx]{5}{2}
			\draw[ultra thick,cyan] (6.5,0)--(6.5,3);			
			\qgateCNC[ibmqx]{b}{5}{3}
			\qgateCNC[ibmqx]{b}{5}{1}
			\qgateU[ibmqxA]{5}{0}{+1}
			\qwire[ibmqx]{6}{2}
			\draw[ultra thick,cyan] (7.8,0)--(7.8,3);			
			\qgateCNC[ibmqx]{b}{6}{3}
			\qgateCNC[ibmqx]{b}{6}{1}
			\qgateU[ibmqxA]{6}{0}{+1}
			\qnode{5.8}{1.3}{$1$}
			\qwire[ibmqx]{7}{2}
			\draw[ultra thick,cyan] (9.1,0)--(9.1,3);			
			\qgateCNC[ibmqx]{b}{7}{3}
			\qgateCNC[ibmqx]{b}{7}{1}
			\qgateU[ibmqxA]{7}{0}{+1}
			\qnode{6.8}{3.3}{$1$}
			\qnode{7.7}{3}{$D$}
			\qnode{7.7}{2}{$g_1$}
			\qnode{7.7}{1}{$g_2$}
			\qnode{7.85}{0}{$b_{out}$}
		\end{tikzpicture}
		\captionof{figure}{Ternary full subtractor circuit realization.}
		\label{FullSub}
	\end{minipage}
	\begin{minipage}{0.25\textwidth}
		\scalebox{0.8}[0.8]{
			\begin{tabular}{|>{\columncolor{green!20}}c|>{\columncolor{green!20}}c|>{\columncolor{green!20}}c||>{\columncolor{orange!40}}c|>{\columncolor{orange!20}}c|}
				\hline
				\rowcolor{cyan!50}
				A & B & C & D & $b_{out}$\\
				\hline\hline
				0 & 0 & 0 & 0 & 0\\
				0 & 0 & 1 & 2 & 1\\
				0 & 0 & 2 & 1 & 1\\
				0 & 1 & 0 & 2 & 1\\
				0 & 1 & 1 & 1 & 1\\
				0 & 1 & 2 & 0 & 1\\
				0 & 2 & 0 & 1 & 1\\
				0 & 2 & 1 & 0 & 1\\
				0 & 2 & 2 & 2 & 2\\
				
				1 & 0 & 0 & 1 & 0\\
				1 & 0 & 1 & 0 & 0\\
				1 & 0 & 2 & 2 & 1\\
				1 & 1 & 0 & 0 & 0\\
				1 & 1 & 1 & 2 & 1\\
				1 & 1 & 2 & 1 & 1\\
				1 & 2 & 0 & 2 & 1\\
				1 & 2 & 1 & 1 & 1\\
				1 & 2 & 2 & 0 & 1\\
				
				2 & 0 & 0 & 2 & 0\\
				2 & 0 & 1 & 1 & 0\\
				2 & 0 & 2 & 0 & 0\\
				2 & 1 & 0 & 1 & 0\\
				2 & 1 & 1 & 0 & 0\\
				2 & 1 & 2 & 2 & 1\\
				2 & 2 & 0 & 0 & 0\\
				2 & 2 & 1 & 2 & 1\\
				2 & 2 & 2 & 1 & 1\\
				\hline
		\end{tabular}}
		\captionof{table}{Truth table for ternary full subtractor.}
		\label{TruthFullSub}
	\end{minipage}	
\end{figure}

\subsection{Ternary Multiplier}
In a two-qutrit multiplier $A_0B_0 \cross A_1B_1$, each digit of the first number is multiplied by each digit of the second number. Then all the partial products are added in the way shown in Fig. \ref{2-Multiply}. Therefore, we need ternary partial product generation (TPPG) circuits and adder circuits for the two-qutrit multiplier. This kind of multiplier is discussed in Ref. \cite{panahi2019novel}. The outputs of the two numbers are $P_0$, $P_1$, $P_2$, and $P_3$ and the carry is $c_{out}$.
\begin{figure}[h!]
	\centering
	\begin{tabular}{ccccc}
		\rowcolor{green!20}
		&&&$A_1$&$A_0$\\
		\rowcolor{green!20}
		&&$\cross$&$B_1$&$B_0$\\
		\hline
		\rowcolor{cyan!20}
		&&$c_0$&&\\
		\rowcolor{cyan!20}
		&$c_2$&$cp_1$&$cp_0$&\\
		\rowcolor{cyan!20}
		&$c_1$&$cp_2$&$A_1B_0$&$A_0B_0$\\
		\rowcolor{cyan!20}
		+&$cp_3$&$A_1B_1$&$A_0B_1$&0\\
		\hline
		\rowcolor{orange!20}
		$c_{out}$&$P_3$&$P_2$&$P_1$&$P_0$
	\end{tabular}
	\caption{Two-qutrit multiplication.}
	\label{2-Multiply}
\end{figure}

The carries produced by addition are represented as $c_i$, whereas the $cp_i$ are the carries we get as a result of one-digit multiplication. The first digit of multiplication is $P_0=A_0B_0$ and its carry is represented by $cp_0$. To compute $P_1,P_2,P_3$ and $c_{out}$, adder circuits are needed. As each of the digits $A_0,B_0,A_1,B_1$ can have values $0,1,2$, not all the partial products produce carries in multiplication. We do not need extra input lines and gates corresponding to the carries, and therefore, it is cheaper to implement the circuit without full adders. Instead, we are using the Adder Blocks as in Panahi \cite{panahi2019novel}. The one-digit TPPG circuit is shown in Fig. \ref{TPPG} and the truth table is shown in Fig. \ref{TruthTPPG}. The carry-out appears only when both the input values are at value $2$ in the one-digit multiplication. The existing realization in Ref. \cite{panahi2019novel} has 13 MS and shift gates.
\eq{P_0&=A_0B_0\nonumber\\
	P_1&=cp_0+A_1B_0+A_0B_1\nonumber\\
	P_2&=c_0+cp_1+cp_2+A_1B_1\nonumber\\
	P_3&=c_2+c_1+cp_3}

\begin{figure}[h!]
	\centering
	\begin{minipage}{0.5\textwidth}
		\centering
		\begin{tikzpicture}[scale=1]
			\draw[fill=red!20,opacity=0.5] (2,0.4) rectangle (5.8,3.6);
			\qnode{1.2}{3}{$A$}
			\qnode{1.2}{2}{$B$}
			\qnode{1.2}{1}{$0$}
			\qnode{1.2}{0}{$0$}
			\draw[ultra thick,cyan] (2.6,1)--(2.6,2);
			\qwire[ibmqx]{2}{3}
			\qgateU[ibmqxD]{2}{1}{+1}
			\qgateCNC*[ibmqx]{b}{2}{2}
			
			\qwire[ibmqx]{3}{2}
			\draw[ultra thick,cyan] (3.9,1)--(3.9,3);
			\qgateCNC[ibmqx]{b}{3}{3}
			\qgateU[ibmqxA]{3}{1}{12}
			\qwire[ibmqx]{4}{2}
			\draw[ultra thick,cyan] (5.2,1)--(5.2,3);		
			\qgateCNC[ibmqx]{b}{4}{3}
			\qgateCNC*[ibmqx]{b}{4}{2}
			\qgateU[ibmqxA]{4}{1}{+2}
			\qnode{3.8}{3.3}{$0$}
			
			\qwire[ibmqx]{5}{1}
			\draw[ultra thick,cyan] (6.5,0)--(6.5,3);					
			\qgateCNC[ibmqx]{b}{5}{3}
			\qgateCNC[ibmqx]{b}{5}{2}
			\qgateU[ibmqxA]{5}{0}{+1}
			\qnode{5.8}{3}{$A$}
			\qnode{5.8}{2}{$B$}
			\qnode{5.8}{1}{$P$}
			\draw[ultra thick] (1.95,0)--(6,0);
			\qnode{5.8}{0}{$cp$}
		\end{tikzpicture}
		\captionof{figure}{Circuit design of the TPPG component.}
		\label{TPPG}
	\end{minipage}
	\begin{minipage}{0.3\textwidth}
		\centering
		\begin{tabular}{|>{\columncolor{green!20}}c|>{\columncolor{green!20}}c||>{\columncolor{orange!40}}c|>{\columncolor{orange!20}}c|}
			\hline
			\rowcolor{cyan!50}
			A & B & $P$ & $cp$\\
			\hline\hline
			0 & 0 & 0 & 0\\
			1 & 0 & 0 & 0\\
			2 & 0 & 0 & 0\\
			0 & 1 & 0 & 0\\
			1 & 1 & 1 & 0\\
			2 & 1 & 2 & 0\\
			0 & 2 & 0 & 0\\
			1 & 2 & 2 & 0\\
			2 & 2 & 1 & 1\\
			\hline
		\end{tabular}
		\captionof{table}{Truth table for a two-qutrit partial product.}
		\label{TruthTPPG}
	\end{minipage}	
\end{figure}

From the Fig. \ref{2-Multiply}, we have $P_1=cp_0+A_1B_0+A_0B_1$. Adder Block 1 adds the partial products $A_1B_0,A_0B_1$ and the carry $cp_0$ of the partial products. As we can see from the truth table \ref{TruthTPPG} of the partial product, the carry in the first partial product is never 2. The input values can be $0$, $1$ and $2$ but carry is only $0$ or $1$. That is, the $cp_0$, $cp_1$, $cp_2$, and $cp_3$ are 1 and the carry is only for the case when inputs are at values 2. The truth table for the Adder Block 1 is shown in Table \ref{TruthBlock1}, and the implementation is shown in Fig. \ref{Block1}. The input lines with label $0$ are ancilla lines, and $g$ are the garbage outputs.
Adder Block 2 adds three qutrits $A_1B_1$, $cp_1$ and $cp_2$ to get $P_2$. The sum of these three qutrits will be added to $c_0$ using Block 4. The first input has values $0,1,2$, but the second and third inputs have values $0,1$. The partial product $A_1B_1$ is added to the two input carry numbers, which have values $0,1$. The truth table and implementation of Block 2 is shown in Table \ref{TruthBlock2} and Fig. \ref{Block2}.
Adder Block 3 adds qutrits $c_1$, $c_2$ and $cp_3$ to get $P_3$. The truth table is shown in Table \ref{TruthBlock3} and the implementation is shown in Fig. \ref{Block3}. All the inputs have values $0$ or $1$. The output carry appears only when all the inputs have values $1$.
Adder Block 4 is used to add the sum of Adder Block 2 and $c_0$. The half adder is used for its implementation. The output carry is only for one of the cases. The input carry $c_{in}$ is either $0$ or $1$ and input $A=0,1,2$. The truth table of Block 4 is shown in Table \ref{TruthBlock4} and its implementation is shown in Fig. \ref{Block4}. The Block 1 is designed with two Clifford and 4 non-Clifford gates, Block 2 has 2 Clifford and 2 non-Clifford gates, Block 3 consists of 2 Clifford and one non-Clifford gates, whereas Block 4 is implemented with only one Clifford gate and 1 non Clifford gate.

\begin{figure}[h!]
	\centering
	\begin{minipage}{0.6\textwidth}
		\centering
		\begin{tikzpicture}[scale=1]
			\draw[fill=red!20,opacity=0.5] (4.6,0.5) rectangle (7.1,3.6);
			\qnode{0.2}{3}{$A$}
			\qnode{0.2}{2}{$B$}
			\qnode{0.2}{1}{$c_{in}$}	
			\qnode{0.2}{0}{$0$}
			\qwire[ibmqx]{1}{3}
			\qwire[ibmqx]{1}{1}
			\draw[ultra thick,cyan] (1.3,0)--(1.3,2);			
			\qgateCNC[ibmqx]{b}{1}{2}
			\qgateU[ibmqxA]{1}{0}{+1}
			\qwire[ibmqx]{2}{1}
			\draw[ultra thick,cyan] (2.6,0)--(2.6,3);			
			\qgateCNC[ibmqx]{b}{2}{3}
			\qgateCNC[ibmqx]{b}{2}{2}
			\qgateU[ibmqxA]{2}{0}{+1}			
			\qnode{1.8}{2.3}{$1$}
			\qwire[ibmqx]{3}{1}
			\draw[ultra thick,cyan] (3.9,0)--(3.9,3);			
			\qgateCNC[ibmqx]{b}{3}{3}
			\qgateCNC[ibmqx]{b}{3}{2}
			\qgateU[ibmqxA]{3}{0}{01}			
			\qnode{2.8}{3.3}{$0$}
			\draw[ultra thick,cyan] (5.2,2)--(5.2,3);
			\qgateU[ibmqxD]{4}{3}{+1}
			\qgateCNC*[ibmqx]{t}{4}{2}
			\qwire[ibmqx]{4}{1}
			\qwire[ibmqx]{4}{0}
			\qwire[ibmqx]{5}{2}
			\qwire[ibmqx]{5}{0}
			\draw[ultra thick,cyan] (6.5,1)--(6.5,3);
			\qgateU[ibmqxD]{5}{3}{+1}
			\qgateCNC*[ibmqx]{t}{5}{1}
			\qwire[ibmqx]{6}{2}
			\draw[ultra thick,cyan] (7.8,0)--(7.8,3);			
			\qgateCNC[ibmqx]{b}{6}{3}
			\qgateCNC[ibmqx]{b}{6}{1}
			\qgateU[ibmqxA]{6}{0}{+1}			
			\qnode{5.8}{1.3}{$1$}
			\qnode{5.8}{3.3}{$0$}
			\qnode{6.9}{3}{$Sum$}
			\qnode{6.8}{2}{$g_0$}
			\qnode{6.8}{1}{$g_1$}	
			\qnode{6.85}{0}{$c_{out}$}
		\end{tikzpicture}
		\captionof{figure}{Implementation for the ternary Block 1.}
		\label{Block1}		
	\end{minipage}
	\begin{minipage}{0.3\textwidth}
		\centering
		\scalebox{0.9}[0.9]{
			\begin{tabular}{|>{\columncolor{green!20}}c|>{\columncolor{green!20}}c|>{\columncolor{green!20}}c||>{\columncolor{orange!40}}c|>{\columncolor{orange!20}}c|}
				\hline
				\rowcolor{cyan!50}
				A & B & $c_{in}$ & $Sum$ & $c_{out}$\\
				\hline\hline
				0 & 0 & 0 & 0 & 0\\
				1 & 0 & 0 & 1 & 0\\
				2 & 0 & 0 & 2 & 0\\
				0 & 1 & 0 & 1 & 0\\
				1 & 1 & 0 & 2 & 0\\
				2 & 1 & 0 & 0 & 1\\
				0 & 2 & 0 & 2 & 0\\
				1 & 2 & 0 & 0 & 1\\
				2 & 2 & 0 & 1 & 1\\
				0 & 0 & 1 & 1 & 0\\
				1 & 0 & 1 & 2 & 0\\
				2 & 0 & 1 & 0 & 1\\
				0 & 1 & 1 & 2 & 0\\
				1 & 1 & 1 & 0 & 1\\
				2 & 1 & 1 & 1 & 1\\
				0 & 2 & 1 & 0 & 1\\
				1 & 2 & 1 & 1 & 1\\
				2 & 2 & 1 & 2 & 1\\
				\hline
		\end{tabular}}
		\captionof{table}{Truth table for ternary Adder Block 1.}
		\label{TruthBlock1}
	\end{minipage}	
\end{figure}


\begin{figure}[h!]
	\centering
	\begin{minipage}{0.5\textwidth}
		\centering
		\begin{tikzpicture}[scale=1]
			\draw[fill=red!20,opacity=0.5] (4.6,0.5) rectangle (7,3.7);
			\qnode{2.2}{3}{$A$}
			\qnode{2.2}{2}{$B$}
			\qnode{2.2}{1}{$c_{in}$}	
			\qnode{2.2}{0}{$0$}
			\qwire[ibmqx]{3}{1}
			\draw[ultra thick,cyan] (3.9,0)--(3.9,3);			
			\qgateCNC[ibmqx]{b}{3}{3}
			\qgateCNC[ibmqx]{b}{3}{2}
			\qgateU[ibmqxA]{3}{0}{+1}			
			\qnode{2.8}{2.3}{$1$}
			\draw[ultra thick,cyan] (5.2,2)--(5.2,3);
			\qgateU[ibmqxD]{4}{3}{+1}
			\qgateCNC*[ibmqx]{t}{4}{2}
			\qwire[ibmqx]{4}{1}
			\qwire[ibmqx]{4}{0}
			\qwire[ibmqx]{5}{2}
			\qwire[ibmqx]{5}{0}
			\draw[ultra thick,cyan] (6.5,1)--(6.5,3);
			\qgateU[ibmqxD]{5}{3}{+1}
			\qgateCNC*[ibmqx]{t}{5}{1}
			\qwire[ibmqx]{6}{2}
			\draw[ultra thick,cyan] (7.8,0)--(7.8,3);			
			\qgateCNC[ibmqx]{b}{6}{3}
			\qgateCNC[ibmqx]{b}{6}{1}
			\qgateU[ibmqxA]{6}{0}{+1}			
			\qnode{5.8}{3.3}{$0$}
			\qnode{5.8}{1.3}{$1$}
			\qnode{6.9}{3}{$Sum$}
			\qnode{6.8}{2}{$g_2$}
			\qnode{6.8}{1}{$g_3$}	
			\qnode{6.9}{0}{$c_{out}$}
		\end{tikzpicture}
		\captionof{figure}{Implementation of Block 2.}
		\label{Block2}		
	\end{minipage}
	\begin{minipage}{0.4\textwidth}
		\centering
		\scalebox{0.9}[0.9]{
			\begin{tabular}{|>{\columncolor{green!20}}c|>{\columncolor{green!20}}c|>{\columncolor{green!20}}c||>{\columncolor{orange!40}}c|>{\columncolor{orange!20}}c|}
				\hline
				\rowcolor{cyan!50}
				A & B & $c_{in}$ & $Sum$ & $c_{out}$\\
				\hline\hline
				0 & 0 & 0 & 0 & 0\\
				1 & 0 & 0 & 1 & 0\\
				2 & 0 & 0 & 2 & 0\\
				0 & 1 & 0 & 1 & 0\\
				1 & 1 & 0 & 2 & 0\\
				2 & 1 & 0 & 0 & 1\\
				0 & 0 & 1 & 1 & 0\\
				1 & 0 & 1 & 2 & 0\\
				2 & 0 & 1 & 0 & 1\\
				0 & 1 & 1 & 2 & 0\\
				1 & 1 & 1 & 0 & 1\\
				2 & 1 & 1 & 1 & 1\\
				\hline
		\end{tabular}}
		\captionof{table}{Truth table for ternary Adder Block 2.}
		\label{TruthBlock2}
	\end{minipage}	
\end{figure}

\begin{figure}[h!]
	\centering
	\begin{minipage}{0.5\textwidth}
		\centering
		\begin{tikzpicture}[scale=1]
			\draw[fill=red!20,opacity=0.5] (0.6,0.5) rectangle (3.2,3.6);
			\qnode{0.2}{3}{$A$}
			\qnode{0.2}{2}{$B$}
			\qnode{0.2}{1}{$c_{in}$}	
			\qnode{0.2}{0}{$0$}
			\draw[ultra thick,cyan] (1.3,2)--(1.3,3);
			\qgateU[ibmqxD]{1}{3}{+1}
			\qgateCNC*[ibmqx]{t}{1}{2}
			\qwire[ibmqx]{1}{1}
			\qwire[ibmqx]{1}{0}
			
			\qwire[ibmqx]{2}{2}
			\qwire[ibmqx]{2}{0}
			\draw[ultra thick,cyan] (2.6,1)--(2.6,3);
			\qgateU[ibmqxD]{2}{3}{+1}
			\qgateCNC*[ibmqx]{t}{2}{1}
			
			\qwire[ibmqx]{3}{2}
			\draw[ultra thick,cyan] (3.9,0)--(3.9,3);			
			\qgateCNC[ibmqx]{b}{3}{3}
			\qgateCNC[ibmqx]{b}{3}{1}
			\qgateU[ibmqxA]{3}{0}{+1}			
			\qnode{2.8}{1.3}{$1$}
			\qnode{2.8}{3.3}{$0$}
			
			\qnode{3.9}{3}{$Sum$}
			\qnode{3.8}{2}{$g_4$}
			\qnode{3.8}{1}{$g_5$}	
			\qnode{3.85}{0}{$c_{out}$}
		\end{tikzpicture}
		\captionof{figure}{Implementation of ternary Adder Block 3.}
		\label{Block3}
	\end{minipage}
\begin{minipage}{0.4\textwidth}		
	\centering	
	\begin{tabular}{|>{\columncolor{green!20}}c|>{\columncolor{green!20}}c|>{\columncolor{green!20}}c||>{\columncolor{orange!40}}c|>{\columncolor{orange!20}}c|}
		\hline
		\rowcolor{cyan!50}
		A & B & $c_{in}$ & $Sum$ & $c_{out}$\\
		\hline\hline
		0 & 0 & 0 & 0 & 0\\
		1 & 0 & 0 & 1 & 0\\
		0 & 1 & 0 & 1 & 0\\
		1 & 1 & 0 & 2 & 0\\
		0 & 0 & 1 & 1 & 0\\
		1 & 0 & 1 & 2 & 0\\
		0 & 1 & 1 & 2 & 0\\
		1 & 1 & 1 & 0 & 1\\
		\hline
	\end{tabular}
	\captionof{table}{Truth table for ternary Adder Block 3.}
	\label{TruthBlock3}
\end{minipage}	
\end{figure}


\begin{figure}[h!]
	\centering
	\begin{minipage}{0.4\textwidth}
		\centering
		\begin{tikzpicture}[scale=1]
			\draw[fill=red!20,opacity=0.6] (2,0.5) rectangle (3.2,2.6);
			\qnode{0.2}{2}{$A$}
			\qnode{0.2}{1}{$c_{in}$}	
			\qnode{0.2}{0}{$0$}
			\draw[ultra thick,cyan] (1.3,0)--(1.3,2);
			\qwire[ibmqx]{1}{2}
			\qgateCNC[ibmqx]{b}{1}{1}
			\qgateCNC[ibmqx]{b}{1}{2}
			\qgateU[ibmqxA]{1}{0}{+1}
			\qnode{0.8}{1.3}{$1$}
			\draw[ultra thick,cyan] (2.6,1)--(2.6,2);	
			\qwire[ibmqx]{2}{0}
			\qgateCNC*[ibmqx]{t}{2}{1}
			\qgateU[ibmqxD]{2}{2}{+1}
			\qnode{2.9}{2}{$Sum$}
			\qnode{2.8}{1}{$g_6$}
			\qnode{2.9}{0}{$c_{out}$}
		\end{tikzpicture}
		\captionof{figure}{Implementation of Block 4.}
		\label{Block4}
	\end{minipage}
	\begin{minipage}{0.4\textwidth}		
		\centering
		\scalebox{1}[1]{
			\begin{tabular}{|>{\columncolor{green!20}}c|>{\columncolor{green!20}}c||>{\columncolor{orange!40}}c|>{\columncolor{orange!20}}c|}
				\hline
				\rowcolor{cyan!50}
				A & $c_{in}$ & $Sum$ & $c_{out}$\\
				\hline\hline
				0 & 0 & 0 & 0\\
				1 & 0 & 1 & 0\\
				2 & 0 & 2 & 0\\
				0 & 1 & 1 & 0\\
				1 & 1 & 2 & 0\\
				2 & 1 & 0 & 1\\
				\hline
		\end{tabular}}
	\captionof{table}{Truth table for ternary Adder Block 4.}
		\label{TruthBlock4}
	\end{minipage}	
\end{figure}

Now we will combine the TPPG circuits and the Adder Blocks to get a full two-digit multiplier circuit. From the two-digit qutrit multiplication in Fig. \ref{2-Multiply}, the first digits of two numbers give the first partial product and first multiplication digit $P_0$. The second multiplication digit is the addition of partial products $A_0B_1$ and $A_1B_0$ and added to the carry from the first partial product. These partial products also create the carries. The carries from these partial products and the carries of additions are added to the partial product $A_1B_1$. These give the third digit of multiplication. The additions generate two carry digits.

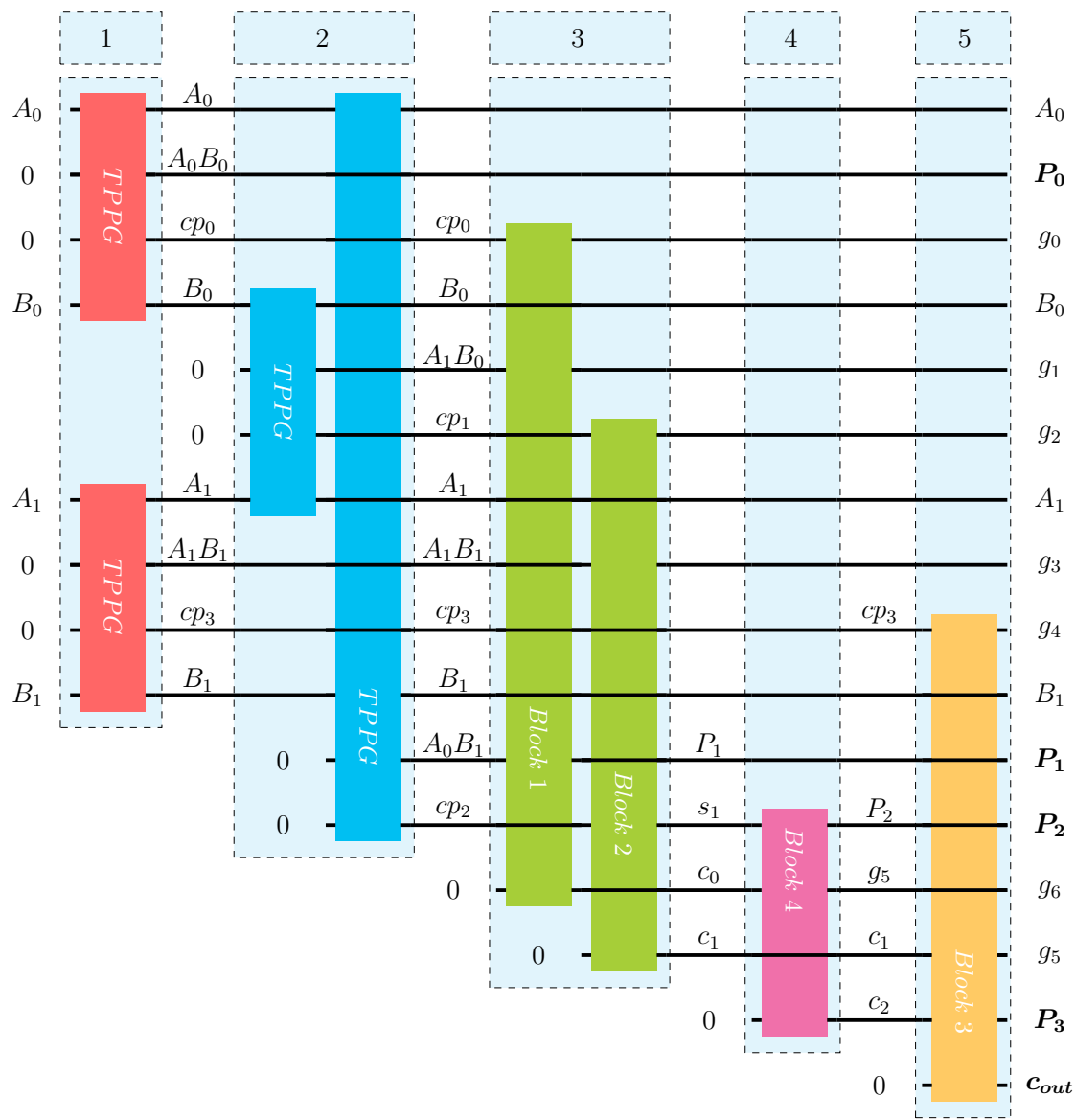
\begin{figure}[t!]
	\centering
\scalebox{0.9}{
	\begin{tikzpicture}[scale=1]
		\draw[dashed,fill=cyan!10] (1.8,9.7) rectangle (3.35,10.5);
		\draw[dashed,fill=cyan!10] (4.45,9.7) rectangle (7.2,10.5);
		\draw[dashed,fill=cyan!10] (8.35,9.7) rectangle (11.1,10.5);
		\draw[dashed,fill=cyan!10] (12.25,9.7) rectangle (13.7,10.5);
		\draw[dashed,fill=cyan!10] (14.85,9.7) rectangle (16.3,10.5);
		\node at (2.5,10.1) {$1$};
		\node at (5.8,10.1) {$2$};
		\node at (9.7,10.1) {$3$};
		\node at (12.95,10.1) {$4$};
		\node at (15.6,10.1) {$5$};
		\draw[dashed,fill=cyan!10] (1.8,-0.5) rectangle (3.35,9.5);
		\draw[dashed,fill=cyan!10] (4.45,-2.5) rectangle (7.2,9.5);
		\draw[dashed,fill=cyan!10] (8.35,-4.5) rectangle (11.1,9.5);
		\draw[dashed,fill=cyan!10] (12.25,-5.5) rectangle (13.7,9.5);
		\draw[dashed,fill=cyan!10] (14.85,-6.5) rectangle (16.3,9.5);
		\qnode{1}{9}{$A_0$}
		\qnode{1}{8}{$0$}	
		\qnode{1}{7}{$0$}
		\qnode{1}{6}{$B_0$}
		\qnode{1}{3}{$A_1$}
		\qnode{1}{2}{$0$}
		\qnode{1}{1}{$0$}
		\qnode{1}{0}{$B_1$}
		\qgateUu[ibmqxE]{2}{8}{}
		\qgateUuu[ibmqxE]{2}{7}{{\small \begin{turn}{-90}$TPPG$\end{turn}}}
		\qgateUu[ibmqxE]{2}{2}{}
		\qgateUuu[ibmqxE]{2}{1}{{\small \begin{turn}{-90}$TPPG$\end{turn}}}
		\qnode{3}{9.23}{$A_0$}
		\qnode{3}{8.23}{$A_0B_0$}	
		\qnode{3}{7.2}{$cp_0$}
		\qnode{3}{6.23}{$B_0$}
		\qnode{3}{5}{$0$}
		\qnode{3}{4}{$0$}
		\qnode{3}{3.23}{$A_1$}
		\qnode{3}{2.23}{$A_1B_1$}
		\qnode{3}{1.2}{$cp_3$}
		\qnode{3}{0.23}{$B_1$}
		\qwire[ibmqx]{3}{9}
		\qwire[ibmqx]{3}{8}
		\qwire[ibmqx]{3}{7}
		\qwire[ibmqx]{3}{6}
		\qwire[ibmqx]{3}{3}
		\qwire[ibmqx]{3}{2}
		\qwire[ibmqx]{3}{1}
		\qwire[ibmqx]{3}{0}
		\qwire[ibmqx]{4}{9}
		\qwire[ibmqx]{4}{8}
		\qwire[ibmqx]{4}{7}
		\qwire[ibmqx]{4}{6}
		\qwire[ibmqx]{4}{3}
		\qwire[ibmqx]{4}{2}
		\qwire[ibmqx]{4}{1}
		\qwire[ibmqx]{4}{0}
		\qnode{4}{-1}{$0$}
		\qnode{4}{-2}{$0$}
		\qgateUu[ibmqxD]{4}{5}{}
		\qgateUuu[ibmqxD]{4}{4}{{\small \begin{turn}{-90}$TPPG$\end{turn}}}
		\qgateUuu[ibmqxD]{5}{8}{}
		\qgateUuu[ibmqxD]{5}{6}{}
		\qgateUuu[ibmqxD]{5}{4}{}
		\qgateUuu[ibmqxD]{5}{2}{}
		\qgateUu[ibmqxD]{5}{0}{}
		\qgateUuu[ibmqxD]{5}{-1}{{\small \begin{turn}{-90}$TPPG$\end{turn}}}
		\qwire[ibmqx]{5}{8}
		\qwire[ibmqx]{5}{7}
		\qwire[ibmqx]{5}{6}
		\qwire[ibmqx]{5}{5}
		\qwire[ibmqx]{5}{4}
		\qwire[ibmqx]{5}{3}
		\qwire[ibmqx]{5}{2}
		\qwire[ibmqx]{5}{1}
		\qwire[ibmqx]{6}{9}
		\qwire[ibmqx]{6}{8}
		\qwire[ibmqx]{6}{7}
		\qwire[ibmqx]{6}{6}
		\qwire[ibmqx]{6}{5}
		\qwire[ibmqx]{6}{4}
		\qwire[ibmqx]{6}{3}
		\qwire[ibmqx]{6}{2}
		\qwire[ibmqx]{6}{1}
		\qwire[ibmqx]{6}{0}
		\qwire[ibmqx]{6}{-1}
		\qwire[ibmqx]{6}{-2}
		\qnode{6}{7.23}{$cp_0$}
		\qnode{6}{6.23}{$B_0$}
		\qnode{6}{5.2}{$A_1B_0$}
		\qnode{6}{4.2}{$cp_1$}
		\qnode{6}{3.23}{$A_1$}
		\qnode{6}{2.23}{$A_1B_1$}
		\qnode{6}{1.23}{$cp_3$}
		\qnode{6}{0.23}{$B_1$}
		\qnode{6}{-0.77}{$A_0B_1$}
		\qnode{6}{-1.77}{$cp_2$}
		\qnode{6}{-3}{$0$}
		\qgateUuu[ibmqxC]{7}{6}{}
		\qgateUuu[ibmqxC]{7}{4}{}
		\qgateUuu[ibmqxC]{7}{2}{}
		\qgateUuu[ibmqxC]{7}{1}{}
		\qgateUu[ibmqxC]{7}{-3}{}		
		\qgateUuu[ibmqxC]{7}{-1}{{\small \begin{turn}{-90}$\ \ Block \ 1$\end{turn}}}
		\qwire[ibmqx]{7}{9}
		\qwire[ibmqx]{7}{8}
		\qwire[ibmqx]{7}{6}
		\qwire[ibmqx]{7}{4}
		\qwire[ibmqx]{7}{3}
		\qwire[ibmqx]{7}{2}
		\qwire[ibmqx]{7}{1}
		\qwire[ibmqx]{7}{0}
		\qwire[ibmqx]{7}{-2}
		\qnode{7}{-4}{$0$}
		\qgateUu[ibmqxC]{8}{3}{}
		\qgateUuu[ibmqxC]{8}{2}{}
		\qgateUuu[ibmqxC]{8}{0}{}
		\qgateUu[ibmqxC]{8}{-4}{}
		\qgateUuu[ibmqxC]{8}{-2}{{\small \begin{turn}{-90}$\ \ Block \ 2$\end{turn}}}
		\qwire[ibmqx]{8}{9}
		\qwire[ibmqx]{8}{8}
		\qwire[ibmqx]{8}{7}
		\qwire[ibmqx]{8}{6}
		\qwire[ibmqx]{8}{5}
		\qwire[ibmqx]{8}{3}
		\qwire[ibmqx]{8}{1}
		\qwire[ibmqx]{8}{0}
		\qwire[ibmqx]{8}{-1}
		\qwire[ibmqx]{8}{-3}
		\qwire[ibmqx]{9}{9}
		\qwire[ibmqx]{9}{8}
		\qwire[ibmqx]{9}{7}
		\qwire[ibmqx]{9}{6}
		\qwire[ibmqx]{9}{5}
		\qwire[ibmqx]{9}{4}
		\qwire[ibmqx]{9}{3}
		\qwire[ibmqx]{9}{2}
		\qwire[ibmqx]{9}{1}
		\qwire[ibmqx]{9}{0}
		\qwire[ibmqx]{9}{-1}
		\qwire[ibmqx]{9}{-2}
		\qwire[ibmqx]{9}{-3}
		\qwire[ibmqx]{9}{-4}
		\qnode{9}{-0.77}{$P_1$}
		\qnode{9}{-1.77}{$s_1$}
		\qnode{9}{-2.77}{$c_0$}
		\qnode{9}{-3.77}{$c_1$}
		\qnode{9}{-5}{$0$}
		\qgateUu[ibmqxF]{10}{-5}{}
		\qgateUuu[ibmqxF]{10}{-3}{{\small \begin{turn}{-90}$\ Block \ 4$\end{turn}}}
		\qwire[ibmqx]{10}{9}
		\qwire[ibmqx]{10}{8}
		\qwire[ibmqx]{10}{7}
		\qwire[ibmqx]{10}{6}
		\qwire[ibmqx]{10}{5}
		\qwire[ibmqx]{10}{4}
		\qwire[ibmqx]{10}{3}
		\qwire[ibmqx]{10}{2}
		\qwire[ibmqx]{10}{1}
		\qwire[ibmqx]{10}{0}
		\qwire[ibmqx]{10}{-1}
		\qwire[ibmqx]{10}{-4}
		\qwire[ibmqx]{11}{9}
		\qwire[ibmqx]{11}{8}
		\qwire[ibmqx]{11}{7}
		\qwire[ibmqx]{11}{6}
		\qwire[ibmqx]{11}{5}
		\qwire[ibmqx]{11}{4}
		\qwire[ibmqx]{11}{3}
		\qwire[ibmqx]{11}{2}
		\qwire[ibmqx]{11}{1}
		\qwire[ibmqx]{11}{0}
		\qwire[ibmqx]{11}{-1}
		\qwire[ibmqx]{11}{-2}
		\qwire[ibmqx]{11}{-3}
		\qwire[ibmqx]{11}{-4}
		\qwire[ibmqx]{11}{-5}
		\qnode{11}{1.23}{$cp_3$}
		\qnode{11}{-1.77}{$P_2$}
		\qnode{11}{-2.77}{$g_5$}
		\qnode{11}{-3.77}{$c_1$}
		\qnode{11}{-4.77}{$c_2$}
		\qnode{11}{-6}{$0$}
		\qgateUu[ibmqxB]{12}{0}{}
		\qgateUuu[ibmqxB]{12}{-1}{}
		\qgateUuu[ibmqxB]{12}{-3}{}
		\qgateUuu[ibmqxB]{12}{-5}{{\small \begin{turn}{-90} $ Block \ 3$\end{turn}}}
		\qwire[ibmqx]{12}{9}
		\qwire[ibmqx]{12}{8}
		\qwire[ibmqx]{12}{7}
		\qwire[ibmqx]{12}{6}
		\qwire[ibmqx]{12}{5}
		\qwire[ibmqx]{12}{4}
		\qwire[ibmqx]{12}{3}
		\qwire[ibmqx]{12}{2}
		\qwire[ibmqx]{12}{0}
		\qwire[ibmqx]{12}{-1}
		\qwire[ibmqx]{12}{-2}
		\qwire[ibmqx]{12}{-3}
		\qnode{13}{9}{$A_0$}
		\qnode{13}{8}{$\bm{P_0}$}
		\qnode{13}{7}{$g_0$}
		\qnode{13}{6}{$B_0$}
		\qnode{13}{5}{$g_1$}
		\qnode{13}{4}{$g_2$}
		\qnode{13}{3}{$A_1$}
		\qnode{13}{2}{$g_3$}
		\qnode{13}{1}{$g_4$}
		\qnode{13}{0}{$B_1$}
		\qnode{13}{-1}{$\bm{P_1}$}
		\qnode{13}{-2}{$\bm{P_2}$}
		\qnode{13}{-3}{$g_6$}
		\qnode{13}{-4}{$g_5$}
		\qnode{13}{-5}{$\bm{P_3}$}
		\qnode{13}{-6}{$\bm{c_{out}}$}
	\end{tikzpicture}}
	\caption[Ternary two-qutrit multiplier is implemented using the Adder Blocks and TPPG circuits.]{Ternary two-qutrit multiplier is implemented using the Adder Blocks and TPPG circuits \cite{panahi2019novel}.}
	\label{MultiFull}
\end{figure}

The full two-digit qutrit multiplication circuit is shown in Fig. \ref{MultiFull}. When the line goes on the top of a Block, then it is non-interacting, but when a line goes below the Block, then it is given to that Block as the input line.
In the Fig., the upper TPPG at stage 1 gives a partial product $P_0=A_0B_0$ and the carry $cp_0$.
The lower TPPG at the stage 1 computes $A_1B_1$, and we have the carry $cp_3$.
At stage 2, the first TPPG computes partial product $A_1B_0$ and produces the carry $cp_1$, whereas the second TPPG computes $A_0B_1$ and produces the carry $cp_2$. Inputs of the first TPPG are $B_0$ and $A_1$ and two ancillae, whereas the inputs of the second TPPG are $A_0$ and $B_1$.
The stages 3, 4 and 5 consist of Adder Blocks. As we can see in Fig. \ref{2-Multiply}, we need to add $cp_0$, $A_0B_1$ and $A_1B_0$ to get $P_1$ and the carry $c_0$. This is done by Block 1 at stage 3. At this stage, the Adder Block 2 adds $cp_1,A_1B_1$ and $cp_2$. The output of this block is $s_1$ and the carry is $c_1$. At stage 4, the sum $s_1$ is added to $c_0$ by using the Block 4 and the output $P_2$ is obtained. The Block 4 has a carry $c_2$. At stage 5, $cp_3$, $c_1$ and $c_2$ are added using the Block 3 to get an output qutrit $P_3$ and the carry $c_{out}$.

	\chapter{Conclusion}\label{Conclusion}
	Topological quantum computing is a promising candidate for fault-tolerant quantum computing. The main focus of this dissertation was to find the ternary arithmetic circuit implementation in topological quantum computation, and the fundamentals on which topological quantum computation is based. It is also explained how to build fault-tolerant gates by using the ideas based on quantum topology and braiding of anyons.

Topological quantum computation is a wide field of research. The areas of topology and knot theory, topological phases of matter, quantum field theory, and quantum computing are merged in this field. The modular categories and the quantum group are the algebraic mathematical models for the topological phases of matter. To make this dissertation self-contained, we discussed all of these areas and made their connections with quantum computing.

First, we discussed the quantum binary and ternary logic design in conventional quantum computing and how the quantum computer can work better in terms of complexity. 
Topology and knot theory as a mathematical background were presented. The physical insight, behind the knots and braids, is explained through the geometric phases in quantum physics. These phases also explain the existence of the degenerate state in the topological phases of matter and the evolution of the system in these states.
There are several proposals for the physical systems for topological quantum computation. Among these, the non-Abelian anyons in the quantum Hall effect and Majorana fermions in topological superconductors are the most popular. We discussed the quantum Hall effect in detail and its similarities and differences to the other topological materials like the topological insulator. Superconductivity and the appearance of Majorana fermions in topological superconductors were explained. The Chern-Simons theory is used as an effective field theory to explain the charge-flux composites as the non-Abelian anyons in such systems and their connection to knots and braids.
The fusion and braiding matrices, topological spin, and Hilbert space are discussed in terms of topological quantum field theory and category theory.

Both binary and ternary logic gates design are presented. Theoretical models for a binary logic design are based on the Fibonacci and Ising anyons. For ternary logic design, metaplectic anyons are used. The basic ternary logic gates are reproduced. The topological data for metaplectic anyons are obtained by the quantum deformation of recoupling theory. The metaplectic anyons are defined in terms of the category theory that also provides their topological data. The fusion matrices $F$ and braiding matrices $R$ are discussed in category theory and quantum deformation of the recoupling theory. 

Our arithmetic circuits are based on a combination of Clifford gates and non-Clifford gates. The quantum cost of the implementations of the ternary adder, ternary subtractor, and ternary multiplier circuits is smaller than the existing realization. We proposed the implementation of the arithmetic circuits by using the topological charge measurement gates $C_c(X)$ and the braiding gates $SUM$ and $SWAP$. We modified the existing realizations to have more controlled increment gates than the controlled permutation gates so that these circuits can be realized by braiding. 

	\cleardoublepage
	\phantomsection
	\addcontentsline{toc}{part}{\numberline{}Appendices}

	\cleardoublepage
	\appendix
	\chapter{Abstract Algebra}\label{AbsAlg}
\section{Group}
The group theory is indispensable in physics for the study of symmetry. But group representation theory is of direct use in quantum physics and other branches of physics. 

A set $G$ is a \textbf{group} when it fulfills the following axioms:
\begin{itemize}
	\item Closure: For any two elements $A$ and $B$ of $G$, their composition $A\circ B$ is also in $G$. Here, $\circ$ is addition or multiplication. 
	\item Associativity: The composition of any three elements $A,B,C$ $\in$ $G$, is associative. That is, $A\circ(B\circ C) = (A\circ B)\circ C$
	\item Identity: There exist an identity element $I$ $\in$ $G$, such that, for any element $A$ $\in$ $G$, $I\circ A = A \circ I = A$. The identity element is $0$ for a group under addition, and $1$ for a group under multiplication. 
	\item Inverse: For any element $A$ $\in$ $G$, there exist a unique element $A^{-1}$ $\in$ $G$ such that $A\circ A^{-1} = A^{-1}\circ A = I$. The additive inverse is the negative of an element, whereas the multiplicative inverse is the reciprocal of an element of the group. 
\end{itemize}

The matrix groups are of particular interest in physics. The set of all matrices form a group under addition. The null matrix is the additive identity. The additive inverse of a matrix $A$ is $-A$. The group is non-commutative when the order of the composition matters, that is, when $A\circ B \neq B \circ A$. The commutative group is called an \textbf{Abelian} group, whereas the non-commutative group is called the \textbf{non-Abelian} group. Matrices form a \textbf{non-Abelian} group under multiplication, as the matrix multiplication is not commutative in general.

An \textbf{ordered pair} is a pair of objects in which one element of the pair is designated as the first entry and the other element is designated as the second entry. It is denoted as $(a,b)$. The order of entries matters in the ordered pair, that is $(a,b)\ne (b, a)$. The \textbf{Cartesian product} of two sets is denoted as $A\cross B=\left\{(a,b)| a\in A \ \text{and} \ b\in B\right\}$. It is the set of all possible ordered pairs such that the first entry is from the set $A$ and the second entry is from the set $B$.

The number of elements of a group is called the \textbf{order} of the group. A group with a finite number of elements is called a \textbf{finite group} and the group with an infinite number of elements is called an \textbf{infinite group}. The infinite group may be \textbf{discrete} if the number of elements is enumerably infinite, or \textbf{continuous} if the number of elements is non-enumerably infinite. The set of integers is a discrete infinite group, whereas the set of real numbers is a continuous infinite group. The Lie group is a continuous group. 

Two elements $A$ and $B$ of a set are equivalent if there is some kind of equivalence relation between them. The \textbf{equivalence relation} is written as $A\sim B$, and it should have three properties; reflexive, symmetric, and transitive. The reflexive property means that any element is equivalent to itself, $A \sim A$. The symmetric property implies, if $A$ is equivalent to $B$ then $B$ is also equivalent to $A$. That is, $A\sim B \Rightarrow B \sim A$. The transitive property tells that if $A$ is equivalent to $B$ and $B$ is equivalent to $C$ then $A$ is equivalent to $C$. That is, if $ A \sim B \ \ and  \  \ B \sim C$ then $A \sim C$. All those elements which have the common property can be put into one class called the \textbf{equivalence class}. A set can be \textit{partitioned} into the set of equivalence classes that are disjoint, and their union is the parent set. One of the examples of the equivalence classes is the \textbf{conjugacy class}. Two elements $A$ and $B$ of a group are said to be in the same conjugacy class if they are related as $C^{-1} A C = B$, where $C$ is also an element of the group. 

\textbf{Homomorphism} is a structure preserving mapping from one set to another. Let $f$ be a mapping from a set $A$ to a set $B$ represented as $f: A\rightarrow B$. Let us consider an element $a$ of the set $A$. An element $b$ of $B$ is obtained by mapping the element $a$ to the element $b$ in $B$. That is  
$$b = f(a), \qquad \text{where} \ a \ \in \ A \ \text{and} \ b \ \in \ B.$$
Therefore, $b= \text{Im} f$ is the image of $f$ from $A$ in $B$. 
A mapping is \textbf{onto or surjective} when each element in the image is mapped from at least one element in the domain. The mapping is called \textbf{one-to-one or injective} when exactly one element in the domain is mapped to some element in the image. A mapping is \textbf{onto and one-to-one} or \textbf{bijective} when exactly one element in the domain is mapped to exactly one element in the image. The bijective mapping is reversible.
The \textbf{group homomorphism} is a mapping from one group to another which respects the composition law. The same set of axioms and the rules of composition hold in the image as in the domain. It can be written as 
$$f(a_1 a_2) = f(a_1)f(a_2).$$

\subsection{Subgroups}

The subset $H$ of a group $G$ is called a \textbf{subgroup} of $G$ if the elements of $H$ also fulfill the same composition axioms as $G$. Any group has two \textit{trivial subgroups}, the identity and the group itself. If a group has no other subgroups other than itself and the identity element, then it is called a \textbf{simple} group.

Consider a group $G$ of the order $n$ and its subgroup $H = (e, h_1, h_2,...h_m)$ of order $m$. Let us take an element $x$ of $G$ and take its composition with the elements of $H$ denoted by $xh$, such that
$$ xh = (xe, xh_1, xh_2,...,xh_m).$$
Here $xh$ is $x\circ h$, but we omit $\circ$ to avoid cluttering. The $xh$ should be assumed as a composition of $x$ and $h$ and this composition can be addition or multiplication. 
Now if $x$ is also an element of $H$ then the $xh$ must be the $H$ itself because the composition of $x$ with $h$ would be just a rearrangement of the elements of $h$. Then we can write $xh = H$. On the other hand, if $x$ is not in $H$ then the elements $xh$ do not belong to $H$. But this set $xh$ of course belongs to $G$. In this case, we say that the $xh$ and $H$ are \textbf{disjoint sets} and written as $H\cap (xh) = \emptyset $. 

We can similarly compose $x$ and $h$ as $hx = (ex, h_1x,h_2x,...,h_mx)$. The set $xh$ is called a \textbf{left coset} and $hx$ called a \textbf{right coset} of $H$ in $G$ with respect to $x$.
The subgroup $H$ of a group $G$ is called a \textbf{normal subgroup} or an \textbf{invariant subgroup} of $G$ if left and right cosets of a subgroup $G$ are the same. More specifically, every element of $xh$ is equal to some element of $hx$, such that
$$ xh_i = h_j x \implies x^{-1}h _j x = h_i,$$
From the above expression, we can see that $h_i$ and $h_j$ are in the same conjugacy class. \textit{$H$ is a normal subgroup if the subgroup $H$ consists of complete conjugacy classes of the group $G$}.
If $H$ is a normal subgroup of $G$, the set of all distinct cosets of $H$ in $G$ is called a \textbf{quotient group} or the \textbf{factor group} of $G$ with respect to $H$. It is denoted as 
$$K = G/H.$$
If the order of $G$ is $n$ and the order of $H$ is $m$ then the order of $K$ is $n/m$.
The quotient group is the group of equivalence classes of $G$ with respect to $H$. It is the set of cosets and this set is also a group. 

\subsection{Representation of a Group}
A group homomorphism from the group elements to the matrices which obey the same group composition rules as the group itself is called the representation of the group.
Let a group $G = \left\{e, a, b, c,... \right\}$ be a finite group, and let $D= \left\{D(e), D(a), D(b) , D(c),...\right\}$ be a collection of nonsingular matrices having the property
$$D(ab) = D(a)D(b),$$
then the collection of matrices is called representation of the group.
For example, if the group composition is $ab=c$, then $D(a)D(b) = D(c)$. As $ea=ae= a$, the representation implies that $D(e)D(b) =D(a)D(e) = D(a)$. The representation for the expression $a^{-1}a = e$ should be $D(a^{-1}a) =D(a^{-1})(a) = D(e)= e$, which implies $D(a^{-1})=[D(a)]^{-1}$
The dimension of the representation is the dimension of space on which it acts.

Two representations $D_1(a)$ and $D_2(a)$ are equivalent if there exist a non-singular matrix $S$ such that $D_1(a)$ and $D_2(a)$ are related as $$D_1(a)=S^{-1}D_2(a)S.$$ This kind of transformation is called a \textbf{similarity transformation}. 
If the action of a representation $D(a)$ on a vector in a subspace results in the elements of the same subspace, then the subspace is called an \textbf{invariant subspace}, otherwise, it is \textbf{irreducible}. A representation is reducible if it has an invariant subspace. A \textbf{completely reducible} representation can be written as a matrix in block diagonal form, which can also be written as a \textbf{direct sum} of the subrepresentation. If all the elements can be represented by distinct matrices, then the representation is called a \textbf{faithful representation}. 

\section{Ring, Field, Vector Space and Module}
A set $\mathcal{A}$ is called a ring if it fulfills the following axioms.
\begin{itemize}
	\item $\cal A$ is an additive Abelian group. 
	\item $\cal A$ is associative:
	$$a\cross(b\cross c) = (a\cross b)\cross c.$$
	\item Distributive law holds, That is, for $a,b,c \ \in \ \cal A,$
	$$a\cross(b+c) = a\cross b+a\cross c,$$ and  $$(b+c)\cross a = b\cross a +c \cross a.$$  
\end{itemize}
When a set is a group under both the operations of multiplication and addition and also holds the distributive law, then it is called a \textbf{field}. The set of real numbers $\mathbb{R}$ and the set of complex numbers $\mathbb{C}$ are examples of fields.

A \textbf{vector space V} over the field is a set on which two operations, addition and multiplication, are defined. The elements of a vector space are usually called \textit{vectors} and the field are called \textit{scalars}. Let $\textbf{u}$, $\textbf{v}$ and $\textbf{w}$ be vectors in a vector space $\textbf{V}$ and $a$ and $b$ be scalars in a field $K$, then a vector space would hold the following axioms 
\begin{enumerate}
	\item $\bold {u+v=v+u}$
	\item $\bold {(u+v)+w=u+(v+w)}$
	\item There exist the zero vector $\bold 0$, such that $\bold{u+0=u}$ 
	\item There exist the inverse vector $\bold{-u}$ such that $\bold{u+(-u)= 0}$
	\item $a\bold{(u+v)}=a\bold u+a \bold v$
	\item $(a+b)\bold u=a\bold u+b\bold u$
	\item $(ab)\bold u=a(b\bold u)$
	\item There exist a unit element such that $\bold {1u=u}$.
\end{enumerate}

The set of vectors $ \{\bold v_i \}$ are \textbf{linearly independent} if for some number $x_i$ it is true that
$x_1 v_1 + x_2 v_2 + ... + x_kv_k = 0$ only when $x_i = 0$,
otherwise, the set of vectors are linearly dependent. A vector space has the \textbf{basis} vectors in terms of which any other vector is written uniquely, e.g.
$$\bold{v} = v^1 \bold{e_1} + v^2 \bold{e_2} +... +    v^i \bold{e_i},$$
where $v^i \in K$ are the components of a vector $\bold{v}$ with respect to $\{\bold e_i\}$. The number of basis is the \textbf{dimension} of the vector space. A vector space over the ring is called a \textbf{module}.  A vector space equipped with the bilinear product is called an \textbf{algebra}. A bilinear product is the one by which we combine two vector spaces and get the third one when this product is linear. A \textbf{tensor product} $u\otimes v \in U\otimes V$, when $u\in U$ and $v\in V$, is a product when each element of the first vector space is multiplied with all the elements of the second vector space.
A mapping of a vector space to itself is called the \textbf{endomorphism}. If this endomorphism is reversible then it is called the \textbf{automorphism}.

There is a \textbf{dual} vector space that fulfills the axioms of the vector space, and when taking the inner product with the vectors in vector space we get numbers. This also means that the inner product of a basis vector with its dual gives identity. We can say that a dual vector space is a linear functional from a vector space to the real numbers. A vector space equipped with an inner product is called a \textbf{Hilbert space}.

\section{Lie Group and Lie Algebra}

The Lie group is a continuous group which is also a differentiable manifold. It has a smooth structure and so the differential calculus can be performed on it. The group multiplication and inverses can be taken on the manifold, therefore the Lie group provides a natural framework for a continuous symmetry in physics. 
Any closed subgroup of the $GL(n,\mathbb{C})$, $n\cross n$ matrices with entries in $\mathbb{C}$, is a Lie group. The special linear group $SL(n,\mathbb{R})$ or $SL(n,\mathbb{C})$ consists of $n\cross n$ matrices with the entries in $\mathbb{R}$ or $\mathbb{C}$ and all matrices having a determinant one. The unitary groups and special unitary groups, $U(n)$ and $SU(n)$, also have $n \cross n$ matrices with the property that $U^{-1} = U^*$ and determinant one in case of $SU(n)$.
The orthogonal groups $O(n)$ and the special orthogonal groups $SO(n)$, consist of $n\cross n$ matrices with $R^T = R^{-1}$ and $\det(R) = 1$ in case of $SO(n)$.

Let us suppose a group $G$ whose elements $g(\alpha)$ depend on some parameter $\alpha$.  Here we consider that this dependence is smooth and continuous. That is, if two group elements are close together, the corresponding values of the parameter are also close.
Let the identity be an element for $\alpha = 0$, that is, $g(\alpha)|_0 = e$ then we have $D(\alpha)|_0 = 1$.

As the $d\alpha$ is infinitesimally small, in the neighborhood of the identity, taking only the first term in the Taylor expansion of $D(\alpha)$ we have
\begin{align}
	D(d\alpha) = 1+ id\alpha_a X_a + ...,
\end{align}
where $X_a$ are called the generators of the group and written as
\begin{align}
	X_a \equiv -i \frac{\partial}{\partial \alpha_a}D(\alpha)|_{\alpha = 0},
\end{align}
where $a= 1,...,N$.
For the representation to be unitary, $X_a$ would be Hermitian operators.
Parameterizing the group by going away from the identity, we get another group element $D(d\alpha) = 1+ id\alpha_a X_a$. For a finite $\alpha$, we define the representation of the group element 
\begin{align}
	D(d\alpha) = \lim_{k \to \infty} (1+ i\alpha_a X_a/k)^k = e^{i \alpha_a X_a}.
\end{align}
This is called the \textit{exponential parameterization}. Thus, we can write the group elements in terms of the generators. These generators form a vector space. 
The one parameter group is of the form $U(\lambda) = e^{i\lambda \alpha_a X_a}$. The group multiplication is $U(\lambda_1)U(\lambda_2) = U(\lambda_1 + \lambda_2)$. As we know that $e^{i\alpha_a X_a}e^{i \beta_a X_a} \ne e^{i(\alpha_a + \beta_a)X_a}$, but for the representation of the group, the product of two exponentials should be another exponential as
\begin{align} \label{Generators1}
	e^{i\alpha_a X_a}e^{i \beta_a X_a} = e^{i\delta_a X_a}.
\end{align}
We will find out what makes this equation to be true by equating the powers on both sides of the equation. We will do this by taking logarithm on both sides. Therefore, we have
\begin{align} \label{Generators2}
	i\delta_a X_a = \ln{(e^{i\alpha_a X_a}e^{i \beta_a X_a})}.
\end{align}
Now expanding the exponentials in parentheses,
\begin{align} \label{GenrtrExpnd1}
	e^{i\alpha_a X_a}e^{i \beta_a X_a} &= [1+ i \alpha_a X_a - \frac{1}{2}(\alpha_a X_a)^2 + ...]\nonumber \\
	&[1+ i \beta_b X_b - \frac{1}{2}(\beta_b X_b)^2 + ...] \nonumber \\
	&= 1 + i \alpha_a X_a + i \beta_b X_b - \alpha_a X_a \beta_b X_b \nonumber \\
	&- \frac{1}{2}(\alpha_a X_a)^2 - \frac{1}{2}(\beta_a X_a)^2 + ... .
\end{align}
Rewriting the Eq. \ref{Generators2} in the form as
\begin{align} \label{Generators3}
	i\delta_a X_a = \ln{(1+ e^{i\alpha_a X_a}e^{i \beta_a X_a} - 1)},
\end{align}
where we added and subtracted 1. The advantage of writing like this is that we can expand this as a series of $\ln{(1+K})$, where $K$ is $e^{i\alpha_a X_a}e^{i \beta_a X_a} - 1$. Now using Eqs. \ref{GenrtrExpnd1} in \ref{Generators3} we get
\begin{align}\label{GenrtrExpnd2}
	i\delta_aX_a &= \ln{(1+K)} \nonumber\\
	&= K - \frac{1}{2}K^2 +... \nonumber \\
	& = i \alpha_a X_a + i \beta_b X_b - \alpha_a X_a \beta_b X_b - \frac{1}{2}(\alpha_a X_a)^2 \nonumber \\
	&- \frac{1}{2}(\beta_a X_a)^2 + \frac{1}{2}( \alpha_a X_a + \beta_b X_b) +...\nonumber\\
	&= i \alpha_a X_a + i \beta_b X_b - \frac{1}{2}[\alpha_a X_a , \beta_b X_b] + ... .
\end{align}
\begin{align}
	\Rightarrow [\alpha_a X_a , \beta_b X_b]= -2i(\delta_c - \alpha_c - \beta_c)X_c + ...  \equiv i\gamma_c X_c.
\end{align}
The Lie algebra is obtained by taking an infinitesimal transformation of the Lie group. Lie groups are manifolds that have tangent spaces at the identity elements of the group. The elements of the tangent spaces give the Lie algebra. 

\textbf{Definition:} A Lie algebra $\mathfrak{g}$ consists of a vector space $\mathfrak{g}$ equipped with a product $[.,.]:\mathfrak{g} \otimes \mathfrak{g} \to \mathfrak{g}$, that satisfies the following 
\begin{itemize}
	\item The anti-symmetry relation
	\eq{[X,Y] = -[Y,X]}
	\item Jacobi identity 
	\eq{&[X,[Y,Z]]+ [Y,[Z,X]] + [Z,[X,Y]] = 0, \\ &\text{for all }X,Y \in \mathfrak{g}}.
\end{itemize}
The product in a Lie algebra is denoted by square brackets and is called the \textbf{Lie bracket}. The matrices are not algebra under multiplication. Lie algebra is an algebra made by Lie bracket. 
$\mathfrak{sl}_n(\mathbb{K})$ is the Lie algebra consisting of a vector space of $n\cross n$ matrices over $\mathbb{K}$ that have trace zero and Lie bracket given by $[A,B]= A.B - B.A$. $\mathfrak{sl}_2$ is a vector space over $\mathbb{C}$ with the $X,Y,H$ with the Lie bracket given as
\eq{[H,X] = 2X, \ [H,Y] = -2Y,\ [X,Y] = H}

Conventionally, capital letters are written for the names of the Lie groups, and small curvy letters are used for the Lie algebras. The Lie algebra $\mathfrak{sl}(2)$ over a complex field $\mathbb{C}$ consists of $2\cross2$ matrices.
Bases of this algebra are chosen as 
\eq{X = \begin{bmatrix}
		0 & 1 \\
		0 & 0
	\end{bmatrix}, \ Y= \begin{bmatrix}
		0 & 0 \\
		1 & 0
	\end{bmatrix}, \ H = \begin{bmatrix}
		1 & 0 \\
		0 & -1
	\end{bmatrix}, \ I= \begin{bmatrix}
		1 & 0 \\
		0 & 1
\end{bmatrix} }
We can see that 
\eq{&[X,Y] = H, \ [Y,X] = -H, \\
	&[X,H] = -2X, \ [H,X] = 2X, \\
	&[Y,H] = 2Y, \ [H,Y] = -2Y, \\
	&[X,I]= [I,X]= [Y,I]= [I,Y] = [I,H]= [H,I]=0.}
$X,Y,H$ have the property that their trace is zero. The matrices $\left\{X,Y,H\right\}$ make the bases of subspace $\mathfrak{sl}(2)$ of $\mathfrak{gl}(2)$.

The Lie bracket is not associative. Universal enveloping algebras $U(\mathfrak{sl}_2)$ are introduced to recover the associativity. $U(\mathfrak{sl}_2)$ is generated by the same elements as $\mathfrak{sl}_2$, but the difference is that the elements like $H^2X$ are also included in this algebra. The quantum group is the q-analog of unital associative universal enveloping algebra. We used the universal enveloping algebra in Appendix \ref{HopfAlg} for computing the knot invariant.
	\chapter{Topology and Differential Geometry}\label{Top}

\section{Topology}
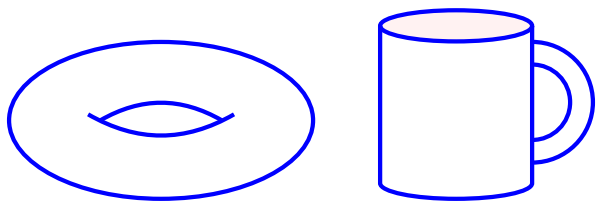
\begin{figure}[h!]
	\centering
	\begin{tikzpicture}[scale=0.8]
		\draw[bend left,ultra thick,blue] (-1,-0) to (1,-0);
		\draw[bend right,ultra thick,blue] (-1.2,0.1) to (1.2,0.1);
		\draw[rotate=0,ultra thick,blue] (0,0) ellipse (2.5 and 1.3);
	\end{tikzpicture}\qquad
	\begin{tikzpicture}
		\draw[rotate=0,ultra thick,blue] (1,0.2) ellipse (0.8 and 0.8);
		\draw[rotate=0,ultra thick,blue] (1,0.2) ellipse (0.5 and 0.5);
		\node [cylinder,draw=black,ultra thick,blue,aspect=1.5,minimum height=2.5cm,minimum width=2cm,shape border rotate=90,cylinder uses custom fill,cylinder end fill=red!5] at (0,0){};
	\end{tikzpicture}
	\caption{A doughnut is homeomorphic to a coffee cup.}
	\label{coffeecup}
\end{figure}

In Euclidean geometry, we take the distances between the points and the angles between lines for describing and comparing spaces. But in topology, the distances and angles have no significance, but some other properties that remain invariant under continuous deformation of the spaces.
Think of the spaces as made up of some rubber material that can be smoothly transformed from one shape to the other. Squeezing, pulling, or twisting would not change topology unless we tear it apart. The two shapes in Fig.  \ref{coffeecup} are considered the same in topology because one can be continuously transformed to the other. Tearing the hole apart or making a new hole would change the topology. A circle has the same topology as any closed curve. More examples are given in Chapter \ref{Knot}.

\section{Topological Spaces}
An \textbf{open interval} in ${\mathbb R}$ is written as $(a,b)$ which means that the $a$ and $b$ are not included, but all real numbers between them are included in the interval. A \textbf{closed interval} is written as $[a,b]$, means that $a$, $b$ and all numbers between them are included. In a Euclidean space $\mathbb R^n$, all points $p \ \in \ \mathbb R^n$ are included, except the boundary, which is called an open ball. The union of open sets is open, and the intersection of open sets is open. The empty set $\emptyset$ and universal set $X= \mathbb R^n$ are open and closed at the same time. 

Let $X$ be a set and $T$ is a collection of subsets of $X$ that satisfies the following requirements: 
\begin{itemize}
	\item The $\emptyset$ and $X$ are in $T$,
	\item The union of sets of $T$ are in $T$,
	\item The intersection of sets from $T$ are in $T$,
\end{itemize}
then $T$ is called a \textbf{topology} on the set $X$. The ordered pair $(X,T)$ is called a \textbf{topological space} and the elements (sets) of $T$ are open sets. The elements of a subset $B$ of $T$ that is, $B \subset T$, are called the basis of the topological space if all open subsets of $T$ can be expressed as unions of the elements of the $B$. 

A map from $X$ to $Y$ is \textbf{continuous} if the inverse image of the $Y$ in $X$ is open. 
Consider a point $x \ \in \ X$. An open set $N$ is a \textbf{neighborhood} of a point $x$ if $N$ is subset of $X$ and $x$ belongs to an open subset of $N$.
Let a topological space $X$ be a family of open subsets $\mathcal{F}$ whose union is $X$ then the $\mathcal{F}$ is called an \textbf{open cover} of $X$. $\mathcal{F}'$ is a \textbf{subcover} of $\mathcal{F}$ if $\cup \mathcal{F}' = X$. A subspace is \textbf{bounded} if the open cover has a finite number of subcovers. A \textbf{compact} topological space is closed and bounded. The corresponding surface is not extended to infinity, and also the boundary points are all included. 
$x$ is an \textbf{interior} point of the $X$ if there is $\epsilon>0$ such that the ball $B(x)$ around $x\in X$ has the property that $B(x) \subset X$. $y$ is an \textbf{exterior} point of $X$ if there is $\epsilon>0$ such that the ball $B(y)$ has the property that $B(y) \cap X = \emptyset$. $z$ is the \textbf{boundary} point of $X$ if there is $\epsilon>0$ such that the ball $B(x)$ intersects both $X$ and $X^c$. The \textbf{connectedness} is a property of a space when it is all in one piece. If we can connect all the piece by a path, the space is called \textbf{path connected}.

A \textbf{metric} on a set $X$ is a map $d:X\cross X \rightarrow \mathbb{R}$ tha satisfies the three conditions:
\begin{itemize}
	\item $d(x,y) = d(y,x)$,
	\item $d(x,y) \ge 0$ zero for $x=y$,
	\item $d(x,y) = d(x,z) + d(z,y)$,
\end{itemize}
where $x,y,z \in X$. The $X$ or sometime $(x,d)$ is called a \textbf{metric space}. As an example, let be an open set $U$ in $\mathbb R^n$ and $x \ \in \ U$ and all the points sufficiently close to $x$ also be in $U$. 
Let us define a continuous function on a topological space. A function is continuous if it sends nearby points to nearby points. More specifically, if a function (also called a map) $f: X_1 \rightarrow X_2$, and $(X_1, T_1)$ and $(X_2, T_2)$ are two topological spaces, and if any open subset $O_2 \subset X_2$ the inverse image $f^{-1}(O_2) \subset X_1$ is an open subset of $X_1$, then $f$ is said to be \textbf{continuous}. 
A function which is a bijection, that is, a one-to-one correspondence between $X_1$ and $X_2$ then the function is called the \textbf{homeomorphism}. The \textbf{homeomorphism} is a map $f: X\rightarrow Y$ which is onto, one-to-one and has a continuous inverse. Homeomorphic spaces are in the same equivalence class. 
If $f$ is a homeomorphism then $f^{-1}$ is too.

\section{Topological Invariants}
The distances and angles are irrelevant in topology.
How to characterize the equivalence classes? The idea is to find the properties that do not change under homeomorphism and uniquely specify the equivalence classes. These properties are called topological invariants. Different invariants have their advantages and limitations. The invariants can be numbers, or certain properties of the topological spaces like connectedness, compactness, homotopy group, homology group, or cohomology group. Fundamental group and knot invariants are already discussed in Chapter \ref{Knot}.

Homeomorphism is the study of shapes that can be continuously transformed from one to another. Topology is changed when there is discontinuity, like tearing apart or welding. The continuous deformation should be done in such a way that the dimension of the diagram should not be changed. A circle is a one-dimensional object $S^1$ called 1-sphere, a boundary of a ball or a hollow cube is a 2-sphere $S^2$, and a torus is $T^2 = S^1\cross S^1$.

The \textbf{genus} is the handle in a space. In the famous coffee cup and doughnut example, both have one handle, so they are homeomorphic, but if we tear apart the handle then the two spaces are topologically not the same anymore, see \ref{coffeecup}. So a genus is also a topological invariant but not strong enough. The \textbf{Euler number} is another topological invariant calculated using the formula
\eq{V-E+F = 2-2g,}
where $g$ is the genus, $V$, $E$, and $F$ are the number of vertices, edges, and faces of a topological space. These are obtained by the triangulation of the space. The idea is to imagine the space as consists of simplicial complexes. A point is a 0-simplex, a closed interval is a 1-simplex, a triangle is a 2-simplex, and a tetrahedron is a 3-simplex. A topological space can be obtained by gluing several simplices along with their faces. See \cite{armstrong2013basic} for the discussion on triangulation.  

A very important theorem, known as the Gauss-Bonnet theorem, relates the geometry and topology
\eq{ \int_M K dA = 2\pi \chi,}
where $\chi = 2(1-g)$ is the Euler characteristic, and $g$ is the genus, $A$ is area of a region in $M$. $K$ is the Gaussian curvature given as $K=k_1k_2$. Where $k_1$ is a curvature when going in one direction and $k_2$ is for the other direction. A flat surface has Gaussian curvature zero. A saddle has Gaussian curvature negative, as one of the $k_1$ or $k_2$ is negative. The sphere has positive Gaussian curvature. A torus has a negative curvature on some points and a positive on some other points. 

\section{Manifolds} 
A manifold is a topological space that is Euclidean space locally. The neighborhood of each point of the manifold is homeomorphic to the Euclidean space, but globally it can have the structure that is not homeomorphic to the Euclidean space $\mathbb{Z}^n$. It is obtained by patching together the open sets of $\mathbb R^n$. Each patch looks like $\mathbb R^n$. The manifold is locally homeomorphic to open subsets of $\mathbb R^n$. If we take a small area around a point on some surface or a shape, this area is tangent to the surface. For example, a sphere $x^2 + y^2 +z^2 = r$ looks like a flat surface or a plane in $\mathbb R^2$ if we consider only a small patch of it. The sphere is a 2-dimensional manifold and is called \textbf{2-sphere}.
Based on the classification of manifold, different types of topology are studied, such as point-set topology, algebraic topology, differential topology, geometric topology, combinatorics topology, general topology.

A \textbf{differentiable or differential manifold} is a manifold that locally looks like a linear space, and hence calculus can be done on it. A \textbf{diffeomorphism} is a smooth invertible map from one differentiable manifold to the other. We can divide a manifold into patches. A transition map is a coordinate transformation between these patches. The collection of all the patches is called the \textbf{atlas}. We can define a tangent vector on every point of the differentiable manifold. The collection of all the tangent vectors is called the \textbf{tangent bundle}. Similarly, we can have a cotangent bundle that consists of the collection of all the dual vectors on the differential manifold.

\begin{figure}[h!]
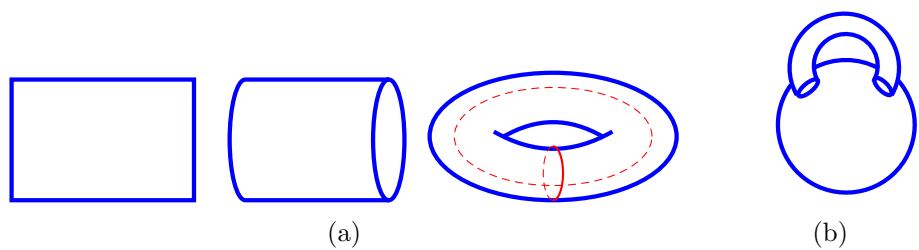

	\centering
	\begin{subfigure}[t]{0.6\textwidth}
		\centering
		\tik{[scale=0.8]\draw [ultra thick,blue] (0,0.1) rectangle (3,2.1);} \ \
		\tik{\node [cylinder,draw=black,ultra thick,blue,aspect=1.5,minimum height=2.3cm,minimum width=1.6cm,shape border rotate=0,cylinder uses custom fill] at (0,0){};} \
		\tik{[scale=0.65]\draw[bend left,ultra thick,blue] (-1,-0) to (1,-0);
			\draw[bend right,ultra thick,blue] (-1.2,0.1) to (1.2,0.1);
			\draw[rotate=0,ultra thick,blue] (0,0) ellipse (2.5 and 1.3);
			\draw[densely dashed,red] (0,0) ellipse (2 and 1);
			\draw[densely dashed,red] (0,-1.3) arc (270:90:.2 and 0.550);
			\draw[thick,red] (0,-1.3) arc (-90:90:.2 and .550);}
		\caption{}
	\end{subfigure}\qquad
	\begin{subfigure}[t]{0.1\textwidth}
		\centering
		\tik{[scale=0.6]\draw[ultra thick,blue] (0,0) arc (-30:210:1.2);
			\draw[ultra thick,blue] (-0.4,0.3) arc (-30:210:0.7);
			\draw[rotate=40,ultra thick,blue] (-1.3,1.3)  ellipse (0.27 and 0.1);
			\draw[rotate=-45,ultra thick,blue] (-0.27,0)  ellipse (0.25 and 0.1);
			\draw[ultra thick,blue] (0.12,0.35) arc (42:-216:1.5);
			\draw[ultra thick,blue] (-0.3,0.6) arc (62:118:1.5);}
		\caption{}
	\end{subfigure}
	\caption[Gluing and surgery of topological spaces.]{(a) A rectangular sheet wrapped to get a cylinder which further can be bent and make torus. (b) An example of surgery; a torus is the result of gluing a cylinder to a sphere.}
	\label{TopCylinder}
\end{figure}

A \textbf{product space} is the Cartesian product $X\cross Y$ of topological spaces. In Fig.  \ref{TopCylinder} (a), a rectangular sheet is a product space of two intervals or lines, the cylinder is the product space of a circle and an interval, and the torus has the product topology of two circles. A topological space that locally has the product topology, but globally it may have a different structure, is called a \textbf{fiber bundle}. A cylindrical hoop and the M\"obius strip look the same locally but are different globally. Fiber bundles are very important concepts in topology and quantum physics, we will discuss them in the next section. 

Topology can be understood through the study of surfaces. Surfaces can be identified and glued together to get diverse topological shapes. For example, we can think of a cylinder made up of a sheet wrapped in a way that opposite sides are identified.
The cylinder can further be identified on two ends so that we get a torus. The torus has one hole in it, so it is a genus-one surface, as in Fig.  \ref{TopCylinder} (a).

\begin{figure}[h!]
	\centering
	\begin{tikzpicture}[scale=0.8]
		\begin{axis}[
			hide axis,
			view = {40}{40}
			]
			\addplot3 [color=blue,thick,    
			domain     = 360:720,samples y=0,
			] (
			{(1+0.5*0.5*cos(x/2)))*cos(x)},
			{(1+0.5*0.5*cos(x/2)))*sin(x)},
			{0.5*0.5*sin(x/2)}
			);
			\addplot3 [
			surf,
			colormap={blackwhite}{gray(0cm)=(1); gray(1cm)=(0.7)},
			shader     = faceted interp,opacity = 0.7,
			point meta = x,
			samples    = 40,
			samples y  = 4,
			z buffer   = sort,
			domain     = 0:360,
			y domain   =-0.5:0.5
			] (
			{(1+0.5*y*cos(x/2)))*cos(x)},
			{(1+0.5*y*cos(x/2)))*sin(x)},
			{0.5*y*sin(x/2)}
			);
			\addplot3 [color=blue,ultra thick,    
			domain     = -140:497.5,samples y=0,samples=(640/360)*24+1,
			] (
			{(1+0.5*0.5*cos(x/2)))*cos(x)},
			{(1+0.5*0.5*cos(x/2)))*sin(x)},
			{0.5*0.5*sin(x/2)}
			);
		\end{axis}
	\end{tikzpicture}
	\caption{M\"obius strip is obtained by wrapping sheet of paper after giving it one twist.}
	\label{MobiusSheet}
\end{figure}
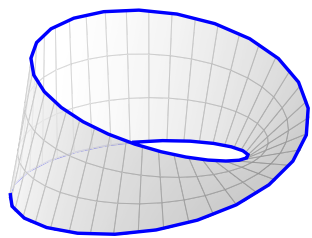

If we make one twist of the sheet before identifying, then we get a M\"obius band shown in Fig.  \ref{MobiusSheet}. This is topologically not the same as a cylindrical shape because of the \textbf{orientability}. Let us take a vector on the cylinder pointing in some specific direction. If we move it around in a full cycle, then the arrow would point in the same direction after the full cycle. But on M\"obius strip, we get a direction opposite to that of the original one. Furthermore, there are two circles on each end of the cylinder, but the whole M\"obius strip is only one circle, and there is no inner or outer side on the M\"obius trip. A cylinder is orientable but a M\"obious band is not orientable. The orientability is also a topological invariant, but it has limitations.
Topological surfaces can also be understood through gluing the smaller surfaces. This is called the \textbf{surgery}. For example, a torus shape can be obtained by gluing a cylinder to the sphere, as in Fig.  \ref{TopCylinder} (b). 

\section{Fiber Bundle}
A fiber bundle is a topological space $T$ which is a product space locally, but not necessarily be a product space globally. It is a way to take a product of spaces. A fiber bundle which is a product space, both locally and globally, is called the \textbf{trivial bundle}. The fiber bundle is a general term, some specific sorts of bundles are vector bundles, tangent bundles, principal bundles.

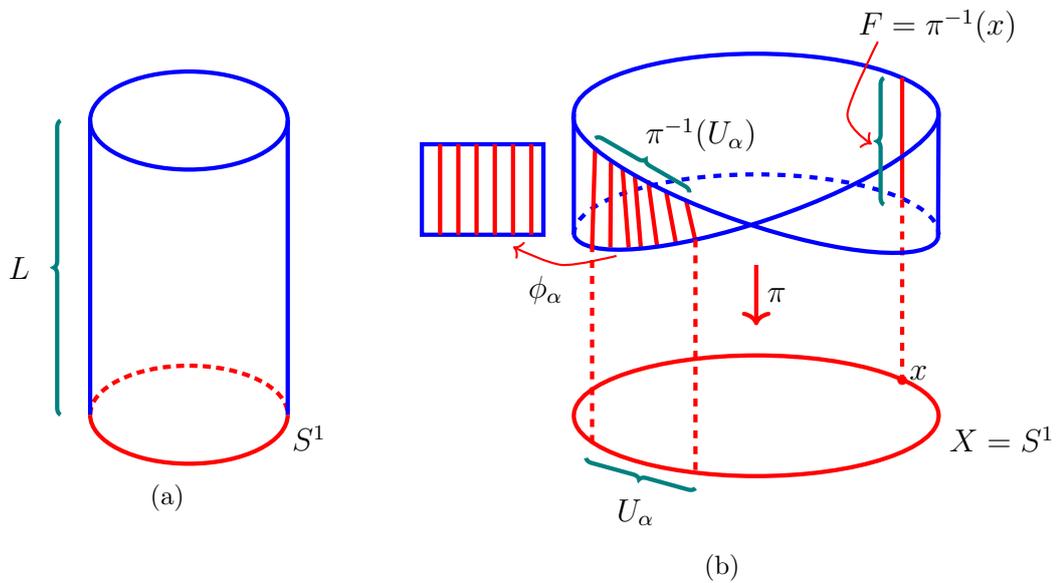
\begin{figure}[h!]
	\centering
	\begin{subfigure}{0.4\textwidth}
		\centering
		\begin{tikzpicture}[scale=1.3]
			\draw[ultra thick,blue] (0,1.5) ellipse (1 and 0.5);
			\draw[ultra thick,densely dashed,red] (-1,-1.5) arc (180:0:1 and 0.5);
			\draw[ultra thick,red] (-1,-1.5) arc (-180:0:1 and 0.5);
			\draw[ultra thick,blue] (-1,-1.5)--(-1,1.5);
			\draw[ultra thick,blue] (1,-1.5)--(1,1.5);
			\draw [decorate,decoration={brace},ultra thick,teal] (-1.3,-1.5) -- (-1.3,1.5);
			\node[left] at (-1.5,0) {$L$};
			\node[below left] at (1.5,-1.5) {$S^1$};
		\end{tikzpicture}
		\caption{}
	\end{subfigure}
	\begin{subfigure}{0.5\textwidth}
		\centering
		\begin{tikzpicture}[scale=0.8]
			\draw [ultra thick,blue] (0,2) arc(180:0:3 and 1);
			\draw [ultra thick,dashed,blue] (0,0) arc(180:0:3 and 1);
			\draw[ultra thick,blue] (0,0)--(0,2);
			\draw[ultra thick,blue] (5.99,-0.06)--(6,2);
			\draw[ultra thick,red] (5.4,0.6)--(5.4,2.6);
			\draw[ultra thick,dashed,red] (5.4,-2.45)--(5.4,0.6);
			\filldraw[red] (5.4,-2.41) circle (2pt);
			\draw[ultra thick,red,->] (3,-0.5)--(3,-1.5);
			\node[right] at (3,-1) {$\pi$};
			
			\draw[ultra thick,red] (0.3,-0.2)--(0.35,1.45);
			\draw[ultra thick,red] (0.6,-0.21)--(0.61,1.2);
			\draw[ultra thick,red] (0.9,-0.21)--(0.8,1.1);
			\draw[ultra thick,red] (1.1,-0.2)--(1,1);
			\draw[ultra thick,red] (1.37,-0.2)--(1.22,0.9);
			\draw[ultra thick,red] (1.68,-0.15)--(1.52,0.7);
			\draw[ultra thick,red] (2,-0.1)--(1.85,0.55);
			\draw [ultra thick,blue,rotate=19] (0,0) arc(-180:0:3.165 and 0.8);
			\draw [ultra thick,blue,rotate=-19] (-0.65,1.9) arc(-180:0:3.165 and 0.8);
			\draw[ultra thick,red] (3,-3) ellipse (3 and 1);
			
			\draw[ultra thick,blue] (-2.5,0) rectangle (-0.5,1.5);
			\draw[ultra thick,red] (-2.2,0)--(-2.2,1.5);
			\draw[ultra thick,red] (-1.9,0)--(-1.9,1.5);
			\draw[ultra thick,red] (-1.6,0)--(-1.6,1.5);
			\draw[ultra thick,red] (-1.3,0)--(-1.3,1.5);
			\draw[ultra thick,red] (-1,0)--(-1,1.5);
			\draw[ultra thick,red] (-0.7,0)--(-0.7,1.5);
			
			\draw[ultra thick,dashed,red] (0.3,-3.45)--(0.3,-0.1);
			\draw[ultra thick,dashed,red] (2,-3.9)--(2,-0.05);
			\node[above right] at (1,1.2) {$\pi^{-1}(U_\alpha)$};
			\node[above right] at (4.5,3) {$F=\pi^{-1}(x)$};
			\node[below left] at (0,-0.5) {$\phi_\alpha$};
			\node[below left] at (6,-2) {$x$};
			\node[below right] at (6,-3) {$X=S^1$};
			\node[below left] at (1.5,-4.2) {$U_\alpha$};
			
			\draw [decorate,decoration={brace},ultra thick,teal] (0.35,1.6) -- (1.9,0.7);
			\draw [decorate,decoration={brace},ultra thick,teal] (5.1,0.5) -- (5.1,2.6);
			\draw [decorate,decoration={brace,mirror},ultra thick,teal] (0.2,-3.7) -- (2,-4.2);
			\draw [red,thick,<-] plot [smooth] coordinates {(4.9,1.5) (4.5,2) (5,3.2)};
			\draw [red,thick,->] plot [smooth] coordinates {(0.7,-0.35) (-0.2,-0.5) (-1,-0.2)};
		\end{tikzpicture}
		\caption{}
	\end{subfigure}
	\caption{A cylinder is a trivial fiber bundle but M\"obious strip is not.}
	\label{FiberMobious}
\end{figure}

The following data are included in a fiber bundle:
\begin{enumerate}[i]
	\item A topological space $E$ called the \textit{total space}.
	\item A topological space $X$ called the \textit{base space}, and \textit{projection} $\pi: E \rightarrow X$ of $E$ on $X$.
	\item A topological space $F$ called the \textit{fiber} 
	\item A \textit{group} $G$ of homomorphism of fiber $F$
	\item A set of open coordinate neighbourhood $U_\alpha$ covering of $X$ which reflect the local triviality of the bundle $E$. 
	\eq{\phi_\alpha : \pi^{-1}(U_\alpha)\rightarrow U_\alpha \cross F,}
	where $\phi_\alpha ^{-1}$ is such that $\pi\phi_\alpha ^{-1}(x,f)=x$, with $x$ $\in$ $U_\alpha$, $f$ $\in$ $F$.
\end{enumerate}
A cylinder is a trivial fiber bundle, it is a product space both locally and globally. It is written as $T=L\cross S^1$, where $L$ is a line segment and $S^1$ is a circle, as shown in Fig.  \ref{FiberMobious} (a). But a M\"obious strip is a product space or a trivial bundle only locally and looks like a cylinder if we see the small rectangular segment of a M{\"o}bius strip shown in Fig.  \ref{FiberMobious} (b), but globally the two spaces are different. 

Let the total space $E$ be a M{\"o}bious strip whose base space $X$ is a circle $S^1$, the fiber bundle $F$ on the base space are line segments. Fiber on a point $x$ of $S^1$ is $\pi^{-1}(x)$ as shown in Fig.  \ref{FiberMobious} by red lines on the total space.
The action of the projection $\pi$ is shown by dotted lines. Let a patch be the open set $U_\alpha$ of $X$ and its projection $\pi^{-1}(U_\alpha)$, then homeomorphism $\phi_\alpha$ untwist the $\pi^{-1}(U_\alpha)$ into the product $U_\alpha \cross F$. This is called the \textbf{trivialization}.
This map is the \textbf{transition function} and is a homeomorphism of $F$. The set of all these homeomorphisms for the choices of $U_\alpha, \phi_\phi$ form a group which is called \textbf{structure group} of the fiber on the fiber bundle $E$. If we move an arrow on the two spaces, it will get rotated on M\"obious strip but will stay in the same direction on the cylinder. The trivialization would let us see the amount of rotation. 

A \textbf{section} is a map of a segment of base space into the total space. The section is written as $\sigma_\alpha : U_\alpha \rightarrow \pi^{-1}(U_\alpha)$. A fiber bundle is trivial if a single section can be obtained for the whole fiber bundle. The section introduces a local coordinate system for the portion of the total space above the coordinate patch $U_\alpha$ by designating a definite element in each fiber as the identity element. 

Let, we are studying the tangent spaces at different points on a manifold. The tangent space on the total space can have a horizontal and vertical component. The structure group acts only on the vertical component. This is the difference between the ordinary Cartesian product space and the fiber. If we compare the tangent space at the total space with that on the base space, the rule of moving up or down on a vertical subspace is provided by the connection. Formally, the \textbf{connection} maps the tangent spaces $T_x(X)$ over the base space $X$ to the tangent spaces $T_\theta(E)$ in the total space $E$ over a point $\theta$ in the fiber $\psi(x)$ above $x$. Thus, the connection is a rule that splits the tangent space at each point $\theta$ into vertical and horizontal subspaces \cite{auyang1995quantum}. 

\section{Parallel Transport, Connection, and Curvature}
Let us differentiate a vector field on a manifold. If we are working in the Cartesian coordinates, then we do not need to take the derivative of the unit vectors because they will remain the same in magnitude and direction. 
However, if our problem involves the curvilinear (non-orthonormal) coordinates, we need to compensate for the change of the bases vectors from point to point. Therefore, things get a bit complicated when working on curved spaces, non-Euclidean or Riemannian geometry.

To take the derivative on curved spaces, we need to compare a vector at one point with the one at a nearby point. But the tangent spaces at two different points are not related due to length and direction change at different locations. To take a derivative in this case, we will first discuss a concept called the \textbf{parallel transport} of a vector. The parallel transport is moving a vector on a manifold in such a way that the direction of the vector stays constant. It is for the observer who is moving the vector. In a flat space, the parallel transport will not change the direction of the vector at the final point, comparing with the one at the starting point. The shortest path on a curved space is called the \textbf{geodesic}, which is a straight line in a flat space.

The covariant derivative will help us to differentiate the vector field on a curved space. Let the derivative of a vector in a curved space be written as
\eq{\partial_\mu \bm{A} = \partial_\mu (A^\mu \bm{e_\mu}) = [\partial_\mu A^\mu + A^\mu \partial_\mu] \bm{e_\mu}.}
The summation convention is assumed. The second term is zero in a flat space. The quantity in the bracket is the \textbf{covariant derivative}.
It is a compensation for the change in basis vector along with the motion. 
The covariant derivative is zero for a parallel transported vector, similar as the ordinary derivative is zero for a straight line parallel to the horizontal axis. 

When we parallel transport a tangent vector along the geodesic, the twist of the tangent space around the geodesics is called the \textbf{torsion}. The change of direction of a tangent vector to left-right or up-down is due to \textbf{curvature}. The curvature is how much a line deviated from the straight line and a plane deviated from the flat surface. Therefore, the curvature is the non-commutativity of the covariant derivative.  
\eq{F(v,w) = D_vD_w - D_wD_v,}
where $D_{\textbf{w}}\textbf{v}$ is the covariant derivative of \textbf{v} in the direction \textbf{w}. 
The curvature is provided by the \textit{connection}. The connection is explained in the section on fiber bundle. The connection is \textbf{flat connection} for vanishing curvature.
The \textbf{holonomy} is the twist of a vector moving in a loop on a curved space. In general relativity, the second term involves the Christoffel symbols. Its failure to be zero is the Riemann curvature tensor. Ricci tensor is the contribution from the curved space to an area or a volume from point to point on curved spacetime. This contribution is added to the flat derivative. 

Now on the fiber bundle, the total space may not be flat, so at a point on the total space above the base point, the tangent space may not be parallel to the tangent space at the base space. 
In $U(1)$ gauge theory, the second term is due to the gauge transformation at every point. To compensate for this contribution, we need to change the connection. In this context, the connection is the vector potential, the curvature is the field strengths, and holonomy is the geometric phase. The Berry phase, Berry connection, and Berry curvature are calculated on the same idea in momentum space on the Brillouin zone. 

\section{Tensors}
A \textbf{tensor} is an algebraic object used to describe the physical quantities. The tensor can be thought of as a multidimensional array of numbers.
A scalar is a zero-dimensional array. It has only one component, the magnitude. A scalar is called a rank-0 tensor. Its components can be written as $3^0 = 1$. 
The rank-1 tensor is a vector. It has a one-dimensional array.
A vector has three components, magnitude and direction in each dimension. Its components are given as $3^1 = 3$.
The rank-2 tensor is a two-dimensional array, looks like matrices. The rank-2 tensor has $3\cross 3 = 3^2 =9$ components. The rank-3 tensor is a three-dimensional array, looks like a cube, and has components $3^3 = 27$.  
In relativity, we use space and time on equal footings. Therefore, instead of 3, we need 4-component tensors corresponds to $t,x,y,z$, one-time, and three space coordinates.
\textit{Tensors are the generalization of the vectors for which every possible combination of the product of the basis vectors are taken as its components.}
The vectors are invariant, but the components are not invariant. If a tensor (a vector, for example) represents a length in the $n$-dimensional space, then the length remains the same if we change our coordinate system.
\textit{A tensor is an object that is invariant under the change of coordinates and has components that change especially and predictably.}

Let us multiply two vectors $U$ and $V$ terms by term that can be written as $UV$,
\eq{UV = \mu_{11} ii + \mu_{12}ij + \mu_{13}ik + \mu_{21}ji +....}
It is not a dot product or a cross product. 
If the vectors $U$ and $V$ have three components each, then $UV$ is $3\cross 3$ array given by
$$\begin{matrix}
	\mu_{11} & \mu_{12} & \mu_{13}\\
	\mu_{21} & \mu_{22} & \mu_{23}\\
	\mu_{31} & \mu_{32} & \mu_{33}
\end{matrix}$$
\textbf{Einstein's summation convention} is used in above equations. Let us have the sum as 
$\sum_{i}^{3} A_iB^i = A_1B^1+A_2B^2+A_3B^3$.
In the Einstein's summation convention, this sum is written as $A_iB^i$. That is, when an index is repeated, then the summation is assumed on that index. A vector would be written as $V= V^ae_a$. So we can write the general tensor  
\eq{T = T^{lmn...}_{abc...}w^aw^bw^c...e_le_me_n....}
Taking the example of a vector in two dimensions,
\eq{ds^2 &= d\textbf{r}\cdot d\textbf{r} = \frac{\partial r}{\partial y^i} dy^i + \frac{\partial r}{\partial y^j}dy^j \nonumber\\
	&=e_i\cdot e_j dy^idy^j \nonumber\\
	&= g_{ij} dy^idy^j.}
The $g_{ij}$ is called a \textbf{metric tensor}.

We can represent a vector on a coordinate basis. When we use the longer unit vectors then the components decrease in lengths and when we use the shorter unit vectors then the components increase accordingly. Such components which transform oppositely to the unit vectors are called the \textbf{contravariant} components of the vectors. We can define the components in another way. Take the projections of the vector on a coordinate basis. These projections are the components described by the length of its dot products with each of the basis vectors. In this case, if we increase the length of the basis vector, the dot products also increase and vice versa. These components are \textbf{covariant} components of a vector. The covariant vectors are sometimes called the covectors. 
The covariant components are written with subscript, whereas the superscripts are written with the contravariant components. 
A general tensor is the collection of vectors and covectors combined using the tensor product.

An example of the contravariant basis and covaraint basis is as below
\eq{e^1 = \partial r/\partial u, \ e^2 = \partial r/\partial v, \ e^3 = \partial r/\partial w \qquad e_1 = \nabla u, \ e_2 = \nabla v, \ e_3 = \nabla w.}
Now, let us consider the transformation between the basis $(x^1,x^2,x^3) \rightarrow (y^1,y^2,y^3)$. 
The contravariant tensor will transform as 
\eq{T^i(y^1,...,y^N) = \frac{\partial y^i}{\partial x^j}T^j(x^i,...,x^N),}
whereas the covariant transforms according to rules 
\eq{T_i(y^1,...,y^N) = \frac{\partial x^j}{\partial y^i}T_j(x^i,...,x^N).}
The contravariant and contravariant transformation properties are related to the \textit{push forward and pull back maps} in differential geometry.

An example of rank-2 tensor is \textit{energy-momentum tensor} that can be written in the form of $4\cross 4$ matrix as
\eq{T^{\mu\nu} = \begin{pmatrix}
		T^{00}&T^{01}&T^{02}&T^{03}\\
		T^{10}&T^{11}&T^{12}&T^{13}\\
		T^{20}&T^{21}&T^{22}&T^{23}\\
		T^{30}&T^{31}&T^{32}&T^{33}
	\end{pmatrix},}
where zero index is for the time component of a particle's momentum and the indices $1,2,3$ are for three spatial components. It can be explained as follows. $T^{\mu\nu}$ is a momentum component $\mu$ flows in the direction of the component $\nu$. The time component of the momentum is energy, so the first row is the energy of the particle. So $T^{00}$ is energy density, and $T^0k$ with $k=1,2,3$ are the energy flux. The first column $T^{k0}$ with $k=1.2,3$, excluding $T^{00}$, is the momentum at constant time so it is the momentum density. The remaining 9 components are the spatial momentum in the spatial direction, that is momentum flux. Three diagonal components are called pressure and other off-diagonal components are called shear stress.

The electromagnetic field tensor can be written as
\eq{F_{\mu\nu} = \begin{pmatrix}
		0&E_x/c&E_y/c&E_z/c\\
		-E_x/c&0&-B_z&B_y\\
		-E_y/c&B_z&0&-B_x\\
		-E_z/c&-B_y&B_x&0
	\end{pmatrix}.}
Since electric and magnetic fields can be written in terms of the four-components vector potential, the EM tensor can be calculated using the relation 
\eq{F_{\mu\nu} = \partial_\mu A_\nu -\partial_\nu A_\mu}
where $A$ is a spacetime vector potential with $A_0$ as a scalar potential and $A$ with three spatial components is a usual vector potential.

	\chapter{Link Invariants from TQFT and Quantum Group}\label{HopfAlg}
	\section{Link Invariant from TQFT}
Let a 3-manifold $M$ be a connected sum of two submanifolds $M_1$ and $M_2$. They are connected along the boundary $S^2$ as shown in Fig. \ref{ConnectedSum} (a) \cite{witten1989quantum}.
When there is no knot in this manifold, the Feynman path integral or partition function on this manifold is given by
\eq{\frac{Z(M)}{Z(S^3)} = \frac{Z(M_1)}{Z(S^3)}.\frac{Z(M_2)}{Z(S^3)}.}	
If the two submanifolds are copies of $S^3$ and have a knot in them, then the ratio of partition functions in the above equation becomes a knot invariant, which is the Jones polynomial. The knot invariants are multiplicative when taking the sum of knots. 
A physical Hilbert space can be associated with $S^2$, the common boundary of $M_1$ and $M_2$.
Let Feynman path integral on $M_1$ determines a vector $\chi$ in $H$, and a dual vector space $H'$. $M_2$ is dual to $M_1$ with the opposite boundary orientation. The path integral on $M_2$ gives a vector $\psi \in H'$. The partition function of $M$ can be written as 
$$Z(M) = \bra{\chi}\ket{\psi}.$$
$S^2$ separates $S^3$ into two three-balls, $B_L$ and $B_R$. The path integrals on $B_L$ and $B_R$ give the vectors $v$ and $v'$,
$$Z(S^3) = \bra{v}\ket{v'}.$$
As the $H$ is one-dimensional, $v$ is a multiple of $\chi$, similarly $v'$ is a multiple of $\psi$. A fact from algebra is
$$\bra{\chi}\ket{\psi}.\bra{v}\ket{v'} = \bra{\chi}\ket{v'}.\bra{v}\ket{\psi}.$$
On the right side, the two factor are actually partition functions $Z(M_1)$ and $Z(M_2)$. When there are links in the $S^3$, with $S$ unlinked and unknotted circles $C_i$, with each circle there is associated representation $R_i$. The partition function with collection of the Wilson lines is $Z(S^3; C_1,...,C_s)$, so we have
$$\frac{Z(S^3;C_1,...,C_s)}{Z(S^3)} = \prod_{k=1}^{s}\frac{Z(S^3;C_k)}{Z(S^3)}.$$
The normalized expectation value of a link $L$ is
$\langle L \rangle = Z(S^3;L)/Z(S^3)$ then for any collection of the unknotted and unlinked Wilson lines shown in Fig. \ref{ConnectedSum} (b),
$$ \langle C_1,...C_s \rangle = \prod_{k}\langle C_k \rangle .$$
\begin{figure}[h!]
	\centering
	\begin{subfigure}{0.4\textwidth}
		\centering
		\begin{tikzpicture}[use Hobby shortcut,scale=1]
			\begin{knot}[
				consider self intersections=true,
				ignore endpoint intersections=false,
				flip crossing=2,
				only when rendering/.style={}]
				\draw[ultra thick,densely dashed,blue] (3,-0.7) arc (270:90:.1 and 0.7);
				\strand [red,ultra thick,blue](0.4,-0.76)..(0,0).. (1,1).. (3,0.7)..(5,1)..(6,0)..(5,-1)..(3,-0.7)..(1,-1)..(0.4,-0.76);
				\draw[bend left,ultra thick,red] (1,0) to (1.8,0);
				\draw[bend right,ultra thick,red] (0.8,0) to (2,0);
				\draw[bend left,ultra thick,red] (4.5,0.3) to (5,0.3);
				\draw[bend right,ultra thick,red] (4.3,0.3) to (5.2,0.3);
				\draw[bend left,ultra thick,red] (4.5,-0.3) to (5,-0.3);
				\draw[bend right,ultra thick,red] (4.3,-0.3) to (5.2,-0.3);	
				\draw[ultra thick,blue] (3,-0.7) arc (-90:90:.1 and 0.7);
				\node at (1,-1.5) {$M_1$};
				\node at (4.5,-1.5) {$M_2$};
			\end{knot}
			\path (0,0);
		\end{tikzpicture}
		\caption{}
	\end{subfigure}
	\begin{subfigure}{0.3\textwidth}
		\centering
		\begin{tikzpicture}[scale=0.8]
			\draw[ultra thick,densely dashed,knot=blue] (-2,0) arc (180:0:2 and 0.5);		
			\draw[ultra thick, ->,red] (-1,1) arc (90:-360:0.5);
			\draw[ultra thick, ->,red] (-0,0.3) arc (90:-360:0.5);
			\draw[ultra thick, ->,red] (1,-0.3) arc (90:-360:0.5);
			\draw [ultra thick,blue] (0,0) circle (2);
			\draw[ultra thick,blue] (-2,0) arc (-180:0:2 and 0.5);
			\node [above] at (-1,0.9) {$ C_1 $};
			\node [] at (-0.2,-1) {$ C_2 $};
			\node [] at (0.6,-1.5) {$ C_3 $};
		\end{tikzpicture}
		\caption{}
	\end{subfigure}	
	\caption[Unknotted Wilson links on a connected sum of $M_1$ and $M_2$ joined at the boundary $S^2$.]{(a) The manifold $M$ is connected sum of $M_1$ and $M_2$ joined at the boundary $S^2$, (b) Unknotted Wilson links in a three-manifold \cite{witten1989quantum}.}
	\label{ConnectedSum}
\end{figure}
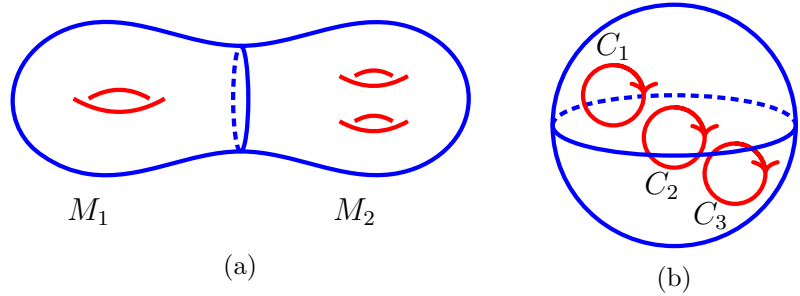

\begin{figure}[h!]
	\centering
	\begin{tikzpicture}[scale=1.5,rotate=-90]
		\draw[ultra thick,knot=teal] (-1,0.6)--(0,0.7);
		\draw[ultra thick,knot=teal] (-1,0.3)--(0,0.3);
		\draw[ultra thick,knot=teal] (-1,-0.3)--(0,-0.3);
		\draw[ultra thick,knot=teal] (-1,-0.6)--(0,-0.7);	
		\draw[ultra thick,blue] (0,-1) .. controls(-3,-1) and (-3,1)..(0,1);
		\draw [ultra thick,knot=red] (0,0) ellipse (0.3 and 1);
		\draw [ultra thick,fill=gray!20] (-1.2,0) ellipse (0.6 and 0.7);
		\fill[orange] (0,0.7) circle(2pt);
		\fill[orange] (0,0.3) circle(2pt);
		\fill[orange] (0,-0.3) circle(2pt);
		\fill[orange] (0,-0.7) circle(2pt);
		\draw[teal,->,ultra thick] (-0.4,0.65)--(-0.5,0.65);
		\draw[teal,->,ultra thick] (-0.4,0.3)--(-0.5,0.3);
		\draw[teal,->,ultra thick] (-0.5,-0.3)--(-0.4,-0.3);
		\draw[teal,->,ultra thick] (-0.5,-0.65)--(-0.4,-0.65);
		\node [] at (-1,-1.2) {$M_L$};	
	\end{tikzpicture}\\
	\begin{tikzpicture}[scale=1.5,rotate=-90]
		\draw[ultra thick,knot=teal](0,0.7) .. controls(1.5,0.4) and (1.5,0.1)..(0,-0.3);
		\draw[ultra thick,knot=teal](0,0.3) .. controls(1.5,0.1) and (1.5,-0.3)..(0,-0.7);	
		\draw[ultra thick,blue] (0,-1) .. controls(2,-0.7) and (2,0.7)..(0,1);
		\draw [ultra thick,knot=red] (0,0) ellipse (0.3 and 1);
		\fill[orange] (0,0.7) circle(2pt);
		\fill[orange] (0,0.3) circle(2pt);
		\fill[orange] (0,-0.3) circle(2pt);
		\fill[orange] (0,-0.7) circle(2pt);
		\draw[teal,->,ultra thick] (0.5,0.58)--(0.45,0.59);
		\draw[teal,->,ultra thick] (0.5,0.21)--(0.45,0.22);
		\node [] at (0.5,-1.3) {$M_R$};		
	\end{tikzpicture}
	\begin{tikzpicture}[scale=1.5,rotate=-90]
		\draw[ultra thick,knot=teal](0,0.3) .. controls(1,0.2) and (1,-0.2)..(0,-0.3);
		\draw[ultra thick,knot=teal](0,0.7) .. controls(1.5,0.3) and (1.5,-0.3)..(0,-0.7);		
		\draw[ultra thick,blue] (0,-1) .. controls(2,-0.7) and (2,0.7)..(0,1);
		\draw [ultra thick,knot=red] (0,0) ellipse (0.3 and 1);
		\fill[orange] (0,0.7) circle(2pt);
		\fill[orange] (0,0.3) circle(2pt);
		\fill[orange] (0,-0.3) circle(2pt);
		\fill[orange] (0,-0.7) circle(2pt);
		\draw[teal,->,ultra thick] (0.5,0.54)--(0.45,0.55);
		\draw[teal,->,ultra thick] (0.5,0.2)--(0.45,0.21);
		\node [] at (1.6,0.5) {$X_1$};
	\end{tikzpicture}
	\begin{tikzpicture}[scale=1.5,rotate=-90]
		\draw[ultra thick,knot=teal](0,0.3) .. controls(1.5,0.1) and (1.5,-0.3)..(0,-0.7);
		\draw[ultra thick,knot=teal](0,0.7) .. controls(1.5,0.4) and (1.5,0.1)..(0,-0.3);		
		\draw[ultra thick,blue] (0,-1) .. controls(2,-0.7) and (2,0.7)..(0,1);
		\draw [ultra thick,knot=red] (0,0) ellipse (0.3 and 1);
		\fill[orange] (0,0.7) circle(2pt);
		\fill[orange] (0,0.3) circle(2pt);
		\fill[orange] (0,-0.3) circle(2pt);
		\fill[orange] (0,-0.7) circle(2pt);
		\draw[teal,->,ultra thick] (0.5,0.58)--(0.45,0.59);
		\draw[teal,->,ultra thick] (0.5,0.21)--(0.45,0.22);
		\node [] at (1.6,0.5) {$X_2$};
	\end{tikzpicture}
	\caption[A link $C$ in a general three-manifold.]{A link $C$ in a general three-manifold \cite{witten1989quantum}.}
	\label{ManifoldWithLink}
\end{figure}
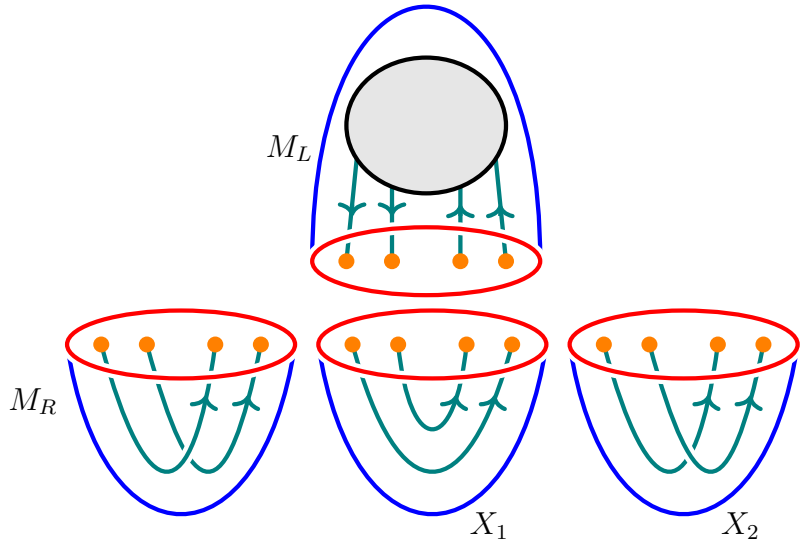
Let there be a link $L$ on the manifold $M$. The components of the knots have associated representation of some group $G$. Let $Z(L)$ be the partition function of the knot, and the Wilson lines are in a gauge group $G=SU(N)$. And let us cut $M$ into two pieces, $M_R$ and $M_L$ for more study. Here, the physical Hilbert spaces $H_R$ and $H_L$ are two-dimensional. The Feynman path integrals on $M_R$ and $M_L$ give $\ket{\psi} \in H_R$ and $\ket{\chi} \in H_L$. The vector spaces $H_R$ and $H_L$ are dual to each other, hence
\eq{Z(L) = \bra{\chi}\ket{\psi}.}
In two dimensions, there is a linear relation among vectors. Let there be two other vectors in $H_R$, we can write their relation as
\eq{\alpha \ket{\psi} +\beta \ket{\psi_1} +\gamma \ket{\psi_2} = 0,}
where the weights $\alpha,\beta,\gamma$ are complex numbers add up to zero. 
Let $X_1,X_2$ be the same manifolds as $M_R$, but with different braids. Connecting the points on the boundary we get
\eq{\alpha \bra{\chi}\ket{\psi} + \beta \bra{\chi}\ket{\psi_1} + \gamma \bra{\chi}\ket{\psi_2} = 0.}
These two manifolds are the same from outside but are different from inside as in Fig. \ref{ManifoldWithLink}. Gluing $X_1$ and $X_2$ is the same as gluing $M_R$ and $M_L$, but the link $L$ is replaced by $L_1$ and $L_2$. The expectation value would be given as
\eq{\alpha Z(L) + \beta Z(L_1)+ \gamma Z(L_2) = 0.}

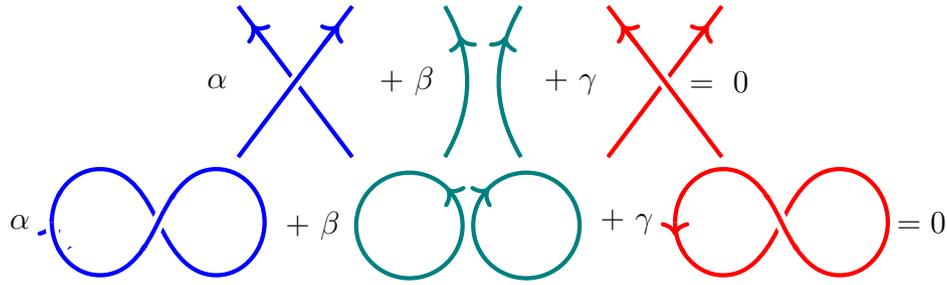
\begin{figure}[h!]
	\centering
	\begin{tikzpicture}[scale = 0.5]
		\node [left] at (0,2) {$\alpha$};
		\draw[blue,knot,ultra thick] (3,0) -- (0,4);
		\draw[blue,knot,ultra thick] (0,0) -- (3,4);
		\draw[blue,->,ultra thick] (0.75,3)--(0.35,3.5);
		\draw[blue,->,ultra thick] (2.25,3)--(2.65,3.5);
	\end{tikzpicture}
	\begin{tikzpicture}[scale = 0.5]
		\node [left] at (0,2) {$+ \ \beta $};
		\draw[teal,bend left,ultra thick] (2,0) to (2,4);
		\draw[teal,knot,bend right,ultra thick] (0,0) to (0,4);
		\draw[teal,->,ultra thick] (0.45,3)--(0.4,3.2);
		\draw[teal,->,ultra thick] (1.55,3)--(1.6,3.2);
	\end{tikzpicture}
	\begin{tikzpicture}[scale = 0.5]
		\node [left] at (0,2) {$+ \ \gamma$};
		\draw[red,knot,ultra thick] (0,0) -- (3,4);
		\draw[red,knot,ultra thick] (3,0) -- (0,4);
		\node [left] at (4,2) {$= \ 0$};
		\draw[red,->,ultra thick] (0.75,3)--(0.35,3.5);
		\draw[red,->,ultra thick] (2.25,3)--(2.65,3.5);
	\end{tikzpicture}\\
	\begin{tikzpicture}[use Hobby shortcut,scale=0.7]
		\begin{knot}[
			consider self intersections=true,
			ignore endpoint intersections=false,
			flip crossing=1,
			only when rendering/.style={}]
			\node [left] at (-0.2,0) {$\alpha $};
			\strand [blue,ultra thick,->] ([closed]0,0) .. (1,-1) .. (2,0) .. (3,1) .. (4,0) .. (3,-1) .. (2,0)..(1,1)..(0,0);
		\end{knot}
		\path (0,0);
	\end{tikzpicture}
	\begin{tikzpicture}
		\begin{knot}[flip crossing=2,scale=0.7]
			\node [left] at (-2.3,0) {$+ \ \beta $};
			\strand [teal, ultra thick](1,0) circle[radius=1];
			\strand[teal,ultra thick] (-1.2,0) circle[radius=1];
			\draw[teal,->,ultra thick] (0.2,0.6)--(0.3,0.7);
			\draw[teal,->,ultra thick] (-0.4,0.6)--(-0.5,0.7);
		\end{knot}
	\end{tikzpicture}
	\begin{tikzpicture}[use Hobby shortcut,scale=0.7]
		\begin{knot}[consider self intersections=true,
			ignore endpoint intersections=false,
			flip crossing=1,
			only when rendering/.style={}]
			\node [left] at (-0.2,0) {$+ \ \gamma$};
			\strand [red,ultra thick,-<] ([closed]0,0) .. (1,1) .. (2,0) .. (3,-1) .. (4,0) .. (3,1) .. (2,0)..(1,-1)..(0,0);
			\node [left] at (5.3,0) {$ =0$};
		\end{knot}
		\path (0,0);
	\end{tikzpicture}
	\caption{Skein relations for a knot}
	\label{WittenSkein}
\end{figure}

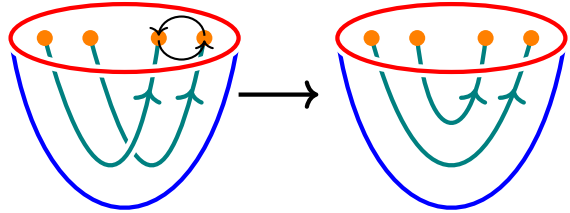
\begin{figure}
	\centering
	\begin{tikzpicture}[scale=1.5,rotate=-90]
		\draw[ultra thick,knot=teal](0,0.7) .. controls(1.5,0.4) and (1.5,0.1)..(0,-0.3);
		\draw[ultra thick,knot=teal](0,0.3) .. controls(1.5,0.1) and (1.5,-0.3)..(0,-0.7);	
		\draw[ultra thick,blue] (0,-1) .. controls(2,-0.7) and (2,0.7)..(0,1);
		\draw [ultra thick,knot=red] (0,0) ellipse (0.3 and 1);
		\fill[orange] (0,0.7) circle(2pt);
		\fill[orange] (0,0.3) circle(2pt);
		\fill[orange] (0,-0.3) circle(2pt);
		\fill[orange] (0,-0.7) circle(2pt);
		\draw[thick,->] (0.02,0.3) arc (-90:90:0.17 and 0.2);
		\draw[thick,->] (-0.02,0.7) arc (90:270:0.17 and 0.2);
		\draw[teal,->,ultra thick] (0.5,0.58)--(0.45,0.59);
		\draw[teal,->,ultra thick] (0.5,0.21)--(0.45,0.22);
		\draw[->,ultra thick] (0.5,1)--(0.5,1.7);
	\end{tikzpicture}
	\begin{tikzpicture}[scale=1.5,rotate=-90]
		\draw[ultra thick,knot=teal](0,0.3) .. controls(1,0.2) and (1,-0.2)..(0,-0.3);
		\draw[ultra thick,knot=teal](0,0.7) .. controls(1.5,0.3) and (1.5,-0.3)..(0,-0.7);		
		\draw[ultra thick,blue] (0,-1) .. controls(2,-0.7) and (2,0.7)..(0,1);
		\draw [ultra thick,knot=red] (0,0) ellipse (0.3 and 1);
		\fill[orange] (0,0.7) circle(2pt);
		\fill[orange] (0,0.3) circle(2pt);
		\fill[orange] (0,-0.3) circle(2pt);
		\fill[orange] (0,-0.7) circle(2pt);
		\draw[teal,->,ultra thick] (0.5,0.54)--(0.45,0.55);
		\draw[teal,->,ultra thick] (0.5,0.2)--(0.45,0.21);		
	\end{tikzpicture}
	\caption[Half-monodromy operation on $S^2$.]{Half-monodromy operation on $S^2$ \cite{witten1989quantum}.}
	\label{HalfMonodromy}
\end{figure}

The knot projection has $p$ crossings, where $p=0$ for unknot. If $\beta = 0$ then the number of crossings reduced giving a new link expectation value. Now, one could at each crossing, replace an overcrossing by an under crossing, that is, pass the two strands through each other with a factor of $-\gamma/\alpha$.  In this case, it is then possible to untie all knots. The skein relations in Fig. \ref{WittenSkein} are used to untie the knot.	
The first and the third diagrams have one unknotted circle, the second has two circles. The partition function $Z(S^3,C)$ for unlinked and unknotted circles in $SU(N)$ representation is now written as
\eq{(&\alpha + \gamma) Z(S^3,C) + \beta Z(S^3,C) = 0, \nonumber\\
	&\langle C \rangle = -\frac{\alpha + \gamma}{\beta}. \label{CExpectation}}
For a knot invariant which includes the self linking number, we need to pick a \textit{framing}. The partition function with the insertion of the Wilson loop $W_R(C)$, in a representation $R$, is not a well-defined holonomy operator. This operator transforms under the change of framing.
The mapping class group of the boundary act naturally on $H_R$. The diffeomorphism is performed before gluing. The \textit{Dehn twist} is as twisting the $S^3$ by one full twist before gluing. It would look like the same picture, but the framing is shifted. The diffeomorphism picked is the Dehn twist acts on $H_R$, causes a phase of $e^{2\pi i h_R}$. This factor is multiplied when straightening a twist in an oriented strand, where $h_R$ is the \textit{conformal dimension or conformal weight}. It is the \textit{topological spin} of a particle in $R$ representation. The $h_R$ is integer then the phase factor $e^{2\pi i h_R}$ is identity, and when $h_R$ is half-integer then the phase $e^{2\pi i h_R}$ is $-1$. For vacuum, $h_R$ is zero. For the anyons, it can have an arbitrary value.
The diffeomorphism of $S^2$ in Fig. \ref{HalfMonodromy} is the \textit{half-monodromy}, depends on the framing. The monodromy is the phase accumulated when one coordinate of a wave function moved around the other. The total phase a wave function can get, on the exchange of two particles, is the Berry phase plus the monodromy.

\begin{figure}[h!]
	\centering
	\begin{tikzpicture}[scale = 0.5]
		\node [left] at (0,2) {$-q^{N/2}$};
		\draw[blue,knot,ultra thick] (3,0) -- (0,4);
		\draw[blue,knot,ultra thick] (0,0) -- (3,4);
		\draw[blue,->,ultra thick] (0.75,3)--(0.35,3.5);
		\draw[blue,->,ultra thick] (2.25,3)--(2.65,3.5);
	\end{tikzpicture}
	\begin{tikzpicture}[scale = 0.5]
		\node [left] at (0,2) {$+ \ (q^{1/2} - q^{-1/2}) $};
		\draw[teal,bend left,ultra thick] (2,0) to (2,4);
		\draw[teal,knot,bend right,ultra thick] (0,0) to (0,4);
		\draw[teal,->,ultra thick] (0.45,3)--(0.4,3.2);
		\draw[teal,->,ultra thick] (1.55,3)--(1.6,3.2);
	\end{tikzpicture}
	\begin{tikzpicture}[scale = 0.5]
		\node [left] at (0,2) {$+ \ q^{-N/2}$};
		\draw[red,knot,ultra thick] (0,0) -- (3,4);
		\draw[red,knot,ultra thick] (3,0) -- (0,4);
		\node [left] at (4,2) {$= \ 0$};
		\draw[red,->,ultra thick] (0.75,3)--(0.35,3.5);
		\draw[red,->,ultra thick] (2.25,3)--(2.65,3.5);
	\end{tikzpicture}
	\caption{Skein relations for Wilson lines in $S^3$ in defining $n$-dimensional representation of $SU(N)$.}
	\label{SkeinFraming}
\end{figure}
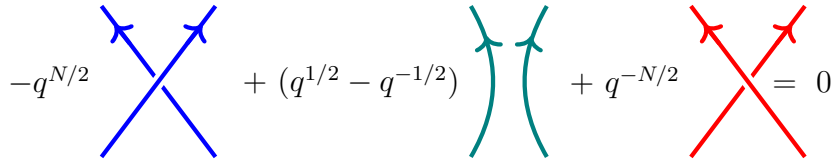

The values of $\alpha,\beta,\gamma$ depend on the framing of knots. Here some knowledge of the conformal field theory (CFT) will be used, see \cite{witten1989quantum} for further detail. Let $H_R$ be the conformal weight of the conformal primary field, transforming as $R$. The primary fields can be thought of as the quasiparticles on the two-dimensional surface. Let $E_i$ be the irreducible representations of $SU(N)$ appearing in the decomposition of $R\otimes R$, and let $h_{E_i}$ be conformal weights of the corresponding primary fields. Let braid $B$ in CFT be such that $\ket{\psi_1} = B\ket{\psi}$ and $\ket{\psi_2} = B^2\ket{\psi}$, then eighenvalues of the $B$ are 
\eq{\lambda_i = \pm \exp(i\pi(2h_R - h_{E_i})),}
where $+,-$ for whether $E_i$ appearing symmetrically or anti-symmetrically in $R\otimes R$. If $R$ is N-dimensional representation of $SU(N)$ \cite{witten1989quantum} then
\eq{ \lambda_1 = \exp(\frac{i\pi(-N+1)}{N(N+k)}), \qquad \lambda_2 = -\exp(\frac{i\pi(N+1)}{N(N+k)}),}
$$h_R = \frac{N^2 -1}{2N(N+k)},\qquad h_{E_1} = \frac{N^2 +N -2}{N(N+k)}, \qquad h_{E_2} = \frac{N^2 - N-2}{N(N+k)}.$$
It is the same framing as in Fig. \ref{HalfMonodromy}, but when glued left and right, then we do not have canonical framing for each link. In order to agree with the knot theory literature, multiply $\beta$ by $\exp(-2\pi i h_R)$ and $\gamma$ by $\exp(-4\pi i h_R)$, we have
\eq{\alpha &= - \exp(\frac{2\pi i}{N(N+k)}), \\
	\beta &= -\exp(\frac{i \pi (2 - N - N^2)}{N(N+k)}) + \exp(\frac{i \pi(2+ N - N^2)}{N(N+k)}), \\
	\gamma &= \exp(\frac{2\pi(1 - N)}{N(N+k)}).}
Let $q = \exp(\frac{2 \pi i}{N+k})$ and common factor $\exp(\frac{i\pi (N^2 -2)}{N(N+k)})$ can be taken out we get
\eq{-q^{N/2} L_+ + (q^{1/2} - q^{-1/2})L_0 + q^{-N/2} L_- =0.}
Compare this relation with that found in Chapter \ref{Knot} for $q=t$. Here $L_i$ is used instead of $Z(L_i)$ and is shown pictorially in Fig. \ref{SkeinFraming}.  
Therefore, using the Eq. \ref{CExpectation}, we obtain the relation for the unknotted Wilson line on $S^3$
\eq{\langle C \rangle = \frac{q^{N/2} - q^{-N/2}}{q^{1/2} - q^{-1/2}}.}
This is the invariant of an unknotted Wilson line in $S^3$ \cite{witten1989gauge}. The Kauffman bracket of a link is evaluated, taking $A= q^{1/4}$.
Compare this result with the knot invariant calculated using the Hopf algebra discussed in the next section.

Now we will write the partition function in terms of the $S$-matrix. The partition function of a Wilson line in $S^3$ is given as \cite{witten1989gauge}
\eq{\langle \tik{\draw (0,0) circle (3pt);} \rangle = \frac{Z_{L_{1,0}}}{Z_{S^3}} = \frac{S_{10}}{S_{00}}, \qquad \langle \text{Hopf} \rangle = \frac{Z_{L_{1,1}}}{Z_{S^3}} = \frac{S_{11}}{S_{00}},}
where $S_{00}$ is the expectation value of the vacuum when no link is present, and $S_{10}$ is the expectation value of one link. The expectation value of a Hopf link is $S_{11}$. $S_{a,b}$ are the matrix elements of the $S$-matrix correspond to the modular transformation. See the section on modular transformation in Chapter \ref{Cat}.
	
\section{Link Invariant from Quantum Group}
 
In quantum mechanics, the commuting classical observables are replaced with the noncommuting Hermitian operators, hence we can say that we deform a classical algebra and the deformation parameter is the Planck constant $h$. 
The noncommutativity of the variables $X$ and $Y$ in the deformed space is written as 
\eq{XY = qYX,}
where $q$ is a complex number in general. It is called the \textit{deformation parameter}. Let $q$ be a number different from $1$ and $h$ is a number different from $0$. If we take $x= qx_0$ or $x=x_0+h$ we can have classical values when $h\to 0$ or $q\to 1$. These two are related by $q= e^h$ \cite{jaganathan2000introduction}. For $q \to 1$, we would get back the classical commuting variables. The quantum group is q-analog of the Hopf algebra.

\subsection{Quantum Deformation of an Algebra}
To describe the quantum deformation of an algebra \cite{jaganathan2000introduction}, let us consider a classical two-dimensional vector space with elements $x,y$. In matrix form
$$\begin{pmatrix}
	x  \\
	y
\end{pmatrix},$$
where $x,y$ are real and commute with each other. Their partial derivatives satisfy the differential calculus, so $$xy = yx, \implies [x,y] = 0,$$ and 
$$ \ [\partial/\partial x, \partial/\partial y] = 0, \ [\partial/\partial x , y] = 0,$$ 
$$ [\partial/\partial y , x] = 0, \ [\partial/\partial x , x] = 1, \ [\partial/\partial y , y] = 1,$$
where the commutator is defined as $[A,B] = AB-BA$. The linear transformation of the vector is given by
\eq{\begin{pmatrix}
		x^{'}\\
		y^{'}
	\end{pmatrix} = M\begin{pmatrix}
		x\\
		y
	\end{pmatrix} = \begin{pmatrix}
		a & b \\
		c & d
	\end{pmatrix} \begin{pmatrix}
		x\\
		y
	\end{pmatrix}.}
The entries of the $M$ are real and satisfy $ad -bc = \det M = 1$. This transformation is an element of the group $SL(2,R)$. The partial derivatives in new coordinates would satisfy the same relations
\eq{\begin{pmatrix}
		\partial/ \partial x^{'}\\
		\partial/ \partial y^{'}
	\end{pmatrix} = \tilde{M}^{-1}\begin{pmatrix}
		\partial/ \partial x\\
		\partial/ \partial y
	\end{pmatrix} = \begin{pmatrix}
		d & -c \\
		-b & a
	\end{pmatrix} \begin{pmatrix}
		\partial/ \partial x\\
		\partial/ \partial y
	\end{pmatrix},}
where $\tilde{M}$ is a transpose of $M$. The noncommutativity of the variables $X$ and $Y$ in the deformed space is written as 
\eq{XY = qYX.}
For the deformation parameter $q \to 1$, we would get back the classical commuting variables. 
The differential calculus on the quantum plane have the following relations
\eq{&XY = qYX, \qquad \frac{\partial}{\partial X}\frac{\partial}{\partial X} = q^{-1} \frac{\partial}{\partial Y}\frac{\partial}{\partial X},\\
	&\frac{\partial}{\partial X} Y = q Y \frac{\partial}{\partial X}, \qquad \frac{\partial}{\partial Y} X = q X \frac{\partial}{\partial Y},\\
	&\frac{\partial}{\partial X}X - q^2 X \frac{\partial}{\partial X} = 1 + (q^2 -1) Y \frac{\partial}{\partial Y},\\
	&\frac{\partial}{\partial Y} Y - q^2 Y \frac{\partial}{\partial Y} = 1.}
The differential calculus on the quantum plane is covariant under the transformations
\eq{\begin{pmatrix}
		X^{'}\\
		Y^{'}
	\end{pmatrix} = T\begin{pmatrix}
		X\\
		X
	\end{pmatrix} = \begin{pmatrix}
		A & B \\
		C & D
	\end{pmatrix} \begin{pmatrix}
		X\\
		Y
	\end{pmatrix},}

\eq{\begin{pmatrix}
		\frac{\partial}{\partial X^{'}}\\
		\frac{\partial}{\partial Y^{'}}
	\end{pmatrix} = \tilde{T}^{-1}\begin{pmatrix}
		\frac{\partial}{\partial X}\\
		\frac{\partial}{\partial Y}
	\end{pmatrix} = \begin{pmatrix}
		D & -qC \\
		-q^{-1}B & A
	\end{pmatrix} \begin{pmatrix}
		\frac{\partial}{\partial X}\\
		\frac{\partial}{\partial Y}
	\end{pmatrix},}
$A,B,C,D$ commute with $X,Y$. Now we have
\eq{&AB = qBA, \ CD = qDC, AC = qCA, \ BD = qDB, \nonumber \\
	&BC = CB, \ AD-DA = (q-q^{-1})BC, \nonumber \\
	&AD - qBC = \det(T)_q = 1,}
\eq{T^{-1} = \begin{pmatrix}
		D & -q^{-1}B\\
		-qC & A
	\end{pmatrix}, \qquad TT^{-1} = \mathbb{I}.}
Let $T_1 = \begin{pmatrix}
	A_1 & B_1\\
	C_1 & D_1
\end{pmatrix} ,\ \ T_2 = \begin{pmatrix}
	A_2 & B_2\\
	C_2 & D_2
\end{pmatrix}$,\\ 
$T_1,T_2$ should satisfy conditions in above equation
\eq{\Delta_{12}(T) &= T_1 \otimes T_2 = \begin{pmatrix}
		A_1 & B_1\\
		C_1 & D_1
	\end{pmatrix} \otimes \begin{pmatrix}
		A_2 & B_2\\
		C_2 & D_2
	\end{pmatrix} \nonumber\\
	&=\begin{pmatrix}
		A_1 \otimes A_2 + B_1 \otimes C_2 & A_1 \otimes B_2 + B_1 \otimes D_2\\
		C_1 \otimes A_2 + D_1 \otimes C_2 & C_1 \otimes B_2 + D_1 \otimes D_2
	\end{pmatrix} =\begin{pmatrix}
		\Delta_{12}(A) & \Delta_{12}(B)\\
		\Delta_{12}(C) & \Delta_{12}(D)
	\end{pmatrix}.}
The product $\Delta$ is called \textit{coproduct or comultiplication}. The matrix $T$ corresponds to the fundamental, irreducible representation of $SL_q(2)$. It can be generalized to a higher dimensional algebra. 

\subsection{Hopf Algebra}
The Hopf algebra not only has unit and bilinear product, but also counit and coproduct. Therefore, it can accommodate the creation and annihilation of particles. The quasitriangular Hopf algebra is called the quantum group. The knot invariant can be computed by it \cite{ohtsuki2002quantum,jackson2019introduction,sawin1996links}.
Quantum groups have asserted their influence in such areas as category theory, representation theory, topology, combinatorics, non-commutative geometry, symplectic geometry, and knot theory to name a few. For detailed discussion on Hopf algebra and quantum group, see \cite{chari1995guide,majid2000quantum,majid2000foundations}.
Before defining the Hopf algebra, we have to define bialgebra and coalgebra. The arrows below are maps from left to right. These arrows can also be interpreted as the morphisms in the categories of quantum group \cite{chari1995guide}.

If $R$ is a ring and $A$ is \textbf{algebra} $(A,\mu,\eta)$ over $R$, $A$ is a module together with the morphisms

\begin{tikzcd}
	\mu: A \otimes A \arrow[r,thick,blue] & A, \qquad \eta: R \arrow[r,thick,blue] & A,
\end{tikzcd}

such that $A$ has 

associativity

\begin{tikzcd}
	A \otimes A \otimes A 
	\arrow[r,thick,blue,shift left, "\mu \otimes 1"] 
	\arrow[r,thick,blue,shift right, "1 \otimes \mu" below]  
	& A \otimes A \arrow[r,thick,blue, "\mu"] 
	& A,
\end{tikzcd}

identity

\begin{tikzpicture}[->]
	\node (a) [left] at (0,0) {A};
	\node (b) [right] at (2,0) {A $\otimes$ A};
	\node (c) [right] at (5,0) {A.};
	\node [above] at (1.3,0) {$\eta \otimes 1$};
	\node [below] at (1.3,0) {$1 \otimes \eta$};
	\node [above] at (4,0) {$\mu$}; 
	\node [above] at (2.5,-0.8) {$1$};
	\path [transform canvas={yshift=0.5ex},thick,blue] (a) edge (b);%
	\path [transform canvas={yshift=-0.5ex},thick,blue] (a) edge (b);
	\path [thick,blue] (b) edge (c);
	\path [bend right=25,thick,blue] (a) edge (c);
\end{tikzpicture}

$C$ is a \textbf{coalgebra} $(C,\Delta,\varepsilon)$ over ring $R$ together with the module morphism

\begin{tikzcd}
	\Delta: C \arrow[r,thick,blue] & C \otimes C, \qquad  \varepsilon: C \arrow[r,thick,blue] & R
\end{tikzcd}

such that 

\begin{tikzcd}
	C \arrow[r,thick,blue, "\Delta"]
	& C \otimes C 
	\arrow[r,thick,blue, shift left, "\Delta \otimes 1"] 
	\arrow[r,thick,blue, shift right, "1 \otimes \Delta" below]  
	& C \otimes C \otimes C,
\end{tikzcd}

\begin{tikzpicture}[->]
	\node (a) [left] at (0,0) {C};
	\node (b) [right] at (1.5,0) {C $\otimes$ C};
	\node (c) [right] at (5,0) {C,};
	\node [above] at (1,0) {$\Delta$};
	\node [below] at (3.7,0) {$1 \otimes \varepsilon$};
	\node [above] at (3.7,0) {$\varepsilon \otimes 1$}; 
	\node [above] at (2.5,-0.8) {$1$};
	\path [transform canvas={yshift=0.5ex},thick,blue] (b) edge (c);%
	\path [transform canvas={yshift=-0.5ex},thick,blue] (b) edge (c);
	\path [thick,blue] (a) edge (b);
	\path [bend right=25,thick,blue] (a) edge (c);
\end{tikzpicture}

where $\Delta$ is \textbf{comultiplication} and $\varepsilon$ is \textbf{counit}.
Let $R$ be a commutative ring, the $R$-module $B$ is \textbf{bialgebra} $(B,\mu,\eta,\Delta,\varepsilon)$ with algebra and coalgebra structures

\begin{tikzcd}
	\mu: B \otimes B \arrow[r,thick,blue] & B, \qquad \eta: R \arrow[r,thick,blue] & B,\\
	\Delta: B \arrow[r,thick,blue] & B \otimes B, \qquad  \varepsilon: B \arrow[r,thick,blue] & R.
\end{tikzcd}

$R$-\textbf{Hopf} algebra is an $R$-bialgebra $H$ together with $R$-module morphism

\eq{S: H \rightarrow H \nonumber}
called the \textbf{antipode} satisfies the following diagram

\begin{tikzpicture}[->]
	\node (a) [left] at (0,0) {H};
	\node (b) [right] at (1.2,0) {H $\otimes$ H};
	\node (c) [right] at (4.2,0) {H $\otimes$ H};
	\node (d) [right] at (6.5,0) {H,};
	\node [above] at (0.7,0) {$\Delta$};
	\node [above] at (3.2,0) {$S \otimes 1$};
	\node [below] at (3.2,0) {$1 \otimes S$}; 
	\node [above] at (6,0) {$\mu$};
	\path [thick,blue] (a) edge (b);	
	\path [transform canvas={yshift=0.5ex},blue,thick] (b) edge (c); 
	\path [transform canvas={yshift=-0.5ex},blue,thick] (b) edge (c);
	\path [thick,blue] (c) edge (d);
	\node (e) [below] at (3.2,-0.8) {$R$};
	\node [below] at (2,-0.7) {$\varepsilon$};
	\node [below] at (4.5,-0.7) {$\eta$};
	\draw [thick,blue,->] (a) -- (e);
	\draw [thick,blue,->] (e) -- (d);
\end{tikzpicture}

which is $\mu \circ (S\otimes id) \circ \Delta = \mu \circ (id \otimes S) \circ \Delta = \eta \circ \varepsilon$. An antipode plays a similar role as the inverse in a group. The Hopf algebra can naturally be extended to braided monoidal categories where $(H,\mu,\eta)$ is a monoid. Let $H$ be an object in category $C$, the Hopf algebra is a sextuple $(H, \mu, \eta, \Delta, \varepsilon, S)$, where $\mu,\eta,\Delta,\varepsilon, S$ are respectively the multiplication, unit, comultiplication, counit, and antipode. $\Delta$ maps a strand to a union of two strands, $S$ takes the strand to the same strand with opposite orientation with additional curves at its ends, and the $\varepsilon$ removes a strand. The antipode can be realized as in the Fig. \ref{Antipode}.

\begin{figure}[h!]
	\centering
	\tik{\draw[ultra thick,blue,->] (0,0) -- (0,2);
		\node at (1,1.3) {$S$};
		\draw[ultra thick,red,->] (0.5,1) -- (1.5,1);}
	\begin{tikzpicture}[use Hobby shortcut]
		\begin{knot}[consider self intersections=true,ignore endpoint intersections=false,flip crossing=0,only when rendering/.style={ultra thick,blue}]
			\strand[ultra thick,blue] (0,0)..(0.5,-0.5)..(1,0)..(1,1)..(1.5,1.5)..(2,1);
			\draw[ultra thick,blue,->] (1,0.6) -- (1,0.5);
		\end{knot}
		\path (0,0);
	\end{tikzpicture}
	\caption{Antipode}
	\label{Antipode}
\end{figure}
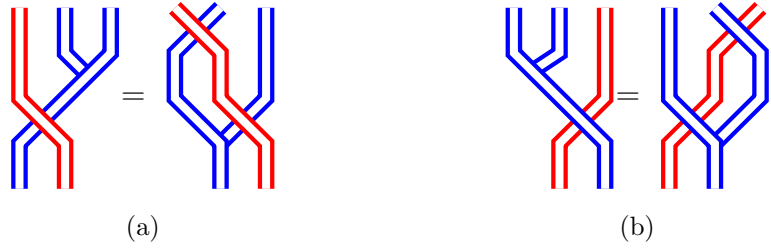
\begin{figure}[h!]
	\centering
	\begin{subfigure}{0.4 \textwidth}
		\centering
		\begin{tikzpicture}[scale=0.6]
			\draw[double distance = 3,ultra thick,blue](0.5,2.5)--(0,3)--(0,4);
			\draw[double distance = 3,ultra thick,blue](-1,0)--(-1,1)--(1,3)--(1,4);
			\node at (1.5,2) {$=$};
			\draw[double distance = 3,ultra thick,red](0,0)--(0,1)--(-1,2)--(-1,4);
		\end{tikzpicture}
		\begin{tikzpicture}[scale=0.6]
			\draw[double distance = 3,ultra thick,blue](0,1)--(1,2)--(1,3)--(1,4);
			\draw[double distance = 3,ultra thick,blue](0,0)--(0,1)--(-1,2)--(-1,3)--(0,4);
			\draw[double distance = 3,ultra thick,red](1,0)--(1,1)--(0,2)--(0,3)--(-1,4);
		\end{tikzpicture}
		\caption{}
	\end{subfigure}
	\begin{subfigure}{0.4\textwidth}
		\centering
		\begin{tikzpicture}[scale=0.6]
			\draw[double distance = 3,ultra thick,red](0,0)--(0,1)--(1,2)--(1,4);
			\draw[double distance = 3,ultra thick,blue](-0.6,2.6)--(0,3)--(0,4);
			\draw[double distance = 3,ultra thick,blue](1,0)--(1,1)--(-1,3)--(-1,4);
			\node at (1.5,2) {$=$};	
		\end{tikzpicture}
		\begin{tikzpicture}[scale=0.6]
			\draw[double distance = 3,ultra thick,red](0,0)--(0,1)--(1,2)--(1,3)--(2,4);
			\draw[double distance = 3,ultra thick,blue](1,1)--(2,2)--(2,3)--(1,4);
			\draw[double distance = 3,ultra thick,blue](1,0)--(1,1)--(0,2)--(0,4);
		\end{tikzpicture}
		\caption{}
	\end{subfigure}
	\caption{The second and third relations for quasitriangular Hopf algebra (a) $(\Delta \otimes id_A)(R) = R_{13}R_{23}$, (b) $(id_A) \otimes \Delta)(R) = R_{13}R_{12}$}
	\label{Quasi12}
\end{figure}

The Hopf algebra $H$ is \textbf{quasitriangular Hopf algebra} if there exist an element $R \in H\otimes H$, such that it has the properties
\eq{&\Delta^{op}(x) =\tau_{H,H}\circ \Delta(x) = R\Delta(x)R^{-1} \ \text{ for all } x\in H,\nonumber \\	 
	&(\Delta \otimes id_A)(R) = R_{13}R_{23}, \nonumber\\
	&(id_A) \otimes \Delta)(R) = R_{13}R_{12}, \label{QuasiTri}}
where $R_{12} = R\otimes I, \ R_{23} = I \otimes R,$ and $R_{13}= (\tau_{A,A} \otimes id_A)(I\otimes R)$.\\
The $R$ matrix is called the  \textit{universal R-matrix}. It should also satisfy the Yang-Baxter equation
\eq{(R\otimes id_V)(id_V\otimes R)(R\otimes id_V)= (id_V\otimes R)(R\otimes id_V)(id_V\otimes R).}
The second and the third properties in Eq. \ref{QuasiTri} implies that the braiding of two particles around the third and then fusing is the same as fusing the two particles then braid around the third as in Fig. \ref{Quasi12}.

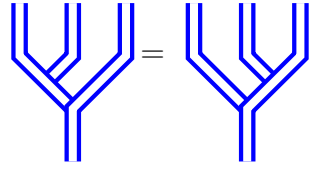
\begin{figure}[h!]
	\centering
	\begin{tikzpicture}[scale=0.7]
		\draw[double distance=3,ultra thick,blue](-0.5,1.5)--(0,2)--(0,3);
		\draw[double distance=3,ultra thick,blue](0,1)--(-0.5,1.5)--(-1,2)--(-1,3);
		\draw[double distance=3,ultra thick,blue](0,0)--(0,1)--(1,2)--(1,3);
		\node at (1.5,2) {$=$};
	\end{tikzpicture}
	\begin{tikzpicture}[scale=0.7]
		\draw[double distance=3,ultra thick,blue](0.5,1.5)--(0,2)--(0,3);
		\draw[double distance=3,ultra thick,blue](0,1)--(-0.5,1.5)--(-1,2)--(-1,3);
		\draw[double distance=3,ultra thick,blue](0,0)--(0,1)--(1,2)--(1,3);
	\end{tikzpicture}
	\caption{$(\Delta \otimes id_A)\Delta = (id_A\otimes \Delta) \Delta$}
	\label{Coasso}
\end{figure}

If the $R$-matrix exists, the braiding is defined in terms of the $R$-matrix as
$$\sigma_{V,W}(v\otimes w) = \tau_{V,W}(R(v\otimes w)),$$
where $\tau_{V,W}(v\otimes w) = w\otimes v$ is a flip map or permutation.
A \textbf{ribbon Hopf algebra} $(A,m,\Delta,u,\varepsilon,S,R,v)$ is a quasitriangular Hopf algebra which possesses an invertible element $\nu$ commonly known as the ribbon element, such that the following hold:
\eq{v^2 = S(u)u, S(v) = v, \varepsilon(v) =1, \nonumber \\
	\Delta(v) = (v \otimes v)(R_{21}R_{12})^{-1},}
where $u= m(S\otimes id)(R_{21})$. $u$ exist for any quasitriangular Hopf algebra, and $uS(u)$ must satisfy 
\eq{(\varepsilon\otimes id_A)(R) = (id_A \otimes \varepsilon)(R).}
Moreover, if it has an antipode then $R^{-1} = (S\otimes id_A)(R)$.
The action of ribbon algebra on Hilbert space is a rotation by $2\pi$ in clockwise direction. 
If the bialgebra has the following properties
\eq{(\Delta \otimes id_A)\Delta &= (id_A\otimes \Delta) \Delta,\nonumber \\
	\Delta^{op} &= \Delta,}
where $\Delta^{op} = \tau_{A,A} \Delta$. The first property is called  coassociativity as in Fig. \ref{Coasso}, whereas the second is called cocommutativity. 
We also have 
$(\varepsilon \otimes id_A)\Delta = (id_A \otimes \varepsilon)\Delta = id_A$,
which is the same as F-symbol

\begin{figure}[h!]
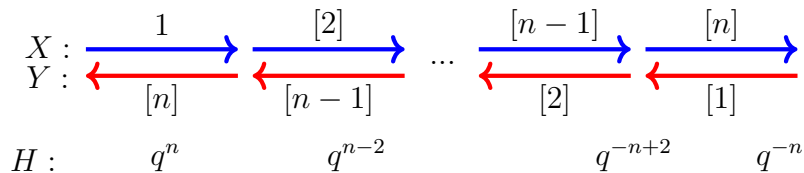

	\centering
	\tik{
		\node at (-0.5,0.2) {$X:$};
		\node at (-0.5,-0.2) {$Y:$};
		\draw[ultra thick, blue, ->, yshift= 5] (0,0)--(2,0);
		\draw[ultra thick, red, ->, yshift = -5] (2,0)--(0,0);
		\node at (1,0.5) {$1$};
		\node at (1,-0.5) {$[n]$}; 
	}
	\tik{
		\draw[ultra thick, blue, ->, yshift= 5] (0,0)--(2,0);
		\draw[ultra thick, red, ->, yshift = -5] (2,0)--(0,0);
		\node at (1,0.5) {$[2]$};
		\node at (1,-0.5) {$[n-1]$};
		\node at (2.5,0) {$...$};
	}
	\tik{
		\draw[ultra thick, blue, ->, yshift= 5] (0,0)--(2,0);
		\draw[ultra thick, red, ->, yshift = -5] (2,0)--(0,0);
		\node at (1,0.5) {$[n-1]$};
		\node at (1,-0.5) {$[2]$}; 
	}
	\tik{
		\draw[ultra thick, blue, ->, yshift= 5] (0,0)--(2,0);
		\draw[ultra thick, red, ->, yshift = -5] (2,0)--(0,0);
		\node at (1,0.5) {$[n]$};
		\node at (1,-0.5) {$[1]$};
	}\\
	\tik{\node at (0,0) {$H:$};} \qquad 
	\tik{\node at (0,0) {$q^n$};}\qquad \qquad
	\tik{\node at (0,0) {$q^{n-2}$};}\qquad \qquad \qquad
	\tik{\node at (0,0) {$q^{-n+2}$};}\qquad
	\tik{\node at (0,0) {$q^{-n}$};}
	\caption{The operators in Lie algebra $\mathfrak{sl_2}$}
	\label{GenLie}
\end{figure}

Now we turn our attention to the quantum group. The quantum group is not a group but a Hopf algebra. The term quantum group is reserved for the Hopf algebra, which is neither commutative nor non-cocommutative. 
We have a \textit{modular Hopf algebra} when there are finitely many irreducible representations, including the trivial one. $U_q(\mathfrak{g})$ is a modular Hopf algebra when $q$ is a root of unity. When $q^n=1$ then we say that the $q$ is the $n$th root of unity.
The quantum group is a ribbon Hopf algebra, which has the additional property that the framed link invariant can be constructed from it. The R-matrix is derived using the two-dimensional representation of $U_q(sl_2)$ at a root of unity, where $U_q(sl_2)$ is the quantum deformation of the universal enveloping Lie algebra $U(sl_2)$. The generators of $U(sl_2)$ are written as
\eq{\qquad [X,H] = 2X, \qquad [Y,H] = 2Y, \ [X,Y] = H.}
Deforming the $U(sl_2)$ to get $U_q(sl_2)$, we get 
\eq{[X,H] = 2X,\ [Y,H] = 2Y, \ [X,Y] =  \frac{q^H -q^{-H}}{q-q^{-1}}.}
The right-hand side is equal to $H$ when $q\to 1$. These operators can be visualized comparing with the raising and lowering operators in the theory of angular momentum. See Fig. \ref{GenLie}.
\\
The Hopf algebra structure on $U_q(sl_2)$ is defined as
\eq{\Delta(H) &= 1\otimes H + H\otimes 1,\nonumber \\
	\Delta(X) &= X\otimes q^{H/2} +q^{-H/4} \otimes X,\nonumber\\
	\Delta(Y) &= Y\otimes q^{H/2} +q^{-H/4} \otimes Y,\nonumber\\
	\Delta(1) &= 1\otimes 1,\nonumber\\
	\epsilon(1) &= 1,\ \epsilon(H) = 0, \ \epsilon (X) = 0,\ \epsilon(Y) = 0, \nonumber \\
	S(1) &= 1, \ S(H) = -H, \ S(X) = -q^{-1/4}X,\ S(Y)= -q^{1/4}Y.}
The representation of the algebra is $\rho(A):U(\mathfrak{g}) \to End(V)$.  Now 1, 2 and 3 dimensional representations of $U_q(sl_2)$ are the as follows
\eq{&X=Y=0, \ H=1,\nonumber\\
	&X= \begin{pmatrix}
		0&1\\
		0&0
	\end{pmatrix}, \ Y = \begin{pmatrix}
		0&0\\
		1&0
	\end{pmatrix}, \ H = \begin{pmatrix}
		q&0\\
		0&q^{-1}
	\end{pmatrix}, \nonumber \\
	&X= \begin{pmatrix}
		0&[2]&0\\
		0&0&[1]\\
		0&0&0
	\end{pmatrix}, \ Y = \begin{pmatrix}
		0&0&0\\
		[1]&0&0\\
		0&[2]&0
	\end{pmatrix}, \ K = \begin{pmatrix}
		q&0&0\\
		0&1&0\\
		0&0&q^{-1}
	\end{pmatrix}.}

\subsection{Link Invariant}
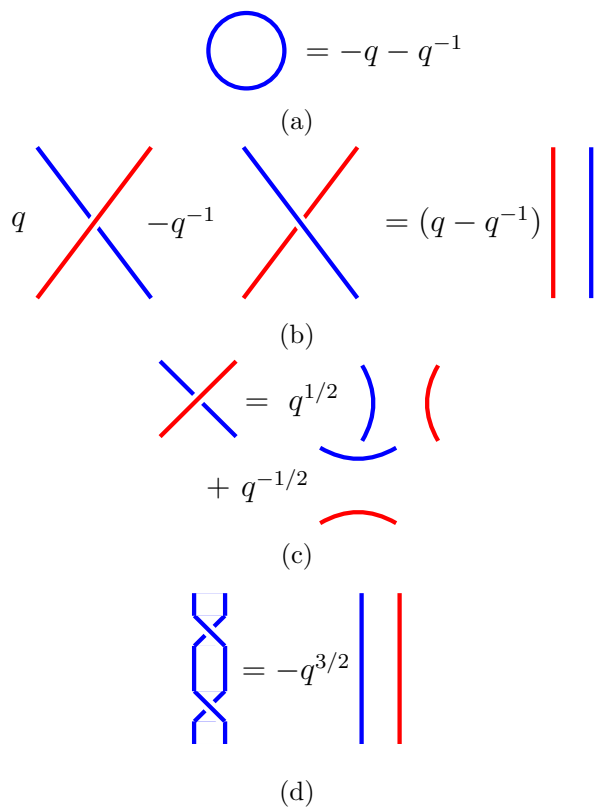
\begin{figure}
	\centering
	\begin{subfigure}{0.15 \textwidth}
		\centering
		\tik{
			\draw [blue, ultra thick] (0,0) circle (0.5);
			\node [left] at (3,0) {$=-q-q^{-1}$};}
		\caption{}
	\end{subfigure}\\
	\begin{subfigure}{0.5\textwidth}
		\centering
		\begin{tikzpicture}[scale = 0.5]
			\node [left] at (0,2) {$q$};
			\draw[blue,knot,ultra thick] (3,0) -- (0,4);
			\draw[red,knot,ultra thick] (0,0) -- (3,4);
			\node [left] at (5,2) {$ - q^{-1}$};
		\end{tikzpicture}
		\begin{tikzpicture}[scale = 0.5]
			\draw[red,knot,ultra thick] (0,0) -- (3,4);
			\draw[blue,knot,ultra thick] (3,0) -- (0,4);
		\end{tikzpicture}
		\begin{tikzpicture}[scale = 0.5]
			\node [left] at (0,2) {$ = (q - q^{-1})$};
			\draw[blue,ultra thick] (1,0) -- (1,4);
			\draw[red,knot,ultra thick] (0,0) -- (0,4);
		\end{tikzpicture}
		\caption{}
	\end{subfigure}\\
	\begin{subfigure}{0.4\textwidth}
		\centering
		\begin{tikzpicture}[scale = 1] 
			\draw[blue,knot,ultra thick] (1,0) -- (0,1);
			\draw[red,knot,ultra thick] (0,0) -- (1,1);
			\node [left] at (2.5,0.5) {$= \ q^{1/2}$};
		\end{tikzpicture}
		\begin{tikzpicture}[scale = 1]
			\draw [blue,ultra thick,bend right] (0,0) to (0,1);
			\draw [red,ultra thick,bend left] (1,0) to (1,1);
		\end{tikzpicture}
		\begin{tikzpicture}[scale = 1]
			\node [left] at (0,0.5) {$ + \ q^{-1/2}$};
			\draw [blue,ultra thick,bend right] (0,1) to (1,1);
			\draw [red,ultra thick,bend left] (0,0) to (1,0);
		\end{tikzpicture}
		\caption{}
	\end{subfigure}\\
	\begin{subfigure}{0.2\textwidth}
		\centering
		\begin{tikzpicture}
			\draw[ultra thick, blue,double distance=10pt] (0,0) -- (0,0.3);
			\draw[ultra thick,blue] (-0.2,0.3) -- (0.2,0.7);	
			\draw[ultra thick,knot=blue] (0.2,0.3) -- (-0.2,0.7);		
			\draw[ultra thick, blue,double distance=10pt] (0,0.7) -- (0,1.3);
			\draw[ultra thick,blue] (-0.2,1.3) -- (0.2,1.7);
			\draw[ultra thick,knot=blue] (0.2,1.3) -- (-0.2,1.7);
			
			\draw[ultra thick, blue,double distance=10pt] (0,1.7) -- (0,2);
			\node [left] at (2,1) {$ = -q^{3/2}$};
			\draw[blue,ultra thick] (2,0) -- (2,2);
			\draw[red,knot,ultra thick] (2.5,0) -- (2.5,2);
		\end{tikzpicture}
		\caption{}
	\end{subfigure}
	\caption{Skein relations in quantum group representation}
	\label{SkeinQG}
\end{figure}

In order to get the link invariant from the quantum group, we will first drive the $R$-matrix for a vector space, then take that vector space in the representation of the quantum group. For more technical discussion, see \cite{ohtsuki2002quantum,jackson2019introduction}. For the 2-dimensional vector space with a basis $R(e_i\otimes e_j) = \sum R^{kl}_{ij} e_k \otimes e_l$, 
due to the property of the conservation of charge, we have $\sum R^{kl}_{ij} = 0$ unless $i+j=k+l$.
The $R$-matrix with respect to the basis $\left\{e_0\otimes e_0, e_0\otimes e_1, e_1\otimes e_0,e_1\otimes e_1 \right\}$ of $V\otimes V$ can be written in the form 
\eq{R= \begin{pmatrix}
		a&0&0&0\\
		0&b&c&0\\
		0&d&e&0\\
		0&0&0&f
	\end{pmatrix}.}
Since $R\otimes id_V$ and $id_V\otimes R$ preserve the four subspaces of $V\otimes V\otimes V$ spanned by the basis $\left\{e_0\otimes e_0\otimes e_0\right\}$, 
$\left\{e_0\otimes e_0\otimes e_1,e_0\otimes e_1\otimes e_0,e_1\otimes e_0\otimes e_0\right\}$,\\
$\left\{e_0\otimes e_1\otimes e_1,e_1\otimes e_0\otimes e_1,e_1\otimes e_1\otimes e_0\right\}$, $\left\{e_1\otimes e_1\otimes e_1\right\}$, $V\otimes V\otimes V$ is the direct sum of four subspaces, 
\eq{R\otimes id_V = (a) \oplus \begin{pmatrix}
		a&0&0\\
		0&b&c\\
		0&d&e
	\end{pmatrix}\oplus \begin{pmatrix}
		b&c&0\\
		d&e&0\\
		0&0&f
	\end{pmatrix}\oplus(f),}
\eq{id_V\otimes R = (a) \oplus \begin{pmatrix}
		b&c&0\\
		d&e&0\\
		0&0&a
	\end{pmatrix} \oplus \begin{pmatrix}
		f&0&0\\
		0&b&c\\
		0&d&e
	\end{pmatrix} \oplus(f).}
To get the Yang-Baxter equation,
\eq{&(R\otimes id_V)(id_V\otimes R)(R\otimes id_V) =\nonumber\\
	&(a^3) \oplus \begin{pmatrix}
		a^2b&abc&ac^2\\
		abd&b^2e+acd&bce+ace\\
		ad^2&bde+ade&cde+ae^2
	\end{pmatrix} \oplus \begin{pmatrix}
		b^2f+bcd&bcf+bce&c^2f\\
		bdf+bde&cdf+be^2&cef\\
		d^2f&def&ef^2
	\end{pmatrix} \oplus(f^3),}
\eq{&(id_V\otimes R)(R\otimes id_V)(id_V\otimes R)= \nonumber \\
	&(a^3) \oplus \begin{pmatrix}
		ab^2+bcd&abc+bce&ac^2\\
		abd+bde&acd+be^2&ace\\
		ad^2&ade&a^2e
	\end{pmatrix} \oplus \begin{pmatrix}
		bf^2&bcf&c^2f\\
		bdf&b^2e+cdf&bce+cef\\
		d^2f&bde+def&cde+e^2f
	\end{pmatrix} \oplus(f^3).}
The $R$-matrix should satisfy the Yang-Baxter equation, therefore we can equate the elements of the above two equations
\eq{&b(cd+ab-a^2)=0, \ b(cd+bf-f^2)=0,\nonumber\\
	&e(cd+ae-a^2)=0, \ e(cd+ef-f^2) = 0,\nonumber\\
	&bce=0, \ bde=0, \ be(b-e)=0.}
Let us take $b=0, e\ne 0$
$a^2-ae = cd=f^2-ef$, hence $e= a-\frac{cd}{a}, \ (a-f)(a+f-e)=0$.
For $a-f =0$, we obtain the following $R$ matrix for Jones polynomial
\eq{R= \begin{pmatrix}
		a&0&0&0\\
		0&0&c&0\\
		0&d&a-cd/a&0\\
		0&0&0&a
	\end{pmatrix}.}
After some normalization, it can be written in somewhat nicer form
\eq{R= \begin{pmatrix}
		t^{1/2}&0&0&0\\
		0&0&t&0\\
		0&t&t^{1/2}-t^{3/2}&0\\
		0&0&0&t^{1/2}
	\end{pmatrix} \in End(V\otimes V).}
The trace of a braid gives an invariant of a link. The isotopy invariance demands that the trace should be invariant under all three Reidemeister moves, and so $trace_2(R^{\pm}) =1$. But the computation shows that this equality does not hold. Therefore, consider the following modification \cite{ohtsuki2002quantum}. Let us consider a map
\eq{h= \begin{pmatrix}
		t^{-1/2}&0\\
		0&t^{1/2}
	\end{pmatrix} \in End(V).}
Now $trace(h^{\otimes n}.\psi_n(b))$ is invariant under all the three Reidemeister moves, where $\psi_n(b)$ is the representation of the braid. Charge conjugation property is given by $R\cdot(h\otimes h) = (h\otimes h)\cdot R$.
Now the Skein relation for Jones polynomial (see Chapter \ref{Knot}), we talked about in the knot theory, can be written as 
\eq{t^{-1}R-tR^{-1}&= t^{-1}\begin{pmatrix}
		t^{1/2}&0&0&0\\
		0&0&t&0\\
		0&t&t^{1/2}-t^{3/2}&0\\
		0&0&0&t^{1/2}
	\end{pmatrix}\nonumber\\
&-t \begin{pmatrix}
		t^{-1/2}&0&0&0\\
		0&t^{-1/2}-t^{-3/2}&t^{-1}&0\\
		0&t^{-1}&0&0\\
		0&0&0&t^{-1/2}
	\end{pmatrix}
	&= (t^{-1/2} -t^{1/2})id_V.}
Here $R,R^{-1}$ and $id_V$ stand for positive crossing, negative crossing and parallel strand respectively.
Analogously, the Kauffman bracket as an operator invariant can be written as
\eq{R= \begin{bmatrix}
		A&0&0&0\\
		0&0&A^{-1}&0\\
		0&A^{-1}&A-A^{-3}&0\\
		0&0&0&A
	\end{bmatrix},}
\eq{n= \begin{bmatrix}
		0 & A & -A^{-1} & 0
	\end{bmatrix} \in Hom(V\otimes V \to \mathbb{C}), \ \
	u= \begin{bmatrix}
		0 \\ -A \\ A^{-1}\\ 0
	\end{bmatrix} \in Hom(\mathbb{C}\to V\otimes V),}
and
\eq{
	&\begin{tikzpicture}[scale = 0.5]
		\node [left] at (0.5,0) {$A \Bigg[$};
		\draw [blue,ultra thick] (1,-0.8) to [curve through={(1.3,0)}](1.1,0.8);
		\draw [red,ultra thick] (1.9,-0.8) to [curve through={(1.7,0)}](1.9,0.8);
		\node [left] at (3.5,0) {$\Bigg]$};
	\end{tikzpicture}
	\begin{tikzpicture}[scale = 0.5]
		\node [left] at (0.5,0) {$ +A^{-1} \Bigg[$};
		\draw [blue,ultra thick] (0.8,0.5) to [curve through={(1.4,0.2)}](2.2,0.5);
		\draw [red,ultra thick] (0.8,-0.5) to [curve through={(1.4,-0.2)}](2.2,-0.5);
		\node [left] at (3.5,0) {$\Bigg]$};
	\end{tikzpicture}\\
	&=A(id_{V \otimes V})+ A^{-1} (n\cdot u)
	= A\begin{bmatrix}
		1&0&0&0\\
		0&1&0&0\\
		0&0&1&0\\
		0&0&0&1
	\end{bmatrix}+A^{-1}\begin{bmatrix}
		0&0&0&0\\
		0&-A^2&1&0\\
		0&1&-A^2&0\\
		0&0&0&0
	\end{bmatrix}\\
	&\begin{tikzpicture}[scale = 0.5] 
		\node at (0,0) {$=R=$};
		\draw[blue,knot,ultra thick] (1.9,-0.8) -- (1.1,0.8);               
		\draw[red,knot,ultra thick] (1.1,-0.8) -- (1.9,0.8);		
\end{tikzpicture}}
\eq{\tik{\node [left] at (1.7,0) {$\Bigg[$};
		\draw [blue, ultra thick] (2.1,0) circle (0.4);
		\node [left] at (3,0) {$\Bigg]$};
		\node [left] at (7,0) {$=n\cdot u=-A^2 -A^{-2}.$};}}
When we have strands on the representation of quantum group then we will take the $q$-deformed $R$-matrix and operators correspond to positive and negative crossings are given by
\eq{Q(L_+) = \begin{pmatrix}
		q^{1/2}&0&0&0\\
		0&0&q^{-1/2}&0\\
		0&q^{-1/2}&q^{1/2}-q^{-3/2}&0\\
		0&0&0&q^{1/2}
\end{pmatrix},}
\eq{Q(L_-) = [Q(L_+)]^{-1} = \begin{pmatrix}
		q^{-1/2}&0&0&0\\
		0&q^{1/2}-q^{-3/2}&q^{1/2}&0\\
		0&q^{1/2}&0&0\\
		0&0&0&q^{-1/2}
\end{pmatrix}.}
The skein relation for Jones polynomial now becomes
\eq{&q^{1/2} \begin{pmatrix}
		q^{1/2}&0&0&0\\
		0&0&q^{-1/2}&0\\
		0&q^{-1/2}&q^{1/2}-q^{-3/2}&0\\
		0&0&0&q^{1/2}
	\end{pmatrix}\nonumber \\
	&-q^{-1/2} \begin{pmatrix}
		q^{-1/2}&0&0&0\\
		0&q^{1/2}-q^{-3/2}&q^{1/2}&0\\
		0&q^{1/2}&0&0\\
		0&0&0&q^{-1/2}
	\end{pmatrix} = (q-q^{-1})\bm{I}.}
It can be written as 
\eq{q^{1/2}Q(L_+)-q^{-1/2}Q(L_-) = (q-q^{-1})Q(L_0).}
The skein diagrams in the representation of quantum group are given in Fig. \ref{SkeinQG}. The elementary tangle diagrams in Fig. \ref{ElTangle} would become the oriented tangle diagrams as in Fig. \ref{ElTangleColored}. In the case of the ribbon Hopf algebra, the basic units of the tangle diagrams can be drawn yet in another way, as in Fig. \ref{RibbonQG}. 

\begin{figure}
	\centering	
	\begin{tikzpicture}[use Hobby shortcut,scale=0.7]
		\draw[ultra thick, blue] (0,0)--(0,1.5);
		\draw[ultra thick, red] (1,0)..controls(1,2) and (2,2)..(2,0);
		\draw[ultra thick, blue] (3,1.5)..controls(3,-0.5) and (4,-0.5)..(4,1.5);
		\braid[ultra thick,above=43,right=115, style strands={1}{red},style strands={2}{blue},style strands={3}{green}]
		s_1^{-1};
		\braid[ultra thick,above=43,right=170, style strands={1}{red},style strands={2}{blue},style strands={3}{green},style strands={4}{green}]
		s_1;
	\end{tikzpicture}
	\caption{The elementary tangle diagrams}
	\label{ElTangle}
\end{figure}
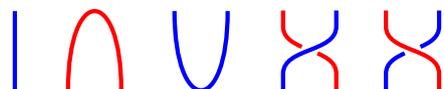

\begin{figure}
	\centering	
	\begin{tikzpicture}[use Hobby shortcut,scale=0.7]
		\draw[ultra thick, blue,->] (-5,0)--(-5,1.5);
		\draw[ultra thick, blue,->] (-4,1.5)--(-4,0);		
		\draw[ultra thick, red,->] (-2,0)..controls(-2,2) and (-3,2)..(-3,0);
		\draw[ultra thick, blue,->] (0,1.5)..controls(0,-0.5) and (-1,-0.5)..(-1,1.5);		
		\draw[ultra thick, red,->] (1,0)..controls(1,2) and (2,2)..(2,0);
		\draw[ultra thick, blue,->] (3,1.5)..controls(3,-0.5) and (4,-0.5)..(4,1.5);
		\braid[ultra thick,above=43,right=115, style strands={1}{red},style strands={2}{blue},style strands={3}{green}]
		s_1^{-1};
		\braid[ultra thick,above=43,right=170, style strands={1}{red},style strands={2}{blue},style strands={3}{green},style strands={4}{green}]
		s_1;
		\draw[ultra thick, red,->] (5,1.5)--(5,1.7);
		\draw[ultra thick, blue,->] (6,1.5)--(6,1.7);
		\draw[ultra thick, red,->] (7,1.5)--(7,1.7);
		\draw[ultra thick, blue,->] (8,1.5)--(8,1.7);
	\end{tikzpicture}
	\caption{The elementary oriented tangle diagrams}
	\label{ElTangleColored}
\end{figure}
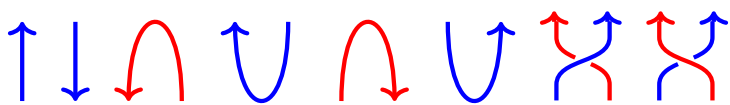

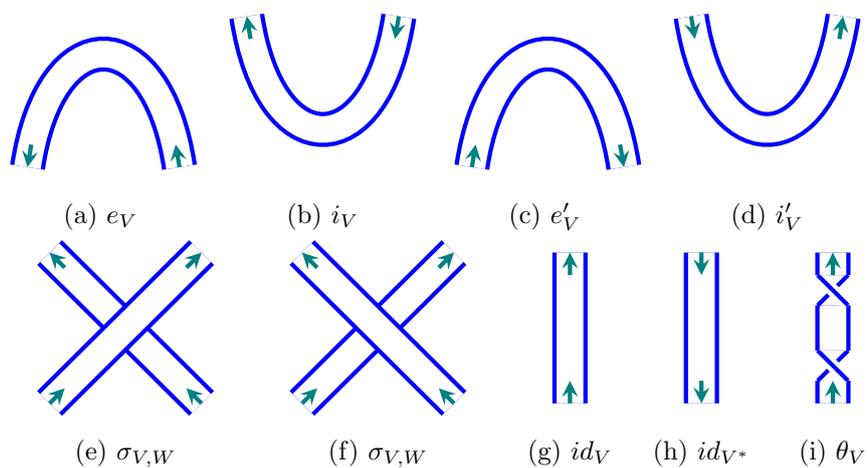
\begin{figure}
	\centering
	\begin{subfigure}{0.15\textwidth}
		\centering
		\begin{tikzpicture}
			\draw[ultra thick, blue,double distance=10pt] (0,0) .. controls (0.3,2) and (1.7,2) .. (2,0);
			\draw[ultra thick,teal,->,>=stealth] (2,0) -- (1.95,0.3);
			\draw[ultra thick,teal,->,>=stealth] (0.05,0.3) -- (0,0);	
		\end{tikzpicture}
		\caption{$e_V$}
	\end{subfigure}
	\begin{subfigure}{0.2\textwidth}
		\centering
		\begin{tikzpicture}
			\draw[ultra thick, blue,double distance=10pt] (0,0) .. controls (0.3,-2) and (1.7,-2) .. (2,0);
			\draw[ultra thick,teal,->,>=stealth] (0.05,-0.3) -- (0,0);
			\draw[ultra thick,teal,->,>=stealth] (2,0) -- (1.95,-0.3);	
		\end{tikzpicture}
		\caption{$i_V$}
	\end{subfigure}
	\begin{subfigure}{0.15\textwidth}
		\centering
		\begin{tikzpicture}
			\draw[ultra thick, blue,double distance=10pt] (0,0) .. controls (0.3,2) and (1.7,2) .. (2,0);
			\draw[ultra thick,teal,->,>=stealth] (0,0) -- (0.05,0.3);
			\draw[ultra thick,teal,->,>=stealth] (1.95,0.3) -- (2,0);	
		\end{tikzpicture}
		\caption{$e'_V$}
	\end{subfigure}
	\begin{subfigure}{0.2\textwidth}
		\centering
		\begin{tikzpicture}
			\draw[ultra thick, blue,double distance=10pt] (0,0) .. controls (0.3,-2) and (1.7,-2) .. (2,0);
			\draw[ultra thick,teal,->,>=stealth] (0,0) -- (0.05,-0.3);
			\draw[ultra thick,teal,->,>=stealth] (1.95,-0.3) -- (2,0);	
		\end{tikzpicture}
		\caption{$i'_V$}
	\end{subfigure}\\
	\begin{subfigure}{0.2\textwidth}
		\centering
		\begin{tikzpicture}
			\draw[ultra thick, blue,double distance=10pt] (0,2) -- (2,0);
			\draw[ultra thick, blue,double distance=10pt] (0,0) -- (2,2);
			\draw[ultra thick,teal,->,>=stealth] (0,0) -- (0.2,0.2);
			\draw[ultra thick,teal,->,>=stealth] (2,0) -- (1.8,0.2);
			
			\draw[ultra thick,teal,->,>=stealth] (1.8,1.8) -- (2,2);
			\draw[ultra thick,teal,->,>=stealth] (0.2,1.8) -- (0,2);	
		\end{tikzpicture}
		\caption{$\sigma_{V,W}$}
	\end{subfigure}
	\begin{subfigure}{0.2\textwidth}
		\centering
		\begin{tikzpicture}
			\draw[ultra thick, blue,double distance=10pt] (0,0) -- (2,2);
			\draw[ultra thick, blue,double distance=10pt] (0,2) -- (2,0);
			
			\draw[ultra thick,teal,->,>=stealth] (0,0) -- (0.2,0.2);
			\draw[ultra thick,teal,->,>=stealth] (2,0) -- (1.8,0.2);
			
			\draw[ultra thick,teal,->,>=stealth] (1.8,1.8) -- (2,2);
			\draw[ultra thick,teal,->,>=stealth] (0.2,1.8) -- (0,2);	
		\end{tikzpicture}
		\caption{$\sigma_{V,W}$}
	\end{subfigure}
	\begin{subfigure}{0.1\textwidth}
		\centering
		\begin{tikzpicture}
			\draw[ultra thick, blue,double distance=10pt] (0,0) -- (0,2);
			\draw[ultra thick,teal,->,>=stealth] (0,0) -- (0,0.3);
			\draw[ultra thick,teal,->,>=stealth] (0,1.7) -- (0,2);		
		\end{tikzpicture}
		\caption{$id_{V}$}
	\end{subfigure}
	\begin{subfigure}{0.1\textwidth}
		\centering
		\begin{tikzpicture}
			\draw[ultra thick, blue,double distance=10pt] (0,0) -- (0,2);
			\draw[ultra thick,teal,->,>=stealth] (0,0.3) -- (0,0);
			\draw[ultra thick,teal,->,>=stealth] (0,2) -- (0,1.7);		
		\end{tikzpicture}
		\caption{$id_{V^*}$}
	\end{subfigure}
	\begin{subfigure}{0.1\textwidth}
		\centering
		\begin{tikzpicture}
			\draw[ultra thick, blue,double distance=10pt] (0,0) -- (0,0.3);	
			\draw[ultra thick,blue] (-0.2,0.3) -- (0.2,0.7);	
			\draw[ultra thick,knot=blue] (0.2,0.3) -- (-0.2,0.7);		
			\draw[ultra thick, blue,double distance=10pt] (0,0.7) -- (0,1.3);
			\draw[ultra thick,blue] (-0.2,1.3) -- (0.2,1.7);
			\draw[ultra thick,knot=blue] (0.2,1.3) -- (-0.2,1.7);
			\draw[ultra thick, blue,double distance=10pt] (0,1.7) -- (0,2);
			\draw[ultra thick,teal,->,>=stealth] (0,0) -- (0,0.3);
			\draw[ultra thick,teal,->,>=stealth] (0,1.7) -- (0,2);
		\end{tikzpicture}
		\caption{$\theta_V$}
	\end{subfigure}
	\caption{The oriented ribbon elementary tangle diagrams.}
	\label{RibbonQG}
\end{figure}

	\chapter{Quantum Field Theory}\label{QFT}
	
\section{Introduction}
Quantum field theory is the most successful theory in physics and is widely used in the philosophy of science. Three of the four forces of nature are unified using the quantum field theory (QFT), therefore it is synonymous to the \textit{standard model} of fundamental particles and their interactions. Classical mechanics is used to studying objects of large sizes at low speeds. Special relativity deals with the objects at a speed comparable to the speed of light. Quantum mechanics is the physics of very small objects. The quantum field theory becomes the theory of very small objects and at very high speeds or energies.
The quasiparticles in condensed matter physics are not elementary particles, but their theory can be built similarly as the quantum field theory is formulated. The quasiparticles excitations are the amplitudes of some kind of quantum probability liquid in some material.  

Fields are some entities that permeate the spacetime and have wavelike oscillations whose amplitude could be a scalar, a vector, a complex number, or a tensor. The classical electromagnetic fields are continuous objects, but quantum mechanics gives the concept of the photon. How to reconcile the two? The missing link is the quantum field. The classical field theory, quantum mechanics, and special relativity are combined in QFT.
The special relativity theory describes the motion of particles at velocities comparable to the speed of light. The Schr\"odinger equation can be used to deal with several particles, but the types and number of particles must be fixed. According to Einstein's mass-energy equation $E=mc^2$, mass can be created from energy. Therefore, at high enough energy, particle number is not fixed but different types of particles and antiparticles are created and destroyed. Non-relativistic quantum mechanics cannot handle these facts. 

To merge relativity and quantum mechanics, an attempt was made in an equation known as the \textit{Klein-Gordon equation}.
Schr\"odinger derived this equation before his famous equation, but he rejected it because it gave the wrong fine structure of the hydrogen atom. Also, it gave the negative probabilities and allows negative energy states. These problems are resolved and made compatible with special relativity by considering $\phi$ as a field instead of describing a relativistic single-particle wave function. 
Paul Dirac also derived an equation for the relativistic massive particle by making use of the relation $E^2 = m^2c^4 + p^2c^2$ from the relativity into the Schr\"odinger equation.
There is no negative probability, but this equation gave negative energy states. It allows particles creation and annihilation. For example, an electron-positron pair can be annihilated to photons, or a photon can create a particle-antiparticle pair. Later we will see that by using the concept of field, the negative energy states are interpreted as the antiparticles or particles moving backward in time. But the concept of time is different in quantum mechanics than in relativity. 

Quantum mechanics deals differently with space and time. Space is an operator, but time is a parameter. But in the relativity theory, time and space are taken at equal footings. Hence, it is impossible to combine relativity with quantum mechanics. We need to make either time as an operator or make the position as a parameter. 
If we make time an operator, then things would get complicated because there is more than one notion of time. The one is the proper time that moves with the frame. The other notion of time is the coordinate on the clock in the stationary frame, which is then promoted to an operator. 
A more convenient approach is demoting the position coordinate to the parameter, which is a label of an operator. This operator is called the \textit{quantum field operator}. These field operators are obtained by quantization of the classical fields.

The classical field theory is based on the \textit{Lagrangian formalism}. To formulate the QFT, we start with the classical fields and then quantize them. This procedure is called the \textit{second quantization}. In quantum mechanics, the solutions to the relevant wave equations are states or particles. In QFT, the solutions are field operators that create or destroy states or particles. These operators are analogous to the creation and annihilation operators in a simple harmonic oscillator. Every fundamental particle has a corresponding field whose excitation is that particle. 
The quantized field concept enables us to write a many-body system in terms of a single object in space-time with the \textit{infinite degrees of freedom} so-called field oscillators. The field corresponding to the field operators are called free fields without interactions.

The interaction involves the product of operators that create and annihilate the virtual particles which mediate the forces. 
The scattering amplitudes in the interactions can intuitively be represented by the \textit{Feynman diagrams}. There can be contributions from the creation and annihilation of particles at an intermediate stage of the process. These contributions sometimes give diverging amplitudes. The method used to remove these contributions is called the \textit{renormalization}. Not all QFTs are renormalizable.

As in relativity, time is dealt on equal footings as space dimension, therefore we use four spacetime coordinates $(t,\bm{x}) = (t,x,y,z)=(x^0,x^1,x^2,x^3)$, three space and one time. The observables quantities are defined as the four-vectors. For example, $x=(t,\bm{x}),\ p^\mu=(E,\bm{p}), \ A^\mu= (\phi,\bm{A}), \ j^\mu = (\rho,\bm{j})$, and $\partial_\mu \equiv \partial/\partial x^\mu = (\partial/\partial t,\nabla)$, where the boldface letter means the three space components.

We are dealing with the relativistic object, it is convenient if we use units such that the most frequently used universal constants of nature $\hbar $ and $c$ are unity, so they would not appear in the equations. Such units are called the \textit{natural units}. Take $c=1$ and $\hbar = 1$ then derive other units from these. These two constants are also taken as dimensionless with similar reasoning.
Using these units, mass turns out to have the same unit as energy and momentum.

There are two approaches to QFT; canonical quantization or second quantization, and path integral. In the following sections, we present the basics of both the approaches without going into the tedious details of solving integrals. The canonical quantization and interactions are explained only for the case of the scalar fields. The section on the Lagrangian is needed for the both approaches, whereas the simple harmonic oscillator treatment, from quantum mechanics, is discussed for better understanding of the canonical quantization. The section on interactions is based on canonical quantization, but independent of the path integral and gauge theory.
For basic knowledge on quantum field theory, see \cite{huang2007fundamental,tom2015quantum,mcmahon2008quantum,klauber2013student}, for more detailed foundation, see \cite{peskin1995quantum,ryder1996quantum,weinberg1995quantum}, and for conceptual understanding, see \cite{auyang1995quantum,brooks2011fields,lederman2011symmetry,weinberg1977search}
	\section{Lagrangian}

Classically, the motion of a particle of mass $m$ under the influence of force $F$ is described by Newton's law given by
\eq{F=m\ddot {x} \label{Newton},}
where $x$ can be found by integrating this equation.
This equation is not quantum-friendly. The alternate approach is due to Joseph-Louis Lagrange and William Rowan Hamilton. Total energy, which is the sum of kinetic and potential energies, remains the same, but potential and kinetic energies may vary during the motion. We will calculate how kinetic energy and potential energy vary along the trajectory. Kinetic and potential energies are written as 
\eq{\bar{T} = \frac{1}{\tau} \int_{0}^{\tau} \frac{1}{2} m[\dot{x}(t)]^2dt, \qquad \bar{V} = \frac{1}{\tau} \int_{0}^{\tau} \frac{1}{2} V[x(t)]dt.}
We will treat average kinetic energy and average potential energy as a functional, which depends on the function $x(t)$. Any function $f(x)$ returns a value when we put some value of $x$. A functional $F[f(x)]$ returns some value when we put a function into it. The derivative of a function is written as $\frac{df}{dx}$ and the corresponding functional derivative is written as $\frac{\delta F}{\delta f(x)}$.

Now we will see how average kinetic energy and average potential energy  changes during the trajectory. Taking the derivatives of the functional $\bar{T}$ and $\bar{V}$ we get
\eq{\frac{\delta \bar{T}[x]}{\delta x(t)} = -\frac{m \ddot{x}}{\tau}, \qquad  \frac{\delta \bar{V}[x]}{\delta x(t)} = \frac{V^{'}[x(t)]}{\tau}.}
Since the force $F= -\frac{dV}{dt}$, the Eq. \ref{Newton} would become
$m\ddot {x}=-\frac{dV}{dt}$,
which implies 
\eq{\frac{\delta \bar{T}[x]}{\delta x(t)} =\frac{\delta \bar{V}[x]}{\delta x(t)}, \qquad \implies \frac{\delta}{\delta x(t)}(\bar{T}[x]- \bar{V}[x])=0 \label{diff}.}
This shows that the classical trajectory is the one for which the difference of kinetic and potential energies is stationary. If we deviate from the classical trajectory, then both average KE and PE increase by the same amount. The difference of kinetic energy (KE) and potential energy (PE) is \textit{Lagrangian} $L=T-V$, and the integral of the Lagrangian is called \textit{action}. It has the dimensions of energy$\cross$time, (Joule-second), the same units as $\hbar$ has. The action $S$ can be written as
\eq{S= \int_{0}^{\tau} Ldt = \int_{0}^{\tau} (T-V) dt= \tau (\bar{T}[x]-\bar{V}[x]),}
using the Eq \ref{diff} we get
\eq{\frac{\delta S}{\delta x(t)}=0 \label{leastaction}.}
This is known as the \textit{principle of least action}. It states that, the action is stationary for the trajectory followed by the particle. 
Lagrangian $L$ can be written as a function of position $x(t)$ and velocity $\dot{x}(t)$ then the variation of $S$ with position and velocity can be written as 
\eq{\frac{\delta S}{\delta x(t)} &= \int du \Big[ \frac{\delta L}{\delta x(u)} \frac{\delta x(u)}{\delta x(t)} + \frac{\delta L}{\delta \dot{x}(u)} \frac{\delta \dot{x}(u)}{\delta x(t)}\Big] \nonumber \\
	&= \int du \Big[ \frac{\delta L}{\delta x(u)}\delta (x-t) + \frac{\delta L}{\delta \dot{x}(u)} \frac{d}{dt}\delta (u-t)\Big] \nonumber\\
	&= \frac{\delta L}{\delta x(t)}+ \Big[\delta (x-t)\frac{\delta L}{\delta \dot{x}(u)}\Big]_{t_i}^{t_f} - \int du \delta(u-t)\frac{d}{dt}\frac{\delta L}{\delta \dot{x}(u)} \nonumber \\
	&= \frac{\delta L}{\delta x(t)} -\frac{d}{dt}\frac{\delta L}{\delta \dot{x}(u)}. \nonumber}
By using Eq. \ref{leastaction} we get the \textit{Euler-Lagrange equation} as below
\eq{\frac{\delta L}{\delta x(t)} -\frac{d}{dt}\frac{\delta L}{\delta \dot{x}(u)} = 0. \label{EL}}

In QFT, in terms of the four spacetime dimensions, the \textit{Lagrangian density} $\cal L$ is used that can be written as $L= \int dx \cal L$. In terms of the Lagrangian density, the action $S$ is written as $S= \int dt L = \int dt dx \cal L$.
Let the Lagrangian density $\cal L$ depends on a function $\phi(x)$ (we will take it as a field later) of a point $x$ in spacetime. 
\eq{S=\int d^4 x \cal L (\phi, \partial_\mu \phi) \label{Lfield}}
Here we are using four-vector notation, $\mu = 0,1,2,3$, and $x=(t,\textbf{x})=(x^0,x^1,x^2,x^3)$. A four-vector version of the Euler-Lagrange equation for this function would be 
\eq{\frac{\delta S}{\delta \phi} = 0, \qquad \frac{\partial \cal L}{\partial \phi} - \partial_\mu \Big(\frac{\partial \cal L}{\partial(\partial_\mu \phi)}\Big) = 0.\label{ELField}}
This equation is used to study the dynamics of the fields in quantum field theory.

\subsection{Continuum Mechanics}
The Hamiltonian is related to the energy of the system. It is conserved when the Lagrangian does not change with time. As the Lagrangian is a function of generalized coordinate and velocity, $L= L(q,\dot{q})$, we have 
\eq{\frac{dL}{dt} = \frac{\partial L}{\partial q_i} \dot{q}_i + \frac{\partial L}{\partial \dot{q}_i}\ddot{q}_i.}
The summation is assumed on the index $i$. 
Using the Euler Lagrange equation in the above equation, we get
\eq{\frac{dL}{dt} = \frac{d}{dt}\Big(\frac{\partial L}{\partial q_i}\Big) \dot{q}_i + \frac{\partial L}{\partial \dot{q}_i}\ddot{q}_i = \frac{d}{dt}\Big(\frac{\partial L}{\partial \dot{q}_i}\dot{q}_i \Big), 	\Rightarrow \ \frac{d}{dt}(p_i \dot{q}_i - L) = 0,}
where $p_i = \frac{\partial L}{\partial \dot{q}_i}$. 
Defining the Hamiltonian as 
\eq{H= p_i \dot{q}_i - L, \ \text{where} \ \frac{dH}{dt} = 0. \label{Hamiltonian}}
Now we will talk about another important concept \textit{Poisson bracket} that is given by
\eq{\left\{A,B\right\} = \frac{\partial A}{\partial q_i}\frac{\partial B}{\partial p_i} - \frac{\partial A}{\partial p_i}\frac{\partial B}{\partial q_i}.}
The rate of change of a function $F(q_i, p_i)$ can be written as
\eq{\frac{dF}{dt} &= \frac{\partial F}{\partial t} + \frac{\partial F}{\partial q_i}\dot{q}_i + \frac{\partial F}{\partial p_i}\dot{p}_i \nonumber \\
	& = \frac{\partial F}{\partial t} + \frac{\partial F}{\partial q_i}\frac{\partial H}{\partial p_i} - \frac{\partial F}{\partial p_i}\frac{\partial H}{\partial q_i} = \frac{\partial F}{\partial t} + \left\{F,H\right\}.}
Thus, if the function $F$ itself does not depend on time, then 
\eq{\frac{dF}{dt} =  \left\{F,H\right\}.}
It implies that if $ \left\{F,H\right\} = 0$ then $F$ is a constant of motion. There is a connection between the conservation laws with the Poisson bracket of Hamiltonian and $F$ being zero. In quantum mechanics, the Poisons bracket is replaced by a commutator.
The rate of change of expectation value of an operator $\hat{F}$ is
\eq{\frac{d\langle \hat{F} \rangle}{dt} = \frac{1}{i\hbar}\langle [\hat{F} , \hat{H}] \rangle .}
Poisson brackets from classical and the commutator in quantum mechanics are related as
\eq{\left\{F,H\right\} \rightarrow \frac{1}{i\hbar}\langle [\hat{F} , \hat{H}] \rangle .}
By using Eq. \ref{Hamiltonian}, we have
\eq{\delta H = p_i \delta \dot{q}_i + \delta p_i \dot{q}_i - \frac{\partial L}{\partial q_i} \delta q_i - \frac{\partial L}{\partial \dot{q}_i}\delta \dot{q}_i.}
Since $p_i = \frac{\partial L}{\partial \dot{q}_i}$, first term would cancel with last term, so that 
\eq{\delta H = \delta p_i \dot{q}_i - \frac{\partial L}{\partial q_i} \delta q_i \label{Heq1}.}
Since $H$ is the function of $q_i$ and $p_i$, we can also write
\eq{\delta H =  \frac{\partial H}{\partial q_i} \delta q_i + \frac{\partial H}{\partial p_i} \delta p_i \label{Heq2}.}
From the last two equations, $\frac{\partial H}{\partial q_i} = - \frac{\partial L}{\partial q_i}$. But from taking derivative of $p_i = \frac{\partial L}{\partial \dot{q}_i}$ we get $\dot{p}_i =\frac{d}{dt}\Big(\frac{\partial L}{\partial \dot{q}_i}\Big)$. Using the Euler-Lagrange equation here we get $\dot{p}_i = - \frac{\partial L}{\partial q_i}$. Comparing Eqs. \ref{Heq1} and \ref{Heq2} we get
\eq{\dot{q}_i = \frac{\partial H}{\partial p_i} \qquad
	\dot{p}_i = -\frac{\partial H}{\partial q_i}.}
These equations are called \textit{Hamilton's equations of motion}.
	\section{Symmetry}
Symmetry is one of the most important concepts in physics, described by group theory. When a system looks the same before and after some kind of transformation, then we say that the system has symmetry.
Symmetry can be global or local. Global symmetry is the one possessed by the entire system as a whole and applied to every point. But when a system transforms differently for different points on it, then the system has a local symmetry. The gauge symmetry is internal. The spacetime does not involve internal symmetry.
The global symmetry in quantum physics means something that cannot be measured. The local symmetry preserves causality, as the charge cannot be disappeared at one place and reappear at another place. It should travel from the former point to the latter.
The conservation laws are related to symmetries by the Lagrangian formalism. There is a transformation corresponding to every symmetry. For example, an invariance under the unitary transformation $U(1)$ means that by changing $\phi(x) \to \phi'(x)=e^{-i\theta}\phi(x)$ does not change the action. Therefore, symmetry leaves the equation of motion invariant. The translation, rotation, time, etc. are examples of continuous symmetries.
The Lagrangian is a scalar. It is unaffected by the Lorentz transformation that ensures that the action is invariant.
By using the Euler-Lagrange equation $$\frac{\partial L}{\partial x^\mu} - \frac{d}{dt}\Big(\frac{\partial L}{\partial \dot{x}^\mu}\Big)=0,$$ 
we get $p_\mu = \partial L/\partial \dot{x}^\mu$ and $\dot{p}_\mu= \partial L/ \partial x^\mu$.
If the Lagrangian $L$ does not depend on coordinates $x^\mu$, that is, when the Lagrangian is invariant with respect to the translation of components of the spacetime, then $\partial L/ \partial x^\mu =0$ and $p^\mu$ remains constant. We say that the momentum is a conserved quantity in this case. The Hamiltonian is a conserved quantity when the Lagrangian does not depend on time, and the energy of the system remains constant. 
When symmetry is broken, new phases appear. The new phase is described by a smaller group, comparing with that of the unbroken phase. The \textit{spontaneous symmetry breaking} is a process in which a symmetric state ends up in an asymmetric state. 

\subsection{Noether Theorem}
Symmetry and the conserved quantities are related according to Noether's theorem.
When something looks the same after rotation, then angular momentum is conserved. When something looks the same after some time translation, then energy is conserved. If there is invariance under the translation, then the linear momentum is conserved. The electric charge has the same value in all inertial frames, hence it is a Lorentz invariant. There is some kind of operators corresponding to the conserved quantities.

Let our spacetime coordinates change as $x^\mu \to x^\mu + a^\mu$. Under this transformation the field and the Lagrangian change as $\phi \to \phi(x+a) = \phi(x)+\delta \phi(x)$, and $\mathcal{L}=\mathcal{L}+\delta \mathcal{L}$. We want $\delta \mathcal{L} =0$. Since, 
\eq{\delta \mathcal{L} &= \frac{\partial \mathcal{L}}{\partial \phi}\delta \phi + \frac{\partial \mathcal{L}}{\partial (\partial_\mu \phi)}\partial_\mu (\delta \phi)\\
	&= \partial_\mu \Bigg(\frac{\partial \mathcal{L}}{\partial [\partial_\mu\phi]}\Bigg)\delta \phi + \frac{\partial \mathcal{L}}{\partial (\partial_\mu \phi)}\partial_\mu (\delta \phi)= \partial_\mu \Bigg(\frac{\partial \mathcal{L}}{\partial [\partial_\mu\phi]}\delta \phi\Bigg),}
where the Euler-Lagrange equation is used for the first term, suppose the change in a field is not changing the Lagrangian, this leads to the result
\eq{\partial_\mu \Bigg(\frac{\partial \mathcal{L}}{\partial [\partial_\mu\phi]}\delta \phi\Bigg) = 0, \implies \partial_\mu J^\mu =0,}
where $J$ is called the \textit{conserved current} and $Q$ is called the conserved charge written as
\eq{J^\mu =\frac{\partial \mathcal{L}}{\partial [\partial_\mu\phi]}\delta \phi, \ \implies \ Q = \int d^3xJ^0.}
These are the conserved quantities in a transformation. The energy-momentum tensor is a special case of the Noether theorem when the conserved charges are energy and momentum. The time component $T^0_0$ is the Hamiltonian density, which is the conservation of energy $\partial_0 T^0_0=0$. It is the energy density. The momentum density $T^0_i$, $i$ are three spatial components. The energy-momentum tensor is discussed in the Appendix \ref{Top}. 

\subsection{Discrete Symmetry}
The symmetry of the continuously varying parameters, like translation and rotation, are called \textit{continuous symmetry}. The parity, charge, and time-reversal symmetries are of fundamental importance in QFT. These are called \textit{discrete symmetries}. When a system is having symmetry, it means that the laws of nature are the same in the system transformed by the symmetry transformation.

If $\psi(x)$ is a solution to the Schr\"odinger equation, so is the $\psi(-x)$ with the same eigenvalue. They are related as $\psi(x) = \alpha \psi(-x)$. We say that the wave function has the \textit{parity} symmetry. The system has an even parity when $\alpha=1$ such that $\psi(-x)= \psi(x)$, and an odd parity when $\alpha = -1$ such that $\psi(-x)=-\psi(x)$. It is the reflection around the y-axis. The parity operator $P$ is defined as $P^2= I$, $P\psi(x) = \pm \psi(x)$. The spin-1/2 particle like quarks and electrons have positive parity, whereas the spin-1/2 antiparticles have negative parity. The parity is conserved in the electromagnetic and strong interactions but violated in the weak interaction. The violation of parity means the mirror image universe does not behave the same way as our universe.
The \textit{charge conjugation} operator $C$ changes a particle into an antiparticle and vice versa. As $C^2\ket{\psi} = C\ket{\bar{\psi}}= \psi$, so its eigenvalues are $\pm 1$. It is not conserved in weak interactions.
The system has the \textit{time reversal} symmetry when it remains the same while changing $\ket{\psi}$ to $\ket{\psi'}$, that is, the system evolves time flowing in the negative direction. For time-reversal symmetry, linear and angular momentum change the sign.

After the discovery of parity violation in 1960, scientists hoped that the CP will be preserved for the weak interactions. If the mirror symmetry is violated, then matter is also needed to be changed into antimatter. But CP violation was also observed for the weak interaction in 1964 in the $K$-meson decay. It means the antimatter universe is not the same as this universe. The CP violation might explain why there is more matter in the universe than antimatter.
The \textit{$CPT$ theorem} says that when $C, P$, and $T$ are taken together, the laws of nature remain invariant. The universe would be indistinguishable if parity is reversed, change particle to antiparticle and take momentum in the reverse direction. If one of the symmetries is violated, then the others are violated too.
	\section{Dirac Equation}
In Schr\"dinger equation, there is a second-order space derivative on the left side and the first-order time derivative on the right side,
\eq{\partial_t\psi(t,\bm{x})=-\frac{1}{2m}\nabla^2 \psi(t,\bm{x}).}
The natural units are used $\hbar=c=1$.
Relativity deals with space and time as a whole. In the relativity theory, we have the relation $E^2=m^2c^4 + p^2c^2$, which becomes $E^2=m^2 + p^2$ in the natural units.  
Putting the operator equivalent of this relation in Schr\"dinger equation, we get the \textit{Klein-Gordon equation},
\eq{(\partial^2 +m^2)\psi(x) = 0.}
The Klein-Gordon equation describes the motion of spin-0 fields, but it had the problems of negative energy and negative probability. The concept of quantum fields removed these problems, as we already discussed.

The Dirac equation describes the motion of a relativistic Fermi particle. Dirac wanted to make both the time and space derivative operators first order. Let us take the square root of the operator in Klein-Gordon equation as $\partial^2+m^2 = (\sqrt{\partial^2}+im)(\sqrt{\partial^2}-im)$. But the square root of an operator is meaningless. Dirac defined four vector $\gamma^\mu$ with the properties $\gamma^0 =1, \ \gamma^1 =-1, \ \gamma^2 =-1, \ \gamma^3 =-1$, and they anticommute $\left\{\gamma^\mu,\gamma^\nu\right\} = \gamma^\mu\gamma^\nu+\gamma^\mu\gamma^\nu=0$.
Now we can write
$(\partial^2+m^2) =((\gamma^\mu\partial_\mu)^2+m^2)= (\gamma^\mu\partial_\mu+im)(\gamma^\mu\partial_\mu-im)$.
If we take the factor with the plus sign and treat it as an operator as $(\gamma^\mu\partial_\mu+im)\psi(x)=0$, we get the Dirac equation from this as
\eq{(i\gamma^\mu \partial_\mu -m)\psi(x) = 0.}
Summation convention is assumed according to which there is a summation on $\mu = 0,1,2,3$. The $\gamma^\mu = (\gamma ^0,\gamma ^1,\gamma ^2,\gamma ^3) = (\gamma ^0,\bm{\gamma})$ is a four-vector. Its components are four by four covariant matrices, and the $\psi(x)$ of the Dirac equation must have the four-components wave function. Since $E_{\textbf{p}}^2=(\textbf{p}^2+m^2)$, the Dirac equation admits the negative energy solutions. 
The matrices $\gamma^\mu$ are not unique, but one way of writing these matrices is as follows
\begin{align}
	\gamma^0 = \begin{pmatrix}
		0 & I \\
		I & 0  
	\end{pmatrix}, \qquad \bm{\gamma} = \begin{pmatrix}
		0 & \sigma_i \\
		-\sigma_i & 0 
	\end{pmatrix},
\end{align}
where $\sigma_i$ are three \textit{Pauli matrices}.
The four-component wave function $\psi(x) = \begin{pmatrix} &\psi_L(x) \\ &\psi_R(x) \end{pmatrix}$ is called \textit{Dirac Spinors}. 
$\psi_R$ and $\psi_L$ are positive and negative frequency solutions of the Dirac equation. These two-component objects are called the \textit{Weyl spinors}. The positive frequency solution describes Fermi particles and negative frequency solution are describes Fermi antiparticles. Feynman's interpretation of the negative energy solution is that these are particles moving backward in time. These solutions can also be written as \cite{tom2015quantum}
\begin{align}
	u(p)e^{-ip\cdot x} = \begin{pmatrix} u_L(p) \\ u_R(p) \end{pmatrix} e^{-i p\cdot x}, \qquad v(p)e^{ip\cdot x} = \begin{pmatrix} v_L(p) \\ v_R(p) \end{pmatrix} e^{i p\cdot x}.
\end{align}  
When a particle's spin is in the same direction as its direction of motion, we say that it has a positive \textit{helicity}, otherwise, the particle has a negative helicity. $\psi_L$ and $\psi_R$ are the eigenstates of the helicity operator, with eigenvalues $+1$ and $-1$ respectively.  
The right-handed particles have positive helicity and the left-handed have negative helicity. But for the left-handed antiparticles, the helicity is positive and right-handed antiparticles have negative helicity. The system also has a property called \textit{chirality} when the system is not equal to its mirror image. It is a property described with respect to the inertial frames of reference. The \textit{parity} operator corresponds to the symmetry transformation between these two states. The four components in Dirac equations are because the Dirac particles have the same properties with parity reversed. The chirality and helicity are in general different, but they are the same for a massless particle.
The discovery of antiparticles predicted by the Dirac equation was a success story of merging relativity with quantum mechanics. The antiparticles in the Dirac equation have the same mass and spin, but have the opposite charge.

The Dirac equation is the Euler-Lagrange equation of the Lagrangian $\mathcal{L}= \bar{\psi}(i\gamma^\mu \partial_\mu -m)\psi(x)$, where $\bar{\psi}=\psi^\dagger\gamma^0$. We are not going into the details of the canonical quantization of the Dirac field. 

	\section{Simple Harmonic Oscillator} \label{SHO}
The theory of simple harmonic oscillator explains oscillatory motions in quantum mechanics. Assume a mass $m$ hanging with a spring with spring constant $K$ and displacement from the origin is $x$. This particle has momentum $p=m\dot{x}$, KE $=p^2/2m$ and PE $=\frac{1}{2}Kx^2$. Position and momentum are operators in quantum mechanics.
Then Schr\"odinger equation is given by
\eq{\hat{H}\ket{\psi} = \Big(-\frac{\hbar^2}{2m}\frac{\partial^2}{\partial x^2}+ \frac{1}{2}Kx^2 \Big) \ket{\psi} = E\ket{\psi}. \label{SchSHO}}
The solution of this equation is obtained by series-solution method which is somewhat involved and can be found in books on quantum mechanics. Energy values obtained from the solution have the form
\eq{E_n=\big(n+\frac{1}{2}\big)\hbar\omega, \label{EnergySHO}}
where $\omega = \sqrt{K/m}$, so $K= m\omega^2$. It can be seen from this equation that the energy levels for simple harmonic oscillator are \textit{equally-spaced}. We can write Hamiltonian as
\begin{align}
\hat{H} = \frac{\hat{p}^2}{2m} + \frac{1}{2} m\omega^2 \hat{x}^2.
\end{align}
We can make this Hamiltonian more useful by thinking of factorizing it.
\begin{align}
\hat{H} &= \frac{\hat{p}^2}{2m} + \frac{1}{2} m\omega^2 \hat{x}^2 - \frac{\hbar \omega}{2} + \frac{\hbar \omega}{2} \nonumber \\
& = \frac{1}{2}m \omega^2 \Big(\hat{x} -\frac{i}{m\omega} \hat{p}\Big)\Big(\hat{x} + \frac{i}{m\omega}\hat{p}\Big) + \frac{1}{2}\hbar \omega \label{H}.
\end{align}
We added and subtracted $\hbar \omega /2$, because when the factors in parenthesis are multiplied we get extra term due to commutator of $\hat{x}$ and $\hat{p}$ that is $[\hat{x}, \hat{p}] = i\hbar$.

Factors in above equation are adjoint of each other. In non-relativistic quantum mechanics, problem of the simple harmonic oscillator is solved by the introduction of creation and annihilation operator for quanta. We name them as $\hat{a}$ and $\hat{a}^\dagger$ with some constants in front which are useful later calculations. 
\begin{align}
\hat{a}= \sqrt{\frac{m\omega}{2 \hbar}}\Big(\hat{x} + \frac{i}{m\omega}\hat{p}\Big), \qquad \hat{a}^\dagger = \sqrt{\frac{m\omega}{2 \hbar}} \Big(\hat{x} - \frac{i}{m\omega}\hat{p}\Big).
\end{align}
Since $x$ and $p$ are operators, $a$ and $a^{\dagger}$ also operators and their commutation relation is written as
\begin{align}
[\hat{a},\hat{a}^{\dagger}] &= \frac{m\omega}{2 \hbar} \Big(-\frac{i}{m\omega} [\hat{x}, \hat{p}] +  \frac{i}{m\omega}[\hat{p},\hat{x}]\Big) = 1.
\end{align}
These can be inverted and get $\hat{x}$ and $\hat{p}$,
\begin{align}
\hat{x} = \sqrt{\frac{\hbar}{2m \omega}}(\hat{a} + \hat{a}^\dagger) \qquad \hat{p} = - i \sqrt{\frac{\hbar m\omega}{2}}(\hat{a}-\hat{a}^\dagger).
\end{align}
Putting $\hat{a}$ and $\hat{a}^\dagger$ in \eqref{H}, we get Hamiltonian in the form,
\begin{align}
\hat{H} = \hbar \omega \Big(\hat{a}^\dagger \hat{a} + \frac{1}{2}\Big).
\end{align}

The product $\hat{a}^\dagger a$ is called \textit{number operator} by comparison with \eqref{EnergySHO}. Let $\hat{a}^\dagger a$ has eigenstate $\ket{n}$ with eigenvalue $n$ then $\ket{n}$ also the eigenstate of $\hat{H}$ with eigenvalue $ \hbar \omega (n + \frac{1}{2})$. Therefore, we can define a number operator and its eigenvalue equation as
$$ \hat{n} = \hat{a}^\dagger \hat{a}, \qquad \hat{n} \ket{n} = n \ket{n}.$$
Here $n$ is the energy level or number of quanta with energy $\hbar \omega$ each. Now Hamiltonian would become
\begin{align}
\hat{H} = \hbar \omega \Big(\hat{n} + \frac{1}{2}\Big) \qquad
\hat{H}\ket{n} = \Big(n + \frac{1}{2}\Big) \hbar \omega \ket{n}.
\end{align}
How $\hat{a}$ and $\hat{a}^\dagger$ act on the state $\ket{n}$? To see this we operate the number operator $\hat{n}$ on the state $\hat{a}^\dagger \ket{n}$, 
$$ \hat{n}\hat{a}^\dagger \ket{n} = \hat{a}^\dagger a \hat{a}^\dagger \ket{n}.$$
Since $[\hat{a}, \hat{a}^\dagger] = 1$ then $\hat{a}\hat{a}^\dagger - \hat{a}^\dagger \hat{a} =1$, which implies $\hat{a}\hat{a}^\dagger = 1 + \hat{a}^\dagger \hat{a}$, therefore,
\begin{align}
\hat{n}\hat{a}^\dagger \ket{n} = (n+1) \hat{a}^\dagger \ket{n}.
\end{align}
On similar lines we found that
$$ \hat{n}\hat{a} \ket{n} = (n-1)\hat{a} \ket{n}.$$
It implies $\hat{a}^\dagger$ and $\hat{a}$ have the effect of adding one quantum of energy and subtracting one quantum of energy. For this reason $\hat{a}^\dagger$ is called \textit{raising operator}, and $\hat{a}$ is called \textit{lowering operator}.

Normalization of $\hat{a}$ and $\hat{a}^\dagger$ is as follows:
As $\hat{a}$ is lowering operator we can write $\hat{a}\ket{n} = k \ket{n-1}$, and so
\begin{align}
\abs{\hat{a}\ket{n}}^2 = \abs{k}^2 \bra{n-1}\ket{n-1} = \abs{k}^2.
\end{align}
But
\begin{align}
\abs{\hat{a}\ket{n}}^2 = \bra{n}\hat{a}\hat{a}^\dagger \ket{n}
= \bra{n}\hat{n} \ket{n} = n
\end{align}
Comparing these two equations we get $k = \sqrt{n}$. On similar lines
\begin{align}
\abs{\hat{a}^\dagger \ket{n}}^2 = \abs{c}^2 \bra{n+1}\hat{a}\hat{a}^\dagger \ket{n+1} = \abs{c}^2.
\end{align}
But
\begin{align}
\abs{\hat{a}^\dagger\ket{n}}^2 = \bra{n}\hat{a}\hat{a}^\dagger \ket{n}  = \bra{n}(1+ \hat{a}^\dagger\hat{a} )\ket{n} = \bra{n}(1+ \hat{n} )\ket{n} = n+1.
\end{align}
Therefore, we get $c = \sqrt{n+1}$. Now $\hat{a}$ and $\hat{a}^\dagger$ are applied to a number state as
\begin{align}
\hat{a}\ket{n} = \sqrt{n}\ket{n-1} \qquad
\hat{a}^\dagger\ket{n} = \sqrt{n+1}\ket{n+1}.
\end{align}
From the equations above we see that applying $\hat{a}$ lowers the quanta, but what when we get $\ket{0}$ state. Here the state is annihilated, that is $\hat{a}\ket{0} = 0$. This state $\ket{0}$ is called ground state of harmonic oscillator. Now we will see the effect of Hamiltonian on this state
\begin{align}
\hat{H}\ket{0} = \hbar \omega \big(\hat{n} + \frac{1}{2}\big)\ket{0} = \frac{1}{2}\hbar \omega \ket{n}.
\end{align}
The eigenvalue of ground state is non-zero and non-integral. This energy $\frac{1}{2}\hbar \omega$ of ground state is called \textbf{zero point energy}.

On the other hand if we keep applying raising operator on ground state $\ket{0}$ we will get states $\ket{n}$ with more quanta,
\begin{align}
\hat{a}^\dagger\ket{0} &= \ket{1}, \nonumber \\
\hat{a}^\dagger\ket{1} &= \sqrt{2}\ket{2} \Rightarrow \ket{2} = \frac{\hat{a}^\dagger}{\sqrt{2}}\ket{1} = \frac{(\hat{a}^\dagger)^2}{\sqrt{2}}\ket{0}, \nonumber \\
\hat{a}^\dagger\ket{2} &= \sqrt{3}\ket{3} \Rightarrow \ket{3} = \frac{\hat{a}^\dagger}{\sqrt{3}}\ket{2} = \frac{(\hat{a}^\dagger)}{\sqrt{3}}\frac{(\hat{a}^\dagger)^2}{\sqrt{2}}\ket{0} = \frac{(\hat{a}^\dagger)^3}{\sqrt{3\cdot 2}} \ket{0}, \nonumber   \\
\Rightarrow \ket{n} &= \frac{(\hat{a}^\dagger)^n}{\sqrt{n!}} \ket{0}.
\end{align}
From this discussion, we can say that $\hat{a}^\dagger$ creates the one quantum of energy $\hbar \omega$ and $\hat{a}$ annihilates the one quantum of energy or a quanta, which behave like particles. Due to this the operator $\hat{a}^\dagger$ also called \textit{creation operator} and $\hat{a}$ is called \textit{annihilation operator}. Analogously, we will discuss creation and annihilation operators for fields in the next section.

	\section{Second Quantization}

The dynamical variables from classical mechanics are promoted to operators in quantum mechanics. This is the first quantization.
The dynamical variables are quantized to operators which act on the wave function. As the wave functions are obtained in first quantization, and we are quantizing it again, hence the name second quantization. The first quantization is associated with particles and the second quantization is associated with the fields. The classical particles are total values, whereas the classical fields are density values. As we already saw in one of the previous sections, the Poisson bracket from classical mechanics is promoted to the commutator in quantum mechanics. Analogously, in the second quantization, we will promote classical fields values to the commutators of fields and their conjugate values.

Here we will consider the steps to second quantize the fields \cite{tom2015quantum}, and later we will interpret it to make sense of it. Let us take an example of the massive free scalar field. The Euler-Lagrange equation for this Lagrangian leads to the Klein-Gordon equation $(\partial^2 + m^2)\phi = 0$.

\textbf{Step I:} First write the Lagrangian density for the field as
\begin{align}
	{\cal{L}} &= \frac{1}{2}[\partial_\mu \phi(x)]^2 - \frac{1}{2} m^2 [\phi(x)]^2 = \frac{1}{2}[(\partial_0 \phi)^2 - (\nabla \phi)^2 - m^2 \phi^2].
\end{align}
The Hamiltonian density can be written as 
\begin{align}
	{\cal{H}} &= \Pi^0(x) \partial_0\phi(x) - {\cal{L}} \nonumber\\
	& = \partial^0\phi(x)\partial_0\phi(x) - {\cal{L}} \nonumber \\
	& = \frac{1}{2} [(\partial_0 \phi(x))^2 + (\nabla \phi)^2 + m^2 (\phi(x))^2].
\end{align}

\textbf{Step II:} At the second step, momentum density is written as
\begin{align}
	\Pi^\mu(x) = \frac{\partial {\cal{L}}}{\partial (\partial_\mu \phi(x))} = \partial^\mu \phi(x).
\end{align}
The time component of the momentum density is $\Pi^0(x) = \pi(x) = \partial^0 \phi(x)$.

\textbf{Step III:} Next step is turning the fields into operator-valued fields, by taking $\phi(x)$ to $\hat{\phi}(x)$ and $\Pi^0(x)$ to $\hat{\Pi}^0(x)$. As $[\hat{x}, \hat{p}]$ is the single particle commutator in quantum mechanics, by the same analogy, we can define the \textit{equal-time commutator} for the field operators,
\begin{align}
	[\hat{\phi(t, \bm{x})}, \hat{\Pi^0(t, \bm{y})}] = i \delta^{(3)}(\bm{x}-\bm{y}).
\end{align}
The operators at different time commute,
\begin{align}
	[\hat{\phi(x)}, \hat{\phi(x)}] = [\hat{\Pi^0(x)}, \hat{\Pi^0(x)}]= 0.
\end{align}

\textbf{Step IV:} There is an analogy between quanta in oscillators and particles in momentum space. We will expand the field in terms of these creation and annihilation operators. The field operators is written as
\begin{align}
	\hat{\phi}(\bm{x}) = \int \frac{d^3 p}{(2\pi)^{3/2}}\frac{1}{(2E_p)^{1/2}} (\hat{a}_p e^{i\bm{p}\cdot \bm{x}} + \hat{a}^\dagger e^{-i\bm{p}\cdot \bm{x}}).
\end{align}
This expansion is sometimes called the \textit{mode expansion}. Here, the creation and annihilation operators have the relation $[\hat{a}. \hat{a}^\dagger]= \delta^{(3)}(\bm{x}-\bm{y})$, and $E_{\bm{p}} = \sqrt{\bm{p}^2+m^2}$. The choice of the normalizing factor is to make the integral Lorentz-invariant \cite{tom2015quantum}.

\subsubsection{Interpretation}
Now we will discuss what does it all means. The first quantization says that particles behave like waves described by the Schr\"odinger equation.  
The first quantization promotes the Poisson bracket to the commutator
\eq{\left\{x^i,p_j\right\} = \delta^i_j \ \underrightarrow{1st \ quantization} \ [x^i,p_j] = i\hbar \delta^i_j. \nonumber}
The quantization of classical theory of fields to quantum theory of field is the second quantization
\eq{\left\{\phi^r(x,t),\pi_s(y,t)\right\} = \delta^r_s \delta(x-y)\ \underrightarrow{2nd \ quantization} \ [\phi^r(x,t),\pi_s(y,t)] = i\hbar \delta^r_s \delta(x-y). \nonumber}
This implies that the wave function is imagined as a field, and we are quantizing this field to get the operators. This wave function is quantized to the field operators. 
As canonical variables are used, hence both quantizations are also called \textit{canonical quantizations}. The dynamics of fields are studied by the Lagrangian formulation, in which the generalized coordinate is replaced by the field. 

We can look at the fields as a time-dependent mapping from each point $x$ to $\phi(x,t)$ so, it has an infinite number of degrees of freedom. This mapping is a real number in the case of a classical field and an operator in the case of a quantum field. The fields are something different from matter and energy. The momentum can be ascribed to them even in the absence of matter. 

The wave function in quantum mechanics maps the position to the probability of finding a particle at position $x$. For QFT, analogous to positions are classical fields $\phi(x)$, and analogous to the wave function, we have wave functional $\psi[\phi(x)]$. It maps functions to the numbers. Now $\abs{\psi[\phi(x)]}^2$ is interpreted as the probability that we will find $\psi(x)$ when measured. The quantum mechanical superposition of particle states is analogous to a superposition of classical field configurations.

The Hamiltonian density is the sum of kinetic and potential energies of the field. Kinetic energy is given by the term that corresponds to the change in configuration space in time and potential energy term reflecting the energy cost for spatial changes in the field and is related with the mass term.
The commutation relation in quantum mechanics has objects with three degrees of freedom. The field operators create and destroy particles at each spacetime point, therefore the commutation relation in QFT has infinite degrees of freedom. Each degree of freedom is with 4-tuple spacetime. 

For a simple harmonic oscillator we discussed in the last section
\eq{\ket{n_1...n_N} = \frac{1}{\sqrt{n_1!...n_N!}}(a^\dagger)^{n_1}...(a^\dagger)^N\ket{0...0}.}
In a similar way, the field operators are applied on the Hilbert space built on the vacuum. Vacuum is a state with zero particle. There is a harmonic oscillator at every spacetime point.
The quantum wave function is acted upon by operators, whereas the field operators itself act on the state of space. Analogous to the zero-point energy in a simple harmonic oscillator, we can write the energy of the vacuum as
\eq{\hat{H}= \frac{1}{2}\int d^3p E_p(\hat{a}\hat{a}^\dagger_p+\hat{a}^\dagger_p \hat{a}), \implies E=\int E_p(\hat{a}^\dagger_p \hat{a}+\frac{1}{2}\delta^3(0)).}
The second term integral gives infinity contribution to the total vacuum energy, which is nonsensical. To avoid this infinity, we need normal ordering of the operators in the Hamiltonian above. The \textit{normal ordering} is arranging all the creation operators to the left. It will give us the difference of energy between the configurations instead of the total energy. 

\subsubsection{So what is QFT?} 
It is all about choosing a suitable Lagrangian that includes the free field terms and the interaction terms. This field is quantized by second quantization and the interaction is calculated by the $S$-matrix discussed in the next section. The simplest Lagrangian is the mass-less \textit{scalar} field. The spin-0 fields are scalar fields, spin-1/2 are \textit{spinors}, spin-1 are \textit{vector} and spin-2 fields are \textit{tensor} fields. \textit{Real} fields are charge-less, but the \textit{complex} fields are charged. With the increase of the parameters of the field, the form of Lagrangian and solutions change. On similar lines, more terms are added to the Lagrangian. The Lagrangian of the grand unified theory (GUI) or the standard model has the terms of fields and their interaction terms related to the three interactions QED, QCD, and weak interaction. The fourth interaction, quantum gravity, is not yet unified with the other three forces. The interactions involve the exchange of some \textit{virtual} particles. This exchange may change the properties of the particle involved in the interaction. So we call the interacting particle before the exchange of the virtual particles as the \textit{incoming} particles, and after the exchange of the virtual particles as the \textit{outgoing} particles.

We have negative and positive solutions in the mode expansion. As we will discuss at the end of this section, the solution $e^{-ip.x}$ is a plane wave corresponds to the incoming particles. It is annihilated in the system. In relativity, $E^2=p^2c^2 + m^2c^4$, which is equal to $E^2=p^2+m^2$ in natural units $\hbar=c=1$. Therefore, $E_{\bm{p}} = \pm \sqrt{\bm{p}^2+m^2}$ and we cannot just ignore the negative energy solution. The mode expansion includes this negative energy solution as
\eq{\hat{\phi}(x) =\sum_{\bm{p}} \text{positive }E_{\bm{p}} \text{ mode} + \sum_{\bm{p}} \text{negative }E_{\bm{p}} \text{ mode}.}
According to the Feynman interpretation of the negative energy states, these are outgoing antiparticles. These outgoing antiparticles are created and have a factor $e^{ip.x}$. 
\eq{\hat{\phi}(x) =&\sum_{\bm{p}} \text{incoming positive energy }E_{\bm{p}} \text{ particle annihilated} + \nonumber \\
	&\sum_{\bm{p}} \text{outgoing positive energy }E_{\bm{p}} \text{ antiparticle created}.}
Now the mode expansion can be written as
\begin{align}
	\hat{\phi}(\bm{x}) = \int \frac{d^3 p}{(2\pi)^{3/2}}\frac{1}{(2E_p)^{1/2}} (\hat{a}_p e^{i\bm{p}\cdot \bm{x}} + \hat{b}^\dagger e^{-i\bm{p}\cdot \bm{x}}).
\end{align}
Here $\hat{a}$ and $\hat{a}^\dagger$ annihilate and create particles, whereas $\hat{b}$ and $\hat{b}^\dagger$ annihilate and create antiparticles. When the spin is zero, then particles are their own antiparticles.

The integrals related to the incoming, outgoing and the propagation of the virtual particles, are computed here.
These integrals will be useful in the next section.
Since $\hat{a}^\dagger \ket{0} = \ket{\bm{p}}$, we have
\eq{\hat{\phi}(x) \ket{0} = \int \frac{d^3p}{(2\pi)^{3/2}(2E_{\bm{p}})^{1/2}}e^{ip\cdot x}\ket{\bm{p}}.}
This is an outgoing superposition of states. The relativistic normalized state can be written as $\bra{q} = (2\pi)^{3/2}(2E_{\bm{p}})^{1/2}\bra{\bm{q}}$. It implies that
\eq{(2\pi)^{3/2}(2E_{\bm{p}})^{1/2}\bra{\bm{q}}\hat{\phi(x)}\ket{0} = \int d^3 p e^{ip\cdot x}\bra{\bm{q}}\ket{\bm{p}} = \int d^3xe^{ip\cdot x}\delta^(3)(\bm{q-p}) = e^{iq\cdot x}. \label{WaveSolution}}
Similarly, we can calculate for the incoming states as $e^{-iq\cdot x}$.
The virtual particle is represented by the \textit{Feynman propagator} and is given by 
\eq{\bra{0}T[\phi(x)\phi(y)]\ket{0} = \Delta(x-y) = \lim_{\epsilon \to 0} \int \frac{d^4k}{(2\pi)^4}\frac{e^{-ik\cdot(x-y)}}{k^2-m^2 +i\epsilon}.\label{FeynmanProagator}}
where $T$ is for time ordering. It is computed using the contour integral. See \cite{klauber2013student} or \cite{tom2015quantum} for detailed calculation.
	\section{Interactions}
By canonical quantization, quantum field theory is exactly solvable only for free particles. For the interactions, some approximate methods, like perturbation theory, are used. Particles are created and destroyed in the interactions. Free particles are fired on each other, they interact for a short time, then they go far apart. This interaction is described by the S-matrix.

The Hamiltonian can be written as the sum of free part and interaction part $\hat{H}= \hat{H_0} + \hat{H}^{'}$. The free part of the Hamiltonian describes non-interacting particles.
In the Heisenberg picture, state-vectors are independent of time, and in the Schr\"odinger's picture, states evolve with time, but operators are independent of time. Whereas, in \textit{interaction picture}, both the operators and states evolve with time. The operators evolve by the free part of the Hamiltonian $\hat{\phi_I(t)}= e^{i\hat{H}_0} \hat{\phi}e^{-i\hat{H}_0t}$, but states evolve according to the interaction part of the Hamiltonian as $i\frac{\partial}{\partial t}\ket{\psi(t)}_I = \hat{H}_I(t)\ket{\psi(t)}_I$, where $\hat{H}_I(t) = e^{i\hat{H}_0}\hat{H'}e^{-i\hat{H}_0}$. 
When particles are far apart at the start and at the end, then the interaction part $\hat{H}^{'}$ is zero. When $\hat{H}_I = 0$ then we have the Heisenberg picture, and we have free part $\hat{H}_0$.
States at the start and at the end are represented as $\ket{\phi} = \ket{\phi_I(\pm \infty)}$. These states are the eigenstates of $\hat{H}_0$. The free part allows particles from vacuum $\ket{0}$ by using creation and annihilation operators. This vacuum is the vacuum of $\hat{H}_0$. It is called free, non-interacting, or bare vacuum. 
When the particles interact, $\hat{H}_I$ is nonzero and states evolve. Operators also evolve with time according to $\hat{H}_0$. We will use the operator $\hat{\phi}(x)$ and its expansion in terms of creation and annihilation operators. 

Let, we have two particles in a momentum state $\ket{\psi} = \ket{p_2p_1}$. This is the state when two particles are far apart $(t\rightarrow - \infty)$. We call this 'in' state and write as $\ket{\psi} = \ket{p_1p_2}_{in}$. After the interaction, two particles again get apart at $t\rightarrow \infty$ then we will call this state 'out' state $\ket{q_2q_1}_{out}$. The amplitude of scattering with particles starting at state $\ket{p_2p_1}_{in}$ and ending at state $\ket{q_2q_1}_{out}$ is given by
\begin{align}
	{\cal{A}} = _{out}\bra{q_1q_2}\hat{S}\ket{p_2p_1}_{in}.
\end{align}
Now we will go behind the $\hat{S}$-operator. At $t=0$ we will write the free part of operators and states
\begin{align}
	\bra{\phi}\hat{S}\ket{\psi} = _{out}\bra{\phi}\ket{\psi}_{in} = \bra{\phi_I(0)}\ket{\psi_I(0)}.
\end{align}
Let $\hat{U}_I(t_2,t_1)$ be the time evolution operator in the interaction picture, then
\begin{align}
	\bra{\phi}\hat{S}\ket{\psi} = \bra{\phi_I(\infty)}\hat{U}_I(\infty,-\infty)\ket{\psi_I(-\infty)}= \bra{\phi}\hat{U}_I(\infty, -\infty)\ket{\psi}.
\end{align}
The interaction Hamiltonian at different times may not commute. So we need to define \textit{time-ordered product} written as $T[\hat{A}_1(t_1)\hat{A}_2(t_2) ...\hat{A}_n(t_n)]$.
\textit{Time ordering} keep operators on the right which are earlier in time and keeps operator on the left which are later in time. It is not an operator, but just to tell how we should arrange the strings. Time ordering makes things in time ordering commute. It is analogours to the \textit{normal ordering} $\bra{0}N[\hat{A}\hat{B}\hat{C}...\hat{Z}]\ket{0}$ which places the creation operators to the left and the annihilation operators to the right. As the unitary operator can be written in terms of the Hamiltonian as 
\begin{align}
	\hat{U}_I(t_2,t_1) = T\big[e^{-i\int_{t_1}^{t_2} dt \hat{H}_I(t)}\big].
\end{align}
Now $\hat{S}$-operator in the limit of $t\rightarrow \infty$ after replacing Hamiltonian with the Hamiltonian density can be written as
\begin{align}
	\hat{S} = T\big[e^{-i\int_{-\infty}^{\infty} dx^4 \hat{{\cal{H}}}_I(t)}\big].
\end{align}
This integral cannot be easily done. A more useful form is when we expand the exponential
\begin{align}
	\hat{S} = T\Big[ 1- i \int d^4z \hat{{\cal{H}}}_I(z) + \frac{(-1)^2}{2!}\int d^4y d^4 w \hat{\cal{H}}_I (y) \hat{\cal{H}}_I(w) + ...\Big].
\end{align}
This is called \textit{Dyson expansion}. This expansion is valid when $\hat{\cal{H}}_I(x)$ is small. We integrate over different spacetime for each term. 

\subsection{Wick's Theorem}
When \textit{vacuum expectation value} (VEV) of the operators is evaluated, we encounter a kind of correlator  $\bra{0}T[\hat{A}\hat{B}\hat{C}...\hat{Z}]\ket{0}$. This is the time-ordered operator product, and it is difficult to calculate comparing with the normal ordered products.  
Wick's theorem is a way to relate the time ordering with the normal ordering to evaluate it easily.
The mode expansion of a field operator $\hat{\phi}$ has two parts: creation and annihilation. Let us call those parts briefly as $\hat{\phi}^+$ and $\hat{\phi}^-$. We have $\bra{0}\hat{\phi}^+ = 0$ and $\hat{\phi}^- \ket{0}=0$.
Let us compute the product of two field operators $\hat{A}$ and $\hat{B}$ as
$$\hat{A}\hat{B} = (\hat{A}^+ + \hat{A}^-)(\hat{B}^+ + \hat{B}^-) = \hat{A}^+\hat{B}^+ + \hat{A}^-\hat{B}^- + \hat{A}^+\hat{B}^- + \hat{A}^-\hat{B}^,.$$
$$	N[\hat{A}\hat{B}] = \hat{A}^+\hat{B}^+ + \hat{A}^-\hat{B}^- + \hat{A}^+\hat{B}^- +\hat{B}^+\hat{A}^-.$$
We can see that only the last term of the above equations is different. So we can write
$$	\hat{A}\hat{B} - N[\hat{A}\hat{B}] = \hat{A}^-\hat{B}^+ - \hat{B}^+\hat{A}^- = [\hat{A}^- , \hat{B}^+].$$
Since 
$$	T[\hat{A}(x)\hat{B}(y)] =  
\begin{cases}
	& \hat{A}(x)\hat{B}(y) \ \ x^0 > y^0 \\
	& \hat{B}(y)\hat{A}(x) \ \ x^0 < y^0
\end{cases}$$
is the Feynman propagator $\bra{0}T[\phi(x)\phi(y)]\ket{0} = \Delta(x-y)$ as in Eq. \ref{FeynmanProagator},
we can also write time ordering for the fields $\hat{A}$ and $\hat{B}$ with respect to spacetime coordinates $x^0$ and $y^0$ as
$$	T[\hat{A}(x)\hat{B}(y)] - N[\hat{A}\hat{B}] =  
\begin{cases}
	& [\hat{A}^-(x),\hat{B}^+(y)] \ \ x^0 > y^0 \\
	& [\hat{B}^-(y),\hat{A}^+(x)] \ \ x^0 < y^0
\end{cases}.
$$
The normal ordered products have VEV zero, we can write
$$	\bra{0}T[\hat{A}(x)\hat{B}(y)]\ket{0} = 
\begin{cases}
	& \bra{0}[\hat{A}^-(x),\hat{B}^+(y)]\ket{0} \ \ x^0 > y^0 \\
	& \bra{0}[\hat{B}^-(y),\hat{A}^+(x)]\ket{0} \ \ x^0 < y^0
\end{cases}.$$
Define the Wick contraction as
\eq{\contraction{}{\hat{A}}{\hat{B}}{} \hat{A}\hat{B} = T[\hat{A}\hat{B}] - N[\hat{A}\hat{B}].}
This Wick's contraction is a \textit{c-number} and have the value 
\eq{T[\hat{A}\hat{B}] = N[\hat{A}\hat{B} + \contraction{}{\hat{A}}{\hat{B}}{}\hat{A}\hat{B}] = N[\hat{A}\hat{B}] + \contraction{}{\hat{A}}{\hat{B}}{}\hat{A}\hat{B}.}
When generalized to many operators yield Wick theorem,
\eq{T[\hat{A}\hat{B}...\hat{Z}] = N[\hat{A}\hat{B}...\hat{Z} + \text{all contractions of }\hat{A}\hat{B}...\hat{Z}].}
When applied only to free field we get
\eq{\bra{0}\hat{\phi}(x_1)\hat{\phi}(x_2)\hat{\phi}(x_3)\hat{\phi}(x_4) \ket{0} &= \Delta(x_1-x_3)\Delta(x_3-x_4) 
	+ \Delta(x_1-x_3)\Delta(x_2-x_4) \nonumber \\ 
	&+ \Delta(x_1-x_4)\Delta(x_2-x_3).}
The time ordering is equal to the normal ordering plus contraction. The contraction is a commutator or propagator of VEV.
	\subsection{Feynman Diagrams}
The scattering process can be shown pictorially by Feynman diagrams. These diagrams are not only more intuitive, but we can also solve the scattering problem using Feynman rules without going into complicated mathematics. Suppose time is in the upward direction. Each term in the perturbation expansion of the S-matrix is represented by a diagram. Particles are represented by arrows in the direction of time, as in Fig. \ref{FeynmanDiagrams} (a), and antiparticles are represented by arrows in the downward direction, as in Fig. \ref{FeynmanDiagrams} (b). Antiparticles are equivalent to particles moving backward in time. The filled circles correspond to the scattering of particles. 

By convention for interaction diagrams, $\psi$ is represented as a line with an arrow toward the vertex of the interaction, whereas $\psi^\dagger$ represented by an arrow going out of the vertex. $g$ is called \textit{coupling constant} gives the strength of interaction. 
In Fig. \ref{FeynmanDiagrams} (c), a particle-antiparticle pair creation at time $t=t_0$ and Fig. \ref{FeynmanDiagrams} (d) is the annihilation of particles and antiparticles. In Fig. \ref{FeynmanDiagrams} (e), a particle-antiparticle pair is created at $t=t_2$ and the antiparticle is annihilated with the incoming particle. The net result is moving the particle from an initial to final position in spacetime. The source term is represented by the diagram as in Fig. \ref{FeynmanDiagrams} (f). The Figs. \ref{FeynmanDiagrams} (g), (h), and (i) are two-particle interactions in electromagnetic, weak, and strong force respectively. When taking the time in upward direction, two particles coming close to each other, interact and exchange some virtual particles then scattered off. We can equally interpret it as; a pair of a particle and an antiparticle came close, fused to make a gauge particle, and then at another spacetime point, the pair is created back. 
\begin{figure}[h!]
	\centering
	\begin{subfigure}{0.1\textwidth}
		\centering
		\tikzset{middlearrow/.style={decoration={markings,mark= at position 0.5 with {\arrow{#1}},},postaction={decorate}}}
		\begin{tikzpicture}
			\draw[ultra thick,blue,middlearrow={>}] (0,0)--(0,2);
		\end{tikzpicture}
		\caption{}
	\end{subfigure}
	\begin{subfigure}{0.1\textwidth}
		\centering
		\tikzset{middlearrow/.style={decoration={markings,mark= at position 0.5 with {\arrow{#1}},},postaction={decorate}}}
		\begin{tikzpicture}
			\draw[ultra thick,blue,middlearrow={<}] (0,0)--(0,2);
		\end{tikzpicture}
		\caption{}
	\end{subfigure}
	\begin{subfigure}{0.1\textwidth}
		\centering
		\tikzset{middlearrow/.style={decoration={markings,mark= at position 0.5 with {\arrow{#1}},},postaction={decorate}}}
		\begin{tikzpicture}
			\draw[ultra thick,blue,middlearrow={>}] (-1,1)--(0,0);
			\draw[ultra thick,blue,middlearrow={>}] (0,0)--(1,1);
			\draw[dashed,thick,blue] (-0.5,0)--(0.5,0);
			\node[below] at (0,0) {$t=t_0$};
		\end{tikzpicture}
		\caption{}
	\end{subfigure}
	\begin{subfigure}{0.2\textwidth}
		\centering
		\tikzset{middlearrow/.style={decoration={markings,mark= at position 0.5 with {\arrow{#1}},},postaction={decorate}}}
		\begin{tikzpicture}
			\draw[ultra thick,blue,middlearrow={>}] (-1,0)--(0,1);
			\draw[ultra thick,blue,middlearrow={>}] (0,1)--(1,0);
			\draw[dashed,thick,blue] (-0.5,1)--(0.5,1);
			\node[above] at (0,1) {$t=t_1$};
		\end{tikzpicture}
		\caption{}
	\end{subfigure}
	\begin{subfigure}{0.2\textwidth}
		\centering
		\tikzset{middlearrow/.style={decoration={markings,mark= at position 0.5 with {\arrow{#1}},},postaction={decorate}}}
		\begin{tikzpicture}[use Hobby shortcut,scale=1]
			\begin{knot}[
				consider self intersections=true,
				ignore endpoint intersections=false,
				flip crossing=2,
				only when rendering/.style={}]
				\strand [ultra thick,blue,middlearrow={>},->](0.2,0)..(0.3,0.33)..(0.45,0.67)..(1,1).. (1.5,0.5)..(2.2,1)..(2.4,1.6);
			\end{knot}
			\draw[ultra thick,blue,dashed] (0.3,1.01)--(1.5,1.01);
			\draw[ultra thick,blue,dashed] (1,0.47)--(2.3,0.47);
			\node[above] at (1,1) {$t=t_3$};
			\node[below] at (1.5,0.5) {$t=t_2$};
		\end{tikzpicture}
		\caption{}
	\end{subfigure}
	\begin{subfigure}{0.2\textwidth}
		\centering
		\begin{tikzpicture}
			\draw[ultra thick,blue] (0,0)--(2,0);
			\filldraw[teal] (0,0) circle (5pt);
			\node[below] at (0,-0.2) {$J$};
			\node[below] at (1,0) {$\phi(z)$};
		\end{tikzpicture}
		\caption{}
	\end{subfigure}
	
	\begin{subfigure}{0.3\textwidth}
		\centering
		\tikzset{middlearrow/.style={decoration={markings,mark= at position 0.5 with {\arrow{#1}},},postaction={decorate}}}
		\begin{tikzpicture}
			\draw[ultra thick,blue,middlearrow={>}] (1,1)--(0,2);
			\draw[ultra thick,blue,middlearrow={>}] (0,0)--(1,1);	
			\draw [decorate,ultra thick,blue, decoration={snake}] (1,1) --(3,1);
			\draw[ultra thick,blue,middlearrow={>}] (4,0)--(3,1);
			\draw[ultra thick,blue,middlearrow={>}] (3,1)--(4,2);
			\filldraw[red] (1,1) circle (3pt);
			\filldraw[red] (3,1) circle (3pt);
		\end{tikzpicture}
		\caption{}
	\end{subfigure}
	\begin{subfigure}{0.3\textwidth}
		\centering
		\tikzset{middlearrow/.style={decoration={markings,mark= at position 0.5 with {\arrow{#1}},},postaction={decorate}}}
		\begin{tikzpicture}
			\draw[ultra thick,blue,middlearrow={>}] (1,1)--(0,2);
			\draw[ultra thick,blue,middlearrow={>}] (0,0)--(1,1);	
			\draw [decorate,ultra thick,blue, dashed] (1,1) --(3,1);
			\draw[ultra thick,blue,middlearrow={>}] (4,0)--(3,1);
			\draw[ultra thick,blue,middlearrow={>}] (3,1)--(4,2);
			\filldraw[red] (1,1) circle (3pt);
			\filldraw[red] (3,1) circle (3pt);
		\end{tikzpicture}
		\caption{}
	\end{subfigure}
	\begin{subfigure}{0.3\textwidth}
		\centering
		\tikzset{middlearrow/.style={decoration={markings,mark= at position 0.5 with {\arrow{#1}},},postaction={decorate}}}
		\begin{tikzpicture}
			\draw[ultra thick,blue,middlearrow={>}] (1,1)--(0,2);
			\draw[ultra thick,blue,middlearrow={>}] (0,0)--(1,1);	
			\draw [decorate,ultra thick,blue, snake=coil] (1,1) --(3,1);
			\draw[ultra thick,blue,middlearrow={>}] (4,0)--(3,1);
			\draw[ultra thick,blue,middlearrow={>}] (3,1)--(4,2);
			\filldraw[red] (1,1) circle (3pt);
			\filldraw[red] (3,1) circle (3pt);
		\end{tikzpicture}
		\caption{}
	\end{subfigure}
	
	\begin{subfigure}{0.2\textwidth}
		\centering
		\begin{tikzpicture}
			\draw[ultra thick,blue] (0,0)--(0,3);	
			\draw[ultra thick,blue] (0.5,1.5) ellipse (0.5 and 0.3);
			\filldraw[red] (0,1.5) circle (3pt);
			\node[below] at (1,1.5) {$\hat{\phi}\hat{\phi}$};
			\node[below] at (0,0) {$\overbracket{\hat{a}_q\hat{\phi}}$};
			\node[above] at (0,3) {$\overbracket{\hat{\phi}\hat{a}^\dagger}$};
		\end{tikzpicture}
		\caption{}
	\end{subfigure}
	\begin{subfigure}{0.1\textwidth}
		\centering
		\begin{tikzpicture}
			\draw[ultra thick,blue] (0,0)--(0,3);	
			\node[below right] at (0,0) {$\overbracket{\hat{a}\hat{a}^\dagger}$};
		\end{tikzpicture}
		\caption{}
	\end{subfigure}
	\begin{subfigure}{0.2\textwidth}
		\centering
		\begin{tikzpicture}
			\draw[ultra thick,blue] (0,1.5) ellipse (0.3 and 0.5);
			\draw[ultra thick,blue] (0,0.5) ellipse (0.3 and 0.5);
			\filldraw[red] (0,1) circle (3pt);
			\node[below] at (0,0) {$\overbracket{\hat{\phi}\hat{\phi}}$};
			\node[above] at (0,2) {$\overbracket{\hat{\phi}\hat{\phi}}$};
		\end{tikzpicture}
		\caption{}
	\end{subfigure}
	\begin{subfigure}{0.2\textwidth}
		\centering
		\begin{tikzpicture}
			\draw[ultra thick,blue] (-1,-1)--(1,1);
			\draw[ultra thick,blue] (1,-1)--(-1,1);
			\filldraw[red] (0,0) circle (3pt);
			\node[above] at (-1,1) {$\phi$};
			\node[above] at (1,1) {$\phi$};
			\node[below] at (-1,-1) {$\phi$};
			\node[below] at (1,-1) {$\phi$};
			\node[right] at (0,0) {$-i\lambda$};
		\end{tikzpicture}
		\caption{}
	\end{subfigure}
	\caption{Feynman diagrams.}
	\label{FeynmanDiagrams}
\end{figure}

\subsubsection{The Example of $\phi^4$ Theory}
The full procedure for calculating the S-matrix for $\phi^4$ theory is as follows. The Lagrangian density is written as
\begin{align}
	{\cal L} = \frac{1}{2}[\partial_\mu \phi(x)]^2 - \frac{m^2}{2}\phi(x)^2 - \frac{\lambda}{4!}\phi(x)^4.
\end{align}
The first two terms together make a free part and give rise to a free Hamiltonian. The interaction part is given by ${\cal L} = - \frac{\lambda}{4!}\phi(x)^4$ and $\hat{{\cal H}}_I= \frac{\lambda}{4!}\phi(x)^4$. We work in interaction picture when we talk about $\hat{S}$-operator. The field operators evolved by free Hamiltonian but state evolve via interaction Hamiltonian. 
We will find S-matrix by the following steps. 

\textbf{Step I:} The element of S-matrix to be calculated is written as the vacuum expectation value of free vacuum $\ket{0}$ and the Wick's theorem is used to make it simple. The amplitude of the in-state in momentum state $p$ and the out-state with momentum $q$ is given by
\begin{align}
	{\cal A} = \bra{q}\hat{S} \ket{p} = (2\pi)^3 (2 E_q)^{1/2}(2 E_p)^{1/2}\bra{0}\hat{a}_q\hat{S}\hat{a}^\dagger_p \ket{0}.
\end{align}

\textbf{Step II:}
Expanding the $\hat{S}$-operator using the Dyson's expansion, we have
\begin{align}
	\hat{S} &=T \big[\exp\big(-i\int d^4z \hat{{\cal H}}_I(z) \big) \big] \nonumber \\
	& = T\big[ 1- i\int d^4z \hat{{\cal H}}_I(z) + \frac{(-i)^2}{2} \int d^4y d^4w \hat{{\cal H}}_I(y)\hat{{\cal H}}_I(w) +...\big] \nonumber \\
	& = T\big[ 1- \frac{i\lambda}{4!} \int d^4 z \hat{\phi}(z)^4
	+ \frac{(-i)^2}{2!}\big(\frac{\lambda}{4!}\big)^2 \int d^4y d^4w \hat{\phi}(y)^4 \hat{\phi}(w)^4 + ...\big]
\end{align}

\textbf{Step III:} Now $\hat{S}$-matrix element are written as
\begin{align}
	{\cal A } &= \bra{q}\hat{S} \ket{p} \nonumber \\
	& = (2\pi)^3 (2 E_q)^{1/2)}(2 E_p)^{1/2)}T\big[\bra{0}\hat{a}_q\hat{a}^\dagger_p \ket{0} + \big(\frac{-i\lambda}{4!}\big) \int d^4z \bra{0}\hat{a}_q\hat{\phi}(z)^4\hat{a}^\dagger_p \ket{0} \nonumber \\
	& \ + \frac{(-i)^2}{2!}\big(\frac{\lambda}{4!}\big)^2 \int d^4y d^4w \bra{0} \hat{a}_q\hat{\phi}(y)^4 \hat{\phi}(w)^4 \hat{a}^\dagger_p + ...\big]
\end{align}
Above equation can be written as ${\cal A} = {\cal A}^{(0)} + {\cal A}^{(1)}+ {\cal A}^{(2)} + ...,$ where ${\cal A}^{(n)}$ proportional to $\lambda^n$

\textbf{Step IV:}
Consider ${\cal A}^{(1)}$, the first order term. \\ As $\bra{0}\hat{a}_q \hat{\phi}(x)^4 \hat{a}^\dagger_p \ket{0} \equiv \bra{0}\hat{a}_q\hat{\phi}(z)\hat{\phi}(z)\hat{\phi}(z)\hat{\phi}(z)\hat{a}^\dagger_p \ket{0}$, we can have contractions by using Wick's theorem. 
\begin{align}
	\bra{0}\overbracket{\hat{a}_q\overbracket{\hat{\phi}(z)\hat{\phi}(z)}\overbracket{\hat{\phi}(z)\hat{\phi}(z)}\hat{a}^\dagger_p} \ket{0} = \bra{0}\hat{a}_q\hat{a}^\dagger_p\ket{0}\bra{0}T\hat{\phi}(z)\hat{\phi}(z)\ket{0}\bra{0}T\hat{\phi}(z)\hat{\phi}(z)\ket{0}
\end{align}
There are three ways of contraction to get this term, and there are twelve ways to contract $\hat{a}$-operators and $\hat{\phi}$ operators. For example 
\begin{align}
	\bra{0}\overbracket{\hat{a}_q\hat{\phi}(z)}\overbracket{\hat{\phi}(z)\hat{\phi}(z)}\overbracket{\hat{\phi}(z)\hat{a}^\dagger_p }\ket{0} = \bra{0}\hat{a}_q\hat{\phi}(z)\ket{0}\bra{0}T\hat{\phi}(z)\hat{\phi}(z)\ket{0}\bra{0}\hat{\phi}(z)\hat{a}^\dagger_p\ket{0}
\end{align}
Now ${\cal A}^{(1)}$ would be written as
\begin{align}
	{\cal A}^{(1)} = \big(\frac{-i\lambda}{4!}\big) \int d^4z \big[3\bra{0}\hat{a}_q\hat{a}^\dagger_p\ket{0}\bra{0}T\hat{\phi}(z)\hat{\phi}(z)\ket{0}\bra{0}T\hat{\phi}(z)\hat{\phi}(z)\ket{0} \nonumber \\
	+ 12 \bra{0}\hat{a}_q\hat{\phi}(z)\ket{0}\bra{0}T\hat{\phi}(z)\hat{\phi}(z)\ket{0}\bra{0}\hat{\phi}(z)\hat{a}^\dagger_p\ket{0} \big]
\end{align}
Some remarks about this equation are as follows. The contraction between the field operators is a free propagator $\overbracket{\hat{\phi}(y)\hat{\phi}(z)} = \bra{0}T\hat{\phi}(y)\hat{\phi}(z) \ket{0} = \Delta(y-z)$. The Eq. \ref{FeynmanProagator} will be used for this. 
The contraction between the creation operator and the field is given by $\overbracket{\hat{\phi}(z)\hat{a}^\dagger_p} = \bra{0} \hat{\phi}(z)\hat{a}^\dagger_p\ket{0}$ is calculated by using \ref{WaveSolution}.  
The factor $e^{-ip\cdot z}$ corresponds to the incoming particle. Similarly, the factor results from the contraction $\overbracket{\hat{a}_q\hat{\phi(x)}}$ give $e^{ip\cdot x}$ corresponds to the outgoing particle. The contraction $\overbracket{\hat{a}_q\hat{a}^\dagger}$ gives the delta function $\delta^{(3)}(\bm{q}-\bm{p})$.

\textbf{Step V:}
Feynman diagrams represent amplitudes in the expansion of the $S$-matrix. The number of interactions is the same as the order of expansion. The interaction vertices are drawn in the diagram for the interactions. The legs of the diagrams are uncontracted field operators. These legs are joined with each other or to the external particles by Wick's contraction. \\

\textbf{Rules for Feynman diagrams for $\phi^4$ theory in position space.}
\begin{itemize}
	\item Draw interaction vertices. Each vertex has a contribution $-i\lambda$.
	\item Connect incoming line to one of the legs which corresponds to the contraction $\overbracket{\hat{\phi(x)} \hat{a}\dagger_p}$. Each incoming line gives the contribution $e^{-ip\cdot x}$. 
	\item The field-field contraction $\overbracket{\hat{\phi(x)} \hat{\phi}(y)}$ gives the propagators, which are drawn as lines linking the points. These are internal to the diagrams and can be thought of as virtual particles. Each line gives a propagator $\Delta(x-y)$.
	\item The outgoing lines are drawn as the contraction $\overbracket{\hat{a}_q \hat{\phi}(x)}$. Each outgoing line gives the contribution $e^{ip\cdot x}$.
\end{itemize}
The position of vertices is integrated over all spacetime. The diagrams related to the Feynman rules are given in Fig. \ref{FeynmanDiagrams} (j), (k), (l), (m). Each term is divided by some factor $D$ to get the right answer.
The factor $D$ is the number of ways of arranging the propagators and vertices through contraction to get a particular diagram.
Sometimes it is easier to do the calculations in momentum space, the Feynman rules are defined for momentum space in that case. 
We draw the diagram and write the equation related to the diagram. From the diagrams, we know the contraction and interactions, but we do not need to go through all the expansion. The amplitude ${\cal A} = _{out}\bra{q}\ket{p}_{in} = \bra{q}\hat{S}\ket{p}$ is written as the sum of diagrams, with each diagram as an integral.
The incoming lines are particles entering the process, and the outgoing lines are particles leaving the process. The internal lines represent the virtual particles exchange. 
They appear not to be connected to anything. The vacuum diagrams have no external lines and are not connected to the incoming or outgoing particles, hence they don't affect transition probabilities. 
Self-energy diagrams are the ones to describe how interactions affect the amplitudes of single particles. 

\subsubsection{Renormalization}
When there are loop diagrams on the internal lines, the propagator integral diverges. These loop diagrams are self-energy diagrams. For these diagrams, there can be an infinite number of ways the particle's momentum is split into several particles. The process of removing these infinities is called \textit{renormalization}.
The idea of renormalization is as follows. The limit of integration is taken to some large but a finite momentum $\Lambda$. The interactions change particles to a dressed particle which behave differently than the \textit{bare particles}. We need to take some kind of pragmatic measure by shifting masses $m$ and coupling constants $\lambda$ in the perturbation expansion of the theory to some physical parameters $m_P$ and $\lambda_P$. It is done by adding some counter-terms to the Lagrangian. These terms correspond to our ignorance below the cut-off scale $1/\Lambda$ and cancel the dependence on the scale $\Lambda$. When these counter-terms can be found then the theory is called \textit{renormalizable}, otherwise it is non-renormalizable. We can think of this process as we were using the wrong parameters before, but now we are using the real-life or physical parameters that nature gave us. For renormalizable theories, some terms can cancel all infinities, but an infinite number of counterterms are needed for non-renormalizable theories. This procedure for solving the problem of divergent amplitudes is questionable, but it is very successful.
	\section{Path Integral}

In contrast to classical mechanics, the particles' motion is described by the wave function in quantum mechanics. A quantum particle traversed all possible paths between the two points, and each path has some phase associated with it. The paths of different phases interfere destructively, hence cancel out. The stationary paths correspond to the constructive interference.
This approach leads to the Feynman path integral formalism of the quantum mechanics discussed in this section. 

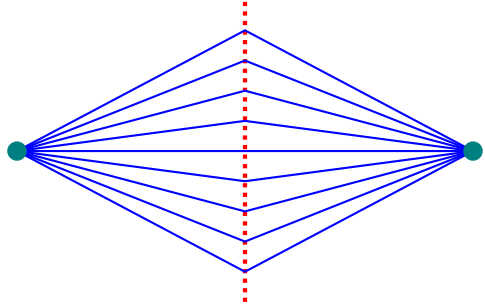
\begin{figure}[h!]
	\centering
	\begin{tikzpicture}	
		\draw[dotted,ultra thick,red] (0,-2)--(0,2);
		\draw[thick,blue] (-3,0)--(0,1.6)--(3,0);
		\draw[thick,blue] (-3,0)--(0,1.2)--(3,0);
		\draw[thick,blue] (-3,0)--(0,0.8)--(3,0);
		\draw[thick,blue] (-3,0)--(0,0.4)--(3,0);
		\draw[thick,blue] (-3,0)--(0,0)--(3,0);
		\draw[thick,blue] (-3,0)--(0,-0.4)--(3,0);
		\draw[thick,blue] (-3,0)--(0,-0.8)--(3,0);
		\draw[thick,blue] (-3,0)--(0,-1.2)--(3,0);
		\draw[thick,blue] (-3,0)--(0,-1.6)--(3,0);
		\draw [ultra thick,teal,fill=teal] (-3,0) ellipse (0.1 and 0.1);
		\draw [ultra thick,teal,fill=teal] (3,0) ellipse (0.1 and 0.1);
	\end{tikzpicture}
	\caption{A quantum particle travels all possible paths between two points.}
	\label{Paths}
\end{figure}

The state vector and the time evolution of a wave function in quantum mechanics is written as
\eq{\ket{\psi(x,t)} = \bra{x}\ket{\psi,t}=\bra{x,t}\ket{\psi}, \ \ket{\phi(x,t)}= \hat{U}\ket{\phi(x)} = e^{i\hat{H}t}\ket{\phi(x)}, \ \ket{x,t}= e^{i\hat{H}t}\ket{x}. \label{StateVector1}}
Since the projectors and inner products in quantum mechanics are written as
\eq{\int dx' \ket{x'}\bra{x'} = 1, \qquad \bra{x}\ket{x'} = \delta(x-x'),}
therefore by inserting the projectors in state vector we get
\eq{ \bra{x}\ket{\psi} = \int dx' \bra{x}\ket{x'}\bra{x'}\ket{\psi}.}
The wave function of the particle at position $x_f$ and at time $t_f$ can be written in terms of the wave function $\psi(x_i,t_i)$ at an earlier position $x_i$ and $t_i$,
\eq{\psi(x_f,t_f) &= \int dx_i \bra{x_f,t_f}\ket{x_i,t_i}\bra{x_i,t_i}\ket{\psi}\\
	&= \int dx_i \bra{x_f,t_f}\ket{x_i,t_i}\psi(x_i,t_i).}
The integral is for the sum of all different paths a particle can take from the initial to the final position, as shown in Fig. \ref{Paths}. The quantity $\bra{x_f,t_f}\ket{x_i,t_i}$ is called the \textit{propagator}, or the \textit{correlation function}. It is the probability amplitude when $\psi(x_f,t_f)$ is decomposed in terms of $\psi(x_i,t_i)$. If we know the initial position of the particle, that is the $\psi(x_i,t_i)$ is delta function, then the probability of finding the particle at later time $t_f$ at position $x_f$ is 
\eq{P = \abs{\bra{x_f,t_f}\ket{x_0,t_0}}^2.}

Now let us divide the path from the initial position to the final position into two parts. Let the particle went from the position $x_i$ at time $t_i$ to the position $x_1$ at time $t_1$ and then in the next step it traveled to $x_f$ at $t_f$. It can be written as 
\eq{\bra{x_f,t_f}\ket{x_i,t_i}=\int dx_1 \bra{x_f,t_f}\ket{x_1,t_1}\bra{x_1,t_1}\ket{x_i,t_i}.}
We can further divide the path from the initial position to the final position into $n$ equal steps so that
\eq{\bra{x_n,t_n}\ket{x_0,t_0}=\int dx_1...dx_{n-1} \bra{x_n,t_n}\ket{x_{n-1},t_{n-1}}...\bra{x_1,t_1}\ket{x_0,t_0}.}
All steps are multiplied to get the whole path. Using the Eq. \ref{StateVector1}, we get
\eq{\bra{x_f,t_f}\ket{x_i,t_i} &= \bra{x_f}\hat{U}(t_i,t_f)\ket{x_i} = \bra{x_f}e^{-i\hat{H}(t_f-t_i)}\ket{x_i}.}
Again we can divide the time into small steps $\Delta t$ and write the unitary time evolution operator for each step as 
$\hat{U}(t_f,t_i)= \hat{U}(t_f,t_1)\hat{U}(t_1,t_i) = e^{-i\hat{H}(t_f-t_1)}e^{-i\hat{H}(t_1-t_i)}$. The propagator now becomes
\eq{\bra{x_n,t_n}\ket{x_0,t_0}= \int d[x]\bra{x_f}(e^{-iH(\Delta t)})^n\ket{x_i},}
\eq{\bra{x_n,t_n}\ket{x_0,t_0}&=\int dx_1...dx_{n-1} \bra{x_n}e^{-i\hat{H}(t_n-t_{n-1})}\ket{x_{n-1}}...\bra{x_1}e^{-i\hat{H}(t_1-t_0)}\ket{x_0}\\
	& = \int dx_1...dx_{n-1} \bra{x_n}e^{-i\hat{H}(\Delta t)}\ket{x_{n-1}}...\bra{x_1}e^{-i\hat{H}(\Delta t)}\ket{x_0}.}
After inserting the Hamiltonian $\hat{H}= \frac{\hat{p}^2}{2m} + \hat{V}(x)$, we reach at the relation
\eq{\bra{x_n,t_n}\ket{x_0,t_0} = \int {\mathcal{D}}[x] \exp[i\int_{t_0}^{t_n} L(x,\dot{x})dt] =\int {\mathcal{D}}[x] e^{iS[x]/\hbar},}
where $L= \frac{m\dot{x}}{2}- V(q)$ and $\mathcal{D}[x]= dx_1...dx_n$. In this last equation, the Gaussian integral is used after applying the operators. For the detail of this integral, see \cite{tom2015quantum,zee2010quantum}.

The last equation is the path integral formulation of quantum mechanics. The left-hand side can also be written as $\bra{x_n,t_n}\ket{x_0,t_0} = \bra{x_n}\hat{U}(t_n-t_0)\ket{x_0} = \bra{x_n}e^{-i\hat{H}\Delta t}\ket{x_0}$. Single-particle quantum mechanics is not enough for a multiparticle state, so we use fields $\phi(x)$ in place of dynamical variable $x$. Now the integral is over the spacetime. By promoting the position $x$ to the field $\phi(x)$ and the Lagrangian to the Lagrangian density, we get 
\eq{\bra{\phi_2}e^{-iH\Delta t}\ket{\phi_1} = \int {\mathcal{D}}[\phi] \exp[i\int_{t_0}^{t_n} \mathcal{L}(\phi,\dot{\phi})d^4x] =\int {\mathcal{D}}[\phi] e^{iS[\phi]/\hbar}.}
When we have $\phi_1 = \phi_2 = \phi$, the vacuum field in the above equation, then it is vacuum to vacuum transition amplitude. We can find the energy of the ground state in this scenario.  
Let us disturb the vacuum for something exciting. It would lead to generating functional. It is conventional to represent this amplitude by $Z[J]$. This $Z[J]$ goes with the names; \textit{partition function, generating functional, path integral with source, and state sum}. 
Finding generating functional $Z[J]$ which has the Green's function information in it, we start with no particle and end with no particle in the presence of the source. The generating functional is written in terms of the functional integral as
\eq{Z[J] = \int {\cal D}[\phi(x)] e^{i\int d^4x \big({\cal L}[\phi(x)] + J(x)\phi(x)\big) } = \int {\cal D}[\phi(x)] e^{iS[\phi]+\int d^4x J(x)\phi(x)}. \label{FieldIntegral}} 
We can split the Lagrangian into the free part and the interaction part, ${\cal L} = {\cal L}_0 + {\cal L}_I$. The free part is solvable by canonical quantization, but the interaction part isn't solvable by canonical quantization.
By taking the functional derivative with respect to $J$, we get
\eq{\frac{1}{Z[0]}\frac{\delta Z[J]}{\delta J(x_1)}_{J=0} = \frac{1}{Z[0]} \int d[\phi] \phi(x_1)e^{-iS[\phi]}.}
Analogous to the statistical mechanics, this integral is the expectation value of $\phi$, that is $\langle \phi(x_1) \rangle$. Similarly,
\eq{\frac{1}{Z[0]}\frac{\delta^2 Z[J]}{\delta J(x_1)\delta J(x_2)}\Bigg|_{J=0} = \frac{1}{Z[0]} \int d[\phi] \phi(x_1)\phi(x_2)e^{-iS[\phi]} = \langle \phi(x_1)\phi(x_2) \rangle.}
This is called two-point function, 
\eq{\bra{0}\phi(x_1)\phi(x_2)\ket{0} = \frac{\int {\cal D}\phi \phi(x) \phi(y) e^{{i}\int d^4x {\cal L}_0[\phi]}}{\int {\cal D}\phi e^{{i}\int d^4x {\cal L}_0[\phi]}}.}
On the similar lines, the $n$-point correlation function is given by
\eq{\langle \phi(x_1)\phi(x_2)...\phi(x_n) \rangle = \frac{\int {\cal D}[\phi] \phi(x)\phi(x_2)...\phi(x_n) e^{{i}\int d^4x {\cal L}_0[\phi]}}{\int {\cal D}\phi e^{{i}\int d^4x {\cal L}_0[\phi]}}.}

The Feynman path integral in quantum field theory has an interpretation that is not identical to that of quantum mechanics. The correlation function, also called the \textit{transition amplitude}, in quantum field theory is the probability of having a final field configuration such that the initial configuration is known. Many processes can occur between the initial state and the outcome. Each distinct history describes a state at any given time, and is a distinct path through the configuration space. Each path has an associated probability amplitude. The probability that a system is in a given configuration is the sum over amplitudes of each path connecting the initial and final configurations. Therefore, the interactions and Feynman diagrams can also be studied by path integral approach, but it is a difficult one comparing with the $S$-matrix approach.
	\section{Gauge Theory}
When we have identical observables for different configurations of the field, we say that there is an internal vagueness. This inherent vagueness is intrinsic. It is sometimes called internal symmetry, but it is not a symmetry, instead, it is our inability to find a unique description \cite{tom2015quantum}. Choosing one particular description is the choice of gauge. For example, the choice of zero potential is arbitrary. In quantum mechanics, the zero of the phase is arbitrary. 
The transformation from one choice to the other is called the gauge transformation, and the underlying invariance is called gauge invariance.

Let the magnetic vector potential be transformed as $\bm{A} \rightarrow \bm{A}+ \nabla \chi$, where $\chi$ is a scalar function. Both of the choices of vector potentials give the same magnetic field $\bm{B}$. 
$\bm{A}(x)$ is chosen with function $\chi (x)$ which may vary with the position, so it needs to distinguish between global and local gauge transformation. 
Let us see the effect of gauge transformation on the complex scalar field. The Lagrangian for the field can be written as
\eq{{\cal L} = (\partial^\mu \psi)^\dagger (\partial _\mu \psi)- {m}^{2} \psi ^\dagger \psi \label{CompField}}
This field has U(1) symmetry. That is, if we replace $\psi(x) \rightarrow \psi(x) e^{i\alpha}$, the Lagrangian and the equation of motion do not change. It is a global gauge transformation because it changes the field by the same amount at every point. Let the phase of the field changed differently at different points in spacetime, which would lead to different transformations at different points and can be written as 
\eq{\psi(x)\rightarrow \psi(x)e^{i\alpha(x)}.}
Here, $\alpha(x)$ depends on the position in spacetime. How can we make the theory invariant under this local phase change?
For the local gauge transformation, the second term in Eq. \ref{CompField} will remain the same, but the derivative term would be changed, because now we have to take the derivative of $\alpha(x)$ too. Consider the first term,
\eq{\partial _\mu \psi(x) &\rightarrow \partial_\mu (\psi(x) e^{i\alpha(x)}) \nonumber \\
	&= e^{i\alpha(x)} \partial_\mu \psi(x) + \psi(x) e^{i\alpha (x)} i \partial_\mu \alpha (x) \nonumber \\
	&= e^{i\alpha(x)} [\partial_\mu + i \partial_\mu \alpha (x)]\psi(x),}
and
\eq{\partial^\mu \psi^{\dagger}(x) \rightarrow e^{i\alpha(x)} [\partial^\mu - i \partial^\mu \alpha (x)]\psi^\dagger(x).}
Multiplying these two equations, the first term in the Lagrangian would become  
\eq{(\partial^\mu \psi^{\dagger})(\partial _\mu \psi) &\rightarrow \Big[e^{-i\alpha} [\partial^\mu - i \partial^\mu \alpha ]\psi^\dagger \Big] \Big[e^{i\alpha} [\partial_\mu + i \partial_\mu \alpha ]\psi\Big] \nonumber \\
	&=(\partial^\mu \psi^{\dagger})(\partial _\mu \psi)+i (\partial^\mu \psi^{\dagger})(\partial _\mu \alpha) \psi -i(\partial^\mu \alpha) \psi^{\dagger}(\partial _\mu \psi) + (\partial^\mu \alpha)(\partial _\mu \alpha) \psi \psi^\dagger,}
where $x$ is omitted in the last equation to avoid cluttering. 
Due to $\alpha (x)$ dependence on $x$, the theory is not invariant under local transformation $U(1)$. To restore the local symmetry, introduce another field $A_\mu$ which varies from point to point in spacetime to cancel the extra terms and make the theory invariant. Theoretically, it is done by introducing another derivative called \textit{covariant derivative} which is written as 
\eq{D_\mu = \partial _\mu +iqA_\mu(x)\label{Covariant}}
If the new field $A_\mu$ transforms in such a way as 
\eq{A_\mu \rightarrow A_\mu -\frac{1}{q}\partial _\mu \alpha (x)}
then the $U(1)$ symmetry can be restored. Here, $q$ is known as the \textit{coupling constant} of the field. It tells how strongly a field interacts with other fields. 
Now, the first term of the Lagrangian density in the Eq. \ref{CompField}  would become
\eq{D_\mu \psi= (\partial _\mu +iq A_\mu) \psi &\rightarrow (\partial_\mu \psi) e^{i\alpha(x)} + i (\partial _\mu \alpha)\psi + iq A_\mu \psi e^{i\alpha} - i(\partial \alpha )\psi \nonumber \\
	&=D_\mu (\psi e^{i\alpha}).}
Therefore, if we write the Lagrangian density with the covariant derivative then it would be invariant under the local gauge transformation 
\eq{{\cal L} = (D^\mu \psi)^\dagger (D_\mu \psi) - m^2 \psi ^\dagger \psi}
In short, the theory is locally invariant if 
\eq{\psi(x) &\rightarrow \psi(x) e^{i\alpha (x)}, \ \text{and} \ \
	A_\mu(x) \rightarrow A_\mu(x) - \frac{1}{q}\partial _\mu \alpha(x).}
The gauge field $A_\mu$ is introduced to make the theory invariant with respect to local transformation. The gauge field has its dynamics. The above discussion is for the Abelian gauge theory. Electromagnetism is an Abelian gauge theory where the gauge field is vector potential. Yang-Mills theory is an example of non-Abelian gauge theory where the gauge field has matrix-valued $SU(2)$ symmetry. Weak and strong interactions are described by the theories based on non-Abelian gauge theories. The spontaneous symmetry breaking of the gauge symmetry leads to the \textit{Higgs mechanism}. It plays a role in unified theories of weak and electromagnetic interactions by the extension to $SU(2)\cross U(1)$ proposed by Salam, Weinberg, and Glashow \cite{kibble2015spontaneous}. We will not discuss the quantization of the gauge fields here.

\subsection{Topology and Gauge Theory}
The gauge theories can intuitively be described by the concept of \textit{fiber bundle} in topology. For the basic understanding of the product spaces and fiber bundle, see the related section in Appendix \ref{Top} and the Fig. \ref{FiberMobious}, and \cite{moriyasu1983elementary,nakahara2003geometry,baez1994gauge} for more in-depth discussion. 

A fiber bundle is a product space locally if we take a small portion of the space, but globally it may have a complicated structure. If the space is a product space both locally and globally, then it is a \textit{trivial bundle}. A cylinder is a trivial bundle, but a M\"obious strip is a non-trivial bundle. To illustrate the fiber bundle, take the following examples. Suppose we want to study the motion of a vector on a spacetime. The spacetime manifold is called the \textit{base space} $M$ and the bundle of all the vectors on these manifolds together make \textit{total space} $D$. The base space and the total space together make a fiber bundle. All those points make one \textit{fiber} $\psi(x)$ which are in the total space at the specific point $x$ of the base manifold. The \textit{structure group} $G$ acts on the element of fiber and gives another element of the same fiber. The \textit{projection map} $\pi$ maps a point of fiber to a point of the base space, and a \textit{section} is an inverse map of a portion of the base space to the total space. The rule for going from one fiber on a particular point of a base space to the nearby fiber on the base space is called \textit{connection} $A$. The connection map the tangent spaces over the base space to the tangent spaces in the total space.   
When each fiber is the structure group itself, then the fiber bundle is called the \textit{principal fiber bundle}. When typical fibers are vector spaces, then we have \textit{vector bundles}.

To describe the gauge theory on a fiber bundle, consider the example of the interaction of a photon with an electron. This is the $U(1)$ gauge theory. The photon is a gauge field or simply the vector potential. We will call it the \textit{interaction field}. It will be represented by the \textit{principal fiber bundle} $(D,M,G, \pi)$. Let the matter field be an \textit{associated vector bundle} $(P,M,G,\pi_D)$. We will call it the \textit{dynamic field}. The matter field is coupled to the interaction field at every point. The structure group becomes a local symmetry group or a gauge group in gauge theory. The vector bundle and principal bundle share the base space and local symmetry group.

\begin{figure}
	\centering
	\includegraphics[scale = 0.7]{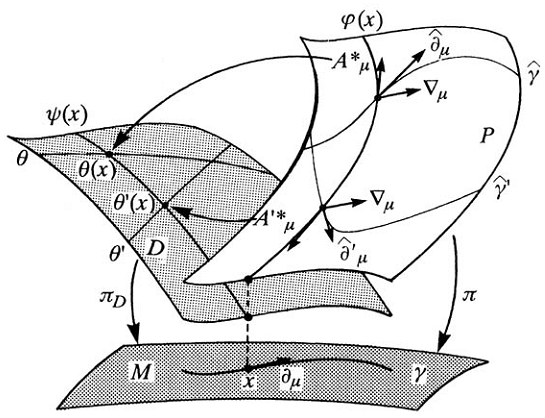}
	\caption[Interaction dynamics of the gauge fields on a fiber bundle.]{Interaction dynamics of the gauge fields on a fiber bundle. The base space $M$ is spacetime, the matter field is $D$ in gray color and the gauge field is $P$ in white color. This diagram is taken directly from the Ref. \cite{auyang1995quantum}.}
	\label{GaugeFieldFiber}
\end{figure}

As shown in \ref{GaugeFieldFiber}, a fiber $\psi(x)$ in the vector bundle has a point of the matter field. The wave function of the matter field is represented by a section of the vector bundle. Let $\theta(x)$ and $\theta(x^{'})$ be phases in the fiber. The actual phase at a point $\psi(x)$ depends on its coupling with the potential of the interaction field. The vector potential is represented by connections over the fiber $\phi(x)$ in the principal fiber bundle. The two fibers $\phi(x)$ and $\psi(x)$ share the base point $x$. This is the idea of point interaction. $M$ is usually called spacetime, and it constitutes the parameter space of the field theory. 

Let a time-like curve $\gamma$ in $M$ picks out some spatio-temporal segment of the dynamical system.
The internal states of the dynamical system are represented by the total space on the bundle. The spatio-temporal change occurs when moving from one point to another in the total space. How are they related? Suppose a section in the total space above $\gamma$, maps $\gamma$ into $\hat{\gamma}$ in total space of principal fiber bundle. The total change of system along $\gamma$ is given by the partial derivative $\hat{\partial}_\mu$ of the section $\hat{\gamma}$. The connection on the principal bundle is given, the derivative is decomposed into two parts
\eq{\hat{\partial}_\mu = \nabla_\mu + A^*_\mu,} 
where $\nabla_\mu$ is a covariant derivative of $\hat{\gamma}$. This is the \textit{horizontal lift} of the partial derivative $\partial_\mu$ of $\gamma$. It is the measure of the change due to spatio-temporal variation by tracking the directional derivative of $\gamma$ in $M$. The difference between derivatives $\nabla_\mu$ and $\hat{\partial}_\mu$ is called the fundamental vector $A^*_\mu$, which determines the dynamical variation of the system. It is related to the interaction potential $A_\mu$. $A^*_\mu$ determines a point $\theta(x)$ in the vector bundle uniquely. Since $\theta(x)$ is the phase of matter field, the interaction field and matter field are coupled at the point $x$.
By picking a different section $\hat{\gamma}^{'}$ we get different fundamental vector $A^{*^{'}}_\mu$ and different phase $\theta^{'}(x)$. Different sections are transformed into each other by the symmetry group $G$. The local symmetry group transforms $\hat{\gamma}$ and $\theta$ simultaneously, so that the interaction potential and the phase of matter field transformed simultaneously, leaving the dynamical system invariant.

Compare the fundamental vector and potential at $x$ and $x^{'}$. The presence of the matter field and variation in its phases induces a variation in the potentials, hence the values of the connection also vary. That leads to nonzero curvature, which is an exterior derivative of the connection. The physical curvature in the principal fiber bundle is the intensity of the interaction field  \cite{auyang1995quantum}.
	\section{Standard Model of Particle Physics}
Quantum field theory is synonymous with the standard model that describes the fundamental particles and their interactions. There are three types of interactions in the standard model; electromagnetic interaction, weak interaction, and strong interaction. These interactions are mediated by the \textit{gauge bosons} with integral spin equal to 1. Gauge bosons are quanta of fields. The number of gauge bosons is equal to the number of generators of the unitary group to describe the symmetry of the field. These forces are of different ranges.

The electromagnetic force is described by the field theory called \textit{electrodynamics}. Its symmetry group is the $U(1)$ gauge group. The quanta are chargeless and massless \textit{photons} with two polarization states.
The weak force has $SU(2)$ symmetry group with three massive generators; $W^+, W^-,Z$. Two of them are charged and one is neutral. 
The strong force is responsible for the binding of nucleons. It is an interaction between quarks and has the gauge group $SU(3)$. It has eight quanta called \textit{gluons}. These particles are massless and carry a color charge. The field theory for strong force is called \textit{quantum chromodynamics}. 

Particles that makeup matter are spin-1/2 particles. The matter particles are divided into two groups; leptons and quarks. Each one of these groups comes in three families. Leptons are charged particles with charge $-1$ that interact via weak and electromagnetic forces and have different masses. They include $e,\mu^-,\tau^-$. Tau $\tau$ and muon $\mu$ are unstable and decay into electron and neutrino. Each of these leptons has their corresponding neutrinos $\nu_e, \nu_\mu, \nu_\tau$ which are neutral particles. Neutrinos interact only through weak interactions.
Each lepton has antilepton with a charge of $+1$ and mass equal to lepton. Positron, antimuaon, and antitau respectively are $e^+, \mu^+,\tau^+$. There are neutral antineutrinos too. Antiparticles have the same properties as particles, but the charge is opposite. 

Other than the charge, leptons carry a quantum number called \textit{lepton number}. It has value; $+1$ for particles and neutrinos, $-1$ for antiparticle and antineutrino, and zero for particles that are not leptons. The lepton number explains why there are antineutrinos in beta decay of neutron $n= p+e+\bar{\nu}_e$. Neutrons and protons have lepton number zero Therefore, on the right side, the lepton number must be zero. And neutrino involved in this decay should be antineutrino with lepton number $-1$ to get canceled with electron with lepton number 1.

Neutrons and protons are made of fundamental particles called \textit{quarks}. They carry the labels called \textit{color}. Quarks have three colors; red, blue, and green. Their antiparticles have colors antired, antiblue, antigreen. They are combined in such a way that their total color must be white.
In addition, quarks have six types of flavors; up, down, strange, charmed, top, bottom. Like leptons, there are three families of quarks $(u,d),(s,c),(t,b)$. If one family member has a charge of $+2/3$, the other has $-1/3$. There is an antiparticle for each quark. The bound state of quarks is called \textit{hadrons}. Hadrons are of two types; \textit{baryons and mesons}. A baryon consists of three quarks or three antiquarks, whereas a meson consists of one quark and one antiquark. For example, protons and neutrons are baryons, and $\pi^0,\pi^+,\pi^-$ are mesons.

Masses of particles are calculated by putting a Higgs field in the Lagrangian. The quanta of the Higgs field is chargeless, spin-0 \textit{Higgs boson}. The Higgs field fills all the space. Larger masses have more interactions with this field, and massless particles do not interact.

The \textit{electroweak theory} is the unification of electromagnetic force and weak force. The \textit{grand unification} schemes unify all the three forces. That means that at high enough energy, these forces would be merged into one single force. 
One of the schemes for GUTs is \textit{supersymmetry}, which proposes a symmetry between bosons and fermions. There is a boson that exists for each fermion with the same mass. These particles are called \textit{superpartners}. The difference in the present masses could be due to the breaking of the supersymmetry. There is no experimental evidence for superpartners yet. \textit{String theory} is another unification scheme, according to which the fundamental particles are not pointlike, but extended strings. The quanta of the gravitational field are spin-2 \textit{graviton}, which naturally arises in string theory.

	\cleardoublepage
	\phantomsection
	\addcontentsline{toc}{part}{Bibliography}
	\bibliographystyle{unsrt}
	\bibliography{zRef}	
	
\end{document}